\documentclass[a4paper,12pt]{book}
\usepackage{amsfonts,wrapfig,amssymb}
\usepackage[latin1,cp1251]{inputenc}
\usepackage[russian]{babel}
\begin{document}
\def\rot{\mathop{\rm rot}\nolimits}
\def\div{\mathop{\rm div}\nolimits}
\def\grad{\mathop{\rm grad}\nolimits}
\def\gv#1{\mathop{\vspace{2pt}\mathfrak  #1}\nolimits}
\def\rv#1{\mathop{\vspace{2pt}\rm \bf #1}\nolimits}
\newcommand{\Arctg}{\mathop{\rm Arctg}\nolimits}
\newcommand{\Arcctg}{\mathop{\rm Arcctg}\nolimits}
\newcommand{\re}{\mathop{\rm Re}\nolimits}
\newcommand{\im}{\mathop{\rm Im}\nolimits}
\newenvironment{fr}[1]{\begin{tabular}{|p{#1}|}\hline}
     {\\\hline\end{tabular}}
\newcommand{\fg}[2]{\par\vspace{-\parskip}\hspace*{-22pt}
     \hangindent=#1\hangafter=#2}
\newcommand{\sign}{\mathop{\rm sign}\nolimits}

\oddsidemargin=-0.4mm \evensidemargin=-0.4mm
\topmargin=-0.4mm
\headsep=7mm
\textheight=231.875mm
\textwidth=160mm
\mathsurround=2.5pt
\unitlength=1mm
%\begin{document}
%\input{macr.tex}
\thispagestyle{empty}
\baselineskip=\normalbaselineskip
%\baselineskip=0.9\normalbaselineskip

\begin{center}\Large{\textit{В.А.Бережной,\hspace{0.05cm} В.Н.Курдюмов}}
\end{center}
\vspace*{55mm}
\begin{center}\LARGE{\bf ЛЕКЦИИ ПО ВЫСОКОЧАСТОТНОЙ}\end{center}
\begin{center}\LARGE{\bf ЭЛЕКТРОДИНАМИКЕ}\end{center}
\vspace*{135mm}
\begin{center}\normalsize{МОСКВА --- 2013}
\end{center}
\newpage\thispagestyle{empty}
УДК 538.566 (075.8)
\begin{center}
            \subsubsection*{\it А\,Н\,Н\,О\,Т\,А\,Ц\,И\,Я}
      \begin{minipage}[c]{0.75\textwidth}
         \parindent0.75cm В  настоящей  книге сделана попытка изложить
         содержание лекций по высокочастотной электродинамике, которые
         читаются   на   четвёртом  курсе  МФТИ;  кратко  представлены
         наиболее   употребительные   в   настоящее    время    методы
         теоретического анализа электромагнитных волн и колебаний.

         \parindent0.75cm <<Лекции>>  в значительной степени соответствуют
         программам   одноимённых  дисциплин,  утверждённых в учебных
         планах  кафедры ФПФЭ <<Фундаментальные взаимодействия и
         космология>>  и  кафедры  радиофизики   ФОПФ   ---   базовыми
         институтами кафедр являются Институт ядерных исследований РАН
         и Московский радиотехнический институт РАН.  Книга может быть
         использована    для   подготовки   студентов   и   аспирантов
         радиофизических и радиотехнических  специальностей,  а  также
         специалистами соответствующего профиля.
      \end{minipage}
\end{center}
\vspace*{10.cm}
%$\copyright$ Авторы
%\footnotesize{\hspace*{9.1cm}
~~~~ISBN 978-5-94274-227-0  \\
\\
$\copyright$\parbox[t]{9cm}{Федеральное государственное \\
             бюджетное учреждение науки\\
Институт  ядерных  исследований\\
Российской академии наук, 2013\\
Institute for Nuclear Research\\
 of the Russian Academy of Sciences, 2013}

\newpage
\vspace*{30mm}
\thispagestyle{empty}
\normalsize{
\begin{center}\subsubsection*{CОДЕРЖАНИЕ}\end{center} \parskip=0.5em

\noindent\hbox to 0.955\textwidth{Предисловие\dotfill 5} \parskip=0em

\noindent\hbox to 0.955\textwidth{Некоторые обозначения\dotfill 7}

\parskip=0.5em
\noindent Г\,Л\,А\,В\,А\,\hbox to  0.84\textwidth{  1.  {\bf  Основные
положения классической электродинамики}\dotfill 9} \parskip=0em

\parindent=1.5cm
   \hspace{1ex}\hbox to 0.85\textwidth{1. Уравнения Максвелла \dotfill
9}

   \hspace{1ex}\hbox to  0.85\textwidth{2.  Монохроматические  поля  и
комплексные амплитуды\dotfill 21}

   \hspace{1ex}\hbox to    0.85\textwidth{3.    Волновое    уравнение.
Потенциалы электромагнитного поля\dotfill 31}

\parskip=0.5em
\noindent Г\,Л\,А\,В\,А\,\hbox  to  0.84\textwidth{  2.  {\bf  Плоские
волны} \dotfill 42} \parskip=0em

\parindent=1.5cm
   \hspace{1ex}\hbox to 0.85\textwidth{4.  Плоские волны в  однородной
безграничной среде\dotfill 42}

   \hspace{1ex}5. Падение   плоской   монохроматической   волны   \par
\hspace{2.4ex} \hbox to 0.825\textwidth{ на  плоскую  границу  раздела
двух сред\dotfill 51}

   \hspace{1ex}\hbox to    0.85\textwidth{6.    Теория   скин-эффекта.
Граничное условие Щукина-Леонтовича\dotfill 62}

\parskip=0.5em
\noindent Г\,Л\,А\,В\,А\,\hbox  to 0.84\textwidth{ 3.  {\bf Волноводы}
\dotfill 72} \parskip=0em

\parindent=1.5cm
   \hspace{1ex}\hbox to 0.85\textwidth{7. Цилиндрические волны\dotfill
72}

   \hspace{1ex}\hbox to 0.85\textwidth{8. Волноводы\dotfill 81}

   \hspace{1ex}\hbox to 0.85\textwidth{9.  Волноводы прямоугольного  и
круглого сечений\dotfill 91}

  1\hbox to   0.85\textwidth{0.   Волноводы   с   двухсвязной   формой
поперечного сечения\dotfill 101}

  1\hbox to 0.85\textwidth{1. Потери в волноводах\dotfill 111}

\parskip=0.5em
\noindent Г\,Л\,А\,В\,А\,\hbox  to  0.84\textwidth{ 4.  {\bf Медленные
волны}\dotfill 121} \parskip=0em

\parindent=1.5cm
   1\hbox to  0.85\textwidth{2.  Диэлектрические замедляющие структуры
\dotfill 121}

   1\hbox to  0.85\textwidth{3.  Медленные   волны   в   периодических
структурах\dotfill 135}

   1\hbox to 0.85\textwidth{4. Диафрагмированный волновод\dotfill 147}

\parskip=0.5em
\noindent Г\,Л\,А\,В\,А\,\hbox to 0.84\textwidth{ 5.  {\bf  Объёмные
резонаторы}\dotfill 156}
\parskip=0em

\parindent=1.5cm
   1\hbox to 0.85\textwidth{5. Резонаторы простых форм \dotfill 156}

   1\hbox to 0.85\textwidth{6. Общая теория резонаторов\dotfill 167}

\parskip=0.5em
\noindent Г\,Л\,А\,В\,А\,\hbox to 0.84\textwidth{ 6. {\bf Сферические
волны и их возбуждение}\dotfill 178}
\parskip=0em

\parindent=1.5cm
   1\hbox to    0.85\textwidth{7.   Излучение   сферических   волн   в
безграничном пустом\hfil}\par\hspace{1.7em}\hbox to 0.81755
\textwidth {пространстве\dotfill 178}

   1\hbox to 0.85\textwidth{8. Лемма Лоренца и её следствия \dotfill
189}

\parskip=0.5em
\noindent Г\,Л\,А\,В\,А\,~7.  {\bf Граничные
задачи и возбуждение ограниченных} \vspace{-0.25cm}
\par\hspace{1.7em}\hbox to 0.81755
\textwidth {{\bf структур}\dotfill 197}
\parskip=0em

\parindent=1.5cm
   1\hbox to 0.85\textwidth{9.  Типы граничных задач в электродинамике
СВЧ\dotfill 197}

   2\hbox to 0.85\textwidth{0. Возбуждение волноводов\dotfill 211}

   2\hbox to 0.85\textwidth{1. Возбуждение резонаторов\dotfill 225}

\newpage\parskip=0.5em
\noindent Г\,Л\,А\,В\,А\,\hbox    to    0.84\textwidth{    8.     {\bf
Нерегулярности в электродинамических структурах}\dotfill 237}
\parskip=0em

\parindent=1.5cm
   2\hbox to 0.85\textwidth{2. Матрица рассеяния\dotfill 237}

   2\hbox to  0.85\textwidth{3.   Нерегулярности   в волноводах\dotfill 255}

\parskip=0.5em
\noindent Г\,Л\,А\,В\,А\,\hbox to 0.84\textwidth{ 9. {\bf Методы ТФКП
в электродинамике}\dotfill 272}
\parskip=0em

\parindent=1.5cm
   2\hbox to   0.85\textwidth{4.   Метод  Винера-Хопфа-Фока    \dotfill 272}

25. Метод задачи Римана-Гильберта. Расчёт дифракционного излу-
\par \hspace{1.3em} \hbox  to 0.82\textwidth {чения сторонних источников
методом Винера-Хопфа-Фока \dotfill 287}

  2\hbox to   0.85\textwidth{6. Дифракция на телах, имеющих форму
кругового цилиндра\dotfill 304}

\parskip=0.5em
\noindent Г\,Л\,А\,В\,А\,\hbox   to   0.8425\textwidth{    10.    {\bf
Квазиоптические структуры} \dotfill 314} \parskip=0em

\parindent=1.5cm
   2\hbox to 0.85\textwidth{7. Квазиоптические линии передачи\dotfill 314}

   2\hbox to 0.85\textwidth{8. Открытые резонаторы\dotfill 330}

\parskip=0.5em
\noindent\hbox to 0.96\textwidth{{\bf  Дополнения}\dotfill 344}
\parskip=0em

\parindent=1.5cm
   \hspace{-0.5em}Д\hbox to     0.86\textwidth{-1.      Интегрирование
неоднородных волновых уравнений \dotfill 344}

   \hspace{-0.5em}Д-2. Кильватерные поля в теории ускорителей
\par     \hspace{1.5em}     \hbox     to    0.82\textwidth    {
заряженных частиц \dotfill 354}

   \hspace{-0.5em}Д\hbox to   0.86\textwidth{-3.   Импеданс связи в
теории ускорителей \dotfill 369}\par \parskip=0.5em

\noindent\hbox to 0.96\textwidth{{\bf Задачи и упражнения}\dotfill 380}
\parskip=0.5em

\noindent\hbox to 0.96\textwidth{Приложения\dotfill 387}
\parskip=0.5em

\noindent\hbox to 0.96\textwidth{Литература\dotfill 405}}

\newpage
\thispagestyle{empty}
%\vspace*{5mm}

 \begin{center}
\normalsize{\rm П\,Р\,Е\,Д\,И\,С\,Л\,О\,В\,И\,Е \,\,К\,О\,\,
    В\,Т\,О\,Р\,О\,М\,У\,\, И\,З\,Д\,А\,Н\,И\,Ю }
\end{center}

      Во втором издании <<Лекций>>  исправлены замеченные опечатки,
допущенные в первом издании. В текст внесены немногочисленные
уточнения, способствующие лучшему пониманию материала.
\par\vspace*{0.5cm}\hspace{7,5cm}{\it В.А. Бережной, В.Н. Курдюмов}\\
Москва, декабрь 2012 г.

\vspace*{10mm}
 \begin{center}
\subsubsection*{\rm П\,Р\,Е\,Д\,И\,С\,Л\,О\,В\,И\,Е \,\,К\,\,
П\,Е\,Р\,В\,О\,М\,У\,\, И\,З\,Д\,А\,Н\,И\,Ю }
\end{center}

     Книга по  высокочастотной  электродинамике  соответствует лекциям,
которые   читаются   студентам    четвёртого    курса    Московского
физико-технического института, обучающимся на кафедре <<Фундаментальные
взаимодействия и космология>>  Факультета Проблем Физики и Энергетики  и
кафедре  радиофизики  Факультета  Общей и Прикладной Физики.  Учитывая
основные направления исследований,  проводимых в базовых  организациях
---  Институте  Ядерных Исследований РАН и Московском Радиотехническом
Институте  РАН,  в  программу  курса  лекций  были  внесены  некоторые
изменения по сравнению со стандартными курсами ВЧ электродинамики.

     В первую  очередь  эти  изменения  касаются  таких вопросов,  как
теория антенн и распространения волн (в  меньшей  степени  ---  теории
дифракции),  которые  описаны относительно неполно.  С другой стороны,
более    подробно,    чем     обычно,     представлено     возбуждение
электродинамических  структур  источниками  поля.  При  этом  основное
внимание  уделено  проблемам,  наиболее  близким  к  теории  ускорения
заряженных частиц.

     К сожалению,  ограниченный  объём  книги  не дал возможности на
достаточно глубоком  уровне  представить  многочисленные  методы
анализа  (особенно  численного)  задач высокочастотной электродинамики
--- даже простое перечисление  этих  методов  потребовало  бы  гораздо
больше  места,  чем  отведено соответствующему курсу.  <<Лекции>>  должны
послужить своего рода <<трамплином>>  для тех,  кто  будет  в  дальнейшем
более глубоко изучать электродинамику СВЧ или близкие к ней вопросы, а
также дать необходимый минимум сведений тем  специалистам,  которые  в
своей    научной    деятельности   лишь   только   <<соприкасаются>>,   с
электродинамикой.

     Изложение материала для студентов четвёртого курса начинается с
напоминания   сведений    об    основных    положениях    классической
электродинамики,  которые  должны быть известны из курса теоретической
физики.  В  первую  очередь  рассматриваются  комплексные   амплитуды,
волновые  уравнения  и  распространение  плоских  волн  в средах,  как
правило,  прозрачных.  Естественным продолжением  этой  темы  является
изложение  таких  вопросов,  как  отражение  плоской  волны от плоской
границы раздела двух сред,  теория скин-эффекта  и  граничные  условия
Щукина-Леонтовича.

     Значительное внимание    уделяется     цилиндрическим     волнам,
направляющим  структурам  типа  гладких  (в том числе диэлектрических)
волноводов,   волноводам   с   двухсвязным   поперечным   сечением   и
периодическим    структурам.    Последние    являются   представителем
электродинамических   систем,   способных   обеспечивать   эффективное
взаимодействие  потока  частиц  с  электромагнитным  полем;  отдельный
раздел    посвящён    широко     используемому     в     ускорителях
диафрагмированному   волноводу.   Довольно   подробно   обсуждаются  и
объёмные  резонаторы,  которые  являются  одной  из  наиболее  часто
применяемых в ускорительной технике структур;  меньше внимания уделено
открытым резонаторам, а также квазиоптическим линиям.

     Изучение сферических волн и их возбуждения позволяет естественным
образом   перейти   в   одном   из  последующих  разделов  к  вопросам
использования для  решения  задач  электродинамики  функции  Грина,  а
теория  возбуждения  волноводов  и  объёмных  резонаторов излагается
после ознакомления студентов с такими темами курса,  как лемма Лоренца
и  типы  граничных  задач  в электродинамике СВЧ.  Изложение материала
ведётся в соответствии с существующей традицией,  однако в  качестве
примеров применения теории возбуждения выбраны такие задачи, в которых
источником поля являются  заряженные  частицы  (или  пучки  заряженных
частиц).  Отдельный  раздел  посвящён  краткому  введению  в  теорию
матрицы рассеяния,  а  неоднородности  в  резонаторах  и  волноводах в
значительной мере рассматриваются с единой точки зрения --- на  основе
использования функционалов сравнения.

     Применение теории  функций   комплексного переменного в электродинамике
СВЧ представлено в   основном   методом Винера-Хопфа-Фока  и  методом  задачи
Римана-Гильберта,  которые в настоящее  время  уже заняли  своё место  в
электродинамике.  Как  и в  предыдущих    разделах,   примеры, иллюстрирующие
применение   этих   методов, подобраны так, чтобы быть максимально связанными
с  ускорительной   проблематикой.

     В заключительной  части <<Лекций>>  некоторое  внимание  уделяется
представлениям   функции   Грина  в  различных  системах  координат  и
применению этих  представлений  для  решения  нескольких  относительно
простых,  но  показательных  задач  возбуждения электромагнитных волн;
кроме этого,  кратко рассматриваются кильватерные поля в металлических
структурах  и  сделано  введение в теорию импеданса.  Раздел <<Задачи и
упражнения>>,   предназначен   преимущественно    для    самостоятельной
проработки студентами.

     В настоящее   время   авторам   неизвестен  учебник,  который  бы
достаточно хорошо соответствовал данному курсу лекций.  Хотя  в  книге
приводится  список  литературы,  представляется затруднительным отдать
предпочтение какой либо из  книг  по  сравнению  с  другими.  Материал
излагается  с  учётом  того,  что  подавляющая  часть  литературы из
приведённого  списка  является  труднодоступной.  Авторы  отдают  себе
отчёт  в том,  что в <<Лекциях>>  имеется много недостатков,  связанных
как с изложением,  так и с  отбором  соответствующего  материала;  они
заранее  благодарны  всем,  кто  выскажет свои замечания и пожелания в
связи с этим.

     Учебник может   быть   использован  для  подготовки  студентов  и
аспирантов радиофизических и радиотехнических специальностей,  а также
в  качестве  пособия для самоподготовки специалистами соответствующего
профиля.
\par\vspace*{0.5cm}\hspace{7.5cm}{\it В.А. Бережной, В.Н. Курдюмов}\\
Москва, ноябрь 2000 г. 

\newpage\thispagestyle{empty}
\normalsize{
\begin{center}
\vspace{25mm}
\centering
     \subsubsection*{\rm Н\,Е\,К\,О\,Т\,О\,Р\,Ы\,Е\,\,
          О\,Б\,О\,З\,Н\,А\,Ч\,Е\,Н\,И\,Я}
\vspace{15mm}
\centering
\baselineskip=\normalbaselineskip
%\baselineskip=1.5\normalbaselineskip
\begin{tabular}{rl}
$\mbox{\boldmath $\gv A$}$,~$\rv A$ ---&вектоpный потенциал,
его комплексная амплитуда\\
$a$ ---&шиpокая стоpона прямоугольного волновода; pадиус круглого
волновода\\
$\mbox{\boldmath $\gv B$}$,~$\rv B$ ---&индукция магнитного поля, её
комплексная амплитуда\\
$b$ ---&узкая стоpона прямоугольного волновода\\
$\beta$ ---&безразмерная скорость\\
$\chi$ ---&магнитная восприимчивость\\
$\mbox{\boldmath $\gv D$}$,~$\rv D$ ---&индукция электpического поля, её
комплексная амплитуда\\
$D$ ---&период структуры\\
$\delta$ ---&толщина скин-слоя\\
$\mbox{\boldmath $\gv e$}$,~$\mbox{\boldmath $\gv E$}$,~$\rv E$ ---
&напряжённость электpического поля, её комплексная амплитуда\\
$\varepsilon$ ---&диэлектpическая пpоницаемость\\
%$f;~f_n,~f_{pq}$ ---&частота; критические частоты\\
$\Phi$ ---&скалярный потенциал\\
$G$ ---&функция Гpина, симметрическая матрица\\
$g$ ---&поперечное волновое число\\
%$\Gamma$ ---&коэффициент отражения\\
$\gamma$ ---&релятивистский фактор\\
$\mbox{\boldmath $\gv h$}$,~$\mbox{\boldmath $\gv H$}$,~$\rv H$ ---
&напряжённость магнитного поля, е\"е комплексная  амплитуда\\
$h$ ---&пpодольное волновое число\\
$\mbox{\boldmath $\gv I$}, \rv I$ ---&поверхностная плотность тока,
её комплексная амплитуда\\
$\mbox{\boldmath $\gv j$},~\rv j$ ---&объёмная плотность тока,
её комплексная амплитуда\\
$J$ ---&полный ток\\
$K,~\rv K$ ---&волновое число и волновой вектор в среде\\
$k,\rv k$ ---&волновое число, волновой вектор в пустоте\\
%$k_{n},~k_{pql}$ ---&критичекое значение волнового числа (критическая
%частота)\\
$\ae$ ---&коэффициент поглощения\\
%$l$ ---&длина; длина дуги; число  вариаций по $z$\\
$\Lambda$ ---&длина волны в волноводе\\
$\lambda$ ---&длина волны; собственное значение\\
$\mu$ ---&магнитная пpоницаемость\\
$n$ ---&показатель пpеломления\\
$\rv n$ ---&нормаль к поверхности\\
$\nu$ ---&эффективная частота столкновений\\
$\mbox{\boldmath $\cal P$}$,~$\mbox{\boldmath$\Pi$}$ ---&вектоp Геpца,
его комплексная амплитуда\\
$Q$ ---&полный заpяд; плотность потерь в веществе; добротность\\
$q$ ---&заpяд частицы\\
$\rho$ ---&объёмная плотность заpяда\\
$\rho_{\mbox{\footnotesize{\it пов}}}$ ---&поверхностная плотность заpяда\\
$\mbox{\boldmath $\gv s$}$,~$\mbox{\boldmath $\gv S$}$,~$\rv S$ ---
&вектоp Умова-Пойнтинга\\
$S$ ---&площадь поверхности; матрица рассеяния\\
$\Sigma$ ---&полный поток энергии\\
$\sigma$ ---&пpоводимость\\
$U$ ---&плотность энергии поля (в сплошной среде)\\
$u$ ---&плотность энергии (микроскопического) поля\\
$V$ ---&объём\\
$v$,~$\rv v;$~$v_{\mbox{\footnotesize{\it ф}}}$,~$v_{\mbox{\footnotesize{\it гр}}}
$ ---&скоpость; фазовая, групповая
     скоpости\\
$W$ ---&волновое сопpотивление среды\\
\end{tabular}
\end{center}
\newpage
\thispagestyle{empty}
%%\end{document}
\begin{center}
\begin{tabular}{rl}
$\gv W_{\parallel}$ ---&продольный кильватерный потенциал\\
$\mbox{\boldmath$\gv W$}_{\perp}$ ---&поперечный кильватерный потенциал\\
$\varsigma$ ---&поверхностный импеданс среды\\
$\omega$ ---&круговая частота\\
$Z_{\parallel}$ ---&продольный импеданс связи\\
$\rv Z_{\perp}$ ---&поперечный импеданс связи
\end{tabular}
\end{center}}

%\end{document}

%\end{document} \%\vspace*{10mm}

\newpage
\oddsidemargin=-0.4mm\evensidemargin=-0.4mm
\topmargin=-0.4mm
\headsep=7mm
\textheight=231.875mm
\textwidth=160mm
\mathsurround=2,5pt
\unitlength=1mm
%\begin{document}
%\input{macr.tex}
\thispagestyle{empty}
%\addtocounter{page}{8}
\baselineskip=\normalbaselineskip
%\baselineskip=1.005\normalbaselineskip

\begin{center}
   \subsubsection*{\rm Г\,Л\,А\,В\,А\, 1}
      \vspace{-1.15em}
      \line(6,0){160}\\
      \vspace{-1em}
      \line(6,0){160}
      \vspace{-1.15em}
   \subsubsection*{ОСНОВНЫЕ ПОЛОЖЕНИЯ КЛАССИЧЕСКОЙ ЭЛЕКТРОДИНАМИКИ}
      \vspace{31mm}
   \subsubsection*{1. Уравнения Максвелла}
\end{center}
\vspace{.5cm}

\markboth{Глава 1. Основные положения классической электродинамики}
{1. Уравнения Максвелла}

\begin{center}\begin{minipage}[c]{0.75\textwidth}
\footnotesize{\parindent=0.5cm
         Уравнения Лоренца-Максвелла   и    основные    энергетические
         соотношения  классической <<микроскопической>>  электродинамики.
         Основные уравнения макроскопической электродинамики  сплошных
         сред.   Феноменологический   подход.   Граничные  условия  на
         поверхности раздела двух сред.  Закон  сохранения  энергии  в
         макроскопической    электродинамике.   Усреднение   уравнений
         Лоренца-Максвелла.  Место электродинамики СВЧ в  классической
         электродинамике.}
\end{minipage}\end{center}
\vspace{.5cm}

     Этой лекцией   начинается  годовой  курс  <<Электродинамика  СВЧ>>.
Термин   <<электродинамика   СВЧ>>  безусловно    требует
уточнения. Электродинамика   в   широком  смысле  этого  слова
является наукой, описывающей взаимодействие всех заряженных
элементарных частиц и  всех заряженных    тел    между    собой
посредством порождаемого   ими электромагнитного поля.
Электромагнитное взаимодействие ---  одно  из известных
на сегодняшний день фундаментальных взаимодействий в природе  и,
несомненно, одно из  наиболее  распространённых  и доступных в
своих проявлениях. Безразмерная  константа связи этого
взаимодействия, характеризующая   его   интенсивность, называется
постоянной тонкой структуры;  она равна $e^2/{\hbar c}\approx
1/137$, где заряд   электрона   $e=4,8\cdot 10^{-10}$~СГС,
$\hbar\approx 10^{-27}$~эрг/c  ---  постоянная Планка,  $c=3\cdot
10^{10}$~см/с
--- скорость  света  в  пустоте. Электрический  заряд   определяет
силу электромагнитного взаимодействия  и  при  этом является
сохраняющейся величиной.  В настоящее время теория этого
взаимодействия  разработана достаточно   полно  и  служит
надежной  основой  широких технических приложений.

     Электромагнитные явления в своём подавляющем  большинстве  либо
периодические,  либо  обладают  временным интервалом,  которому
всегда можно сопоставить спектр частот  с  характерной
центральной  частотой \(\omega\).  Переменные во времени
электромагнитные поля всегда являются и переменными в пространстве
--- частоте $\omega$ соответствует  длина волны \(\lambda=2\pi
c/\omega\).  При достаточно больших расстояниях и временных
интервалах взаимодействие частиц  и  поведение  поля  вполне
удовлетворительно описываются классической электродинамикой, в
которой поле  в  каждой  точке  пространства-времени  определяется
векторами напряжённости  электрического  \(\mbox{\boldmath$\gv
e$}(\rv  r,t)\) и магнитного $\mbox{\boldmath $\gv h$}(\rv  r,t)$
полей.   Истинные или, как  их   часто   называют,
<<микроскопические>> поля удовлетворяют релятивистски инвариантной
системе уравнений Лоренца-Максвелла:
     $$\hspace{-0.95cm}\rot\mbox{\boldmath$\gv e$}\,\!=\!-\displaystyle
       {\frac 1 c  \frac{\partial\mbox{\boldmath $\gv h$}}{\partial t}}\,,
        \eqno(1.1\mbox{\textit а})$$
     $$\rot\mbox{\boldmath $\gv h$}\!=\!\displaystyle{\frac 1 c\frac
       {\partial\mbox{\boldmath $\gv e$}}{\partial t}+{\frac{4\pi} c}
        \mbox{\boldmath $\gv j$}\,,}\eqno(1.1\mbox{\textit б})$$
     $$\hspace{-1.5cm}\div\mbox{\boldmath $\gv e$}\!=\!4\pi\rho\,,
       \eqno(1.1\mbox{\textit в})$$
     $$\hspace{-1.85cm}\div\mbox{\boldmath $\gv h$}\!=\!0\,,
       \eqno(1.1\mbox{\textit г})$$
где $\rho$ и $\mbox{\boldmath$\gv j$}$ --- соответственно плотности
заряда и тока. В микроскопической  электродинамике заряженные  частицы
принципиально являются точечными, так что для системы частиц
     $$\rho=\sum_n q_n\delta(\rv r-\rv r_n(t)),\qquad\mbox{\boldmath
       $\gv j$}=\sum_n q_n\rv v_n\delta(\rv r-\rv r_n(t)),\eqno(1.2)$$
где $\delta(\rv r)$ ---  трёхмерная  дельта-функция  Дирака,  $q_n$,
$\rv  r_n$  и  $\rv  v_n$  ---  соответственно заряд,  радиус-вектор и
скорость $n$-той частицы.

     В уравнениях (1.1),  (1.2) и везде далее,  где это специально  не
оговорено,  используется  абсолютная  система  единиц  СГС (гауссова).
Использование практических систем  единиц,  например, СИ,  приводит  к
ненужному  усложнению  записи  уравнений электродинамики и появлению
в них   величин,   не   имеющих   физического   содержания,    например,
диэлектрической  проницаемости  вакуума.  Гауссова система оказывается
неудобной  лишь   при   вычислении   таких   величин,   как   волновое
сопротивление  кабеля,  сопротивление  излучения  и в некоторых других
случаях.  В Приложении П-1 указан простой способ перехода от гауссовой
к практической системе и наоборот.

     Нетрудно убедиться,   что   плотность   заряда   и   тока   (1.2)
удовлетворяют уравнению непрерывности
     $$\frac{\partial\rho}{\partial t}+\div\mbox{\boldmath $\gv j$}
       =0\,, \eqno(1.3)$$
выражающему собой   закон   сохранения   электрического   заряда.  Это
уравнение  непосредственно  следует  и   из   системы   (1.1).   Сила,
действующая  на  частицу  с  зарядом  $q$,  называемая  силой Лоренца,
определяется через поля $\mbox{\boldmath $\gv e$}$ и $\mbox{\boldmath
$\gv h$}$:
     $$\rv f=q(\mbox{\boldmath $\gv e$}+\frac 1 c[\rv v\mbox{\boldmath
$\gv h $}])\,.\eqno(1.4)$$

     Рассмотрим основные   энергетические   характеристики
электромагнитного   поля,   определяемого  системой  уравнений  (1.1).
Умножая первое уравнение системы на $\mbox{\boldmath $\gv h$}$, а второе
уравнение  на $-\mbox{\boldmath $\gv e$}$, почленно складывая и пользуясь
известной формулой векторного анализа
     $$\div [\rv A\rv B]=\rv B\rot\rv A-\rv A\rot\rv B\,,\eqno(1.5)$$
получаем
     $$\frac{\partial u}{\partial t}+p+\div\mbox{\boldmath $\gv s$}=0\,,
      \eqno(1.6)$$
где
     $$u=\frac{\mbox{\boldmath $\gv e$}^2+\mbox{\boldmath $\gv h$}^2}
         {8\pi}\,,\eqno(1.7)$$
вектор Умова-Пойнтинга
      $$\mbox{\boldmath $\gv s$}=\frac c{4\pi}[\mbox{\boldmath $\gv e$}
        \mbox{\boldmath $\gv h$}]\,,\eqno(1.8)$$
а
     $$p=\mbox{\boldmath $\gv j$}\mbox{\boldmath $\gv e$}.\eqno(1.9)$$

     Проинтегрируем соотношение (1.6) по  некоторому  объёму  $V$  и
применим к слагаемому $\div\mbox{\boldmath $\gv s$}$ теорему Гаусса
     $$\int\limits_V\div\rv A\,dV=\oint\limits_S\rv A\,d\rv S\,,
             \eqno(1.10)$$
где $S$  --  замкнутая  поверхность,   ограничивающая   объём
$V$. Преобразуем второе слагаемое,  учитывая (1.2) и соотношение,
следующее из уравнений движения заряженной частицы в электрическом
поле:  $q\rv v\mbox{\boldmath $\gv e$}=dE_{\footnotesize\textit {кин}}/dt$,
где
$E_{\textit{кин}}$ --- кинетическая энергия частицы. В результате
получим:
   $$\frac{\partial}{\partial t}\left\{\int\limits_V u\,dV+
        \sum_n {E_{\footnotesize\textit {кин}}}\right \}=-\oint\limits_S\mbox
        {\boldmath $\gv s$}\,d\rv S\,,\eqno(1.11)$$
где суммирование производится по всем частицам,  заключённым в
рассматриваемом объёме.  Это соотношение, называемое \textit{
теоремой Умова-Пойнтинга}, представляет собой  закон  сохранения
энергии  и позволяет  называть величину $u$,  определённую
формулой (1.7), \textit {плотностью энергии} электромагнитного
поля, вектор $\mbox {\boldmath $\gv s$}$, определённый формулой (1.8)
---   \textit {плотностью  потока  энергии} (количество  энергии
поля, протекающее в единицу времени  через единичную  площадку  с
нормалью вдоль $\mbox{\boldmath $\gv s$}$), величину $p$,
определённую формулой (1.9)
--- \textit {плотностью работы поля над токами}.

     Совокупность уравнений  (1.1)--(1.4) и (1.6)--(1.9)  представляет  собой  основу
классической  электродинамики,  которая даёт прекрасное совпадение с
опытом.  Она является непротиворечивой теорией вплоть  до  расстояний,
близких    к   так   называемому   классическому   радиусу   электрона
$r_e=e^2/m_ec^2=2,8\cdot 10^{-13}\,\mbox{см}$,  при котором энергия
электромагнитного  поля  электрона  сравнивается  с его энергией покоя
(масса электрона $m_e=9,1\cdot 10^{-28}\,\mbox{г}$). Отметим, что ещё
раньше начинают проявляться квантовые эффекты, и  правильное  описание
электромагнитного взаимодействия может быть получено только  в  рамках
квантовой   электродинамики.  В  этом  курсе  нас  будут  интересовать
взаимодействия  зарядов  на  значительно  больших  расстояниях,  когда
классическая  микроскопическая  электродинамика  при применении её к
реальным телам обладает другим  недостатком  ---  она  даёт  излишне
подробное описание.

     Действительно, при переходе к изучению электромагнитных  полей  в
макроскопических  объёмах  вещества бессмысленно следить за
быстрыми изменениями поля в межмолекулярном пространстве. Дело не
только в том, что  такой  подход  практически  невозможно
реализовать,  так  как  в 1~см$^3$ вещества порядка $10^{21\pm2}$
молекул или  атомов,  но  даже каким-то   образом   полученный
громадный   объём   информации   о мелкомасштабном поведении
поля  мало  чем  помог  бы,  например,  при учёте  влияния
данного  объёма  вещества  на  поведение  поля  на некотором
расстоянии  от   него.   Макроскопическая   электродинамика
оперирует  только величинами,  усреднёнными по <<физически
бесконечно малым>> объёмам $\Delta  V$  и  интервалам  времени
$\delta  t$,  не интересуясь   микроскопическими   изменениями
величин,  обусловленных молекулярным строением вещества.  Вместо
истинного <<микроскопического>>  значения напряжённости
электрического поля $\mbox{\boldmath $\gv e$}$ рассматривается его
усреднённое значение $\overline{\mbox{\boldmath $\gv
e$}}=\mbox{\boldmath $\gv E$}(\rv r,t)=\displaystyle{\frac 1
{\Delta V \delta T}\int\,d^3\xi\int d\tau\mbox{\boldmath $\gv e$}
(\rv r+\vec\xi,t+\tau)}$. Аналогичным образом производится
усреднение напряжённости  магнитного поля:  $\overline
{\mbox{\boldmath $\gv h$}} =\mbox{\boldmath $\gv B$}(\rv r,t)$,
причём по установившейся традиции среднее значение
микроскопического магнитного поля называется магнитной индукцией.

     В результате  усреднения  плотности  тока  и  заряда эти величины
становятся непрерывными функциями координат.  Сама операция
усреднения для реального многообразия существующих в природе
веществ --- процедура сложная  и  не  всегда  однозначная.
Фактически   она   может   быть последовательно  проведена  только
для некоторых упрощённых моделей строения  вещества.  Основная
ценность  такого  подхода   состоит   в получении     причинно
обусловленной    системы    уравнений    для электромагнитных
полей  и  токов,   базирующихся   на   точных  <<микроскопических>>
уравнениях.   Прежде,  чем  приступать  к  выводу уравнений
электродинамики   сплошной   среды   путём усреднения
микроскопических   уравнений   поля   (1.1), остановимся  на исторически
более  раннем  феноменологическом  подходе,  широко распространённом
и  в современных руководствах  по электродинамике СВЧ.

     Утверждается, что  электромагнитное  поле в сплошной материальной
среде определяется четырьмя векторами: $\mbox{\boldmath $\gv E$}$ ---
электрическое поле, $\mbox{\boldmath $\gv H$}$ --- магнитное поле,
$\mbox{\boldmath $\gv D$}$ --- электрическая индукция,  $\mbox{\boldmath
$\gv B$}$ --- магнитная индукция, которые связаны с плотностью зарядов
$\rho$ и токов $\mbox{\boldmath $\gv j$}$ системой уравнений Максвелла:
     $$\hspace{-1.2cm}\rot\mbox{\boldmath $\gv E$}\,\!=\!-\displaystyle
       {\frac 1 c \frac{\partial\mbox{\boldmath $\gv B$}}{\partial t}}\,,
       \eqno(1.12\mbox{\textit а})$$
     $$\rot\mbox{\boldmath $\gv H$}\!=\!\displaystyle{\frac 1 c\frac
        {\partial\mbox{\boldmath $\gv D$}}{\partial t}+{\frac{4\pi} c}
        \mbox{\boldmath $\gv j $}}\,,\eqno(1.12\mbox{\textit б})$$
     $$\hspace{-1.8cm}\div\mbox{\boldmath $\gv D$ }\!=\!4\pi\rho\,,
         \eqno(1.12\mbox{\textit в})$$
     $$\hspace{-2.35cm}\div\mbox{\boldmath $\gv B$}\!=\!0\,.
         \eqno(1.12\mbox{\textit г})$$

     Приведенная система уравнений  пригодна  для  полного  описания
электромагнитного  поля  в  любых средах.  Однако одних этих уравнений
недостаточно для решения конкретных задач,  хотя бы потому,  что число
уравнений в системе меньше числа неизвестных.  Их необходимо дополнить
так  называемыми   {\it   материальными   уравнениями},   учитывающими
поведение  вещества  в  электромагнитном поле и устанавливающими связь
между векторами поля. Простейшие такие уравнения имеют вид:
     $$\hspace{-1.4cm}\mbox{\boldmath $\gv D$}\!=\!\varepsilon\mbox
       {\boldmath $\gv E$},\eqno(1.13)$$
     $$\hspace{-1.25cm}\mbox{\boldmath $\gv B$}\!=\!\mu\mbox{\boldmath
       $\gv H$},\eqno (1.14)$$
     $$\mbox{\boldmath $\gv j$}\!=\!\sigma\mbox{\boldmath $\gv E$}+
       \mbox{\boldmath $\gv j$}^e_{\footnotesize\textit {ст}}\,.\eqno(1.15)$$
Для однородной изотропной среды  $\varepsilon,\,\mu,\, \sigma$
---    постоянные    действительные    величины:   $\varepsilon$   ---
диэлектрическая  проницаемость,  $\mu$  ---  магнитная
проницаемость, $\sigma$ --- проводимость  вещества,
$\mbox{\boldmath $\gv  j$}^e_{\footnotesize\textit {ст}} $
--- плотность стороннего электрического тока.

     Слово <<сторонний>> говорит о  том,  что  этот  ток  не  обусловлен
усреднённым  движением  заряженных  частиц  вещества  под
действием рассматриваемого электромагнитного поля,  а имеет иное
происхождение, например,  представляет  собой  движение  пучка
частиц,  выведенных из ускорителя и пролетающих через вещество.  В
других случаях,  например, при вычислении излучения антенного
вибратора, ток в его элементах тоже рассматривается как
$\mbox{\boldmath $\gv j$}^e_{\footnotesize\textit{ст}}$,
хотя  по  своему происхождению  он несомненно  есть ток
проводимости в металлических элементах вибратора, возбуждённый
тем полем,  с помощью которого к вибратору  подводится
высокочастотная  мощность.  Поле,  возбуждающее вибратор,  тоже
можно рассматривать как $\mbox{\boldmath $\gv
E$}_{\footnotesize\textit {ст}}$, исключив с его помощью из
уравнений $\mbox{\boldmath $\gv j$}^e_{\footnotesize\textit{ст}}$.
Сторонние токи могут быть заданы и в виде петли с током,
и щели в стенке волновода, и иным способом.

     Правильное выделение стороннего тока или поля --- важный  элемент
постановки  электродинамической  задачи для реальной системы.  С одной
стороны,  такое выделение должно максимально упростить задачу для  той
части полной системы,  которая в данной задаче представляет наибольший
интерес,  а   с   другой   ---   не   позволить   пренебречь   важными
взаимодействиями между частями полной системы.  Без первого --- строго
поставленная задача для всей системы может оказаться  недоступной  для
решения, без   второго   ---  решение  может  оказаться  далёким  от
реальности.

     В дальнейшем нижний индекс  ${}_{\textit{ст}}$
у $\mbox{\boldmath $\gv j$}^e_{\footnotesize \textit{ст}}$
опускается.   Верхний индекс ${}^e$ оставляется в тех случаях, когда необходимо
отличить реальный ток, обусловленный движением электрических зарядов,
от фиктивного стороннего магнитного тока, отмечаемого  верхним индексом
${}^m$, который формально вводится в уравнения Максвелла  и позволяет
существенно упростить  решение  ряда электродинамических задач.

     Значения констант  $\varepsilon,\,\mu,\,\sigma$  для конкретных
материалов берутся  из  опыта,  зачастую  проводимого  при  постоянных
полях.  Материальные  уравнения  удовлетворительно описывают поведение
достаточно медленно меняющихся  полей  во  многих  изотропных  средах.
Однако  применение их к быстропеременным полям в ряде случаев приводит
к  бессмысленным  результатам.  Невозможно  придумать  никакой  модели
строения  вещества,  для  которой подобные соотношения были бы верны и
для достаточно быстрых временных изменений поля при строгом усреднении
микроскопических уравнений.

     Говорят, что соотношения (1.13)--(1.15)  не  учитывают  временной
(как,  впрочем,  и  пространственной)  дисперсии  свойств
материальной среды.  Если отвлечься от этого <<небольшого
недостатка>>,  то  система уравнений является замкнутой,
позволяющей математически строго решить большое   количество
практически    важных    задач.    Существенным дополнительным
достоинством  системы  является  возможность на её основе
последовательным способом ввести энергетические  характеристики
поля,  подобно  тому,  как  это было сделано выше для
микроскопических уравнений.

     Прежде чем  переходить к рассмотрению энергетических соотношений,
вытекающих из уравнений (1.12),  остановимся ещё на  так
называемых \textit{граничных условиях} для этих уравнений.
Параметры $\varepsilon,\,\mu,\,\sigma$,  определяющие
электрические  и  магнитные  свойства вещества,  могут,  в случае
неоднородной среды,  произвольным способом изменяться в
пространстве, оставаясь непрерывными функциями координат. Среди
этих  изменений важную роль играют такие,  при которых свойства
среды меняются вблизи  какой-то  поверхности  очень  резко,
например, вблизи поверхности соприкосновения двух сред с
различными свойствами.

     Разумной идеализацией такого поведения параметров, очень полезной
для  решения  электродинамических  задач,  является  замена   его
на изменение  скачком.  При этом,  вообще говоря,  могут терпеть
разрыв и компоненты электромагнитного поля.  Соответствующие
граничные  условия легко  получить из  уравнений Максвелла,
записанных в интегральной форме. Переход к интегральному представлению
уравнений  (1.12)  осуществляется  с помощью  уже использованной
выше теоремы Гаусса (1.10) и теоремы Стокса:
     $$\int\limits_S\rot\rv A\,d\rv S=\oint\limits_C\rv A\,d\rv l\,,
                                         \eqno(1.16)$$
где $S$ -- произвольная поверхность, натянутая на замкнутый контур $C$.
В результате имеем:
     $$\hspace{-0.9cm}\oint\limits_C\!\mbox{\boldmath  $\gv E$}\,d\rv l=-
       \frac 1 c\frac{\partial}{\partial t}\int\limits_S\!\mbox
        {\boldmath $\gv B$}\,d\rv S\,,\eqno(1.12\mbox{\textit а}')$$
     $$\oint\limits_C\!\mbox{\boldmath $\gv H$}\,d\rv l=\frac 1c\frac
       {\partial}{\partial t} \int\limits_S\!\mbox{\boldmath $\gv D$}\,
        d\rv S+\frac{4\pi}c J\,,\eqno(1.12\mbox{\textit б}')$$
     $$\hspace{-3.2cm}\oint\limits_S\!\mbox{\boldmath $\gv B$}\,d\rv S=0\,,
       \eqno(1.12\mbox{\textit в}')$$
     $$\hspace{-2.7cm}\oint\limits_S\!\mbox{\boldmath $\gv D$}\,d\rv
       S=4\pi Q\,, \eqno(1.12\mbox{\textit г}')$$
где $J=\int \mbox{\boldmath $\gv j$}\,d\rv S$
--- полный ток,  охватываемый контуром, $Q$ ---  полный  заряд,
заключённый  внутри замкнутой поверхности. Во избежание
недоразумений отметим, что поверхности интегрирования $S$ во всех
этих  уравнениях произвольные,  но  при  этом в
(1.12$\mbox{\textit а}'$) и (1.12$\mbox{\textit б}'$) они
незамкнутые, а в (1.12$\mbox{\textit в}'$) и (1.12$\mbox{\textit г}'$)
--- замкнутые.

     Уравнения Максвелла  в  интегральной форме полностью эквивалентны
дифференциальным уравнениям (1.12)  и,  следовательно,  описывают  все
явления  электродинамики.  При  конкретных  расчётах полей они часто
оказываются  более  удобными  для  численных  методов  решения   задач
электродинамики. Дифференциальная же форма уравнений имеет несомненное
преимущество при аналитических исследованиях,  особенно  для  описания
волновых процессов.

     Проведём внутри слоя,  в пределах которого  происходит  быстрое
изменение  параметров  среды,  некоторую  поверхность и примем
её за границу  раздела  со  скачкообразным  изменением  свойств
среды.  Для установления  граничных  условий  на  тангенциальные
компоненты  поля построим небольшой прямоугольный контур (рис.1а),
плоскость  которого содержит нормаль к поверхности раздела,
проходящий по обе стороны этой поверхности  и  охватывающий  всю
область,  где   происходит   резкое изменение  свойств  среды. Две
стороны контура параллельны касательной $\rv \tau$  к границе
раздела в точке,  для  которой  ищутся граничные условия. Выпишем
для этого контура уравнения (1.12$\mbox{\textit а}'$)  и
(1.12$\mbox{\textit б}'$). Путём предельного   перехода   к
нулевой толщине   слоя получим соотношения между тангенциальными
компонентами  поля  по обе стороны поверхности раздела:
     $$\hspace{-0.5cm}[\rv n,\mbox{\boldmath $\gv E$}_2-\mbox
       {\boldmath $\gv E$}_1]=0\,,\eqno(1.17)$$
     $$[\rv n,\mbox{\boldmath $\gv H$}_2-\mbox{\boldmath $\gv H$}_1]=
        \frac{4\pi}c \mbox{\boldmath$\gv I$},\eqno(1.18)$$
где нижние индексы у векторов поля соответствуют нумерации сред по обе
стороны границы,  а нормаль $\rv n$ направлена из среды 1 в  среду  2.
Входящая  в  формулу  (1.18)  величина  $\mbox{\boldmath $\gv I$}$ ---
поверхностная плотность тока, представляющая собой предел интеграла от
объёмной плотности тока $\mbox{\boldmath $\gv j$}$ по
толщине слоя, в котором он протекает, при стремлении проводимости слоя
к  бесконечности, а его толщины к нулю.  Такой  предельный  переход
соответствует   модели идеального  проводника  --- среды, которую часто
используют в задачах электродинамики СВЧ как хорошее приближение для
реальных  проводников ввиду  их высокой проводимости  в рассматриваемом
диапазоне частот. Понятие  поверхностного  тока  подробнее  обсуждается
далее  при рассмотрении скин-эффекта; для сред с конечной проводимостью,
в том числе и для непроводящих,  правая часть (1.18) обращается в нуль,
если только $\mbox{\boldmath $\gv I$}$ не задан в  качестве  стороннего.

\begin{picture}(160,55)
\put(5,50){\special{em:graph  fig1-1a.bmp}}
\put(84,50){\special{em:graph fig1-1b.bmp}}
\end{picture}

\begin{center}\begin{minipage}[c]{0.9\textwidth}
\footnotesize{\parindent=0.5cm
Рис.~1.1.~Вспомогательные построения для вывода граничных условий: {\it а)}
---  прямоугольный  контур  для  тангенциальных составляющих $\mbox
{\boldmath $\gv E$}$ и $\mbox{\boldmath $\gv H$}$, {\it б)} --- цилиндр для
нормальных составляющих $\mbox{\boldmath $\gv D$}$,~$\mbox{\boldmath
$\gv B$}$. }\end{minipage}\end{center}\vspace*{0.25cm}

     Отметим, что на границе  двух  сред,  одна  из  которых  обладает
магнитными свойствами ($\mu\ne 1$),  в частности, на границе вакуума с
такой средой,  непрерывными являются  не  тангенциальные  составляющие
истинных  магнитных  полей  (в случае среды речь идет об усреднённом
поле),  а составляющие вектора  $\mbox{\boldmath $\gv H$}$, физический
смысл  которого обсуждается дальше.

     Применяя уравнения (1.12$\mbox{\textit в}'$) и (1.12$\mbox{\textit г}'$)
к небольшому цилиндру, расположенному по обе стороны поверхности
раздела (см.  рис.~1.1б),  и устремляя его высоту к нулю,  получим
граничные условия для нормальных компонент векторов поля:
     $$\hspace{-1.1cm}(\mbox{\boldmath $\gv B$}_2-\mbox{\boldmath
        $\gv B$}_1)\rv n=0\,,\eqno(1.19)$$
     $$(\mbox{\boldmath $\gv D$}_2-\mbox{\boldmath $\gv D$}_1)\rv n=
        4\pi\rho_{\footnotesize\textit {пов}}\,,\eqno (1.20)$$
где $\rho_{\footnotesize\textit {пов}}$ --- поверхностная
плотность заряда.

     Введение усреднённых  полей и плотностей заряда и тока приводит
в областях с непрерывным  изменением  параметров  среды  к
устранению сингулярностей,    присущих    этим   величинам   в
микроскопической электродинамике из-за точечности элементарных
зарядов. Однако введение граничных  поверхностей  раздела,  на
которых скачкообразно изменяются свойства среды,  также может
привести  к  сингулярностям.  Это  имеет место,  например, на
изломах поверхности металлических проводников (на остриях,
рёбрах).  Сингулярности в этом случае имеют совсем  другую
природу,  чем  в  микроскопической  электродинамике:  они
обусловлены некорректностью    моделирования    реальных
поверхностей     тел. Принципиальное отличие между этими двумя
типами сингулярностей состоит в бесконечности энергии поля  вблизи
точечных  зарядов  и  конечности энергии   поля,   сосредоточенной
в   окрестности   особых  точек  в усреднённом поле.  В
большинстве случаев последние сингулярности  не приводят  ни  к
каким принципиальным трудностям как при аналитическом
исследовании,  так и при численном  решении,  хотя  правильный
учёт поведения  поля  вблизи  изломов может существенно
упростить процедуру численных расчётов.  В некоторых же случаях,
например, вблизи кромки бесконечно  тонкой  идеально  проводящей
поверхности,  математика  не позволяет  выбрать  однозначное
решение   уравнений   Максвелла   без дополнительного  условия на
конечность энергии поля вблизи сингулярных точек.  Это так
называемое {\it  условие  на  ребре}  будет  подробнее рассмотрено
при решении дифракционных задач.

     Здесь уместно отметить,  что для переменных во времени  полей,  а
только  такие поля и рассматриваются в электродинамике СВЧ,
уравнения (1.12\textit{в}),  (1.12\textit{г})  являются следствием
уравнений (1.12\textit{а}),  (1.12\textit{б}), поскольку для
произвольного векторного поля $\rv A(\rv r)$ имеет место тождество
$\div\rot \rv A=0$.  В этом случае для определения  полей  и
плотности наведенных  токов достаточно  проинтегрировать уравнения
(1.12\textit{а}) и (1.12\textit{б}), а плотность заряда $\rho$,
как,  в  частности, и поверхностная плотность  заряда
$\rho_{\footnotesize\textit{пов}}$, может быть найдена из
уравнения непрерывности (1.3).  Впрочем, эти величины для СВЧ
полей в большинстве случаев не представляют интереса.

     Закон сохранения   энергии   в  макроскопической  электродинамике
выводится из системы уравнений (1.12) с помощью тех же преобразований,
которые  были  использованы  выше  при  получении  уравнения  (1.6)  в
микроскопической электродинамике; в результате имеем, что
     $$-\frac c{4\pi}\div [\mbox{\boldmath $\gv E\gv H$}]=\frac 1{4\pi}
        \Bigl(\mbox{\boldmath$\gv E$} \frac{\partial \mbox{\boldmath
        $\gv D$}}{\partial t}+\mbox{\boldmath $\gv H$} \frac{\partial
         \mbox{\boldmath $\gv B$}}{\partial t}\Bigr)+\mbox
         {\boldmath$\gv E$}\mbox{\boldmath$\gv j$}.\eqno(1.21)$$
Левую часть этого уравнения можно записать в виде $-\div
\mbox{\boldmath$\gv  S$}$,  где вектор Умова-Пойнтинга в электродинамике
сплошной среды
     $$\mbox{\boldmath $\gv S$}=\frac c{4\pi}[\mbox{\boldmath $\gv E$}
       \mbox{\boldmath $\gv H$}]\eqno(1.22)$$
сохраняет физический смысл вектора $\mbox{\boldmath $\gv s$}$,
определённого  формулой (1.8)   для  пустоты.  Это  следует  из
непрерывности  тангенциальных составляющих полей $\mbox{\boldmath
$\gv E$}$ и  $\mbox{\boldmath $\gv  H$}$  (граничные  условия  (1.17)  и
(1.18), в котором $\mbox{\boldmath $\gv I$}$ следует положить равной нулю).

     Более того,  выражение  (1.22)  сохраняет  смысл  плотности
потока  энергии  в  единицу  времени  и  при учёте дисперсии,  и при
наличии   дополнительного (необусловленного   токами   проводимости)
поглощения   в  среде,  когда  материальные  уравнения  (1.13)--(1.15)
становятся неприменимыми. Таким образом, вектор Умова-Пойнтинга (1.22)
является   наиболее   важной  энергетической  характеристикой  поля  в
сплошной среде.  Обратим внимание, что в среде с магнитными свойствами
он  определяется не усреднённым истинным магнитным полем
$\mbox{\boldmath $\gv B$}$,  а вектором $\mbox{\boldmath $\gv H$}$.
В отличие от вектора Умова-Пойнтинга сила  Лоренца, действующая  на
движущийся  в  веществе  пробный заряд,  определяется усреднёнными
истинными полями:
              $$\rv f=q(\mbox{\boldmath $\gv E$}+\frac 1 c
[\rv v \mbox{\boldmath $\gv B$}]).\eqno(1.23)$$

     Первое слагаемое в  правой  части  уравнения  (1.21)  с  учётом
материальных     уравнений     (1.13),     (1.14)    при    постоянных
$\varepsilon$ и $\mu$ можно  рассматривать  как  производную  по
времени от величины
     $$U=\frac 1{8\pi}(\varepsilon \mbox{\boldmath $\gv E$}^2+
         \mu \mbox{\boldmath $\gv H$}^2),\eqno(1.24)$$
что позволяет при сделанных допущениях  считать  последнюю  плотностью
энергии  электромагнитного  поля в сплошной среде.  Второе слагаемое в
правой части  уравнения  (1.21)  можно  на  основании  формулы  (1.15)
представить  в виде суммы двух членов --- плотности джоулевых потерь в
проводящей среде $\sigma \mbox{\boldmath $\gv E$}^2$ и плотности
работы  поля  над сторонним током $\mbox{\boldmath $\gv j$}^e\mbox
{\boldmath $\gv E$}$.

     Забегая вперёд,  отметим,  что несмотря на формальную строгость
получения  выражений  (1.24)  для  плотности  энергии поля,  они имеют
ограниченную область применимости.  Это обусловлено видом материальных
уравнений    (1.13)--(1.15),    справедливых   лишь   для   достаточно
низкочастотных полей.  Распространение этих уравнений на все временные
процессы приводит,  в частности, к нарушению принципа причинности, что
подробнее обсуждается при рассмотрении монохроматических полей.

     Остановимся теперь  кратко     на     процедуре    усреднения
микроскопических уравнений.  В соответствии с  определением
средних значений  векторов  поля  $\overline {\mbox{\boldmath $\gv
e$}}=\mbox {\boldmath $\gv E$}$ и $\overline{\mbox{\boldmath $\gv
h$}}=\mbox {\boldmath $\gv B$}$ уравнение (1.12\textit{а})
непосредственно следует из уравнения (1.1\textit{а}). Уравнение
(1.1\textit{б}) после усреднения принимает вид
     $$\rot \mbox{\boldmath $\gv B$}=\frac 1c\frac{\partial \mbox
      {\boldmath $\gv E$}}{\partial t}+\frac{4\pi}
         c\overline{\Sigma q\rv v}\,,\eqno(1.25)$$
а результат усреднения микроскопического тока зарядов  вещества  можно
представить в виде суммы трёх членов:
     $$\overline{\Sigma q\rv v}= \mbox{\boldmath $\gv j$}+\frac
       {\partial \mbox{\boldmath $\gv P$}}{\partial t}+c\rot
        \mbox{\boldmath $\gv M$}.\eqno(1.26)$$

     Первое слагаемое справа в этом уравнении связано с  усреднённым
движением  <<свободных>>  зарядов  (обычный  ток в проводниках),
второе слагаемое $\partial\mbox{\boldmath $\gv P$}/\partial t$
называется током поляризации.  Вектор  \textit  {диэлектрической
поляризации} вещества $\mbox{\boldmath $\gv P$}$ связан со средней
плотностью заряда в диэлектрике соотношением
     $$\bar\rho =-\div  \mbox{\boldmath $\gv P$};\eqno(1.27)$$
при этом подразумевается,  что в диэлектрике нет посторонних
зарядов ($\int\bar\rho\,dV$,  взятый по всему объёму тела, равен
нулю) и что $\mbox{\boldmath $\gv P$}=0$ вне диэлектрика (в
пустоте). Определённый таким  образом вектор  $\mbox{\boldmath
$\gv  P$}$  представляет  собой электрический  дипольный момент
единицы объёма диэлектрика. В этом легко убедиться, если
вспомнить, что в теории поля \textit {дипольным моментом} системы
зарядов называется величина
     $$\rv d=\sum_n q_n\rv r_n\,,\eqno(1.28)$$
а из (1.27)  путём  несложных  преобразований  вывести соотношение
  $$\int\rv r\bar\rho\,dV=\int\mbox{\boldmath $\gv P$}\,dV\,.\eqno(1.29)$$
Таким образом, член $\partial \mbox{\boldmath $\gv P$}/\partial t$
представляет собой плотность тока поляризации.

     Третье слагаемое в (1.26) соответствует атомным или  молекулярным
токам  в  веществе  (часто  называемым  токами  Ампера),  для
которых интеграл от  среднего  значения  микроскопической
плотности  тока  по любому  поперечному  сечению  тела  равен
нулю.  Последнее  позволяет представить   соответствующую   часть
$\overline{\Sigma q\rv   v}$ пропорциональной ротору вектора
\textit {намагниченности} $\mbox{\boldmath $\gv M$}$:
     $$\overline{\Sigma q\rv v}=c\rot\mbox{\boldmath
        $\gv M$}\,.\eqno(1.30)$$
Определённый   таким  образом  вектор   $\mbox{\boldmath $\gv
M$}$ при дополнительном условии,  что  он отличен от нуля только в
объёме тела, представляет  собой магнитный момент  единицы
объёма  вещества. Напомним, что  в теории  поля {\textit
магнитный момент} системы зарядов $\rv  m$ определяется как
     $$\rv m={\frac 1 {2c}}\sum_n q_n[\rv r_n\rv v_n]\,,\eqno(1.31)$$
а из  (1.30)  путём  простых  векторных  преобразований  может  быть
выведено соотношение
     $$\frac 1{2c}\int[\rv r\overline{\Sigma q\rv v}]\,dV=\int\mbox
        {\boldmath $\gv M$} \,dV\,.\eqno(1.32)$$

     Если ввести  вектор  электрической  индукции  $\mbox{\boldmath
$\gv  D$}$  и  вектор напряжённости магнитного поля $\mbox{\boldmath
$\gv H$}$ c помощью соотношений
       $$\mbox{\boldmath $\gv D$}=\mbox{\boldmath $\gv E$} +4\pi
         \mbox{\boldmath $\gv P$},\eqno(1.33)$$
       $$\mbox{\boldmath $\gv B$}=\mbox{\boldmath $\gv H$}+ 4\pi
         \mbox{\boldmath $\gv M$},\eqno(1.34)$$
то уравнение (1.25) перейдёт в уравнение (1.12{\textit б}).

     Все приведенные  выше  преобразования  микроскопического   тока
являются  формальными,  пока  не  получены выражения для векторов
$\mbox{\boldmath $\gv P, \gv M, \gv j $}$ в функции
$\mbox{\boldmath $\gv E, \gv B$}$. При этом надо сказать, что в
переменных   полях разделение  зарядов  вещества  на  <<свободные>>
движением которых обусловлен  ток  проводимости,  и  на
<<связанные>>, движение которых определяет ток поляризации,
процедура, вообще говоря, неоднозначная.  Дело в том,  что при
очень  быстрых  изменениях  полей электромагнитные свойства
веществ,  резко различающиеся  в постоянных и медленно
изменяющихся полях (как, например, свойства  металлов и
диэлектриков), становятся почти идентичными. Соотношения же (1.29)
и (1.31),  связывающие между собой векторы  поля  и плотности
электрического дипольного момента и магнитного момента среды
--- величины,  однозначно определённые для микроскопических полей  и
зарядов, проясняют физический смысл векторов
 $\mbox{\boldmath $\gv D$}$  и $\mbox{\boldmath $\gv H$}$.

     В самом  общем  виде связь между векторами $\mbox{\boldmath $\gv P,
~\gv M,~\gv j$}$ и полями может быть записана в некотором функциональном
виде
    $$\mbox{\boldmath $\gv P$}=\mbox{\boldmath $\gv P$}\{\mbox
    {\boldmath $\gv E$}\},\quad\mbox{\boldmath $\gv j$}=\mbox{\boldmath
    $\gv j$}\{\mbox{\boldmath $\gv E$}\},\quad\mbox{\boldmath $\gv M$}=
     \mbox{\boldmath $\gv M$}\{\mbox{\boldmath $\gv H$}\},\eqno(1.35)$$
который не  подразумевает  ни локальности этих связей, ни их
одновременности.  Принцип причинности требует лишь того, чтобы значения
всех этих векторов в данный момент времени определялись  значениями
полей только в  предшествовавшие  моменты  (последовательнее  было  бы
рассматривать зависимость $\mbox{\boldmath $\gv M$}$ от $\mbox{\boldmath
$\gv B$}$, однако такова традиция). При  не  очень  сильных  полях,
которые  в дальнейшем  только и рассматриваются,  эта связь должна быть
линейной, а в случае медленных изменений поля эта линейная зависимость
может быть представлена в виде
     $$\mbox{\boldmath $\gv P$}=\alpha\mbox{\boldmath $\gv E$},\quad
       \mbox{\boldmath $\gv j$}=\sigma\mbox{\boldmath $\gv E$},\quad
       \mbox{\boldmath $\gv M$}=\chi\mbox{\boldmath $\gv H$}.\eqno(1.36)$$

     Напомним, что    $\alpha$    называется    \textit  { коэффициентом
поляризуемости}    вещества,    а    $\chi$    ---  \textit{
магнитной восприимчивостью}.  Они связаны с введёнными раньше
диэлектрической проницаемостью   $\varepsilon$   и   магнитной
проницаемостью  $\mu$ соотношениями, очевидными из определений
(1.33), (1.34):
     $$\alpha=\frac{\varepsilon-1}{4\pi},\quad\chi=\frac{\mu-1}
         {4\pi}\,.\eqno(1.37)$$
В быстропеременных полях из-за инерции  зарядов  в  веществе  процессы
установления  тока  проводимости,  поляризации  и  намагниченности  не
успевают  следовать  за  изменениями  полей.  Поэтому  по  аналогии  с
соотношениями (1.36) эту связь  представляют в операторном виде
     $$\mbox{\boldmath $\gv P$}=\hat\alpha\mbox{\boldmath $\gv E$},
       \quad\mbox{\boldmath $\gv j$}=\hat\sigma\mbox{\boldmath $\gv E$},
       \quad\mbox{\boldmath $\gv M$}=\hat\chi\mbox{\boldmath $\gv H$}.
       \eqno(1.38)$$

     В случае   локальной   связи   (нелокальность  связи  приводит  к
пространственной дисперсии характеристик вещества,  которая  в  данном
курсе  не  рассматривается)  общий  вид  такого  оператора  может быть
представлен  на  примере  оператора  $\hat\varepsilon$,   связывающего
векторы $\mbox{\boldmath $\gv D$}$ и $\mbox{\boldmath $\gv E$}$:
     $$\mbox{\boldmath $\gv D$}(\rv r,t)=\mbox{\boldmath $\gv E$}
       (\rv r,t)+\int\limits_0^\infty f(\tau)\mbox{\boldmath $\gv E$}
       (t-\tau)\,d\tau=\hat\varepsilon\mbox{\boldmath $\gv E$}(\rv r,t)
       \,.\eqno(1.39)$$
Ядро этого интегрального оператора, функция $f(\tau)$, и описывает все
диэлектрические свойства  среды.  Очевидно,  что  описание  с  помощью
функции $f(\tau)$  даёт  значительно  больше  возможностей  передать
свойства   вещества,   чем   это   может   сделать   одна   постоянная
$\varepsilon$.  С  другой  стороны,  очевидно,  что  решать конкретные
задачи с операторной связью типа (1.39) в общем  виде  крайне  трудно.
Приходится  прибегать к спектральному разложению всех полей и работать
с гармоническим составляющими.  Возникающие при этом особенности будут
рассмотрены в следующем разделе.

     В заключение уточним место электродинамики  СВЧ  в  общем  здании
классической   электродинамики   сплошных   сред.   В  зависимости  от
соотношения  между  характерной  длиной  волны   $\lambda$   волнового
процесса  и  размером тел $L$,  участвующих в нем,  можно выделить три
случая:  1) $\lambda \gg L$, когда волновые свойства электромагнитного
поля  не  проявляются,  эффекты  запаздывания  из-за конечной скорости
распространения света не играют роли --- это  область  электростатики,
магнитостатики  и теории квазистационарного поля;  2) $\lambda \ll L$,
область оптических явлений,  когда  в  основном  используется  лучевое
описание поля и от уравнений Максвелла переходят к уравнению эйконала;
3) $\lambda \simeq L$,  промежуточный случай,  при котором  приходится
использовать  уравнения Максвелла в самом общем виде --- это  и
есть область электродинамики СВЧ, для которой характерны длины волн от
$1$~м  до  $1$~мм.  Решаемые электродинамикой СВЧ задачи имеют широкое
приложение в радиосвязи, радиолокации, ускорителях элементарных частиц
и различных приборах научного и прикладного значения.

     Отметим ещё   раз,   что   используемое   в  курсе  приближение
макроскопической   электродинамики,    естественно,    не    учитывает
дискретности  зарядов (заряд распределён в пространстве непрерывно),
параметр $L$ много больше  размеров  атомов  и  молекул  (то  есть  не
учитывается  дискретность  вещества),  а время изучения объектов много
больше характерных ядерных (атомных) времён.  Под  точечным  зарядом
понимается  совокупность  элементарных  зарядов,  размер которой много
меньше размера изучаемых объектов $L$.

%\end{document}

\newpage
\oddsidemargin=-0.4mm \evensidemargin=-0.4mm
\headsep=7mm
\textheight=231.875mm
\textwidth=160mm
\mathsurround=2,5pt \unitlength=1mm
%\begin{document}
%\input{macr.tex}
\thispagestyle{empty}
%\addtocounter{page}{20}
\baselineskip=\normalbaselineskip
%\baselineskip=1.075\normalbaselineskip

\begin{center}
\subsubsection*{2. Монохроматические поля и комплексные амплитуды}
\end{center}
\vspace*{0.5cm}

\markboth{Глава 1. Основные положения классической электродинамики}
{2. Монохроматические поля и комплексные амплитуды}

\begin{center}\begin{minipage}[c]{0.75\textwidth}
\footnotesize{\parindent=0.5cm
         Особая роль  гармонических   процессов   в   электродинамике.
         Представление   гармонических  полей  с  помощью  комплексных
         амплитуд.  Уравнения  Максвелла  для  комплексных   амплитуд.
         Комплексные   диэлектрическая   и   магнитная   проницаемости
         вещества.  Дисперсионные соотношения. Вычисление квадратичных
         по полю величин.  Комплексная теорема Умова-Пойнтинга. Потери
         энергии в среде.  Энергия поля  в  диспергирующей  среде  без
         поглощения.
}\end{minipage}\end{center}
\vspace*{0.5cm}

 Векторы электромагнитного поля $\mbox{\boldmath $\gv E,~\gv D, ~\gv B,~\gv
H$}$ и  плотность тока  $\mbox{\boldmath $\gv j$}$  являются
функциями радиуса-вектора точки наблюдения  $\rv  r$  и  времени
$t$. Среди  всевозможных  временных процессов в электродинамике
особую роль играют гармонические или,  как их ещё часто
называют, монохроматические,  при  которых  компоненты всех полей
и токов меняются по синусоидальному закону:
     $$\mbox{\boldmath $\gv E$}(\rv r,t)=\mbox{\boldmath
       $\gv E$}(\rv r) \cos(\omega t-\alpha).\eqno(2.1\mbox{\textit а})$$
Всякая гармоническая  зависимость  характеризуется  тремя  величинами:
амплитудой (в данном случае $\mbox{\boldmath $ \gv E$}(\rv r)$),
частотой $\omega$ и фазой $\alpha$.

     Можно указать   несколько   причин,   по   которым   рассмотрение
гармонической  зависимости  от  времени  представляет  особый интерес.
Во-первых,  всякая временная зависимость реальной физической  величины
может быть представлена в виде интеграла (или ряда) Фурье:
     $$F(t)=\frac 1{2\pi}\int\limits_{-\infty}^\infty F(\omega)\,
         e^{-i\omega t}\, d\omega,\qquad F(\omega)=\int\limits_
         {-\infty}^\infty F(t)\,e^{i\omega t}\,dt,\eqno(2.2)$$
то есть  в  виде интеграла от комплексной гармонической функции.  Хотя
вычисление   интеграла   по   $\omega$   может    оказаться    сложной
математической операцией, но никаких принципиальных трудностей на этом
пути нет.  Во-вторых,  многие реальные устройства излучают поля, очень
близкиe  к  гармоническим,  что  обусловлено  не  в  последнюю очередь
свойствами одного из основных элементов техники  СВЧ  ---  объёмного
резонатора.  В  третьих,  в  силу  линейности  уравнений Максвелла все
входящие в них векторы изменяются с  одной  и  той  же  частотой,  что
позволяет   с  помощью  несложного  математического  приёма  полностью
исключить из уравнений время.  При этом  сложная  интегральная  операторная
связь (1.39) между векторами  $\mbox{\boldmath $\gv D$}$
и~ $\mbox{\boldmath $\gv E$}$, имеющая место  при произвольном временном
изменении этих величин, для их гармонических компонент сводится   к
алгебраической.

     Этот математический  приём  состоит в использовании комплексных
амплитуд  для  вещественных  векторов  поля  и  тока,  так  что  любая
компонента  вектора монохроматического поля вида (2.1\textit а)
представляется в виде
     $$\mbox{\boldmath $ \gv E$}(\rv r,t)=\re[\rv E(\rv r)
       e^{-i\omega t}],\eqno(2.3)$$
где $\rv E(\rv r)$ --- комплексная амплитуда поля (вектор),  зависящая
только  от  координат.  Поскольку на протяжении всего курса в основном
рассматриваются   монохроматические   зависимости,   то   именно   для
комплексных  амплитуд  векторов поля и плотности тока выбраны наиболее
употребительные обозначения $\rv E$,~$\rv B$,~$\rv D$,~$\rv H$ и~ $\rv
j$. Запись вида (2.3) позволяет после подстановки всех векторов поля в
уравнения Максвелла опустить общий множитель $e^{-i\omega  t}$  и  тем
самым фактически избавиться от времени $t$. Однако теперь все величины
становятся комплексными и при переходе к истинным физическим величинам
приходится выделять действительную часть.  Отметим, что это усложнение
не  столь  велико,  как  при  альтернативном   способе   представления
гармонической величины (2.1а) в виде
     $$\mbox{\boldmath $\gv E$}(\rv r,t)=\rv E_s(\rv r)\sin{\omega t}+
       \rv E_c(\rv r)\cos{\omega t}\,,\eqno(2.1\mbox{\textit б})$$
где $\rv E_s(\rv r)=\mbox{\boldmath $\gv E$}(\rv r)\sin\alpha$,~
     $\rv E_c(\rv r)=\mbox{\boldmath $ \gv E$}(\rv r)  \cos\alpha  $
---  величины  действительные.  Поскольку множители $\sin{\omega t}$ и
$\cos{\omega t}$ линейно независимы,  то из каждого уравнения системы
(1.12) можно сделать два,  разбив члены уравнения на две группы в
зависимости от того,  какой множитель они содержат. Из-за того,  что
операция дифференцирования по $t$ переводит члены из одной группы  в
другую,  усложнения  в  выкладках  оказываются  несравненно большими,
чем при выделении реальной части комплексного числа.

     Уравнения Максвелла  для  комплексных  амплитуд,   определённых
формулой   (2.3),  получаются  из  (1.12)  путём  формальной  замены
$\partial /{\partial t \rightarrow -i \omega}$ и  опускания  множителя
$e^{-i\omega t}$. После этих несложных преобразований получаем:
     $$\left.\begin{array}{rcl}\rot \rv H &\!=\!&-ik\rv D+
         \displaystyle{\frac{4\pi}c\rv j},\\[0.25cm]\rot \rv E &\!=
         \!&ik\rv B,\\[0.25cm]\end{array}\right\}\eqno(2.4)$$
где
     $$k=\frac\omega c\eqno(2.5)$$
--- так  называемое  {\it  волновое  число}.  В  дальнейшем,  говоря о
частотной  зависимости  какой-то  величины,  будем  часто   при   этом
подразумевать не $\omega$, а пропорциональную ей величину $k$. Отметим
также, что линейная связь между полями приводит к линейной связи между
комплексными    амплитудами;    $\re$   ---   линейный   оператор   и,
следовательно, коммутирует с  другими  линейными   операторами   вида
$\rot\rot$,~$\grad\div$,   ~$\Delta$,   присутствующими  в  уравнениях
электродинамики.

     В случае  простейших  материальных  уравнений  (1.13),  (1.14)  и
(1.15) уравнения (2.4) можно записать в виде
     $$\left.\begin{array}{rcl}\rot \rv H &\!=\!&-\displaystyle
         {ik(\varepsilon+i\frac{4\pi\sigma}\omega) \rv E
         +\frac{4\pi}c\rv j^e},\\[0.25cm]\rot \rv E &\!=\!&ik\mu
         \rv H\,.\\[0.25cm]\end{array}\right\}\eqno(2.6)$$
Если ввести  теперь  {\it  комплексную  диэлектрическую проницаемость}
$\tilde{\varepsilon}$ c помощью соотношения
     $$\tilde\varepsilon=\varepsilon+i\frac{4\pi\sigma}\omega,
         \eqno (2.7)$$
то уравнения (2.6) примут окончательный вид
     $$\left.\begin{array}{rcl}\rot \rv H &\!=\!&-ik\varepsilon \rv E
         +\displaystyle{\frac{4\pi}c\rv j^e},\\[0.25cm]\rot \rv E &\!
         =\!&ik\mu\rv H\,,\end{array}\right\}\eqno(2.8)$$
в котором  они  будут  применяться на протяжении большей части данного
курса.  Здесь  и  в  дальнейшем  везде,  где  это  не  может   вызвать
недоразумений,   знак  $\tilde{\phantom{\varepsilon}}$  у  комплексной
диэлектрической проницаемости $\tilde\varepsilon$ опущен.

     За счёт   введения  комплексной  диэлектрической  проницаемости
$\varepsilon=\varepsilon'+i\varepsilon''$ из уравнений Максвелла (2.6)
исчез  в  явном  виде ток проводимости.  При этом действительная часть
$\varepsilon'$   обусловлена   током   смещения,   а   мнимая    часть
$\varepsilon''=4\pi\sigma/\omega  $  ---  током проводимости.  Обратим
внимание,  что  теперь  даже  для  простейших  материальных  уравнений
(1.13),  (1.15)  диэлектрическая  проницаемость  (точнее  её  мнимая
часть) является функцией $\omega$.

     Зависимость диэлектрической  проницаемости от частоты имеет более
глубокий физический смысл,  чем  результат  перегруппировки  членов  в
уравнении. Как было сказано в предыдущем разделе, при не очень сильных
полях, когда связь между векторами поля остаётся линейной, соотношение
между  векторами  $\mbox{\boldmath $\gv  D$}$ и $\mbox{\boldmath
$\gv E$}$ в произвольном временном процессе имеет   интегральный
операторный   характер.   Напомним,   что   при пренебрежении
пространственной дисперсией самый общий вид соотношения между этими
векторами есть (1.39).  Если применить это  соотношение  к
монохроматической  компоненте  Фурье-разложения (2.2),  то связь между
$\rv D$ и $\rv E$ приобретает вид
     $$\rv D=\varepsilon(\omega)\rv E\,,\eqno(2.9)$$
где, вообще   говоря,   комплексная   функция    $\varepsilon(\omega)$
определяется через ядро $f(\tau)$ интегрального оператора (1.39):
     $$\varepsilon(\omega)=1+\int\limits_0^\infty f(\tau)e^{i\omega
         \tau}\,d\tau.\eqno(2.10)$$
Таким образом, для гармонических полей связь между векторами $\rv D$ и
$\rv   E$   линейная   алгебраическая,  но  соответствующий  множитель
$\varepsilon$ является функцией частоты. О зависимости диэлектрической
проницаемости от частоты говорят как о законе её {\it дисперсии}.

     Функция $\varepsilon(\omega)$ определена  соотношением  (2.10)  и
для отрицательных значений $\omega$, причём
     $$\varepsilon(-\omega)=\varepsilon^*(\omega),\eqno(2.11)$$
где значком   $\phantom{a}^*$   здесь   и   везде  далее  обозначается
комплексное   сопряжение.   Для   действительной   и   мнимой    части
$\varepsilon$ получаем отсюда
     $$\varepsilon'(-\omega)=\varepsilon'(\omega),\qquad\varepsilon''
         (-\omega)=-\varepsilon''(\omega).\eqno(2.12)$$

     Одного представления функции  $\varepsilon(\omega)$ в виде (2.10)
оказывается достаточным,  чтобы выяснить некоторые её  общие
свойства.   Напомним,   что   (2.10)   по  существу  выражает  принцип
причинности:  значение вектора $\mbox{\boldmath $\gv D$}$ в  данный
момент  времени  $t$ может   определяться    значениями   вектора
$\mbox{\boldmath $\gv E$}$   только в предшествовавшие моменты времени.
Не будем здесь  останавливаться  на этом подробнее,  а приведём лишь
окончательный результат, известный как {\it дисперсионные соотношения},
связывающие   между   собой действительную  и мнимую части
$\varepsilon(\omega)$.  Для диэлектрика эти соотношения имеют вид
     $$\varepsilon'(\omega)=1+\frac 1\pi V.p.\!\!\int\limits_{-\infty}
         ^\infty\frac{\varepsilon''(x)}{x-\omega}\,dx,\eqno(2.13)$$
     $$\varepsilon''(\omega)=-\frac 1\pi V.p.\!\!\int\limits_{-\infty}
         ^\infty\frac{\varepsilon'(x)-1}{x-\omega}\,dx,\eqno(2.14)$$
где $V.p.$  означает,  что интеграл следует понимать в смысле главного
значения.  Дисперсионные соотношения показывают,  что  не  может  быть
причинно   обусловленной   модель   вещества,   которая   приводит   к
диэлектрической проницаемости, не зависящей от частоты и не обладающей
мнимой частью,  с которой,  как будет видно дальше, связано поглощение
энергии электромагнитного поля  в  среде.  Но  эти  соотношения    не
препятствуют и  тому,   чтобы   в   достаточно  широкой  полосе  частот
существовали {\it области прозрачности},  где поглощение  пренебрежимо
мало.   При  этом  из  соотношения  (2.13)  следует,  что  в  областях
прозрачности     $$\frac{d\varepsilon}{d\omega}>0\,;\eqno(2.15)$$
в этом же диапазоне частот выполняется и неравенство
     $$\frac{d\varepsilon}{d\omega}>\frac{2(1-\varepsilon)}\omega\,,
         \eqno(2.16)$$
которое при  $\varepsilon<1$ является более сильным,  чем (2.15).  Эти
два  неравенства,  как  будет  показано  чуть   дальше,   обеспечивают
положительность  плотности энергии электромагнитного поля в прозрачной
среде и то,  что распространение возмущений и перенос  энергии  в  ней
происходят со скоростью, меньшей скорости света в пустоте.

     Обратим внимание на то,  что в  диэлектриках  при  $\omega\to  0$
функция  $\varepsilon(\omega)$  стремится,  очевидно,  к  статическому
значению  диэлектрической  постоянной,  а   первый   член   разложения
$\varepsilon''(\omega)$ по частоте пропорционален $\omega$.  При малых
частотах  можно  рассматривать  функцию  $\varepsilon(\omega)$   и   в
металлах,   если   заранее   оговорить,   что   уравнение   $\rot
\mbox{\boldmath $\gv H$}=\displaystyle{\frac 1 c}\displaystyle{\frac
{\partial\mbox{\boldmath $\gv  D$}}{\partial t}}$   при $\omega\to 0 $
переходит   в  уравнение  $\rot  \mbox{\boldmath $ \gv H$}=4\pi\sigma
\mbox{\boldmath $\gv E$}/c$;  для монохроматических полей из  сравнения
этих соотношений следует,  что $\varepsilon(\omega)=4\pi i\sigma/\omega$,
где $\sigma$ --- проводимость по отношению к постоянному  току.  Таким
образом,  если  речь идёт о проводнике,  то в правой части выражения
(2.14) для  $\varepsilon''(\omega)$  должен  появиться  дополнительный
член  $4\pi\sigma/\omega$ (соответствующий полюсу в точке $\omega=0$ в
выражении для $\varepsilon(\omega)$).

     Если в  среде  имеются потери на перемагничивание,  то аналогично
комплексной диэлектрической проницаемости необходимо ввести  магнитную
комплексную  проницаемость $\mu = \mu'+i\mu''$;  её действительная и
мнимая  части  также   удовлетворяют   соответствующим   дисперсионным
уравнениям.

     Рассмотрим для  примера  простейшую  модель   вещества,   которая
правильно  передаёт свойства функции $\varepsilon(\omega)$ при очень
больших значениях $\omega\to\infty$ для всех  веществ  и  используется
наряду  с  этим  для вычисления диэлектрической проницаемости холодной
плазмы. В этой модели все электроны считаются свободными, совершающими
колебания в однородном поле электромагнитной волны.  При очень высоких
частотах такое предположение оправдывается тем,  что скорости движения
электронов в атомах малы по сравнению со скоростью света,  а амплитуды
высокочастотных колебаний много меньше длины волны.  Поэтому уравнение
движения электрона имеет вид:
     $$m_e\frac{d\rv v}{dt}=e\mbox{\boldmath $\gv E$}=e\rv E e^
       {-i\omega t},\eqno(2.17)$$
где $\rv  v$  --- дополнительная скорость,  приобретаемая электроном в
поле волны.  В  результате $\rv v=ie\mbox{\boldmath $\gv E$}/m_e\omega$,
а   смещение электрона под  действием  поля  $\rv  r=-e\mbox{\boldmath
$\gv E$}/m_e\omega^2$.  Вектор поляризации  $\mbox{\boldmath $\gv  P$}$
вещества  есть  дипольный  момент  единицы  его объёма. В  результате
суммирования по всем электронам получаем для комплексных амплитуд:
     $$\rv P=\sum\limits_{i=1}^n e\rv r_i=-\frac{e^2}{m_e\omega^2}
           n\rv E\,,\eqno(2.18)$$
где $n$   ---   число  электронов  в  единице  объёма  вещества.  По
определению  электрической  индукции  $\rv  D=\varepsilon  \rv   E=\rv
E+4\pi\rv P$ и поэтому
     $$\varepsilon(\omega)=1-\frac {\omega_p^2}{\omega^2}\,,
         \eqno(2.19)$$
где $\omega_p=\sqrt{4\pi  ne^2/m_e}$  ---  так  называемая электронная
плазменная частота.  Полученное  выражение  для  $\varepsilon$  даёт
правильное описание поведения вещества в области высоких частот, но не
удовлетворяет дисперсионным уравнениям (что  ясно  уже  из  того,  что
$\varepsilon(\omega)$ --- величина действительная),  а потому не может
быть верным в достаточно широкой  области  частот,  в  частности,  при
малых частотах.

     Отметим ещё, что даже для этой простейшей модели разбиение тока
на  поляризационный  и  ток  проводимости процедура в переменных полях
неоднозначная.  Так,  с равным основанием можно считать все  свободные
электроны  электронами  проводимости и тогда,  полагая множитель между
комплексной  амплитудой  плотности  тока  $\rv  j=\sum   e\rv   v$   и
электрического     поля     $\rv    E$    комплексной    проводимостью
$\sigma(\omega)$, получим для неё следующее выражение:
     $$\sigma(\omega)=i\frac{ne^2}{m_e\omega}\,.\eqno(2.20)$$
Поскольку рассматриваемая  модель  среды  не  обладает  потерями,   то
проводимость оказывается чисто мнимой величиной.

     Если же в эту простейшую модель среды ввести <<трение>> электронов,
обусловленное их    столкновениями    с    неподвижными   ионами   или
кристаллической решеткой твёрдого тела,  и  считать,  что  электроны
осциллируют с некоторой собственной (резонансной) частотой $\omega_0$,
то уравнение движения преобразуется к виду
     $$\frac{d^2\rv r}{dt^2}+\omega_0^2\rv r+\nu\frac{d\rv r}{dt}=
        \frac{e\rv E}{m_e}\, e^{-i\omega t},\eqno(2.21)$$
где $\nu$  ---  эффективная  частота столкновений (константа затухания
осциллятора). В результате

     $$\rv r=\frac{e\rv E}{m_e}\cdot\frac{e^{-i\omega t}}{\omega^2_0-
         \omega^2-i\omega\nu}\eqno(2.22)$$
и диэлектрическая проницаемость становится комплексной:

     $$\varepsilon (\omega)=1+\omega_p^2\cdot\frac{\omega_0^2-\omega
         ^2+i\nu\omega}{(\omega_0^2-\omega^2)^2+\nu^2\omega^2}\,.
         \eqno(2.23)$$
Нетрудно убедиться,    что    полученное    выражение    удовлетворяет
дисперсионным соотношениям (2.13),  (2.14);  очевидно,  что  уравнения
(2.21) описывают систему осцилляторов с затуханием.

     При вычислениях  величин,  в  которые поля и токи входят линейным
образом,  использование комплексных амплитуд не может  привести  ни  к
каким  недоразумениям,  если  при  переходе к истинным величинам взята
действительная часть выражения.  Несколько осторожнее надо подходить к
вычислению   квадратичных  по  полю  величин,  каковыми  являются  все
энергетические характеристики электромагнитного  поля.  Действительно,
пусть  имеются две  монохроматические  действительные величины,
изменяющиеся с одной частотой, представленные в виде
     $$A(t)=\re({\rm A}\,e^{-i\omega t}),\qquad  B(t)=\re({\rm B}\,e^
         {-i\omega t})\eqno(2.24)$$
с комплексными  амплитудами   ${\rm   A}=ae^{i\alpha}$,~   ${\rm   B}=
be^{i\beta}$,  где  $a$,~  $b$,~$\alpha$,~  $\beta$ --- действительные
числа. Тогда их произведение
     $$A(t) B(t)=\frac 12 ab[\cos(\alpha-\beta)+\cos(2\omega t-\alpha
         -\beta)]\eqno(2.25)$$
содержит постоянную часть и часть,  осциллирующую с двойной  частотой,
которая   при  усреднении  по  периоду  обращается  в  нуль.  Усреднив
выражение (2.25), получим
     $$\overline{A(t) B(t)}=\frac12 ab\cos(\alpha-\beta),\eqno(2.26)$$
что иначе можно записать в виде
     $$\overline{A(t) B(t)}=\frac 12 \re(\rm A \rm B^*)=\frac 12 \re
         (\rm A^* \rm B).\eqno(2.27)$$
При $A(t)=B(t)$ из этой  формулы  получается  выражение  для  среднего
квадрата:
     $$\overline{A^2(t)}=\frac 12 {\rm A}{\rm A^*}=\frac 12 |{\rm
         A}|^2=\frac 12a^2.\eqno(2.28)$$

     Отметим, что  в  высокочастотной  электродинамике  в  подавляющем
большинстве случаев интерес представляют  именно  средние  по  времени
квадратичные величины,  которые выражаются через комплексные амплитуды
при помощи  формулы  (2.27),  а  осциллирующие  части  используются  в
основном  для контроля правильности вычислений;  осциллирующая --- или
колеблющаяся --- часть равна  $\re  [{\rm  A\rm  B}\cdot\exp(-2i\omega
t)]/2$.  Подчеркнём ещё раз, что при вычислении нелинейных величин
пользоваться комплексными  амплитудами  нельзя  и  надо  переходить  к
физическим  полям  --- однако средние по периоду значения квадратичных
величин вычисляются по комплексным амплитудам.

     Важнейшей энергетической характеристикой  электромагнитного  поля
является вектор Умова-Пойнтинга, определяемый формулой
     $$\mbox{\boldmath $\gv S$}=\frac c{4\pi}[\mbox{\boldmath $\gv E$}
       \mbox{\boldmath $\gv H$}],\eqno(1.22)$$
который сохраняет  свой  смысл  плотности  потока  энергии  в  единицу
времени  для материальных уравнений самого общего вида и произвольного
изменения   полей   во   времени.   Для    монохроматического    поля,
представленного  в  виде  (2.3),  средняя  по времени плотность потока
энергии в соответствии с формулой (2.27)  определяется  действительной
частью комплексного вектора Умова-Пойнтинга:
     $$\overline {\mbox{\boldmath $\gv S$}}=
       \re \rv S_{\mbox{\it\footnotesize  ком}}=\frac
       c{8\pi}\re[\rv E \rv H^*].\eqno(2.29)$$
Выразим дивергенцию вектора $\rv S_{\mbox{\it\footnotesize ком}}$ с помощью системы  уравнений
(2.8). Для этого произведём в первом уравнении операцию комплексного
сопряжения и умножим результат на $\rv E$, второе уравнение умножим на
$-\rv  H^*$  и  сложим результаты преобразований.  Пользуясь известной
формулой векторного анализа (1.5), имеем:
     $$-\div \rv S_{\mbox{\it\footnotesize ком}}=\frac {i\omega}{8\pi}(\varepsilon^*|\rv E|^2-
         \mu|\rv H|^2)+\frac 12 \rv E\rv j^{e*}.\eqno(2.30)$$
Это уравнение     часто    называют    {\it    комплексной    теоремой
Умова-Пойнтинга} (или {\it теоремой о комплексной мощности}).

     Проинтегрировав (2.30) по некоторому  объёму,  воспользовавшись
теоремой Гаусса (1.10) и взяв действительную часть,  получаем
соотношение
     $$-\oint\overline{\mbox{\boldmath $\gv  S$}}\rv n\,dS=\int\frac
        \omega{8\pi}(\varepsilon''|\rv E|^2+\mu'' |\rv H|^2)\,dV+\int
        \frac 12 \re\rv E\rv j^{e*}\,dV,\eqno(2.31)$$
которое в соответствии с физическим  смыслом  вектора  $\overline{\mbox
{\boldmath $\gv S$}}$  следует  понимать  как  баланс  энергии
электромагнитного поля в среднем  по  периоду  колебания:  энергия,
поступающая  в   некоторый объём (знак -- в левой части связан с тем,
что $\rv n$ -- внешняя нормаль),  расходуется  на  поглощение  в веществе
и на работу поля над сторонними  токами.  Такая  трактовка  соотношения
(2.31)   позволяет назвать величину
     $$Q=\frac\omega{8\pi}(\varepsilon''|\rv E|^2+\mu'' |\rv H|^2)
         \eqno(2.32)$$
средней плотностью потерь в веществе,  а величину $\displaystyle \frac
12\re  \rv  E\rv  j^{e*}$ --- средней плотностью работы электрического
поля над сторонним током.  Таким образом,  потери  энергии  поля  $Q$,
выделяемые  в  веществе  в  виде тепла,  выражаются через мнимые части
комплексных   диэлектрической   и   магнитной   проницаемостей.    Это
накладывает  определённые  ограничения на значения $\varepsilon''$ и
$\mu''$: в любой модели пассивного вещества на всех  частотах   должны
выполняться неравенства
     $$\varepsilon''>0\,,\qquad\mu''>0\,.\eqno(2.33)$$
Из   (2.31)  следует,  что    при  строгом   подходе  макроскопическая
электродинамика не даёт возможности различить, чем обусловлены потери:
электрическим гистерезисом или протеканием тока проводимости в среде.

     Выше (см.  (2.31))  была  подсчитана  реальная  часть дивергенции
вектора $[\rv  E\rv  H^*]$;  получившееся  выражение  иногда  называют
теоремой  об активной мощности в отличие от теоремы о мнимой мощности,
которой соответствует мнимая часть выражения для $\div[\rv E\rv H^*]$.
Если  проделать  аналогичные  выкладки и найти $\div[\rv E\rv H]$,  то
получающиеся при этом соотношения  называют  теоремой  о  колеблющейся
мощности.  Необходимо  подчеркнуть,  что  только  теорема  об активной
мощности,  то есть (2.31),  имеет конкретный физический  смысл,  в  то
время   как   теоремы   о   мнимой   и  колеблющейся  мощностях  носят
вспомогательный характер, хотя иногда и могут оказаться полезными.

     Заметим теперь, что в средний по периоду баланс энергии (2.31) не
входит  сама  величина   энергии   электромагнитного   поля:   это   и
естественно,  поскольку  строго  монохроматическое  поле  через каждый
период возвращается к своему  исходному  значению,  и,  следовательно,
энергия электромагнитного поля в среднем не изменяется. Чтобы получить
выражение для средней плотности энергии  поля,  следует  рассматривать
временной  процесс,  при  котором  амплитуда  монохроматического  поля
медленно изменяется со временем.  Такое поле уже  не  является  строго
монохроматическим, но его разложение в интеграл Фурье содержит заметно
отличные от нуля компоненты  только  в  узкой  области  частот  вокруг
$\omega$.  Поскольку теперь амплитуда поля изменяется со временем,  то
изменяется и его  средняя  энергия.  Внимательный  анализ  показывает,
однако,   что  при  наличии  поглощения  в  среде  получить  выражение
плотности электромагнитной  энергии  через комплексные $\varepsilon$ и
$\mu$ не представляется возможным.  Это удаётся сделать  только  для
конкретных  моделей  вещества.  Однако для области прозрачности среды,
где можно пренебречь мнимыми частями $\varepsilon$ и $\mu$  и  считать
эти  величины  действительными функциями $\omega$,  существует общее
выражение для  средней  энергии  монохроматического  поля.  Перейдём
теперь к его выводу.

     При отсутствии поглощения в среде, а значит, в частности, и токов
проводимости,  а также сторонних токов,  изменение  плотности  энергии
поля  в  среде  в единицу времени вычисляется как $\div\mbox{\boldmath
$\gv S$}$.  В этом случае из уравнений Максвелла (1.12) следует:
     $$-\div\mbox{\boldmath $ \gv S$}=\frac 1{4\pi}\Bigl(\mbox
        {\boldmath $\gv E$}\frac{\partial\mbox{\boldmath $ \gv D$}}
        {\partial t}+\mbox{\boldmath $ \gv H$}\frac{\partial\mbox
        {\boldmath $ \gv B$}}{\partial t}\Bigr).\eqno(2.34)$$
Для вычисления  плотности энергии поля необходимо величину,  стоящую в
правой части уравнения  (2.31),  представить  в  виде  производной  по
времени  от  какой-нибудь  величины.  Последнюю  и можно будет считать
плотностью  энергии  поля.  Напомним,  что  в   отсутствие   дисперсии
$\varepsilon$ и $\mu$ --- действительные числа, вектора $\mbox{\boldmath
$ \gv D$}$ и~$\mbox{\boldmath $ \gv E$}$,~$\mbox{\boldmath $ \gv B$}$ и
~$\mbox{\boldmath $ \gv H$}$ связаны материальными уравнениями (1.13),
(1.14) и, следовательно, в качестве   плотности   энергии   поля  можно
рассматривать величину
     $$U=\frac 1{8\pi}(\varepsilon \mbox{\boldmath $ \gv  E$}^{\,2}+
         \mu\mbox{\boldmath $ \gv  H$}^{\,2}).
         \eqno(2.35)$$
Это выражение перестаёт быть адекватным  энергии  поля  при  наличии
дисперсии  и  даёт  бессмысленный  результат  в области частот,  где
$\varepsilon<0$,  в  частности,  в  случае   плазмы,   диэлектрическая
проницаемость $\varepsilon$ для которой определяется формулой (2.19).

     Чтобы вычислить   среднюю  энергию  монохроматического  поля  при
наличии дисперсии, рассмотрим гармоническое поле с медленно меняющейся
амплитудой
     $$\mbox{\boldmath $ \gv E$}(t)=\re\{\rv E(t)\,e^{-i\omega t}\},
       \eqno (2.36)$$
где комплексную амплитуду $\rv E(t)$ разложим в интеграл Фурье:
     $$\rv E(t)=\frac 1{2\pi}\int \limits _{-\infty}^\infty \rv E
          (\alpha)\,e^{-i\alpha t}\,d\alpha.\eqno (2.37)$$
Запомним, что в этом интеграле функция $\rv E(\alpha)$ отлична от нуля
лишь при  $\alpha\ll\omega$,  что  и  означает  медленность  изменения
функции $\rv E(t)$ за период колебания. Подставим теперь выражение для
поля (2.36) в уравнение (2.34)  и  вычислим  среднее  по  времени  его
правой  части,  а  для  сокращения  записи  рассмотрим  только  первое
слагаемое  в  скобках  ---  все  преобразования  со  вторым  слагаемым
производятся аналогично.

     Индукция $\mbox{\boldmath $ \gv D(t)$}$ и её производная
$  \partial\mbox{\boldmath $ \gv  D$}/\partial t$ связаны с $\mbox
{\boldmath $ \gv E$}(t)$ операторными соотношениями
     $$\mbox{\boldmath $ \gv D$}(t)=\hat\varepsilon \mbox{\boldmath
$ \gv E$}(t),\qquad \frac{\partial\mbox{\boldmath $ \gv D$}}
         {\partial t}=\hat {f} \mbox{\boldmath $\gv E$}(t),\eqno(2.38)$$
где оператор  $\hat  \varepsilon$ определён формулой (1.39), а $\hat
f=\displaystyle{\frac{\partial \hat{\varepsilon}}{\partial t}}$. В случае
строго  монохроматического поля (амплитуда $\rv E$ в (2.36) не зависит
от $t$) действие операторов $\hat\varepsilon$ и $\hat{f}$  на  функцию
$\mbox{\boldmath $ \gv E$}(t)$ сводится к операциям умножения:
     $$\hat{\varepsilon}\mbox{\boldmath $ \gv E$}(t)=\varepsilon(\omega)
       \mbox{\boldmath $ \gv E$}(t),\qquad \hat f \mbox{\boldmath
       $ \gv E$}(t)=f(\omega)\mbox{\boldmath $ \gv E$}(t),\eqno(2.39)$$
причём функция $\varepsilon(\omega)$ определена формулой (2.10), а
     $$f(\omega)=-i\omega\varepsilon(\omega).\eqno(2.40)$$
При действии  оператора  $\hat   f$   на   гармоническое   поле   $\rv
E(\alpha)e^{-i(\omega+\alpha)t}$
 получаем:
     $$\hat f\rv E(\alpha)\,e^{-i(\omega+\alpha)t}=f(\omega+\alpha)
         \rv E(\alpha)\,e^{-i(\omega+\alpha)t}.\eqno (2.41)$$

     Разложим теперь  функцию  $f(\omega+\alpha)$  в  ряд  по   малому
параметру
     $\alpha$ $$f(\omega+\alpha)=f(\omega)+\alpha\frac{d  f(\omega)}{d
     \omega}+\dots\eqno (2.42)$$
и с учётом  этого  проинтегрируем  (2.41)  по  $\alpha$,  тем  самым
суммируя все гармонические составляющие; тогда
     $$\hat f\mbox{\boldmath $ \gv E$}(t)=f(\omega)\mbox{\boldmath
 $\gv E$}(t)+i\frac{df(\omega)}{d\omega}
         \frac{\partial\rv E}{\partial t}\,e^{-i\omega t}\eqno (2.43)$$
или
     $$\frac{\partial\mbox{\boldmath $ \gv D$}}{\partial t}=
       -i\omega\varepsilon(\omega) \mbox{\boldmath $ \gv E$}(t)+
         \frac{d(\omega\varepsilon)}{d\omega}\frac {\partial\rv E}
         {\partial t}\,e^{-i\omega t}.\eqno (2.44)$$

     Наконец, вычислим с помощью (2.27) среднее значение  произведения
$\mbox{\boldmath $ \gv  E$}(t)\displaystyle{\frac{\partial  \mbox
{\boldmath $ \gv D$}}{\partial t}}$ за период
колебания,  считая,  что  за  этот  промежуток   времени   комплексные
амплитуды обоих сомножителей не успевают измениться; поэтому
     $$\overline{\mbox{\boldmath $ \gv E$}(t)\frac{\partial \mbox
       {\boldmath $ \gv D$}}{\partial t}}=\frac 12 \re\{\rv E[i\omega
        \varepsilon^*(\omega)\rv E^*+\frac{d(\omega\varepsilon^*)}
        {d\omega}\frac{\partial\rv E^*}{\partial t}]\}.\eqno(2.45)$$
Отсюда, с  учётом  действительности  функции  $\varepsilon(\omega)$,
получим  для  среднего  значения  величины,  стоящей  в  правой  части
уравнения (2.34), выражение
     $$ \frac1{4\pi}\overline{\Bigl(\mbox{\boldmath $ \gv E$}\frac
       {\partial\mbox{\boldmath $ \gv D$}}{\partial t}+\mbox
       {\boldmath $ \gv H$}\frac{\partial\mbox{\boldmath $ \gv B$}}
       {\partial t}\Bigr)}=  \frac{d\overline{U}}{dt},\eqno (2.46)$$
где
     $$\overline{U}=\frac 1{16\pi}\left\{\frac{d(\omega\varepsilon)}{d
         \omega}|\rv E|^2+\frac{d(\omega\mu)}{d\omega}|\rv H|^2\right
         \}\eqno(2.47)$$
и есть искомое выражение для средней энергии монохроматического поля в
области  прозрачности  диспергирующей  среды.  Условие положительности
средней энергии (2.47) накладывает ограничения на допустимые  значения
$\varepsilon$   и   $\mu$   ---   эти  величины  должны  удовлетворять
неравенствам
     $$\frac{d(\omega\varepsilon)}{d\omega}>0\,,\qquad\frac{d(\omega
         \mu)}{d\omega}>0\,.\eqno(2.48)$$
Нетрудно видеть,  что  они  всегда  выполняются  как  следствие  более
сильных неравенств (2.15) и (2.16).

     Вычислим в  заключение  среднюю энергию монохроматического поля в
плазме,  диэлектрическая проницаемость которой  определяется  формулой
(2.19):
     $$\overline{U}=\frac 1{16\pi}\left(\varepsilon+\omega\frac{d
         \varepsilon}{d\omega}\right)|\rv E|^2=\frac 1{16\pi}\left(1+
         \frac {\omega_p^2}{\omega^2}\right)|\rv E|^2.\eqno (2.49)$$
Полученное выражение  позволяет  рассматривать  среднюю энергию поля в
среде как сумму энергии поля  в  пустом  пространстве  и  кинетической
энергии  движения  зарядов,  обусловленного полем.  Если же пренебречь
дисперсией и воспользоваться для  энергии  поля  формулой  (2.35),  то
получается, что при $\omega<\omega_p$ энергия поля отрицательна.

%\end{document}

\newpage
\oddsidemargin=-0.4mm \evensidemargin=-0.4mm
\topmargin=-0.4mm
\headsep=7mm
\textheight=231.875mm
\textwidth=160mm
\unitlength=1mm
\mathsurround=2.5pt
%\begin{document}
%\input{macr.tex}
\thispagestyle{empty}
%\addtocounter{page}{30}

\begin{center}
\subsubsection*{3. Волновое уравнение. Потенциалы электромагнитного
     поля}\end{center} \vspace*{0.5cm}

\markboth{Глава 1. Основные положения классической электродинамики}
{3. Волновое уравнение. Потенциалы электромагнитного поля}

\begin{center}\begin{minipage}[c]{0.75\textwidth}
\footnotesize{\parindent=0.5cm
         Волновое уравнение  в пространственно-временном представлении
         и    волновое    уравнение    для    комплексных     амплитуд
         монохроматического  поля.  Скалярный  и векторный потенциалы.
         Калибровочные   преобразования.   Волновые   уравнения    для
         потенциалов;   потенциалы   Герца.   Волновое  уравнение  для
         скалярной   функции   (уравнение    Гельмгольца).    Типичные
         постановки  задач  в  электродинамике  СВЧ.  Проблема  выбора
         единственного решения.
}\end{minipage}\end{center}
\vspace*{0.5cm}

     Уравнения Максвелла  (1.12)  содержат  в   качестве   неизвестных
шестнадцать  скалярных  функций (каждая векторная функция определяется
тремя  скалярными  составляющими   вектора).   В   случае   простейших
материальных уравнений (1.13)--(1.15)
в системе  уравнений (1.12)   остаются   шесть  неизвестных  функций,
в  качестве  которых  традиционно  используются составляющие векторов
$\mbox{\boldmath $\gv E$}$ и $\mbox{\boldmath $\gv H$}$.  Эти
составляющие  должны  быть найдены из системы двух векторных уравнений
(то есть шести скалярных уравнений) в частных производных  первого
порядка
     $$\left.\begin{array}{rcl}\rot\mbox{\boldmath $\gv H$}&\!=\!&
       \phantom{-}\displaystyle{\frac{\varepsilon} c \frac{\partial
       \mbox{\boldmath $\gv E$}}{\partial t}+\frac{4\pi\sigma} c
       \mbox{\boldmath $\gv E$}+\frac{4\pi} c\mbox{\boldmath $\gv j$}^
         {\,e}},\\[0.25cm]\rot\mbox{\boldmath $\gv E$}&\!=\!&-
       \displaystyle{\frac{\mu} c \frac{\partial\mbox{\boldmath $\gv H$}
       }{\partial t}}\\[0.25cm]\end{array}\right\}\eqno(3.1)\vspace{1cm}$$
с соответствующими граничными и (или)  начальными  условиями.  Отметим
здесь,   что  для  переменных  во времени полей (а только такие поля в
дальнейшем и рассматриваются)   уравнение   $ \div \mbox{\boldmath
$\gv B$}=0 $     ввиду тождества $\div\rot \rv A \equiv 0  $  является
следствием  уравнения
(1.12a)  и  может  быть  опущено.  Уравнение  (1.12в)  в  этом  случае
фактически  определяет  плотность  заряда  $\rho$  ---  величину,   не
представляющую  интереса  в  большинстве  задач  электродинамики  СВЧ.
Впрочем,  она может быть найдена по  известному  $\mbox{\boldmath
$\gv  E$}$  с  учётом (1.15) и из уравнения непрерывности.

      При решении конкретных задач почти  всегда  удобнее  работать  с
возможно  меньшим  числом неизвестных.  В однородной среде,  в которой
$\varepsilon$,  $\mu$ и $\sigma$ не зависят от координат,  из  системы
(3.1)  нетрудно  исключить  один  из  векторов $\mbox{\boldmath
$\gv E$}$ или $\mbox{\boldmath $\gv H$}$.  В основе такой процедуры
лежит известная формула векторного анализа:
     $$\rot\rot \rv A=\grad\div \rv A-\Delta \rv A.\eqno(3.2)$$
 В декартовой  системе  координат каждая составляющая  вектора $\Delta\rv A$
определяется  соответствующей составляющей вектора $\rv A$ по формулам
$  (\Delta  A)_{\alpha}  =  \Delta  (A_{\alpha}),\,\alpha= x,y,z$, где   $\Delta=
\partial^2/\partial   x^2+\partial^2/\partial  y^2+\partial^2/\partial
z^2$ --- оператор Лапласа. Составляющие вектора $\Delta\rv A$ в других
системах координат приведены в Приложении П-4.

      Для исключения вектора $\mbox{\boldmath $\gv H$}$ из системы
(3.1) вычислим  $\rot$ от  обеих  частей  второго  уравнения  и  с
помощью (3.2) в результате несложных преобразований получим  уравнение
для  вектора  $\mbox{\boldmath $\gv  E$}$, которое  при отсутствии
в рассматриваемой области сторонних токов ($\mbox{\boldmath $\gv j$}^
{\,e}=0$) имеет вид
     $$\Delta \mbox{\boldmath $\gv E$}-\frac{\varepsilon\mu}{c^2}
       \frac{\partial^2\mbox{\boldmath $\gv E$}}{\partial t^2}-
       \frac{4\pi\sigma\mu}{c^2}\frac{\partial\mbox{\boldmath $\gv E$}}
         {\partial t}=0\,.\eqno(3.3)$$
При наличии сторонних токов уравнение становится  неоднородным  ---  в
правую  часть входят пространственные производные $\mbox{\boldmath
$\gv j$}^e(\rv r)$.  В таком виде уравнение обычно не используется,
поскольку в этом  случае намного удобнее соответствующее уравнение для
потенциалов,  включающее в себя сами токи.  Вектор $\mbox{\boldmath
$\gv H$}$ может быть исключен из системы (3.1) и  в  том  случае, когда
среда  неоднородна,  однако  получающееся в результате уравнение для
$\mbox{\boldmath $\gv E$}$ используется очень редко.

     Если же  из системы уравнений (3.1) аналогичным образом исключить
$\mbox{\boldmath $\gv E$}$, то для $\mbox{\boldmath $\gv H$}$ получим
точно такое же уравнение, как и (3.3):
     $$\Delta \mbox{\boldmath $\gv H$}-\frac{\varepsilon\mu}{c^2}
       \frac{\partial^2\mbox{\boldmath $\gv H$}}{\partial t^2}-
       \frac{4\pi\sigma\mu}{c^2}\frac{\partial\mbox{\boldmath $\gv H$}}
         {\partial t}=0\,.\eqno(3.4)$$
Отметим, что  уменьшение  числа  неизвестных   достигается   за   счет
повышения порядка уравнения,  а также то, что нет необходимости решать
оба уравнения (3.3) и (3.4).  Достаточно решить одно из  них  ---  то,
которое  удобнее  в  данной конкретной задаче,  а составляющие другого
вектора искать непосредственно из уравнений Максвелла.

     Для прозрачной однородной среды, когда потерями можно пренебречь,
и,  следовательно,  ток  проводимости  отсутствует   ($\sigma=0$),   а
$\varepsilon$,   ~$\mu$   ---   действительные   постоянные  величины,
уравнения (3.3) и (3.4) упрощаются:
     $$\Delta \mbox{\boldmath $\gv E$}-\frac{\varepsilon\mu}{c^2}
        \frac{\partial^2\mbox{\boldmath $\gv E$}}
         {\partial  t^2}=0\eqno(3.5)$$
и
     $$\Delta \mbox{\boldmath $\gv H$}-\frac{\varepsilon\mu}{c^2}
       \frac{\partial^2\mbox{\boldmath $\gv H$}}
         {\partial  t^2}=0\,.\eqno(3.6)$$
Уравнения (3.5) и (3.6) являются уравнениями гиперболического  типа  и
их принято называть {\it волновыми уравнениями}.

     В декартовых координатах  однородные  векторные  уравнения  (3.3)
--(3.6)  в  силу  свойства вектора $\Delta\rv A$ распадаются каждое на
три скалярных волновых уравнения,  которые, вообще говоря, не являются
независимыми,  поскольку  могут  быть  связаны  между собой требуемыми
граничными условиями. Скалярное волновое уравнение
     $$\Delta f-\frac 1{v^2}\frac{\partial^2 f}{\partial  t^2}=0
         \eqno(3.7)$$
встречается во многих разделах математической физики и хорошо
изучено. В частности,  показано,  что входящая в (3.7) постоянная
$\,v\,$  есть скорость распространения начального возмущения.
Поскольку эта скорость в электродинамике всегда должна быть меньше
скорости света  в  пустоте $\,c\,$,  то  уравнения  (3.5),  (3.6)
имеют смысл только для значений $\varepsilon$ и $\mu$,
удовлетворяющих условию $ \varepsilon  \mu>1$. При этом всегда
следует помнить,  что эти уравнения получены для среды без
дисперсии,  и потому   необходимо быть осторожными  при
использовании  общих решений  уравнения  (3.7). Многие из них
получены без использования разложения поля на  гармонические
составляющие. Например,  в  случае зависимости  только  от  одной
пространственной координаты $x$ решение скалярного волнового
уравнения
     $$\frac{\partial^2 f}{\partial x^2}-\frac 1{v^2}\frac{\partial
         ^2 f}{\partial t^2}=0\eqno(3.7\mbox{\textit а})$$
есть
     $$f(x,t)=f_1(x-vt)+f_2(x+vt),\eqno(3.7\mbox{\textit б})$$
где $f_1$  и  $f_2$  ---  произвольные  функции своего аргумента.  Для
реальной  сплошной  среды такое решение может не иметь смысла,  если в
спектральное  разложение  $f(x,t)$   входят   частоты,   для   которых
дисперсией свойств среды пренебречь нельзя.

     В электродинамике  СВЧ  практически всегда имеют дело с волновыми
уравнениями для комплексных амплитуд монохроматического поля,  которые
выводятся  либо  из  уравнений  (3.5)--(3.6) путём формальной замены
$\partial  /\partial  t   \to   -i\omega$   и   введения   комплексной
диэлектрической  проницаемости,  либо  исключением  одного из векторов
поля из уравнений Максвелла  (2.8).  При  отсутствии  сторонних  токов
однородные волновые уравнения приобретают вид:
     $$\Delta\rv E+k^2\varepsilon\mu\rv E=0\,,\eqno(3.8)$$
     $$\Delta\rv H+k^2\varepsilon\mu\rv H=0\,.\eqno(3.9)$$
Эти уравнения   имеют  значительно  более  широкую
область   применения,  поскольку  комплексные проницаемости
 $\varepsilon$  и  $\mu$  позволяют  помимо дисперсии учесть ещё
 и потери в среде,  в  частности,  обусловленные  током проводимости.
 При этом они заметно проще,  чем уравнения (3.3), (3.4).

     В случае  использования  волновых  уравнений,  будь  то уравнения
(3.5),  (3.6)  или  (3.8),  (3.9),  следует  помнить,   что   они   не
эквивалентны  уравнениям  Максвелла  (в  данном  случае  без сторонних
токов).  Это следует хотя бы из того, что уравнения (3.8), (3.9) имеют
решение  $\rv  H\equiv  0$,  $\rv  E\ne  0$,  которое не удовлетворяет
уравнениям (2.8) при $\rv j^e=0$. При этом любое гармоническое решение
однородных   уравнений  Максвелла  удовлетворяет  волновым  уравнениям
(3.8), (3.9).

     При наличии  сторонних  токов  более  удобным способом уменьшения
числа  неизвестных  в  системе  уравнений  Максвелла  (1.12)  является
введение потенциалов поля.  {\it Векторный потенциал} $\mbox{\boldmath
$\gv A$}$ вводится как некоторая векторная функция, которая связана с
магнитной индукцией $\mbox{\boldmath $\gv B$}$ соотношением
     $$\mbox{\boldmath $\gv B$}=\rot \mbox{\boldmath $\gv A$}.
       \eqno(3.10)$$
Согласно векторному  тождеству $\div\rot\rv A\equiv 0$ уравнение
$\div \mbox{\boldmath $\gv B$}=0$ выполняется при любом выборе
функции $\mbox{\boldmath $\gv A$}$.  Подстановка (3.10) в
(1.12\textit а) приводит к соотношению
     $$\rot {\left (\mbox{\boldmath $\gv E$}+\frac 1c \frac{\partial
       \mbox{\boldmath $\gv A$}}{\partial t}\right)}=0\,,\eqno(3.11)$$
что на  основании   векторного   тождества   $\rot\grad\Phi\equiv   0$
позволяет выразить электрическое поле в виде
     $$\mbox{\boldmath $\gv E$}= -\grad\Phi-\frac 1c \frac{\partial
       \mbox{\boldmath $\gv A$}}{\partial t}\,,\eqno(3.12)$$
где $\Phi$  ---  {\it  скалярный  потенциал}.  Отметим,  что  введение
потенциалов   посредством  соотношений  (3.10)  и  (3.12)  приводит  к
автоматическому удовлетворению тем двум уравнениям Максвелла,  которые
получаются  непосредственной  заменой  полей  $\mbox{\boldmath $\gv e$}$
и  $\mbox{\boldmath $\gv h$}$  в микроскопических уравнениях
Лоренца-Максвелла   их    усреднёнными значениями.  Поэтому  такое
определение потенциалов никак не связано с видом материальных уравнений
и не зависит от токов и зарядов.

     Введение потенциалов сразу уменьшает число неизвестных функций  с
шести (составляющие   векторов   $\mbox{\boldmath $\gv E$}$  и
$\mbox{\boldmath $\gv  B$}$)  до  четырёх (составляющие вектора
$\mbox{\boldmath $\gv  A$}$  и  скаляр  $\Phi$).  Естественно, это
достигается   за   счет   повышения  порядка  уравнений,  которым
эти потенциалы должны удовлетворять,  в  сравнении  с  порядком
уравнений Максвелла.   При   отсутствии   токов   проводимости  и
использовании материальных  уравнений  (1.13),  (1.14)  уравнения
для   потенциалов получаются  путем  подстановки  (3.10)  и (3.12)
в (1.12\textit б) и (1.12\textit в) в виде
     $$\left.\begin{array}{rcl}\displaystyle{\Delta\mbox{\boldmath
       $\gv A$}-\frac{\varepsilon\mu}{c^2}\frac{\partial^2
        \mbox{\boldmath $\gv A$}}{\partial t^2}-\grad\Bigl(\div
        \mbox{\boldmath $\gv A$}+\frac{\varepsilon\mu} c\frac
        {\partial\Phi}{\partial t}\Bigr)}&\!=\!&\displaystyle
        {-\frac{4\pi} c\mu\mbox{\boldmath $\gv j$}^{\,e},}\\[0.25cm]
        \displaystyle{\Delta\Phi+\frac 1 c\frac{\partial}{\partial t}
        (\div\mbox{\boldmath $\gv A$})}&\!=\!&\displaystyle{-\frac{4\pi
         \rho^e}{\varepsilon}}\,.\\[0.25cm]\end{array}\right\}
         \eqno(3.13)$$

     Необходимо сказать,  что соотношениями (3.10) и (3.12) потенциалы
$\mbox{\boldmath $\gv A$}$ и $\Phi$ определены неоднозначно. Они допускают
так называемые {\it калибровочные преобразования}
     $$\left.\begin{array}{rcl}\displaystyle{\mbox{\boldmath $\gv A$}}
       &\!\longrightarrow \!&\displaystyle{\mbox{\boldmath $\gv A$}+
        \grad \Lambda\,,}\\[0.2cm]\displaystyle {\Phi}&\!\longrightarrow
        \!&\displaystyle{\Phi-\frac 1 c \frac{\partial\Lambda}
        {\partial t}\,,}\\[0.2cm]\end{array}\right\}\eqno(3.14)$$
где $\Lambda$  ---  произвольная  функция  координат  и  времени,  при
которой электрическое поле  $\mbox{\boldmath $\gv E$}$  и  магнитная
индукция  $\mbox{\boldmath $\gv  B$}$ (реальные  физические  величины,
с которыми оперирует электродинамика сплошных сред) остаются неизменными.
Это  позволяет  выбрать  функцию $\Lambda$ таким образом, чтобы
выполнялось {\it условие Лоренца}
     $$\div\mbox{\boldmath $\gv A$}+\frac{\varepsilon\mu} c\frac
       {\partial\Phi}{\partial t} =0\,.\eqno(3.15)$$
В результате система (3.13) сводится к двум {\it неоднородным волновым
уравнениям}  для  потенциалов,  схожим  по  форме  записи и содержащим
каждое лишь один из потенциалов:
     $$\left.\begin{array}{rcl}\displaystyle{\Delta\mbox{\boldmath
       $\gv A$}-\frac{\varepsilon\mu}{c^2}\frac{\partial^2
        \mbox{\boldmath $\gv A$}}{\partial t^2}}&\!=\!&\displaystyle
        {-\frac{4\pi} c\mu\mbox{\boldmath $\gv j$}^{\,e},}\\[0.35cm]
         \displaystyle{\Delta\Phi-\frac{\varepsilon\mu}{c^2}
         \frac{\partial^2 \Phi}{\partial t^2}}
          &\!=\!&\displaystyle{-\frac{4\pi\rho^e}{
         \varepsilon}}\,.\\[0.25cm]\end{array}\right\}\eqno(3.16)$$

     Потенциалы, удовлетворяющие условию  Лоренца  (3.15),  все  ещё
содержат   некоторую   неопределённость:  группа  {\it  ограниченных
калибровочных   преобразований}   (3.14)   с    функцией    $\Lambda$,
удовлетворяющей уравнению
     $$\Delta\Lambda-\frac {\varepsilon\mu}{c^2}\frac{\partial^2
         \Lambda}{\partial t^2}=0\,,\eqno(3.17)$$
оставляет условие   Лоренца   (3.15)   без   изменений.   Лоренцовская
калибровка  является  наиболее  употребительной в теории поля, имеющей
дело с уравнениями Максвелла в пустоте ($\varepsilon=1$,~$\mu=1$),
из-за своей релятивистской инвариантности, поскольку в этом случае
потенциалы $\mbox{\boldmath$\gv A$}$ и $\Phi$ образуют 4-х вектор.
В сплошной среде лоренцовская калибровка привлекательна тем, что при
её использовании волновые уравнения (3.16) для потенциалов $\Phi$ и
$\mbox{\boldmath $\gv A$}$  (во  всяком  случае  в  декартовой  системе
координат) одинаковы, хотя она уже не обладает релятивистской
инвариантностью при переходе к  другой инерциальной системе  координат.
%Далее практически везде используется именно калибровка (3.15).

     Помимо лоренцовской  применение  находит  и  так  называемая {\it
кулоновская калибровка}, которая вводится условием $\div\mbox{\boldmath
$\gv A$}=0$. Её особенно  удобно  применять  в  отсутствие  источников
поля --- в этом случае $\Phi=0$, векторный потенциал $\mbox{\boldmath
$\gv A$}$ удовлетворяет однородному волновому уравнению, а поля
определяются формулами
     $$\mbox{\boldmath $\gv E$}=-\displaystyle{\frac 1 c\frac
       {\partial\mbox{\boldmath $\gv A$}}{\partial t}
         }\,,\qquad\mbox{\boldmath $\gv B$}=\rot\mbox{\boldmath $\gv A$}\,.
        \eqno(3.18)$$
Кулоновскую калибровку называют ещё  {\it  поперечной}  калибровкой.
Происхождение  термина  обусловлено  тем,  что всякое
векторное поле $\rv A(\rv r)$ может быть  однозначно  разбито  на  две
части  ---  продольную  $\rv A_l$ и поперечную $\rv A_t$ --- таким образом,
что $\rot \rv A_l=0$,~$\div\rv A_t=0$.  При этом оказывается, что в правую
часть   соответствующего   неоднородного   волнового   уравнения   для
векторного  потенциала  входит  только  поперечная   часть   плотности
стороннего тока $\mbox{\boldmath $\gv j$}^e_t$. Разбиение векторного
поля на продольную и поперечную части ещё встретится нам далее  при
рассмотрении  теории возбуждения резонаторов.

     В электродинамике СВЧ широкое применение находит ещё  один  вид
потенциала --- {\it вектор Герца} $\mbox{\boldmath ${\cal P}$}$,  который
связан с векторным и скалярным потенциалами следующими соотношениями:
     $$\mbox{\boldmath $\gv A$}=\displaystyle{\frac {\varepsilon\mu}{c}
       \frac{\partial{\mbox{\boldmath ${\cal P}$}}}{\partial t}};
       \qquad\Phi=-\div{\mbox{\boldmath ${\cal P}$}}.\eqno(3.19)$$
Нетрудно видеть,   что  при  этом  автоматически  выполняется  условие
Лоренца (3.15),  а в случае отсутствия сторонних токов вектор
$\mbox{\boldmath ${\cal P}$}$ удовлетворяет  тому  же  однородному
волновому уравнению (3.5),  что и поля $\mbox{\boldmath $\gv E$}$ и
$\mbox{\boldmath $\gv H$}$.

     При гармонической  зависимости  от  времени комплексные амплитуды
полей и потенциалов связаны соотношениями
     $$\rv B=\rot\rv A,\qquad\rv E=-\grad\Phi+ik\rv A\,,\eqno(3.20)$$
где $\rv A$ --  комплексная  амплитуда  векторного  потенциала;  для
комплексной амплитуды скалярного потенциала использовано то же
обозначение $\Phi$, что  и  для  самого  потенциала.  Это  не  может
в дальнейшем вызвать недоразумений,  поскольку при калибровке  Лоренца
(3.15)  комплексные амплитуды  скалярного  и  векторного потенциалов
связаны алгебраическим соотношением
     $$\Phi=-\frac {i}{k\varepsilon\mu}\div \rv A\eqno(3.21)$$
и амплитуда $\Phi$  исключается  из  всех  уравнений.

     В результате подстановки соотношений (3.20) в уравнения Максвелла
(2.8) с учётом тождества (3.2) получаем  уравнение  для  $\rv  A$  в
однородной изотропной среде:
     $$\Delta \rv A+k^2\varepsilon\mu\rv A=-\frac{4\pi}c\mu\rv j^e\,.
         \eqno(3.22)$$
Комплексные амплитуды поля $\rv E$ и $\rv H$ в такой среде  выражаются
через $\rv A$ формулами
     $$\rv H=\frac1\mu\rot\rv A\,,\qquad \rv E=ik\rv A+\frac i{k
         \varepsilon\mu}\grad\div \rv A .\eqno(3.23)$$

     Поскольку комплексная амплитуда вектора Герца $\rv \Pi$ связана с
$\rv A$ соотношением
     $$\rv A=-ik\varepsilon\mu\rv \Pi\,,\eqno(3.24)$$
то из (3.22) сразу следует уравнение для $\rv \Pi$:
     $$\Delta \rv \Pi+k^2\varepsilon\mu\rv \Pi=-\frac{4\pi i}
         {\omega\varepsilon}\rv j^e\,,\eqno(3.25)$$
причём поля выражаются через вектор $\rv\Pi$ формулами
     $$\rv H=-ik\varepsilon\rot \rv \Pi,\quad\rv E=k^2\varepsilon\mu
         \rv \Pi+\grad\div\rv \Pi\,.\eqno(3.26)$$
При отсутствии  источников  поля  вектор Герца $\rv \Pi$ удовлетворяет
тому же однородному волновому уравнению (3.8),  которому удовлетворяют
в этом случае векторы поля $\rv E$,~$\rv H$ и векторный потенциал $\rv
A$.

     Введённые до  сих  пор  потенциалы $\rv A$ и $\rv \Pi$ с полным
основанием могут быть названы  {\it  электрическими},  поскольку  и  в
уравнения Максвелла (2.8),  и в неоднородные волновые уравнения (3.23)
и (3.25) входит сторонний электрический ток $\rv j^e$.

     Введём теперь  в  уравнения  (2.8) сторонний магнитный ток $\rv
j^m$ (пока формально для симметризации уравнений),  в результате  чего
они примут вид:
     $$\left.\begin{array}{rcl}\rot\rv H&\!=\!&-ik\varepsilon\rv E+
         \displaystyle{\frac{4\pi} c}\rv j^e,\\[0.25cm]\rot\rv E&\!=
         \!&\phantom{-}ik\mu\rv H-\displaystyle{\frac{4\pi} c}\rv j^m.
         \\[0.25cm]\end{array}\right\}\eqno(3.27)$$
Решение этой системы уравнений всегда можно представить в виде
     $$\rv E=\rv E^e+\rv E^m,\qquad\rv H=\rv H^e+\rv H^m;\eqno(3.28)$$
при этом каждое слагаемое в (3.28)  вследствие  принципа  суперпозиции
удовлетворяет своей системе уравнений, а именно:
     $$\left.\begin{array}{rcl}\rot\rv H^e&\!=\!&-ik\varepsilon\rv E^e
         +\displaystyle{\frac{4\pi} c}\rv j^e,\\[0.25cm]\rot\rv E^e&\!
         =\!&\phantom{-}ik\mu\rv H^e;\end{array}\right\}
         \eqno(3.29)$$
     $$\quad\;\,\left.\begin{array}{rcl}\rot\rv H^m&\!\!=\!\!&-ik
         \varepsilon\rv E^m,\\\rot\rv E^m&\!\!=\!\!&\phantom{-}ik\mu
         \rv H^m-\displaystyle{\frac{4\pi} c}\rv j^m.\\[0.2cm]\end
         {array}\right\}\eqno(3.30)$$
Отметим, что  системы уравнений (3.29) и (3.30) переходят друг в друга
при замене величин
     $$\rv E^m\longleftrightarrow-\rv H^e;\quad\rv H^m
         \longleftrightarrow\rv E^e;\quad\varepsilon
         \longleftrightarrow\mu;\quad\mu\longleftrightarrow
         \varepsilon;\quad\rv j^m\longleftrightarrow\rv j^e,
         \eqno(3.31)$$
а введённые выше  потенциалы  $\rv  A$  и  $\rv  \Pi$  соответствуют
системе  уравнений  (3.29);  будем  в  дальнейшем  там,  где это может
вызвать  недоразумения,  помечать  их  верхним   индексом   ---   $\rv
A^e$,~$\rv\Pi^e$.

     Для системы уравнений (3.30) с помощью замены (3.31) по  аналогии
введём   {\it  магнитный}  векторный  потенциал  $\rv  A^m$  и  {\it
магнитный} вектор Герца $\rv \Pi^m$.  Тогда  вектор  $\rv  A^m$  будет
удовлетворять уравнению
     $$\Delta \rv A^m+k^2\varepsilon\mu\rv A^m=-\frac{4\pi}c
         \varepsilon\rv j^m,\eqno(3.32)$$
а поля выражаться формулами
     $$\rv E^m=-\frac1\varepsilon\rot\rv A^m,\qquad \rv H^m=ik\rv A^m
         +\frac i{k\varepsilon\mu}\grad\div \rv A^m\eqno(3.33)$$
и
     $$\rv E^m=ik\mu\rot\rv\Pi^m,\quad\rv H^m=
         k^2\varepsilon\mu\rv \Pi^m+\grad\div\rv \Pi^m.\eqno(3.34)$$
Таким образом,   решение   системы   уравнений   (3.27)   может   быть
представлено  в виде суммы полей (3.23) и (3.33).  Через векторы Герца
$\rv\Pi^m$ и $\rv\Pi^e$  электрическое  и  магнитное  поля  выражаются
аналогичным образом:
     $$\left.\begin{array}{rcl}\rv E&\!=\!&(\grad\div + k^2\varepsilon
         \mu)\rv\Pi^e+ik\mu\rot\rv\Pi^m,\\[0.25cm]\rv H&\!=\!&(\grad
         \div + k^2\varepsilon\mu)\rv\Pi^m-ik\varepsilon\rot\rv \Pi^e.
         \end{array}\right\}\eqno(3.35)$$
При отсутствии  в рассматриваемой области пространства сторонних токов
потенциалы $\rv A^e$ и  $\rv  A^m$  удовлетворяют  одному  и  тому  же
однородному  волновому уравнению.  Поэтому для любого частного решения
этого уравнения поля,  вычисленные как по формулам (3.23),  так  и  по
формулам (3.33), будут удовлетворять однородным уравнениям Максвелла.

     Введение двух    векторных    потенциалов    является    заведомо
“избыточным”  и  на  первый  взгляд  противоречит  исходной  цели
--- уменьшить число переменных в уравнениях.  Вообще говоря,  это
не  так, поскольку  в  большом  числе  задач отличными от нуля
оказываются лишь некоторые из составляющих, а часто --- и лишь
одна из них, какого-нибудь одного    из    потенциалов.   Выбор
нужного   потенциала   зачастую подсказывается видом  источника
поля,  лежащего  вне  рассматриваемой области, а также видом
граничных условий.

     Заканчивая описание всего многообразия  используемых  потенциалов,
отметим,  что в электродинамике сплошной среды введение потенциалов не
является  столь  принципиальным  моментом,  как  в  теории  поля,  где
векторный  и  скалярный  потенциалы  образуют в совокупности 4-вектор.
Роль потенциалов особенно возрастает в квантовой электродинамике,  где
они  приобретают  вполне  конкретный  физический  смысл.  В  случае же
сплошной среды потенциалы являются вспомогательным средством и  многие
задачи  решаются  и  без них.  Подчеркнём,  однако,  что при наличии
источников неоднородные волновые уравнения для  векторного  потенциала
(3.22)  и  вектора  Герца (3.25) обладают определённым преимуществом
перед другими,  поскольку в них непосредственно входит  сторонний  ток
$\rv j^e (\rv r)$, а не его производные по координатам.

     В заключение   рассмотрения    общих    положений    классической
электродинамики  сплошных  сред  остановимся  на  проблеме  корректной
постановки   электродинамической   задачи   и   выборе    её    {\it
единственного}  решения.  Учитывая,  что  для  электродинамики  СВЧ по
изложенным выше причинам основной интерес представляют  установившиеся
монохроматические  процессы,  в  которых  все  поля  и  токи  являются
гармоническими  функциями  времени,  а  характеризующая   их   частота
$\omega$  ---  действительная  величина,  подробнее  рассмотрим  здесь
именно этот случай.

     Самая общая постановка задачи включает в себя задание в некоторой
ограниченной области пространства сторонних токов и свойств  среды  во
всём неограниченном пространстве. Свойства среды (будем считать её
изотропной) определяются двумя  комплексными  скалярными  непрерывными
функциями   координат   $\varepsilon(\rv   r)$   и   $\mu(\rv  r)$,  а
электромагнитные поля везде удовлетворяют уравнениям Максвелла  (2.8).
В   такой   постановке   задача  не  может  быть  решена  в  замкнутом
аналитическом виде, однако она имеет единственное решение при условии,
что  параметры  среды  удовлетворяют  условиям  причинности  (в данном
случае обе проницаемости должны иметь  в  каждой  точке  положительную
мнимую  часть  или  точно  равняться единице) и что поля нигде,  в том
числе и на бесконечности,  не имеют особенностей, за исключением, быть
может, точек, где расположены сторонние токи.

     Основной практический  интерес  представляют  задачи,  в  которых
проницаемости   $\varepsilon(\rv   r)$   и   $\mu(\rv   r)$   являются
кусочно-постоянными  функциями  координат,  в  результате  чего  всё
пространство  может  быть  разделено  на  несколько  областей  с  {\it
однородными} средами (в частности,  область может быть и  одна,  когда
всё    пространство    представляет    собой    однородную   среду).
Скачкообразное изменение свойств среды несомненно является  упрощающей
идеализацией  реальности,  но  ни к какой некорректности или трудности
выбора единственного решения оно не приводит,  во всяком случае до тех
пор,   пока   постоянные   $\varepsilon$  и  $\mu$  являются  причинно
обусловленными (имеют мнимые части или  тождественно  равны  единице).
При  сделанных  предположениях  в  каждой  области  задача  сводится к
решению  неоднородных  векторных  волновых   уравнений   типа   (3.22)
относительно  полей  или  потенциалов.  Неоднородность  в правой части
уравнения обусловлена возможным присутствием в рассматриваемой области
заданных  сторонних токов.  {\it Декартовы компоненты} искомых величин
(полей или потенциалов) удовлетворяют скалярному волновому уравнению
     $$\Delta \Psi+k^2\varepsilon\mu\Psi=f(\rv r)\,,\eqno(3.36)$$
которое в математической физике называется уравнением Гельмгольца. Это
уравнение  встречается  во  многих  областях  физики,  в частности,  в
акустике,  и его решения хорошо изучены.  Отметим,  что  хотя  искомые
величины $\Psi$ в (3.36) являются декартовыми компонентами,  но решать
эти  уравнения  можно,  а   зачастую   и   целесообразно,   в   других
ортогональных  системах  координат  (например,  в  цилиндрической  или
сферической) в  зависимости  от  структуры  сторонних  токов  и  формы
границы области.

     Все электродинамические задачи можно  разделить  на  два  класса:
скалярные,  которые  могут  быть  сведены  к  решению {\it одного} или
нескольких независимых уравнений вида  (3.36),  и  векторные,  которые
требуют  совместного  решения нескольких таких уравнений.  Несомненно,
что скалярные  задачи  значительно  проще;  более  того,  при  решении
большинства   векторных   задач  удобнее  работать  непосредственно  с
уравнениями Максвелла (2.8), а не с волновыми уравнениями (3.22).

     Простейший случай,   когда   задача   является   скалярной,   ---
неограниченная однородная среда.  При этом  для  выбора  единственного
решения  (необходимость  такого выбора следует из линейности волнового
уравнения --- к частному решению неоднородного уравнения (3.36) всегда
можно  добавить  произвольное решение однородного волнового уравнения)
достаточно потребовать его непрерывности во  всём  пространстве  (за
исключением точек, где расположены сторонние токи) и ограниченности на
бесконечности.  Последнее  требование  в  среде  с  потерями  означает
отсутствие   волн,   идущих   из   бесконечности. В  случае прозрачных
сред и, в частности, пустого пространства требование ограниченности на
бесконечности  недостаточно для   выделения  единственного  решения
---  необходимо  использовать дополнительное ограничение на поведение
поля  на  бесконечности,  так называемое {\it условие излучения
Зоммерфельда} (математическая формулировка его приведена в разделе
17),  которое обеспечивает отсутствие приходящих из бесконечности волн
и в пустоте. Этот же результат может быть получен при использовании
{\it принципа предельного поглощения},  который  гласит:
решение  уравнения  (3.36)  при $\varepsilon, \mu =1$ есть предел
единственного  решения  уравнения  $\Delta\psi_\delta+(k^2+i   \delta)
\psi_\delta=f(\rv   r)$   при  $\delta\to 0$.  Этот  принцип  позволяет
выделить единственное решение уравнения, хотя он при практическом
применении менее удобен, поскольку фактически вводит вместо пустоты
среду с комплексными проницаемостями.

     Если всё  пространство  состоит   из нескольких  однородных  по
электромагнитным свойствам  областей,  то в общем случае  задача
является  векторной  для каждой области.   Это связано с тем,  что на
поверхности,  разделяющей однородные  области,   должны  выполняться
граничные   условия  непрерывности тангенциальных  компонент  поля.
В  результате отдельные  составляющие  электрического или
магнитного  полей в рассматриваемой области оказываются связанными
друг с  другом, чем и определяется  необходимость совместного   решения
нескольких   уравнений типа  (3.36).  Некоторые  области  оказываются
замкнутыми (полностью   ограниченными  поверхностями раздела, при
этом  они  могут  быть  и  многосвязными, например,  шаровой   слой
между   двумя сферическими  поверхностями). Для   них
электродинамическая задача  является  {\it внутренней}.  Другие  области
простираются до бесконечности и  в  этом  случае  говорят  о  {\it внешней}
задаче.

   Начнём обсуждение с внутренней задачи. Она, в свою очередь, может
быть двух видов, в зависимости от того, присутствуют в рассматриваемой
области  или  нет  сторонние  токи,  и,  следовательно,  являются   ли
соответствующие    уравнения    Максвелла   или   волновое   уравнение
неоднородными  или  однородными.  При  отсутствии   источников   можно
утверждать  (и это будет строго доказано в последующих разделах),  что
решение задачи однозначно определяется  заданием  на  границе  области
тангенциальной   компоненты   одного   из  полей:  электрического  или
магнитного (или на части  граничной  поверхности  одного  поля,  а  на
оставшейся  части  ---  другого).  Тогда  поля внутри области нигде не
имеют  особенностей,  непрерывны  внутри  неё  и  на  самой  границе
области.  Требование непрерывности полей при подходе к границе области
зачастую молчаливо подразумевается,  но именно оно позволяет  выделить
из   бесчисленного   числа   решений  однородных  уравнений  (2.8)  то
единственное, которое имеет физический смысл.

     Это решение   определяет   значение   незаданной   тангенциальной
компоненты  поля  на  граничной  поверхности  и  обладает   следующими
свойствами:  если  внутри  области  пустота,  то интеграл по замкнутой
граничной  поверхности  от   реальной   части   комплексного   вектора
Умова-Пойнтинга,   определяющий   средний  по  времени  поток  энергии
электромагнитного поля через поверхность,  равен нулю,  а  если
проницаемости среды $\varepsilon, \mu\ne 1$, то поток направлен внутрь
и по величине равен интегралу по объёму области от плотности потерь,
определяемой  формулой (2.32).  Рассматривая теперь задачу с граничным
значением,  равным найденной тангенциальной компоненте  поля,  получим
однозначное  решение,  которое  для  незаданной  компоненты представит
исходное значение первоначальной задачи.

     Остаются два  вопроса  ---  что произойдёт с решением,  если на
граничной поверхности задать {\it обе} тангенциальные компоненты  поля
произвольным  образом,  и чем определяется в реальной задаче граничное
значение компоненты,  каким оно может быть?  Ответ  на  первый  вопрос
прост: при произвольном задании компонент существует бесконечное число
решений,  удовлетворяющих   уравнениям   Максвелла   внутри   области,
непрерывных в ней,  но не стремящихся к заданным значениям на границе.
Поэтому ни одно из них не может быть признано  физически  осмысленным.
Единственное  разумное  решение задачи получается только в том случае,
когда обе компоненты согласованы друг с другом  и  поэтому  достаточно
задать только одну из них.

     Ответ на второй вопрос  сложнее.  Остановимся  здесь  на  случае,
когда вся внутренняя область представляет собой пустоту. Очевидно, что
граничные  значения  полей  определяются  при   этом   только   токами
(сторонними и наведёнными), существующими во внешних областях. Можно
утверждать,  что ни при каком реальном распределении  этих  токов  (не
равных  нулю)  тангенциальная  компонента электрического поля не может
быть тождественно равна нулю на всей границе рассматриваемой  области.
Другими  словами,  постановка  внутренней задачи с нулевыми граничными
условиями для однородного уравнения  Гельмгольца  с  физической  точки
зрения  является  некорректной  и  поэтому  имеет  только  тривиальное
нулевое решение.

     Но именно  такая  задача возникает при использовании для среды во
внешней области модели идеального  проводника, и  её  решение  имеет
бесконечный   дискретный   набор  действительных  собственных  частот,
которым соответствуют собственные незатухающие колебания  произвольной
амплитуды.  Тем самым на выделенных частотах нарушается единственность
решения --- математически строгое толкование этого  утверждения  можно
найти    в  курсе   математической   физики,   когда   рассматривается
единственность решения  внутренней  задачи  Дирихле  или  Неймана  для
уравнения   Гельмгольца.   Физическая   же  причина  этого  состоит  в
некорректности  модели   идеального   проводника,   для   которой   не
выполняются  дисперсионные  соотношения,  являющиеся прямым следствием
принципа причинности.

     Тем не менее, эти решения играют важнейшую роль в электродинамике
СВЧ и будут рассматриваться в течение всего дальнейшего курса. Дело не
только  в том,  что распределение полей в них очень мало отличается от
имеющего место при реальных значениях проводимости внешней среды, но и
в  том,  что  они  образуют  полную систему ортогональных функций,  по
которой может быть  разложено  произвольное  поле  на  любой  частоте.
Однако при этом всегда надо помнить,  что в реальных задачах трудности
с выбором единственного  решения  возникают  в  первую  очередь  из-за
модели идеальной проводимости.

     Идеальная проводимость приводит к физически некорректному решению
и  во  внутренней  задаче со сторонними токами.  Хотя в этом случае на
всех  частотах,  отличных  от  собственных,  решение  единственно,  но
возбуждаемое поле таково,  что для любых сторонних токов, в частности,
для элементарного диполя, в установившемся режиме не происходит потери
энергии  на  излучение  (бессмысленность  этого результата очевидна на
примере атома, который является устойчивой структурой и в стационарном
состоянии не теряет энергии на электромагнитное излучение, но сам этот
факт может быть объяснён  только  в  рамках  квантовой  теории).  На
собственных   частотах   решение   внутренней   задачи  при  идеальной
проводимости граничной поверхности отсутствует,  поскольку оно во всех
точках области приводит к бесконечному полю.

     Для внешних задач модель  идеальной  проводимости  для  граничных
поверхностей  не  приводит  к трудностям выбора единственного решения.
Единственность решения обеспечивается при тех же условиях,  как и  для
всего   однородного   неограниченного  пространства.  Однако  как  для
внешней,  так и для внутренней задачи существует  ещё  одна  причина
трудности  выбора  единственного решения --- она состоит в использовании
модели проводящего тела с нулевым объёмом,  представляющего  в  этом
случае  незамкнутую  поверхность.  Очевидно,  что  реальное физическое
решение при наличии в области хорошо проводящей очень тонкой  пластины
не  может  зависеть  от её толщины везде,  за исключением небольшого
пространства вблизи кромки.  Поэтому представляется разумным  с  целью
уменьшения числа геометрических параметров задачи устремить толщину пластины
к нулю (при этом её приходится считать  и  идеально  проводящей).  В
результате  кривизна  поверхности  на  кромке становится бесконечной и
одновременно к бесконечности стремятся поля при приближении к  кромке.
Основная  неприятность  состоит  в  том,  что  в этом случае пропадает
единственность  решения  и  необходимо  ставить  дополнительное   {\it
условие  на  ребре},  которое  сводится  к  требованию интегрируемости
квадрата модуля комплексной амплитуды поля по области вблизи кромки.

     Для немонохроматических  полей  единственность  решения во многих
задачах требует учёта дисперсии свойств среды и наличия  поглощения,
следующих из дисперсионных соотношений. В противном случае интеграл по
частотам,  который возникает при использовании для решения  разложения
полей в интеграл Фурье, становится неопределённым или расходящимся.

     Подводя итоги,  можно  сказать,  что  уравнения   электродинамики
сплошной  среды  всегда  имеют  единственное  решение  при  корректной
постановке  задачи  с  учётом  требований,  вытекающих  из  принципа
причинности.

%\end{document}

\newpage
\oddsidemargin=-0.4mm \evensidemargin=-0.4mm
\headsep=7mm
\textheight=231.875mm
\textwidth=160mm
\mathsurround=2.5pt
\unitlength=1mm

%\begin{document}
%\input{macr.tex}
\thispagestyle{empty}
%\addtocounter{page}{41}

\begin{center}
   \subsubsection*{\rm Г\,Л\,А\,В\,А\, 2}
   \vspace{-1.15em}
   \line(6,0){160}\\
   \vspace{-1em}
   \line(6,0){160}
   \vspace{-1.15em}
   \subsubsection*{ПЛОСКИЕ ВОЛНЫ}
      \vspace{35mm}
   \subsubsection*{4.\hspace{1ex}Плоские волны в однородной
         безграничной среде}
\end{center}\vspace*{0.5cm}

\markboth{Глава 2. Плоские волны}
{4. Плоские волны в однородной безграничной среде}

\begin{center}\begin{minipage}[c]{0.75\textwidth}
\footnotesize{\parindent=0.5cm
         Определение плоской   монохроматической   волны   и   понятие
         комплексного волнового  вектора.  Однородные  и  неоднородные
         плоские волны.  Соотношение между компонентами поля в плоской
         волне. Монохроматическая плоская волна в диспергирующей среде
         без  поглощения.  Одномерные  решения  волнового  уравнения в
         однородной среде без дисперсии.
}\end{minipage}\end{center}
\vspace*{0.5cm}

     {\it Плоские   волны}   представляют   собой   простейшее  решение
однородных    уравнений    Максвелла.    Начнём    рассмотрение    с
монохроматического  решения,  в  котором  зависимость  всех  полей  от
времени  определяется  множителем  $e^{-i\omega  t}$.   В   однородной
изотропной  среде,  характеризуемой двумя комплексными проницаемостями
$\varepsilon$ и $\mu$, зависящими, вообще говоря, от частоты $\omega$,
но  не  от  координат,  комплексные  амплитуды полей $\rv E$ и $\rv H$
удовлетворяют уравнениям
     $$\rot\rv E=ik\mu(\omega)\rv H,\quad\rot\rv H=-ik\varepsilon
         (\omega)\rv E\,,\eqno(4.1)$$
из которых  после  исключения  одного  из  векторов  следуют  волновые
уравнения
     $$\triangle\rv E+k^2\varepsilon\mu\rv E=0\,,\quad\triangle\rv H+k^2
         \varepsilon\mu\rv H=0\,.\eqno(4.2)$$

     Плоской монохроматической волной называют решение этих уравнений,
имеющее вид
     $$\rv E(\rv r)=\rv E_0e^{i\rv K\rv r},\quad\rv H(\rv r)=\rv H_0
         e^{i\rv K\rv r},\eqno(4.3)$$
в котором  вся зависимость от координат выражается множителем $e^{i\rv
K\rv r}$, прич\"ем вектор $\rv K$ называется {\it волновым вектором },
а  векторы $\rv  E_0$,~$\rv  H_0$  определяют   амплитуды поля.
Уравнения (4.2) допускают решения и  для  комплексных  волновых
векторов  $\rv  K=\rv K'+i\rv  K^{''}=K'\rv  n'+iK^{''}\rv  n^{''}$,
где $K'$,~$K^{''}$ --- действительные  положительные  числа,
$\rv   n',\;\rv   n^{''}$   --- единичные  векторы.  Подставляя  (4.3)
в  волновое  уравнение  (4.2), получаем,  что для заданной частоты
$\omega$  или, что то  же, для заданного волнового числа $k=\omega/c$,
квадрат вектора $\rv K$ должен быть равен
     $$\rv K^2=K^2= K^{'2} - K^{''2} + 2iK'K''\rv n'\rv n^{''}
           =k^2\varepsilon\mu\,. \eqno(4.4)$$
Величина $K=k\sqrt{\varepsilon\mu}$, называемая {\it волновым числом в
среде},  вообще говоря,  комплексная,  что обусловлено  поглощением  в
среде,  а  сам  волновой  вектор  $\rv  K$  может быть комплексным и в
прозрачной  среде,  в том числе и в  пустоте,  при   ортогональности
векторов $\rv n'$ и $\rv n^{''}$.

     При наличии  поглощения  волновое  число  в  среде  $K$   принято
записывать в виде
     $$K=k\sqrt{\varepsilon\mu}=k(n+i\mbox{\ae}),\eqno(4.5)$$
где действительные величины $n$   и   $\mbox{\ae}$  называют соответственно
 {\it  показателем  преломления} и   {\it коэффициентом поглощения}.
Величины $n$ и  $\mbox{\ae}$  нетрудно  выразить через
действительную и мнимую части  комплексных  проницаемостей
$\varepsilon$ и $\mu$. Во избежание излишней громоздкости
приведём формулы для случая $\mu=1$:
     $$n=\sqrt{\frac{\varepsilon'+\sqrt{\varepsilon^{'2}+\varepsilon
        ^{''2}}}2}\,,\quad\mbox{\ae}=\sqrt{\frac{-\varepsilon'+
       \sqrt{\varepsilon^{'2}+\varepsilon^{''2}}}2}\,.\eqno(4.6)$$

     Величины $K'$,  $K^{''}$  выражаются  через $n$ и $\mbox{\ae}$,  как это
следует из (4.4) и (4.5), следующим образом:
     $$K^{'2}=\frac{k^2} 2\Bigl[n^2-\mbox{\ae}^2+\sqrt{(n^2-\mbox{\ae}^2)^2
       +\Bigl(\frac{2n\mbox{\ae}}{\rv n'\rv n^{''}}\Bigr)^2}\;\Bigr]\;,
        \eqno(4.7)$$
     $$K^{''2}=\frac{k^2} 2\Bigl[\mbox{\ae}^2-n^2+\sqrt{(n^2-\mbox{\ae}^2)^2
       +\Bigl(\frac{2n\mbox{\ae}}{\rv n'\rv n^{''}}\Bigr)^2}\;\Bigr]\;.
        \eqno(4.8)$$

      Если вектор $\rv E_0$ действительный (линейная поляризация), то с
учётом  временного  множителя  $e^{-i\omega t}$ электрическое
поле плоской волны принимает вид
     $$ \gv E(\rv r,t)= \re{\{\rv E_0e^{i(\rv K\rv r-\omega t}}\}=
         \rv E_0 e^{-K{''}\rv n{''}\rv r}\cos{(K'\rv n'\rv r-\omega t)}
         \,.\eqno(4.9)$$
Отсюда следует,  что плоскости,  перпендикулярные  вектору  $\rv  n'$,
являются плоскостями постоянной фазы $\varphi=K'\rv n'\rv r-\omega t$,
а плоскости,  перпендикулярные  вектору  $\rv  n''$,  ---  плоскостями
постоянной амплитуды.  Поверхности же, на которых поле в данный момент
$t$  имеет  одинаковое  значение,  вообще  могут  не  быть   плоскими.
Скорость,  с  которой перемещается в направлении вектора $\rv n'$ фаза
волны $\varphi$, называется {\it фазовой} и равна
     $$v_{\mbox\footnotesize\textit{ф}}=\frac {\omega} {K'}\,,\eqno(4.10)$$
длина волны (пространственный период в распределении поля)
     $$\lambda=\frac{2\pi}{K'}\,.\eqno(4.11)$$
Амплитуда волны  убывает  в  направлении  вектора  $\rv   n^{''}$   по
экспоненциальному закону с показателем экспоненты $\sim K^{''}$. Такое
затухание амплитуды не  обязательно  обусловлено  поглощением  энергии
волны в среде;  как уже отмечалось, оно возможно и в прозрачной среде,
в частности,  в пустоте.  В следующем разделе будет показано,  что при
определённых условиях это имеет место, например, при падении плоской
волны на плоскую границу раздела двух сред.

     Волна (4.9)  в  общем случае произвольных $\rv n'$ и $\rv n^{''}$
называется {\it неоднородной} плоской волной.  В частном случае,
когда векторы   $\rv   n'$   и   $\rv  n^{''}$  совпадают
(обозначим  тогда направляющий вектор просто $\rv n$) волна
называется {\it однородной}. Поля  такой  волны являются функциями
одной координаты,  отсчитываемой вдоль $\rv n$, а плоскости равных
фаз и равных  амплитуд  совпадают  и нормальны  к  $\rv  n$;
формулы  (4.7)  и  (4.8)  для $K'$ и $K^{''}$ существенно
упрощаются и $K'=kn$,~$K^{''}=k\mbox{\ae}$.  В этом случае затухание
амплитуды волны вдоль направления её распространения зависит только
от $\mbox{\ae}$, что позволяет называть эту величину также и
{\it коэффициентом затухания}. В однородной волне комплексным является
не сам волновой вектор $\rv  K$, а только его <<длина>> (волновое число в среде)
$K$.

     Все приведенные  до  сих  выражения  для  компонент   волнового
вектора  $\rv  K$  вытекают  из  требования  того,  что  решение (4.3)
удовлетворяет волновым уравнениям (4.2). Для удовлетворения уравнениям
Максвелла (4.1) необходимо,  помимо этого,  выполнение определённого
соотношения между $\rv E_0$ и $\rv H_0$.

     Подставляя (4.3)  в  уравнения  (4.1),  получаем:
     $$k\mu\rv H_0=[\rv K\rv E_0],\quad k\varepsilon\rv E_0=-
         [\rv K\rv H_0]\,. \eqno(4.12)$$
В результате скалярного умножения этих уравнений на $\rv K$ имеем
     $$\rv K\rv E_0=0\,,\qquad\rv K\rv H_0=0\,,\eqno(4.13)$$
а возводя каждое из уравнений (4.12)  в  квадрат  и  используя  (4.4),
приходим к соотношению
     $$\varepsilon\rv E_0^2=\mu\rv H_0^2.\eqno (4.14)$$
Ввиду комплексности всех трёх векторов $\rv K$,~$\rv E_0$,~$\rv H_0$
соотношениям    (4.12)--(4.14)    трудно   придать   определённый
геометрический  смысл,  в  частности,  нельзя  говорить о поперечности
неоднородной плоской волны.

     В однородной  плоской  волне  вектор $\rv K$ может быть записан в
виде $\rv K=K\rv n$. Поэтому соотношения (4.12) и (4.13) упрощаются:
     $$\rv E_0=-W [\rv n\rv H_0],\qquad
         \rv H_0=\frac 1 W[\rv n\rv E_0]\,,\eqno(4.15)$$
где величина
     $$W=\sqrt\frac\mu\varepsilon\eqno(4.16)$$
называется {\it волновым сопротивлением} среды, и
     $$\rv n\rv E_0=0\,,\qquad\rv n\rv H_0=0\,.\eqno(4.17)$$
Из (4.15) и (4.17) следует, что в однородной плоской волне три вектора
$\rv n$,~$\rv E_0$,~$\rv H_0$  взаимно  ортогональны.  Необходимо  при
этом  отметить,  что  оба  вектора  ($\rv E_0$ и $\rv H_0$) могут быть
комплексными,  то есть представимыми в  виде  $\rv  E_0=\rv  E_0'+i\rv
E_0^{''}$,  $\rv  H_0=\rv  H_0'+i\rv  H_0^{''}$,  причём  угол между
векторами $\rv E_0'$ и $\rv E_0^{''}$ и векторами $\rv  H_0'$  и  $\rv
H_0^{''}$   одинаков.   Комплексность  вектора  $\rv  E_0$  определяет
характер поляризации волны.  В общем случае эллиптической  поляризации
концы векторов электрического и магнитного поля вращаются в плоскости,
нормальной к $\rv n$,  причём концы векторов  описывают  эллипсы.  В
этом  случае  вектор $\rv E_0$ может быть записан в виде $\rv E_0=(\rv
b_1+i\rv  b_2)e^{i\alpha}$,  где   $\rv   b_1$   и   $\rv   b_2$   ---
действительные    ортогональные    векторы   ($\rv   b_1\rv   b_2=0$),
определяющие  ориентацию  и   длину   осей   эллипса,   $\alpha$   ---
действительное  число.  Если  один  из векторов $\rv b_1$,  ~$\rv b_2$
равен нулю, то волна является {\it линейно} поляризованной --- векторы
поля  волны  по  мере  её  распространения  лежат  в  одних и тех же
ортогональных   плоскостях.   Очевидно,   что    любую    эллиптически
поляризованную  волну  можно  представить  в  виде  суммы двух линейно
поляризованных; далее,  если это специально не оговорено,  для волн  в
среде подразумевается линейная поляризация.

     Особенно простые результаты получаются в  случае  плоской  волны,
распространяющейся    в    непоглощающей   среде   ($\varepsilon''=0$,
~$\mu''=0$),    для  чего  необходимо  выполнение   условия
$\varepsilon\mu>0$.   В  этом  случае  волновое  сопротивление  $W$  и
волновое число $K$ --- действительные величины, фазовая скорость
           $$v_{\mbox{\footnotesize\textit ф}}=\frac c{\sqrt{\varepsilon\mu}}\,,\eqno(4.18)$$
длина волны
     $$\lambda=\frac{2\pi}K=
        \frac{2\pi}{k\sqrt{\varepsilon\mu}}\,.\eqno(4.19)$$

     Помимо общего для плоских волн соотношения (4.14) в данном случае
выполняется более определённое:
     $$\varepsilon|\rv E|^2=\mu |\rv H|^2,\eqno(4.20)$$
которое, однако,  для среды с дисперсией  не означает равенства в
волне  плотностей  энергии  электрического и магнитного полей, как
это имеет место в пустоте.  При наличии дисперсии и возможности
пренебречь поглощением  плотность энергии электромагнитного поля в
среде (2.47) с учетом (4.20) и (4.4) сводится к виду
     $$\overline U=\frac {c^2}{16\pi\mu\omega}\frac{dK^2}{d\omega}|
         \rv E|^2=\frac c{8\pi W}\frac{dK}{d\omega}|\rv E|^2.
         \eqno(4.21)$$
Поток энергии в волне направлен вдоль $\rv n$ и его  среднее  значение
даётся вектором Умова-Пойнтинга
     $$\overline {\mbox{\boldmath$\gv S$}}=
       \frac c{8\pi}\re {[\rv E\rv H^*]}=\rv n
         \frac c{8\pi W}|\rv E|^2\,.\eqno(4.22)$$

     Скорость переноса энергии в волне,  по традиции  называемая  {\it
групповой скоростью} $\rv v_{\mbox\footnotesize\textit{гр}}$,
определяется очевидным соотношением между плотностью  энергии
$\overline  U$  и плотностью  её  потока $\overline
{\mbox{\boldmath$\gv S$}}$:
     $$\overline {\mbox{\boldmath$\gv S$}}=
       \rv v_{\mbox\footnotesize\textit{гр}}\overline U.\eqno (4.23)$$
Из соотношений  (4.21) и (4.22) следует,  что в рассматриваемом
случае изотропной среды $\rv v_{\mbox\footnotesize\textit{гр}}$
направлена вдоль $\rv n$ и  её  модуль равен
     $$v_{\mbox\footnotesize\textit{гр}}=\frac {d\omega}{dK}\eqno(4.24)$$
или --- с учётом (4.4) ---
     $$v_{\mbox\footnotesize\textit{гр}}=
     \frac c{\sqrt{\varepsilon\mu}+\displaystyle{
         \frac \omega{2\sqrt{\varepsilon\mu}}{\Bigl(\mu\frac{d\varepsilon}{d
         \omega}+\varepsilon\frac{d\mu}{d\omega}\Bigr)}}}\,.\eqno(4.25)$$
Поскольку в  прозрачной  среде  согласно  условию (2.15),
налагаемому дисперсионными    соотношениями,
$\displaystyle{\frac{d\varepsilon} {d\omega}}>0$  и
$\displaystyle{\frac{d\mu}{d\omega}}>0,$ то
$v_{\footnotesize\textit{гр}}$ меньше
$v_{\footnotesize\textit{ф}}$,  определённой формулой
(4.18), а в силу неравенства (2.16) всегда
$v_{\footnotesize\textit{гр}}<c$.

     Скорость $v_{\footnotesize{\textit{гр}}}$  обязана  своему
названию  тому,  что   с   ней распространяется  в  пространстве  и
группа  монохроматических  волн, образующих так  называемый
волновой  пакет.  Для  любого  одномерного волнового   процесса
в  среде  каждая  компонента  поля  может  быть представлена в
виде интеграла Фурье:
     $$\psi(z,t)=\int\limits_{-\infty}^{\infty}A(K)e^{i[Kz-\omega(K) t
         ]}\,dK.\eqno(4.26)$$
Функция (4.26)   образует   волновой   пакет   в   том  случае,  когда
подынтегральная функция $A(K)$ заметно отлична от нуля только в  узкой
области  $K-\Delta  K<K_0<K+\Delta  K$  вокруг  центрального  значения
$K_0$, так что её можно представить в виде
     $$\psi(z,t)=\int\limits_{K_0-\Delta K}^{K_0+\Delta K}A(K)
         e^{i[Kz-\omega(K) t]}\,dK.\eqno(4.27)$$
Функция $\omega(K)$   определяется   уравнением  (4.4)  и  может  быть
разложена в ряд вблизи $K_0$, то есть в ряд
     $$\omega(K)=\omega(K_0)+\left.\Bigl(\frac{d\omega}{dK}\Bigr)
         \right|_{K=K_0}\cdot(K-K_0)+\cdots,\qquad\omega(K_0)=\omega
         _0\,,\eqno(4.28)$$
в котором ввиду  малости  $\Delta  K$  достаточно  ограничиться  двумя
первыми членами. Тогда (4.26) можно переписать как
     $$\psi(z,t)=e^{i(K_0z-\omega_0t)}\cdot\int\limits_{K_0-\Delta K}
         ^{K_0+\Delta K}A(K)e^{i(K-K_0)\bigl[z-\left.\bigl(\frac{d
         \omega}{dK}\bigr)\right|_{K=K_0}\cdot t\bigr]}\,dK.
         \eqno(4.29)$$

     Интеграл в этой формуле фактически представляет  собой  амплитуду
волнового процесса, которая постоянна на плоскости
     $$z-\left.\biggl(\frac{d\omega}{dK}\biggr)\right|_{K=K_0}t=const\,,
         \eqno(4.30)$$
то есть  скорость  его  перемещения  или  скорость  группы  волн
есть $v_{\footnotesize\textit{гр}}=  d\omega/dK.$  Для
непоглощающей среды $v_{\footnotesize\textit{ф}}=\omega/ K$
и её можно рассматривать как функцию $K$, что позволяет
вычислять $v_{\footnotesize\textit{гр}}$ по формуле
     $$v_{\footnotesize\textit{гр}}=
         \frac{d(Kv_{\footnotesize\textit{ф}})}{dK}=
       v_{\footnotesize\textit{ф}}+
         K\frac{dv_{\footnotesize\textit{ф}}}{dK}\,.\eqno(4.31)$$

     Отметим, что  при  наличии  затухания  в среде волновые пакеты по
мере  своего  распространения  быстро   расплываются   и   определение
групповой  скорости  как  скорости перемещения волнового пакета теряет
свой смысл.  Таким образом,  ввести понятие групповой скорости тем или
иным   способом   можно  только  при  тех  же  условиях,  при  которых
определяется  понятие  средней  энергии   электромагнитного   поля   в
диспергирующей среде.

     Изучим теперь типичные волновые процессы в однородной  изотропной
среде,  которые  могут с полным основанием рассматриваться как плоские
волны,  поскольку все составляющие полей в них зависят только от одной
декартовой  координаты  (примем  её  за  $z$) и времени,  но которые
обладают широким спектром частот.  Исследовать такой процесс  в  общем
виде  удаётся  только  при использовании материальных уравнений вида
(1.13)--(1.15).  В этом случае из уравнений (3.3),  (3.4) следует, что
любая составляющая электромагнитного поля удовлетворяет уравнению
     $$\frac{\partial^2\psi}{\partial z^2}-\frac{\varepsilon\mu}
         {c^2}\frac{\partial^2\psi}{\partial t^2}-\frac{4\pi\sigma\mu}
         {c^2}\frac{\partial\psi}{\partial t}=0\,.\eqno(4.32)$$
Использование этого   уравнения   для   описания   волнового  процесса
допустимо при условии, что разложение (временной зависимости $\psi$) в
интеграл  Фурье содержит только такие частоты,  для которых дисперсией
свойств среды можно пренебречь.

     Общая постановка  задачи  такова:  в  плоскости  $z=0$  считаются
заданными граничные условия
     $$\left.\begin{array}{rcl}\psi(0,t)&\!=\!&f(t)\,,\\[0.25cm]
         \displaystyle{\frac{\partial\psi}{\partial z}}(0,t)&\!=
         \!&F(t)\,,\end{array}\right\}\eqno(4.33)$$
где $F(t)$ и $f(t)$ -- произвольные функции времени.  Требуется  найти
решение уравнения (4.32),  удовлетворяющее этим условиям и описывающее
дальнейшее развитие волнового процесса как функцию $z$ и $t$.

     Любое решение     уравнения    (4.32),    представляющее    собой
гармоническую функцию частоты $\omega$, может быть записано в виде
     $$\psi=(\tilde Ae^{iKz}+\tilde Be^{-iKz})\,e^{-i\omega t}\,,\eqno(4.34).$$
где амплитуды $\tilde A$,~$\tilde B$ следует считать функциями только $\omega$,  $K$
--- комплексное волновое число
     $$K=k\sqrt{\varepsilon\mu+i\displaystyle{\frac{4\pi\sigma
         \mu}\omega}}\,,\eqno(4.35)$$
которое для дальнейшего удобно записать в виде
     $$K={\frac 1 v}\sqrt{\omega^2+2b\omega i}\,,\eqno(4.36)$$
где
     $$v=\frac c{\sqrt{\varepsilon\mu}}\,,\quad b=\frac{2\pi\sigma
         }{\varepsilon}\,.\eqno(4.37)$$
Нетрудно видеть,  что  для  любого  гармонического  решения (4.34) $v$
представляет собой фазовую скорость затухающей волны.

     Согласно теории   интегралов   Фурье  любая  функция,  являющаяся
решением  уравнения  (4.32),  представляет  собой  непрерывный  спектр
плоских волн (4.34):
     $$\psi(z,t)=\frac 1{2\pi}\int\limits_{-\infty}^\infty[A(\omega)
         e^{iKz}+B(\omega)e^{-iKz}]e^{-i\omega t}\,d\omega\,.
         \eqno(4.38)$$
Подставляя <<решение>>  (4.38)  в  граничные  условия  (4.33),
получаем систему уравнений для амплитуд:
     $$\left.\begin{array}{rcl}f(t)&\!=\!&\displaystyle{\frac 1{2\pi}}
         \int\limits_{-\infty}^\infty[A(\omega)+B(\omega)]e^{-i
         \omega t}\,d\omega\,,\\[0.65cm]F(t)&\!=\!&\displaystyle{\frac i
         {2\pi}}\int\limits_{-\infty}^\infty K[A(\omega)-B(\omega)]e^
         {-i\omega t}\,d\omega\,.\\[0.25cm]\end{array}\right\}\eqno(4.39)$$
С помощью  обратного  преобразования Фурье определяем теперь
амплитуды  $A(\omega)$ и $B(\omega)$ в виде следующих интегралов:
     $$\left.\begin{array}{rcl}A(\omega)&\!=\!&\displaystyle{
         \frac 12}\int\limits_{-\infty}^\infty[f(\alpha)+\frac 1{iK}
         F(\alpha)]e^{i\omega\alpha}\,d\alpha\,,\\[0.45cm]B(\omega)&\!=
         \!&\displaystyle{\frac 12}\int\limits_{-\infty}^\infty
         [f(\alpha)-\frac 1{iK}F(\alpha)]e^{i\omega\alpha}\,d\alpha\,.
         \\[0.25cm]\end{array}\right\}\eqno(4.40)$$
Наконец, подставляя   (4.40)  в  (4.38),  получим  выражающееся  через
двойной интеграл решение для $\psi(z,t)$, которое запишем в виде
     $$\left.\begin{array}{rcl}\psi&\!=\!&\psi_1+\psi_2,\\[0.35cm]
         \psi_1&\!=\!&\displaystyle{\frac 1{2\pi}}\int\limits_{-
         \infty}^\infty f(\alpha)\,d\alpha\int\limits_{-\infty}^\infty
         \cos Kz\cdot e^{i\omega(\alpha-t)}\,d\omega,\\[0.45cm]
         \psi_2&\!=\!&\displaystyle{\frac 1{2\pi}}\int\limits_{-
         \infty}^\infty F(\alpha)\,d\alpha\int\limits_{-\infty}^\infty
         \displaystyle{\frac{\sin Kz}{K}}\cdot e^{i\omega(\alpha-t)}
         \,d\omega\,.\\[0.25cm]\end{array}\right\}\eqno(4.41)$$
Изложенный способ   решения   широко  используется  при  исследованиях
распространения  одномерных  импульсов   в   среде.   Фактически   все
преобразования тривиальны, однако они остаются формальными до тех пор,
пока в (4.39) не вычислены интегралы по $\omega$.  Из двух  интегралов
достаточно  суметь вычислить только тот,  который определяет $\psi_2$.
Другой  интеграл   после   этого   легко   находится   как   результат
дифференцирования   по   параметру   $z$  (с  соответствующей  заменой
$F(\alpha)$  на  $f(\alpha)$).   Вся   зависимость   от   $\omega$   в
подынтегральном  выражении (помимо экспоненциального множителя) входит
в неявном виде через волновое число $K$.

     Поэтому задача  состоит  в том,  чтобы представить $\sin{Kz}/K$ в
виде,  удобном для интегрирования по $\omega$.  При $K$,  определяемом
формулой (4.36),  это удаётся сделать с помощью широко используемого
в математической физике интеграла Гегенбауэра.  Приведём без  вывода
окончательный результат:
     $$\frac{\sin Kz} K=\frac v 2\int\limits_{-z/v}^{z/v}e^{i\gamma(
         \omega+i b)}J_0\Bigl({\frac b v}\sqrt{z^2-v^2\gamma^2}
         \Bigr)\,d\gamma\,.\eqno(4.42)$$
После подстановки   (4.42)   в   (4.41)  интеграл  по  $\omega$  легко
вычисляется и сводится к $\delta$-функции в соответствии  с  известным
её представлением:
     $$\delta(x)=\frac 1{2\pi}\int\limits_{-\infty}^\infty e^{-i
         \omega x}\,d\omega\,.\eqno(4.43)$$
Дальнейшее вычисление интеграла по $\gamma$ становится тривиальным и в
результате получаем
     $$\psi_2=\frac v 2\int\limits_{t-z/v}^{t+z/v}F(\alpha)J_0\Bigl(
         \frac b v\sqrt{z^2-v^2(t-\alpha)^2}\Bigr)e^{-b(t-\alpha)}\,d
         \alpha\,.\eqno(4.44)$$
Теперь описанным выше способом находим $\psi_1$:
     $$\begin{array}{l}\psi_1=\displaystyle{\frac 1 2e^{bz/v}f(t+z/v
         )}+\displaystyle{\frac 1 2}e^{-bz/v}f(t-z/v)+\\[0.4cm]\qquad
         +\displaystyle{\frac v 2}e^{-bt}\int\limits_{t-z/v}^{t+z/v}f(
         \alpha)e^{b\alpha}\displaystyle{\frac{\partial}{\partial z}
         J_0\Bigl(\frac b v\sqrt{z^2-v^2(t-\alpha)^2}\Bigr)\,d\alpha}\,.
         \end{array}\eqno(4.45)$$

     Для непроводящей среды $(\sigma=0)$ $b=0$ и решение упрощается:
     $$\psi=\frac 1 2f(t+\frac z v)+\frac 1 2 f(t-\frac z v)+\frac v 2
         \int\limits_{t-z/v}^{t+z/v}F(\alpha)\,d\alpha\,.\eqno(4.46)$$
Если теперь ввести функцию
     $$h(\beta)=-v\int\limits_0^\beta F(\alpha)\,d\alpha\,,\eqno(4.47)$$
то это решение можно переписать в виде
     $$\psi(z,t)=\displaystyle{\frac 12}f(t+z/v)+\frac 12 f(t-z/v)
         -\displaystyle{\frac 12}h(t+z/v)+\frac 1 2 h(t-z/v)\,.
         \eqno(4.48)$$
Заметим, что  при  $\sigma=0$  уравнение  (4.32)  переходит в
волновое уравнение (3.7\textit а) и (4.48) является частным
случаем его общего  решения (3.7\textit б).

     Докажем теперь,  что задания  двух  функций  $f(t)$  и  $F(t)$  в
плоскости  $z=0$  достаточно  для  того,  чтобы  полностью  определить
одномерное электромагнитное поле  во  всем  пространстве  и  для  всех
моментов  времени.  Пусть  электрическое поле поляризовано в плоскости
$xz$. Тогда уравнения Максвелла можно записать в виде
     $$\frac{\partial E_x}{\partial z}+\frac{\mu} c\frac{\partial H_y}
         {\partial t}=0\,,\qquad \frac{\partial H_y}{\partial z}+\frac
         {\varepsilon} c\frac{\partial E_x}{\partial t}=0\,.
         \eqno(4.49)$$
Примем теперь,  что $E_x$ есть  функция  $\psi(z,t)$,
определяемая решением (4.48):
     $$E_x=\displaystyle{\frac 12}[f(t+z/v)+f(t-z/v)-h(t+z/v)+h(t-z/v
         )]\,.\eqno(4.50)$$
С помощью уравнений (4.49)  нетрудно  убедиться,  что  в  этом  случае
     $$H_y=\displaystyle{\frac 12\sqrt{\frac{\varepsilon}{\mu}}}
         [f(t-z/v)-f(t+z/v)+h(t+z/v)+h(t-z/v)]\,.\eqno(4.51)$$
На плоскости $z=0$
     $$\left.\begin{array}{l}E_x=f(t)\,,\qquad\displaystyle{\frac{
         \partial E_x}{\partial z}}=F(t)\,,\\H_y=\displaystyle{\sqrt{
         \frac{\varepsilon}{\mu}}\,h(t)=-\displaystyle\frac c \mu \int
         \limits_0^tF(t)\,dt\,,\qquad\displaystyle{\frac{\partial H_y}{
         \partial z}}=-\frac{\varepsilon} c \frac{df}{dt}}\,.
         \end{array}\right\}\eqno(4.52)$$
Отсюда следует, что электромагнитное поле определено для всех $z,\,t$,
если на плоскости $z=0$ заданы как независимые функции  времени  любые
две  из  величин,  содержащих как $f(t)$,  так и $F(t)$.  В частности,
могут быть заданы либо электрическое поле и  его  производная  по  $z$,  либо
оба поля --- электрическое и магнитное.

     На первый  взгляд  последнее  утверждение  противоречит   условию
единственности  решения электродинамической задачи,  согласно которому
достаточно  задать  только  одну  тангенциальную  компоненту  поля  на
границе области, выбрав в качестве таковой, например, полупространство
$z>0$.  Однако  рассматриваемая  задача  является   внешней   (область
простирается  до  бесконечности)  и  для  решения  (4.50) не выполнено
условие излучения Зоммерфельда:  решение содержит волну, приходящую из
бесконечности  ($f(t+z/v)$  при $z>0$ и $f(t-z/v)$ при $z<0$).  Именно
поэтому на границе области $z=0$ приходится  задавать  две  компоненты
поля. Отметим  также,  что  уравнению  (4.32)  соответствует оператор,
называемый в  математической  физике   {\it   телеграфным}   ---   для
нахождения физически корректного решения тогда необходимо использовать
его фундаментальное   решение,   удовлетворяющее   условию   излучения
Зоммерфельда.

     Скажем ещё несколько слов о физическом смысле решений (4.48)  и
(4.45).  Первое  из  них  описывает  распространениe  импульса,  форма
которого определяется функциями $f(t)$ и $F(t)$:  от плоскости $z=0$ в
обе  стороны распространяются со скоростью ${v= \displaystyle {c/\sqrt
{\varepsilon\mu}}}$  два  импульса,  полностью  сохраняющие  по   мере
продвижения свою первоначальную форму. При распространении импульсов в
поглощающей среде (решение (4.45)) форма импульса меняется по мере его
удаления от исходной плоскости,  уменьшается амплитуда и,  кроме того,
даже после прохождения заднего фронта в среде остается шлейф,  который
также затухает со временем.

%\end{document}

\newpage
\oddsidemargin=-0.4mm \evensidemargin=-0.4mm
\topmargin=-0.4mm
\headsep=7mm
\textheight=231.875mm
\textwidth=160mm
\mathsurround=2.5pt
\unitlength=1mm
%\begin{document}
%\input{macr.tex}
\thispagestyle{empty}
%\addtocounter{page}{50}
\baselineskip=\normalbaselineskip
%\baselineskip=0.955\normalbaselineskip

\begin{center}\subsubsection*{5. Падение плоской монохроматической
волны на плоскую границу раздела двух сред}\end{center}\vspace*{0.5cm}

\markboth{Глава 2. Плоские волны}{5. Падение плоской волны на плоскую
границу раздела}

\begin{center}\begin{minipage}[c]{0.75\textwidth}
     \footnotesize{\parindent=0.5cm
         Представление поля в виде падающей,  отражённой и прошедшей
         плоских волн.  Соотношения между волновыми  векторами  трёх
         волн  --- законы отражения и преломления.  Полное отражение и
         возникновение  поверхностной   волны   у   границы   раздела.
         Соотношения между амплитудами волн (формулы Френеля). Условие
         отсутствия     отражения     при      нормальном      падении
         (радиолокационная   <<невидимка>>).   Соотношения   подобия   в
         электродинамике сплошных сред. Переходный слой.
}\end{minipage}\end{center}\vspace*{0.5cm}

     Рассмотрим отражение   и  прохождение  плоской  монохроматической
волны, падающей на границу раздела двух однородных сред. Пусть
граница раздела  совпадает  с  плоскостью  $z=0$  и нормаль $\rv
n$  к этой плоскости направлена  вдоль  оси  $z$. Пространство
$z<0$  заполнено средой~1  с  проницаемостями $\varepsilon_1$,
$\mu_1$,  пространство $z>0$ --- средой~2 с проницаемостями
$\varepsilon_2$,~$\mu_2$. Среду~1 будем   считать прозрачной,  так
что  $\varepsilon_1$,~$\mu_1$  --- действительные   величины,   и,
кроме    того,    для возможности распространения   волны
потребуем,  чтобы выполнялось  неравенство $\varepsilon_1\mu_1>0$.
Относительно среды~2   таких   предположений делать   не   будем
---  она может  обладать  поглощением,  так  что
$\varepsilon_2$,~$\mu_2$ в общем случае комплексные величины.
Плоская однородная  волна, распространяющаяся  в  среде~1,  падает
на границу раздела,  так что угол  между  её  волновым  вектором
$\rv  K_i$  и нормалью $\rv n$ (угол падения) равен $\varphi_i$.
Полное поле в этой
\begin{wrapfigure}[14]{l}{8cm}
\begin{picture}(80,55)
\put(-4,50){\special{em:graph fig5-1.bmp}}
\end{picture}
\hbox to 8cm{\hfil\footnotesize{Рис.~5.1.~Падение волны на плоскую
границу.}\hfil}
\end{wrapfigure}
среде будем  искать  в  виде  суммы  падающей  и  отражённой
плоских однородных волн,  а поле в среде~2 --- прошедшую волну
---  в  виде, вообще  говоря,  неоднородной плоской волны.
Прошедшую волну называют преломлённой,  поскольку направление её
распространения обычно  не совпадает  с  направлением падающей
волны.  При таком выборе решения уравнения Максвелла будут
удовлетворены во  всем  пространстве  и  нам остаётся обеспечить
выполнение  граничных  условий  на поверхности раздела,  то есть
непрерывность тангенциальных компонент полей $\rv E$ и $\rv H$.

     Комплексные амплитуды любой плоской волны с учётом (4.12) могут
быть записаны в виде:
     $$\rv E(\rv r)=\rv E_0e^{i\rv K\rv r},\quad\rv H(\rv r)=\rv H_0
         e^{i\rv K\rv r},\quad \rv H_0=\frac 1{\mu k}[\rv K\rv E_0]\,,
         \quad\rv E_0=-\frac 1{\varepsilon k}[\rv K\rv H_0]\,,
         \eqno(5.1)$$
где $\rv K$  ---  волновой  вектор.  Все  величины  в  этих  формулах,
относящиеся   к   падающей,  отражённой  и  прошедшей  волнам  будем
отмечать,  соответственно,  нижними индексами  $i$,~$r$  и~$t$.  Ввиду
одинаковой  зависимости  полей  $\rv  E$  и  $\rv  H$ от координат для
полного описания плоской волны достаточно привести выражение для  $\rv
E(\rv  r)$  и соотношение между постоянными векторами $\rv E_0$ и $\rv
H_0$.  В однородной плоской волне волновой вектор $\rv K$  может  быть
записан  в  виде  $\rv  K=K\rv  e_K$,  где  $\rv  e_K$  ---  единичный
направляющий вектор,  а $K$ ---  его  длина
(волновое  число  в  среде),  которая  в  поглощающей среде величина
комплексная.  В неоднородной плоской волне $\rv K$ определяется
двумя   единичными   векторами   $\rv   n'$   и   $\rv  n''$  и  двумя
действительными  положительными  числами  $K'$  и  $K''$:  $\rv  K=\rv
K'+\, i\rv  K''=K'\rv  n'+\, i K''\rv  n''$.  Ввиду  прозрачности  среды  1 $
K_i=K_r=K_1=k\sqrt{\varepsilon_1 \mu_1}$ --- действительное  число,  а
между  определяющими  вектор $\rv K_t$ величинами согласно (4.4) имеет
место соотношение
     $$K_t^{'2}-K_t^{''2}+2iK_t^{'}K_t^{''}\rv n'\rv n''=K_2^2=k^2
         \varepsilon_2\mu_2\,.\eqno(5.2)$$

     Плоскость, в которой лежат векторы $\rv n$ и~$\rv n_i$, примем за
плоскость $xz$,  что позволяет записать вектор $\rv  n_i$  через  свои
компоненты в виде $\rv n_i=\{\sin \varphi_i,0,\cos \varphi_i\}$. Таким
образом, падающая волна может быть представлена следующим образом:
     $$\rv E_i(\rv r)=\rv E_{0i}e^{ikn_1(x\sin\varphi_i+z\cos\varphi_i
         )}\,,\qquad \rv H_{0i}=\frac 1{W_1}[\rv n_i \rv E_{0i}]
          \,, \eqno(5.3)$$
где $n_1=\sqrt{\varepsilon_1\mu_1}$ --- \textit{показатель преломления},  $W_1=
\sqrt{\mu_1/\varepsilon_1}$ --- \textit{волновое сопротивление} среды 1.

     Ввиду полной однородности задачи  в  плоскости  $xy$  зависимость
векторов  поля  от  координат  $x$ и $y$ во всём пространстве должна
быть одинаковой и,  следовательно,  равны компоненты  $K_x$,~$K_y$  во
всех  трёх  волнах.  Поэтому  все волновые векторы лежат в плоскости
$xz$  и  $\rv  n_r=\{\sin   \varphi_r,   0,-\cos   \varphi_r\}$,   где
$\varphi_r$  ---  угол отражения.  Из равенства $x$-компонент волновых
векторов следуют соотношения
     $$K_1\sin \varphi_i=K_1\sin\varphi_r=K_t^{'}n_{tx}^{'},\quad n_
         {tx}^{''}=0\,,\eqno(5.4)$$
которые позволяют  с  помощью  (5.2)  найти  $z$-компоненты:
     $$n_{rz}=-n_{iz}=-\cos\varphi_i,\;
        K_t^{'}n_{tz}^{'}+iK_t^{''}n_{tz}^{''} =\sqrt{K_2^2 -K^{'2}
         _t n^{'2}_{tx}}=k\sqrt{\varepsilon_2\mu_2-\varepsilon_1\mu_1
         \sin^2\varphi_i}\,.\eqno(5.5)$$
Согласно этим формулам $\rv n_r=\{\sin \varphi_i, 0,-\cos \varphi_i\}$
и отражённая волна может быть записана в виде
     $$\rv E_r(\rv r)=\rv E_{0r}e^{ikn_1(x\sin\varphi_i-z\cos\varphi_i
         )}\,,\qquad
           \rv H_{0r}=\frac 1{W_1}[\rv n_r\rv E_{0r}]\,. \eqno(5.6)$$
Направляющие векторы преломлённой волны в  соответствии  с  (5.4)  и
(5.5) равны
     $$\rv n'_t=\{\sin\varphi_t,0,\cos\varphi_t\}\,,\qquad \rv n_t^
       {''}=\{0,0,1\}\,,\eqno(5.7)$$
где $\varphi_t$ --- угол преломления, определяемый равенством
     $$K'_t\sin\varphi_t=k\sqrt{\varepsilon_1\mu_1}\sin{\varphi_i}=
          kn_1\sin\varphi_i\,.\eqno(5.8)$$

     Если использовать для комплексного волнового числа в среде 2
стандартное обозначение (4.5)
     $$K_2=k\sqrt{\varepsilon_2\mu_2}=k(n_2+i\mbox{\ae} _2)\,,\eqno(5.9)$$
то следующая   из   (5.2)   система   уравнений   для  $K_{t}^{'}$
и~$K_{t}^{''}$ записывается в виде
     $$\left.\begin{array}{rcl}K_{t}^{'2} - K_{t}^{''2}&=&k^2(n_2^2-
         \mbox{\ae}_2^2),\\[.2cm]K'_tK_t^{''}\cos\varphi_t
         &=&k^2n_2\mbox{\ae}_2\,,\end{array}\right\}\eqno(5.10)$$
откуда после исключения  $\cos\varphi_t$ с помощью (5.8) получим:
     $$K_{t}^{'2}=\frac{k^2} 2\left[n_2^2-\mbox{\ae}_2^2+n_1^2\sin^2
         \varphi_i+\sqrt{(n_2^2-\mbox{\ae}_2^2-n_1^2\sin^2\varphi_i)^2+4n
         _2^2\mbox{\ae}_2^2}\,\right]\,,\eqno(5.11)$$
     $$K_{t}^{''2}=\frac{k^2} 2\left[\mbox{\ae}_2^2-n_2^2+n_1^2\sin^2
         \varphi_i+\sqrt{(n_2^2-\mbox{\ae}_2^2-n_1^2\sin^2\varphi_i)^2+4n
         _2^2\mbox{\ae}_2^2}\,\right]\,.\eqno(5.12)$$

Обозначив $K^{'}_t=\hat{k} n$, находим для преломлённой волны
следующее выражение:
     $$\rv E_t(\rv r)=\rv E_{0t}e^{ik\tilde n(x\sin\varphi_t +z\cos
         \varphi_t )}e^{-K_{t}^{''}z}\,,\qquad \rv H_{0t}=
         \frac 1{k\mu_2}[\rv K_t \rv E_{0t}]\,,\eqno(5.13)$$
причём соотношение  между  амплитудами  $\rv  E_{0t}$ и $\rv H_{0t}$
существенно зависит от поляризации падающей волны.

     Таким образом, между углами $\varphi_i,\,\varphi_r,\, \varphi_t$,
которые образуют волновые вектора $\rv K_i,\,\rv K_r,  \,\rv  K'_t$  с
нормалью  к  плоскости  раздела  $\rv  n$,  имеют  место  соотношения,
называемые   {\it   законами   отражения   и   преломления}:
     $$\sin\varphi_i=\sin \varphi_r\,,\qquad\sin{\varphi_t}=\frac
         {n_1}{\tilde n}\sin{\varphi_i}\,.\eqno(5.14)$$

     Для достаточно  хорошего  проводника  все   формулы   существенно
упрощаются, так как тогда $\displaystyle{\varepsilon_2\approx
i\frac{4 \pi\sigma_2}\omega}$ и ${\displaystyle{n_2=
\mbox{\ae}_2=\sqrt{  \frac  {2\pi \sigma_2\mu_2}\omega}}}$; в
результате получаем, что
     $$K'_t=K^{''}_t=k\tilde n\,,\qquad \tilde n=\sqrt{ \frac 12\Bigl
         (n^2_1\sin^2\varphi_i+\sqrt{n^4_1\sin^4\varphi_i+\bigl(\frac
         {4\pi\sigma_2\mu_2}{\omega}\bigr)^2}\Bigr)}\,.\eqno(5.15)$$
В случае   таких   проводников,   как   медь,   в   диапазоне  частот,
представляющих   интерес   для    электродинамики    СВЧ,    отношение
$\omega/\sigma$  очень  мало (для меди $\sigma=5\cdot 10^{17} c^{-1}$,
верхняя  граница  диапазона  $\omega$  порядка  $10^{11}  c^{-1}$)  и,
следовательно,  $\tilde  n$  --- большая величина,  которая может быть
записана в виде
     $$\tilde n=n_2\sqrt{\sqrt{1+\alpha^2}+\alpha}\,,\eqno(5.16)$$
где малый параметр
     $$\alpha=\displaystyle{\frac{\omega}{4\pi\sigma_2\mu_2}}
         \,n_1^2\,{\sin^2{\varphi_i}}\,.\eqno(5.17)$$
Угол  преломления, определяемый формулой (5.14), в данном случае зависит
только от $\alpha$:
     $$\sin{\varphi_t}=\sqrt{2\alpha(\sqrt{1+\alpha^2}-\alpha)}\,.
         \eqno(5.18)$$
Он очень  мал,  и  при  любом  угле  падения  $\varphi_i$  в проводнике
фактически  по  нормали  к  его  поверхности  распространяется  быстро
затухающая волна ($K_{t}''$ велико).

    Если среда  2  прозрачная,  то  $\mbox{\ae}_2=0$,~$\tilde  n=n_2=  \sqrt{
 \varepsilon_2 \mu_2}$ и система уравнений (5.10) имеет два решения:
     $$K'_t=kn_2,\quad K^{''}_t=0,\quad \sin\varphi_t=\frac {n_1}{n_2}
         \sin\varphi_i,\quad n'_t=\{\sin\varphi_t,0,\cos\varphi_t\}
         \eqno(5.19)$$
и
     $$K'_t=kn_1\sin\varphi_i\,,\quad \sin\varphi_t=1\,,\quad
         K_t^{''}=k\sqrt{n^2_1\sin^2\varphi_i-n^2_2}\,,\quad n'_t=\{1,
         0,0\}\,.\eqno(5.20)$$
Первое решение соответствует однородной плоской волне
    $$\rv E_t(\rv r)=\rv E_{0t}e^{ikn_2(x\sin\varphi_t+z\cos\varphi
         _t)}\,,\qquad \displaystyle{\rv H_{0t}=\frac 1{W_2}
           [\rv n_t\rv E_{0t}]}\eqno(5.21)$$
и всегда  реализуется  при  падении волны из оптически менее плотной в
более плотную среду ($n_1<n_2$);  в противном случае  ($n_2<n_1$)  оно
пригодно  лишь  при  достаточно  малых углах падения,  удовлетворяющих
условию
     $$\sin\varphi_i<\frac{n_2}{n_1}\,.\eqno(5.22)$$
\begin{wrapfigure}[15]{l}{8cm}
\begin{picture}(80,50)
\put(-1,50){\special{em:graph fig5-2.bmp}}
\end{picture}
\hbox to 8cm{\footnotesize\hfil
{Рис.~5.2.~Полное отражение: сплошные линии --- }}
\hbox to 8cm
{\footnotesize {плоскости равных фаз, штрих-пунктирные ---}\hfil}
\hbox to 8cm
{\footnotesize {плоскости равных амплитуд преломлённой волны.}}
\end{wrapfigure}

     При нарушении  этого  условия реализуется второе решение (5.20) и
преломлённая волна становится неоднородной.  Эта неоднородная  волна
распространяется не вглубь среды,  а вдоль границы раздела,  и поэтому
говорят о полном отражении или о полном  внутреннем  отражении  (менее
удачный и устаревший термин).

     Остановимся на этом случае более  подробно.  Пусть  для  простоты
записи $\varepsilon_2=1$,~$\mu_2=1$,  то есть падение волны происходит
на   границу   среды   с   пустотой,   и   имеет   место   неравенство
$n_1\sin\varphi_i>1$,   противоположное   (5.22).  Тогда  составляющие
действительных  векторов  $\rv  K'_t$  и   $\rv   K''_t$,   образующих
комплексный вектор $\rv K_t$, равны
     $$\rv K_t^{'}=\{kn_1\sin\varphi_i, 0, 0\},\quad\rv K_t^{''}=
         \{0, 0,k\sqrt{n_1^2\sin^2\varphi_i-1}\}\,,\eqno(5.23)$$
и прошедшую волну можно записать в виде
     $$\rv E_t(\rv r)=\rv E_{0t}e^{ikn_1x\sin\varphi_i}e^{-k
         \sqrt{n_1^2\sin^2\varphi_i-1}\,z}\,.\eqno(5.24)$$

     Плоскости равных  амплитуд  в  этой  волне  параллельны плоскости
раздела,  а плоскости равных фаз нормальны и к плоскости падения  и  к
плоскости  раздела.  Если  в  падающей волне магнитное поле направлено
вдоль оси $y$,  то и в прошедшей отлична  от  нуля  только  компонента
$H_{0ty}$,  через  которую  с помощью формул (5.1) и (5.23) выражаются
компоненты электрического поля:
     $$E_{0tx}=i\sqrt{n_1^2\sin^2\varphi_i-1}\,H_{0ty}\,,\quad E_
         {0tz}=-n_1\sin\varphi_i\,H_{0ty}\,.\eqno(5.25)$$
Вектор Умова-Пойнтинга    преломлённой    волны    направлен   вдоль
поверхности раздела, так как
     $$\overline{{\gv S}}_z=\frac{c}{8\pi}\re {E_x H_y^{*}}=\frac c
         {8\pi}\re {i\sqrt{n_1^2\sin^2{\varphi_i}-1}}\,|H_{0ty}|^2=0\,,
         \eqno(5.26)$$
     $$\quad\overline{\gv S}_x=-\frac c{8\pi}\re E_zH_y^{*}=
         \frac c{8\pi}n_1\sin{\varphi_i}|H_{0ty}|^2e^{-2k\sqrt{n_1^2
         \sin^2\varphi_i-1}\,z}\,.\eqno(5.27)$$
\hspace{1mm}

     Отметим характерные для  прошедшей  волны  свойства,  которые  не
представляются  очевидными  для  плоской  волны,  распространяющейся в
пустом полупространстве:

     --- фазовая скорость волны  (вдоль  оси  $x$)
$v_{\footnotesize\textit{ф}}=c/(n_1\sin\varphi_i)<c$ , такая волна называется
 {\it медленной};

     --- поле волны сосредоточено вблизи плоскости $z=0$, такая волна
называется {\it поверхностной};

     --- волна  не  является  поперечной,  поскольку  у  неё имеется
продольная  компонента   электрического   поля   $E_x$.  \\
Волна   с перечисленными    свойствами   может  быть использована
для  ускорения заряженных частиц.

     Найдём теперь  соотношения  между  амплитудами  рассматриваемых
волн, которые обычно называют {\it формулами Френеля}. Эти соотношения
различаются в зависимости от того, лежит ли вектор электрического поля
$\rv   E_{0i}$   в   плоскости   падения   (плоскость   $xz$)   или  в
перпендикулярной к ней плоскости.  Произвольная падающая волна  всегда
может   быть   представлена  в  виде  суперпозиции  двух  таких  волн.
Установившейся терминологии относительно этих двух волн  в  литературе
нет,   поэтому  будем  называть  волной  первой  поляризации  волну  с
компонентами поля  $E_x$,~$E_z$,~$H_y$  и  волной  второй  поляризации
волну с компонентами поля $H_x$,~$H_z$,~$E_y$.

     Для первой поляризации граничные условия равенства тангенциальных
компонент поля с учётом (5.1) и (5.7) сводятся к двум уравнениям
     $$H_{0iy}+H_{0ry}=H_{0ty}\,,\qquad W_1\cos\varphi_i(H_{0iy}-H_
         {0ry})=\frac 1{k\varepsilon_2}(K'_t\cos\varphi_t+iK_t^{''})H
         _{0ty}\,,\eqno(5.28)$$
из которых  с  помощью  равенства $K'_t\cos\varphi_t+iK_t^{''} =k\sqrt
{\varepsilon_2^2    W_2^2-\varepsilon_1^2    W_1^2\sin^2\varphi_i}\,,$
следующего   из   (5.5),   после   несложных  преобразований  получаем
соотношения между амплитудами:
     $$H_{0ry}=\;\frac{\cos
         \varphi_i-\sqrt{\displaystyle{\frac{W_2^2}{W_1^2}-\frac{
         \varepsilon_1^2}{\varepsilon_2^2}\sin^2\varphi_i}}}{\cos
         \varphi_i+\sqrt{\displaystyle{\frac{W_2^2}{W_1^2}-{\frac{
         \varepsilon_1^2}{\varepsilon_2^2}}\sin^2\varphi_i}}}\:
         H_{0iy}\,, \eqno(5.29)$$
     $$H_{0ty}=\;\frac{2\cos \varphi_i}{\cos\varphi_i
         +\sqrt{\displaystyle{\frac{W_2^2}{W_1^2}-{\frac{\varepsilon_
         1^2}{\varepsilon_2^2}}\sin^2\varphi_i}}}\:H_{0iy}\,.
         \eqno(5.30)$$
Аналогично рассматривается и вторая поляризация,  только в этом случае
следует выразить компоненту  поля  $H_x$  через  $E_y$;  в  результате
получим соотношения
     $$E_{0ry}=\;\frac{\cos \varphi_i-\displaystyle{\sqrt{\frac{W_1^2}
         {W_2^2}- \displaystyle{\frac{\mu_1^2}{\mu_2^2}}\sin^2\varphi
         _i}}}{\cos \varphi_i+\displaystyle{\sqrt{\frac{W_1^2}{W_2^2}-
         \frac{\mu_1^2}{\mu_2^2}\sin^2\varphi_i}}}\:E_{0iy}\,,
         \eqno(5.31)$$
     $$E_{0ty}=\;\frac{2\cos \varphi_i}{\cos \varphi_i+\displaystyle
         {\sqrt{\frac{W_1^2}{W_2^2}- \frac{\mu_1^2}{\mu_2^2}\sin^2
         \varphi_i}}}\:E_{0iy}\,.\eqno(5.32)$$

     {\it Коэффициент  отражения}  $R$  определяется   как   отношение
среднего по времени потока энергии в отражённой волне, направленного
от границы,  к потоку в падающей,  направленному к границе. Эти потоки
определяются  $z$-компонентами вектора Умова-Пойнтинга соответствующей
волны и поскольку  обе  волны  распространяются  в  среде  1,  то  это
отношение   равно   отношению  квадратов  модулей   комплексных амплитуд
любой  тангенциальной компоненты поля.  С помощью  формул
(5.29),  (5.31)  для  коэффициентов  отражения  волн  первой  и второй
поляризаций получаем соответственно следующие выражения:
     $$R_1=\left |\frac{\cos \varphi_i-\displaystyle{\sqrt{\frac{W_2
         ^2}{W_1^2}-\frac{\varepsilon_1^2}{\varepsilon_2^2}\sin^2
         \varphi_i}}}{\cos \varphi_i+\displaystyle{\sqrt{\frac{W_2^2}
         {W_1^2}-\frac{\varepsilon_1^2}{\varepsilon_2^2}\sin^2\varphi
         _i}}}\right |^2,\quad R_2=\left |\frac{\cos \varphi_i-
         \displaystyle{\sqrt{\frac{W_1^2}{W_2^2}-\frac{\mu_1^2}{\mu_2
         ^2}\sin^2\varphi_i}}}{\cos \varphi_i+\displaystyle{\sqrt{
         \frac{W_1^2}{W_2^2}- \frac{\mu_1^2}{\mu_2^2}\sin^2\varphi_i
         }}}\right |^2.\eqno(5.33)$$
Очевидно, что оба коэффициента отражения стремятся к единице  в  любом
из   четырёх   предельных   случаев:   $W_1\longrightarrow  0$,~$W_2
\longrightarrow 0$, ~$W_1 \longrightarrow \infty$,~$W_2\longrightarrow
\infty.$  Кроме  этого,  коэффициенты  отражения  $R_1$  и~$R_2$ равны
единице и в том  случае,  когда  обе  среды  прозрачны  и  выполняется
условие
     $$\displaystyle{\sqrt{\frac{\varepsilon_2\mu_2}
         {\varepsilon_1\mu_1}}\leqslant\sin\varphi_i}\,,\eqno(5.34)$$
то есть при рассмотренном выше явлении полного  отражения.

     Если аналогичным образом ввести коэффициент прохождения  $T$,  то
есть определить его как отношение среднего по времени потока энергии в
прошедшей волне,  направленного по нормали к границе,  к  аналогичному
потоку  в  падающей  волне,  то  нетрудно  убедиться,  что выполняется
очевидное  из  соображений  сохранения  энергии  соотношение  $1-R=T$.
Многие   авторы   называют   коэффициентами  отражения  и  прохождения
множители,  связывающие соответствующие  амплитуды  полей  в  формулах
(5.29)--(5.32).  Тогда в случае первой поляризации приходится говорить
о  коэффициенте  отражения  по  магнитному  полю,  в   случае   второй
поляризации  ---  по  электрическому  полю.  Это  представляется менее
удобным,  в частности,  и потому,  что реально  наблюдаемой  на  опыте
величиной почти всегда является поток энергии.

     При нормальном падении волны на границу раздела $(\sin  \varphi_i
=0)$  даже при наличии поглощения в среде 2 прошедшая волна
остаётся плоской   однородной   волной,   затухающей   в
направлении   своего распространения.  При  этом  $\tilde  n  =
n_2$,  $K_t''=k\mbox{\ae}_2$ и формулы~(5.13) сильно упрощаются:
     $$\rv E_t(\rv r)=\rv E_{0t}e^{ikn_2z}e^{-k\mbox{\scriptsize{\ae}}_2z},\qquad
         \rv H_{0t}=\frac 1{W_2}[\rv n \rv E_{0t}]\,.\eqno(5.35)$$
Коэффициенты отражения  в этом случае для обеих поляризаций совпадают,
что  и   естественно   ввиду   полного   равноправия   соответствующих
плоскостей, и становятся равными
     $$R=\left |\frac{W_1-W_2}{W_1+W_2}\right |^2.\eqno(5.36)$$
Если $W_1=W_2$,  то  коэффициент  отражения  равен  нулю.  Соотношение
(5.36)  лежит  в  основе  многочисленных  разработок покрытий летающих
объектов  с  целью  максимального  уменьшения  отражённой   мощности
радиолокационного сигнала.  Поскольку падающий сигнал распространяется
в  воздухе,  где  практически  $\varepsilon=1$,~$\mu=1$,  то  условием
отсутствия отражения является равенство $\varepsilon_2=\mu_2$.  Тело с
таким покрытием обладает свойствами радиолокационной <<невидимки>>.

     При рассмотрении падения плоской волны на границу раздела полезно
использовать  соотношение  подобия,  имеющее  место  в электродинамике
сплошных  сред.  Пусть  нам  известно  решение  задачи  для  какого-то
распределения  в пространстве параметров среды $\widetilde \varepsilon
(x,y,z)$  и  $\widetilde  \mu(x,y,z)$  и  сторонних  токов,  то   есть
найденное решение удовлетворяет системе уравнений
     $$\left.\begin{array}{rcl}\rot\widetilde{\rv E}&\!=\!&\phantom{-}
         i\widetilde{k}\widetilde{\mu}\widetilde{\rv H}-\displaystyle
         {\frac{4\pi}c}\widetilde{\rv j}^m,\\[0.25cm]\rot\widetilde
         {\rv H}&\!=\!&-i\widetilde{k}\,\widetilde{\varepsilon}\,
         \widetilde{\rv E}+\displaystyle{\frac{4\pi} c}\widetilde
         {\rv j}^e\,.\end{array}\right\}\,\eqno(5.37)$$
Возьм\"ем теперь два произвольных  комплексных  числа  $\varepsilon_1$
и~$\mu_1$  и введём новые величины $\rv E$,~$\rv H$,~$\rv j^e$,~$\rv
j^m$,~$ \varepsilon$,~$\mu\,,k$ посредством соотношений
     $$\left.\begin{array}{l}\displaystyle{\rv E=\frac{\widetilde
         {\rv E}}{\sqrt{\varepsilon_1}}\,,\quad\rv H=\frac{\widetilde
         {\rv H}}{\sqrt{\mu_1}}\,,\quad{\rv j}^e=\frac{\widetilde{\rv
         j}^e}{\sqrt{\mu_1}}\,,\quad{\rv j}^m=\frac{\widetilde{\rv j}
         ^m}{\sqrt{\varepsilon_1}}}\,,\\[0.2cm]\displaystyle{\qquad
         \varepsilon=\varepsilon_1\widetilde\varepsilon\,,\qquad\mu=
         \mu_1\widetilde\mu\,,\qquad k=\frac{\widetilde k}{\sqrt{
         \varepsilon_1\mu_1}}}\,.\end{array}\right\}\eqno(5.38)$$
Прямой подстановкой   (5.38)  в  (5.37)  убеждаемся,  что  введённые
величины удовлетворяют системе уравнений
     $$\left.\begin{array}{rcl}\rot\rv E&\!=\!&\phantom{-}ik\mu\rv H-
         \displaystyle{\frac{4\pi} c}\rv j^m,\\[0.25cm]\rot\rv H&\!=\!&
         -ik\,\varepsilon\rv E+\displaystyle{\frac{4\pi} c}\rv j^e\,.
         \end{array}\right\}\eqno(5.39)$$
Поэтому, если  нам  известно решение задачи о падении плоской волны из
среды с параметрами $\varepsilon_1$,~$\mu_1$ на плоскость  раздела  со
средой  с  параметрами  $\varepsilon_2$,~$\mu_2$,  то одновременно нам
известно,  например,  и решение задачи о падении волны из  пустоты  на
плоскую границу среды с параметрами $\tilde \varepsilon =\varepsilon_2
/\varepsilon_1$,~$ \tilde\mu=\mu_2/\mu_1.$

     До сих   пор  распространение  плоской  волны  рассматривалось  в
однородной среде или в среде,  где плоская граничная поверхность делит
все пространство на две однородные по своим электромагнитным свойствам
среды.  Очевидно,  что  полученные  решения  можно
распространить и на случай среды, состоящей из любого числа однородных
слоёв с  параллельными  плоскими  границами.  В  каждом  слое  будет
существовать  прямая  и  обратная  плоская  волна,  амплитуды  которых
необходимо  определить   с   помощью   граничных   условий   на   всех
поверхностях.  Задача  сведётся  к  системе  линейных алгебраических
уравнений относительно неизвестных амплитуд.

     Помимо этого существует ещё один класс задач, который допускает
решение  в  виде   плоских   волн   с   амплитудами,   зависящими   от
пространственной  координаты, и  который  имеет  широкое  приложение к
многочисленным реальным процессам, происходящим в окружающем нас мире.
Эти  задачи  касаются  вопросов  прохождения  плоской  волны  в среде,
свойства которой меняются  вдоль  одной  пространственной  координаты.
Ниже  рассматривается простейший случай прозрачной среды без магнитных
свойств ($\mu=1$), диэлектрическая проницаемость $\varepsilon$ которой
меняется вдоль оси $z$.  Вдоль этой же оси направлен и волновой вектор
плоской  волны.  Пусть  электрическое  поле  в  волне  поляризовано  в
плоскости   $xz$.   В   этом  случае  компонента  $E_x$  удовлетворяет
сравнительно простому волновому уравнению
     $$\frac{d^2E_x}{dz^2}+k^2\varepsilon(z)E_x=0\,,\eqno(5.40)$$
получаемому из уравнений Максвелла (4.1)  с  учётом  того,  что
при выбранной   поляризации   электрического   поля   из уравнения
$\div \varepsilon \rv E=0$ следует уравнение $\div \rv E=0$.  Для
некоторых конкретных функций $\varepsilon(z)$ удаётся получить
строгое решение уравнения (5.40) в  замкнутом аналитическом  виде,
выраженное  через специальные  функции,  но в данном случае нас
больше интересует хоть и {\it  приближённое}, но  {\it   общее
решение},   пригодное   для произвольных функций $\varepsilon(z)$,
принадлежащих к определённому классу. Этот  класс  ограничен
функциями,  которые  достаточно  {\it медленно меняются},  точнее,
их относительное изменение на расстояниях порядка длины волны
много меньше единицы.

     Такое приближённое  общее  решение  можно получить,  представив
искомую функцию $E_x(z)$ в виде
     $$E_x(z)=\Psi(z)e^{i\varphi(z)}\;,\eqno(5.41)$$
подразумевая при   этом,   что   функция   $\Psi(z)$   ---    медленно
изменяющаяся,  так что масштаб её изменения $L$ велик по сравнению с
$1/k$. Подстановка (5.41) в (5.40) приводит к уравнению
     $$\Psi'' +2i\Psi'\varphi'+i\varphi''\Psi+[k^2\varepsilon(z)-
         {\varphi'}^2]\Psi=0\,.\eqno(5.42)$$
Положим теперь
     $$\varphi(z)=\pm k\int\limits_0^z\sqrt \varepsilon(\xi)\,d\xi\,,
         \eqno(5.43)$$
обратив тем самым в нуль четвёртое слагаемое в левой  части
(5.42). Первое слагаемое имеет порядок малости $ 1/L^2$, а два
последующих --- $k/L$,  поскольку согласно (5.43) $\varphi'\sim
k$,  а~$\varphi''\sim k/L$.  На  этом  основании  можно
попытаться решать уравнение (5.42), отбросив первое слагаемое,  то
есть  старшую  производную.  Тогда  для неизвестной  функции
$\Psi(z)$  остается  простое  уравнение  первого порядка
     $$2\varphi'\Psi'+\varphi''\Psi=0\,,\eqno(5.44)$$
которое легко интегрируется:
     $$\Psi(z)=\frac 1{\sqrt[4]{\varepsilon(z)}}\,.\eqno(5.45)$$

     Таким образом, приближённое решение уравнения (5.40) может быть
представлено в виде
     $$E_x(z)=\frac{A}{\sqrt[4]{\varepsilon(z)}}e^{ik\int\limits_0^z\sqrt
         {\varepsilon(\xi)}\,d\xi}+\frac{B}{\sqrt[4]{\varepsilon(z)}}
         e^{-ik\int\limits_0^z\sqrt{\varepsilon(\xi)}\,d\xi}\,,
         \eqno(5.46)$$
где $A$ и~$B$ --- произвольные постоянные.  Аналогичное
приближённое решение  уравнения Шредингера в квазиклассическом
пределе,  получаемое таким  же  способом  отбрасывания  члена   со
старшей   производной, называется решением
Венцеля-Крамерса-Бриллюэна (ВКБ) и находит широкое применение в
квантовой механике.  Здесь  фактически  используются  две идеи:
выделение   в   решении   быстро  осциллирующего  множителя  и
отбрасывание в уравнении для  медленно  изменяющейся  функции
старшей производной  (уравнение  эллиптического  типа превращается
в уравнение параболического типа).

     Первый член в правой части решения (5.46) естественно сопоставить
с  плоской   волной   переменной   амплитуды,   распространяющейся   в
положительном  направлении  оси  $z$,  второй  ---  с такой же волной,
направленной в обратную сторону.  Эти две волны представляют собой два
линейно   независимых   решения   уравнения   (5.40).
\begin{wrapfigure}[14]{l}{8cm}
\begin{picture}(80,55)
\put(-1,50){\special{em:graph  fig5-3.bmp}}
\end{picture}
\hbox to 8cm {\hfil\footnotesize{Рис.~5.3.~Плавный   переходный
слой в диэлектрике.}\hfil}
\end{wrapfigure}
     Посмотрим теперь,  как  же  на  основе  этих двух волн может быть
построен итерационный  процесс  решения  задачи  о  согласующем  слое.
Предположим (см.  рис. 5.3), что слева от плоскости $z=0$ расположена
однородная среда с диэлектрической проницаемостью  $\varepsilon_1$,  а
справа  от  плоскости  $z=l$  ---  однородная  среда c диэлектрической
проницаемостью $\varepsilon_2$.  Внутри же  переходного  слоя  $0<z<l$
функция   $\varepsilon(z)$   плавно  меняется  от  $\varepsilon_1$  до
$\varepsilon_2$.  Пусть слева на переходный слой падает плоская волна,
поляризованная  в  плоскости  $xz$,  с комплексной амплитудой $E_{ix}=
\displaystyle   {\frac    1{    \sqrt[4]    {\varepsilon_1}}e^{ik\sqrt
{\varepsilon_1}z}}$. Если  бы толщина переходного слоя равнялась нулю,
то  согласно (5.1) и формулe  Френеля  (5.29)  амплитуда  отражённой
волны составила бы
     $$E_{rx}=-\displaystyle{\frac{\sqrt{\varepsilon_2}-\sqrt
         {\varepsilon_1}}{\sqrt{\varepsilon_2}+\sqrt{\varepsilon_1}}}
         \,E_{ix}\,.\eqno(5.47)$$

     Будем искать   поле   в   области   $0<z<l$   в   виде  (5.46)  с
коэффициентами $A(z)$ и $B(z)$, являющимися функциями $z$:
     $$E_x(z)=\frac{A(z)}{\sqrt[4]{\varepsilon(z)}}e^{ik\int\limits_0^z
         \sqrt{\varepsilon(\xi)}\,d\xi}+\frac{B(z)}{\sqrt[4]
         {\varepsilon(z)}}e^{-ik\int\limits_0^z\sqrt{\varepsilon(\xi)}
         \,d\xi}\,.\eqno(5.48)$$
Поскольку теперь у нас вместо одной неизвестной функции $E_x(z)$ стало
две --- $A(z)$ и~$B(z)$, то можно связать их дополнительным условием
     $$\frac{dE_x}{dz}=ik\sqrt[4]{\varepsilon(z)}A(z)e^{ik\int\limits_0
         ^z\sqrt{\varepsilon(\xi)}\,d\xi}-ik\sqrt[4]{\varepsilon(z)}
         B(z)e^{-ik\int\limits_0^z\sqrt{\varepsilon(\xi)}\,d\xi}\,,
         \eqno(5.49)$$
позволяющим получить  из  уравнения  (5.40)  наиболее  простую систему
уравнений для функций $A(z)$ и $B(z)$:
     $$\frac{dA}{dz}=\frac 1{4\varepsilon}\frac{d\varepsilon}{dz}B(z)
         e^{-2ik\int\limits_0^z\sqrt{\varepsilon(\xi)}\,d\xi}\,,
         \eqno(5.50\mbox{\textit а})$$
     $$\frac{dB}{dz}=\frac 1{4\varepsilon}\frac{d\varepsilon}{dz}A(z)
         e^{2ik\int\limits_0^z\sqrt{\varepsilon(\xi)}\,d\xi}\,.
         \eqno(5.50{\textit б})$$
Отметим, что до сих пор не совершалось никаких приближённых
операций и поэтому система  уравнений  (5.50)  эквивалентна
уравнению  (5.40); появление  в  формулах  (5.50)  малых
множителей  не  должно вызывать удивления хотя бы потому,  что
именно такого вида коэффициент  следует из   (5.47),   если
положить  $\varepsilon_2  =\varepsilon_1  +\Delta \varepsilon$,~$
\Delta\varepsilon \ll \varepsilon_1.$

     Приближённый характер   приводимых   ниже   решений   связан  с
итерационным процессом их получения.  Предположим на первом шаге этого
процесса,  что  поскольку  производная  функции  $A(z)$,  определяемая
уравнением (5.50\textit{а}), пропорциональна малой величине $\displaystyle\frac
{d\varepsilon}{dz}$,  то во всём слое $A(z)=1$, то есть $A(z)$ равно
своему граничному значению при $z=0$.  Это позволяет на следующем шаге
решить уравнение (5.50\textit{б}) для функции $B(z)$ в квадратурах:
     $$B(z)=\int\limits_l^z\frac 1{4\varepsilon}\frac {d\varepsilon}
         {dz}e^{2ik\int\limits_0^{z'}\sqrt{\varepsilon(\xi)}\,d\xi}\,
         dz'\,,\eqno(5.51)$$
где учтено,  что  при $z=l$ приходящей справа волны нет.  В результате
амплитуда отражённой плоской волны в плоскости $z=0$ составит
     $$E_{rx}=-\frac 1{\sqrt[4]{\varepsilon_1(0)}}\int\limits_0^l
         \frac1{4\varepsilon}\frac{d\varepsilon}{dz}e^{2ik\int\limits
         _0^{z'}\sqrt{\varepsilon(\xi)}\,d\xi}\,dz'\,.\eqno(5.52)$$
На следующем   шаге   можно   получить   уточнённое   значение
для коэффициента  $A(z)$,  соответствующего  распространяющейся  в
прямом направлении волне:
     $$A(z)=1+\int\limits_0^z\frac 1{4\varepsilon}\frac{d\varepsilon}
         {dz}B(z)e^{-2ik\int\limits_0^{z'}\sqrt{\varepsilon(\xi)}\,d
         \xi}\,dz'\,.\eqno(5.53)$$
Поскольку разложение ведётся по малому  параметру,  то
итерационный процесс   быстро   сходится.  В  результате  для
электрического  поля отражённой  волны  в  плоскости  $z=0$
получаем   заметно   меньшее значение,  чем  следующее  из формулы
Френеля  (5.47).  Причём это значение тем меньше,  чем шире
переходный слой;  кроме того, чем более плавным  является
изменение $\varepsilon(z)$ в согласующем слое,  тем более
широкополосным при том же  значении  коэффициента  отражения  он
является.  Как показывают численные расчёты, уже формула (5.52)
даёт решение с хорошей точностью практически для любой достаточно
плавной зависимости $\varepsilon(z)$.

%\end{document}

\newpage
\oddsidemargin=-0.4mm \evensidemargin=-0.4mm
\topmargin=-0.4mm
\headsep=7mm
\textheight=231.875mm
\textwidth=160mm
\mathsurround=2.5pt
\unitlength=1mm
%\begin{document}
%\input{macr.tex}
\thispagestyle{empty}
%\addtocounter{page}{61}
%\baselineskip=\normalbaselineskip
%\baselineskip=0.955\normalbaselineskip

\begin{center}\subsubsection*{6. Теория скин-эффекта. Граничное условие
Щукина-Леонтовича}\end{center}\vspace*{0.5cm}

\markboth{Глава 2. Плоские волны}{6. Теория скин-эффекта. Граничное
условие Щукина-Леонтовича}

\begin{center}\begin{minipage}[c]{0.75\textwidth}
\footnotesize{\parindent=0.5cm
         Скин-эффект при   нормальном   падении   плоской   волны   на
         поверхность   проводника.  Понятие  идеального  проводника  и
         граничные  условия  на  его  поверхности.  Граничные  условия
         Щукина-Леонтовича  и критерии их применимости.  Выражение для
         потока энергии в  металл.  Поверхностный  импеданс  металлов.
         Скин-эффект в цилиндрическом проводе.
}\end{minipage}\end{center}\vspace*{0.5cm}

     Вернёмся к задаче о нормальном падении плоской волны на плоскую
поверхность однородной среды с потерями.  Напомним,  что нормаль $\rv
n$  к  этой  поверхности направлена в сторону среды и совпадает с осью
$z$.  Прошедшая волна в  этом  случае  представляет  собой  затухающую
однородную плоскую волну,  поле которой согласно (5.35) записывается в
виде
     $$\rv E_t(z)=\rv E_{0t}e^{ikn_2z}e^{-k\mbox{\scriptsize{\ae}}_2z}\;,
     \quad\rv H_{0t}=\frac  1 {W_2}[\rv n\rv E_{0t}]\;,\eqno(6.1) $$
где $n_2$,~$\mbox{\ae}_2$  и  $W_2$  соответственно  показатель
преломления, коэффициент поглощения и волновое  сопротивление
среды.  Пусть  волна падает  из пустоты ($\varepsilon_1=\mu_1=1$),
её электрическое поле поляризовано в плоскости $xz$,  а
поглощающая среда  является  хорошим проводником,   для   которого
$\displaystyle{   \varepsilon_2\approx
i\frac{4\pi\sigma}{\omega}}$ (далее  нижний индекс  ${}_{2}$  у  всех
характеристик среды опускается).

     В этом случае
     $$n=\mbox{\ae}=\displaystyle{\sqrt{\frac{2\pi\sigma\mu}{\omega}}},\qquad
         W=\sqrt\frac\mu\varepsilon=(1-i)\sqrt\frac{\mu\omega}{8\pi
         \sigma}\eqno(6.2)$$
и  (6.1)  можно  переписать  в  виде
     $$E_{tx}(z)=E_{0tx}\displaystyle{e^{i\frac z\delta}e^{-\frac  z
         \delta}},\qquad H_{0ty}=\frac 1 W E_{0tx}\,,
         \eqno(6.3)  $$
где введена величина $\delta$, имеющая размерность длины:
     $$\delta=\frac  c{\sqrt{2\pi\sigma\mu\omega}}\;.\eqno(6.4) $$
Эта величина,   называемая   {\it   толщиной   (глубиной)  скин-слоя},
определяет степень убывания поля  вглубь  среды.  Для  металлов  из-за
высокой их проводимости в диапазоне СВЧ она очень мала.  Так, для меди
($\sigma=5\cdot 10^{17}$~c$^{-1}$,  $\mu\approx 1$) при $\lambda=3$~см
толщина  скин-слоя  $\delta=0,7$~мкм.  Быстрое убывание амплитуды поля
при углублении в проводник по нормали  к  его  поверхности  называется
{\it скин-эффектом}.

     Волновое число  в  металле  $K$  однозначно  связано  с  глубиной
скин-слоя:
     $$K=k(n+i\mbox{\ae})=\frac{1+i}{\delta}\,,\eqno(6.5)$$
так что длина волны в металле
     $$\lambda=\frac{2\pi}{\re K}=2\pi\delta\eqno(6.6)$$
составляет несколько скин-слоёв. При этом трудно говорить о волновом
процессе,  поскольку поле затухает практически до нуля на длине волны.
С волновым сопротивлением $W$ глубина скин-слоя связана соотношением
     $$W=\frac{1-i}2 \mu k\delta\,,\eqno(6.7)$$
а амплитуды  поля  прошедшей  и отражённой волн связаны с амплитудой
падающей волны $E_{0ix}=H_{0iy}$  формулами  Френеля  (5.29),  (5.30),
которые в данном случае можно записать в виде
     $$E_{0tx}=\frac{2W}{1+W}E_{0ix},\quad H_{0ty}=\frac 2{1+W}E_{0ix},
       \quad  E_{0rx}=-H_{0ry}=-\frac  {1-W}{1+W}E_{0ix}\,.\eqno(6.8)$$

     Средняя плотность   потока   мощности  через  плоскость  $z=0$  в
проводник составляет
     $$\overline{\gv S}_z=\frac
     c{8\pi}\re(E_{0tx}H_{0ty}^{*})=\frac
        c{8\pi}\re W|H_{0ty}|^2= \frac{\sigma\delta}4|E_{0tx}|^2=
         \frac c{2\pi}\frac{\re W}{|1+W|^2}|E_{0ix}|^2\,.\eqno(6.9)$$
Объёмная плотность тока,  наводимого в  металле,
экспоненциально убывает вглубь проводника:
     $$j_x(z)=\sigma E_{tx}(z)=\sigma E_{0tx}\displaystyle{e^{i\frac z
         \delta}e^{-\frac z\delta}}\,,\eqno(6.10)$$
а полный ток (на единицу длины по  $y$)  по  всей  глубине  проводника
равен
     $$I_x=\int\limits_0^\infty j_x(z)\,dz=(i+1)\frac{\sigma\delta}2
         E_{0tx}=\frac c{4\pi W}E_{0tx}=\frac c{4\pi}H_{0ty}\,.
         \eqno(6.11) $$
Джоулевы потери   в   проводнике,  равные  работе  поля  над  током
проводимости и выделяющиеся  в  виде  тепла,  на  единицу  поверхности
составляют
     $$Q=\frac 12\re\int\limits_0^\infty j_x(z)E_{tx}^{*}(z)\,dz=\frac
         {\sigma} 2 |E_{0tx}|^2\int\limits_0^\infty e^{-\frac{2z}
         \delta}\,dz=\frac {\sigma\delta} 4 |E_{0tx}|^2\eqno(6.12)$$
и, естественно,  равны  плотности  потока  мощности  (6.9) в проводник
через его поверхность.

     Во многих  практических  задачах  распределение поля в металле не
представляет интереса и достаточно знать поле вне его, в данном случае
в  пустоте,  откуда  падает плоская волна и где возникает отражённая
волна.  Поэтому желательно иметь возможность решать задачу для области
вне проводника, не учитывая распределение поля в самом проводнике. Для
этого  необходимо  задать  на  границе   раздела   сред   условие   на
тангенциальные компоненты поля в интересующей нас области, учитывающее
свойства среды,  расположенной по другую сторону  границы.  Простейшим
примером  является  задача о падении плоской волны на поверхность {\it
идеального   проводника}   ---   проводника   с   бесконечно   высокой
проводимостью   $\sigma$  (магнитная  проницаемость  $\mu$  остаётся
конечной).

     Посмотрим, что  происходит  со всеми величинами в полученном выше
решении при $\sigma\longrightarrow\infty$. К нулю стремятся
     $$W\sim\sigma^{-\frac 12}\longrightarrow 0,\quad\delta\sim\sigma^
         {-\frac 12}\longrightarrow 0,\quad E_{0tx}\sim\sigma^
         {-\frac 12}\longrightarrow 0\,,\quad Q=\overline{\gv S}_z\sim
         \sigma^{-\frac 12}\longrightarrow 0\,,\eqno(6.13)$$
а величины  $H_{0ty}$,~$I_x$,~$E_{0rx}$,  как  это  следует из (6.8) и
(6.11), остаются конечными:
     $$H_{0ty}=H_{0iy}+H_{0ry}\longrightarrow 2H_{0iy}\,,\quad
           I_x\longrightarrow\frac c{2\pi} E_{0ix}\,,\quad E_{0rx}
           \longrightarrow -E_{0ix}\,.\eqno(6.14)$$
При этом полный ток (на  единицу  длины  вдоль  $y$),  который  теперь
становится  поверхностным,  по-прежнему связан со значением магнитного
поля на поверхности соотношением (6.11).

     Решение, полученное   предельным  переходом  $\sigma\to  \infty$,
находится без труда,  если на поверхности проводника $z=0$ потребовать
равенства нулю тангенциальной компоненты электрического поля:
     $$[\rv n \rv E]=0\,.\eqno(6.15)$$
В данном случае это условие принимает вид $E_{0ix}+E_{0rx}=0 $, откуда
сразу следуют предельные значения (6.14) для полей.  Поскольку в самом
идеальном  проводнике  зависящее  от  времени  поле  равно  нулю,  что
непосредственно     следует     из     уравнений     Максвелла     при
$\sigma\longrightarrow\infty$,   то,   следовательно,   тангенциальная
составляющая  магнитного  поля  терпит  разрыв.  Согласно   граничному
условию  (1.18)  комплексная амплитуда соответствующего поверхностного
тока равна
     $$\rv I=-\frac c{4\pi}[\rv n \rv H]\eqno (6.16)$$
и совпадает с предельным значением тока в проводнике (6.14).  Отметим,
что условие (6.15) при $\sigma \to \infty$ не столь очевидно,  как это
представляется на первый взгляд,  и во всяком  случае  не  следует  из
равенства  нулю  переменного  электромагнитного поля внутри идеального
проводника.  Так,  в идеальном магнетике ($\mu\to \infty$) поле внутри
тоже  равно  нулю,  но  правильное решение внешней задачи находится из
условия $[\rv n \rv H]=0\,.$

     Вне идеального   проводника   отличие  полей  от  соответствующих
точному решению пропорционально малому параметру  $|W|$,  который  при
$\mu$,  близких  к  единице,  для реальных проводников даже на верхней
границе СВЧ диапазона составляет величину порядка $10^{-4}$.  С  такой
же   относительной   погрешностью   вычислен   и  полный  ток  (6.11),
протекающий в проводнике,  однако в случае идеального проводника  этот
ток  становится  поверхностным.  Основным недостатком рассматриваемого
приближения следует признать отсутствие  потока  энергии  в  идеальный
проводник, который  в  реальном проводнике всегда имеет место и в ряде
случаев существен.

     Этот недостаток   может  быть  устранён  путём  наложения  на
поверхности проводника граничного условия, называемого в литературе по
имени    предложивших    его    авторов    {\it   граничным   условием
Щукина-Леонтовича}, а именно:
     $$\vspace{.2cm}[\rv n\rv E]=-W[\rv n[\rv n\rv H]]\,,\eqno(6.17)$$
которое также  позволяет  ограничиться  решением  задачи  только   для
внешней  области  и  в  то же время вычислить поток мощности в металл.
Нетрудно  убедиться,   что  в  рассматриваемой   простейшей   задаче
использование  граничного  условия (6.17) приводит к точному решению и
для поля вне металла,  и для потока мощности в металл.  Действительно,
достаточно  воспользоваться  равенством  нулю  $x-$  и  $y-$ компонент
волнового  вектора  отражённой  волны,  следующим  из   однородности
распределения  поля в плоскости $xy$,  чтобы из (6.17) получить точное
соотношение амплитуд (6.8).  Для определения поля в металле,  конечно,
приходится   решать   внутреннюю  задачу,  но  в  большинстве  случаев
достаточно знать только поток мощности в металл через его  поверхность
и связанное с этим выделение тепла. Отметим, что полный наведённый в
проводнике ток правильно определяется формулой (6.16).

     Использование граничного  условия (6.17) является частным случаем
общего подхода к решению сложных задач электродинамики  СВЧ  путём
введения  на поверхности,  разделяющей две области,  {\it импедансного
условия}. Для рассматриваемой задачи получено точное решение и поэтому
легко  оценить  ошибку,  которая  возникает при использовании того или
иного приближения.  Отметим,  что  при  решении  задачи  о  нормальном
падении  плоской  волны  с  помощью  граничного  условия (6.17) точное
решение  вне  проводника  получается  без  предположения   о   малости
волнового сопротивления $W$.

     Для более сложной задачи --- наклонного падения плоской волны  из
пустоты на плоскую поверхность проводника --- использование граничного
условия (6.17) не приводит к точному решению и для поля  вне  металла.
Точное  решение внешней задачи даётся общей формулой Френеля (5.29),
которая в данном случае  для  первой  поляризации  ($E_x,\,E_z,\,H_y$)
может  быть  записана  --- с учётом второго из равенств (5.28) --- в
виде
     $$H_{0ry}=\frac{\cos\varphi_i-W\sqrt{1+i\alpha}}{\cos\varphi_i+W
         \sqrt{1+i\alpha}}\;H_{0iy}\;,\qquad E_{0tx}=W\sqrt{1+i\alpha}
         H_{0ty}\,,\eqno(6.18)$$
где безразмерный  параметр  $\alpha$,  введённый  формулой   (5.17),
связан с глубиной скин-слоя $\delta$ соотношением
     $$\alpha=\frac{k^2\delta^2\sin^2{\varphi_i}}2\,.\eqno(6.19)$$
Средний поток  мощности  в  проводник,  нормальный  к его поверхности,
составляет
     $$\overline{\gv S}_z=\frac c{2\pi}\frac {\cos^2{\varphi_i}\re (W
         \sqrt{1+i\alpha})}{|\cos\varphi_i+W\sqrt{1+i\alpha}|^2}|H_
         {0iy}|^2\,,\eqno(6.20)$$
а интегральный по глубине проводника ток равен
     $$I_x=\frac c{4\pi W\sqrt{1+i\alpha}} E_{0tx}e^{ikx\sin\varphi_i}
         \,.\eqno(6.21)$$

     Приближённое решение  на  основе  граничных  условий (6.17) для
первой поляризации поля в падающей волне ($E_x,\,E_z,\,H_y$)  приводит
к следующему соотношению амплитуд:
     $$H_{0ry}=\frac{\cos\varphi_i-W}{\cos\varphi_i+W}H_{0iy}\,,
         \eqno(6.22)$$
так что тангенциальная составляющая электрического поля на поверхности
проводника
     $$E_x\Bigr |_{z=0}=W(H_{0ry}+H_{0iy})e^{ikx\sin\varphi_i}=\frac
         {2W\cos\varphi_i}{\cos\varphi_i+W}E_{0ix}e^{ikx\sin\varphi
         _i}\,,\eqno(6.23)$$
полный ток в проводнике, вычисляемый по формуле (6.16),
     $$I_x=\frac c{4\pi W}(E_{0ix}+E_{0rx})e^{ikx\sin\varphi_i}\,,
         \eqno(6.24)$$
средний поток мощности в металл
     $$\overline{\gv S}_z=\frac c{2\pi}\frac {\cos^2{\varphi_i}\re W
         }{|\cos\varphi_i+W|^2}|H_{0iy}|^2\,.\eqno(6.25)$$
Все величины      $H_{0ry},\,E_x(z=0),\,I_x,\,\overline{\gv S}_z$,
вычисленные на основе граничного условия (6.17),  могут быть  получены
из соответствующих точных значений путём замены в последних величины
$W\sqrt{1+i\alpha}$ на $W$,  то есть  различаются  начиная  с  первого
порядка малости в разложении по параметру $\alpha$.

     Для металлических  проводников  $\alpha$   всегда   очень   малая
величина.  Она мала и в том случае,  когда за счёт большого значения
$\mu$ перестаёт быть малым волновое сопротивление среды --- $\alpha$
при  этом  ещё  более уменьшается.  В соответствии с (6.19) $\alpha$
представляет собой квадрат отношения глубины скин-слоя  (точнее  длины
волны в металле) к длине волны в окружающем пространстве,  домноженный
на $\sin^2{\varphi_i}/2$ --- величину,  всегда меньшую единицы. Именно
этот  множитель  обусловливает  совпадение  приближённого решения на
основе граничного условия (6.17) с точным, поскольку при $\varphi_i=0$
и  $\alpha=0$.  Малость  параметра  $\alpha$ приводит и к малости угла
преломления    $\varphi_t$,     определяемого     формулой     (5.18):
$$\sin\varphi_t=\sqrt{2\alpha(\sqrt{1+\alpha^2}-\alpha)}\,.
         \eqno(6.26)$$
Поэтому при  любом  угле  падения  плоской  волны   поле   в   металле
представляет собой волну,  распространяющуюся практически по нормали к
его поверхности и затухающую на длине волны.

     На практике  для  расчёта  потерь  в  различных  волноводных  и
резонаторных   структурах   широкое   применение   находит   смешанный
приближённый подход к вычислению потока мощности в металл. В нулевом
приближении   на   основе  граничного  условия  (6.15)  для  идеальной
проводимости вычисляется касательная составляющая магнитного поля $\rv
H_0$   на  поверхности  металла,  затем  с  помощью  (6.17)  находится
тангенциальная  составляющая  электрического  поля,  а  по  этим  двум
величинам  определяется  уже  и  средняя  плотность  потока мощности в
металл:
     $$\overline{\gv S}_n=\frac c{16\pi}\mu k\delta|[\rv n\rv H_0]|^2
         =\frac c{8\pi}\re W |[\rv n \rv H_0]|^2.\eqno(6.27)$$
При таком подходе основная погрешность  расчёта  потерь  обусловлена
приближённым вычислением   магнитного  поля (ошибка  порядка $|W|$,)
тогда как погрешность, обусловленная использованием граничного условия
Щукина-Леонтовича,  всегда существенно меньше  (порядка $\alpha\ll |W|$).

     Необходимо сказать   несколько   слов   о   применимости   широко
используемой  в  электродинамике  СВЧ  модели  идеального проводника и
соответствующего  граничного  условия  (6.15).  Без  предположения  об
идеальной  проводимости  стенок  теория таких важнейших элементов СВЧ,
как волновод и резонатор,  утрачивает свою простоту и наглядность и не
может  быть сведена к замкнутым аналитическим выражениям.  Допускаемые
при этом ошибки в значениях полей в большинстве случаев невелики,  что
определяется  малостью  волнового  сопротивления  среды  из-за малости
отношения $\omega/\sigma$ для  металлов  в  рассматриваемом  диапазоне
частот  (при  умеренных  значениях  $\mu$).  Однако  понятие идеальной
проводимости является физически внутренне противоречивым.

     Действительно, в  простейшей  теории  проводимости  металлов  ток
обусловлен  движением  свободных  электронов    и   в   самом   грубом
приближении  $\sigma=ine^2/m_e\nu$,  где  $n$  -- плотность электронов
проводимости,  $\nu$ -- эффективная частота столкновений. При этом для
возможности  пренебрегать током смещения и поляризационным током $\nu$
должна быть много больше частоты внешнего поля  и  собственных  частот
вещества,  в  чём  легко  можно  убедиться,  опираясь  на результаты
раздела 2.  Большие значения $\sigma$ не могут быть получены за счёт
увеличения $n$,  ограниченной плотностью вещества, а предел $\nu\to 0$
вступает в прямое противоречие с исходными предположениями.  В  равной
степени   предположение   о  $\sigma\to  \infty$  (и,  соответственно,
$\varepsilon^{''}\to i\infty$) вступает  в  противоречие  с  принципом
причинности,  находящим  свое  выражение в дисперсионных соотношениях.
Вследствие этого в ряде  случаев  в  теории  возникают  принципиальные
трудности  в  выборе единственного решения задачи,  которые могут быть
устранены  только  путем  учёта  конечной   проводимости   вещества,
приводящей к проникновению в него поля, и отказа от граничного условия
(6.15).

     Между интегральным  током  по  всей глубине плоского проводника и
тангенциальным электрическим  полем  на  его  поверхности  выполняется
соотношение
     $$E_x(0)=\zeta I_x\,,\eqno(6.28)$$
где коэффициент пропорциональности
     $$\zeta=\frac {4\pi}c W=\frac{1-i}{\sigma\delta}\eqno(6.29)$$
называется {\it   поверхностным   импедансом}  металла.  Поверхностный
импеданс,  как и волновое сопротивление,  от  которого  он  отличается
только численным множителем, величина комплексная:
     $$\zeta=\rho-i\xi\,.\eqno(6.30)$$
Его действительную  часть  $\rho$  называют  {\it активным},  а мнимую
часть $\xi$ ---  {\it  реактивным}  поверхностным  сопротивлением.  Из
формулы (6.29) следует, что активное поверхностное сопротивление $\rho
=1/\sigma\delta$ совпадает с  обычным  сопротивлением  при  постоянном
токе   (когда   ток   равномерно   заполняет  все  поперечное  сечение
проводника) пластинки  толщиной  $\delta$.  Это  соотношение  поясняет
смысл  термина \textit{толщина скин-слоя}:  $\delta$ показывает,  какая часть
проводника эффективно проводит переменный электрический ток. Используя
теорему  о  комплексной мощности (комплексная теорема Умова-Пойнтинга,
см.  (2.30)),  легко  можно  убедиться,  что  активное   поверхностное
сопротивление $\rho$ определяет мощность потерь в проводнике,  а $\xi$
--- магнитную энергию, находящуюся в нём.

     В рассмотренных   задачах  высокая эффективность использования
граничного условия Щукина-Леонтовича   связана  с малостью  произведения
$k\delta$,  которое  фактически  представляет  собой отношение толщины
скин-слоя к длине волны в пустоте (в более общем случае к длине  волны
в той среде вне металла,  где ищется поле). В других задачах критерием
применимости  этого  граничного  условия  служит   малость   отношения
$\delta$  к какому-нибудь другому,  наименьшему характерному линейному
размеру задачи,  в качестве  которого  может  быть  размер  проводника
(диаметр  проводника,  толщина  плоскопараллельной  пластины),  радиус
кривизны поверхности или радиус  кривизны  волнового  фронта  падающей
волны.   При   выполнении  этого  критерия  условие  (6.17)  позволяет
значительно расширить класс задач, допускающих аналитическое решение.

     Сам по  себе  скин-эффект  наблюдается  и в таких процессах,  где
волновая природа электромагнитного поля вне проводника не проявляется.
Типичным  примером  служит   задача   о   протекании   переменного
электрического  тока  в  круглом  цилиндрическом  проводе.  Пусть  ось
провода    совпадает   с   осью   цилиндрической   системы   координат
$(r\,,\varphi\,,z)$;  будем считать,  что радиус  провода  равен  $a$,
плотность   тока   и   электрическое   поле  в  проводе  имеют  только
$z$-составляющие, обладают симметрией по азимуту $\varphi$ и однородны
по   $z$,   а   протекающий   по   проводу  полный  ток  имеет  вид  $
J=I_0e^{-i\omega t}\;.  $  Электрическое поле в  проводе  удовлетворяет
уравнению
     $$\Delta E_z+K^2E_z=0\,,\eqno(6.31)$$
где комплексное  волновое  число  в  проводнике $K$ связано с толщиной
скин-слоя  соотношением(6.5).  При сделанных выше
предположениях   $E_z$  зависит  только  от  $r$, и  уравнение  (6.31)
представляет собой обыкновенное дифференциальное уравнение:
     $$\frac{d^2E_z}{dr^2}+\frac 1 r\frac{dE_z}{dr}+K^2E_z=0\;.
         \eqno(6.32)$$

     Решение этого уравнения,  не имеющее особенностей при $r=0$, есть
с точностью до произвольного множителя функция Бесселя от комплексного
аргумента:
     $$E_z=CJ_0(Kr)\,.\eqno(6.33)$$
Постоянная $C$   определяется   с   помощью   граничного   условия  на
поверхности провода.  Из-за отсутствия в данной задаче зависимости  от
угла  $\varphi$ имеется только $\varphi$-составляющая магнитного поля,
которая   связана   с   электрическим   полем   уравнением   Максвелла
$\displaystyle{H_\varphi(r)=\frac  i{k\mu}\frac{dE_z}{dr}},$  то  есть
граничное условие имеет вид
     $$\left.\frac{dE_z}{dr}\right |_{r=a}= - ik\mu H_\varphi (a).
         \eqno(6.34)$$
Значение $H_\varphi (a)$ легко связать с полным током $I_0$ c
помощью уравнения Максвелла $(1.12\mbox{\it{б'}})$, учитывая,
что в нём можно пренебречь током  смещения  из-за  высокой
проводимости металла.  В  результате $H_\varphi (a)=2I_0/ca$ и,
следовательно, постоянная $C$ равна
     $$C=\frac{KI_0}{2\pi\sigma  a J_1(Ka)}\,.\eqno(6.35)$$

     Особенности скин-эффекта в этой  задаче  определяются  поведением
функции
     $$f(r)=\frac{j_z(r)}{j_z(a)}=\frac {E_z(r)}{E_z(a)}=\frac{J_0(Kr)}
         {J_0(Ka)}\,,\eqno(6.36)$$
представляющей собой отношение плотности  тока  на  некотором  радиусе
провода  $r$  к  значению  этой  величины  на поверхности провода.  На
рис~6.1 представлены результаты  вычисления  $|f(r)|$  для  нескольких
значений  параметра $a/\delta$.  При малых значениях $a/\delta$ вплоть
до $a/\delta \approx 0,5$ можно  считать,  что  ток  распределён  по
сечению провода  равномерно,  но  уже  при  значениях  $a/\delta\geqslant 10$
практически весь ток протекает в приповерхностном слое.

     Так как   функция   Бесселя  $J_0(x)$  при  $\im  x\gg  1$  имеет
асимптотическое представление
     $$J_0(x)=\frac 1{\sqrt{2\pi x}}\,e^{-i(x-\pi/4)}\,,\eqno(6.37)$$
\begin{wrapfigure}[14]{l}{8cm}
\begin{picture}(80,50)
\put(-1,52){\special{em:graph Fig6-1.bmp}}
\end{picture}
\hbox to 8cm{\hfil\footnotesize{Рис.~6.1.~Распределение тока
в цилиндрическом}\hfil}
\hbox to 8cm{\hfil\footnotesize{проводе.\hspace{3.5cm}}\hfil}
\end{wrapfigure}
то функция $f(r)$ при больших $Kr$ может быть записана в виде:
     $$f(r)\approx\sqrt{\frac a r}\exp((1-i)(a-r)/\delta)\eqno(6.38)$$
и при $r\approx a$
     $$|f(r)|=\exp(-(a-r)/\delta),\eqno(6.39)$$
что совпадает   с  характером  убывания  поля  вглубь  проводника  при
рассмотренном выше падении плоской волны на его плоскую поверхность.

     Воспользуемся найденным  решением  задачи  для выяснения вопроса,
каким образом внутрь проводника поступает энергия,  выделяющаяся там в
виде  тепла.  Поглощаемая  энергия  может  быть  вычислена как средняя
работа поля над токами проводимости.  Потери на единицу длины  провода
определяются интегралом по его поперечному сечению:
     $$Q=\frac{\sigma}2\re\int\limits _S E_z(r)E_z^*(r)\,dS=\frac
         {\sigma|C|^2}2\re\int\limits_0^{2\pi}\,d\varphi\int\limits_
         0^a J_0(Kr)J_0(K^{*}r)r\,dr\,.\eqno(6.40)$$
Интеграл по  $dr$  легко  вычисляется  в  замкнутом виде и после
несложных преобразований получаем  следующее  выражение  для
$Q$:
     $$Q=\frac{|I_0|^2}{2\pi a^2\sigma}\re F\Bigl(\frac a\delta\Bigr)
         \,,\eqno(6.41)$$
где  комплексная функция
     $$F(x)=\frac 12 \sqrt{2i}x\frac{J_0(\sqrt{2i}x)}{J_1(\sqrt{2i}x)
         }\,.\eqno(6.42)$$

     Поскольку рассматривается    стационарное    решение,    то   для
компенсации потерь должен существовать поток мощности внутрь  провода.
Этот поток не может быть направлен вдоль провода:  в этом случае из-за
поглощения неизбежным было бы уменьшение амплитуды поля вдоль оси $z$,
а в задаче по $z$ всё однородно. Единственная возможность --- приток
энергии через боковую поверхность провода. Вычислим плотность среднего
по времени радиального потока мощности через поверхности $r=a:$
     $$\overline{\gv S}_r=\frac {c}{8\pi}\re \{E_z(a)H^*_{\varphi}(a)
         \}\,.\eqno(6.43)$$
Ввиду азимутальной симметрии  средний  поток  мощности  $\Sigma_r$  на
единицу длины провода легко вычисляется:
     $$\Sigma_r(r=a)=2\pi a\overline{\gv S}_r=\frac{|I_0|^2}{2\pi a^2
         \sigma}\re F\Bigl (\frac a\delta\Bigr )\,\eqno(6.44)$$
и совпадает с джоулевыми потерями в проводе.

     Если воспользоваться  граничным условием Щукина-Леонтовича (6.17)
для вычисления  потока  энергии  внутрь  провода  и  учесть  при  этом
следующее из (6.2) и (6.4) равенство $\re W=c/{4\pi\delta\sigma}$,  то
поток мощности на единицу длины провода равен
     $$\Sigma_r=\frac {|I_0|^2}{4\pi a\delta\sigma}.\eqno(6.45)$$
При больших значениях аргумента  $x$  функция  $F(x)$,  определённая
формулой  (6.42),  асимптотически  стремится  к $x(1-i)/2$, и выражения
(6.44) и  (6.45)  совпадают.  Таким  образом,  критерием  применимости
условия (6.17) в данной задаче является малость отношения $\delta/a$.

\begin{wrapfigure}[15]{l}{8cm}
\begin{picture}(80,55)
\put(-1,52){\special{em:graph Fig6-2.bmp}}
\end{picture}
\hbox to 8cm{\hfil\footnotesize{Рис.~6.2.~Погонный внутренний импеданс
}\hfil}
\hbox to 8cm{\hfil\footnotesize{провода круглого сечения.}\hfil}
\end{wrapfigure}
     Электрическое поле  на  поверхности   проводника
     $$E_z(a)=\frac{I_0}{\pi\sigma a^2}F\Bigl(\frac a{\delta}\Bigr)=
                Z_t I_0\eqno(6.46)$$
пропорционально, очевидно,  комплексной  амплитуде полного тока $I_0$,
протекающего по проводу. Коэффициент пропорциональности $Z_t=R-iX$ ---
величина комплексная и согласно терминологии,  принятой в теории цепей
переменного  тока,  его  следует  назвать  {\it  погонным   внутренним
импедансом}    провода.    При    постоянном    токе   соответствующая
действительная величина называется погонным сопротивлением  провода  и
равна
     $$R_0=\frac 1{\sigma \pi a^2}\,.\eqno(6.47)$$
Из формулы  (6.39)  следует,  что   отношение   действительной   части
импеданса
     $$R=R_0\re F\Bigl(\frac a{\delta}\Bigr)\eqno(6.48)$$
к погонному  сопротивлению  $R_0$  является  функцией одного параметра
$a/\delta$.  График этой функции и график функции $X/R_0$ представлены
на рис.~6.2.

     Мнимая часть (в данном случае $X=-\im Z_t$)  погонного импеданса
     $$X=\frac 1{\sigma\pi a^2}\im F\Bigl(\frac a{\delta}\Bigr
         )\eqno(6.49)$$
и может быть записана в виде $X=\omega  L$,  где  $L$  ---  внутренняя
погонная  индуктивность провода.  Отношение этой величины к внутренней
погонной индуктивности провода на постоянном токе  $L_0=1/2c^2$  также
представлено  на  рис.~6.2;  параметрам  $R$ и~$L$ на основе теоремы о
комплексной мощности можно приписать тот же энергетический смысл,  что
и параметрам $\rho$ и~$\xi$.

     В заключение  отметим  ещё  раз,  что  из-за  малого  волнового
сопротивления проводников скин-эффект проявляется и на таких частотах,
на которых соответствующая длина волны в пустоте много больше размеров
провода.   Так,  для  обычной  сети  переменного  тока  50~Гц  толщина
скин-слоя в медном проводе $\delta=0,95\;$~см.  В частности, поэтому в
силовых  линиях  передач не применяются провода радиусом больше одного
см.  Длина волны для частоты 50~Гц равна 6000~км,  то  есть  близка  к
радиусу Земли.
%\end{document}

\newpage
\oddsidemargin=-0.4mm \evensidemargin=-0.4mm
\topmargin=-0.4mm
\headsep=7mm
\textheight=231.875mm
\textwidth=160mm
\mathsurround=2.5pt
\unitlength=1mm
%\begin{document}
%\input{macr.tex}
\thispagestyle{empty}
%\addtocounter{page}{71}

\begin{center}
  \subsubsection*{\rm Г\,Л\,А\,В\,А\, 3}
     \vspace{-1.15em}
     \line(6,0){160}\\
     \vspace{-1em}
     \line(6,0){160}
     \vspace{-1.15em}
\subsubsection*{ВОЛНОВОДЫ}
     \vspace{35mm}
\subsubsection*{7. Цилиндрические волны}
\end{center}
\vspace*{.5cm}

\markboth{Глава 3. Волноводы}{7. Цилиндрические волны}

\begin{center}\begin{minipage}[c]{0.75\textwidth}
\footnotesize{\parindent=0.5cm

     Определение цилиндрической   волны и условие её существования.
     Магнитные  и  электрические волны в однородной среде.  Векторы
     Герца как  потенциальные  функции  цилиндрических  волн.
     Выражение   составляющих   электромагнитного  поля  через
     потенциальные  функции.  Волновое  уравнение   и  его решение
     методом разделения переменных. Разложение цилиндрической волны
     на плоские волны. Разложение плоской волны по цилиндрическим
     функциям.

}\end{minipage}\end{center}
\vspace*{.5cm}

     До сих пор  нами  рассматривались плоские  волны,  которые
представляют собой простейшие решения однородных (без источников)
уравнений Максвелла
     $$\rot{ \bf E}=ik\mu{\bf H},\quad \rot{\bf H}=-ik\varepsilon
         {\bf E}\eqno(2.8)$$
для комплексных амплитуд монохроматического поля в однородной
($\varepsilon, \mu=const$) среде  или непосредственно
следующего из них волнового уравнения
     $$\Delta {\bf E}+ \varepsilon \mu k^2 {\bf E}=0\eqno(3.8)$$
(такой же вид волновое уравнение имеет и для вектора $\rv H$).
Перейдём теперь  к  более  сложным  решениям --- так называемым {\it
цилиндрическим волнам}.  Как и  плоская  волна,  цилиндрическая  волна
характеризуется  выделенным  направлением  своего  распространения,  в
качестве которого  выберем ось  $z$.  Тогда  поле  цилиндрической
волны может быть записано в виде
     $${\bf E}(x,y,z)={\bf E}(x,y)e^{ihz},\eqno(7.1\mbox{\textit a)}$$
     $${\bf H}(x,y,z)={\bf H}(x,y)e^{ihz}.\eqno(7.1\mbox{\textit б)}$$
Величина $h$ в формулах  (7.1)  называется  {\it  продольным  волновым
числом}  и  является  основной  характеристикой  цилиндрической волны.
Вообще  говоря,  $h$  ---  комплексная  величина:  $h=h'+ih''$,
~$h'\geqslant 0$,~$h''\geqslant 0.$ Компоненты поля $E_z$ и $H_z$
принято называть продольными; компоненты, лежащие в ортогональной к оси
$z$ плоскости --- поперечными.

     Волна (7.1) распространяется в сторону положительной  оси  $z$  c
фазовой скоростью
     $$v_{\mbox{\footnotesize\textit ф}}=\frac{\omega}{h'}=
         c\frac{k}{h'}\,,\eqno(7.2)$$
длиной волны
     $$\Lambda=\frac{2\pi}{h'}=\lambda\, \frac{k}{h'}\eqno(7.3)$$
и затуханием,  определяемым   множителем   $e^{-h''z}$.   Для
волны, распространяющейся  в  сторону  отрицательных  значений
$z$,  в (7.1) вместо $e^{ihz}$ надо подставить $e^{-ihz}$. Если
$h'>k$, то $v_{\mbox{\footnotesize\textit ф}}<c$ и волна  {\it
медленная};  если  же  $h'<k$  и  $v_{\mbox{\footnotesize\textit
ф}}>c$,  то волна {\it быстрая}.

     Решение уравнений (2.8), имеющее структуру цилиндрической волны
(7.1), существует для любой среды,  в которой $\varepsilon$ и $\mu$ не
зависят от $z$ (однородны по $z$), при произвольной зависимости этих
параметров от поперечных координат. В цилиндрической волне,  как и в
плоской, поверхностью  постоянной фазы поля $\varphi=hz-\omega t$
является плоскость, нормальная к направлению  распространения,
однако амплитуда поля в этой плоскости --- в отличие от  случая
плоской однородной волны --- уже не постоянна.  Таким образом, плоская
однородная волна является частным случаем  цилиндрической.
Цилиндрическую волну называют также {\it неоднородной плоской волной},
хотя это понятие намного шире, чем те неоднородные плоские волны,
которые рассматривались в предыдущих разделах в среде, однородной во
всех направлениях.

     Интерес к цилиндрическим волнам обусловлен несколькими причинами.
Одна из них состоит в том,  что поля,  возбуждаемые во многих реальных
конструкциях,  таких, как разнообразные линии передачи, в частности, в
волноводах, имеют структуру поля как раз такого типа. Однако и в целом
ряде других устройств --- в антеннах,  во многих электронных  приборах
---   поле   имеет  структуру,  близкую  к  цилиндрической  волне  или
суперпозиции таких волн.  Другая  причина  заключается  в  возможности
разложить  поле,  произвольным  образом  зависящее от $z$,  в интеграл
Фурье вида $\rv  E(x,y,z)=\int\limits^\infty_{-\infty}  \rv  E(x,y,h)e
^{ihz}   \,dh$,   то  есть  представить  поле  в  виде  разложения  по
цилиндрическим  волнам.  Хотя  вычисление   такого   интеграла   часто
оказывается  непростой  математической  операцией,  сам  поиск решения
уравнений Максвелла с зависимостью искомой  функции  от  $z$  в  виде
множителя  $\displaystyle  e^{ihz}$  несомненно  проще,  чем  в случае
функции трёх переменных произвольного вида.

   Важным моментом, упрощающим работу с цилиндрическими волнами вида
(7.1), является то обстоятельство, что для них все  компоненты  поля
могут  быть  выражены  через  {\it  две скалярные   функции};
другими  словами,  расчёт  полей  сводится  к нахождению двух
потенциальных функций, зависящих лишь от поперечных координат.

     В этом  нетрудно   убедиться,   выписав   поперечные   компоненты
уравнений (2.8) с учётом (7.1) :
     $$\left.\begin{array}{l}\displaystyle{\vspace*{1mm}
         {\partial E_z\over \partial y}-ih E_y=\phantom{-}ik\mu H_x\,,
         \quad-{\partial E_z\over\partial x}+ih E_x=\phantom{-}ik
         \mu H_y\,,}\\[6mm]
         \displaystyle{{\partial H_z\over \partial y}-ih H_y=-ik
         \varepsilon E_x\,,\quad-{\partial H_z\over\partial x}+ih H_x=
         -ik\varepsilon E_y}\,,\end{array}\right\}\eqno(7.4)$$
откуда непосредственно вытекают следующие выражения
поперечных   компонент поля   через  две  продольные  компоненты
$E_z(x,y)$  и $H_z(x,y)$:
     $$\vspace{2mm}\left.\begin{array}{l}\displaystyle{E_x=\frac{i}
         {k^2\varepsilon\mu-h^2}\left (h{\partial E_z\over\partial x}
         +k\mu{\partial H_z\over\partial y} \right)}\,,\quad
         \displaystyle{E_y=\frac{i}{k^2\varepsilon\mu-h^2}\left (h{
         \partial E_z\over\partial y}-k\mu{\partial H_z\over
         \partial x} \right)}\,,\\[5mm]\displaystyle{H_x=\frac{i}{k^2
         \varepsilon\mu-h^2}\left (h{\partial H_z\over\partial x}-k
         \varepsilon{\partial E_z\over\partial y}\right)}\,,\quad
         \displaystyle{H_y=\frac{i}{k^2\varepsilon\mu-h^2}\left (k
         \varepsilon{\partial E_z\over\partial x}+h{\partial H_z\over
         \partial y} \right)}\,.\end{array}\right\}\eqno(7.5)$$

    Видно, что  в качестве    двух    потенциальных    функций
цилиндрической  волны  могут  быть  выбраны  две продольные компоненты
поля $E_z(x,y)$  и $H_z(x,y)$.  Уравнения для этих функций легко могут
быть   получены  подстановкой  поперечных компонент поля (7.5)
в два  оставшихся невыписанными скалярных уравнения, являющихся
проекцией уравнений  (2.8) на ось $z$. При произвольной зависимости
проницаемостей $\varepsilon$ и $\mu$ от поперечных координат
получающаяся  в результате система двух связанных уравнений в частных
производных  слишком сложна, очень редко используется в практических
расчетах и далее не рассматривается.  В электродинамике СВЧ основной
интерес представляют среды, состоящие из однородных областей,
ограниченных цилиндрическими поверхностями с образующими,
параллельными оси $z$. В этом случае для каждой однородной области
получаются два независимых волновых уравнения
     $$\Delta_2 E_z(x,y)+(\varepsilon\mu k^2-h^2)E_z(x,y)=0\,,
         \eqno(7.6)$$
     $$\Delta_2 H_z(x,y)+(\varepsilon\mu k^2-h^2)H_z(x,y)=0\,.
         \eqno(7.7)$$
Двумерный  оператор  $\Delta_2$,  действующий  на скалярную функцию
поперечных координат, в   декартовых координатах имеет вид
     $$\Delta_2={\partial^2\over\partial x^2} + {\partial^2\over
         \partial y^2}\,,\eqno(7.8)$$
в полярных координатах ---
     $$\Delta_2={\partial^2\over\partial r^2}+\frac{1}{r}{\partial \over
         \partial r}+\frac{1}{r^2}{\partial^2\over\partial\varphi^2}\,.
         \eqno(7.9)$$

     Выбор той или иной системы координат в поперечной плоскости
определяется геометрией цилиндрических поверхностей, разделяющих
однородные области, и структурой источников возбуждения. Для
получения аналитических решений пригодны только те системы
координат, в которых переменные разделяются.

     Из приведённых  преобразований  ясно,  что в неограниченной
однородной среде вся совокупность цилиндрических волн (7.1) распадается
на две  независимые системы  волн:  у  одной  из  них  отсутствует
продольная составляющая электрического поля ($E_z=0$) и волны называются
{\it  магнитными ($H$-волнами)}; у  другой   отсутствует   продольная
составляющая магнитного поля ($H_z=0$) и волны  называются  {\it
электрическими ($E$-волнами)}. Такое деление цилиндрических волн
на два типа имеет место и при определённых условиях на границе
однородной области, в частности, при идеальной проводимости
граничной поверхности.

    Характерным свойством как электрических, так и магнитных волн
является ортогональность проекций векторов электрического и магнитного
поля на плоскость $z=const$. В этом легко убедиться, вычислив с помощью
формул (7.5) скалярное произведение $\rv E\rv H= E_xH_x+E_yH_y$,
которое для обоих типов волн оказывается равным нулю.

     В качестве  потенциальных  функций  цилиндрических  волн  удобнее
использовать не $E_z$ и $H_z$,  а $z$-компоненты векторов Герца.  Не в
последнюю  очередь  это  связано  с  тем  обстоятельством,   что   при
выполнении     условия    $h=k\sqrt{    \varepsilon\mu}$    существует
цилиндрическая волна,  у которой и $E_z$, и $H_z$ равны нулю. Эта, так
называемая  {\it TEM-волна},  не может быть описана системой уравнений
(7.5)--(7.7).  Выполнение условия $h=k\sqrt{ \varepsilon\mu}$ приводит
к  превращению  уравнений  Гельмгольца  в  уравнения  Лапласа, и,
следовательно,  распределение  поля  в  плоскости  $z=const$  совпадает
с решением соответствующей статической задачи.

     Введём электрический  и  магнитный векторы Герца с единственной
отличной от нуля $z$-составляющей:
     $$\rv \Pi^e=\{0,\,0,\,\Pi^e(x,y)\,e^{ihz}\}\,,\eqno(7.10)$$
     $$\rv \Pi^m=\{0,\,0,\,\Pi^m(x,y)\,e^{ihz}\}\,.\eqno(7.11)$$
В соответствии  с  формулами  (3.26) и (3.34)  $z$-составляющие  этих
векторов связаны  с  продольными составляющими электрического и
магнитного поля соотношениями
     $$E_z(x,y,z)=(k^2\varepsilon\mu-h^2)\,\Pi^e(x,y)\,e^{ihz}\,,
         \eqno(7.12)$$
     $$H_z(x,y,z)=(k^2\varepsilon\mu-h^2)\,\Pi^m(x,y)\,e^{ihz}\,.
         \eqno(7.13)$$
Функции $\Pi^e(x,y)$  и~$\Pi^m(x,y)$  удовлетворяют  всё  тому   же
скалярному уравнению Гельмгольца, что $E_z(x,y)$ и $H_z(x,y)$:
     $$\Delta_2\Pi^e+(k^2\varepsilon\mu-h^2)\Pi^e=0\,,\eqno(7.14)$$
     $$\vspace*{5mm}\Delta_2\Pi^m+(k^2\varepsilon\mu-h^2)\Pi^m=0\,.
         \eqno(7.15)$$

    Выражения  для  поперечных  компонент  поля  через
потенциальные функции $\Pi^e$ и $\Pi^m$ в декартовой системе координат
непосредственно  следуют  из  формул  (7.5):
     $$\left.\begin{array}{l}E_x(x,y)=\phantom{-}\displaystyle{ih{\partial
         \Pi^e(x,y)\over\partial x}+ik\mu{\partial\Pi^m(x,y)\over
         \partial y}}\,,\\[5mm]E_y(x,y)=\phantom{-}\displaystyle{ih
         {\partial \Pi^e(x,y)\over\partial y}-ik\mu{\partial\Pi^m(x,y)
         \over\partial x}}\,,\\[5mm]H_x(x,y)=\displaystyle{
         -ik\varepsilon{\partial \Pi^e(x,y)\over\partial y}+ih
         {\partial\Pi^m(x,y)\over\partial x}}\,,\\[5mm]H_y(x,y)=\phantom{-}
         \displaystyle{ik\varepsilon{\partial \Pi^e(x,y)\over
         \partial x}+ih{\partial\Pi^m(x,y)\over\partial y}}\,.\end
         {array}\right \}\eqno(7.16)$$
В дальнейшем нам понадобятся также аналогичные  формулы  для  компонент поля
в полярной  системе координат $r,\varphi $:
     $$\left.\begin{array}{l}E_r(r,\varphi)=\phantom{-}\displaystyle
        {ih{\partial \Pi^e(r,\varphi)\over\partial r}+ik\mu\frac{1}{r}
         {\partial\Pi^m(r,\varphi)\over\partial\varphi}}\,,\\
         [5mm]E_\varphi(r,\varphi)=\phantom{-}\displaystyle{ih\frac{1}{r}
         {\partial\Pi^e(r,\varphi)\over\partial\varphi }-ik\mu
         {\partial\Pi^m(r,\varphi)\over\partial r}}\,,\\[5mm]
          H_r(r,\varphi)=\displaystyle{-ik\varepsilon\frac{1}{r}
         {\partial\Pi^e(r,\varphi)\over\partial\varphi}+ih{\partial
          \Pi^m(r,\varphi)\over\partial r}}\,,\\[5mm]H_\varphi
         (r,\varphi)=\phantom{-}
         \displaystyle{ik\varepsilon{\partial\Pi^e(r,\varphi)\over
         \partial  r}+ih\frac{1}{r}{\partial\Pi^m(r,\varphi)\over
         \partial\varphi}}\,.\end{array}\right \}\eqno(7.17)$$

Формулы (7.12)--(7.17)   образуют   основу математического аппарата
 теории  цилиндрических  волн,  который  будет широко использоваться
 в последующих  разделах.

     Основной способ аналитического решения уравнений (7.14),  (7.15)
---  метод  разделения переменных (метод Фурье). В декартовых координатах
функция $\Pi(x,y)$ (здесь у функций $\Pi^e(x,y)$ и $\Pi^m(x,y)$ верхние
индексы опускаются,  поскольку уравнение для обеих функций одно и то же) ищется
в виде произведения  двух  функций,  каждая  из  которых
зависит только от одной переменной:
     $$\Pi(x,y)=X(x)Y(y)\,.\eqno(7.18)$$
Подставляя (7.18) в скалярное волновое уравнение (7.14), получаем:
     $$\frac 1X{d^2X\over\ dx^2}+\frac 1Y{d^2Y\over\ dy^2}=- (K^2-h^2)
     \,;\eqno(7.19)$$
здесь, как обычно,  $K^2= k^2\varepsilon\mu$. Каждое слагаемое слева в
(7.19)  должно  быть  постоянным  числом;  обозначим  их как $g_x^2$ и
$g_y^2$  (величины  $g_x,\,g_y$  часто   называют   {\it   поперечными
волновыми    числами}).    В    результате    получаем    обыкновенные
дифференциальные уравнения
     $${d^2X\over\ dx^2}+g^2_xX=0\,,\eqno(7.20)$$
     $${d^2Y\over\ dy^2}+g^2_yY=0\,,\eqno(7.21)$$
прич\"ем между волновыми числами должно выполняться соотношение
     $$g^2_x+g^2_y=K^2-h^2\,.\eqno(7.22)$$
Решения уравнений (7.20) и (7.21) хорошо известны:
     $$X=A\cos g_xx+B\sin g_xx\,,\quad Y=C\cos g_yy+D\sin g_yy\,.
         \eqno(7.23)$$
Постоянные $A$,~$B$,~$C$,~$D$   и   допустимые   значения   поперечных
волновых чисел определяются из граничных условий каждой задачи.

     В полярных  координатах  $r$,~$\varphi$   решение скалярного
волнового уравнения ищется в виде:
     $$\Pi(r,\varphi)=R(r)\Phi(\varphi)\,.\eqno(7.24)$$
В результате подстановки (7.24) в (7.14) получаем уравнение
     $${1\over R}{1\over r}{d\over dr}\Bigl(r{dR\over dr}\Bigr)+
         {1\over\Phi r^2}{d^2\Phi\over d\varphi^2}+(K^2-h^2)=0\,,
         \eqno(7.25)$$
которое может удовлетворяться для всех $r$,~$\varphi$ только при условии
     $$ {1\over R}r{d\over dr}\Bigl(r{dR\over dr}\Bigr)+(K^2-h^2)r^2
         =-{1\over\Phi}{d^2\Phi\over d\varphi^2}=m^2\,,\eqno(7.26)$$
где $m$  --- постоянное число.  Таким образом,  получаем два отдельных
обыкновенных дифференциальных уравнения для двух искомых функций
одной переменной:
     $$\displaystyle{r{d\over dr}\Bigl(r{dR\over dr}\Bigr)+[(K^2-h^2)
         r^2-m^2]R}=0\,,\eqno(7.27)$$
     $$\displaystyle{{d^2\Phi\over d\varphi^2}}+m^2\Phi=0\,.
         \eqno(7.28)$$
Решение уравнения  (7.28) очевидно:
     $$\Phi=A\cos m\varphi+B\sin m\varphi\,.\eqno(7.29)$$
Из требования  однозначности  решения  при изменении угла $\varphi$ на
$2\pi$,  следует,  что $m$ --- целое число.
Путём  введения  переменной $\rho=\sqrt{K^2-h^2}\; r$
уравнение   (7.27)   для   функции  $R(\rho)$
сводится к известному  уравнению  Бесселя $m$-того порядка:
     $$\displaystyle{{d^2R\over d\rho^2}+{1\over\rho}{dR\over d
         \rho}+\Bigl(1-{m^2\over\rho^2}\Bigr) R}=0\,.
         \eqno(7.30)$$

     Решениями этого уравнения являются так называемые  цилиндрические
функции, среди которых только одна функция Бесселя $J_m(\rho)$ конечна
в нуле и поэтому именно она выбирается в качестве решения  однородного
волнового  уравнения в области,  примыкающей к оси $z$.  При целых $m$
остальные линейно независимые от $J_m(\rho)$ решения уравнения  (7.30)
---   функции   Неймана   $N_m(\rho)$  и  Ханкеля  1-го  и  2-го  рода
$H_m^{(1)}(\rho)$  и  $H_m^{(2)}(\rho)$  --- имеют  в  нуле   особенность.
Функции  Бесселя  и Неймана описывают стоячие в поперечном направлении
волны, а функции Ханкеля --- бегущие:  одна от оси $z$,  другая к оси.
Подробнее  останавливаться  на  этих  хорошо изученных в математике
функциях здесь не имеет смысла --- в Приложении  П-6  приведён  ряд
полезных    соотношений   для   них.

   Частным видом цилиндрической  волны является решение (7.1)
при $h=0$; в этом случае поля однородны по $z$. Характер поля
в поперечном направлении определяется граничными условиями и
структурой источников возбуждения. Вектор Умова-Пойнтинга, определяющий
направление потока энергии в такой волне, может иметь только
поперечные компоненты.

     Существует глубокая  связь между плоскими и цилиндрическими
волнами,  поскольку и  те, и  другие  являются  решениями  однородного
волнового  уравнения.  Любая цилиндрическая волна может быть разложена
по плоским волнам, то есть представлена в виде совокупности
плоских   волн определённого вида. Обозначим произвольную декартову
компоненту поля монохроматической цилиндрической волны   (7.1)
как функцию $\psi(x,y,z)$.   При выполнении хорошо  известных
требований на её аналитические свойства, согласующихся с физическими
ограничениями на поведение компонент поля, эта функция может быть
разложена в двумерный интеграл Фурье по поперечным компонентам волнового
вектора $\rv K$  плоской волны:
$$
     \psi(x,y,z)=e^{ihz}\int\int\limits_{{\hspace{-0.7cm}}-\infty}^
               {{\hspace{-.5cm}}\infty}g(K_x,K_y)
       e^{i(K_x x+K_y y)}\,dK_xdK_y\,.\eqno(7.31)
$$
Для того, чтобы функция $\psi(x,y,z)$ удовлетворяла волновому уравнению
(3.8), между компонентами волнового вектора $\rv K=\{K_x,K_y,h\} $  и
волновым числом в среде $K=k\sqrt{\varepsilon\mu}$ должно иметь место
соотношение
$$ K_x^2+K_y^2+h^2-K^2=0\,.\eqno(7.32) $$
Если ввести используемые в сферической системе координат углы
$\theta,\varphi$ так, что
$$
   K_x=K\sin{\theta}\cos{\varphi}\,,\quad
   K_y=K\sin{\theta}\sin{\varphi}\,,\quad
   h=K\cos{\theta}\,,\eqno(7.33)
$$
то разложение (7.31) можно переписать в виде
$$
     \psi(x,y,z)=e^{ihz}\int\limits_0^{2\pi} g(\varphi)
       e^{iK\sin{\theta}(x\cos{\varphi}+y\sin{\varphi})}\,d\varphi\,,
       \eqno(7.34)
$$
то есть в виде совокупности однородных плоских волн, распространяющихся
под     одним и тем же углом $\theta$  по отношению к оси $z$. Этот
угол согласно (7.33) однозначно определяется волновым числом в среде
$K$  и продольным волновым числом цилиндрической волны $h$.

    Приведённые рассуждения, строго говоря, относятся только к случаю
прозрачной среды, когда $K$ --- действительная величина. При наличии
поглощения разложение цилиндрической волны в общем случае приходится
проводить по неоднородным плоским волнам, так что углы $\theta$ и
$\varphi$ в (7.34) становятся комплексными. При этом контур
интегрирования в комплексной плоскости $\varphi$  может оказаться и
незамкнутым.

    При решении многих задач в круговых цилиндрических координатах
$r,\varphi,z$ полезно знать разложение плоской волны $\psi(\rv r)=
e^{i\rv K \rv r}$ по цилиндрическим функциям (решениям уравнения
7.30)). Представляя векторы в виде
$$\rv r=\{r\sin{\theta}
\cos{\varphi}, r\sin{\theta}\sin{\varphi},z\},\quad \rv K=
\{K\sin{\theta_0}\cos{\varphi_0}, K\sin{\theta_0}\sin{\varphi_0},h\}
\eqno(7.35)$$
получаем для плоской волны следующее выражение:
$$
  \psi(\rv r)=e^{ihz}\psi(r,\varphi)=
e^{ihz}\cdot e^{i\rho\sin{\theta}\cos{(\varphi-\varphi_0)}}
  \;,\eqno(7.36)$$
где $\rho=Kr\sin{\theta_0}$. Разлагая функцию $\psi(r,\varphi)$
в ряд Фурье по углу $\varphi$
$$\psi(r,\varphi)=\sum\limits_{n=-\infty}^{\infty}
       \psi_n (r) e^{in\varphi}\;,\eqno(7.37) $$
где
$$\psi_n(r)=\frac 1{2\pi}\int\limits_0^{2\pi}
   e^{i[\rho\sin{\theta}\cos{(\varphi-\varphi_0)}-n\varphi]}\,
   d\varphi\,,\eqno(7.38)$$
и пользуясь известным интегральным представлением  функции Бесселя
$$
J_n(x)=\frac 1{2\pi}\int\limits_{-\pi}^{\pi}
e^{i(x\sin{\alpha}-n\alpha)}\,d\alpha\,,\eqno(7.39)
$$
в результате несложных преобразований получаем окончательное
разложение плоской волны:
$$
\psi(r,\varphi,z)=e^{ihz}\sum\limits_{n=-\infty}
    ^{\infty} i^n e^{-in\varphi_0} e^{in\varphi}
         J_n(Kr\sin{\theta_0\,\sin{\theta}})\,.\eqno(7.40)$$
Напомним, что  в  формуле  (7.40)  параметр  $r$   равен длине
радиуса-вектора  в трёхмерной (а  не полярной) системе координат.

     Отметим также,  что при  работе  с  цилиндрическими  функциями  в
системе  круговых  цилиндрических  координат большим подспорьем служит
преобразование   Фурье-Бесселя.    Пусть    $f(r,\varphi)$    является
ограниченной  и  однозначной функцией переменных,  кусочно-непрерывной
вместе со своими первыми производными.  Тогда она может быть разложена
в ряд Фурье по $\varphi$
     $$f(r,\varphi)=\sum\limits_{n=-\infty}^{\infty}f_n (r) e^{in
         \varphi}\;,\quad f_n(r)=\frac 1{2\pi}\int\limits_0^{2\pi}
         f(r,\varphi) e^{-in\varphi}\,d\varphi\,,\eqno(7.41) $$
и для коэффициентов $f_n(r)$  имеет место представление Фурье-Бесселя:
     $$f_n(r)\!=\!\displaystyle{\int\limits_0^\infty g_n(\lambda)J_n
         (\lambda r)\lambda\,d\lambda},\qquad g_n(\lambda)\!=\!
         \displaystyle{\int\limits_0^\infty f_n(r)J_n(\lambda r)r
         \,dr}.\eqno(7.42)$$

     В заключение этого раздела  остановимся  кратко  на  особенностях
решения волнового уравнения
     $$\frac{\partial^2\psi}{\partial x^2}+\frac{\partial^2\psi}
         {\partial y^2}+\frac{\partial^2\psi}{\partial z^2}-\frac
         {\varepsilon\mu}{c^2}\frac{\partial^2\psi}{\partial t^2}-
         \frac{4\pi\mu\sigma}{c^2}\frac{\partial\psi}{\partial t}=0\,,
         \eqno(7.43)$$
пригодного для   описания   произвольного   временного   процесса  при
использовании  простейших  материальных  уравнений  (1.13)--(1.15)  со
всеми  присущими  им  ограничениями.  Будем считать,  что на плоскости
$z=0$ функция $\psi$ задана:
     $$\psi(x,y,0,t)=f(x,y,t).\eqno(7.44) $$

     Представим искомое решение  уравнения  (7.43)  в  виде  4-мерного
интеграла  Фурье  по компонентам волнового вектора плоской волны $K$ и
частоте $\omega$.  Аналогично условию (7.32)  между  этими  величинами
должно выполняться соотношение
     $$K_x^2+K_y^2+h^2=\varepsilon\mu\frac{\omega^2}{c^2}+4\pi i\mu
         \sigma\frac{\omega}{c^2}\,,\eqno(7.45)$$
что позволяет записать решение в виде
     $$\psi(x,y,z,t)=\biggl(\frac 1{2\pi}\biggr)^{\frac 32}\int\!\!
         \int\!\!\int\limits_{\hspace{-0.7cm}-\infty}^{\hspace{-0.5cm}
         \infty}g(K_x,K_y,\omega)e^{i(K_xx+K_yy+hz-\omega t)}\,dK_x
         dK_yd\omega,\eqno(7.46)$$
прич\"ем  можно  считать комплексной только величину $h$:
     $$h=\sqrt{\frac{\omega^2}{ c^2}\mu\varepsilon+4\pi i\mu\sigma
         \frac{\omega}{c^2}-K_x^2-K_y^2}\,.\eqno(7.47)$$
Амплитудная функция $g(K_x,K_y,\omega)$ является образом Фурье функции
$f(x,y,t)$ и равна
     $$g(K_x,K_y,\omega)=\biggl(\frac 1{2\pi}\biggr)^{\frac 32}
         \int\!\!\int\!\!\int\limits_{\hspace{-0.7cm}-\infty}^{
         \hspace{-0.5cm}\infty}f(x,y,t)e^{-i(K_xx+K_yy-\omega t)}
         \,dxdydt\,.\eqno(7.48)$$
При $\sigma=0$  каждая  гармоническая  составляющая   распространяется
вдоль    оси    $z$    со   скоростью   $v=\omega/h$,   но   так   как
$h=\pm\sqrt{(\omega/c)^2\varepsilon\mu-K_x^2-K_y^2}$  не  представляет
собой  линейной  комбинации  $\omega$,  $K_x$  и $K_y$,  то заданное в
плоскости $z=0$ возмущение $f(x,y,t)$ не  будет  распространяться  без
искажений  даже в отсутствии дисперсии и диссипации.  Другими словами,
общего решения уравнения (7.43)  вида  $\psi(x,y,z\pm  vt)$,  как  это
имело  место  для  одномерного  волнового  уравнения,  не  существует.
Физически  этот  результат  представляется  очевидным:  первоначальное
возмущение  поля  $\psi(x,y,0,t)$,  зависящее от поперечных координат,
будет распространяться в пространстве и в поперечном направлении,  что
и приводит к его искажению со временем.
%\end{document}

\newpage
\oddsidemargin=-0.4mm \evensidemargin=-0.4mm
\topmargin=-0.4mm
\headsep=7mm
\textheight=231.875mm
\textwidth=160mm
\mathsurround=2.5pt
\unitlength=1mm
%\begin{document}
%\input{macr.tex}
\thispagestyle{empty}
%\addtocounter{page}{80}
%\baselineskip=\normalbaselineskip
\baselineskip=0.94\normalbaselineskip

\begin{center}\subsubsection*{8. Волноводы}\end{center}
\vspace*{.5cm}

\markboth{Глава 3. Волноводы}{8. Волноводы}

\begin{center}\begin{minipage}[c]{0.75\textwidth}
\footnotesize{\parindent=0.5cm
     Классификация линий   передачи.  Граничные  условия  на  идеально
     проводящей поверхности цилиндрического волновода  с  произвольной
     замкнутой    формой    контура    поперечного    сечения.   Волны
     электрического  и  магнитного  типов.  Собственные   значения   и
     собственные    функции   граничных   задач.   Основные   свойства
     волноводных    волн.    Структура     силовых     линий     поля.
     Распространяющиеся   и   нераспространяющиеся   волны.  Диаграмма
     Бриллюэна. Фазовая  и  групповая   скорости волн  в    волноводе.
     Ортогональность  и нормировка волноводных волн.  Поток мощности и
     энергия электромагнитного поля волны.
}\end{minipage}\end{center}
\vspace*{.5cm}

     Рассмотренные в предыдущем разделе цилиндрические волны  являются
теми решениями уравнений Максвелла, которые описывают электромагнитные
поля в так называемых {\it линиях  передачи}.  С  помощью  этих  линий
переносится СВЧ мощность, например, от генератора к передающей антенне
радиолокатора  или  к  ускорительной  станции  ускорителя   заряженных
частиц.  Они используются и на очень малом уровне сигнала для передачи
информации. Разнообразие их конструкций очень велико  ---  часть  этих
линий перечислена   в   Таблице~8.1.   В   диапазоне   СВЧ  наибольшее
распространение имеют линии передачи,  которые  образованы  замкнутыми
металлическими   цилиндрическими  поверхностями  (будем  считать,  что
образующие этих поверхностей направлены вдоль оси $z$).  В зависимости
от   геометрии  поперечного  сечения  эти  линии  можно  разделить  на
замкнутые  (пространство,  где  распространяются   волны,   ограничено
металлическими стенками) и открытые. Каждый из этих типов линий в свою
очередь имеет смысл разделить на два,  в зависимости от того, является
ли  форма  сечения односвязной или многосвязной (обычно в практических
приложениях --- двухсвязной).

     В настоящее  время  для  передачи  энергии (сигнала)  используются
 двухпроводные  линии, коаксиальные    кабели,  металлические   и
диэлектрические  волноводы, волоконно оптические линии (длина волны
около  1 мм),    радиорелейные  линии  (дециметровые, сантиметровые  и
миллиметровые   волны)  и лучеводные линии  (длины волн  от
субмиллиметрового диапазона до светового). В  данном  курсе   из
перечисленных   выше  линий передачи  практически  не
рассматриваются радиорелейные   линии.

     Для передачи  больших  мощностей  наиболее  широкое  применение в
настоящее время находят {\it металлические волноводы} --- так называют
замкнутые линии  передачи,   как   правило,   с   односвязной   формой
поперечного сечения.  Чаще всего используются волноводы прямоугольного
или круглого сечений,  но встречаются  волноводы  и  с  более  сложным
сечением.  Выявленные  выше  свойства  цилиндрических  волн  позволяют
развить теорию волноводов с идеально проводящими  стенками  достаточно
далеко без конкретизации формы поперечного сечения.

     Будем считать,  что материал, заполняющий волновод, однороден, то
есть проницаемости $\varepsilon$,~$\mu$ не зависят от координат, и что
на идеально проводящих стенках волновода выполняется граничное условие
равенства нулю тангенциальной составляющей электрического поля:
     $$E_t=0\,.\eqno (8.1)$$
Сделанные предположения позволяют утверждать,  что в волноводах  могут
распространяться  цилиндрические  волны  (7.1) и,  следовательно,  все
компоненты  электромагнитного  поля  выражаются  через  две  скалярные
потенциальные  функции,  в  качестве которых удобно выбрать продольные
компоненты электрического  и  магнитного  векторов  Герца. В результате
поле в волноводе может быть представлено в виде совокупности электрических
и магнитных волн. Поскольку зависимость всех компонент поля  от
продольной координаты  $z$  согласно (7.1) определяется  множителем  $e^{ihz}$
(отметим, что для электрических и магнитных волн продольное волновое число
$h$, вообще говоря, различно), то в дальнейшем в этом разделе (если это
не будет специально оговорено) для всех компонент поля
речь  идёт о множителях, определяющих их зависимость
от поперечных координат. В качестве потенциальных используются
функции  $\Pi^e(x,y)$  и $\Pi^m(x,y)$, определенные соотношениями (7.10) и
(7.11) и удовлетворяющие  двумерному волновому уравнению (7.14) и (7.15).
\vspace{0.5cm}

\hbox to\textwidth{\hfill\small{Таблица~8.1. Передающие линии\hfil}}
\vspace{-0.3cm}
$$\begin{array}{|c|c|c|}\hline\vspace{-0.2cm}&&\\\vspace{-0.2cm}
\hbox{Тип линии}&\hbox{Длина волны}&\hbox{Примечание}\\&&\\
     \hline\vspace{-0.2cm}&&\\\vspace{-0.2cm}
     \parbox[t]{4.5cm}{ Двухпроводная линия}&
     \parbox[t]{3cm}{$\lambda > 100$ м}&
     \parbox[t]{7.5cm}{ Расстояние  между   проводами   много   меньше
$\lambda$}
     \\&&\\\vspace{-0.25cm}
     \parbox[t]{4.5cm}{ Коаксиальный кабель}&
     \parbox[t]{3cm}{$\lambda>30$ см}&
     \parbox[t]{7.5cm}{ Поперечный   размер  много  меньше  $\lambda$,
длина --- много больше $\lambda$;  применению на частотах выше  1  ГГц
мешает многоволновость}
     \\&&\\\vspace{-0.25cm}
     \parbox[t]{4.5cm}{ Металлический волновод}&
     \parbox[t]{3cm}{ Сантиметровые и миллиметровые волны}&
     \parbox[t]{7.5cm}{ Одномодовым    режимам    работы   стандартных
волноводов соответствуют $\lambda=10$ см,  3,2 см и  8,2  мм;  дальняя
связь    на    $\lambda=8$~мм    ($H_{01}$)   оказалась   экономически
неэффективной}
     \\&&\\\vspace{-0.25cm}
     \parbox[t]{4.5cm}{ Диэлектрический волновод}&
     \parbox[t]{3cm}{ Миллиметровые волны}&
     \parbox[t]{7.5cm}{ Используются      полистирол,      полиэтилен,
фторопласт,     которые     эффективны     и     в    диапазоне    ИКЛ
($\lambda\approx1$~мкм)  ---  тогда  это  волоконные   световоды   или
оптические волокна}
     \\&&\\\vspace{-0.25cm}
     \parbox[t]{4.5cm}{ Радиорелейная линия}&
     \parbox[t]{3cm}{ Дециметровые, сантиметровые и миллиметровые
волны}&
     \parbox[t]{7.5cm}{ Угол диаграммы направленности  примерно  равен
$\lambda/D$, где $D$ --- размер зеркала}
     \\&&\\\vspace{-0.25cm}
     \parbox[t]{4.5cm}{ Квазиоптические (лучеводные) линии}&
     \parbox[t]{3cm}{ $0,76\hbox{ мкм}<\lambda<1\hbox{ мм}$}&
     \parbox[t]{7.5cm}{ Не нащли широкого применения по  экономическим
причинам}
     \\&&\\\vspace{-0.25cm}
     \parbox[t]{4.5cm}{ Волоконно-оптические линии связи (ВОЛС)}&
     \parbox[t]{3cm}{ $\lambda\approx1\hbox{ мкм}$}&
     \parbox[t]{7.5cm}{ Обычно   используется   кварцевое   стекло  на
$\lambda=0,8$~мкм,  1,3  мкм  (минимальная  дисперсия)   и   1,5   мкм
(минимальные  потери);  основное преимущество --- малое затухание (0,2
дБ/км на 1,5 мкм)
% Диаметр сердцевины $\approx2-3\hbox{ мкм}$, второго
%слоя  с  меньшей  $\varepsilon$  ---  $\sim50-70$~мкм,  полный диаметр
%(вместе с защитным слоем) равен примерно 0,5--1 мм
}
     \\&&\\
     \hline\end{array}$$
%\newpage

     Выразим граничное  условие  (8.1)  через значения функций $\Pi^e$
и~$\Pi^m$ на поверхности  металла.  Введём  на  контуре $C$ поперечного
сечения  $S$ волновода  орты $\rv n$ и $\rv s$ таким
образом, что  $\rv n$ направлен по  нормали
в  металл,  а~$\rv  s$  ---  вдоль контура и при этом ~$\rv n,~\rv s$
и единичный орт вдоль оси  $z$  образуют  правую тройку  (см.  рис.~8.1).
Тогда тангенциальные к поверхности металла
компоненты поля на основании формул (7.17) могут быть записаны в виде
     $$E_s=ih\frac{\partial\Pi^e}{\partial s}-ik\mu\frac{\partial\Pi^m}
         {\partial n}\;,\eqno (8.2\mbox{\textit a})$$
     $$E_z=(k^2\varepsilon\mu-h^2)\Pi^e\;.\eqno(8.2\mbox{\textit б})$$
\begin{wrapfigure}[16]{l}{8cm}
\begin{picture}(80,50)
\put(-0,50){\special{em:graph Fig8-1.bmp}}
\end{picture}
\hbox to 8cm{\hfil\footnotesize{Рис.~8.1.~Поперечное
сечение волновода.}\hfil}
\end{wrapfigure}

     Следовательно, если   функции   $\Pi^e$~   и    ~$\Pi^m$    будут
удовлетворять на контуре~$C$ условиям
     $$\left.\frac{\partial\Pi^m}{\partial n}\right|_C=0\,,
     \eqno (8.3\mbox{\textit а})$$
     $$\left.\Pi^e\right|_C=0\,,\eqno (8.3\mbox{\textit б})$$
а внутри контура волновым уравнениям (7.14) и (7.15),  то  поля  будут
удовлетворять однородным уравнениям Максвелла (4.1) и условию (8.1) на
поверхности металла.

     Заметим, что есть ещё одна возможность удовлетворить граничному
условию (8.1). Достаточно, чтобы вместо условия (8.3 \mbox{\textit
 б}) выполнялись два других:
     $$\left.\frac{\partial\Pi^e}{\partial s}\right|_C = 0
       \eqno (8.4\mbox{\textit a})$$
и, одновременно,
     $$h=k\sqrt{\varepsilon\mu}\,.\eqno (8.4\mbox{\textit б})$$
Эта возможность будет изучена в дальнейшем при рассмотрении волноводов
с двухсвязной формой поперечного сечения.

     Из формул  (8.2)  и  уравнений (7.14),  (7.15) следует,  что {\it
функции $\Pi^e$ и $\Pi^m$ независимы друг от друга}. Это означает, что
цилиндрические  волны  в  волноводах  с  идеально проводящими стенками
распадаются на волны двух типов.  В первом типе волн,  называемых {\it
магнитными  волнами},  $\Pi^e=0$  и поля выражаются через одну функцию
$\Pi^m$.  Характерным свойством этих волн является  отсутствие  у  них
компоненты поля $E_z$.  Второй тип волн --- {\it электрические волны}.
Поля этих волн выражаются через функцию $\Pi^e$,  для них $\Pi^m=0$, и
потому  всюду  равна  нулю компонента $H_z$.  Магнитные волны называют
ещ\"е $H$-волнами,  иногда  $TE$-волнами  (поперечными  электрическими
волнами),  электрические  волны  ---  $E$-волнами  (или $TM$-волнами).
Отметим,  что при учёте  конечности  проводимости  стенок  волновода
электродинамические  свойства  среды  в  поперечной  плоскости  нельзя
считать однородными и разделение  волноводных  волн  на  два  типа  по
отличной от нуля продольной компоненте поля уже невозможно --- в обеих
волнах присутствует и~$E_z$, и~$H_z$.

   Волновое уравнение  и  граничное  условие  для  каждого  типа  волн
односвязного закрытого волновода составляют свою граничную задачу:
     $$\Delta_2\Pi^e+g^2\Pi^e=0\,,\qquad\quad\left.\Pi^e\right|_C
         \phantom{=}=0\qquad\mbox{ --- {\it электрические волны;}}
         \eqno (8.5)$$
     $$\hspace{-0.7cm}\Delta_2\Pi^m+g^2\Pi^m=0\,,\qquad\displaystyle
         {\left.\frac{\partial\Pi^m}{\partial n}\right|_C}=0\qquad\mbox
         { --- {\it магнитные волны,}}\eqno (8.6)$$
где $g^2=k^2\varepsilon\mu-h^2$.  Граничные  задачи  (8.5)  и (8.6) об
отыскании  ненулевых  решений  двумерного  волнового  уравнения   были
подробно  исследованы  задолго до развития теории волноводов в связи с
другими физическими проблемами,  в частности,  при изучении  колебаний
мембран.   Поэтому   потенциальные  функции  $\Pi^m$  и  $\Pi^e$  (без
множителя $\displaystyle{e^{ihz}}$) в литературе часто называются {\it
мембранными  функциями}.  Было установлено,  что уравнения (8.5) и (8.6)
имеют ненулевые решения  только  при  некоторых  дискретных  значениях
параметра  $g^2$;  эти  значения  ---  так называемые {\it собственные
значения}  граничной  задачи  ---  образуют   счётную   возрастающую
последовательность вещественных чисел
     $$g_1^2,\,g_2^2,\,\dots,\,g_n^2,\dots\,.\eqno (8.7)$$

     Вообще говоря,  спектр собственных значений граничных задач (8.5)
и (8.6) различен, но для некоторых форм поперечного сечения волноводов
отдельные  собственные  значения  могут  совпадать.  Функция  $\Pi^e$,
удовлетворяющая  системе  (8.5),  или функция $\Pi^m$, удовлетворяющая
системе (8.6) при соответствующем $g=g_n$, называется {\it собственной
функцией}  граничной  задачи  и  обозначается $\Pi^e_n$ или $\Pi^m_n$.
Собственные функции,  соответствующие различным собственным  значениям
$g_n$ и $g_m\;  (g_n\ne g_m)$,  ортогональны между собой в том смысле, что
     $$\int \limits_S\Pi^e_n\Pi^e_m\,dS=0\,,\qquad
       \int \limits_S\Pi^m_n\Pi^m_m\,dS=0\;,\eqno (8.8)$$
где $S$ --- поперечное сечение волновода, ограниченное контуром $C$.

     Дискретность собственных значений $g_n$ приводит  к  дискретности
значений продольных волновых чисел $h$ для заданного $k$:
     $$h_n^2=k^2\varepsilon\mu-g_n^2\;.\eqno (8.9)$$
При сделанном   предположении   об  идеальной  проводимости  стенок  и
отсутствии поглощения в заполняющей волновод среде продольные волновые
числа  либо  действительные  числа,  либо чисто мнимые.  Условимся для
волны,  распространяющейся или затухающей в положительном  направлении
оси $z$,  выбирать значение корня таким образом, что либо ${h_n}'>0$ и
${h_n}''=0$,  либо ${h_n}'=0$ и ${h_n}{''}>0$.  Данному значению $g_n$
соответствует и волна,  распространяющаяся в отрицательном направлении
оси $z$. Компоненты поля и продольное волновое число этой волны будем
отмечать тем же индексом $n$, но со знаком минус. Мембранная функция у
волн обоих направлений одна и та же, однако некоторые компоненты полей
согласно формулам (7.16) различаются знаком, поскольку  $h_{-n}=-h_n$.

\begin{picture}(160,50)
\put(-10,50){\special{em:graph Fig8-2a.bmp}}
\put(76,50){\special{em:graph Fig8-2b.bmp}}
\end{picture}
\begin{center}\begin{minipage}[c]{0.9\textwidth}
\footnotesize{\parindent=0.5cm
Рис.~8.2.~Проекции силовых линий  поля  на плоскость
$z=const$ в магнитной {\it {а)}} и электрической {\it {б)}} волне.
}\end{minipage}\end{center}\vspace*{0.25cm}

     Продольные компоненты   полей   собственных   волноводных    волн
определяются  формулами  (7.12) и (7.13),  а поперечные компоненты ---
формулами (7.16) в декартовой системе координат и формулами (7.17) ---
в цилиндрической. Отметим, что для  обоих  типов  волн  в  любой точке
пространства  электрическое  и  магнитное   поля   собственной   волны
ортогональны друг другу.  Чтобы убедиться в этом, достаточно вычислить
скалярное произведение $\rv E_n\rv H_n=E_{n,x}H_{n,x}+ E_{n,y}H_{n,y}+
E_{n,z}  H_{n,z}$,  которое  для  обоих  типов волн тождественно равно
нулю. Сама же картина силовых линий в плоскости поперечного сечения по
найденным потенциальным функциям $\Pi^m_n$ и $\Pi^e_n$ для магнитных и
электрических волн строится по-разному.  Но в обоих  случаях  начинать
следует   с   построения   линий  постоянного  уровня  соответствующей
собственной функции.  Для магнитных волн линии уровня  $\Pi^m_n=const$
подходят к контуру $C$ под прямым углом и совпадают с силовыми линиями
электрического поля,  которые в этом случае являются плоскими кривыми,
целиком  лежащими  в  плоскости поперечного сечения.  Проводя линии им
ортогональные,  получаем  проекции  магнитных  силовых  линий  на  эту
плоскость  (рис.~8.2{\it а}).  Для  электрических  волн линии уровня функции
$\Pi^e_n =  const$  совпадают  с  силовыми  линиями  магнитного  поля,
лежащими  в  плоскости поперечного сечения,  а семейство ортогональных
кривых даёт картину проекций силовых линий  электрического  поля  на
эту    плоскость   (рис.~8.2{\it б}).   В   продольных   сечениях   проекции
соответствующей компоненты поля в волнах  обоих  типов  повторяются  с
периодом, равным {\it длине волны в волноводе} $\Lambda_n=2\pi/h_n.$

     Итак, вопрос о том,  какими продольными волновыми  числами  могут
обладать волны в волноводе, решён равенством (8.9), точнее, сведён
к нахождению набора собственных значений $g_n$ граничной задачи ~(8.5)
или~(8.6).  Фактическое  определение чисел $g_n$ и функций $\Pi^e_n$ и
$\Pi^m_n$  для  волноводов  двух  наиболее   распространённых   форм
поперечного сечения производится далее;  однако уже сейчас,  на основе
только формулы (8.9),  можно  выяснить  почти  все  основные  свойства
волноводных волн.

     Величины $g_n$ зависят только от типа волны и от формы и размеров
поперечного  сечения  волновода,  но не зависят от электродинамических
характеристик материала заполнения.  Собственное значение $g_n$  часто
называют  {\it поперечным  волновым  числом}, и тогда соотношение (8.9)
может быть сформулировано  таким  образом:  сумма  квадратов
продольного  и поперечного волновых чисел волноводной волны равна квадрату волнового
числа плоской волны в соответствующей однородной неограниченной среде.

     Приступая к анализу формулы (8.9),  примем для  простоты  записи,
что волновод заполнен однородной прозрачной средой.  Напомним, что при
частичном заполнении волновода  формулы  и  результаты  этого  раздела
неприменимы.  Отвлечёмся также от дисперсионных свойств среды, чтобы
отчётливее выявить дисперсию, присущую волноводным волнам даже в
пустоте.

     В бесконечном ряду (8.7) (будем  считать,  что  в  него  включены
собственные   значения  обеих  граничных  задач  (8.5)  и  (8.6))  при
достаточно больших $k$ имеется конечное  число  членов,  меньших,  чем
$k^2  \varepsilon  \mu$  (их нет,  если только $ k^2 \varepsilon \mu $
меньше первого члена ряда (8.7)).  Для соответствующих волн $h_n^2>0,$
то есть $h_n$ --- действительное число, и такие волны распространяются
в волноводе без изменения амплитуды.  Для всех же $g_n^2>k \varepsilon
\mu$,   то   есть   для   бесконечного  числа  членов,  $h_n^2<0$,  и,
следовательно,  $h_n$ --- чисто мнимая величина. Это значит, что волна
не     распространяется,  а множитель   $e^{ih_nz}$   описывает
экспоненциальное убывание поля вдоль $z$.  Таким образом,  из (8.7)  и
(8.9)   следует,   что   в   волноводах  при  заданной  частоте  могут
существовать несколько распространяющихся  волн  и  бесконечное  число
нераспространяющихся.  Отметим здесь, что поскольку стенки волновода в
настоящем рассмотрении считаются идеально проводящими,  а в заполняющей
среде  нет    потерь,  то распространяющиеся волны не
затухают вдоль волновода.  Затухание же нераспространяющейся  волны  в
свою очередь не связано ни с какими потерями энергии.

     Проследим, как эта ситуация изменяется с  {\it  ростом  частоты}.
При  очень  низких частотах,  когда $k^2\varepsilon\mu<g_1^2,$ ни одна
волна не распространяется,  все они затухают.  Правда,  как это  будет
видно  из  следующих  разделов,  существуют  волноводы  с многосвязной
формой поперечного сечения,  для которых $g_1=0$;  о такой возможности
выше  уже  говорилось  в  связи с граничными условиями (8.4).  В таких
волноводах одна волна существует при сколь угодно низких частотах.

     С ростом частоты,  когда $k$ достигает значения $k = g_1/  \sqrt{
\varepsilon \mu}$,  первая волна становится распространяющейся. Каждый
раз затем, когда $k$ становится равным одному из значений $g_n/ \sqrt{
\varepsilon \mu} $, волна $n$-го номера становится распространяющейся.
Частота,  при которой это происходит, называется критической для волны
данного  номера.  Для  упрощения  записи  будем  называть  далее  {\it
критической частотой}  соответствующее  значение
$k_n=g_n/ \sqrt{\varepsilon \mu} $
(напомним,  что частота отличается от $k$ множителем $c$).
Критическая частота  так  же,  как  и  $g_n$  ---
функция  геометрии  поперечного  сечения,  но  зависит  и  от  свойств
материала заполнения.

     С этими обозначениями формула (8.9) принимает вид
     $$h_n^2=\varepsilon\mu(k^2-k_n^2)\,.\eqno (8.10)$$
Чем больше частота $k$,  тем  больше  волн,  для  которых  критическая
частота  $k_n$  меньше  $k$,  то  есть тем больше различных волн может
распространяться без затухания.

     Рассмотрим теперь,  что  происходит  с волной {\it фиксированного
номера} при изменении частоты и сделаем это в  терминах  длины  волны.
Кроме  обычной  длины волны $\lambda=2\pi/k$ и длины волны в волноводе
$\Lambda_n=2\pi/h_n$,  введём ещё {\it  критическую  длину  волны}
$\lambda_n=2\pi/k_n$.  В пустом волноводе наибольшая критическая длина
волны является естественным  масштабом  длины  ---  $\lambda_1$  имеет
порядок размера поперечного сечения волновода.  Для длинных волн, пока
$\lambda>\lambda_n$,  волна экспоненциально  затухает.  При  $\lambda<
\lambda_n$ волна распространяется без затухания и согласно (8.10)
     $$\frac{1}{\Lambda_n^2}=\frac{1}{\lambda^2}-\frac{1}{\lambda_n^2
         }\,,\eqno (8.11)$$
то есть
     $$\Lambda_n=\frac{\lambda}{\sqrt{1-(\lambda/\lambda_n)^2}}\,.
         \eqno (8.12)$$
Всегда $\Lambda_n>\lambda   $;  при  $\lambda\longrightarrow\lambda_n$
длина волны  в  волноводе  $\Lambda_n\longrightarrow\infty$,  то  есть
вблизи  критической частоты поле как бы теряет волновой характер.  При
очень коротких волнах, когда $\lambda\ll\lambda_n$,~$\Lambda_n \approx
\lambda$,  волноводная  длина  волны лишь немного больше длины волны в
свободном пространстве.  Отметим,  что хотя конкретное значение  $k_n$
зависит  от  номера волны и формы сечения волновода,  формулы (8.10) и
(8.12)  носят  универсальный   характер.   Все   изложенные   свойства
волноводных  волн  хорошо  иллюстрируются  графиком зависимости $k$ от
продольного волнового числа $h$. Такой график для пустого волновода $(
\varepsilon=1,  \mu=1)$ представлен на рис.~8.3; в литературе он часто
называется диаграммой Бриллюэна.  Изображённая  на  нём  пунктиром
биссектриса    первого    квадранта   соответствует   плоской   волне,
распространяющейся в пустоте  со  скоростью  $c$  ---  именно  к  этой
биссектрисе   стремятся  кривые  для  всех  волноводных  волн  при  $h
\rightarrow \infty$.

\begin{wrapfigure}[13]{l}{7.5cm}
\begin{picture}(75,50)
\put(-3,50){\special{em:graph Fig8-3.bmp}}
\end{picture}
\hbox to 7.5cm{\hfil\footnotesize{Рис.~8.3.~Диаграмма
Бриллюэна.}\hfil}
\end{wrapfigure}

     Для распространяющихся  в  волноводе  волн   нетрудно   вычислить
фазовую  и  групповую  скорости.  Фазовая скорость
$v_{\mbox{\footnotesize\textit ф}}=ck/h_n$ даже в пустом волноводе
в отличие  от  скорости  плоской  волны  в  свободном пространстве
зависит  от  частоты  и  всегда  больше фазовой скорости плоской
волны в однородной неограниченной среде $c/ \sqrt{ \varepsilon
\mu}$:
     $$v_{\mbox{\footnotesize\textit ф}}=
         \frac{c}{\sqrt{\varepsilon\mu}}\frac 1{\sqrt{1-(k_n/k)^2}}
         \,.\eqno (8.13)$$
Зависимость $v_{\mbox{\footnotesize\textit ф}}$ от частоты  или,
другими  словами,  {\it  дисперсия фазовой  скорости}  волноводных
волн приводит к расплыванию волнового пакета.  При этом центр
пакета  распространяется  вдоль  волновода  с групповой скоростью,
равной,  как известно,
     $$v_{\mbox{\footnotesize\textit{гр}}}=
         c(dk/dh_n)=\frac{ch_n}{k\varepsilon\mu}\,=
          \frac c{\sqrt{\varepsilon\mu}}
          \sqrt{1-(k_n/k)^2}\,.\eqno (8.14)$$
Нетрудно видеть,  что для  волноводных  волн  существует  удобная  для
запоминания связь между скоростями:
     $$v_{\mbox{\footnotesize\textit ф}} v_{\mbox{\footnotesize\textit{гр}}}
        =\frac{c^2}{\varepsilon\mu}\,.\eqno (8.15)$$

     При учёте   дисперсии  в  среде  заполнения
выражение  для  групповой скорости усложняется
(для про\-сто\-ты за\-пи\-си счи\-та\-ем $\mu=\mbox{\it const}$):
     $$ v_{\mbox{\footnotesize\textit{гр}}}=
         \frac c {\sqrt{\varepsilon\mu}}\sqrt{1-(k_n/k)^2}\;
         \displaystyle{\frac 1 {1+\displaystyle{\frac {\omega}{2
         \varepsilon}\frac{d\varepsilon}{ d\omega}}}}\,,\eqno (8.16)$$
и, разумеется,   всегда   $v_{\mbox{\footnotesize\textit{гр}}}<c$.

     Отметим здесь,  что  групповую  скорость  легко оценить с помощью
приведенной на рис.~8.3  диаграммы  Бриллюэна  (в  данном случае
для пустого волновода): $v_{\mbox{\footnotesize\textit{гр}}}$ для
любой распространяющейся волноводной волны равна скорости света
$c$ ,  умноженной на тангенс  угла  наклона касательной  к
соответствующей  кривой при интересующем нас значении $h_n$;  в
частности,  хорошо видно, что при  малых  $h_n$  групповая
скорость  стремится к нулю,  а при больших $h_n$ --- к скорости
света $c$.  С  групповой скоростью распространяется  любое
возмущение электромагнитного поля, например, импульс в волноводе;
но как и в любой другой среде, предвестник в волноводе всегда
идёт со скоростью  $c$.

\begin{wrapfigure}[14]{l}{7.5cm}
\begin{picture}(80,45)
\put(-6,50){\special{em:graph Fig8-4.bmp}}
\end{picture}
\hbox to 7.5cm{\hfil\footnotesize{Рис.~8.4.~Фазовая
и групповая скорости} \hfil}
\hbox to 7.5cm{\hfil\footnotesize{волны в волноводе.}\hfil}
\end{wrapfigure}

     Зависимость $v_{\mbox{\footnotesize\textit ф}}$  и
$v_{\mbox{\footnotesize\textit{гр}}}$ от частоты (от волнового числа $k$)
представлена на рис.~8.4. Единственным параметром на этом графике
является критическое значение  $k_n$;  в  остальном  этот  рисунок,
как  и формулы (8.13) и (8.14),  универсален,  то есть относится к
волноводам любого сечения и ко  всем  волнам.  Вблизи критической
частоты распространяющаяся волна имеет своеобразные характеристики
--- групповая  скорость  её  очень мала,  а  длина  волны  очень
велика.  Кроме того,  волна в этой
области частот  в  реальных  условиях  неидеально проводящих
стенок сильно затухает,  её трудно возбудить,  так что в этой
области обычно стараются не работать.

     При $k\gg  k_n$  фазовая и групповая скорости близки к $c/ \sqrt{
\varepsilon \mu}$, как у волны в безграничном однородном пространстве;
при этом и {\it поля близки к полю плоской волны}.  Действительно, так
как  $h_n\approx  k\sqrt{\varepsilon\mu}$,  то  согласно  (7.16)  $E_x
\approx  \sqrt{  \mu/  \varepsilon}  H_y,  E_y  \approx  - \sqrt{ \mu/
\varepsilon} H_x$,  а согласно (7.5), (7.12) и (7.13) эти поперечные компоненты
значительно (по порядку величины в $k/k_n$ раз) больше продольных,  то
есть возникают те же соотношения, что и в плоской волне.

     {\it Поперечные компоненты} всей совокупности
электрических и магнитных волн (в неё  должны быть включены
волны   обоих   направлений)   образуют   полную систему
ортогональных  векторных  полей  --- поперечные компоненты
любого  поля,  удовлетворяющего  уравнениям  Масвелла  и
граничному    условию    $\left.E_t\right|_C=0,$    могут   быть
разложены в ряд по полям  этой совокупности.  Полнота
системы  следует  из  полноты  системы скалярных собственных функций
граничных задач (8.5) и (8.6), а их ортогональность понимается в том
смысле, что  интеграл
     $$I=\int\limits_S\bigl[\rv E_n\rv H_m\bigr]_z\,dS\,,  \eqno (8.17)$$
взятый по произвольному поперечному сечению волновода $S$, отличен от нуля
только для полей волн одного типа (электрических или магнитных) и при
условии  $|n|=|m| $. В этом легко убедиться, выразив  компоненты поля
$E_{n,x}$,    ~$E_{n,y}$, ~$H_{m,x}$,  ~$H_{m,y}$,  входящие в интеграл
(8.17), через мембранные функции $\Pi^m$ и~$\Pi^e$.

Для волн разного типа (пусть, например, $\rv E_n$ -- поле электрической,
а   $\rv  H_m$  --  поле магнитной волны)
   $$ I=-h_n h_m
     \int\limits_S \Bigl(\frac{\partial\Pi^e_n}{\partial y}
     \frac{\partial\Pi^m_m}{\partial x}-\frac{\partial\Pi^e_n}{\partial x}
     \frac{\partial\Pi^m_m}{\partial y}\Bigr)\,dS=-h_n h_m
     \int\limits_S \rot_z(\Pi^e_n \nabla_2\Pi_m^m)\,dS\,,\eqno(8.18)$$
где $\nabla_2$ --- двумерный оператор, имеющий в прямоугольной системе
координат вид $\nabla_2=\rv i {\partial / \partial x}+ \rv j {\partial
/\partial  y}$. В силу теоремы Стокса (1.16) и граничных условий (8.5) и (8.6)
интеграл (8.18) равен нулю при всех $m$ и $n$.

   Если оба поля в интеграле (8.17) представляют волны одного типа, то
для электрических и магнитных волн получаем соответственно
       $$I=-h_n k\varepsilon\int\limits_S\nabla_2\Pi^e_n\nabla_2
          \Pi^e_m\,dS\quad\mbox{и}\quad I=-h_m k\mu\int\limits_S
          \nabla_2\Pi^m_n\nabla_2\Pi^m_m\,dS\,.\eqno(8.19)$$
Из двумерной первой формулы Грина
     $$\int\limits_S\{\varphi\Delta_2\psi+\nabla_2\varphi\cdot\nabla_2
         \psi\}\,dS=\oint\limits_C\varphi\frac{\partial\psi}{\partial n}
         \,dl\,,\eqno (8.20)$$
и условий граничных задач (8.5) и (8.6) следует, что
     $$\int\limits_S\nabla_2 \Pi^{e,m}_n\nabla_2 \Pi^{e,m}_m\,dS=
        g_n^2\int\limits_S\Pi^{e,m}_n\Pi^{e,m}_m\,dS=
        g_m^2\int\limits_S\Pi^{e,m}_n\Pi^{e,m}_m\,dS\,.\eqno(8.21)$$
Последнее равенство доказывает ортогональность (8.8) мембранных функций,
а из первого с учётом (8.19) следует ортогональность полей волноводных
волн в сформулированном выше виде.

     Для мембранных функций удобно ввести следующую нормировку:
     $$\int\limits_S(\nabla_2\Pi^m_n)^2\,dS=1\,,\qquad\int\limits_S(
         \nabla_2\Pi^e_n)^2\,dS=1\,.\eqno (8.22)$$
Тем самым устанавливается нормировка и для полей:
     $$\int\limits_S\bigl[\rv E_n\rv H_{-n}\bigr]_z\,dS=-\int\limits_
         S\bigl[\rv E_{-n}\rv H_n\bigr]_z\,dS=-\varepsilon kh_n\mbox
         { --- \it электрические волны,}\eqno(8.23)$$
     $$\int\limits_S\bigl[\rv E_n\rv H_{-n}\bigr]_z\,dS=-\int\limits_
         S\bigl[\rv E_{-n} \rv H_n\bigr]_z\,dS=
         \phantom{-}k\mu h_n\mbox{ ---
         \it магнитные волны.\phantom{ааа}}\eqno(8.24)$$
Интегралы (8.23) и (8.24) сохраняют свои  значения  и  в  том  случае,
когда  поля  $\rv E_n$,~$\rv H_n$ включают в себя множитель $e^{ihz}$,
и, следовательно, они не зависят от $z$ (если в формуле (8.17) индексы
$m$ и~$n$  равны  друг  другу,  то  интегрирование следует проводить в
плоскости $z=0$).

     После введения  определённой  нормировки   можно   говорить   в
дальнейшем  об  амплитудах  возбуждаемых  в  волноводе  волн.  Нередко
используется и нормировка, отличная от (8.22); в задачах о возбуждении
волновода  какими-либо  источниками  это изменяет амплитуды волн,  но,
естественно, не изменяет значения самих полей.

     Обратим особо  внимание  на  то,   что   {\it   полнота   системы
волноводных  волн}  имеет  место только для поперечных компонент поля.
Если найдены коэффициенты  разложения  поперечных  компонент  искомого
поля  по  поперечным  компонентам  волноводных волн,  то выражения для
$E_z$,~$H_z$ в той области пространства, где находятся источники, надо
затем искать из уравнений Максвелла --- они,  вообще говоря,  не будут
равны сумме продольных компонент волноводных волн,  взятых с  теми  же
коэффициентами.   Этот   вопрос  более  подробно  будет  рассмотрен  в
дальнейшем при рассмотрении задач о возбуждении волноводов.

     Ортогональность полей   волноводных   волн   приводит   к  важным
следствиям,    упрощающим   расчёты   энергетических   характеристик
электромагнитного поля,  в частности, потока мощности вдоль волновода.
Средняя по времени плотность потока  мощности
определяется $z$-компонентой вектора Умова-Пойнтинга и для
$n$-й  электрической волны  составляет
     $$\overline{\gv S}_{n,z}=\frac{c}{8\pi}\re\{{[\rv E_n\rv H_n^*]
         }_z\}=\frac{ckh_n\varepsilon}{8\pi}{(\nabla_2\Pi^e)^2}.
         \eqno(8.25)$$
Полный поток   $\overline{\Sigma}_{n}$    находится    интегрированием
$\overline{\gv S}_{n,z}$ по поперечному сечению волновода;  с учётом
нормировки (8.22)
     $$\overline{\Sigma}_{n}=\frac{ckh_n\varepsilon}{8\pi}\,.
         \eqno(8.26)$$
Поток для  магнитной волны даётся аналогичным выражением,  в котором
$\varepsilon$ заменена  на  $\mu$.  Если  же  в  волноводе  возбуждено
несколько  волн,  то  из-за их ортогональности перекрёстные члены от
разных волн не дают вклада в суммарный поток и он равен  просто  сумме
потоков отдельных волн.

     Этим же  свойством  обладает  средняя  энергия   $\overline{U}_n$
электромагнитного  поля  волноводной  волны,  приходящаяся  на единицу
длины волновода; для электрической волны
     $$\begin{array}{l}\overline{U}_{n}=\displaystyle{\frac{1}{16\pi}
         \int\limits_S(\varepsilon\rv E_n\rv E_n^*+\mu\rv H_n\rv H_n
         ^*)}\,dS=\\[.3cm]=\displaystyle{\frac{1}{16\pi}\int\limits_S
         \left\{\varepsilon[h_n^2(\nabla_2\Pi_n^e)^2+g_n^4(\Pi_n^e)
         ^2]+\mu \varepsilon^2 k^2(\nabla_2\Pi_n^e)^2\right\}\,dS}\,.
         \end{array}\eqno(8.27)$$
Можно показать, что из соотношений (8.21) при $m=n$ следует равенство
     $$\varepsilon\int\limits_S\rv E_n\rv E_n^*\,dS=\mu\int\limits_S
         \rv H_n\rv H_n^*\,dS\,,\eqno(8.28)$$
так что с учётом нормировки (8.22)
     $$\overline{U}_{n}=\frac{\varepsilon^2\mu k^2}{8\pi}\,.
         \eqno(8.29)$$
Для энергии  магнитной  волны  получается  аналогичное  выражение,   в
котором $\varepsilon$ и $\mu$ просто следует поменять местами. Обратим
внимание,  что при выбранной нормировке энергия не зависит  от  номера
волны.

     Сопоставив формулы (8.14),  (8.26)  и  (8.29), находим, что между
групповой  скоростью волны,  потоком мощности по волноводу и энергией
на единицу длины имеет место привычное соотношение
     $$\overline{\Sigma}_{n}=v_{\mbox{\footnotesize\textit{гр}}}
       \overline{U}_{n};\eqno(8.30)$$
оно сохраняется  и  при учёте дисперсии свойств заполняющей волновод
среды.  В этом случае групповая скорость определяется формулой (8.16),
а для энергии волны получаем следующее выражение:
     $$\overline{U}_{n}=\displaystyle{\frac{1}{16\pi}\int\limits_S
         \left\{\frac{d{(\omega\varepsilon)}}{d\omega}\rv E_n\rv E_n^*
         +\mu\rv H_n\rv H_n^*)\right\}}\,dS=\frac{\varepsilon^2\mu k^
         2}{8\pi}\left(\displaystyle{1+\displaystyle{\frac{\omega}
         {2\varepsilon}\frac{d\varepsilon}{d\omega}}}\right)\,,
         \eqno(8.31)$$
в котором для сокращения записи учтена дисперсия только  $\varepsilon$,
а при выводе последнего равенства  использована формула (8.28).

%\end{document}

\newpage
\oddsidemargin=-0.4mm \evensidemargin=-0.4mm
\topmargin=-0.4mm
\headsep=7mm
\textheight=231.875mm
\textwidth=160mm
\mathsurround=2.5pt
\unitlength=1mm
%\begin{document}
%\input{macr.tex}
\thispagestyle{empty}
%\addtocounter{page}{90}
\baselineskip=\normalbaselineskip

\begin{center}\subsubsection*{9. Волноводы прямоугольного и круглого
         сечений}\end{center}\vspace*{0.5cm}

\markboth{Глава 3. Волноводы}{9. Волноводы прямоугольного и круглого
         сечений}

\begin{center}\begin{minipage}[c]{0.75\textwidth}
\footnotesize{\parindent=0.5cm
         Собственные значения  и  собственные  функции  прямоугольного
         волновода.  Структура силовых линий основных волн. Волновод с
         круговым   поперечным  сечением.  Поляризационное  вырождение
         волноводных волн. Разложение волноводных волн на плоские.
}\end{minipage}\end{center}\vspace*{0.5cm}

     При исследовании волн, которые могут быть возбуждены в волноводе с
{\it   прямоугольной   формой}   поперечного   сечения,    естественно
использовать  декартову  систему  координат.  Пусть поперечное сечение
волновода расположено так,  как показано на рис.~9.1. При таком выборе
контур  $C$,  представляющий собой прямоугольник,  состоит из отрезков
координатных  линий.  Будем  предполагать  ---  как  это   принято   в
электродинамике  СВЧ  --- что $a>b$,  то есть широкая стенка волновода
помещена  вдоль  оси  $x$,  и  пусть  волновод  заполнен  веществом  с
проницаемостями  $\varepsilon,\,\mu$.  Волноводные  волны определяются
решением краевых задач (8.5)  и  (8.6),  в  которых  необходимо  найти
собственные  значения  $g_n$  и  явный вид собственных функций функций
$\Pi^m_n(x,y)$ и $\Pi^e_n(x,y)$.

\begin{wrapfigure}[10]{l}{7.5cm}
\begin{picture}(80,55)
\put(3,45){\special{em:graph fig9-1.bmp}}
\end{picture}
\hbox to 7.5cm{\hfil\footnotesize{Рис.~9.1.~Прямоугольный волновод.}
\hfil}\end{wrapfigure}
     В прямоугольной системе координат краевая задача (8.5)  для  {\it
магнитных волн} сводится к уравнению
     $$\displaystyle{\frac{\partial^2\Pi^m}{\partial x^2}+\frac{\partial^2
         \Pi^m}{\partial y^2}}+g^2\Pi^m=0\eqno(9.1{\it a})$$
с граничными условиями (условия Ней\-мана)
     $$\left.\begin{array}{rcl}\displaystyle{\frac {\partial\Pi^m}
         {\partial x}}&=&0\quad\mbox{при } $x\,=\,0$\; \mbox{и}\; $x\,=\,a\,,$\\[.4cm]
         \displaystyle{\frac {\partial\Pi^m}{\partial y}}&=&0\quad
         \mbox{при } $y\,=\,0$\; \mbox {и}\;$y\,=\,b\,.$\end{array}\right\}\eqno(9.1\mbox{\textit б})$$

     Решение уравнения  (9.1\mbox{\textit а})  методом разделения переменных
(методом Фурье) уже было приведено ранее (формулы (7.18), (7.23)), так что
     $$\Pi^m(x,y)=(A\cos g_xx+B\sin g_xx)(C\cos g_yy+D\sin g_yy)\,,
         \eqno(9.2)$$
где $g^2_x+g^2_y=g^2=k^2\varepsilon\mu-h^2$. Постоянные
$A$,~$B$,~$C$,~$D$   и   допустимые   значения   поперечного
волнового  числа  $g$ определяются из граничных условий
(9.1\mbox{\textit б}). Из условий на стенках $x=0$ и~$y=0$
следует, что $B=0$ и~$D=0$, а из условий на стенках $x=a$ и~$y=b$
находятся значения
     $$g_x=\frac{\pi p}{a}\,,\qquad g_y=\frac{\pi q}{b}\,,\eqno(9.3)$$
где $p$,~$q$ --- целые  числа.  Окончательно,  с  учётом  нормировки
(8.22),  собственную  функцию  $\Pi^m(x,y)$  для  магнитных волн можно
записать в виде
     $$\Pi^m_{pq}=A_{pq}\cos{\frac{\pi p}{a}\,x}\,\cos{\frac{\pi q}
         {b}\,y}\,,\eqno(9.4)$$
где
     $$A_{pq}=\frac{2}{g_{pq}}\,\sqrt{\frac{1}{ab(1+\delta_
         {0p})(1+\delta_{0q})}}\,,\qquad p,\,q=0,1,2,\dots\,,
         \eqno(9.5)$$
а $\delta_{pq}$ --- символ Кронекера.

     Итак, формула  (9.4)  представляет  собой  общее  решение краевой
задачи (9.1), собственные значения которой
     $$g_{pq}=\sqrt{\frac{\pi^2 p^2}{a^2}+\frac{\pi^2 q^2}{b^2}}
         \eqno(9.6)$$
определяются двумя  числами  $p$  и~$q$.  Каждое  из  этих чисел может
принимать значение $0$,  но нe оба  одновременно.  Волноводная  волна,
определяемая  потенциальной  функцией  (9.4)  и  собственным значением
(9.6), обозначается как $H_{pq}$; её продольное волновое число равно
     $$\vspace{.3cm}h_{pq}=\displaystyle{\sqrt{k^2\varepsilon\mu
         -\frac{\pi^2 p^2}{a^2}-\frac{\pi^2 q^2}{b^2}}}\,.\eqno(9.7)$$
Продольная компонента поля $H_z$ магнитной волны определяется формулой
(7.13),  а все поперечные компоненты поля вычисляются с помощью  (9.4)
по  формулам  (7.16)  (при  этом  необходимо  в  соответствии  с (7.1)
добавить множитель $\displaystyle {e^{ih_{pq}z}}$). Если одно из чисел
$p$  или  $q$  равно  нулю,  то у волны отличны от нуля три компоненты
поля, во всех остальных случаях --- пять компонент.

     Наибольшее практическое   применение   в  технике  находит  волна
$H_{10}$.  Поле этой волны не зависит от координаты  $y$.  Отличны  от
нуля следующие компоненты поля:
     $$\left.\begin{array}{l}E_y=\displaystyle{ik\mu\,\sqrt{\frac{2}{ab
         }}\,\sin{\frac{\pi x}{a}}\,{e^{ih_{10}z}}}\,,\\[.5cm]H_x=
         \displaystyle{-ih_{10}\,\sqrt{\frac{2}{ab}}\,\sin{\frac{\pi
         x}{a}}\,{e^{ih_{10}z}}}\,,\\[.5cm]H_z=\displaystyle{\frac{\pi}
         {a}\,\sqrt{\frac{2}{ab}}\,\cos{\frac{\pi x}{a}}\,{e^{ih_{10}z}}}\,,
         \end{array} \right\} \eqno (9.8)$$
а критическая  частота   $k_{10}=\pi   /a\sqrt{\varepsilon\mu}$.   Для
пустого  волновода  ($\mu=\varepsilon  =1$)  критическая длина волны
равна удвоенному размеру широкой стенки волновода $2a$.

     Из формул  (9.8)  видно,  что  электрическое  поле волны $H_{10}$
направлено  по  оси  $y$,  причем  наибольшая  густота  силовых  линий
достигается   при   $x=a/2$.   Магнитные   силовые   линии   лежат   в
горизонтальных  плоскостях,  их  проекции  на  плоскость   поперечного
сечения являются горизонтальными прямыми, изображёнными на рис.~9.2{\it a}
штриховыми линиями.  В плоскости  продольного  сечения  $x,z$  силовые
линии  магнитного  поля  распространяющейся  волны являются замкнутыми
кривыми,  повторяющимися с периодом,  равным длине волны  в  волноводе
$\Lambda$ (рис.~9.2{\it б}).

     Структура поля  волны  $H_{p0}$  представляет  собой  $p$-кратное
повторение структуры волны $H_{10}$ по оси $x$.  Структура поля  волны
$H_{01}$  та же,  что и у волны $H_{10}$,  но с заменой оси $x$ на ось
$y$ и наоборот.  Поле волны $H_{0q}$ является $q$-кратным  повторением
поля волны $H_{01}$ по оси $y$ в уменьшенном масштабе.

\begin{picture}(10,55)
\put(5,50){\special{em:graph fig9-2a.bmp}}
\put(70,50){\special{em:graph fig9-2b.bmp}}
\end{picture}
\begin{center}\begin{minipage}[c]{0.9\textwidth}
\footnotesize{\parindent=0.5cm
\begin{center}Рис.~9.2.~Волна $H_{10}$ в прямоугольном волноводе.
\end{center}
}\end{minipage}\end{center}\vspace*{0.25cm}

     Структура силовых  линий  поля  волны  $H_{11}$  представлена  на
рис.~9.3.  Эту  картину  легко  построить,  если  учесть,  что функция
$\Pi^m_{11}$,  уровни постоянного значения которой и определяют  форму
силовых  линий  электрического  поля,  достигает своего экстремального
значения  в  каждой  вершине  прямоугольника.  Поэтому  вблизи  каждой
вершины  прямоугольника  в пределах его внутреннего угла силовые линии
близки к четверти эллипса,  в чём легко убедиться,  разложив функцию
(9.4) в ряд по степеням $x$ и~$y$.  При удалении от вершины угла форма
линий изменяется и  они  становятся  более  похожими  на  гиперболы  с
асимптотами  $x=a/2$,~$y=b/2$.  Проекции ортогональных к электрическим
магнитных силовых линий на плоскость поперечного сечения изображены на
рис.~9.3{\it {а}}   штриховой   линией.   Магнитные   силовые  линии  являются
пространственными кривыми  и  их  проекции  на  плоскость  продольного
сечения  $(y,z)$  для  распространяющейся волны $H_{11}$ изображены на
рис.~9.3{\it {б}}.

     Структура электромагнитного   поля   волны   $H_{pq}$  получается
разбиением сечения на $p\times q$ прямоугольников, в каждом из которых
размещается  в  соответственно  уменьшенном  виде структура поля волны
$H_{11}$.

     Перейдем теперь к рассмотрению электрических волн. Краевая задача
(8.6) в декартовой системе координат имеет вид
     $$\displaystyle{\frac{\partial^2\Pi^e}{\partial x^2}+\frac{\partial^2
         \Pi^e}{\partial y^2}+g^2\Pi^e=0}\,\eqno(9.9\mbox{\textit а})$$
с граничными условиями (условиями Дирихле)
     $$\displaystyle{\left.\Pi^e\right|_{x=0}=\left.\Pi^e\right|_{x=
         a}=\left.\Pi^e\right|_{y=0}=\left.\Pi^e\right|_{y=b}=0}\,.
         \eqno(9.9\mbox{\textit б})$$
Решение уравнения  (9.9а) имеет тот же вид (9.2),  как и для магнитных
волн,  однако теперь граничные условия на стенках волновода определяют
другую собственную функцию
     $$\Pi^e_{pq}=A_{pq}\sin{\frac{\pi p}{a}x}\,\sin{\frac{\pi q}
         {b}y}\eqno(9.10)$$
с тем же выражением (9.5) для $A_{pq}$ и теми  же  собственными  значениями
(9.6) как и для  магнитных волн; единственное  отличие    состоит в том,
что для электрических волн числа  $p$  и~$q$  не  могут  быть  нулями,
поскольку для этих значений функция (9.10) обращается в нуль.

\begin{picture}(160,55)
\put(0,50){\special{em:graph fig9-3a.bmp}}
\put(65,50){\special{em:graph fig9-3b.bmp}}
\end{picture}

\begin{center}\begin{minipage}[c]{0.9\textwidth}
\footnotesize{\parindent=0.5cm
\begin{center}Рис.~9.3.~Волна $H_{11}$ в прямоугольном волноводе.
\end{center}
}\end{minipage}\end{center}\vspace*{0.1cm}

     Структура электромагнитного поля в волне $E_{pq}$ устанавливается
путём анализа поля  простейшей  электрической  волны  $E_{11}$,  для
которой собственная функция имеет вид
     $$\Pi^e_{11}(x,y)=\frac 2 \pi \sqrt{\frac{ab}{a^2+b^2}}\,
         \displaystyle{\sin{ \frac {\pi x}{a}}\,\sin{\frac{\pi y}
         {b}}}\,.\eqno(9.11)$$
\vspace{2mm}
Эта функция  принимает  максимальное  значение в центре прямоугольника
при $x=a/2$ и~$y=b/2$, вблизи которого линии уровня имеют вид эллипсов
с  этим  же  центром,  а оси направлены вдоль осей $x$ и $y$.  По мере
увеличения эллипса форма его  искажается  и  он  все  больше  начинает
походить  на  прямоугольник со скруглёнными углами.  Наконец,  линия
уровня   $\Pi^e_{11}=0$    совпадает    с    самим    прямоугольником.
Изображённые   на   рис.~9.4{\it a}   штриховые  линии  $\Pi^e_{11}=const$
согласно  общим  соображениям,  изложенным   в   предыдущем   разделе,
представляют  собой  силовые линии магнитного поля,  а ортогональные к
ним сплошные кривые --- силовые линии электрического поля.

\begin{picture}(160,50)
\put(0,45){\special{em:graph fig9-4a.bmp}}
\put(65,45){\special{em:graph fig9-4b.bmp}}
\end{picture}
\begin{center}\begin{minipage}[c]{0.9\textwidth}
\footnotesize{\parindent=0.5cm
\begin{center}Рис.~9.4.~Волна $E_{11}$ в прямоугольном волноводе.
\end{center}
}\end{minipage}\end{center}\vspace*{0.1cm}

     Cиловые линии   электрического   поля   распространяющейся  волны
$E_{11}$ в продольном сечении  $(y,z)$  изображены  на  рис.~9.4{\it б},  из
которого   видно,  что  они  начинаются  на  стенках  и  сначала  идут
перпендикулярно к ним в плоскости поперечного сечения.  Приближаясь  к
оси волновода, силовые линии отходят от этого сечения, идут в основном
параллельно оси волновода и кончаются опять на стенке,  подходя к  ней
под  прямым  углом.  Картина  силовых  линий  в  плоскости продольного
сечения обладает периодом, равным длине волны в волноводе $\Lambda$.

     Структура электромагнитного  поля  волны  $E_{pq}$  в  поперечном
сечении находится разбиением его  на  $p\times  q$  прямоугольников  и
размещении в каждом из них в соответственно уменьшенном виде структуры
поля волны $E_{11}$.

     Сравним между   собой   критические  длины  волн  $\lambda_{pq}$,
которые   определяются    поперечным    волновым    числом    $g_{pq}$
соответствующей  волны  в  волноводе  по  формуле  $\lambda_{pq} =2\pi
/g_{pq}$. Волна $H_{10}$ имеет самую большую длину $(\lambda_{10}=2a)$
среди  всех  $\lambda_{pq}$  волн  обоего типа и поэтому её называют
{\it основной} в прямоугольном  волноводе.  Имеется  большой  диапазон
длин волн $\lambda$,  меньших $2a$, для которых прямоугольный волновод
работает в одномодовом режиме. Следующую критическую длину волны будет
иметь либо  волна  $H_{01}$~($ \Pi^m_{01} =\displaystyle {\frac{2b}
{\pi^2a}\cos{\frac{\pi  y}{b}}}, ~\lambda_{01}=2b)$, электрическое поле
которой   параллельно   большей  стороне  прямоугольника,  либо  волна
$H_{20}$~($  \Pi^m_{20  }=\displaystyle {\frac a{2\pi^2b}\cos{\frac{2\pi
 x}{a}}}, ~\lambda_{20}=a$), поле которой содержит те же компоненты (9.8),
что и поле волны $H_{10}$, но в которой $E_y$ обращается в нуль не только на
вертикальных стенках, но и на средней линии. Если $a>2b$, то следующей
за основной будет волна $H_{20},$ и диапазон одномодового режима будет
$a<\lambda<2a$.  Если  $a<2b$,  то  следующей  за основной будет волна
$H_{01}$, и волновод будет пропускать только  одну  волну  в  диапазоне
$2b<\lambda<2a.$  Для волн $H_{11}$ и~$E_{11}$ критическая длина волны
одинакова  $(\lambda_{11}=2/\sqrt{1/a^2+1/b^2})$   и   всегда   меньше
соответствующей  длины  по  крайней мере для двух волн.  Все остальные
волны имеют ещё  меньшее значение критической длины волны.

     Как правило,  передача  электромагнитной  энергии по волноводам с
помощью нескольких распространяющихся волн  представляет  значительные
технические неудобства, поэтому для заданного диапазона частот размеры
прямоугольного волновода подбираются  таким  образом,  чтобы  в  нём
могла распространяться только волна $H_{10}$,  а все остальные были бы
затухающими.  Этим  же  (и  габаритными  соображениями)   определяется
преимущественное     использование    прямоугольных    волноводов    в
сантиметровом диапазоне.

     Остановимся ещё  на  проблеме  вырождения  волн в прямоугольном
волноводе, которое состоит в том, что одному собственному значению
краевой  задачи  соответствует  две или больше собственных функций.  В
прямоугольном волноводе с соизмеримыми сторонами $a$ и~$b$  вырождение
всегда  имеет место.  В частности,  в квадратном волноводе (при $a=b$)
волны $E_{pq}$ и~$E_{qp}$~$ (p\neq q)$  имеют  одинаковые  собственные
значения, хотя распределения их полей в пространстве не совпадают.

     Для прямоугольного  волновода  характерно особое вырождение,  так
называемое вырождение $E-H$. Дело в том, что поперечные волновые числа
(9.6)  для  волн  $E_{pq}$  и~$H_{pq}$  совпадают,  и, таким образом, все
собственные значения являются вырожденными.  Невырожденные собственные
значения   имеют   только   волны   $H_{p0}$   и~$H_{0q}$,   поскольку
электрические волны  с  такими  индексами  отсутствуют.  В  частности,
невырожденным   является   собственное   значение  $g_{10}=\pi/a$  для
основной волны прямоугольного волновода $H_{10}$ при $a>b$.

     Вырождение $E-H$  в прямоугольном волноводе приводит к тому,  что
вместо волн $E_{pq}$ и~$H_{pq}$ в качестве двух независимых типов волн
можно брать линейные комбинации этих волн.  Например,  можно составить
такие линейные комбинации,  для одной из которых $E_x\neq 0$,~$H_x=0$,
а  для  другой  $H_x\neq  0$,~$  E_x=0$.  Эти  новые волны также можно
назвать {\it электрическими} и {\it магнитными},  но теперь эта  классификация
производится  относительно  оси  $x$, и эти волны являются продольными.
Можно показать, что они могут быть получены с помощью электрического и
магнитного   векторов   Герца,  имеющих  по  одной  отличной  от  нуля
составляющей $\Pi_x^e$ и~$\Pi_x^m$.

     Рассмотрим теперь электрические и магнитные волны в  волноводе  с
круговым  поперечным  сечением  радиуса  $a$  или,  кратко,  в круглом
волноводе.  Для  этого  естественно   воспользоваться   цилиндрической
системой  координат $r$,~$\varphi$,~$z$,  ось которой совпадает с осью
волновода. Краевая задача для  магнитных волн записывается в
полярных координатах в виде уравнения
     $$\displaystyle{\frac{\partial^2 \Pi^m}{\partial r^2}+\frac{1}{r}
         \frac{\partial\Pi^m}{\partial r}+\frac{1}{r^2}\frac{\partial^2\Pi^m
         }{\partial \varphi^2}+g^2\Pi^m =0}\,\eqno(9.12\mbox{\textit a})$$
с граничным условием
     $$\displaystyle{\left .\frac{\partial\Pi^m}{\partial r}\right|_{r=a}
         =0}\,.\eqno(9.12\mbox{\textit б})$$
Решение уравнения (9.12{\it {а}}) методом Фурье уже было получено в разделе 7.
Ограниченное в нуле решение запишем в виде
     $$\Pi^m(r,\varphi)= A J_p(g r)\cos {p(\varphi-\varphi_
         0)}\,,\eqno(9.13)$$
где $J_p(g r)$ --- функция Бесселя порядка $p$ $(p=0,1,2,\dots)$.
Граничное  условие  (9.12{\textit б})  определяет дискретный набор
собственных значений задачи:
     $$g_{pq}=\frac{\mu_{pq}}{a}\,,\qquad q=1,2,\dots\,,\eqno(9.14)$$
где $\mu_{pq}$ --- корни  производной  функции  Бесселя:  $J'_p  (\mu_
{pq})=0$.  Наименьшие  из  них  $\mu_{11}\approx 1,841$,  ~$\mu_{01}\approx 3,832$
понадобятся нам в дальнейшем. Как и в случае прямоугольного волновода,
собственные значения определяются парой чисел $p,\,q$.

     С учётом нормировки (8.22) мембранная  функция магнитных волн
в круглом волноводе может быть представлена в виде
     $$\Pi^m_{pq}= A_{pq} J_p(\frac{\mu_{pq}}{a}r)\cos{p(\varphi
         -\varphi_0)},\quad p=0,1,2,\dots,\quad q=1,2,\dots\,,
         \eqno(9.15)$$
где
     $$A_{pq}=\frac 1{J_p(\mu_{pq})}\,\sqrt{\frac{2}{\pi(1+\delta_{0p})
         (\mu_{pq}^2-p^2)}}\,.\eqno(9.16)$$
Произвольность угла $\varphi_0$ приводит к тому, что для любого $p\neq
0$ существуют два линейно независимых решения, отличающиеся значениями
$\varphi_0$,  например,  решения  с  множителями  $\cos{p\varphi}$   и
$\sin{p\varphi}$;   решение   с   любым   другим   $\varphi_0$   может
рассматриваться как комбинация этих двух.
Любое    собственное    значение,    соответствующее    несимметричной
электрической  или   магнитной   волне,   то   есть   имеющее   индекс
$p=1,2,3,\dots$,  является  вырожденным,  так как ему соответствуют по
крайней  мере  две  собственные  функции.  Это  вырождение  называется
поляризационным  вырождением  и является следствием того,  что круглый
волновод обладает симметрией вращения.  В  некоторых  случаях  уже  на
первоначальной    стадии   расчётов   электродинамических   структур
приходится    предусматривать    специальные    меры    для     снятия
поляризационного вырождения.

     Волна с мембранной  функцией  (9.15)  называется  волной  типа
$H_{pq}$,  её  продольное  волновое  число  равно
     $$h_{pq}=\sqrt{k^2\varepsilon\mu-\frac{\mu^2_{pq}}{a^2}}\,,
         \eqno(9.17)$$
а продольная компонента магнитного поля
     $$H_z(r,\varphi,z)=g^2_{pq}\,\Pi^m_{pq}\;e^{ih_{pq}z}\,;\eqno(9.18)$$
поперечные компоненты   вычисляются  по  формулам  (7.17),  в  которых
функция  $\Pi^m$  определена  выражением  (9.15),  а  функцию  $\Pi^e$
следует положить равной нулю.

\begin{picture}(160,55)
\put(0,50){\special{em:graph fig9-5a.bmp}}
\put(55,50){\special{em:graph fig9-5b.bmp}}
\put(110,50){\special{em:graph fig9-5c.bmp}}
\end{picture}
\begin{center}\begin{minipage}[c]{0.9\textwidth}
\footnotesize{\parindent=0.5cm
\begin{center}Рис.~9.5.~Силовые линии волн круглого волновода.
\end{center}
}\end{minipage}\end{center}\vspace*{0.25cm}

     Наибольший практический  интерес   среди   всех   волн   круглого
волновода представляет волна $H_{11}$~$(g_{11} \approx 1,841/a)$. Линии уровня
$\Pi^m_{11} = const$  изображены на рис.~9.5{\it {а}}  сплошными  линиями  для
$\varphi_0=0$.  Эти линии нормальны к граничной окружности поперечного
сечения и  совпадают  с  электрическими  силовыми  линиями.  Магнитные
силовые  линии  ортогональны  электрическим  и изображены на рис.~9.5{\it а}
штриховыми линиями.  Картина распределения электрических силовых линий
волны  $H_{11}$  в  круглом волноводе похожа на ту же картину для волн
$H_{10}$  в  прямоугольном  волноводе  (рис.~9.2{\it {а}}):  в  обоих  случаях
имеется  пучок  электрических  силовых  линий,  идущих  в вертикальном
направлении  вдоль  оси  симметрии  волновода.  Для  другого  значения
$\varphi_0$  картина  силовых  линий получается поворотом рис.~9.5{\it {а}} на
этот угол.

     Волна $H_{11}$  является {\it основной} в круглом волноводе.  Она
имеет {\it наименьшее} собственное значение $g_{11}$ не  только  среди
всех $g_{pq}$,  определяемых для $H$-волн формулой (9.14), но и среди
собственных значений для $E$-волн,  о которых речь  впереди.  Как  уже
отмечалось,  это  означает,  что в некотором диапазоне длин волн волна
$H_{11}$  ---  единственная,  распространяющаяся  без  затухания  (при
идеально   проводящих   стенках).   Её   критическая   длина   волны
$\lambda_{11}=3,41a$,  а ближайшей к ней по  критической  длине  волны
оказывается    волна    $E_{01}$,    для    которой   $g_{01}=2,405/a$
и~$\lambda_{01}= 2,61a $.  Таким образом, для круглого волновода режим
одноволновости   выполняется   для  длин  волн,  лежащих  в  диапазоне
$2,61a<\lambda<3,41a$.

     Поле несимметричных $H$-волн $(p\neq 0)$, в частности, поле волны
$H_{11}$,  содержит  пять  компонент.  В  особом  положении  находятся
симметричные   волны   $H_{0q}$   ---  их  поля  содержат  только  три
компоненты: $E_\varphi$,~$H_r$,~$H_z$.

     До сих  пор  при  исследовании  волн  в  прямоугольном  и круглом
волноводах основное внимание обращалось на электромагнитные поля  этих
волн.  Остановимся  теперь  кратко  на  токах,  наводимых  на  стенках
волноводов.  Знание тангенциальных  составляющих  магнитного  поля  на
идеально  проводящих  стенках волновода позволяет с помощью граничного
условия (1.18) вычислить поверхностную плотность  тока  ${\rv  I}$  на
стенках:
     $${\rv I}=-\frac{c}{4\pi}[\rv n\rv H]\,.\eqno(9.19)$$
Для электрических волн $H_z=0$, и поэтому токи имеют только  продольную
составляющую $I_z$.

     Для магнитных волн $H_z\neq 0$,  поэтому поверхностная  плотность
тока  магнитных  волн  имеет,  помимо  продольной,  также  и  поперечную
составляющую.  В частности,  поперечными составляющими  тока  обладают
рассмотренные  выше  волны  ---  $H_{10}$  в прямоугольном волноводе и
$H_{11}$  в  круглом;  эти  волны  несут  и  продольные  токи.  А  вот
симметричные магнитные волны $H_{0q}$ в круглом волноводе возбуждают в
стенках только поперечные (азимутальные) токи.

     Забегая вперёд,   сделаем    здесь    маленькое    отступление.
Предположим,  что  для  какой-то  волноводной  волны с помощью формулы
(9.19) вычислена  плотность  поверхностного  тока.  Забудем  теперь  о
существовании   металлических   стенок  волновода  и  будем  решать  в
неограниченном   однородном   пространстве   неоднородные    уравнения
Максвелла  (2.8),  в  которых  в  качестве  стороннего  тока $\rv j^e$
возьмём найденный  поверхностный  ток  $I$  (неоднородные  уравнения
будут  рассмотрены отдельно).  В результате внутри волновода получится
поле той самой волноводной волны, для которой вычислялся поверхностный
ток, а вне волновода --- нуль.

     Вернёмся к симметричным магнитным  волнам.  Отсутствие  осевого
тока  делает  эти  волны  малочувствительными  к  азимутальным щелям в
стенках волновода (для остальных  волн  азимутальные  щели  перерезают
линии  тока  и  потому вызывают существенное искажение поля).  Поэтому
волна  $H_{01}$  находит   применение   во   вращающихся   сочленениях
радиолокаторов.  Это  же  обстоятельство является причиной уникального
свойства волны $H_{01}$ ---  затухание  этой  волны  (о  котором  речь
пойдет ниже) падает с ростом частоты; поэтому она способна переносить
электромагнитную энергию на большие расстояния с малыми потерями.

     Теория электрических  волн  в  круглом  волноводе  отличается  от
изложенной теории $H$-волн только граничным условием для потенциальной
функции:
     $$\left.\Pi^e\right |_{r=a}=0\,.\eqno(9.20)$$
Собственные значения  краевой  задачи  определяются  теперь   условием
{$J_p(g_{pq} a)=0$}, так что
     $$g_{pq}=\frac{\nu_{pq}}{a},\qquad q=1,\,2,\dots\,,\eqno(9.21)$$
где $\nu_{pq}$ --- корни функции Бесселя:{ $J_p(\nu_{pq})=0$}.  Вообще
говоря,  они отличаются от корней производной этой функции $\mu_{pq}$,
а нам понадобятся в дальнейшем два наименьших корня:  $\nu_{01}\approx 2,405$
и $\nu_{11}\approx 3,832$.

     Cимметричные волны  $E_{0q}$  содержат   три   компоненты   полей
$E_z$,~$E_r$,~$H_\varphi$  и  интересны  тем,  что  в  отличие от всех
других волноводных волн имеют продольную компоненту $E_z$, отличную от
нуля  на оси волновода.  Поэтому волна $E_{01}$ или
её аналог в более  сложных  симметричных  структурах  применяется  в
электронных  приборах,  где надо обеспечить эффективное взаимодействие
электромагнитного поля с  осевым  потоком  электронов.  Силовые  линии
электромагнитного  поля этой волны изображены на рис.~9.5{\it {б}}.  Следующей
(в последовательности критических частот)  за  волной  $E_{01}$  стоит
волна $E_{11}$.

     Интерес к  ней  обусловлен  тем,  что она имеет ту же критическую
частоту,  а потому и ту же фазовую скорость,  что и  волна  $H_{01}$~$
(\nu_{11}\!=\!\mu_{01} $,  тaк  как  $J_1(x)=-J^\prime_0(x))$.  Это
вырождение (совпадение критических частот) волн  $H_{01}$  и  $E_{11}$
оказывается существенным в проблеме передачи волны $H_{01}$ на большие
расстояния, поскольку из-за совпадения фазовых скоростей волн на любых
нерегулярностях  тракта  они  легко  преобразуются  друг  в  друга,  а
затухание волны $E_{11}$ с ростом частоты,  как это будет  показано  в
дальнейшем,  не спадает.  Точно также вырождены все пары волн $E_{1q}$
и~$H_{0q}$; электрическим волнам в круглом волноводе, как и магнитным,
присуще  поляризационное  вырождение,  ибо  оно  обусловлено  не типом
волны, а симметрией вращения.

     Свойства электромагнитных  волн  в  волноводах  становятся  более
наглядными,  если  рассмотреть  любопытную  связь,  существующую между
этими волнами и  плоскими  волнами,  распространяющимися  в  свободном
пространстве. Впервые на эту связь обратил внимание Бриллюэн. Выше уже
говорилось о  возможности  разложить  любую  цилиндрическую  волну  по
плоским  волнам.  Начнём  с  несложных алгебраических преобразований
формул  (9.8)  для  составляющих  поля  основной  волны   $H_{10}$   в
прямоугольном   волноводе.  Используя  формулу  Эйлера  для
тригонометрических функций, перепишем (9.8) в виде
     $$\left.\begin{array}{l}E_y=\displaystyle{\frac{k\mu}{\sqrt{2ab}}\,
         [e^{ik(z\cos\theta+x\sin\theta)}-e^{ik(z\cos\theta-x\sin
         \theta)}]}\,,\\[.5cm]H_x=-\displaystyle{\frac{k}{\sqrt{2ab}}
         \cos\theta\,[e^{ik(z\cos\theta+x\sin\theta)}-e^{-ik(z\cos\theta
         -x\sin\theta)}]}\,,\\[.5cm]H_z=\displaystyle{\frac{k}{\sqrt {2ab}}\sin
         \theta\,[e^{ik(z\cos\theta+x\sin\theta)}+e^{-ik(z\cos\theta-x
         \sin\theta}]}\,,\end{array}\right\}\eqno(9.22)$$
где угол $\theta$ введён с помощью соотношения
     $$\cos\theta=\frac{h_{10}}{k}\,,\eqno(9.23)$$
так что $\sin \theta=g_{10}/k$.

     Первое слагаемое  в  каждой  из  этих  формул  представляет собой
компоненту  поля  плоской  волны,  распространяющейся  в  направлении,
которое  составляет  угол  $\theta$ с осью $z$,  второе --- компоненту
поля такой же волны, волновой вектор которой составляет  с  осью  $z$  угол
$-\;\theta$.   Действительно,   три   первых   слагаемых   удовлетворяют
уравнениям Максвелла,  амплитуды полей  постоянны,  фазовый  множитель
имеет    вид    $ik(z\cos\theta+x\sin\theta)$,   так   что   плоскость
$z\cos\theta+x\sin\theta=const$, в которой, как легко убедиться, лежат
векторы $\bf E$ и $\bf H$, совпадает с поверхностью равных фаз, и волна
является поперечной.  Таким образом,  волна $H_{10}$ состоит  из  двух
плоских   волн   (волн   Бриллюэна),   распространяющихся   под  углом
$\pm\;\theta$ к  оси  $z$.  Интерференция  этих  волн  образует  узловые
плоскости,  совпадающие  с  боковыми  стенками  волновода,  на которых
$E_y=0$.  Каждую из этих волн можно представлять себе как образованную
отражением другой волны в боковой стенке волновода.

     Такое же  преобразование  возможно и для более сложных полей волн
$H_{mq}$ и~$E_{mq}$.  Оно также состоит  в  замене  тригонометрических
функций  суммой двух экспонент;  все компоненты полей представляются в
виде суммы четырёх слагаемых, так что всё поле записывается в виде
суммы четырёх плоских волн. Существенно, что формула (9.23) для угла
$\theta$,  составляемого с осью $z$ нормалями к плоскостям равных  фаз
каждой  из этих волн,  сохраняется.  Она имеет универсальный характер.
Поскольку любая цилиндрическая волна может быть разложена  по  плоским
волнам, то нетрудно показать, что и в волноводах круглого сечения поле
любой волны представимо в виде суперпозиции плоских волн (на этот  раз
---  бесконечного  множества),  и  что нормали к волновым фронтам этих
волн образуют с осью волновода угол, определяемый формулой (9.23).

     В заключение  представляется  целесообразным  ещё раз отметить,
что волна типа $H_{01}$ в круглом волноводе является неустойчивой  ---
даже  при  малой  эллиптичности поперечного сечения она превращается в
волну $E_{11}$,  обладающую той же критической частотой,  но  большими
потерями. Для обеспечения устойчивости волны $H_{01}$ круглый волновод
делают,  например,  из  изолированных  колец   ---   это   не   мешает
распространению  $H_{01}$-волны,  а  волна  $E_{11}$ в таком волноводе
существовать не может  (электрические  силовые  линии  волны  $H_{01}$
представляют   собой  окружности,  а  продольные  электрические  токи,
характерные для волны $E_{11}$,  в волноводе с изолированными кольцами
существовать  не  могут).  В  целом  проблема  устойчивости типа волны
(колебания) в электродинамических структурах часто представляет  собой
достаточно сложную техническую проблему.

%\end{document}

\newpage
\oddsidemargin=-0.4mm \evensidemargin=-0.4mm
\topmargin=-0.4mm
\headsep=7mm \textheight=231.875mm \textwidth=160mm
\mathsurround=2.5pt \unitlength=1mm
%\begin{document}
%\input{macr.tex}
\thispagestyle{empty}
%\addtocounter{page}{100}

\begin{center}\subsubsection*{10. Волноводы с двухсвязной формой
         поперечного сечения}\end{center}\vspace*{0.5cm}

\markboth{Глава 3. Волноводы}{10. Волноводы с двухсвязной формой
         сечения}

\begin{center}\begin{minipage}[c]{0.75\textwidth}
\footnotesize{\parindent=0.5cm
         Критические длины  волн в волноводах с односвязным поперечным
         сечением.  $TEM$-волна в волноводах с многосвязным  сечением  и
         её   основные  свойства.  Высшие  типы  волн  в  плоском  и
         коаксиальном   волноводах.   Аналогия   поля   $TEM$-волны    с
         электростатическим      полем.     Волновое     сопротивление
         коаксиального волновода. Телеграфные уравнения.
}\end{minipage}\end{center}\vspace*{0.5cm}

     В предыдущем разделе было показано,  что в волноводах  простейшей
формы (прямоугольном и круглом) критическая длина основной волны имеет
тот же порядок,  что и наибольший поперечный  размер  волновода.  Так,
значение  $\lambda_{10}$  для волны $H_{10}$ в прямоугольном волноводе
вдвое превышает  размер  широкой  стенки  волновода  $a$,  а  значение
$\lambda_{11}$  для  волны  $H_{11}$  в  круглом волноводе в 1,71 раза
превышает диаметр волновода.  Того же порядка и ширина диапазона  длин
волн, в котором соблюдается режим одноволновости. Однако выбирая форму
поперечного сечения  надлежащим  образом,  можно  добиться  того,  что
критическая   длина  основной  волны  волновода  будет  во  много  раз
превосходить его поперечные габариты.  Проще всего это представить  на
примере   прямоугольного   волновода  с  очень  малой  узкой  стенкой.
Деформируя контур поперечного сечения по спирали или по ломаной  линии
таким образом,  чтобы он вписался в прямоугольник,  наибольшая сторона
которого  много  меньше  широкой  стенки  исходного   сечения,   можно
убедиться, что критическая длина волны останется приблизительно равной
длине  спирали  или  ломаной  и,  следовательно,   будет   существенно
превосходить наибольший поперечный размер свёрнутого волновода.

     Теория волноводов  со  сложной  односвязной  формой   поперечного
сечения  сводится  к  решению тех же двумерных граничных задач (8.5) и
(8.6) и не содержит принципиально новых физических результатов. Всегда
имеется критическая частота, ниже которой распространяющихся волн нет,
и во всех волнах имеется продольная компонента электромагнитного поля.
Существует, однако, широкий класс волноводов с двухсвязными и, вообще,
многосвязными сечениями,  в которых  может  распространяться  волна  с
качественно  отличными  свойствами.  В  поле  этой  волны  отсутствует
продольная компонента,  а её критическая длина равна  бесконечности,
то  есть  распространение  волны  имеет  место при сколь угодно низкой
частоте.  Простейшими примерами таких структур являются: {\it  плоский
волновод},    представляющий    собой    пространство    между   двумя
неограниченными  параллельными  проводящими  плоскостями  и  ---   что
особенно важно --- часто используемый для теоретического моделирования
сложных процессов,  {\it коаксиальный волновод  (кабель)},  образуемый
пространством    между   двумя   круглыми   соосными   цилиндрическими
металлическими поверхностями,  и {\it двухпроводная линия},  состоящая
из двух параллельных круглых цилиндрических проводников.

     Для перечисленных   выше   типов   волноводов   реализуется    именно та
-- дополнительная -- возможность удовлетворения граничному условию
(8.1) на стенках волновода,  о которой уже упоминалось.  Она
состоит в том, что для электрических волн касательные составляющие
электрического поля на стенках волновода равны нулю не только при
условии (8.3\mbox{\textit б}),  но  и  при менее строгом условии
     $$\left.\frac{\partial\Pi^e}{\partial s}\right|_C=0\,,\eqno(8.4\mbox{\textit a})$$
то есть $\left .\Pi^e \right |_C=const$, если одновременно
выполняется и условие (8.4\mbox{\textit б}):
$h^2=k^2\varepsilon\mu$. Но при последнем  условии волновое
уравнение  (7.14)  переходит в двумерное уравнение Лапласа,
которое,  как известно,  при постоянном значении  функции  на
границе односвязной  области имеет  только тривиальное решение
$\Pi^e=const$. Однако в случае многосвязных областей функция,
будучи  постоянной  на каждой граничной поверхности,  может иметь
на них разные значения, и у уравнения  Лапласа  появляются
решения,  которые  и  реализуются, в частности, в перечисленных
волноводах.

     Итак, в  волноводах,  поперечное  сечение  которых   состоит   из
нескольких    замкнутых   контуров,   возможна   волна,   определяемая
потенциальной  функцией  $\Pi^e$,  удовлетворяющей  следующей  краевой
задаче:
     $$\Delta\Pi^e=0\,,\qquad\left.\Pi^e\right|_{C_j}=B_j\,,\eqno(10.1)$$
где $j$ --- номер контура,  $B_j$ --- постоянные, причём хотя бы две
из них разные. Поскольку данная краевая задача соответствует волновому
уравнению  (8.9)  для  $\Pi^e$  с  собственным  значением $g=0$,  то в
цилиндрической волне (7.1)
     $$h=K=k\sqrt{\varepsilon\mu}\,.\eqno(10.2)$$

     Структура поля этой волны значительно проще,  чем у рассмотренных
ранее  волноводных волн.  Прежде всего эта волна поперечная:  согласно
(7.12),  (7.13) в ней отсутствуют продольные компоненты поля  $H_z$  и
$E_z$,  а, как следует из формул (7.16), поперечные компоненты связаны
между собой теми же соотношениями, что и в плоской волне:
     $$E_x=\sqrt{\frac{\mu}{\varepsilon}} H_y\,,\qquad E_y=-\sqrt{\frac{\mu}
         {\varepsilon}} H_x\,.\eqno(10.3)$$

\vspace*{-0.6cm}
\begin{picture}(160,50)
\put(0,45){\special{em:graph fig10-1a.bmp}}
\put(55,45){\special{em:graph fig10-1b.bmp}}
\put(110,45){\special{em:graph fig10-1c.bmp}}
\end{picture}
\begin{center}\begin{minipage}[c]{0.9\textwidth}
\footnotesize{\begin{center}
Рис.~10.1.~Волноводы с двухсвязной формой поперечного сечения.
\end{center}}\end{minipage}\end{center}

     Фазовая скорость  волны  $v_{\mbox{\footnotesize\textit ф}}
=c/\sqrt{\varepsilon\mu}$  такая же, как и у плоской  волны;  и
так  же,  как  плоская  волна,  она  может распространяться  при
любых  самых  низких  частотах  и  не  обладает дисперсией (кроме
обусловленной зависимостью $\varepsilon$ и $\mu$  от $\omega$).
Однако в отличие от плоской волны компоненты (10.3), вообще
говоря,  в поперечном сечении не постоянны,  то есть зависят от
$x$  и $y$.  Имея  в  виду  поперечность  волны,  её называют
$TEM$-волной. Другие используемые в литературе термины --- {\it
главная  волна}  или {\it кабельная волна}.

     Рассмотрим структуру $TEM$-волны и сравним её со структурой волн
электрического и магнитного  типов  в  конкретных  перечисленных  выше
волноводах.  Начнём  с  плоского  волновода (рис.~10.1\textit{а },  расстояние
между   параллельными   проводящими   плоскостями   $a$)  и    будем
рассматривать  поля,  все  компоненты  которых  зависят  лишь от одной
поперечной координаты $x$ и не зависят  от  $y$  (структура  однородна
вдоль   оси   $y$).   В   таком   случае   поля  цилиндрической  волны
представляются в виде
     $$\rv E(x,y,z)= \rv E(x)e^{ihz}, \qquad \rv H(x,y,z)=\rv H(x)e^
         {ihz}\,.\eqno(10.4)$$

     Краевая задача    для    потенциальной    функции     $TEM$-волны
     $$\frac{d^2\Pi^e}{dx^2}=0\,,\quad \left.\Pi^e\right|_{x=
         0}=B_1,\quad\left.\Pi^e\right|_{x=a}=B_2\eqno(10.5)$$
имеет решение: $\Pi^e=\displaystyle{\frac {B_2-B_1}a x}+B_1$. Нормируя
его  условием  $\int\limits_0^a  (\nabla  \Pi^e)^2\,dx=1,$  найдем  по
формулам (7.16) отличные от нуля компоненты поля:
     $$E_x=i\frac{K}{\sqrt{a}} e^{iKz},\qquad H_y=\sqrt{\frac
         {\varepsilon}{\mu}}E_x\,,\eqno(10.6)$$
которые с   точностью   до   нормировочного   множителя   совпадают  с
компонентами поля однородной плоской волны (4.15),  распространяющейся
вдоль оси $z$. Таким образом, $TEM$-волна плоского волновода совпадает
с  плоской  волной  в  неограниченном  пространстве.  Физически   этот
результата  очевиден:  помещение неограниченных проводящих плоскостей,
нормальных к вектору электрического поля плоской  волны,  не  искажает
её структуры.  Отметим,  что поверхностный ток в идеально проводящих
ограничивающих   волновод   плоскостях   течёт   вдоль   направления
распространения  волны,  то  есть  в  данном случае вдоль оси $z$ (это
следует из формулы (9.19)).  Поскольку  и  на  верхней,  и  на  нижней
пластине нормаль направлена в металл,  то токи на них в каждом сечении
текут в противоположные стороны.  Средний поток мощности вдоль оси $z$
на единицу длины по $y$ составляет
     $$\overline{\Sigma}=\frac{ck^2\varepsilon\sqrt{\varepsilon\mu}}
         {8\pi a}\,.\eqno(10.7)$$

     В плоском волноводе  могут  распространяться  и  волны,  присущие
волноводам  с  односвязным поперечным сечением.  Краевые задачи для их
потенциальных  функций  $\Pi^e$  (электрические   волны)   и   $\Pi^m$
(магнитные волны) имеют вид:
    $$\begin{centering}{\textit {Электрические волны} }\end{centering}$$
    $$\left. \begin{array}{l}
         \displaystyle{\frac
         {d^2\Pi^e}{dx^2}+g^2\Pi^e=0}\,,\\[0.4cm]
         \left.\Pi^e\right |_{x=0}=\left.\Pi^e\right|_{x=a}=0\,,
       \end{array} \right\}\eqno(10.8\mbox{\textit{а}})$$
   $$\begin{centering}{\textit {Магнитные волны} }\end{centering}$$

    $$\left. \begin{array}{l}
         \displaystyle{\frac{d^2\Pi^m}{dx^2}+g^2\Pi^m=0}\,,\\[0.4cm]
        \displaystyle{\left.\frac{d\Pi^m}{dx}\right|_{x=0}=
        \left.\frac{d\Pi^m}{dx}\right|_{x=a}=0}\,,
        \end{array} \right\}\eqno(10.8\mbox{\textit{б}})$$

Задачи (10.8)  имеют одинаковый  для  обоих  типов  волн  набор
собственных значений $g_p$,
     $$g_p=\frac{\pi p}{a}\,,\qquad p=1,2,\dots\,,\eqno(10.9)$$
и  соответствующих продольных волновых чисел
     $$h_p=\sqrt{k^2\varepsilon\mu-(\pi p/a)^2}\,.\eqno(10.10)$$

     Потенциальные функции и компоненты полей у этих волн различаются:
      $$\begin{centering}{\textit {Электрические волны} }\end{centering}$$
      $$\displaystyle{\Pi^e_p=\frac{\sqrt{2a}}{\pi  p}\,\sin{\frac{\pi p}{a}x}}\,,
           \eqno(10.11)$$
     $$\left.\begin{array}{l}\displaystyle{H_y=i\sqrt{\frac 2 a}\,k
         \varepsilon \, \cos{\frac{\pi p}{a}x}\; e^{ih_p z}}\,,\\[0.4cm]
         \displaystyle{E_x=i\sqrt {\frac  2 a}\,h_p\,\cos{\frac{\pi p}{a}x}\;e^{ih_p   z}}\,,
         \\[0.4cm]\displaystyle{E_z=\sqrt {\frac 2 a}\,\frac{\pi  p}a\,\sin{\frac{\pi  p}{a}x}
         \; e^{ih_p z}}\,,\end{array}\right\}\eqno(10.12)$$

     $$\begin{centering}{\textit {Магнитные волны}}\end{centering}$$
     $$\displaystyle{\Pi_p^m=\frac{\sqrt{2a}}{\pi p}\,\cos{\frac{\pi p}{a}x}}\,,
           \eqno(10.13)$$
     $$\left.\begin{array}{l}\displaystyle{E_y=i\sqrt{\frac 2 a}\,k\mu\,
          \sin{\frac{\pi p}{a}x}\; e^{ih_p z}}\,,\\[0.4cm]
          \displaystyle{H_x=i\sqrt{\frac 2  a}\,h_p\,\sin{\frac{\pi  p}{a}}
          \; e^{ih_p z}}\,,\\[.4cm]
          \displaystyle{H_z=\sqrt{\frac 2 a}\,\frac{\pi p}a\,\cos{\frac {\pi p}{a}x}
          \; e^{ih_p z}}\,.\end{array}\right\}\eqno(10.14)$$
\vspace{0.2cm}

Как и потенциальная функция $TEM$-волны, потенциальные функции (10.11)
и (10.13) нормированы не условием (8.22),  где интегрирование производится
по всему поперечному сечению волновода,  а условием
 $\int\limits_0^a  (\nabla\Pi)^2\,dx=1$, то  есть  интегрирование  производится
на  единицу  длины  по $y$.  С точностью до  нормирующего  множителя  поля
магнитных  волн  плоского волновода совпадают с волнами $H_{p0}$
прямоугольного волновода.

     Краевая задача  для  $TEM$-волны коаксиального  волновода,  изображённого
на рис.~10.1{\textit{б}},  включает в себя: скалярное  уравнение  Гельмгольца
(7.14),  которое  в  полярных  координатах  для  потенциальной функции
$\Pi^e(r,\varphi)$ записывается в виде
     $$\frac{\partial^2\Pi^e(r,\varphi)}{\partial r^2}+\frac{1}{r}\frac{
         \partial\Pi^e(r,\varphi)}{\partial r}+\frac{1}{r^2}\frac{
         \partial^2\Pi^e(r,\varphi)}{\partial\varphi^2}+(K^2-h^2)
         \Pi^e(r,\varphi)=0\,,\eqno(10.15) $$
дополнительное соотношение (10.2), которое сводит уравнение (10.15)
к уравнению Лапласа, и граничные условия
     $$\left.\Pi^e\right|_{r=a}=B_1,\qquad \left.\Pi^e\right|_
         {r=b}=B_2\,.\eqno(10.16)  $$
Решая задачу   методом   разделения   переменных,   то   есть  полагая
$\Pi^e(r,\varphi)=R(r)\Phi(\varphi)$,  убеждаемся,  что  при   условии
(10.2) она имеет только азимутально симметричное решение
     $$\Pi^e=(B_1-B_2)\ln{\frac r a}+B_2\,.
         \eqno(10.17) $$
Нормируя его  согласно  формуле  (8.22),  найдём  отличные  от  нуля
поперечные компоненты поля:
     $$E_r=\frac{iK}{\sqrt{2\pi\ln{b/a}}}\frac { e^{iKz}}r\,,\qquad
         H_{\varphi}=\sqrt{\frac{\varepsilon}{\mu}} E_r.\eqno(10.18)$$
Полный ток на обоих цилиндрах имеет только продольную составляющую,  в
каждом сечении $z= const$  внешнего и внутреннего проводника направлен
в противоположные стороны и с точностью до знака равен
     $$J_z=\frac{ca H_\varphi(a)}{2}=\frac{i\omega
          \varepsilon}{\sqrt{8\pi\ln{b/a}}}e^{iKz}\,.\eqno(10.19)$$

     Средний поток мощности  вдоль  коаксиального  волновода  для
$TEM$-волны составляет
     $$\overline{\Sigma}=\frac{ck^2\varepsilon\sqrt{\varepsilon\mu}}
         {8\pi}\,,\eqno(10.20)$$
что при учёте условия (10.2) совпадает  с  потоком  мощности  (8.26)
электрических волн в односвязном волноводе.

     В коаксиальной   линии   могут   распространяться   и   обычные
волноводные волны с отличными от нуля значениями  критических  частот.
Краевые  задачи  для  потенциальных  функций электрических и магнитных
волн включают в себя одинаковое волновое уравнения  вида  (10.15),  но,
как и всегда, различаются граничными условиями;  для магнитных
волн эти условия имеют вид
     $$\left.\frac{\partial\Pi^m(r,\varphi)}{\partial r}\right|_{r=b}=
         \left.\frac{\partial\Pi^m(r,\varphi)}{\partial r}\right
         |_{r=a}=0\,,\eqno(10.21) $$
для электрических ---
     $$\Pi^e(b,\varphi)=\Pi^e(a,\varphi)=0\,.\eqno(10.22)$$

     При решении уравнения (10.15)  методом  разделения  переменных  в
полярных  координатах  (формулы  (7.24)--(7.30))  в  качестве  функции
$R(r)$ необходимо взять  два  линейно  независимых  решения  уравнения
Бесселя   (7.30),  чтобы  иметь  возможность  удовлетворить  граничным
условиям (10.21) или (10.22) и при $r=b$,  и при $r=a$. Помимо функции
Бесселя  в качестве второго решения можно взять функцию Неймана;  хотя
она имеет особенность в нуле,  но в данном случае это не  существенно,
поскольку точка $r=0$ не входит в область,  где ищется решение.  Итак,
функция $R(r)$ ищется в виде
     $$R(r)=AJ_p(gr)+BN_p(gr)\,,\eqno(10.23) $$
где $p$ --- число вариаций поля по азимуту  $\varphi$,  а
$g=\sqrt{K^2- h^2}$ --- поперечное волновое число.

     В случае  магнитных  волн  граничные  условия  (10.21) приводят к
системе двух уравнений относительно коэффициентов $A$ и $B$:
     $$ \left.\begin{array}{l} AJ'_p(gb)+BN'_p(gb)=0\,,\\[.3cm]AJ'_p
         (ga)+BN'_p(ga)=0\,.\end{array}\right\}\eqno(10.24)$$
Условие существования  у этой однородной системы нетривиального решения,  то
есть равенство  нулю  определителя  соответствующей  матрицы,  является
характеристическим уравнением   для   поперечного   волнового   числа   $g$:
     $$J'_p(ga)N'_p(gb)-J'_p(gb)N'_p(ga)=0\,.
         \eqno(10.25)$$
Характеристическое   уравнение   для   $E$-волн  выводится аналогично;
в тех же обозначениях имеем
     $$J_p(ga)N_p(gb)-J_p(gb)N_p(ga)=0\,.
         \eqno(10.26) $$
При $a= 0$  уравнения (10.25) и (10.26) переходят
соответственно в уравнения $J'_p(gb)=0$ и~$J_p(gb)=0$,  определяющие
собственные значения для магнитных  и  электрических  волн  в  круглом
волноводе.

      Оба характеристических  уравнения  (10.25)   и   (10.26)   имеют
бесконечный   набор  различающихся  между  собой  решений  $g_{pq}$,
которые  представляют  собой  собственные   значения   соответствующей
краевой  задачи.  Их точные значения $g_{pq}$,  являющиеся функциями
параметра $a/b$,  могут быть найдены  только  численными  методами,
однако нетрудно получить их приближенные значения для больших значений
индекса $q$.  Воспользовавшись асимптотическими разложениями  функций
$J_p(x)$ и $N_p(x)$ при больших значениях аргумента
     $$J_p(x)=\sqrt{\frac 2{\pi x}}\Bigl[\cos{\Bigl(x-\frac{p\pi} 2-
         \frac \pi 4\Bigr)}+O\Bigl(\frac 1 x\Bigr)\Bigr]\,,
         \eqno(10.27)$$
     $$N_p(x)=\sqrt{\frac 2{\pi x}}\Bigl[\sin{\Bigl(x-\frac{p\pi} 2-
         \frac \pi 4\Bigr)}+O\Bigl(\frac 1 x\Bigr)\Bigr]\,,
         \eqno(10.28)$$
после несложных преобразований тригонометрических  выражений  получаем
для электрических волн
     $$g_{pq}=\frac {\pi q}{b-a}\qquad(q\gg p)\,,\eqno(10.29)$$
для магнитных волн
     $$g_{pq}=\frac {\pi (q+1/2)}{b-a}\qquad(q\gg p)\,.
         \eqno(10.30)$$
Таким образом, последовательные дальние собственные значения для обоих
типов волн расположены друг от друга на  одинаковом  расстоянии  и  не
зависят  от индекса $p$,  определяющего число вариаций поля по азимуту
$\varphi$.  Спектр  собственных  значений  и,  следовательно,   спектр
критических частот коаксиала оказывается гуще соответствующего спектра
для  круглого  волновода  того  же  радиуса,  что  и  внешний   радиус
коаксиала.

     Из всех  собственных  значений  для  обоих типов волн минимальным
является $g_{11}$ для магнитной волны.  При $a\approx b$ его  нетрудно
вычислить  путём  разложения  всех  бесселевых  функций,  входящих в
уравнение  (10.25),  в  ряд  Тейлора  вокруг  точки  $r_0=(a+b)/2$.  В
результате получаем
     $$g_{11}\approx\frac   2{a+b}\,,\eqno(10.31)$$
так что  на  средней  окружности поперечного сечения пустого коаксиала
укладывается  самая  длинная   критическая   волна.   Для   коаксиала,
заполненного веществом с проницаемостями $\varepsilon,\;\mu$,  верхняя
граница  полосы   одноволновости   $TEM$-волны   определяется   условием
обращения   в   нуль   продольного   волнового  числа  $h_{11}=\sqrt{k^2
\varepsilon\mu-g^2_{11}}$.

     Собственные функции  краевых  задач   $\Pi^e_{pq}(r,\varphi)$   и
$\Pi^m_{pq}(r,\varphi)$    для   коаксиального   волновода   находятся
следующим образом.  С помощью любого из уравнений системы  (10.24)  по
известным собственным значениям $g_{pq}$ один из коэффициентов $A$ или
$B$ выражается через другой и подставляется в (10.23).  В  результате,
например, для магнитных волн получаем:
     $$ \Pi^m_{pq}=C\bigl [ N'_p(g_{pq}b)J_p(g_{pq} r ) -
        J'_p(g_{pq}b)N_p(g_{pq} r)
         \bigr]\cos p(\varphi-\varphi_0)\,.\eqno(10.32)$$
Произвольный амплитудный коэффициент $C$ удобно определить  с  помощью
нормировочного   условия  (8.22).  Соответствующее  выражение  слишком
громоздко и здесь не  приводится.  Произвольный  параметр  $\varphi_0$
обусловлен  поляризационным  вырождением,  присущим  всем  азимутально
симметричным структурам.  Собственная функция для  электрических  волн
$\Pi^e_{pq}$  отличается от (10.32) заменой всех бесселевых функций со
штрихом,   обозначающим    дифференцирование    по    аргументу,    на
соответствующую функцию без штриха.

     Поперечные компоненты электромагнитного поля волноводных  волн  в
коаксиале находятся с помощью формул (7.17). Отметим, что совокупность
поперечных компонент поля электрических и магнитных волн  в  коаксиале
образуют  полную систему ортогональных функций,  по которым могут быть
разложены  поперечные  компоненты  произвольного  решения   однородных
уравнений  Максвелла,  но  в  эту совокупность обязательно должно быть
включено поле $TEM$-волны.

     В завершение   этого   раздела   рассмотрим  несколько  подробнее
свойства  $TEM$-волн.  Как  видно   из   формул   (7.16)   и   (10.1),
электрическое  поле $TEM$-волны в поперечном сечении $z=0$ совпадает с
{\it электростатическим полем},  которое возникает между проводниками,
заряженными    до   потенциала   $\Phi_j=-iKB_j$.   Электростатический
потенциал  этого  поля  может  быть  найден  для  очень  сложных  форм
поперечного  сечения  методами  двумерной теории потенциала,  в первую
очередь методом конформного отображения.

     Для $TEM$-волн  можно  естественным  путем  определить  обычные в
теории низкочастотных цепей понятия напряжения,  тока и сопротивления.
Под  {\it  напряжением}  понимается  интеграл  от электрического поля,
взятый между проводами вдоль любой линии,  расположенной  в  плоскости
сечения;  для  линии  передачи  из  двух  отдельных  проводников  (см.
рис.~10.1\textit{в}) напряжение
     $$V=-iK(B_2-B_1).\eqno(10.33) $$
Ток по проводам течёт только в продольном направлении,  и  под  {\it
полным  током}  понимается  интеграл от плотности поверхностного тока,
взятый вдоль контура сечения  любого  проводника.  Магнитное  поле  на
поверхности   идеального   проводника   имеет   только  тангенциальную
составляющую.  Проекция этой  составляющей  на  плоскость  поперечного
сечения согласно формулам (7.17) определяется на контуре как
     $$\left.H_s=ik\varepsilon\frac{\partial\Pi^e}{\partial n}
         \right|_C\,,\eqno(10.34) $$
а полный ток, текущий по проводнику, есть
     $$J_z=\frac{ik\varepsilon c}{4\pi} \oint\limits_C\frac
         {\partial\Pi^e}{\partial n}\,ds\,.\eqno(10.35)$$

     {\it Волновым   сопротивлением   линии} называется   отношение
напряжения  между проводами к току в каждом из них:
     $$Z=\frac{4\pi}c\sqrt{\frac\mu\varepsilon} \,
         (B_1-B_2)\Biggm/\oint\limits_C\frac{\partial\Pi^e}
         {\partial  n}\,ds\,.\eqno(10.36)  $$
Эта величина  является  существенной  характеристикой линии передачи и
важна при согласовании линий или линии с оконечным  устройством.  Если
из     правой     части    формулы    (10.36)    выделить    множитель
$\sqrt{\varepsilon\mu}/c $,  то  оставшееся  выражение  имеет  простое
электростатическое  истолкование:  оно  равно  $1/C$,  где $C$ --- так
называемая погонная ёмкость линии, то есть отношение ёмкости между
двумя проводами очень длинной линии к её длине. Действительно, заряд
на каждом проводнике равен нормальной компоненте электрического  поля,
делённой на $4\pi$,  так что знаменатель в (10.36) только множителем
$ik$   отличается   от   полного   заряда   на   единицу    длины    в
электростатической задаче.  Числитель этим же множителем отличается от
разности потенциалов, откуда и следует, что
     $$Z=\frac{\sqrt{\varepsilon \mu} }{cC}\, .\eqno (10.37)$$

     Формула (10.37) принадлежит к тем нескольким  формулам  в  теории
электродинамики СВЧ, для которых используемая в книге гауссова система
единиц неудобна.  Единица  сопротивления  в  этой  системе  в  $9\cdot
10^{11}$   раз   больше   Ома,  погонная  ёмкость  ---  безразмерна.
Подставляя численное значение $c=3\cdot 10^{8}$~м/сек,  получим  для
волнового сопротивления линии в единицах СИ:
     $$Z=30\frac{\sqrt{\varepsilon\mu}}C\quad\hbox{[Ом]}\,.
         \eqno(10.38)$$
Для коаксиальной линии, как видно из формулы (10.17),
     $$Z=60\sqrt{\frac \mu \varepsilon}\,\ln{\frac b a}\quad\hbox
         {[Ом]}\,.\eqno(10.39)$$

     Скажем ещё  несколько  слов  о  так называемых {\it телеграфных
уравнениях},  которые нашли широкое применение в теории кабельных  волн
намного  раньше,  чем  была  создана  теория  волноводов.  Сейчас  эти
уравнения  уже  в  значительной  степени  утратили   свое   прикладное
значение, но всё-таки заслуживают упоминания.

     Электромагнитное поле в телеграфных уранениях  не  фигурирует,  а
вместо  него для каждого поперечного сечения $z$ данной линии передачи
и каждого момента времени $t$ вводятся  две  величины  ---  напряжение
$V=V(z,t)$ и ток $J=J(z,t)$.  Каждый однородный отрезок линии передачи
характеризуется четырьмя параметрами:  $R$ --- погонное  сопротивление
линии,   $L$  ---  погонная  индуктивность  линии,  $C$  ---  погонная
ёмкость линии, $G$ --- погонная утечка.

     Напряжение и  ток  связаны  между  собой  уравнениями  в  частных
производных   первого  порядка:
     $$-\frac{\partial V}{\partial z}=RJ+L\frac{\partial J}{
         \partial t},\quad-\frac{\partial J}{\partial z}=GV+C\frac
         {\partial V}{\partial t}.\eqno(10.40)$$

   При применении телеграфных  уравнений к реальным линиям нужно иметь
в виду, что, например,  погонное сопротивление  линии  $R$  вследствие
скин-эффекта зависит от частоты и поэтому его нельзя рассматривать как
константу для  волн  произвольной  формы.  В  этом  случае  необходимо
ограничиться   рассмотрением  монохроматических  волн,  то  есть  волн
определенной частоты. Для таких колебаний, как и для монохроматических
электромагнитных полей,  удобно использовать комплексные обозначения и
брать временной  множитель  $e^{-i\omega  t}$.  Обозначая  комплексные
амплитуды  напряжения  и  тока  теми же буквами,  из уравнений (10.40)
получим телеграфные уравнения в комплексной форме:
     $$-\frac{dV}{dz}=(R-i\omega L)J\,,\quad-\frac{dJ}{dz}=(G-i\omega C)V\,.
         \eqno(10.41)$$

     Исключая из этих уравнений ток $J$, переходим к уравнению второго
порядка:
     $$\frac{d^2V}{dz^2}+h^2V=0\,,\eqno(10.42)$$
где
     $$h^2=(\omega L+iR)(\omega C+iG)\,.\eqno(10.43)$$
Обозначая  через
     $$h=\sqrt{(\omega L+iR)(\omega C+iG)}\eqno(10.44)$$
корень, удовлетворяющий  условию  $\im  h>0$,  получим  общее  решение
уравнения (10.40) в виде
     $$V=Ae^{ihz}+Be^{-ihz}\eqno(10.45)$$
и, подставляя это выражение в (10.41), найдём ток
     $$J=\frac 1 Z (Ae^{ihz}-Be^{-ihz})\,,\eqno(10.46)$$
где
     $$Z=\sqrt{\frac{\omega L+iR}{\omega C+iG}}\,.\eqno(10.47)$$
Иногда вводится так называемый  погонный  внутренний  импеданс  линии,
определяемый как коэффициент пропорциональности между компонентой поля
$E_z$ на поверхности и током; он равен $(1-i)/2\pi a\sigma\delta$, где
$a$ -- радиус проводника, $\delta$ -- толщина скин-слоя. Величина $h$,
определяемая формулой (10.44), называется комплексным волновым числом.

     Итак, общее решение телеграфных уравнений для однородного отрезка
линии передачи представляет собой  сумму  двух  волн  с  произвольными
комплексными амплитудами: одна волна бежит в положительном направлении
(слагаемое,  пропорциональное  $e^{ihz}$),  другая   в   отрицательном
направлении (слагаемое, пропорциональное $e^{-ihz}$).

     Можно выделить   два    круга    вопросов,    для которых  с помощью
телеграфных уравнений проще получить результаты,  чем
электродинамическими методами.  Во-первых, в
этих  уравнениях  легче  учесть  конечную   проводимость   металла   и
проводимость   среды,  заполняющей  пространство  между  проводниками.
Во-вторых,  телеграфные уравнения позволяют  приближённо  определить
отражение кабельной волны от нерегулярных участков линии, включённых
неоднородностей, и решить ряд других  важных  задач.  Необходимо  также
иметь   в   виду,  что  при  высоких  частотах  телеграфные  уравнения
неприменимы --- длина волны всегда должна быть много больше расстояния
между проводниками линии.

%\end{document}

\newpage
\oddsidemargin=-0.4mm \evensidemargin=-0.4mm
\topmargin=-0.4mm
\headsep=7mm \textheight=231.875mm \textwidth=160mm
\mathsurround=2,5pt \unitlength=1mm
%\begin{document}
%\input{macr.tex}
\thispagestyle{empty}
%\addtocounter{page}{110}

\begin{center}\subsubsection*{11. Потери в волноводах}\end{center}
     \vspace*{0.5cm}

\markboth{Глава 3. Волноводы}{11. Потери в волноводах}

\begin{center}\begin{minipage}[c]{0.75\textwidth}
\footnotesize{\parindent=0.5cm
         Затухание волноводных   волн   из-за   потерь   в    веществе
         заполнения.   Учёт  конечной  проводимости  стенок  методом
         теории возмущений.  Вычисление  затухания  волноводных  волн.
         Зависимость   затухания  от  частоты.  Коэффициент  затухания
         основных   волн   прямоугольного,   круглого   волноводов   и
         коаксиального кабеля. Поправка к фазовой скорости волны.
}\end{minipage}\end{center}\vspace*{0.5cm}

     До сих   пор   изучение   волн   в   волноводах   проводилось   в
предположении,  что  их  стенки  идеально  проводящие,  а   внутреннее
пространство  если  и  заполнено  веществом (например,  чтобы получить
возможность передавать более  длинные  волны  или  замедлить  их),  то
таким,  в  котором  не происходит поглощения энергии электромагнитного
поля.  Преимущественно  рассматривались  пустые  волноводы.  При  этих
условиях  в  зависимости  от  частоты  либо  продольное волновое число
данной волны $h_n$ действительная величина,  волна распространяется по
волноводу с постоянной амплитудой и в любом поперечном сечении средний
по времени поток энергии отличен  от  нуля  и  постоянен,  либо  $h_n$
мнимое число, волна экспоненциально затухает по мере удаления от места
возбуждения и в среднем в любом сечении поток энергии отсутствует.  Но
в обоих случаях диссипации энергии поля не происходит.

     Потери энергии могут быть обусловлены поглощением как в веществе,
заполняющем  волновод,  так  и  в  стенках  волновода  из-за конечного
значения их  проводимости.  Любое  поглощение  приводит  к  тому,  что
продольное волновое число становится комплексным:
     $$h_n=h_n'+ih_n''\,.\eqno(11.1)$$
В результате  пропадает  резкое  деление  волн на распространяющиеся и
затухающие: распространяющаяся волна испытывает затухание амплитуды, а
затухающим  волнам  становится  возможным приписать фазовую скорость, и
они уже переносят активную мощность,  которая и поглощается. Поскольку
амплитуда  поля в волне убывает $\sim e^{-h_n''z}$,  то за коэффициент
затухания волноводной  волны  естественно  принять  величину  $h_n''$.
Расчёт  потерь в волноводе сводится к вычислению этого коэффициента,
имеющего размерность обратной длины. Практический интерес представляет
случай,  когда  потери  невелики.  Это  позволяет  искать  коэффициент
затухания для каждого вида потерь независимо и  использовать  при  его
вычислении  разложение по соответствующему малому параметру,  принимая
за исходное приближение решение при отсутствии потерь.

     Начнём с  учёта потерь в заполняющем волновод веществе,  считая
стенки волновода идеально проводящими.  В этом  случае  поглощение  не
отменяет деления всей совокупности волноводных волн на электрические и
магнитные,  а  продольное  волновое  число  остаётся   связанным   с
собственными значениями $g_n$ соответствующей краевой задачи привычным
соотношением
     $$h_n=\sqrt{k^2\varepsilon\mu-g_n^2}\eqno(11.2)$$
и является комплексной величиной из-за наличия у $\varepsilon$ и $\mu$
мнимых  частей,  через  которые  и  выражаются согласно формуле (2.32)
потери  энергии  поля  в  веществе.  Для  упрощения  записи  будем   в
дальнейшем  считать  $\mu$  величиной  действительной и полагать,  что
только $\varepsilon$ имеет заметную мнимую часть $\varepsilon''$ --- в
данном  случае  формула  для  коэффициента  затухания  непосредственно
следует из (11.2). Действительно, введём обозначение
     $$ h_{0n}^2=k^2\varepsilon'\mu-g_n^2\eqno(11.3) $$
и запишем $h_n$ в виде
     $$h_n=\sqrt{h_{0n}^2+i\varepsilon''\mu k^2}\,;\eqno(11.4)$$
тогда с учётом малости $\varepsilon''$
     $$h_n=h_{0n}+i\varepsilon''\mu k^2/2h_{0n}\,.\eqno(11.5)$$
Если $h_{0n}^2>0$,  то поправочный член в этой формуле чисто мнимый  и
именно  он  и  есть  коэффициент  затухания  амплитуды волны.  Если же
$h_{0n}=i|h_{0n}|$,  то  поправочный  член  представляет  собой  малую
вещественную часть продольного волнового числа,  обусловливает перенос
энергии вдоль волновода в среднем по  времени  и  определяет  конечную
фазовую скорость.

     Обратим внимание,   что  формула  (11.5)
применима только при $\varepsilon''\mu k^2\ll h_{0n}^2$.  Вблизи
критической частоты $k_n$,  определяемой теперь условием
$h_{0n} = 0$, формула (11.5) приводит к бесконечно большому затуханию,
в то время как формула~(11.4)  применима  при  любых  частотах  и  при
$h_{0n}=0$ дает простой результат:  $h_n=k_n\sqrt{i\mu\varepsilon''}.$
На этом примере хорошо видно,  как наличие потерь в заполняющей  среде
делает  скачкообразный  переход  от  распространения  к  затуханию при
критической частоте данной волны непрерывным,  размытым;  само понятие
критической  частоты  теряет свой абсолютный смысл.  Обратим внимание,
что вблизи критической частоты затухание волны  при  том  же  значении
$\varepsilon''$  оказывается  большим,  чем  вдали  от неё в области
частот,  где волна  является  распространяющейся.  В  непосредственной
близости    от    $k_n$    коэффициент   затухания   $h_n''\approx   k
\sqrt{\mu\varepsilon''/2}$,  а вдали,  когда применима формула (11.4),
$h_n''\approx \mu\varepsilon''k^2/2h_{0n}$.  При малых $\varepsilon''$
второе выражение имеет существенно меньшее значение.

     При учёте  конечной  проводимости  стенок  волновода так просто
вычислить коэффициент затухания уже не удаётся.  Строгий расчёт  с
учётом  проникновения  поля  вглубь стенок является примером решения
задачи о распространении цилиндрической волн в поперечно  неоднородной
среде.  Даже  в  кусочно  однородной по своим свойствам среде в волне,
вообще говоря, отличны от нуля все шесть компонент поля и деление волн
на  магнитные  и электрические уже не имеет места.  Краевая задача для
потенциальных функций становится очень громоздкой. Однако для металлов
в сантиметровом диапазоне волн, где в основном используются волноводы,
волновое сопротивление $W$ очень малая по модулю комплексная величина,
скин-эффект сильно выражен и конечная проводимость с хорошей точностью
может быть учтена с помощью граничного условия Щукина-Леонтовича.  При
этом поле в стенках волновода не представляет интереса,  а поле внутри
волновода  определяется  двумя  потенциальными  функциями  $\Pi^e$   и
$\Pi^m$,  для которых остаются в силе те же уравнения (7.14) и (7.15),
что и при идеальной проводимости стенок.  Однако  эти  две  функции  в
каждой   точке   контура   $C$,   ограничивающего  поперечное  сечение
волновода,  оказываются теперь связанными граничным  условием  (6.17),
которое с учётом формул (7.16) записывается в виде
     $$\left.\begin{array}{l}\displaystyle{ih\frac{\partial \Pi^e}{
         \partial s}-ik\mu\frac{\partial \Pi^m}{\partial n}=
         W(k^2\varepsilon\mu-h^2)}\Pi^m\,,\\[.4cm]\displaystyle{(k^2
         \varepsilon\mu-h^2)\Pi^e=-iW\Bigl(k\varepsilon\frac{\partial\Pi^e
         }{\partial n}+h\frac{\partial\Pi^m}{\partial s}\Bigr)}\,.\\
         [.2cm]\end{array}\right\}\eqno(11.6)$$

     Граничные условия (11.6) содержат малый параметр $|W|$,  так  что
возникшая краевая задача отличается от краевой задачи для волноводов с
идеально проводящими стенками  возмущением  в  граничных  условиях.  С
помощью  методов  теории возмущения можно найти поправки к собственным
значениям и собственным функциям идеального волновода.  Будем помечать
все   величины,   относящиеся   к   невозмущённому  решению,  нижним
индексом~0;  тогда с точностью до членов  первого  порядка  по  малому
параметру,  которые пометим нижним индексом~1,  решение краевой задачи
можно искать в виде
     $$\left.\begin{array}{l}\displaystyle{\Pi^e(x,y,z)=[\Pi^e_0(x,y)
        +W\Pi^e_1(x,y)]e^{ihz}}\,,\\[.3cm]\displaystyle{\Pi^m(x,y,z)=
        [\Pi^m_0(x,y)+W\Pi^m_1(x,y)]e^{ihz}}\,,\\[.3cm]h=h_0+Wh_1\,,\\
        [.2cm]g=g_0+Wg_1\,.\end{array}\right\}\eqno(11.7)$$
Если $W=0$,   то   (11.6)   переходит  в  граничное  условие  $E_t=0$,
использованное в рассмотренной  ранее  теории  волноводов  с  идеально
проводящими  стенками.  Малость  параметра $W$ позволяет ожидать,  что
структура   электромагнитного   поля   в   волноводе    с    реальными
металлическими стенками мало отличается от структуры поля в волноводе,
стенкам  которого  приписывается  идеальная  проводимость.  В   первом
порядке    основной   интерес   представляют   величины,   которые   в
невозмущённом решении равны нулю.  К таковым, в частности, относится
мнимая  часть  продольного волнового числа $h''$ в той области частот,
где волна в идеальном волноводе является распространяющейся.

     Рассмотрим простейший пример.  Пусть невозмущённое решение есть
симметричная $H$-волна круглого волновода радиуса $a$.  В соответствии
с формулой (9.15) потенциальная функция для волны $H_{0q}$ есть
     $$\Pi^m_0=AJ_0(g_0r);\quad g_0=\mu_{0q}/a\,.\eqno(11.8)$$
Для упрощения записи здесь у величин $\Pi^m_0$ и $g_0$ и далее у
$g$ и $h$ опущены  индексы,  определяющие  число  вариаций  поля
по  $r$  и $\varphi$. Другая потенциальная функция $\Pi^e$ в этом
частном примере остается равной нулю  во  всех  приближениях  по
параметру  $W$,  что непосредственно следует из второго условия
(11.6),  так как на контуре $C$   производная
$\displaystyle{\frac{\partial\Pi^m_0}    {\partial s}}=0$.
Решение  волнового  уравнения  (7.15),  регулярное  в  нуле,
независимо от возмущения граничных условий есть $\Pi^m=AJ_0(gr)$,
так что первая формула (11.6) принимает вид
     $$ik\mu J_1(ga)=WgJ_0(ga)\,.\eqno(11.9)$$
Здесь $\mu$  ---  магнитная  проницаемость материала заполнения и её
нельзя путать с корнями производной функции Бесселя $\mu_{0q}$,  через
которые   выражаются   собственные  значения  невозмущённой  задачи.
Представляя $g$ в виде (11.7),  раскладывая обе части (11.9) по малому
параметру  $W$  и  учитывая,  что  $J_1(g_0a)=0$,  найдём поправку к
собственному значению:
     $$g_1=-\frac{ig_0}{\mu ka}\,.\eqno(11.10)$$
Таким образом, в   том   же   приближении, что и   формулы   (11.7),
     $$g^2=g_0^2+2Wg_0g_1=g_0^2\Bigl(1-\frac{2iW}{\mu ka}\Bigr),
         \eqno(11.11)$$
     $$h=\sqrt{h_0^2+i\frac{2Wg_0^2}{\mu ka}}=h_0+i\frac{Wg_0^2}{\mu
         kah_0}\,.\eqno(11.12)$$

     Вблизи критической частоты волны,  определяемой условием $h_0=0$,
последнее равенство в формуле (11.12) неверно, и следует  пользоваться
предшествующим  выражением  с  радикалом,  которое пригодно и на самой
критической частоте.  В  результате  коэффициент  затухания  вдали  от
критичеcкой частоты равен
     $$h''=\frac{\re W g_0^2}{\mu kah_0}\,,\eqno(11.13)$$
а вблизи неё
     $$h''=g_0\sqrt{\frac{|W|}{\mu ka}}\,.\eqno(11.14)$$
Как и   в  случае  потерь  из-за  поглощения  в  веществе  заполнения,
коэффициент затухания волны вблизи  критической  частоты  при  той  же
проводимости стенок существенно выше, чем для более высоких частот.

     Приведенные выкладки показывают,  что даже в простейшем  случае
волноводной  волны,  определяемой  и при учёте конечной проводимости
стенок лишь одной потенциальной функцией, прямой расчёт коэффициента
затухания  по методу теории возмущений достаточно громоздкая операция.
Однако существует способ,  существенно упрощающий  расчет  поправок  к
продольному   волновому   числу,   позволяющий   к  тому  же  записать
коэффициент  затухания  в  первом   порядке   по   $W$   в   замкнутом
аналитическом   виде   без  конкретизации  формы  поперечного  сечения
волновода.

     Пусть $\overline  \Sigma(z)$  ---  средняя  по  времени мощность,
переносимая  волной  через  поперечное  сечение   волновода,   которая
вычисляется   интегрированием   по   поперечному   сечению  продольной
компоненты усреднённого вектора Умова-Пойнтинга
     $$\overline{ \gv S}=\frac{c}{8\pi}\re{[{\bf {EH^*}}]}\,.
         \eqno(11.15)$$
Для упрощения  записи здесь и в последующих формулах у всех физических
величин,   присущих   волноводной   волне,   опущен   нижний   индекс,
определяющий  номер собственного значения краевой задачи.  Так как при
наличии  затухания  амплитуда   всех   компонент   поля   для   волны,
распространяющейся   в  положительном  направлении  оси  $z$,  убывает
$\sim\displaystyle{e^{-h''z}}$, то
     $$\overline\Sigma(z)=\overline\Sigma(0) e^{-2h''z}\,,
         \eqno(11.16)$$
и в результате  справедливо следующее дифференциальное соотношение:
     $$\frac{d\overline\Sigma}{dz}=-2h''\overline\Sigma\,.\eqno(11.17)$$

     В соответствии  с  законом  сохранения  энергии  разность потоков
мощности через сечение $z$ и сечение $z+\Delta z$ обусловлена потерями
энергии   на   отрезке   волновода   длиной  $\Delta  z$.  Эти  потери
складываются из потерь  в  заполняющем  волновод  веществе  и  потоков
мощности,  поступающей  из  объёма волновода в металлические стенки.
Объёмная плотность  потерь  в  веществе  $Q$  определяется  формулой
(2.32).  Для  вычисления  потерь  в  выделенном  отрезке волновода это
выражение надо проинтегрировать по поперечному сечению и домножить  на
$\Delta  z$.  Мощность,  поступающая  в стенки волновода,  вычисляется
интегрированием нормальной  компоненты  среднего  по  времени  вектора
Умова-Пойнтинга  $\overline{\gv  S}_n$  по контуру поперечного сечения
$C$ и домножением на $\Delta z$. В пределе $\Delta z \to 0$ приходим к
дифференциальному соотношению, из которого с учётом (11.16) получаем
выражение для коэффициента затухания:
     $$h''= \frac{\displaystyle{\int\limits_S Q\,dS}+
         \displaystyle{\oint\limits_C\overline {\gv S}_n\,ds}}
         {2\overline \Sigma} \,.\eqno(11.18)$$
Эта формула является точной, если в правую часть подставлены величины,
полученные   в   результате   строгого  решения  задачи  о  волноводе,
заполненном веществом с поглощением и стенки которого  рассматриваются
как  отдельная среда с проникающим в них электромагнитным полем.  Но в
таком случае ценность формулы не велика --- она может служить лишь для
контроля  правильности  расчёта,  поскольку  если  получено  строгое
решение задачи,  то известно и  комплексное  $h$.  Однако  соотношение
(11.18) позволяет вычислить $h''$ методом последовательных приближений
в рассматриваемом случае относительно небольших потерь.

     В частности, с помощью (11.18) может быть определён коэффициент
затухания,  обусловленный потерями в веществе заполнения при  идеально
проводящих  стенках  волновода  (в  этом  случае  второе  слагаемое  в
числителе (11.18) равно нулю). Однако для этого формула мало пригодна,
так  как  лишь  усложняет  выкладки  по  сравнению  с прямым способом,
использованным выше.  Кроме того,  из неё непосредственно не следует
выражение для затухания в окрестности критической частоты.  Но формула
(11.18) становится незаменимой для вычисления коэффициента затухания в
том  случае,  когда  для  расчёта потока мощности в стенки волновода
можно пользоваться граничным условием Щукина-Леонтовича, и по сравнению
с этим потоком можно пренебречь объёмными потерями в среде. При этом
правая часть  имеет  общим  множителем  малый  параметр  $|W|$  и  для
вычисления  $h''$  в  первом  порядке теории возмущения все величины в
числителе и  знаменателе  (11.18)  достаточно  подставить  из  решения
задачи  для  волновода  с  идеальными стенками.  Именно таким способом
производится  практический  расчёт  коэффициента  затухания  волн  в
волноводах.

     Стоящий в знаменателе  формулы  (11.18)  средний  поток  мощности
вдоль  волновода $\overline \Sigma$ был вычислен ранее.  При выбранной
нормировке (8.22) в односвязном волноводе он  для  электрических  волн
определяется формулой (8.26), то есть
     $$\overline\Sigma =\frac{ck \varepsilon \re h}{8\pi}\,;
                \eqno(11.19)$$
для магнитных волн
     $$\overline\Sigma=\frac{ck \mu \re h}{8\pi}\,,\eqno(11.20)$$
а для ТЕМ волны в замкнутой двухсвязной линии
     $$\overline\Sigma=\frac{ck^2 \varepsilon \sqrt{\varepsilon\mu}}
       {8\pi}\,.\eqno(11.21)$$

     Потоки мощности  в стенки волновода для электрических и магнитных
волн  существенно  различаются.  В  электрических  волнах,  как  и
в $TEM$-волне,  отсутствует  продольная  составляющая  магнитного
поля и весь поток согласно формуле (6.27) определяется касательной
к  контуру поперечного сечения $C$ компонентой:
     $$\overline\Sigma_n=\frac c{8\pi}\re W\oint\limits_C|H_s|^2\,ds\,,
         \eqno(11.22)$$
где $W$ -- волновое сопротивление материала стенки.  Выражая  $H_s$  с
помощью формул (7.16) через потенциальную функцию $\Pi^e_0$, а $\re W$
через толщину скин-слоя $\delta$  (полагая  для  металлической  стенки
$\mu=1$), находим коэффициент затухания электрических волн:
     $$h''=\frac{k^2\delta}{4h_0 \varepsilon}\oint\limits_C\left|
         \frac{\partial\Pi^e_0}{\partial n}\right|^2 \,ds\,.
         \eqno(11.23)$$
Для ТЕМ-волны  контур  интегрирования в этом выражении включает в себя
контуры сечения обоих проводников.

     В случае  магнитных  волн  в  подынтегральное  выражение  (11.22)
входят обе тангенциальные составляющие магнитного поля:
     $$\overline\Sigma_n=\frac c{8\pi}\re W\oint\limits_C (|H_s|^2+
         |H_z|^2)\,ds=\frac c{8\pi}\re W\oint\limits_C\biggl[h_0^2\left|
        \frac{\partial\Pi^m_0}{\partial s}\right|^2+g_0^4|\Pi^m_0|^2
        \biggr]\,ds\,,\eqno(11.24)$$
В результате для коэффициента затухания магнитных волн получаем:
     $$h''=\frac{\delta}{4h_0 \mu}\oint\limits_C\biggl[h_0^2\left|
     \frac{\partial \Pi^m_0}{\partial s}\right|^2 +g_0^4|\Pi
         ^m_0|^2\,\biggr] ds\,.\eqno(11.25)$$

     Примеры расчётов  коэффициента  затухания  для  важнейших  волн
круглого и прямоугольного волноводов будут рассмотрены чуть  позже,  а
сейчас  исследуем  в общем виде зависимость затухания от частоты волны
для  волновода  произвольного  сечения.  Удобно  ввести   безразмерную
величину  $\nu$,  представляющую  собой  отношение  рабочей  частоты к
критической    частоте    волны:    $\nu=k/k_0$    ($k_0=g_0/    \sqrt
{\varepsilon\mu}$)   и   рассматривать   её   значения  в  диапазоне
$1<\nu<\infty$,  в котором данная волна в идеальном волноводе является
распространяющейся.  Из-за  конечной  проводимости  волна  приобретает
затухание (11.23) или  (11.25),  для  вычисления  которого  необходимо
знать лишь функции $\Pi^e_0$ или $ \Pi^m_0$. Эти собственные функции и
соответствующие  им  собственные  значения  $g_0$  зависят   лишь   от
геометрических параметров волновода и не зависят от частоты,  в данном
случае от $\nu$.  От частоты зависят лишь следующие  величины:  $k\sim
\nu$, $h\sim \sqrt{\nu^2-1}$ и $\delta\sim 1/\sqrt{\nu}$.

     Поэтому для   электрических   волн  коэффициент  затухания  можно
представить в виде
     $$h''=\frac{C_0\nu\sqrt\nu}{\sqrt{\nu^2-1}},\eqno (11.26)$$
а для магнитных
     $$h''=\frac{C_1(\nu^2-1)+C_2}{\sqrt{\nu(\nu^2-1)}},
         \eqno (11.27)$$
где положительные коэффициенты $C_0, C_1$ и $C_2$ от $\nu$ не зависят.
Из  этих  формул  видно,  что  при $\nu\to 1$,  то есть при стремлении
частоты к критической,  затухание волн неограниченно возрастает. Такой
вывод  является ошибочным и объясняется неприменимостью формул (11.26)
и (11.27) вблизи критической частоты.  Однако  при  $\nu$,  близких  к
единице,  затухание  действительно  сильно  возрастает  и во много раз
больше затухания при более высоких частотах.

     При $\nu\to\infty$  формулы  (11.26)  для  электрических  волн  и
(11.27)  для  магнитных   стремятся   к   одному   виду:   $h''\approx
C_0\sqrt\nu$   и  $h''=C_1\sqrt\nu$.  Таким  образом,  при  достаточно
высоких частотах  коэффициент   затухания   растёт   пропорционально
квадратному  корню  из  частоты.  Фактически  оказывается,  что в этом
случае  зависимость  затухания  от   частоты   полностью   обусловлена
зависимостью  от  $\nu$  малого  параметра --- волнового сопротивления
материала  стенок  $  W  $.  Зависимость  $\re  W\sim\nu^{1/2}$
характерна  для  металлов при комнатных температурах;  при температуре
жидкого азота $\re W\sim\nu^{2/3}$,  а для сверхпроводящих  материалов
$\re  W\sim\nu^2$.  При  этом, несмотря  на  более быстрый рост с
частотой, при низких температурах затухание в широком диапазоне частот
существенно  меньше  из-за малости коэффициентов $C_0$, $C_1$ и $C_2$.

\begin{wrapfigure}[14]{l}{7.5cm}
\begin{picture}(80,55)
\put(-5,50){\special{em:graph fig11-1.bmp}}
\end{picture}
\hbox to 7.5cm{\hfil\footnotesize{Рис.~11.1.~Затухание волноводных
волн.}\hfil}
\end{wrapfigure}
     Вследствие возрастания $h''$  при  $\nu\to  1$  и  $\nu\to\infty$
зависимость коэффициента затухания от частоты в общем случае имеет вид
кривой $A$ на рис.~11.2 с одним характерным минимумом, вблизи которого
и целесообразно выбирать рабочую частоту. Затухание электрических волн
всегда  представляется  кривыми  вида  кривой  $A$.  Однако  затухание
некоторых  магнитных  волн  может  происходить  и  по кривой вида $B$,
соответствующей случаю $C_1=0$, когда формула (11.27) упрощается:
     $$h''=\frac{C_2}{\sqrt{\nu(\nu^2-1)}}.\eqno(11.28)$$
В этом случае с увеличением  частоты  затухание  быстро  спадает:  при
$\nu\gg  1$  имеем $h''=C_2\nu^{-3/2}$.  Равенство $C_1=0$ имеет место
при условии
     $$\oint\limits_C\left |\frac{\partial\Pi^m_0}{\partial s}\right|^2
         \,ds=0\,,\eqno(11.29)$$
для чего необходимо
     $$\frac{\partial \Pi^m_0}{\partial s} = 0 \quad (\mbox{или}\;
         \Pi^m_0=const)\quad\mbox{на}\quad C\,.\eqno(11.30)$$
Таким образом,  затухание ведет себя согласно кривой $B$ на  рис.~11.2
лишь  для  магнитной  волны,  потенциальная  функция $\Pi^m_0$ которой
постоянна на контуре $C$.  В прямоугольном волноводе согласно  формуле
(9.4) таких волн нет,  а в круглом они существуют: это волны $H_{0q}$.
Отметим, что условие (11.30) приводит к отсутствию продольных токов на
стенках,  с  чем  и  связано  уменьшение  затухания этих волн с ростом
частоты.

     Здесь имеет   смысл   сделать  небольшое  замечание  относительно
возможности практического использования  этой  закономерности.  Долгое
время  волну  $H_{01}$  в  круглом  волноводе  пытались  применить для
передачи  больших  СВЧ  мощностей   на   значительные   расстояния   с
минимальными  потерями,  что  возможно  лишь  при  достаточно  высоких
частотах.  Однако в этом случае волновод перестаёт быть одноволновым,
и  основные трудности обусловлены возникновением из-за неоднородностей
тракта помимо  рабочей  волны  ещё  и  паразитных  волн,  обладающих
большим затуханием. Избавиться от этой неприятности за счет уменьшения
радиуса   волновода   нельзя,   поскольку   при   сохранении    режима
одноволновости  существенно  увеличивается коэффициент $C_2$ в формуле
(11.28), так что в результате $h''\sim \sqrt{k}$.

     Перейдём теперь  к  расчёту  коэффициентов затухания основных
волн наиболее распространённых волноводов. Начнём с основной волны
$H_{10}$ в прямоугольном волноводе, широко применяемом в сантиметровом
диапазоне волн. Для этого подставим в (11.24)
     $$g_0=\frac{\pi} a\quad \mbox{и}\quad \Pi^m_0=
           \frac{2a}{\pi^2b}\cos\frac{\pi x}{a}
         \eqno(11.31)$$
в соответствии  с  формулами  (9.6)  и  (9.4)  для  магнитных волн.  В
результате несложных выкладок получим для коэффициента затухания волны
$H_{10}$:
     $$h''=\frac{\delta}{h_0a}\left(k^2\frac{a}{2b}+\frac{\pi^2}{a^2}
         \right).\eqno(11.32)$$
Напомним, что в формулах (11.31) и (11.32) $a$ ---  широкая,  $b$  ---
узкая стенка волновода;  для того,  чтобы вычислить затухание основной
волны в плоском  волноводе,  достаточно  в  (11.32)  устремить  $b$  к
бесконечности  (при  этом  следует иметь в виду,  что основная волна в
плоском  волноводе   соответствует   волне   $H_{10}$   прямоугольного
волновода  с  широкой стенкой $b$ и узкой $a$).  В результате $h''\sim
k^{-3/2}$;  такая зависимость обусловлена тем,  что продольные токи  в
случае  $H_{10}$-волны  в  прямоугольном  волноводе  текут  только  по
широкой стенке,  но при $b\to\infty$  их  вкладом  в  затухание  можно
пренебречь.

     Коэффициент затухания основной волны  круглого  волновода $H_{11}$
легко вычисляется с помощью формул (9.15) и (11.25). В результате
     $$h''=\frac{\delta}{2(\mu_{11}^2-1)h_0a}\left[k^2+\frac{\mu_{11}
         ^2(\mu_{11}^2-1)}{a^2}\right],\quad   \mu_{11}\approx 1,841 .
         \eqno(11.33)$$
Аналогично вычисляется    коэффициент    затухания   $TEM$   волны   в
коаксиальном кабеле:
     $$h''=k\delta\frac{a+b}{4ab\ln b/a}\,,\eqno(11.34)$$
где $a$  и~$b$  ---  соответственно  радиусы  внутреннего  и  внешнего
проводников;  зависимость $h''\sim\sqrt{k}$ имеет место  и  для  $TEM$
волны в плоском волноводе.

     Порядок величины  затухания  в сантиметровой области для наиболее
употребительных волноводов --- сотые доли децибела на метр.  Например,
для  основной  волны прямоугольного волновода со сторонами $a=7,2$см,
$b=3,4$см,  при длине волны $\lambda$=10~см и  $\delta=1$мк  (медь),
коэффициент  затухания  $h''=1,9  \cdot  10^{-5}$см$^{-1}$,  то  есть
затухание составляет 0,015~дБ/м.  В коротковолновой области  затухание
становится  значительно большим и одноволновые прямоугольные волноводы
для $\lambda \approx2$~мм имеют затухание порядка 5~дБ/м. Значительное
затухание  ---  одна  из  основных  причин,  ограничивающих применение
волноводов на длинах волн порядка 1--2~мм и меньше.

     Малость параметра   $|W|$   для  металлических  стенок  волновода
позволяет с помощью теории возмущений  связать  коэффициент  затухания
$h''$  с  поправкой  к действительной части $h$,  определяющей фазовую
скорость (длину волны в волноводе).  Для этого представим $h$  в  виде
(11.7)  и  поскольку  $h_0$  ---  величина  действительная  (в области
частот, где волна распространяется), то
     $$h''=\re W\im h_1+\im W\re h_1.\eqno(11.35)$$
Величина $h''$ уже была  вычислена  ранее,  и  обе  формулы  (11.23)  и
(11.25) можно представить в виде
     $$h''=\re W\cdot w,\eqno(11.36)$$
где величина  $w$  определяется соответствующей потенциальной функцией
идеального волновода и, следовательно, не зависит от свойств материала
стенок. Сопоставляя формулы (11.35) и (11.36), получаем:
     $$\re h_1=0;\quad \im h_1=w\,.\eqno(11.37)$$
Таким образом, в первом порядке теории возмущения
     $$h=h_0+iW\cdot w,\eqno(11.38)$$
и поскольку для металлов $\im W=-\re W$, то
     $$h'=h_0+\re W\cdot w=h_0+h''\,,\eqno(11.39)$$
и фазовая скорость записывается в виде
     $$v_{\mbox{\footnotesize\textit{ф}}}= c\;\frac k {h_0+h''}\,.\eqno(11.40)$$

     В заключение  следует  сказать,  что конечная проводимость стенок
волновода снимает вырождение волноводных  волн,  поскольку  они  имеют
теперь разную фазовую скорость и разный коэффициент затухания.  Так, в
частности,  снимается вырождение  волн  $H_{01}$--$E_{11}$  идеального
круглого волновода.  Напомним ещё раз,  что в неидеальном волноводе,
волна,  вообще  говоря,  является  гибридной,  имеет  обе   продольные
составляющие  поля,  и делить волны на электрические и магнитные можно
только условно,  имея ввиду близость полей той или иной волны к  полям
соответствующей волны идеального волновода.

%\end{document}

\newpage
\oddsidemargin=-0.4mm \evensidemargin=-0.4mm
\topmargin=-0.4mm
\headsep=7mm
\textheight=231.875mm
\textwidth=160mm
\mathsurround=2.5pt
\unitlength=1mm
%\begin{document}
%\input{macr.tex}
\thispagestyle{empty}
%\addtocounter{page}{120}

\begin{center}
     \subsubsection*{\rm Г\,Л\,А\,В\,А\, 4}
         \vspace{-1.15em}
         \line(6,0){160}\\
         \vspace{-1em}
         \line(6,0){160}
         \vspace{-1.15em}
     \subsubsection*{МЕДЛЕННЫЕ ВОЛНЫ}\vspace{35mm}
     \subsubsection*{12. Диэлектрические замедляющие структуры}
\end{center}\vspace*{0.5cm}

\markboth{Глава 4.  Медленные  волны}{12.  Диэлектрические замедляющие
структуры}

\begin{center}\begin{minipage}[c]{0.75\textwidth}
\footnotesize{\parindent=0.5cm
         Медленные и быстрые цилиндрические волны.  Медленные волны  в
         диэлектрическом слое.  Поверхностный характер медленных волн.
         Ограниченность спектра медленных волн.  Разложение  медленных
         волн  на плоские.  Физическая природа медленных поверхностных
         волн.  Медленные  волны   в   структурах   с   цилиндрической
         симметрией.
}\end{minipage}\end{center}\vspace*{0.5cm}

     Цилиндрические волны  делятся на волны {\it быстрые} и волны {\it
медленные} в зависимости от того,  больше или меньше скорости
света  в пустоте  $c$  их  фазовая скорость
$v_{\mbox{\footnotesize\textit{ф}}}=ck/h$; поэтому волна является
быстрой при $h<k$ и медленной при  $h>k$. Рассмотренные  до  сих
пор цилиндрические волны в пустых (незаполненных диэлектриком)
волноводах были волнами  быстрыми, поскольку  их  продольное
волновое   число $h=\sqrt{k^2-g^2} <k.$ Однако в ряде электронных
приборов СВЧ, таких, как лампа бегущей волны,  и в ускоряющих
станциях линейных ускорителей заряженных  частиц  требуются
волноводные системы,  в которых фазовая скорость меньше скорости
света.  Это обусловлено тем,  что длительное эффективное
взаимодействие частицы с электромагнитным полем может быть только
при условии синхронизма волны и частицы, а скорость  движения
частиц всегда меньше $c$.

     Существует много электродинамических структур, в которых возможно
распространение  медленных волн.  Эти структуры могут быть разделены
на два класса,  различающиеся  физическим  принципом,  обусловливающим
замедление  волны.  К  первому классу относятся структуры,  однородные
вдоль направления распространения цилиндрической волны вида (7.1),  то
есть   вдоль   оси   $z$.  Будем  называть  их  гладкими  замедляющими
структурами.  В  них  обязательно  должно  присутствовать  вещество  с
$|\varepsilon|>1$,  внутри  которого  сосредоточена существенная часть
электромагнитного поля.  При этом поле имеется и  в  примыкающем
пустом  пространствe,  где  волна  неоднородна  в поперечной плоскости
(амплитуда  её   резко   спадает   при   удалении   от   поверхности
диэлектрика), но скорость перемещения  фазы и там меньше $c$.

     Если электрическими  потерями   в   веществе   можно   пренебречь
($\varepsilon$   --   действительная  величина),  то  медленная  волна
является незатухающей;  при наличии у $\varepsilon$  небольшой  мнимой
части   появляется  затухание  волны  вдоль  $z$,  но  структура  поля
искажается несущественно.  Медленные волны возможны и в  структурах  с
гладкими    металлическими    проводниками,    которым   соответствует
комплексная диэлектрическая проницаемость (2.7) с существенной  мнимой
частью.   Но  и  в  этом  случае  весь  эффект  замедления  обусловлен
проникновением поля волны  внутрь  вещества  ---  в  гладких  идеально
проводящих структурах медленных волн нет.

     Ко второму   классу   следует   отнести   такие структуры,     в     которых
замедление    волны обусловлено  периодической  пространственной   неоднородностью
вдоль  направления распространения волны.  При этом проникновение поля
внутрь   образующих   структуру   элементов    может    играть    лишь
вспомогательную роль --- сам эффект замедления присутствует и в случае
идеально  проводящих  ограничивающих  поверхностей  и  обусловлен   их
геометрией.

     Собственная волна в структурах этого класса  не  является  строго
цилиндрической  волной  вида (7.1),  а представляет собой совокупность
бесконечного   числа   таких   волн,   называемых    пространственными
гармониками  и  обладающих  разными  продольными  волновыми числами и,
следовательно,  фазовыми  скоростями.   При   определённой   частоте
какая-нибудь  из  гармоник  всегда окажется в синхронизме с движущейся
вдоль структуры заряженной частицей. Второй класс замедляющих структур
находит   основное   применение   в  мощных  сильноточных  электронных
устройствах СВЧ и в  ускорителях;  он  будет  рассмотрен  в  следующем
разделе, а здесь основное внимание уделяется гладким структурам.

\begin{wrapfigure}[15]{l}{7.2cm}
\begin{picture}(80,55)
\put(0,50){\special{em:graph fig12-1.bmp}}
\end{picture}
\hbox to 7.2cm{\hfil\footnotesize{Рис.~12.1.~Плоскопараллельный
слой}\hfil}
\hbox to 7.2cm{\hfil\footnotesize{диэлектрика.}\hfil}
\end{wrapfigure}
    Замедление электромагнитной волны имеет  место в любой среде
с $\varepsilon \mu >1$.  В неограниченной однородной среде с такими
свойствами фазовая скорость обычной однородной  плоской волны
$v_{\mbox{\textitф}}=c/\sqrt{\varepsilon \mu}<c$, следовательно, волна ---
медленная. Плоская волна и неограниченная среда являются
идеализацией, а  для практических применений интерес представляют
свойства структур, неоднородных  в поперечной  плоскости. Простейшим
 примером   такой структуры является  круговой диэлектрический
 цилиндр в пустоте.  В  то  же время он  является   хорошей  моделью
 широко используемой на практике волоконной оптической линии связи.

     Но прежде   чем   исследовать   систему    медленных    волн    в
диэлектрическом  цилиндре,  изучим  ещё  более  простую  замедляющую
структуру,  в которой основные физические  результаты  не  заслоняются
громоздкой  математикой  ---  неограниченный  {\it  плоскопараллельный
диэлектрический слой},  представленный на рис.~12.1. Ось $x$ совпадает
с   нормалью   к  поверхности  слоя,  который  расположен  симметрично
относительно  плоскости  $x=0$;  вне  слоя   толщиной   $2a$ ---  пустота
$(\varepsilon=\mu=1)$,   а   сам   он   характеризуется  вещественными
проницаемостями $\varepsilon$,~$\mu$.

     Рассмотрим двумерные цилиндрические волны,  в которых зависимость
от координаты $z$ определяется множителем $e^{ihz}$,  а зависимость от
координаты   $y$   отсутствует.   В   этом   случае  ---  несмотря  на
неоднородность свойств  среды  вдоль  поперечной  координаты  $x$  ---
сохраняется деление волн на магнитные и электрические. Соответствующие
потенциальные   функции   $\Pi^m(x)$   и   $\Pi^e(x)$   внутри    слоя
удовлетворяют одному и тому же волновому уравнению
     $$\frac{d^2\Pi}{dx^2}+(k^2\varepsilon \mu-h^2)\Pi=0\,,
     \eqno(12.1)$$
имеющему частные решения
     $$\Pi(x)=A\sin{g x}\eqno(12.2\mbox{\textit a})$$
и
     $$\Pi(x)=B\cos{g x},\eqno(12.2\mbox{\textit б})$$
где $g$ --- поперечноe волновоe число и $g^2=k^2\varepsilon\mu-h^2$.

     Составляющие электромагнитного поля определяются через потенциалы
по  формулам  (7.12),  (7.13)  и (7.16);  например,  для электрической
волны,  соответствующей решению (12.2а),  отличны от  нуля  компоненты
     $$E_x=ihg A \cos{g x}\; e^{ihz},\qquad E_z=g^2 A \sin{g x}\; e^
         {ihz},\qquad H_y=ikg\varepsilon A\cos{g x}\; e^{ihz}.
         \eqno(12.3)$$
Вне слоя волновое уравнение для $\Pi(x)$ целесообразно записать в виде
     $$\frac{d^2\Pi}{dx^2}-p^2\Pi=0\,,\eqno(12.4)$$
где $p^2=h^2-k^2$; такая запись при условии $p^2>0$ подразумевает, что
поле ищется в виде медленной волны,  у которой  $h>k$.  Убывающее  при
$x\to\infty$  решение  уравнения  (12.4)  есть  $\Pi^e=Ce^{-px}$  (при
$p>0$), поэтому в области $x>a$ для электрических волн имеем следующие
выражения для отличных от нуля компонент поля:
     $$E_x=-iphCe^{-px+ihz},\qquad E_z=-p^2 C e^{-px+ihz},\qquad
         H_y=-ipk Ce^{-px+ihz}.\eqno(12.5)$$

     Приравнивая на   поверхности   раздела    $x=a$    тангенциальные
составляющие  полей  $E_z$  и  $H_y$,  получим  систему  двух линейных
однородных уравнений относительно коэффициентов $A$ и $C$ вида
     $$-p^2Ce^{-pa}=g^2 A \sin{g a}\,,\qquad-pCe^{-pa}=\varepsilon g A
         \cos{g a}\,,\eqno(12.6)$$
условие совместности    которой   (обращение   в   нуль   детерминанта
соответствующей матрицы) определяет в неявном виде  характеристическое
уравнение волны:
     $$p=\frac{1}{\varepsilon}\;g\tg{g a}\,.\eqno(12.7)$$
Если в это уравнение подставить $g$ и $p$, выраженные через $k$ и $h$,
то  получим  характеристическое  уравнение,  из  которого  может  быть
найдено продольное волновое число $h$ как  функция  $k$, и  тем  самым
определена  дисперсия  фазовой скорости волны.  В таком виде уравнение
пригодно лишь для построения зависимости $h(k)$  численными  методами.
Для   качественного  графического  анализа  системы  собственных  волн
структуры  удобнее  уравнение  (12.7),   которое   следует   дополнить
соотношением
     $$p^2+g^2=k^2(\varepsilon \mu-1)\,,\eqno (12.8)$$
являющимся прямым следствием того факта, что продольное волновое число
$h$ одно и то же для волны вне и внутри слоя.

     Для потенциальной  функции  (12.2\textit {б})  характеристическое уравнение
выводится точно таким же способом и имеет вид
     $$p=-\frac{1}{\varepsilon}\; g \ctg{ga}\,.\eqno(12.9)$$
Волны, определяемые  функциями  (12.2\textit а)  и (12.2\textit б),
имеют разную симметрию   относительно   плоскости   $x=0$.
Электрическую волну  с компонентами поля (12.3),  у которой
чётной  функцией $x$  являются $E_x$ и $H_y$,  принято называть
{\it чётной}; волну, определяемую решением (12.2\mbox{\textit
б}), {\it нечётной}. Заметим, что в соответствии  с чётностью
или нечётностью волн граничные условия для компонент полей  на
границе  $x=-a$ удовлетворяются автоматически,  если  они
выполнены при $x=a$.

     Наряду с электрическими волнами вдоль слоя могут распространяться
и  магнитные  волны,  описываемые  потенциальной  функцией $\Pi^m(x)$.
Характеристическое уравнение для магнитных волн получается по  той  же
схеме и имеет вид для чётных волн
     $$p=\frac{1}{\mu}\;g\tg{g a}\,,\eqno(12.10)$$
а для нечётных ---
     $$p=-\frac{1}{\mu}\;g \ctg{g a}\,.\eqno(12.11)$$
Нетрудно видеть,  что  характеристические уравнения   для   магнитных
волн  отличаются от соответствующих  уравнений  для электрических
волн заменой $\varepsilon$ на $\mu$.

    Проведём теперь  графический  анализ  решений характеристических
уравнений  (12.7)  и   (12.9) в случае  электрических   волн, для чего
целесообразно  домножить обе части этих уравнений на $a$,  чтобы иметь
дело с безразмерными переменными $pa$,~$ka$ и~$ga$.  Построим  графики
правых частей уравнений в функции переменной $ga$, имея в виду, что по
смыслу решения (12.5) интерес  представляют  только  значения  $pa>0$;
тогда  на  рис.  12.2  чётным  волнам соответствуют кривые 1,  3,  а
нечётным --- 2, 4.

\begin{wrapfigure}[15]{l}{6.5cm}
\begin{picture}(80,55)
\put(0,50){\special{em:graph fig12-2.bmp}}
\end{picture}
\hbox to 7.0cm{\hfil\footnotesize{Рис.~12.2.~Графическое решение
}\hfil}
\hbox to 7.0cm{\hfil\footnotesize{характеристического уравнения.}\hfil}
\end{wrapfigure}
     Поскольку переменные  $pa$ и $ga$ в соответствии с (12.8) связаны
между собой уравнением окружности радиуса
     $$R=ka\sqrt{\varepsilon\mu-1}\,\eqno(12.12)$$
с центром в начале координат,  то искомые  корни  уравнений  (12.7)  и
(12.9)  лежат на пересечении кривых,  изображающих на рис.~12.2 правые
части уравнений,  с этой окружностью.  Так как окружность пересекается
лишь  с  конечным  числом  кривых на рис.~12.2,  то уравнения (12.7) и
(12.9) имеют конечное число  корней,  и,  следовательно,  на  заданной
частоте  в  слое  может распространяться лишь конечное число медленных
электрических волн.  При $R<\pi/2 $ есть только один корень,  которому
соответствует  волна  $E_{00}$ (значения нижних индексов соответствуют
числу вариаций поля по осям  $x,\,y$);  при  $\pi/2<R<\pi$  появляется
второй корень,  соответствующий волне $E_{10}$,  и так далее. Частоты,
на которых появляются новые медленные волны,  называются критическими;
они соответствуют значениям $R=m\pi/2\; (m=0,1,\dots).$

     Характеристические уравнения (12.10) и (12.11) для магнитных волн
исследуются    аналогичным   образом.   Характер   поведения   кривых,
определяющих зависимость $pa$ от $ga$,  такой же, как у сответствующих
кривых на рис.~12.2,  поскольку правые части уравнений для магнитных и
электрических волн различаются лишь постоянным множителем. Пересечение
кривой,   соответствующей  медленной  волне  $H_{m0}$,  с  окружностью
радиуса  (12.8)  имеет  место  только  при   $R>m\pi/2$.   Поэтому   в
диэлектрическом  слое  критические частоты у магнитных и электрических
волн с одинаковыми индексами совпадают.

     Все выявленные  в результате графического анализа медленные волны
соответствуют значениям $p>0$.  В соответствии с формой записи решения
уравнения   (12.4)   все  компоненты  поля  убывают  при  удалении  от
поверхности диэлектрика по экспоненциальному закону.  Поле  волны  вне
диэлектрика   оказывается   оказывается сосредоточенным вблизи  его
поверхности.  Поэтому рассматриваемые медленные волны  называют  также
{\it  поверхностными волнами}.

\begin{wrapfigure}[14]{l}{7.0cm}
\begin{picture}(80,50)
\put(-5,50){\special{em:graph fig12-3.bmp}}
\end{picture}
\hbox to 7.5cm{\hfil\footnotesize{Рис.~12.3.~Фазовые скорости
волн}\hfil}
\hbox to7.5cm{\hfil\footnotesize{в диэлектрическом слое $(\mu=1)$.}\hfil}
\end{wrapfigure}
     Из рис.~12.2  видно,  что  для  любой  медленной  волны (например,
$E_{00}$)   при   неограниченном возрастании радиуса окружности $R$
---  это эквивалентно  неограниченному возрастанию   частоты  ---
величина    $ga$ стремится  к конечному пределу,  а величина $pa$
неограниченно   возрастает.   Поэтому  при частоте,    значительно
превышающей критическую частоту  данной  волны,  её  поле  в
основном сосредоточено в диэлектрике,  а в  окружающее   пустое
пространство просачивается   слабо.   Действительно,  условие
$pa\gg   1$    приводит   к  тому, что  поперечная протяжённость  поля
в  пустоте  существенно меньше $a$.

     Формула $h=\sqrt  {k^2 \varepsilon \mu-g^2}$ показывает,  что при
высоких  частотах   продольное   волновое   число   $h\approx   k\sqrt
{\varepsilon\mu}$,  то  есть  фазовая  скорость волны такая же,  как в
безграничном диэлектрике, и, наоборот, при критической частоте $p=0$ и
$h=k$.   Это   значит,  что  при  частоте,  лишь  немного  превышающей
критическую,  электромагнитное поле волны сосредоточено в  основном  в
пустоте;  скорость  волны  близка  к  $c$.  На  рис.~12.3 представлено
замедление первых четырёх электрических волн в функции частоты для
случая $\mu=1$.

     Характеристические уравнения   (12.7),   (12.9)--(12.11)   помимо
корней, выявленных проведенным графическим анализом, имеют и другие.
Так,  для каждой дисперсионной кривой на частотах выше критической,
помимо корня $p>0$,  всегда  имеется  и  отрицательный  корень.  В
некотором интервале частот,  меньших критической, таких действительных
отрицательных  корней  два.   Существование   этих   корней   нетрудно
установить  после  добавления  к  каждой  кривой  на рис.12.2 той её
части,   которая   соответствует    отрицательным    значениям    $p$.
Соответствующие   медленные   волны,    называемые  в  литературе
{\it антиповерхностными},  физического смысла не имеют,  поскольку их поле
экспоненциально нарастает  при  удалении от поверхности диэлектрика, и,
следовательно, они не могут быть реально возбуждены в эксперименте.

     При ещё  более  низких  частотах  у  уравнения  появляется пара
комплексных корней с разными знаками мнимой части.  Обе эти волны  вне
диэлектрика  являются  неоднородными  быстрыми  плоскими  волнами  (им
соответствует  $h'<k$),  действительная  часть  их  волнового  вектора
направлена под некоторым углом к поверхности,  а фаза волны у одной из
них движется от поверхности диэлектрика,  у другой --- к  поверхности.
Их   соответственно   называют  {\it втекающей}  и  {\it вытекающей}  волнами,
причём последняя наблюдается в эксперименте.

     Все медленные  волны  в идеальном диэлектрическом слое без потерь
являются незатухающими вдоль оси $z$ и число этих волн тем больше, чем
выше  частота.  В этом отношении диэлектрический слой подобен обычному
плоскому волноводу с идеально проводящими металлическими стенками,  за
что его часто так и называют:  {\it плоскопараллельный диэлектрический
волновод}.

     Однако аналогия  между  диэлектрическим  и  обычным волноводом не
такая уж глубокая. Различие обусловлено тем, что электромагнитное
поле в диэлектрическом волноводе проникает за пределы слоя.  В
нём всякая новая волна,  возникающая при своей критической
частоте,  имеет $h=k$, то есть распространяется со скоростью
света,  в то время как в обычном волноводе при критической частоте
$h=0$  и  $v_{\mbox{\footnotesize\textit{ф}}}\to\infty$.  Другое
важное отличие состоит в том,  что на любой частоте ниже
критической в обычном    волноводе    каждая    собственная волна
становится нераспространяющейся   и  не  переносит энергию  вдоль
волновода.  В диэлектрическом слое в некотором интервале  частот
ниже  критической соответствующей цилиндрической  волны  нет:  она
появляется лишь при ещё более низких частотах в виде быстрой
волны,  переносящей энергию от поверхности   слоя   и   затухающей
вследствие  этого  по  мере распространения вдоль оси $z$.

     Кроме этого,  существует принципиальное различие в математических
свойствах системы  волн  в  этих  двух  типах  волноводов.  В  обычном
волноводе с идеально проводящими стенками при любой частоте существует
бесконечная система собственных  волн  (из  них  лишь  конечное  число
распространяющихся),  образующих  полную  систему функций,  по которым
может быть разложено  произвольное  поперечное  поле  в  волноводе;  в
диэлектрическом  волноводе  (в  том числе и в диэлектрическом стержне,
рассматриваемом ниже)  совокупность  цилиндрических  волн  (включая  и
быстрые)  не  образует  полной  системы.  В результате при возбуждении
структуры произвольным  распределением  сторонних  токов  внутри  слоя
помимо цилиндрических будут излучаться и сферические волны.

     Физический смысл явлений в диэлектрическом  волноводе  становится
более  ясным,  если разложить электромагнитное поле в нём на плоские
волны,  как  это  уже  делалось  раньше  с  полем  основной  волной  в
прямоугольном  волноводе.  Составляющие  поля  (12.3) с помощью формул
Эйлера могут быть представлены в виде:
     $$E_x=\displaystyle{\frac{iAhg}2[e^{i(gx+
         hz)}\!+e^{i(-gx+hz)}]},\, E_z=-\displaystyle{\frac{iAg^2}2
         [e^{i(g x+hz)}\!-e^{i(-g x+hz)}]},\, H_y=\frac{k\varepsilon}h
         E_x.\eqno(12.13)$$

     Так как  волновые  числа  $g$  и  $h$  удовлетворяют  соотношению
$h^2+g^2=k^2\varepsilon\mu$, то можно ввести угол $\theta$, такой, что
     $$g=k \sqrt{\varepsilon\mu} \sin \theta\,,\qquad  h=k\sqrt
         {\varepsilon\mu}\cos\theta\,.\eqno(12.14)$$
После этого нетрудно заметить,  что поле внутри диэлектрического  слоя
представляет    собой    сумму    двух   плоских   волн,   направление
распространения которых образует с осью $z$ углы $\theta$ и $-\theta$.
Угол  падения этих волн на границу диэлектрика --- плоскость $x=a$ или
$x=-a$,  очевидно, равен (рис.~12.4) $\varphi=\pi/2-\theta$. Для того,
чтобы  рассмотренные  выше  волны  были  медленными  волнами,  то есть
выполнялось условие
     $$h=k\sqrt{\varepsilon\mu}\cos\theta=k\sqrt{\varepsilon\mu}\sin
         \varphi>k,\eqno(12.15)$$
угол падения $\varphi$ должен удовлетворять неравенству
     $$\sin\varphi>1/\sqrt{\varepsilon\mu}\,,\eqno(12.16)$$
при котором имеет место полное отражение.

\begin{wrapfigure}[14]{l}{7.2cm}
\begin{picture}(80,50)
\put(-5,50){\special{em:graph fig12-4.bmp}}
\end{picture}
\hbox to 7.5cm{\hfil\footnotesize{Рис.~12.4.~Разложение волны в
диэлектри-}\hfil}
\hbox to 6.9cm{\hfil\footnotesize{ческом слое на плоские волны.}\hfil}
\end{wrapfigure}
     Таким образом, волноводные свойства диэлектрического   слоя
обусловлены   явлением полного отражения. Электромагнитное  поле
в окружающем диэлектрик пространстве возникает благодаря
просачиванию поля   через   поверхность   раздела   сред,  причём
просачивающееся поле убывает при  удалении  от этой    поверхности
по    экспоненциальному закону.  В результате  электромагнитная  энергия
волны   локализована   в   основном   в   пределах диэлектрика.  Чем  ближе
частота  к критической,  тем меньше показатель экспоненты и тем
равномернее распределяется энергия  волны  в пространстве.

     Для частот,  значительно  превышающих  критическую частоту данной
волны,  угол  $\theta$  мал  и  плоские   волны   распространяются
в диэлектрике   почти   по   оси   $z$,   чему  соответствует
$h\approx k\sqrt{\varepsilon\mu}$,  а  вне  слоя   волна
является   медленной неоднородной  плоской  волной,  бегущей вдоль
оси $z$.  При уменьшении частоты угол $\theta$ увеличивается (но
не  до  значения  $\pi/2$),  а угол $\varphi$ уменьшается,  что
имеет место и в обычном волноводе. На критической частоте
     $$\cos\theta=\sin\varphi=1/\sqrt{\varepsilon\mu}
         \eqno(12.17)$$
и в окружающем пустом пространстве поле представляет собой  однородную
плоскую волну, распространяющуюся, как ей и положено, со скоростью $c$
вдоль  оси  $z$.  При  более  низких   частотах   появляется   быстрая
{\it вытекающая}  неоднородная   плоская  волна,  распространяющаяся  под
некоторым углом к поверхности.

     Особое положение   занимают   поверхностные   волны  $E_{00}$  и
$H_{00}$,  для которых критической частоты не существует, и поэтому они
могут   распространяться   при  сколь  угодно  низких  частотах.  Если
выполняется условие $R\ll 1$,  то оно согласно формуле (12.12)  влечет
за  собой  неравенства $pa\ll 1$ и $g a\ll 1$.  Это значит,  что как в
диэлектрике,  так и над ним,  поле становится практически  поперечным:
вдоль слоя распространяется плоская волна, слегка возмущённая слоем.

     Остановимся ещё   на   системе   медленных  поверхностных  волн
диэлектрического слоя, лежащего на идеально проводящей плоскости $x=0$
---   на  ней  должна  быть  равна  нулю  тангенциальная  составляющая
электрического поля. В связи с этим отметим, что в плоскости симметрии
$x=0$  диэлектрического  слоя  без  подложки  в чётных электрических
волнах  $E_{00}$,  $E_{20}$,  $E_{40},   \dots$   обе   тангенциальные
компоненты  $E_y$ и $E_z$ равны нулю (см.  формулы (12.3));  они равны
нулю   в   этой   плоскости   и   в   нечётных   магнитных    волнах
($H_{10}$,~$H_{30},\dots$).  Поэтому  в  слое на подложке толщиной $a$
существуют  такие  и  только  такие  медленные  поверхностные   волны,
структура  поля  которых  совпадает  со структурой полей перечисленных
волн в слое без подложки толщиной $2a$ при $x>0$.

     Волноводные  свойства  диэлектрического  слоя
обусловлены отражением распространяющихся  в  нём  плоских  волн  от
обеих  поверхностей.  Вопрос  о  существовании  медленной волны вблизи
поверхности,  отделяющей полупространство,  заполненное веществом,  от
пустоты,  требует дополнительного исследования.  Из приведенной выше
физической  интерпретации  медленной  волны  как  результата   полного
отражения  плоской волны от поверхности раздела двух сред
следует, что в этом случае она может существовать только при затухании
отражённой  волны вглубь среды.  Пусть  полупространство $x<0$
заполнено веществом с проницаемостями $\varepsilon,\mu$;  тогда поле в
пустоте $(x>0)$ определяется формулами (12.5),  а цилиндрическая волна
в веществе описывается $z$-компонентой вектора Герца,  удовлетворяющей
уравнению (12.1). Решение этого уравнения, затухающее вглубь вещества,
есть
     $$\Pi^e(x)=Be^{-\tilde p x},\eqno(12.18)$$
где
     $$\tilde p^2=h^2-\varepsilon\mu k^2\;,\eqno(12.19)$$
при условии $\tilde p^2>0$ и $\tilde p<0$. Компоненты полей в веществе
даются формулами,  получаемыми из (12.5) заменой $p$ на $\tilde  p$  и
$C$ на $B$.

     Условие равенства  тангенциальных  составляющих  полей на границе
$x=0$ приводит к системе линейных однородных алгебраических  уравнений
относительно  коэффициентов  $B$  и  $C$.  Равенство нулю детерминанта
соответствующей матрицы определяет соотношение между $p$ и $\tilde p$:
     $$\tilde p=\varepsilon p\;.\eqno(12.20)$$
Из (12.20), (12.19) и соответствующего выражения для $p$ следует:
     $$p^2=\frac{1-\varepsilon\mu}{\varepsilon^2-1}k^2\;.
         \eqno(12.21)$$
При действительных  значениях  $\varepsilon$,~$\mu$,   то   есть   при
отсутствии  поглощения  в  веществе,  $p^2$ и $\tilde p^2$ оказываются
действительными числами, большими нуля, при условии $\varepsilon<-1$ и
$\mu>0$.   Классическим  примером  подобного  вещества  (имеющего  для
рассматриваемой задачи чисто теоретический интерес) является  холодная
плазма в области низких частот, таких, что $\omega^2< \omega_p^2/2$.

     Для комплексных значений $\varepsilon$, в частности, в проводящей
среде,  отражённая от границы с пустотой плоская волна всегда  будет
затухать  вглубь  среды  и поэтому вторая граница слоя при достаточной
его толщине не будет влиять на формирование медленной волны;  при этом
ни   при  каких  условиях  нет  и  полного  отражения.  В
результате медленная  волна  всегда  будет  затухающей  в  направлении
своего   распространения   вдоль   граничной   поверхности.  Затухание
обусловлено как джоулевыми  потерями  в  среде,  так  и  излучением  в
свободное  пространство,  поскольку  $p $ в этом случае,  как видно из
формулы  (12.21),  величина  комплексная  и  есть  поток  мощности  от
границы.   Если  $|\varepsilon|\gg  1$,  то  уравнение  (12.21)  можно
записать в виде:
     $$p=iWk\,,\eqno(12.22)$$
где $W=\sqrt{\mu/\varepsilon}$   ---   волновое  сопротивление  среды.
Последнее  уравнение  часто  называют  характеристическим   уравнением
поверхностной  волны  Ценнека.  На  заре  развития радиотехники именно
этими   поверхностными   волнами   пытались   объяснить    особенности
распространения  радиоволн  вдоль  земной  поверхности.  Такой  подход
оказался ошибочным:  при тех высоких значениях  проводимости  почвы  и
морской  воды,  которые  реально  имеют место,  $|W|$ малая величина, и
согласно  формуле  (12.22)  вертикальная  протяжённость   поля   над
проводящей  средой  намного  больше  длины  волны.  Поэтому передающие
антенны  радиостанций  поверхностной  волны  Ценнека  практически   не
возбуждают.

     Перейдём теперь к рассмотрению медленных волн в диэлектрических
структурах,  ограниченных  в  поперечной плоскости пустотой и при этом
обладающих осью симметрии. Простейшим примером служит круглый стержень
из    однородного   диэлектрика.   Физика   явления,   приводящего   к
возникновению медленной поверхностной волны,  остаётся той же самой:
полное   отражение  при достаточно больших углах падения плоской волны
из оптически более плотной среды на границу с менее плотной,
однако математический аппарат заметно усложняется.

     Пусть радиус  стержня  $a$,  его  проницаемости  $\varepsilon$  и
$\mu$. Исследование системы цилиндрических волн в такой неоднородной в
поперечной плоскости структуре естественно проводить в  цилиндрической
системе координат $r,\varphi,z$.  Согласно общей теории цилиндрических
волн  поля  в  этом  случае  могут  быть  найдены   с   помощью   двух
потенциальных  функций,  в  качестве которых удобно выбрать продольные
компоненты электрического и  магнитного  векторов  Герца  $\Pi^e_z$  и
$\Pi^m_z$.

     В отличие от обычных волноводов с идеально  проводящими  стенками
система волн в рассматриваемой структуре не распадается на независимые
электрические и магнитные волны,  а каждая собственная волна  в  общем
случае является гибридной,  то есть содержит все шесть компонент поля,
которые выражаются через $\Pi^e_z$ и $\Pi^m_z$ с помощью общих  формул
(7.12),  (7.13) и (7.17). Оба вектора Герца в каждой из областей $r<a$
(диэлектрик)  и  $r>a$  (пустота)  удовлетворяют  одинаковым  волновым
уравнениям (7.14),  (7.15), которые в цилиндрической системе координат
имеют вид
     $$\frac{\partial^2\Pi_z^{e,m}}{\partial r^2}+\frac{1}{r}
         \frac{\partial \Pi_z^{e,m}}{\partial r}+\frac{1}{r^2}
         \frac{\partial^2\Pi_z^{e,m}}{\partial\varphi^2}+
         (k^2\varepsilon\mu-h^2)\Pi_z^{e,m}=0\,;\eqno(12.23)$$
$\varepsilon=\mu=1$  при  $r>a$.

     Решение этих уравнений в диэлектрике должно удовлетворять условию
конечности  поля  при  $r=0$,  а  в  пустоте представлять собой волну,
распространяющуюся (или затухающую) от  оси  структуры.  Имея  в  виду
последующее приравнивание тангенциальных компонент поля по обе стороны
границы  $r=a$,  целесообразно  записать  соответствующие  решения   в
следующем виде:
     $$\left.\begin{array}{rcl}\Pi_z^e&\!=\!&AJ_m(g r)\sin(m\varphi+
         \varphi_0)\; e^{ihz}\\[0.15cm]\Pi_z^m\!&\!=\!&BJ_m(g r)\cos(m
         \varphi+\varphi_0)\; e^{ihz}\\\end{array}\right\}\quad\hbox{при
         $r<a$,}\eqno(12.24)$$
     $$\left.\begin{array}{rcl}\Pi_z^e&\!=\!&CK_m(pr)\sin(m\varphi+
         \varphi_0)\; e^{ihz}\\[0.15cm]\Pi_z^m\!&\!=\!&DK_m(pr)\cos(m
         \varphi+\varphi_0)\; e^{ihz}\\\end{array}\right\}\quad\hbox{при
         $r>a$,}\eqno(12.25)$$
причём нижний   индекс   $m$   определяет  число  вариаций  поля  по
азимутальному  углу  $\varphi$.  Нас  интересует  диапазон  частот,  в
котором   поперечное  волновое  число в  диэлектрике  $g=  \sqrt  {k^2
\varepsilon \mu-h^2}$ --- величина действительная, а поскольку решение
ищется  в  виде медленной волны,  для которой $h>k$,  то для избежания
мнимых величин при $r>a$ вместо мнимого  поперечного  волнового  числа
введена     величина    $p=\sqrt{h^2-k^2}$,    а    функция    Ханкеля
$H^{(1)}_m(\sqrt{k^2-h^2}r)$  от   мнимого   аргумента   заменена   на
модифицированную   функцию   Бесселя   $K_m(pr)$   от  действительного
аргумента, экспоненциально убывающую при больших его значениях.

     Исходя из формул (12.24) и (12.25) для векторов  Герца  находятся
компоненты  поля  и из требования непрерывности $E_\varphi$,  ~$E_z$,~
$H_\varphi$,  ~$H_z$ получается система линейных однородных  уравнений
относительно   коэффициентов   $A$,~$B$,~$C$,~$D$.   Условие   наличия
ненулевого решения  у  этой  системы  приводит  к  характеристическому
уравнению следующего вида:
     $$\left(\varepsilon f_m-F_m\right)\left(\mu f_m-F_m\right)=m^2
       \left[\frac 1{(g a)^2}+\frac 1{(pa)^2}\right]\left[\frac
  {\varepsilon\mu}{(g a)^2}+\frac 1{(pa)^2}\right],\eqno (12.26)$$
где
     $$f_m=\frac{J_m'(g a)}{g aJ_m(g a)},\qquad F_m=\frac{K_m'(pa)}
         {pa K_m(pa)}\eqno(12.27)$$
(штрихом обозначено  дифференцирование  по  аргументу  соответствующей
функции,  а между входящими в уравнение (12.26) величинами имеет место
дополнительное  соотношение (12.8)).  Из характеристического уравнения
определяется  продольное  волновое  число  $h$  как  функция  $k$.  При
произвольных значениях $m$ анализ этого уравнения весьма затруднён и
получить  аналитические  выражения  хотя  бы  для  критических  частот
гибридных   медленных  волн  не  удается.  Их,  как  и  сами  решения,
приходится искать численными методами,  но  принципиальных  трудностей
при современном развитии вычислительной техники на этом пути нет.

     При $m=0$ характеристическое уравнение (12.26) распадается на два
независимых   уравнения   ---   отдельно   для    симметричных    {\it
электрических}   волн   $(\varepsilon   f_0=F_0)$   и   отдельно   для
симметричных {\it магнитных} волн $(\mu f_0= F_0)$.  Эти уравнения уже
намного   проще   и   по  своей  структуре  схожи  с  соответствующими
уравнениями для диэлектрического слоя. Остановимся кратко на свойствах
симметричных электрических волн.

     Характеристическое уравнение   в этом случае  может быть записано
в следующем виде:
     $$pa\frac{K_0(pa)}{K_1(pa)}=-\frac{ga J_0(ga)}{\varepsilon J_1
         (ga)}.\eqno(12.28)$$
Левая и  правая  части  последнего  уравнения по характеру поведения в
функции  своего  аргумента  очень  близки  к  соответствующим   частям
уравнения  (12.7):  левая часть заметно отличается от линейной функции
лишь при малых значениях $pa$, а положительные значения правой части в
функции  аргумента  $ga$  графически очень сходны с системой кривых на
рис.~12.2  для  нечётных  электрических  волн   в   диэлектрической
пластинке. Основное различие заключается лишь в том, что каждая кривая
лежит между корнями нулевой и  первой  функций  Бесселя,  а  не  между
нулями косинуса и синуса. Правая часть уравнения (12.28) не обращается
в нуль при $g=0$ и поэтому при низких частотах медленной волны нет.

     Поскольку в отличие от  уравнения  (12.7)  уравнение  (12.28)  не
позволяет  в  явном  виде  выразить  $pa$ как функцию от $ga$,  то при
использовании описанного выше графического  метода  решения  уравнения
для   диэлектрической   пластины  в  случае  диэлектрического  стержня
требуется  дополнительное  графическое  построение:  строятся  графики
левой  и  правой  части  уравнения  (12.28) соответственно как функции
переменных $pa$ и $ga$ и уже с их помощью на плоскости этих переменных
наносятся  кривые,  представляющие  функцию $pa(ga)$.

     Пары переменных $pa$ и $ga$,  определяющих для заданного значения
$k$ собственные  медленные  волны,  находятся  как  точки  пересечения
построенных  кривых с окружностью радиуса $R$,  определяемого формулой
(12.12) и пропорционального  $k$.  Дальнейшее  вычисление  продольного
волнового числа $h$ и фазовой скорости не представляет труда.  Из этих
построений становится очевидным, что при малых значениях $k$ медленная
симметричная  волна  в  диэлектрическом  стержне отсутствует.  В целом
система  симметричных  медленных  волн   в   круглом   диэлектрическом
волноводе   качественно   не  отличается  от  соответствующей  системы
нечётных медленных волн в диэлектрическом слое.

     Можно показать,  что уравнение (12.28) имеет только такие решения
для продольного волнового числа, которые удовлетворяют неравенствам
     $$k<h<k\sqrt{\varepsilon\mu},\eqno(12.29)$$
поэтому   фазовые   скорости   медленных   волн  лежат  в
интервале $c>v_{\mbox{\footnotesize\textit{ф}}}>c/
\sqrt{\varepsilon\mu}$.

     Среди гибридных несимметричных волн $(m\geqslant 1)$ диэлектрического
стержня особое положение занимает {\it  основная}  или  {\it  главная}
волна  с  $m=1$  --- она существует при любой частоте.  Основную волну
можно рассматривать как плоскую волну,  распространяющуюся  вдоль  оси
$z$   и   слегка  возмущённую  диэлектрическим  стержнем.  Благодаря
замедлению в стержне волна приобретает поверхностный характер,  однако
при    низких   частотах   соответствующие   значения   $p$   малы   и
экспоненциальное затухание поля в окружающем  пространстве  происходит
очень   медленно.   Также  малы  и  продольные  компоненты  поля,  и  в
значительной части пространства волна не отличается по своей структуре
от плоской.

     Наряду с медленными  волнами  в  диэлектрических  стержнях  могут
распространяться  быстрые  волны:  ниже  критической частоты медленная
волна преобразуется в быструю, затухающую из-за излучения в окружающее
стержень   пространство.   В  определённой  области  параметров  это
затухание  невелико,  и  такие  волны   наблюдаются   в   эксперименте.
Качественно всё происходит так же, как и в диэлектрическом слое.

     Пустой канал  в неограниченном  диэлектрике  в  отличие  от
диэлектрического стержня в пустоте не является направляющей линией
для цилиндрических волн.  Формально это следует из-за отсутствия
решений у уравнения  (12.26)  при  $\varepsilon\leqslant 1$.
Физическое  же различие в условиях распространения волны вдоль
диэлектрического стержня и канала в  диэлектрике  объясняется
явлением  полного   отражения, имеющим место лишь при
распространении  падающей  волны  в  оптически более плотной
среде. Отсутствие цилиндрических волн в канале означает, что
диполь,  помещённый внутрь  него,  возбудит  только  сферическую
волну,  поле  которой спадает при удалении от источника значительно быстрее,
чем у цилиндрической.

     Картина, однако,  резко  меняется,  если  диэлектрик,  в  котором
проделан канал, помещён внутрь проводящей металлической трубы. Такая
структура  представляет  собой обычный волновод,  частично заполненный
диэлектриком,  и в ней уже появляется система цилиндрических  волн,  в
том  числе и медленных.  В частности,  такая структура  может быть использована для
ускорения заряженных частиц, поскольку в медленной волне на оси канала
имеется  продольная  составляющая  электрического поля.  Отметим,  что
пустой канал принципиально важен для ускорения частиц.

     При однородном   заполнении   волновода  диэлектриком  продольное
волновое число $h=\sqrt{\varepsilon  k^2  -  g^2}$,  так  что
фазовая скорость
$v_{\mbox{\footnotesize\textit{ф}}}=c/\sqrt{\varepsilon-(g/k)^2}$
может  быть сколь угодно снижена путем выбора диэлектрика с
достаточно  большим  $\varepsilon$. Замедление  волны  в  этом
случае никак не связано с явлением полного отражения,  а
обусловлено тем,  что при достаточно большом $\varepsilon$ обычное
замедление  плоской  волны  в  неограниченной однородной среде с
$\varepsilon\mu>1$ преобладает над характерным  для пустого
волновода  фазовым убыстрением волны.  При наличии приосевого
пустого канала  в  диэлектрике  полное   отражение  на его границе
играет отрицательную роль.  Помимо уменьшения замедления волны ее
поверхностный характер внутри канала приводит к  быстрому спаданию
продольного электрического поля на оси. Удовлетворительные
результаты могут быть получены только при достаточно малом радиусе
канала. В этом случае,  используя малость этого  параметра,
удаётся получить для основной электрической волны $E_{01}$
аналитические  оценки  полей  и её замедления.

     Рассмотрим кратко характеристики симметричной электрической волны
$E_{01}$   в   предположении   идеальной  проводимости  стенок  трубы,
соосности трубы и канала и малости радиуса канала $a$ по  сравнению  с
радиусом   трубы   $b$:  $a\ll  b$.  Заметим,  что  в  рассматриваемой
структуре,  как и в диэлектрическом  стержне,  симметричные  волны  не
являются   гибридными,   а  распадаются  на  систему  электрических  и
магнитных волн.

     Решение волнового   уравнения  (12.23)   для компоненты электрического
вектора  Герца  $\Pi^e_z$ в канале и диэлектрике целесообразно искать в виде
     $$\hspace*{-0.2cm}\Pi^e_z=A I_0(pr)\qquad\qquad\qquad\quad\;\mbox
         {при $r<a$},\eqno(12.30)$$
     $$\Pi^e_z=B J_0(gr)+ C N_0(gr)\qquad\mbox{при $r>a$},
         \eqno(12.31)$$
где величины  $p$  и  $g$  определены  теми же выражениями,  что и для
диэлектрического стержня.  Решение внутри канала ищется в виде функции
$I_0(pr)$,   так   как   другое   решение  уравнения  $K_0(pr)$  имеет
особенность при $r=0$.  Компоненты поля в  обеих  областях  выражаются
через    $\Pi^e_z$   общими   формулами   (7.12),   (7.13)   и~(7.17).
Характеристическое   уравнение   выводится    стандартным    способом,
неоднократно   применявшимся   выше   ---   из  граничных  условий  на
тангенциальные   составляющие   поля   получается   система   линейных
однородных   уравнений  относительно  коэффициентов  $A$,~$B$,~$C$.  В
данном случае граничные  условия  требуют  равенства  нулю  компоненты
$E_z$  при  $r=b$ и непрерывности составляющих $E_z$ и~$H_\varphi$ при
$r=a$.  Условие наличия у однородной  системы  нетривиального  решения
приводит к характеристическому уравнению
     $$pa\frac{I_0(pa)}{I_1(pa)}=\frac{ga}{\varepsilon}\frac{J_0(ga)
         N_0(gb)-J_0(gb)N_0(ga)}{J_1(ga)N_0(gb)-J_0(gb)N_1(ga)}\;.
         \eqno(12.32)$$

     Решения этого уравнения очевидны в двух предельных случаях: $a=0$
(волновод с однородным диэлектрическим заполнением)  и  $a=b$  (пустой
волновод).  При  произвольном  значении  $a$  можно  получить  решение
уравнения (12.32)  описанным  выше  графическим  методом,  однако  для
использования  рассматриваемой структуры в качестве ускоряющей интерес
представляют узкие каналы:  $a\ll b$.  В этом  случае  нетрудно  найти
аналитическое решение разложением по малому параметру $a/b$. Для волны
$E_{01}$ в нулевом приближении  $g=\nu_{01}/b$  $(\nu_{01}=2,4048\dots
)$  и  поэтому  величины  $ga$ и $pa$ --- малые.  Используя разложения
функций Бесселя при малых значениях аргументов
     $$J_0(x)\approx1-\frac{x^2} 4,\quad J_1(x)\approx \frac x2,\quad
         N_0(x)\approx-\frac 2 {\pi}\ln{\frac 2 {1.7811x}},\quad N_1(x)
         \approx-\frac 2{\pi x},\eqno(12.33)$$
запишем уравнение (12.32) в первом порядке  по  малому  параметру:
     $$ J_0(gb)\Bigl[\frac {4\varepsilon}{\pi ga}-ga\frac 2 \pi\ln{
         \frac2{1,7811 ga}}\Bigr]=ga N_0(gb)(1-\varepsilon)\;.
         \eqno(12.34)$$

     Величину $gb$ удобно искать в виде
     $$gb\approx \nu_{01} (1+\delta)\qquad (\delta\ll 1);\eqno(12.35)$$
тогда $J_0(gb)\approx  -\delta\nu_{01}  J_1(\nu_{01})\quad   (J_1(\nu_
{01})=  0,5192\dots)  $.  Поскольку правая часть уравнения (12.34) уже
пропорциональна малой величине $ga$,  то функцию $N_0(gb)$ достаточно
взять    в   нулевом   приближении:   $N_0(gb)\approx   N_0(\nu_{01})=
2/{\pi\nu_{01} J_1(\nu_{01})}$. В результате получается, что
     $$\delta=\frac 1{2 J_1^2(\nu_{01})}\left (1- \frac 1{\varepsilon}
         \right )\left(\frac a b\right)^2\;.\eqno(12.36)$$

     По найденной  величине  $\delta$  легко  находятся  все  основные
характеристики волны,  которые,  как и следовало  ожидать,  при  малых
значениях  $a/b$  лишь  незначительно  отличаются  от  соответствующих
характеристик при однородном заполнении. Так, фазовая скорость волны
     $$v_{\mbox{\footnotesize\textit{ф}}}=
          \frac c{\sqrt{\varepsilon\mu-\displaystyle{\left(\frac
         {\nu_{01}}{kb}\right)^2}(1+2\delta)}}\;,\eqno(12.37)$$
а её критическая частота
     $$k_{\mbox{\footnotesize\textit{кр}}}=
          \frac{\nu_{01}}{b\sqrt{\varepsilon\mu}}(1+\delta)\;.
         \eqno(12.38)$$

\begin{wrapfigure}[16]{l}{8.5cm}
\begin{picture}(70,60)
\put(0,65){\special{em:graph fig12-5.bmp}}
\end{picture}
\hbox to 8.5cm{\hfil\footnotesize{Рис.~12.5.~Диаграмма Бриллюэна
для волновода,}\hfil}
\hbox to 8.5cm{\hfil\footnotesize{ заполненного диэлектриком с каналом.}\hfil}
\end{wrapfigure}

     Из (12.37) следует, что $v_{\mbox{\footnotesize\textit{ф}}}<c$ при
     $$k>\frac{\nu_{01}}{b\sqrt{\varepsilon\mu-1}}(1+\delta)\;,
         \eqno(12.39)$$
а само  волновое число $k$ и продольное волновое число $h$ при $\mu=1$
связаны соотношением
     $$(kb)^2=\frac 1 \varepsilon [(hb)^2+\nu_{01}^2(1+\delta)^2],
         \eqno(12.40)$$
на основании  которого  построена диаграмма Бриллюэна,  приведенная на
рис.~12.5.  Штриховкой указана область параметров,  в которой возможно
ускорение заряженных частиц.

     В заключение   этого   раздела   отметим,  что  решение  задач  с
диэлектрическими волноводами начинается,  как правило, с предположения
о  структуре поля в каждой отдельной области пространства с однородным
заполнением и написания явных выражений  для  него  через  специальные
функции  и  неопределённые коэффициенты;  затем из граничных условий
устанавливается  однородная  система  линейных  уравнений   для   этих
коэффициентов  и  из  условия  наличия  у неё нетривиального решения
выводится  характеристическое  уравнение.  Этот  подход  типичен   для
решения  задач  о  свободных  колебаниях  и соответствует тому случаю,
когда отсутствует общая теория,  сводящая задачу о полях любых волн  к
двумерному уравнению Гельмгольца.
%\end{document}

\newpage
\oddsidemargin=-0.4mm \evensidemargin=-0.4mm
\topmargin=-0.4mm
\headsep=7mm
\textheight=231.875mm
\textwidth=160mm
\mathsurround=2.5pt
\unitlength=1mm
%\begin{document}
%\input{macr.tex}
\thispagestyle{empty}
%\addtocounter{page}{134}

\begin{center}\subsubsection*{13. Медленные   волны   в  периодических
         структурах}\end{center}
\vspace*{0.5cm}

\markboth{Глава 4.    Медленные    волны}{13.    Медленные   волны   в
         периодических структурах}

\begin{center}\begin{minipage}[c]{0.75\textwidth}
\footnotesize{\parindent=0.5cm
         Трансляционная симметрия периодических структур.  Набег  фазы
         на  периоде --- основная характеристика волны в периодических
         структурах. Волна как совокупность пространственных гармоник.
         Простейшие примеры периодических структур.  Медленная волна в
         открытой гребенчатой  структуре.  Метод  частичных  областей.
         Простейшее решение в приближении частой гребенки. Импедансные
         граничные условия.  Волны вдоль ребристого круглого  стержня.
         Спиральная замедляющая линия.
}\end{minipage}\end{center}\vspace*{0.5cm}

     В предыдущем разделе были рассмотрены однородные (вдоль оси  $z$)
и  неоднородные  в  поперечной  плоскости  структуры,  в которых
могут распространяться  медленные  цилиндрические  волны
($v_{\mbox{\footnotesize\textit{ф}}}<c$).   Такие гладкие
структуры   обязательно   включают  в  себя  диэлектрические
материалы  или  проводники  с  конечной   проводимостью,   которые
в высокочастотных   полях  можно  считать  диэлектриками  с
комплексной диэлектрической проницаемостью.  Все медленные волны в
примыкающей  к поверхности  диэлектрика пустоте являются
поверхностными --- их поле в направлении нормали к поверхности
спадает по экспоненциальному  закону и  не  носит  волнового
характера.  Только  при  этом условии фазовая скорость волны в
пустоте может быть меньше $c$.

     В ряде  приборов  СВЧ,  в  которых  требуется  замедление  волны,
гладкие структуры не находят применения из-за присущих им
недостатков. В  частности,  поскольку  фазовая  скорость  волны в
гладкой структуре всегда находится в диапазоне
$c/\sqrt{\varepsilon\mu}< v_{\mbox{\footnotesize\textit{ф}}}<c$,
то для таких   нерелятивистских  устройств,  как  лампа  бегущей
волны, где требуется очень большое замедление,  невозможно
подобрать материал  с подходящим   значением   $\varepsilon$.   В
мощных   приборах   СВЧ, использующих сильноточные электронные
пучки,   прохождение  пучка вблизи поверхности диэлектрика
вызывает накопление на ней  рассеянного заряда, что приводит к
искажению поля и пробоям.

     От этих   недостатков  свободны  замедляющие  структуры,  которые
состоят только из металлических  элементов  и  в  которых
собственные медленные волны можно изучать в приближении идеальной
проводимости. Такие структуры не могут  быть  однородными  в
направлении  распространения волны    ---    вдоль    гладкой
идеально   проводящей   поверхности распространение медленных волн
в  пустоте  невозможно:  в  замкнутых структурах (металлические
волноводы) $v_{\mbox{\footnotesize\textit{ф}}}>c$,  в  открытых
$v_{\mbox{\footnotesize\textit{ф}}}=c$. Эффект замедления может
быть достигнут  только  за  счёт усложнения формы металлической
поверхности.  Наибольший интерес, во всяком случае для
теоретического    рассмотрения, представляют     бесконечные
периодические  структуры,  обладающие трансляционной симметрией:
при смещении вдоль оси на период $D$ структура  совпадает  сама  с
собой. Простейшими  примерами периодических структур,  о которых
ниже пойдет речь,  служат: плоская гребенчатая  структура  или
гофра  (рис.~13.1), круглый ребристый  стержень (рис.~13.2) и
диафрагмированный волновод, который рассматривается в следующем
разделе.

     Цилиндрические волны в  том  виде,  как  они  были  определены  в
разделе  7,  в  периодических  структурах  распространяться  не  могут
(напомним,  что в цилиндрической волне зависимость полей от координаты
вдоль  оси  распространения  даётся  множителем $e^{ihz}$, и условием
её существования является однородность структуры вдоль этой оси).
Вследствие трансляционной симметрии периодической структуры комплексные
амплитуды  полей   её   собственной   волны   должны   удовлетворять
определённому  соотношению,  которое,  например,  для  вектора Герца
может быть записано в виде
     $$ \rv \Pi (x,y,z+nD)=e^{in\Psi}\rv \Pi(x,y,z)\,,\qquad  n=1,2\dots,
         \eqno(13.1)$$
где параметр  $\Psi$  представляет собой набег фазы волны на периоде и
является основной характеристикой волны в периодической структуре. Как
и в любом гармоническом волновом процессе, фаза определена с точностью
до  значения,  кратного  $2\pi$.  В  случае  периодической   структуры
диапазон изменения фазы целесообразно определить неравенствами
     $$-\pi<\Psi<\pi\;.\eqno(13.2)$$
Тогда при положительных значениях  $\Psi$  естественно  считать  волну
распространяющейся  вдоль  оси  $z$,  при отрицательных --- в обратном
направлении.

     Традиционно, по аналогии с обычной цилиндрической  волной  (7.1),
собственную  волну  периодической  структуры  принято  характеризовать
продольным волновым числом $h_0$, определяемым соотношением
     $$ h_0 D=\Psi\,,\eqno(13.3) $$
что позволяет записать вектор Герца волны в виде
     $$\rv\Pi(x,y,z)=\rv\Pi^0(x,y,z)\,e^{ih_0z}\,,\eqno(13.4)$$
где $\rv\Pi^0(x,y,z)$  ---  периодическая функция $z$;  действительно,
     $$\rv\Pi^0(x,y,z+D)=\rv\Pi(x,y,z+D)\,e^{-ih_0(z+D)}=\rv\Pi(x,y,z)\,e^
         {-ih_0z}=\rv\Pi^0(x,y,z) \,.\eqno(13.5)$$
Периодическую функцию  можно разложить в ряд Фурье; тогда
     $$\rv\Pi^0(x,y,z)=\sum_{n=-\infty}^\infty\rv\Pi_n(x,y)\,e^{2\pi inz
         /D},\eqno(13.6)$$
что позволяет представить вектор Герца волны в виде
     $$\rv\Pi(x,y,z)=\sum_{n=-\infty}^{\infty}\rv\Pi_n(x,y)\, e^{ih_nz},
         \qquad h_n=h_0+\frac{2\pi n}D\,.\eqno(13.7)$$
Формулы (13.1)  и  (13.7) являются прямым следствием теоремы Флоке для
дифференциальных уравнений с периодическими коэффициентами.  В  теории
ускорителей  эти формулы принято называть {\it обобщённой теоремой Флоке},
в  физике твёрдого тела --- теоремой Блоха.

     Представление собственной  волны в периодической структуре в виде
(13.7) называется разложением  на  {\it  пространственные
гармоники}. Каждая  такая  гармоника является обычной
цилиндрической волной (7.1), но при этом всегда надо иметь в виду,
что отдельные гармоники не могут сами  по себе существовать в
структуре:  их поле  не удовлетворяет либо уравнениям Максвелла,
либо граничным условиям, либо и тому, и другому. Реальностью
является   лишь   совокупность   всех  гармоник  ---  их
 амплитуды жёстко связаны между собой  и определены
с точностью  до  общего  множителя,  который  и характеризует
амплитуду волны.  Возможность  разложения   на гармоники
свидетельствует   о неопределённости   понятия фазовой  скорости
для  собственных  волн периодической структуры: каждая  из
гармоник  имеет  свою  {\it фазовую} скорость  и,  более того,  у
разных гармоник эти скорости могут быть направлены в
противоположные стороны.  Говоря о фазовой скорости волны в этом
случае,  обычно подразумевают величину $v_{\mbox{\footnotesize\textit{ф}}}=
ch_0/k=c\Psi/kD$
--- фазовую скорость основной (нулевой)  гармоники.  Другое  дело  ---
групповая   скорость  волны,  которая  одинакова  для  всех  гармоник,
поскольку все производные $\partial \omega / \partial h_n$ равны между
собой.   Интересной  особенностью  собственных  волн  в  периодических
структурах  является  возможность  несовпадения  у  некоторых  из  них
направлений фазовой и групповой скоростей волны.  Такие волны называют
{\it  обратными},  они  находят  важное  практическое  применение,   в
частности,  в  мощном  генераторе  СВЧ ---  так называемой лампе обратной
волны или карсинотроне.

     Начнём рассмотрение  с  простейшей открытой  периодической  структуры     ---
плоской  гребёнки,  обозначение  всех  размеров которой   ясно  из  рис.~13.1.
 Для  вывода характеристического уравнения, определяющего
собственные волны гребёнки,  воспользуемся одним из основных методов
решения электродинамических  задач  для  структур,  обладающих сложной
геометрией, --- {\it методом частичных областей}.  Этот метод особенно
эффективен в тех случаях, когда все ограничивающие исследуемый объём
поверхности совпадают с  частью  одной  из  координатных  поверхностей
такой  системы координат,  в которой разделяются переменные в волновом
уравнении. Тогда весь объём структуры может быть разбит на некоторое
число  сравнительно  простых  областей,  в  каждой  из которых решение
волнового уравнения может быть найдено методом разделения переменных с
точностью до какого-то числа коэффициентов.

\begin{wrapfigure}[13]{l}{7.5cm}
\begin{picture}(80,50)
\put(-1,50){\special{em:graph fig13-1.bmp}}
\end{picture}
\hbox to    7.5cm{\hfil\footnotesize{Рис.~13.1.~Плоская   гребенчатая
структура.
}\hfil}
\end{wrapfigure}
     Неопределённость, присущая   данному методу,  обусловлена  тем,
что какие-то из границ этих   частичных областей
являются вспомогательными поверхностями,
проходящими внутри структуры и разделяющими соседние  области.  На
этих  вспомогательных  границах производится <<сшивание>>  решений,
полученных в каждой отдельной простой области.     Сшивание
состоит  в приравнивании
тангенциальных  составляющих  полей, в результате  чего  получаются пары
функциональных  уравнений.   Переход от этих уравнений  к   системе
линейных  однородных алгебраических  уравнений  для  постоянных
коэффициентов    осуществляется  путем  переразложения собственных
функций  одной  частичной  области по собственным функциям соседней.

     Само разбиение    структуры   на   частичные   области   является
неоднозначной процедурой. В случае рассматриваемых далее периодических
структур  основными  конкурирующими  способами  разбиения  на  области
являются два.  В первом способе  частичные  области  выбираются  таким
образом,  чтобы  вспомогательная  граница  между  областями  проходила
параллельно  направлению  распространения   волны,   во   втором   ---
перпендикулярно этому направлению.  Каждый из способов разбиения имеет
свои достоинства и  недостатки.  При рассмотрении  плоской  гребенки  воспользуемся
первым  способом  разбиения,  а  для  диафрагмированного волновода ---
вторым.

     При выводе  характеристического  уравнения  гребёнки достаточно
рассмотреть  один  период  структуры  $0<z<D$.  Областью  1  назовём
пространство  над  гребенкой  ($x>0$),  областью  2  --- прямоугольную
область,   ограниченную   плоскостями   $x=0$,~$x=-a$,   ~$z=0$,~$z=d$
(см.~рис.~13.1).   Во  избежание  излишней  громоздкости  всех  формул
ограничимся рассмотрением   электрических  волн,  у которых в
данной  геометрии  структуры у всех компонент поля отсутствует
зависимость  от  координаты  $y$ .  В  этих  волнах  отличны  от   нуля
составляющие поля $E_z$,~$E_x$,~$H_y$. Поле в области 1 будем искать в
виде совокупности   медленных   поверхностных   волн   (12.5).   Тогда
единственная  отличная от нуля $z$-компонента вектора Герца может быть
с учётом общих свойств волн в периодической структуре представлена в
виде
     $$\Pi_z^e(x,z)=-e^{i\Psi z/D}\sum_{n=-\infty}^{\infty}\frac {C_n}
         {p_n}\,e^{-p_nx+i2\pi n z/D}\;,\eqno(13.8)$$
где $p_n^2=(\Psi+2\pi   n)^2/D^2-k^2$.   Поскольку   все   $p_n$   при
достаточно  низких  частотах  растут почти $\sim n$,  то на расстоянии
порядка $D$ от  поверхности  гребёнки  в  поле  представлена  только
основная   (нулевая)   гармоника.   Фактически  происходит  усреднение
мелкомасштабных вариаций поля по периоду структуры.

     Согласно общим  формулам (3.33) и (3.34) из представления вектора
Герца (13.8) получаем следующие выражения для тангенциальных компонент
поля,  необходимых  для  сшивания решений в областях 1 и 2 в плоскости
$x=0$:
     $$E_z^{(1)}=e^{i\Psi z/D}\sum_{n=-\infty}^{\infty}p_nC_n\,e^{-p_nx+
         i2\pi n z/D}\;,\eqno(13.9)$$
     $$H_y^{(1)}=ik e^{i\Psi z/D}\sum_{n=-\infty}^{\infty}C_n\,e^{-p_nx+
         i2\pi n z/D}\;.\eqno(13.10)$$

     Поле в  области  2  представим  в виде совокупности электрических
волн плоского волновода,  распространяющихся вдоль оси $x$. С учётом
обращения  в  нуль  тангенциальной составляющей электрического поля на
стенке $x=-a$ имеем:
     $$E_z^{(2)}=\sum_{n=0}^{\infty}\tilde h_nA_n\cos{n\pi \frac z d}\;
         \sin \tilde h_n(x+a)\;,\eqno(13.11)$$
     $$H_y^{(2)}=ik\sum_{n=0}^{\infty} A_n\cos{n\pi \frac z d}\;\cos
         \tilde h_n(x+a)\;,\eqno(13.12)$$
где $\tilde h_n^2=k^2-(\pi  n/d)^2$.  Граничные  условия  в  плоскости
$x=0$ записываются как
     $$E_z^{(1)}=\left\{\begin{array}{l}E_z^{(2)}\qquad\mbox{при}
         \quad0<z<d\,,\\[.1cm]\phantom{m}0\phantom{,}\qquad\mbox{при}
         \quad d<z<D\,,\end{array}\right.\eqno(13.13)$$
     $$\hspace{-0.2cm}H_y^{(1)}=H_y^{(2)}\quad\qquad\mbox{при}\quad
         0<z<d\;\eqno(13.14)$$
и тогда с помощью формул (13.9)--(13.12) уравнения (13.13)  и  (13.14)
приводятся к следующему виду:
     $$e^{i\Psi z/D}\sum_{n=-\infty}^{\infty}C_n p_n \,e^{i2\pi n z/D}=
         \left \{\begin{array}{l}\displaystyle{\sum_{n=0}^{\infty}
         \tilde{h_n}A_n\cos{n\pi\frac z d}}\;\sin{\tilde{h_n}a}\,,\quad 0<z<d,\\
         [.6cm] 0 \;,\qquad d<z<D,\end{array}\right .\eqno(13.15)$$
     $$\hspace{-0.8cm}e^{i\Psi z/D}\sum_{n=-\infty}^{\infty}C_n \,e^{i2
         \pi n z/D}=\displaystyle{\sum_{n=0}^{\infty} A_n\cos{n\pi\frac
          z d}}\;\cos{\tilde{h_n} a}\;.\eqno(13.16)$$

  Переход от  этих  функциональных  уравнений  к  системе
линейных  алгебраических  производится  следующим  образом.  Уравнение
(13.15)  домножается  на  $\displaystyle{e^{-i(\Psi+2\pi  m)  z/D}}$ и
интегрируется в пределах от 0 до $D$ (в правой части  уравнения  (13.15)
интегрирование  производится  фактически  в
пределах от 0 до  $d$).  Уравнение  (13.16)  домножается  на
$\displaystyle{\cos{m\pi  z/d}}$  и  интегрируется  в пределах от 0 до
$d$. В результате получаем:
     $$Dp_mC_m=\sum_{n=0}^{\infty}\tilde{h_n}A_n I_{mn}^{(-)}\,\sin
         {\tilde h_na},\qquad m=0,\pm 1,\pm 2,\dots\;,\eqno(13.17)$$
     $$\frac d 2 (1+\delta_{0m}) A_m\cos{\tilde h_m a}=\sum_{n=-
         \infty}^{\infty} C_n I_{mn}^{(+)},\qquad m=0, 1, 2,\dots\;,
         \eqno(13.18)$$
где
     $$\displaystyle{I_{mn}^{(\pm)}}=\mp \,iD\frac{\Psi+2\pi m}{(\pi n
         D/d)^2-(\Psi+2\pi m)^2}\left [1-(-1)^n \,\displaystyle{e^{\pm i
         (\Psi+2\pi m)d/D}}\right ]\;.\eqno(13.19)$$
Исключая из уравнений (13.17) и (13.18) коэффициенты  $A_n$,  получаем
бесконечную  систему  линейных  алгебраических  уравнений относительно
коэффициентов $C_n$:
     $$p_mC_m-\sum_{n=-\infty}^{\infty}\alpha_{mn}C_n=0\,,\qquad m=0,
         \pm 1,\pm 2,\dots\;,\eqno(13.20)$$
где
     $$\alpha_{mn} = \frac 2 {dD}\sum_{s=0}^{\infty} \frac  { I_{ms}^{(-)}\,I_{ns}^{(+)}}{1+\delta _{0s}}
     \,\tilde h_s\tg{\tilde h_s a}\,.\eqno(13.21)$$

     Однородная система  уравнений (13.20) имеет нетривиальное решение
только в случае обращения в нуль  детерминанта  матрицы  с  элементами
$p_m\delta_{mn}-\alpha_{mn}$.       Это       условие       определяет
характеристическое  уравнение  гребёнки,  из  которого  для  каждого
заданного  набега  фазы  $\Psi$  находится  набор  дискретных значений
частот,  соответствующих   собственным   медленным   волнам.   Каждому
собственному значению $k$ (являющемуся функцией $\Psi$) соответствует
вектор решения однородной системы, составленный из коэффициентов $C_n$
и  определяющий  поле  волны с точностью до общего множителя,  который
естественно принять за амплитуду собственной волны.

     Представляется очевидным,  что получить точное характеристическое
уравнение  гребёнки в замкнутом аналитическом виде с помощью системы
(13.20)  невозможно  ---  для  этого  необходимо  было  бы   вычислить
детерминант соответствующей бесконечной матрицы и приравнять его нулю.
Поэтому  собственные  частоты  для  заданного $\Psi$ приходится искать
численными методами,  редуцируя  систему  (13.20),  то  есть  оставляя
конечное число $N$ уравнений и неизвестных коэффициентов.  Для каждого
выбранного набора значений параметров гребёнки ($a$,~$d$,~$D$)  даже
при  оставлении  нескольких уравнений в системе приходится проделывать
большой  объём  трудоёмких  численных  расчётов.  При  некоторых
соотношениях  между параметрами гребёнки коэффициенты $C_n$ медленно
убывают с ростом номера и для получения приемлемой  точности  параметр
редукции  $N$  приходится  выбирать  очень  большим.  Для установления
достаточно общих  закономерностей  требуемый  объём  вычислений  при
расчётах вручную становится практически нереализуемым. Поэтому ещё
несколько десятилетий назад изложенный подход к задаче представлялся в
значительной степени формальным. В настоящее время в связи с развитием
вычислительной  техники  обращение   матрицы   размерностью $100\times   100$   или
вычисление  её детерминанта  является  тривиальной  операцией и все
дисперсионные  характеристики  гребёнки  для  произвольного   набора
значений  её  параметров  легко  и  быстро могут быть получены путем
решения системы (13.20). Отметим, что такое решение, не будучи, строго
говоря,  аналитическим,  всё  же  сохраняет в значительной степени
физическую картину явления,  чем выгодно отличается от  прямолинейного
численного решения методом сеток.

     Не останавливаясь   на    результатах    численных    расчётов,
рассмотрим,   при  каких  значениях  параметров  могут  быть  получены
приближенные аналитические решения системы (13.20) и какой  они  имеют
физический  смысл.  Самым  грубым  приближением,  сохраняющим,  тем не
менее,  целый  ряд  существенных  характеристик  гребёнки,  является
пренебрежение в системе (13.20) всеми коэффициентами,  кроме $C_0$,  и
всеми уравнениями,  кроме соответствующего $m=0$.  При этом в сумме  в
правой  части  формулы  (13.21)  тоже  учитывается  только  один член,
соответствующий $s=0$.  Такое  приближение  равносильно  представлению
поля в канавках гребёнки в виде суммы $TEM$ волн плоского волновода,
а поля выше гребёнки --- в виде (12.5),  и тем  самым  соответствует
полному пренебрежению всеми мелкомасштабными эффектами, обусловленными
периодичностью структуры. Если ввести параметр
     $$\alpha=\frac{\Psi d}{2D}=\frac{h_0d} 2\;,\eqno(13.22)$$
определяющий набег фазы волны на полуширине канавки, то
     $$\displaystyle{I_{00}^{(\pm)}=d\;e^{\pm i\alpha}\;\frac{\sin{
         \alpha}}{\alpha}}\;.\eqno(13.23)$$
Учитывая, что  $\tilde  h_0=k$,  и   переходя   согласно   (13.13)   к
традиционно  используемому  в этом подходе продольному волновому числу
$h_0$, получим:
     $$\sqrt{h_0^2-k^2}=\displaystyle{\frac d D\left (\frac {\sin{
         \alpha}}{\alpha}\right )^2\;k\tg{ka}}\;.\eqno(13.24)$$

     Пренебрежение эффектами,      обусловленными     пространственной
периодичностью поля,  имеет смысл  только  при  условии  $h_0D\ll  1$,
обеспечивающим  малость  набега  фазы  на  периоде структуры.  Поэтому
множитель в скобках мало отличается от  единицы  и  уравнение  (13.24)
сводится к виду
     $$\sqrt{h_0^2-k^2}=\displaystyle{\frac d D\;k\tg{ka}}\;.
         \eqno(13.25)$$
В случае   очень  тонких  рёбер  гребёнки  $d\approx  D$  и  (13.25)
переходит в уравнение,  часто использумое в  литературе  для  описания
дисперсии гребёнки:
     $$\sqrt{h_0^2-k^2}=k\tg{ka}\;.\eqno(13.26)$$
Отсюда непосредственно  следуют  выражения  для  продольного волнового
числа и фазовой скорости  (которая  в  этом  приближении  определяется
однозначно) как функции частоты:
     $$h_0=\frac k {\cos{ka}}\;,\qquad
      v_{\mbox{\footnotesize\textit{ф}}}=c\;\cos{ka}\;.\eqno(13.27)$$

     Для значений  $ka>\pi/2$  уравнение  (13.26)  решений  не имеет:
правая часть становится отрицательной,  а левая всегда положительна из
условия  затухания  поля  над  гребёнкой  (при  $ka>\pi$  решения  у
уравнения  опять  появляются,  но  они  интереса  для  приложений   не
представляют).  При  приближении $ka$ к $\pi/2$ фазовая скорость волны
становится всё меньше, и волна всё сильнее  поджимается  к  границе
гребёнки.  Однако  при  этом  нарушается условие малости набега фазы
волны на периоде структуры  $h_0D\ll1$,  поскольку  $h_0\to\infty$.  В
результате  длина  волны  $\Lambda$  (пространственный  период  поля в
области над гребёнкой) становится меньше  пространственного  периода
структуры  и  учёт  высших гармоник как в поле над гребёнкой,  так и
внутри канавок, становится обязательным. Результаты численного решения
системы  (13.20)  показывают,  что   волна перестаёт существовать при более низкой
 частоте,  а замедление волны оказывается  меньшим, чем это следует из
 уравнения (13.26). При этом замедление волны тем меньше и тем
раньше (по  частоте)  она перестаёт существовать,  чем  меньше  отношение
глубины канавки $a$ к периоду структуры $D$.

     Уравнение (13.26),  справедливое  лишь  при  условии  $h_0D\ll1$,
может  быть  получено  не  только  методом сшивания решений в соседних
областях,  а и более простым,  широко используемым методом,  в  основе
которого лежат так называемые импедансные граничные условия. Напомним,
что ранее (раздел  6)  использовались  импедансные  граничные  условия
Щукина-Леонтовича (6.17),  позволяющие решать многие задачи только для
внешней области.  Физическое обоснование этих  условий  заключается  в
том,  что  поле  в металле практически не зависит от структуры поля во
внешней области.  Это обусловлено очень высоким значением  комплексной
диэлектрической   проницаемости   проводников:   любая   падающая   на
поверхность плоская волна  (а  любое  поле  может  быть  разложено  по
системе  таких  волн)  независимо  от  угла падения распространяется в
металле почти по нормали к его поверхности  в  виде  плоской  волны,  и
соотношение  между  электрическим и магнитным полем в этой волне равно
волновому  сопротивлению   металла,   которое   является   комплексной
величиной  (волновое  сопротивление может быть и действительным,  если
рассматривается граница раздела двух  сред,  для  которых  справедливо
неравенство      $\varepsilon_2\mu_2\gg\varepsilon_1\mu_1$).     Ввиду
непрерывности тангенциальных  составляющих  поля  на  границе  раздела
компоненты   поля   во  внешней  среде  связаны  между  собой  тем  же
соотношением.
\begin{wrapfigure}[14]{l}{8.5cm}
\begin{picture}(80,55)
\put(0,55){\special{em:graph fig13-2.bmp}}
\end{picture}
\hbox to 7.5cm{\hfil\footnotesize{Рис.~13.2.~Круглый ребристый
стержень.}\hfil}
\end{wrapfigure}
     Если поле   в   канавках,   которые   можно   рассматривать   как
короткозамкнутые  отрезки  плоского  волновода,  представляется в виде
совокупности только $TEM$ волн (прямой и обратной), то соотношение между
магнитным  и электрическим полем в плоскости $x=0$ также не зависит от
координаты $z$ и от вида поля в области $x>0$.  Можно считать,  что  в
плоскости $x=0$ выполняются граничные условия
     $$E_z=\zeta_1 H_y\;,\quad E_y=-\zeta_2 H_z\;,\eqno(13.28)$$
причём поверхностный   импеданс   является  анизотропным.  Поскольку в  первом
приближении   $\zeta_1=-id/D\tg{ka}$,~$\zeta_2=0$, то  анизотропия
обусловлена поляризацией $TEM$ волны:  электрическое поле имеет только
нормальную  к  плоскости  ребра  составляющую.  Импедансные  граничные
условия позволяют получить приближённые аналитические решения целого
ряда задач для периодических структур,  которые  при  строгом  подходе
требуют численных методов.

     Основные результаты,  полученные для  плоской  гребёнки,  легко
обобщаются   на   симметричные   электрические волны,
распространяющиеся  вдоль  круглого   ребристого  стержня,  иначе
называемого гофрированным цилиндром (рис.~13.2).  При
выводе  характеристического  уравнения  в  этом   случае   ограничимся
импедансным  приближением.  Поле  волны во внешнем пространстве $r>a$,
имеющее   составляющие   $E_z$,~$E_r$    и~$H_\varphi$,    описывается
потенциальной  функцией  $\Pi^e_z(r)e^{ih_0z}$,  которая удовлетворяет
уравнению
     $$\frac{d^2\Pi_z^e}{dr^2}+\frac 1 r\frac{d\Pi_z^e}{dr}-p^2\Pi^e_
         z=0\;,\eqno(13.29)$$
где $p^2=h_0^2-k^2  $.  Убывающее  на  бесконечности   решение   этого
уравнения   есть  $\Pi^e_z(r)=CK_0(pr)$;  другое  линейно  независимое
решение --- $I_0(pr)$ --- на бесконечности неограниченно возрастает  и
поэтому  нам не подходит.  Согласно общим формулам (7.12) и (7.17) для
тангенциальных составляющих поля на поверхности $r=a$ имеем:
     $$E_z=-p^2CK_0(pa)\,e^{ih_0z},\qquad H_\varphi=-ikpCK_1(pa)\,e^{ih_0z}
         \;.\eqno(13.30)$$
Отношение этих величин  должно  быть  равно  поверхностному  импедансу
$\zeta$ структуры:
     $$\frac{pK_0(pa)}{ikK_1(pa)}=\zeta\;.\eqno(13.31)$$

     Для вычисления поверхностного импеданса при $r=a$ представим поле
в  области  между  ребрами  стержня   в   виде   симметричной   волны,
тангенциальная  компонента  электрического  поля которой не зависит от
$z$.  Эта волна является аналогом $TEM$  волны  в  плоском  волноводе,
имеет  две  компоненты  поля  $E_z$ и $H_\varphi$ и е\"е потенциальная
функция $\Pi_z^{(e)}$ удовлетворяет уравнению
     $$\frac{d^2\Pi_z^e}{dr^2}+\frac 1 r\frac{d\Pi_z^e}{dr}+k^2\Pi^e_z
         =0\;,\eqno(13.32)$$
общее решение которого
     $$\Pi_z^e=AJ_0(kr)+BN_0(kr)\;.\eqno(13.33)$$
Отличные от нуля составляющие поля согласно формулам (7.12) и (7.17) в
данном случае связаны с $\Pi_z^e$ следующими соотношениями:
     $$E_z=k^2\Pi_z^e\;,\qquad H_\varphi=ik\frac{d\Pi_z^e}{dr}\;
         \eqno(13.34)$$
и с  учётом граничного условия $E_z=0$ при $r=b$ могут быть записаны
в виде
     $$\hspace{-0.6cm}E_z=k^2\tilde A[J_0(kr)N_0(kb)-N_0(kr)J_0(kb)]\;,
         \eqno(13.35)$$
     $$H_\varphi=-ik^2\tilde A[J_1(kr)N_0(kb)-N_1(kr)J_0(kb)]\;.
         \eqno(13.36)$$

     Отношение величин   (13.35),   (13.36)  при  $r=a$  и  определяет
поверхностный  импеданс  $\zeta$,  подстановка  которого   в   (13.31)
приводит к характеристическому уравнению
     $$p\frac{K_0(pa)}{K_1(pa)}=-k\frac{J_0(ka)N_0(kb)-N_0(ka)J_0
         (kb)}{J_1(ka)N_0(kb)-N_1(ka)J_0(kb)}\;;\eqno(13.37)$$
при $a-b\ll b $ уравнение заметно упрощается:
     $$p\frac{K_0(pa)}{K_1(pa)}=-k\tg{k(a-b)}\;.\eqno(13.38)$$
Всякое действительное  положительное  решение  $p(k)$  этих  уравнений
определяет  медленную  поверхностную  волну,  распространяющуюся вдоль
стержня.  Характер поведения правой и левой части уравнений (13.37)  и
(13.38)  такой  же,  как  у  уравнения (13.26) для плоской гребёнки.
Поэтому и  области  значений параметров,   при  которых  могут
распространяться  медленные  волны, качественно одинаковы.

     Ребристые стержни находят применение в качестве антенн метровых и
дециметровых  волн;   правда,   для   этого   используется   излучение
несимметричной волны, уравнение для которой имеет более громоздкий вид
и здесь не приводится.  Попутно отметим,  что  симметричные  магнитные
волны в подобных структурах не существуют, так как у них $E_z\equiv0$, и
гофрированный цилиндр подобен цилиндру из идеального металла.

     Импедансные граничные  условия  позволяют  получить  сравнительно
просто   характеристическое   уравнение   медленных  волн  и  в  такой
замедляющей структуре, как спиральный волновод (или спиральная линия),
представляющий собой винтовую намотку металлическим ленточным проводом
на цилиндрическую поверхность (рис.~13.3{\it а}).  Такая линия применяется в
приборах типа лампы бегущей волны. Самое простое физическое объяснение
замедления электромагнитной волны состоит в следующем.  Вдоль  прямого
идеально  проводящего  провода  волна  бежит  со  скоростью  $c$, и это
свойство в значительной  степени  сохраняется  и  для  искривлённого
провода;   скорость  же  волны  вдоль  оси  винтовой  линии  при  этом
естественно  оказывается  заметно  меньшей.  Она   (в   самом   грубом
приближении) определяется чисто геометрическим фактором,  а именно ---
углом  намотки  $\theta$,  связанным  с  шагом  намотки  $D$   (период
структуры)  и  радиусом  цилиндра $a$ соотношением $\tg{\theta}=D/2\pi
a$; при этом очевидно, что
     $$v_{\mbox{\footnotesize\textit{ф}}}=c\,\sin{\theta}\,.\eqno(13.39)$$
Согласно этой  формуле  фазовая скорость не зависит от частоты.  Более
точные расчёты показывают,  что дисперсия скорости всё же имеет место, хотя
и   невелика.   Простейший   приближённый   способ   получения
характеристического  уравнения  для   спиральной   линии   состоит   в
использовании импедансных граничных условий.

\begin{picture}(160,55)
\put(0,50){\special{em:graph  fig13-3a.bmp}}
\put(80,50){\special{em:graph fig13-3b.bmp}}
\end{picture}
\begin{center}\begin{minipage}[c]{0.9\textwidth}
\footnotesize{\begin{center}
Рис.~13.3.~Спиральный волновод {\it а)} и решётка из  металлических  лент
{\it б)}.
\end{center}}\end{minipage}\end{center}\vspace*{0.25cm}

     Для вывода  этих условий рассмотрим плоскую решётку,  состоящую
из узких  параллельных  металлических  лент  (рис.13.3{\it б}).   Из   опыта
известно,  что  если  на  такую решётку падает волна с электрическим
полем,  поляризованным  вдоль  проводников,  то  она  почти  полностью
отражается  от  решётки;  если  же поле поляризовано перпендикулярно
проводникам, то волна проходит сквозь неё почти без отражения. Этому
легко   дать   простое  объяснение:  в  первом  случае  волна  наводит
значительные токи вдоль лент,  поле которых складывается  с  падающим,
результатом  чего  является  наблюдаемый эффект (именно интерференцией
падающего поля и поля наведённых токов объясняется полное  отражение
волны  от  идеально  проводящей  плоскости).  В поперечном направлении
из-за узости лент наводится слабый ток, возбуждаемое им поле мало, и в
результате  волна проходит сквозь решётку практически без искажений.
Если направление вдоль проводников  обозначить  вектором  $\rv  s$,  а
направление, перпендикулярное к ним, вектором $\rv t$, то решётка из
проводящих  лент  может  быть   описана   приближёнными   граничными
условиями
     $$E^{(1)}_s=E^{(2)}_s=0\,, \qquad E_t^{(1)}=E_t^{2}\,,
               \qquad H_s^{(1)}=H_s^{(2)}\,,  \eqno(13.40)$$
где индексами (1) и (2)  обозначены  поля  по  обе  стороны  плоскости
решётки.  Условие  на магнитное поле следует из отсутствия на лентах
поперечных токов.  В случае падения плоской волны эти условия приводят
к  полному отражению и полному прохождению волны в зависимости от её
поляризации.  Очевидно,  что они не учитывают мелкомасштабные вариации
поля,  сопоставимые с периодом ленточной решётки,  и, следовательно,
применимы только при условии $kD \ll 1$,  которое позволяет  показать,
что  в примыкающей к решётке области с размерами порядка длины волны
можно пренебречь в волновом  уравнении  членом  с  $k^2$.  Тогда  поля
определяются  уравнением  Лапласа  с  обычными граничными условиями на
металле, из решения которого и следуют приближённые условия (13.40).
Эти  условия  можно уточнить,  решая электростатическую задачу методом
конформных отображений,  но они становятся слишком громоздкими и  здесь
не приводятся.

     Рассмотрим теперь кратко вывод характеристического уравнения  для
спиральной  линии  на основе граничных условий (13.40).  В этом случае
спиральную  линию  можно  рассматривать  как   круглый   металлический
волновод радиуса винтовой спирали с анизотропной проводимостью стенок.
Проводимость  в  направлении,  определяемом  углом  намотки  $\theta$,
идеальная,  в поперечном направлении --- равна нулю. Из-за анизотропии
проводимости в структуре могут  существовать  медленные  поверхностные
волны, поле которых частично сосредоточено и вне спирали.

     Все такие   волны,   даже    простейшая    симметричная    волна,
рассматриваемая ниже, являются гибридными, и их поля определяются двумя
потенциальными  функциями.  Для   симметричной   волны   эти   функции
$\Pi_z^e=\Pi^e(r)e^{ih_0z}$        и       $\Pi_z^m=\Pi^m(r)e^{ih_0z}$
удовлетворяют одному и тому же уравнению  вне и внутри спирали:
     $$\frac{d^2\Pi^{e,m}}{dr^2}+\frac 1 r \frac{d\Pi^{e,m}}{dr}-p^2
         \Pi^{e,m}=0\;,\eqno(13.41)$$
где $p^2=h_0^2-k^2>0$.  Из общих формул  (7.12),  (7.13)  и  (7.17)  в
случае   азимутальной   симметрии  поля  для  интересующих  нас  далее
компонент поля следуют выражения:
     $$\left.\begin{array}{l}E_\varphi=-ik\displaystyle{\frac{d\Pi^m}
         {dr}}\,e^{ih_0z}\;,\qquad H_\varphi=ik\displaystyle{\frac{d\Pi^e}
         {dr}}\,e^{ih_0z}\;,\\[.5 cm]E_z=-p^2\Pi^e \,e^{ih_0z}\;,\qquad H_z
         =-p^2\Pi^m \,e^{ih_0z}\;.\end{array}\right \}\eqno(13.42)$$

     Решения уравнения  (13.41)  в  области вне спирали ($r>a$) должны
удовлетворять требованию затухания поля  на  бесконечности,  а  внутри
спирали   ($r<a$)  ---  условию  конечности  поля  при  $r=0$.  Такими
решениями являются
     $$\left.\begin{array}{l}\Pi^e=AI_0(pr)\;,\qquad\quad\Pi^m=BI_0
         (pr)\qquad \mbox{при $r\leqslant a$}\;,\\[.3cm]\Pi^e=CK_0(pr)\;,
         \;\qquad \Pi^m=DK_0(pr)\qquad \mbox{при $r\geqslant a$}\;;\end
         {array}\right \}\eqno(13.43)$$
исходя из   них,  с   помощью  формул  (13.42)  легко  вычислить
тангенциальные  компоненты   поля   по   обе   стороны
проводящей      поверхности      $r=a$.      Компоненты     поля
$E_s$,~$E_t$,~$H_s$,  входящие  в  приближённые  граничные   условия
(13.40), связаны с ними очевидными соотношениями
     $$E_s=E_\varphi \cos{\theta}+E_z\sin{\theta},\quad H_s=H_\varphi
         \cos{\theta}+H_z\sin{\theta},\quad E_t=-E_\varphi \sin{\theta}
         +E_z\cos{\theta}.\eqno(13.44)$$
Таким образом,  граничные условия (13.40) приводят к системе четырёх
линейных    алгебраических    уравнений    относительно     постоянных
$A$,~$B$,~$C$,~$D$,  условие  наличия нетривиального решения которой и
определяет  характеристическое  уравнение  медленной  волны.   Опуская
несложные выкладки, запишем окончательный результат:
     $$p^2=k^2\ctg^2{\theta}\frac {I_1(pa)K_1(pa)}{I_0(pa)K_0
         (pa)}\,.\eqno(13.45)$$

     Нетрудно найти решение этого уравнения при $pa\gg 1$.  В  этом
случае дробь в правой части (13.45) обращается в единицу, так что
     $$p=k\ctg{\theta}\;,\qquad h_0=k/\sin{\theta}\;,
           \qquad v_{\mbox{\footnotesize\textit{ф}}}=
           c\sin{\theta}\;,\eqno(13.46)$$
и, следовательно, выполняется и условие  $ka\gg\tg{\theta}$. Поэтому
при   достаточно   высоких   частотах
оказывается  справедливым элементарное решение (13.39).  Отметим,  что
область  частот,  для  которых  справедливо   приближённое   решение
(13.46),  ограничена  и  сверху  условием малости периода структуры по
сравнению  с  длиной  волны   $\Lambda=2\pi/h_0$,   что   сводится   к
неравенству  $ka\ll 1$.  В реальных приборах,  таких как лампа бегущей
волны,  желательно работать при  достаточно  высоких  значениях  $pa$,
чтобы  обеспечить лишь небольшое провисание поля на оси,  где проходит
электронный поток;  вместе с тем при больших $pa$ поле в  значительной
степени  проникает  в  область  вне  спирали,  что приводит к излишним
потерям и паразитным взаимодействиям.

     В пределе  низких  частот фазовая скорость растёт и стремится к
$c$, что нетрудно видеть из уравнения (13.45), считая $pa\ll 1$.
Тогда дробь   в   правой   части   (13.45)  будет  малой
величиной  порядка $1/\ln{pa}$, так что $pa\ll ka\ctg{\theta}$.
При достаточно малом $ka$ величина $pa$ также мала,
следовательно,  $h\to k$ и $v_{\mbox{\footnotesize\textit{ф}}}\to
c$.  Для промежуточных значений частоты фазовая скорость  с
удовлетворительной для практики точностью может быть найдена с
помощью численного решения уравнения (13.45).

%\end{document}
Забегая вперед,  отметим ещё, что при заданном значении
набега  фазы  $\Psi$  каждая собственная волна имеет свою определённую
частоту.  Все собственные частоты данной волны  лежат  в  ограниченном
диапазоне,  называемом  полосой  пропускания  для данной волны.  Число
собственных  волн  определённой  симметрии  в  замнутой  периодической
структуре  неограничено,  в  открытой структуре существует обычно лишь
одна волна. Полосы пропускания собственных волн в замкнутых структурах
типа   диафрагмированного   волновода   для   нижних   типов  волн  не
перекрываются между собой.  В результате в таком волноводе  существуют
полосы частот, в которых не может распространяться ни одна собственная
волна структуры --- их называют полосами непропускания.

\newpage
\oddsidemargin=-0.4mm \evensidemargin=-0.4mm
\topmargin=-0.4mm
\headsep=7mm
\textheight=231.875mm
\textwidth=160mm
\mathsurround=2.5pt
\unitlength=1mm
%\begin{document}
%\input{macr.tex}
\thispagestyle{empty}
%\addtocounter{page}{146}

\begin{center}\subsubsection*{14. Диафрагмированный волновод}
     \end{center}\vspace*{0.5cm}

\markboth{Глава 4.~Медленные волны}{14.~Диафрагмированный волновод}

\begin{center}\begin{minipage}[c]{0.75\textwidth}
\footnotesize{\parindent=0.5cm
         Расчёт характеристик  диафрагмированного  волновода методом
         частичных областей.  Система собственных волн и  их  основные
         свойства. Диаграмма Бриллюэна. Групповая скорость собственной
         волны.  Прямые и обратные волны. Разложение собственной волны
         на  сумму  цилиндрических волн (гармоник) в канале волновода.
         Приближённое  характеристическое  уравнение  для   основной
         гармоники.
}\end{minipage}\end{center}\vspace*{0.5cm}

     Диафрагмированный волновод,  изображённый  на
рис.~14.1,  можно  рассматривать  как  обычный
гладкий круглый волновод  радиуса $b$, к стенке
которого на равном   расстоянии    $D$ (период
структуры) прикреплены  плоские  перегородки   толщиной $D-d$
с   круглым  осесимметричным  отверстием радиуса   $a$
(диафрагмы).    Такой  волновод  представляет   большой  интерес   и в
теоретическом  отношении  как  классическая замкнутая  (в  поперечных
направлениях)  периодическая структура,  обладающая  полной системой
собственных  волн,  так и в практическом плане --- как основной
ускоряющий элемент линейных электронных  ускорителей.

\begin{wrapfigure}[15]{l}{7.5cm}
\begin{picture}(80,55)
\put(4,50){\special{em:graph fig14-1.bmp}}
\end{picture}
\hbox to 7.5cm{\hfil\footnotesize{Рис.~14.1.~Круглый диафрагмированный
}\hfil}
\hbox to 7.5cm{\hfil\footnotesize{ волновод.}\hfil}
\end{wrapfigure}

     В предположении идеальной проводимости металлических поверхностей
в диафрагмированном волноводе, как и в гладком, могут распространяться
незатухающие собственные волны.  Однако системы собственных волн  этих
волноводов  существенно  различаются  между  собой.  Основное  отличие
состоит не столько в более сложном пространственном распределении поля
волны в диафрагмированном волноводе, а в наличии у каждой волны помимо
нижней критической частоты (как в гладком волноводе), ещё и верхней.
В  результате всякая собственная волна может распространяться только в
ограниченной  полосе  частот.  Между  полосами  распространения   двух
соседних  (по  шкале  частот)  собственных  волн  может быть зазор,  в
результате чего волновод обладает свойствами полосового фильтра.

     Электромагнитные поля    всей    системы   собственных   волн   в
диафрагмированном   волноводе   определяются   двумя    потенциальными
функциями     ---     векторами    Герца    с    одной    составляющей
$\Pi_z^{e,m}(r,\varphi,z)$.  Из-за  азимутальной  симметрии  структуры
система  волн распадается на сумму независимых гармоник,  поля которых
пропорциональны $e^{ip\varphi}$.  Потенциальные функции таких гармоник
удовлетворяют одинаковому волновому уравнению
     $$\frac{\partial^2 \Pi^{e,m}}{ \partial r^2}+\frac 1 r
        \frac{\partial \Pi^{e,m}}{\partial r}+
        \frac{\partial^2 \Pi^{e,m}}{\partial
         z^2}+\biggl(k^2-\frac {p^2}{r^2}\biggr)\Pi^{e,m}=0\;.
         \eqno(14.1)$$
Несимметричные волны $(p\ne 0)$ являются  гибридными  и  содержат  все
шесть  компонент электромагнитного поля,  которые вычисляются по общим
формулам  (3.35).

     Система симметричных   волн   $(p=0)$   распадается   на    волны
электрические, описываемые потенциалом $\Pi^e(r,z)$, и волны магнитные
---   потенциал   $\Pi^m(r,z)$.   Ниже   будет   рассмотрена   система
электрических  волн,  представляющих  особый  интерес для практических
приложений,   поскольку   только   у   них   продольная   составляющая
электрического  поля  $E_z$ отлична от нуля на оси.  Компоненты поля в
этом  случае  выражаются  через   потенциальную   функцию   следующими
соотношениями:
     $$E_r=\frac{\partial^2\Pi^e}{\partial r\partial z}\;,\qquad E_z=
         \left (\frac{\partial^2}{\partial z^2}+k^2 \right )\Pi^e\;,
         \qquad H_\varphi=ik \frac{\partial \Pi^e}{ \partial r}\,;
         \eqno (14.2)$$
остальные составляющие  поля  равны  нулю.  Важнейшей  характеристикой
собственной волны во всякой  периодической  структуре  является  набег
фазы  на  периоде  $\Psi$,  который  связывает значение полей в данной
точке структуры и точке, отстоящей на период:
     $$\rv E(r,z+D)=\rv E(r,z)\,e^{i\Psi}\;,\qquad\rv H(r,z+D)=\rv H
         (r,z)\,e^{i\Psi}\;.\eqno(14.3)$$

     Из-за сложности граничной поверхности структуры уравнение  (14.1)
приходится  решать  методом  частичных областей.  Ввиду трансляционной
симметрии достаточно  рассмотреть  один  период  структуры,  например,
определяемый   условием  $-d/2<z<D-d/2$.  Используем  здесь  разбиение
периода на простые области граничной плоскостью,  перпендикулярной оси
волновода.  При таком разбиении (в отличие от использованного выше для
гребенчатой структуры) поле в приосевой области уже не  представляется
набором  цилиндрических волн (гармоник),  имеющих единое представление
для всех значений координаты $z$.  Обе  выделенные  в  рассматриваемом
периоде  структуры  области  представляют  собой  отрезки  регулярного
круглого   волновода:    область~1    определяется    неравенствами
$0<r<b$,~$-d/2<z<d/2$,   область~ 2  ---  неравенствами  $0<r<a$,~$d/2
<z<D-d/2$.  В силу полноты системы собственных волн гладкого  круглого
волновода  произвольное  электромагнитное  поле  (в  том  числе и поле
собственной волны диафрагмированного волновода) в каждой  из  областей
может быть представлено в виде суммы
     $$\Pi^e_i(r,z)=\sum_{q=1}^{\infty}\frac{J_0(\nu_{0q}r/R_i)}
         {\nu_{0q}J_1(\nu_{0q})}\left [C_{qi}\,e^{ih_{qi}z}+C_{-qi}\,e^
         {-ih_{qi}z}\right ]\;,\eqno(14.4)$$
где индекс  $i=1,2$  соответствует номеру области,  $R_1=b$,~$R_2=a$,~
$C_{\pm q1}= B_{\pm q}$,~$C_{\pm q2}=A_{\pm  q}$,~  $h_{qi}=\sqrt{k^2-
(\nu_{0q}/R_i)^2}$.

     При записи  решения  в  виде (14.4) автоматически удовлетворяются
граничные  условия  на  боковых   стенках обеих областей:
$E_z^{(1)}(b,z)=0$,~$E_z^ {(2)}(a,z)=0$, где здесь и далее в формулах (14.5)--(14.7)
верхний индекс у компонент поля соответствует номеру частичной области.
За счёт выбора коэффициентов  $A_{\pm  q}$  и~$B_{\pm q}$   необходимо
дополнительно  удовлетворить  условию идеальной проводимости
стенок диафрагм
     $$E_r^{(1)}(r,-d/2)=E_r^{(1)}(r,d/2)=0 \qquad \mbox{при}\qquad
         a<r<b\;,\eqno(14.5)$$
условию непрерывности тангенциальных составляющих поля на общей границе
($z=d/2,\,0<r<a$) частичных  областей
     $$E_r^{(1)}(r,d/2)=E_r^{(2)}(r,d/2),\qquad H_\varphi^{(1)}(r,d/2)
         =H_\varphi^{(2)}(r,d/2)\qquad \mbox{при~$0<r<a$}\;,
         \eqno(14.6)$$
а также обеспечить требуемый набег  фазы на периоде:
     $$\left.\begin{array}{l} E_r^{(2)}(r,D-d/2)=
         e^{i\Psi} E_r^{(1)}(r,-d/2)\,,\\[.4cm]
         H_\varphi^{(2)}(r,D-d/2)=e^{i\Psi}H_\varphi^{(1)}(r,-d/2)\,.
         \end{array}\right\}\qquad\mbox{при}\qquad 0<r<a\;.
         \eqno(14.7)$$

     Выражая поля  через  потенциал  (14.4),  получим   функциональные
уравнения;  в  частности,  из  условий  (14.5)  и (14.6) на радиальную
составляющую электрического поля следует, что
     $$\frac a b\sum_{q=1}^\infty\frac{J_1(\nu_{0q}r/b)}{J_1(\nu_{0q}
         )}h_{q1}\left [B_qe^{ih_{q1}d/2}-B_{-q}e^{-ih_{q1}d/2}\right
         ]=\\[.4cm]$$
     \vspace{-1.0cm}$$\eqno(14.8)$$\vspace{-0.75cm}
     $$\hspace{1cm} =\left\{\begin{array}{ll}0&\mbox{при~$a<r<b\;;$}
         \\\displaystyle{\sum_{q=1}^\infty\frac{J_1(\nu_{0q}r/a)}{J_1
         (\nu_{0q})}}h_{q2}\left [A_q e^{ih_{q2}d/2}-A_{-q}e^{-ih_{q2}
         d/2}\right ]&\mbox{при~$0<r<a\;.$}\end{array}\right.$$

Из условия (14.6) на магнитное поле получаем:
     $$\frac a b\sum_{q=1}^\infty \frac{J_1(\nu_{0q}r/b)}{J_1(\nu_{0q
         })}\left [B_qe^{ih_{q1}d/2}+B_{-q}e^{-ih_{q1}d/2}\right ]=\\
         [.4cm]$$
     \vspace{-1.0cm}$$\eqno(14.9)$$\vspace{-0.75cm}
     $$=\displaystyle{\sum_{q=1}^\infty\frac {J_1(\nu_{0q}r/a)}{J_1(
         \nu_{0q})}}\left [A_q e^{ih_{q2}d/2}+A_{-q}e^{-ih_{q2}d/2}
         \right ]\qquad\mbox{при}\qquad 0<r<a\;.$$
Аналогично записываются и фазовые соотношения (14.7):
     $$\frac a be^{i\Psi}\sum_{q=1}^\infty \frac{J_1(\nu_{0q}r/b)}{J_1
         (\nu_{0q})}h_{q1}\left [B_qe^{-ih_{q1}d/2}-B_{-q}e^{ih_{q1}d
         /2}\right ]=\\[.4cm]$$
     \vspace{-1.0cm}$$\eqno(14.10)$$\vspace{-0.75cm}
     $$=\left \{\begin{array}{ll} 0&\mbox{при~$a<r<b\;,$}\\
         \displaystyle{\sum_{q=1}^\infty\frac {J_1(\nu_{0q}r/a)}{J_1(
         \nu_{0q})}}h_{q2}\left [A_q e^{ih_{q2}(D-d/2)}-A_{-q}e^{-ih_
         {q2}(D-d/2)}\right ]&\mbox{при~$0<r<a$}\;;\end{array}\right.
         \;$$
\vspace{.3cm}
     $$\sum_{q=1}^\infty\frac {J_1(\nu_{0q}r/a)}{J_1(\nu_{0q})}\left
         [A_q \,e^{ih_{q2}(D-d/2)}+A_{-q}\,e^{-ih_{q2}(D-d/2)}\right ]=\\
         [.4cm]$$
     $$=\frac a b\, e^{i\Psi}\sum_{q=1}^\infty \frac{J_1(\nu_{0q}r/b)}
         {J_1(\nu_{0q})}\left [B_q\,e^{-ih_{q1}d/2}+B_{-q}\,e^{ih_{q1}d/2
         }\right ]\qquad\mbox{при~$0<r<a$}\;.\eqno(14.11)$$

     Переход от  функциональных  уравнений  к   бесконечным   системам
линейных  алгебраических уравнений производится путем умножения (14.8)
и (14.10) на $J_1(\nu_{0m}r/b)$ и интегрирования с весом $r$ от  0  до
$b$,  а  (14.9) и (14.11) --- на $J_1(\nu_{0m}r/a)$ и интегрирования с
тем же весом от 0 до $a$. Используя известные интегралы
     $$\int\limits_0^R\;J_1(\nu_{0n}r/R)\,J_1(\nu_{0m}r/R)\,r\,dr=
         \displaystyle{\frac {R^2}2 J_1^2(\nu_{0n})}\,\delta_{mn}
         \;,\eqno(14.12)$$
     $$\int\limits_0^a\,J_1(\nu_{0n}r/a)\,J_1(\nu_{0m}r/b)\,r\,dr=
         \frac {a^3} b \frac {\nu_{0m} J_1(\nu_{0n})J_0(\nu_{0m}a/b)}
         {\nu_{0n}^2-(\nu_{0m} a/b)^2}\;,\eqno(14.13)$$
получаем из (14.8), что
     $$h_{m1}\left [B_me^{ih_{m1}d/2}-B_{-m}e^{-ih_{m1}d/2}\right ]=$$
     \vspace{-1.0cm}$$\eqno(14.14)$$\vspace{-0.75cm}
     $$=2\frac{a^2}{b^2}\frac{\nu_{0m}J_0(\nu_{0m}a/b)}{J_1(\nu_{0m})}
         \sum^\infty_{q=1}\frac{h_{q2}}{\nu_{0q}^2-(\nu_{0m}a/b)^2}
         \left[A_qe^{ih_{q2}d/2}-A_{-q}e^{-ih_{q2}d/2}\right ]\;.$$
\vspace{.2cm}

     Аналогичным образом из (14.9)--(14.11) следуют соотношения
     $$A_me^{ih_{m2}d/2}+A_{-m}e^{-ih_{m2}d/2}=$$
     \vspace{-1.0cm}$$\eqno(14.15)$$\vspace{-0.75cm}
     $$=2\frac{a^2}{b^2}\sum_{q=1}^\infty\frac{\nu_{0q}J_0(\nu_{0q}a/b
         )}{J_1(\nu_{0q})}\frac 1{\nu_{0m}^2-(\nu_{0q}a/b)^2}\left[B_q
         e^{ih_{q1}d/2}+B_{-q}e^{-ih_{q1}d/2}\right ]\;,$$
\vspace{.3cm}
     $$h_{m1}e^{i\Psi}\left [B_me^{-ih_{m1}d/2}-B_{-m}e^{ih_{m1}d/2}
         \right ]=$$
     \vspace{-1.0cm}$$\eqno(14.16)$$
     \vspace{-0.75cm}
     $$=2\frac{a^2}{b^2}\frac{\nu_{0m}J_0(\nu_{0m}a/b)}{J_1(\nu_{0m})}
         \sum^\infty_{q=1}\frac{h_{q2}}{\nu_{0q}^2-(\nu_{0m}a/b)^2}
         \left[A_q\,e^{ih_{q2}(D-d/2)}-A_{-q}\,e^{-ih_{q2}(D-d/2)}\right]
         \;,$$
\vspace{.3cm}
     $$A_m\,e^{ih_{m2}(D-d/2)}+A_{-m}\,e^{-ih_{m2}(D-d/2)}=$$
     \vspace{-1.0cm}$$\eqno(14.17)$$\vspace{-0.75cm}
     $$\vspace{.2cm}=2\frac{a^2}{b^2}e^{i\Psi}\sum_{q=1}^\infty\frac
         {\nu_{0q}J_0(\nu_{0q}a/b)}{J_1(\nu_{0q})}\frac 1{\nu_{0m}
         ^2-(\nu_{0q}a/b)^2}\left[B_q\,e^{-ih_{q1}d/2}+B_{-q}\,e^{ih_{q1}
         d/2}\right ]\;.$$
\vspace{.2cm}

     Каждое из   соотношений   (14.14)--(14.17)   представляет   собой
бесконечную     систему     уравнений     относительно     неизвестных
$A_q$,~$A_{-q}$,~$B_q$,~$B_{-q}$,  поскольку все они  выполняются  при
$m=1,2,\dots$.    Нетрудно    исключить    из    этой   системы   либо
$A_q$,~$A_{-q}$,  либо $B_q$,~$B_{-q}$.  Если выбрать  первый  вариант
(исключение   коэффициентов   $A_q$,~$A_{-q}$),   то   удобно   ввести
вспомогательные неизвестные
     $$\vspace{.2cm}D_q=(B_q+B_{-q})\cos{h_{q1}d/2}\,;
       \quad E_q=(B_q-B_{-q})\sin{h_{q1}d/2}\,.\eqno(14.18)$$
Путём несколько   громоздких   преобразований   из  (14.14)--(14.17)
получается бесконечная система линейных уравнений для этих величин:
     $$\left.\begin{array}{l} D_q h_{q1}\tg{h_{q1}d/2}\,\cos{\Psi/2}-
         E_q h_{q1}\ctg{h_{q1}d/2}\,\sin{\Psi/2}=\\[.2cm]=-
         \displaystyle{\sum_{m=1}^\infty}\alpha_{mq}[D_m\cos{\Psi/2}+
         E_m\sin{\Psi/2}]\;,\\[.4cm]D_q h_{q1}\tg{h_{q1}d/2}\,\sin
         {\Psi/2}+E_q h_{q1}\ctg{h_{q1}d/2}\,\cos{\Psi/2}=\\[.2cm]
         =\displaystyle{\sum_{m=1}^\infty}\beta_{mq}[D_m\sin{\Psi/2}-
         E_m\cos{\Psi/2}]\;,\\[.2cm]\end{array}\right \}\eqno(14.19)$$
где элементы  матриц  $\alpha_{mq}$  и $\beta_{mq}$ имеют сходный вид,
так что их удобно представить столбцом
     $$\left [\begin{array}{l}\alpha_{mq}\\\beta_{mq}\end{array}\right
         ]\;=\;\chi_m\chi_q\sum_{s=1}^{\infty}\left [\begin{array}{l}
         \tg[h_{s2}\displaystyle{(D-d)/2}]\\\ctg[h_{s2}\displaystyle{
         (D-d)/2}]\end{array}\right]\frac{h_{s2}}{[\nu_{0s}^2-(\nu_
         {0m}a/b)^2][\nu_{0s}^2-(\nu_{0q}a/b)^2]}\;,\eqno(14.20)$$
причём в   последнем   соотношении  для  сокращения  записи  введено
обозначение
     $$\chi_n=2\left(\frac a b\right)^2\frac{\nu_{0n}J_0(\nu_{0n}a/b)}
         {J_1(\nu_{0n})}\;.\eqno(14.21)$$
Систему уравнений  (14.19) удобно записать в матричном виде;  при этом
матрица состоит из четырёх симметричных блоков.

     Характеристическое уравнение   для   частот   собственных    волн
диафрагмированного  волновода  в функции набега фазы на периоде $\Psi$
представляет  собой  условие  обращения  в  нуль   детерминанта   этой
четырёхблочной матрицы:
     $$\vspace{.5 cm}\left |\begin{array}{ll} \displaystyle{\sin{\frac
         \Psi 2}}\;\;{\rv P}\quad &\phantom{-}\displaystyle{\cos{\frac
         \Psi 2}}\;\; {\rv Q}\\[.4 cm]\displaystyle{\cos{\frac\Psi 2}}
         \;\; {\rv R}\quad &-\, \displaystyle{\sin{\frac \Psi
         2}}\;\;{\rv S}\end{array}\right |=0\;,\eqno(14.22)$$
где элементы матриц $\rv {P,\;Q,\;R,\;S}$ имеют вид
     $$\left. \begin{array}{lllll}P_{mq}\;&=\;&\delta_{mq}h_{q1}\tg(h_
         {q1}d/2)\;&-\;&\beta_{mq}\;,\\[.4 cm]Q_{mq}\;&=\;&\delta_{mq}
         h_{q1}\ctg(h_{q1}d/2)\;&+\;&\beta_{mq}\;,\\[.4 cm]R_{mq}\;&=
         \;&\delta_{mq}h_{q1}\tg(h_{q1}d/2)\;&+\;&\alpha_{mq}\;,\\
         [.4 cm]S_{mq}\;&=\;&\delta_{mq}h_{q1}\ctg(h_{q1}d/2)\;&-\;&
         \alpha_{mq}\;,\end{array} \right \}\qquad m,q=1,\,2,\,\dots
         \;.\eqno(14.23)$$

     Уравнение  (14.22)  определяет   для   данного
значения $\Psi$ дискретный бесконечный набор собственных частот $k_n$,
которые  можно  пронумеровать  в  порядке  их   возрастания.   Каждому
собственному   значению   $k_n$  соответствует  столбец  коэффициентов
$D_n$,~$E_n$,  который  представляет   собой   нетривиальное   решение
однородной  системы  линейных  алгебраических  уравнений.  Это решение
определено с точностью до произвольного общего  множителя,  который  и
определяет амплитуду возбуждаемой собственной волны. Уравнение (14.22)
содержит в себе детерминант бесконечномерной матрицы и  представляется
очевидным,  что  найти  его  решение  в  замкнутом  аналитическом виде
нельзя.  Единственный способ решения состоит в  редукции  матрицы,  то
есть  оставлении  квадратной  матрицы  размером  $N\times  N$  и  её
численном  обращении.  Такая  процедура   при   современном   развитии
вычислительной   техники   является   стандартной  и  не  представляет
серьёзных затруднений.  При этом необходимо убедиться,  что с ростом
$N$  решение  стремится  к  конечному  пределу,  который  и приходится
принимать за решение для бесконечномерной матрицы. Критерии сходимости
решения  для  бесконечной системы линейных уравнений довольно сложны и
выходят за рамки настоящего  изложения.  Укажем  только,  что  решение
уравнения    (14.22)    методом    редукции    сходится   к   решению,
удовлетворяющему  уравнению  (14.1)  и  всем  граничным  условиям  при
добавлении с ростом $N$ к общей матрице равного числа строк и столбцов
из всех её четырёх блоков.

\begin{wrapfigure}[18]{l}{8.2cm}
\begin{picture}(80,70)
\put(2,70){\special{em:graph fig14-2.bmp}}
\end{picture}
\hbox to 7.5cm{\hfil\footnotesize{Рис.~14.2.~Диаграмма Бриллюэна для }\hfil}
\hbox to 7.5cm{\hfil\footnotesize{диафрагмированного волновода.}\hfil}
\end{wrapfigure}

     Чтобы получить  общее   представление  о  системе  распространяющихся
собственных  волн диафрагмированного волновода,  необходимо
построить  диаграмму  Бриллюэна  на плоскости переменных $\Psi$,~$kD$.
Качественный характер этой диаграммы может быть выявлен в значительной
степени путем анализа структуры детерминанта в (14.22).  Очевидно,  что для
построения  кривых  $k_n(\Psi)$  достаточно  найти  решение уравнения
  (14.22)  только  в  интервале значений набега фазы   $0\leqslant \Psi\leqslant\pi$.
Значения функций $k_n(\Psi)$ в  крайних  точках этого интервала определяются
нулями  детерминантов матриц (14.23):  $k_n(0)$ --- матриц $\rv Q$ или
$\rv R$,  $k_n(\pi)$ --- матриц $\rv P$  или  $\rv  S$.  Из  структуры
матрицы  (14.22)  следует  чётность  функций  $k_n$,  что  позволяет
построить  участки  кривых  в   интервале   $-\pi<\Psi<0$   зеркальным
отражением относительно оси $\Psi=0$.  Исходя из периодичности функций
$k_n$,  кривые для произвольного интервала  значений  $\Psi$  строятся
путём  параллельного  переноса  кривых,  построенных  для  интервала
$-\pi\leqslant \Psi\leqslant\pi$,  вправо  или  влево.  Типичный  вид
диаграммы представлен на рис.~14.2, где построены три нижние кривые для
значений $\Psi>0$;  для  лучшего  понимания  свойств  собственных  волн
кривые следует мысленно продолжить на всю действительную ось $\Psi$.

     Каждая отдельная  непрерывная  периодическая кривая соответствует
одной собственной  волне.  Между  кривыми  имеются  полосы  частот,  в
пределах  которых  распространяющихся  волн  нет,  и  поэтому  отрезок
диафрагмированного  волновода  представляет  собой  хороший  полосовой
фильтр.    Участки   кривой,   изображённые   сплошной линией,  соответствуют
волне,  распространяющейся  в  положительном направлении оси  $z$,
штриховой  линией  ---  в  отрицательном. Говоря  о направлении
распространения волны,  следует иметь ввиду знак её групповой скорости
     $$v_{\mbox{\footnotesize\textit{гр}}}= cD\,\frac{dk_n}{d\Psi}\eqno(14.24)$$
(отношение $v_{\mbox{\footnotesize\textit{гр}}}/c $ совпадает с
тангенсом угла наклона  касательной к   кривым   на   диаграмме).
Отметим,   что   на   границах  полосы распространения каждой
собственной волны $v_{\mbox{\footnotesize\textit{гр}}}=0$.

     В отличие  от групповой скорости привычное для рассмотренных выше
в разделе~7 цилиндрических волн понятие фазовой скорости к собственной
волне  в диафрагмированном волноводе неприменимо и именно потому,  что
эта волна не является цилиндрической вида (7.1). Однако, как следствие
теоремы  Флоке  (13.7),  во всей приосевой области структуры вплоть до
кромок диафрагм потенциал $n$-той волны может быть представлен в  виде
{\it суммы} таких волн (гармоник):
     $$\Pi_n(r,z)=e^{i\Psi z/D}\sum_{m=-\infty}^\infty \Pi_{nm}(r)\,e^
         {i2\pi mz/D},\qquad 0<r<a\;,\eqno(14.25)$$
то есть каждой собственной волне ставится в  соответствие  бесконечный
набор  дискретных  значений  продольных  волновых чисел $h_m=\Psi/D\pm
2\pi m/D,\quad m=0,1,2,\dots$,  а, значит, и фазовых скоростей,
среди которых   есть   бесконечное   число   как  положительных,  так  и
отрицательных  значений.  Подчеркнём  ещё   раз,   что   отдельные
гармоники  из  (14.25)  в  структуре  распространяться  не могут,  они
возбуждаются все одновременно, причём соотношение амплитуд отдельных
гармоник  для  данной  собственной волны при заданных частоте и набеге
фазы всегда постоянно.

     С утратой   понятия   фазовой   скорости  для  собственной  волны
диафрагмированного  волновода  теряют  смысл  и  такие  понятия,   как
медленная  и  быстрая  волны;  они  сохраняются  только по отношению к
отдельным гармоникам. При этом высшие гармоники всегда медленные, а та
часть основной гармоники, которая соответствует малым значениям набега
фазы $\Psi$,  всегда быстрая;  чем больше номер собственной волны, тем
больше быстрых гармоник.  Поскольку в каждой собственной волне имеются
сколь угодно медленные гармоники,  то каждая собственная волна (точнее
какая-нибудь  из  её  гармоник)  оказывается  в  {\it синхронизме} с
равномерно движущейся вдоль оси структуры заряженной частицей при любой
её  скорости  $v$, и  само  понятие  синхронизма  для  периодической
структуры требует уточнения.  Частица  и  гармоника  синхронны  в  том
случае,  если  одну и ту же плоскость во всех ячейках частица проходит
при одном и том же значении поля этой гармоники.  Условие  синхронизма
определяется   точкой  пересечения  кривой  $k_n(\Psi)$  на  диаграмме
рис.~14.2 с прямой линией,  угол наклона $\theta$ которой к оси $\Psi$
определяется   соотношением   $\tg{\theta}=v/c$.  Если  в  этой  точке
групповая скорость окажется отрицательной,  то говорят, что возбуждена
{\it обратная волна}. На диаграмме {\it прямыми} оказались волны $k_1$
и $k_2$,  обратной --- волна $k_3$. Из построения очевидно, что одна и
та  же собственная волна на одной и той же частоте может оказаться как
прямой, так и обратной.

     Отметим ещё    два    свойства   собственных   волн   замкнутой
периодической структуры,  которые близки к аналогичным характеристикам
собственных   волн   гладкого   волновода  и  которые  приведём  без
доказательства:

     1. Средняя  по  времени  энергия  электрического поля собственной
волны,  приходящаяся на период структуры,  равна  средней  по  времени
энергии  магнитного  поля.  Подчеркнем,  что  отдельно  для выделенных
областей 1 и 2 (рис.~14.1) это утверждение несправедливо.

     2. Средний  по  времени поток энергии поля через любое поперечное
сечение структуры одинаков и равен средней энергии поля,  приходящейся
на её период, умноженной на групповую скорость и делённой на длину
периода.  Фактически  это  утверждение  можно   считать   определением
групповой скорости.  Поэтому при её вычислении на данной частоте нет
необходимости  в  соответствии   с   (14.24)   вычислять   производную
дисперсионной  кривой (для чего необходимо производить её расчёт в
окрестности  интересующей  частоты),  а  достаточно  вычислить   поток
энергии и среднюю энергию на период на одной частоте.  Напомним, что в
гладком  волноводе  оба  этих  утверждения  справедливы  для   отрезка
волновода произвольной длины.

     Для полной  характеристики  свойств  системы   собственных   волн
замкнутой  периодической  структуры  следует  сказать несколько слов о
её  полноте,  поскольку  по  этому  вопросу  в  ряде  руководств  по
электродинамике СВЧ имеются неточности. Произвольное монохроматическое
поле, удовлетворяющее уравнениям Максвелла и граничным условиям, может
быть  представлено  в  виде  совокупности  таких  волн.  Но не следует
ограничивать эту совокупность волнами,  представленными  на  диаграмме
Бриллюэна  (рис.~14.2).  В  нижней  части  спектра частот на выбранной
частоте существует  только  одна  (или  ни  одной)  распространяющаяся
волна.  Поле  произвольного  источника,  например,  монохроматического
диполя, не может быть представлено одной этой волной, хотя оно и очень
мало   отличается  от  неё  на  достаточно  больших  расстояниях  от
источника. Полное поле включает в себя и затухающие собственные волны,
свойства которых определяются рассмотренной системой уравнений, но при
мнимых значениях $\Psi$.  Впрочем,  это  замечание  в  равной  степени
относится и к системе собственных волн гладкого волновода.

     В случае часто расположенных диафрагм и  при  малом  набеге  фазы
($\Psi\ll  \pi$)  нетрудно  получить приближённое характеристическое
уравнение  для  основных   гармоник   собственных   волн   с   помощью
импедансного  граничного  условия  на поверхности $r=a$,  как это было
сделано в предыдущем разделе  для  ребристого  стержня.  Такой  подход
соответствует   пренебрежению   в  (14.25)  всеми  гармониками,  кроме
основной $(m=0)$.  При тех значениях переменных $\Psi$ и $k$, при которых
основная  гармоника  собственной  волны  является  медленной,  решение
волнового уравнения (13.29) в области вне диафрагм  ($r<a$)  ищется  в
виде
     $$\Pi^e(r)=CI_0(pr)\,,\eqno (14.26)$$
где $p=   \sqrt{(\Psi/D)^2   -k^2}$  --  положительная  действительная
величина.  Поверхностный импеданс вычисляется по той же схеме,  как и для
ребристого   стержня,  и  в  результате  характеристическое  уравнение
получается в следующем виде:
     $${p}a\frac{I_0({p}a)}{I_1({p}a)}=ka\frac{J_0(ka)N_0
         (kb)-N_0(ka)J_0(kb)}{J_1(ka)N_0(kb)-N_1(ka)J_0(kb)}\;.
         \eqno(14.27)$$
Предположения, которые были использованы при выводе  этого  уравнения,
ограничивают на плоскости переменных $\Psi$, $kD$ область, где решения
имеют физический  смысл, неравенствами  $kD\leqslant \Psi \ll \pi$. Из
анализа  уравнения  нетрудно  установить  те полосы частот,  в которых
решения  могут  иметь  место;  действительно,  левая  часть   является
монотонной  функцией  $pa$,  возрастающей от 2 до $\infty$,  и поэтому
решения могут лежать только в тех интервалах частот,  в которых правая
часть (14.26) больше 2. Из этого следует, что основные гармоники могут
быть медленными только у нескольких первых собственных волн.

     Поскольку фазовая скорость основной гармоники определяется формулой
     $$v_{\mbox{\footnotesize\textit{ф}}}=\frac c {\sqrt{1+p^2/k^2}}\,,\eqno(14.28)$$
то значительное  замедление  может быть достигнуто лишь при
достаточно больших значениях $p$. В этом случае в соответствии с
решением (14.26) продольная  составляющая  поля  $E_z$ существенно
<<провисает>>  на оси структуры  по  сравнению  со значением  вблизи
кромок  диафрагм   и, следовательно, при нерелятивистских скоростях
 эффективное взаимодействие возможно лишь с трубчатыми пучками
 заряженных  частиц. По   этой же  причине диафрагмированные
 волноводы  находят  широкое применение в линейных  ускорителях
 электронов,  где  уже  при малых энергиях скорость  частиц близка
 к $c$ и требуемые значения $p$ малы, и, наоборот, не используются
 на начальном этапе ускорения протонов.

%\end{document}

\newpage
\oddsidemargin=-0.4mm \evensidemargin=-0.4mm
\topmargin=-0.4mm
\headsep=7mm
\textheight=231.875mm
\textwidth=160mm
\mathsurround=2.5pt
\unitlength=1mm
%\begin{document}
%\input{macr.tex}
\thispagestyle{empty}
%\addtocounter{page}{155}

\begin{center}
   \subsubsection*{\rm Г\,Л\,А\,В\,А\, 5}
      \vspace{-1.15em}
      \line(6,0){160}\\
      \vspace{-1em}
      \line(6,0){160}
      \vspace{-1.15em}
   \subsubsection*{ОБЪЁМНЫЕ РЕЗОНАТОРЫ}
      \vspace{35mm}
   \subsubsection*{15. Резонаторы простых форм}
\end{center}\vspace*{0.5cm}

\markboth{Глава 5. Объёмные резонаторы} {15. Резонаторы простых форм}

\begin{center}\begin{minipage}[c]{0.75\textwidth}
\footnotesize{\parindent=0.5cm
         Волноводные резонаторы.  Собственные   частоты.   Круглый   и
         коаксиальный   волноводные  резонаторы.  Основные  колебания.
         Прямоугольный   резонатор.    Квазистатические    собственные
         колебания тороидальных резонаторов и их расчёт.
}\end{minipage}\end{center}\vspace*{0.5cm}

     Объёмным резонатором     называется     часть     пространства,
ограниченная  металлическими  стенками.  В  такой  структуре  возможны
собственные   колебания   электромагнитного   поля,   частоты  которых
определяются формой и размерами  полости.  Под  идеальным  резонатором
будем  понимать  резонатор,  у которого стенки полностью замкнуты (нет
отверстий),  проводимость  стенок  бесконечна,  а  полость   заполнена
веществом без потерь (в частности,  вакуумом). В идеальном резонаторе
имеется бесконечный дискретный набор собственных частот и  собственные
колебания  являются  незатухающими.  Мысленно  убедиться  в реальности
таких колебаний проще всего на примере так называемых {\it волноводных
резонаторов},  представляющих  собой  отрезок  волновода  произвольной
формы (постоянного) поперечного сечения,  ограниченного с обоих концов
торцевыми  металлическими  плоскими  стенками,  перпендикулярными  оси
волновода --- такие резонаторы называют ещё  {\it  цилиндрическими}.
Пусть  в  волноводе  распространяется какая-то собственная волна.  При
падении её на идеальную торцевую стенку возникнет отражённая волна
той  же амплитуды,  которая вместе с падающей образует стоячую волну с
узлами, отстоящими на пол длины волны в волноводе. Если вторая торцевая
стенка  окажется  в  плоскости  одного из этих узлов,  то она никак не
исказит поле в  объёме,  и,  следовательно,  в  резонаторе  возможно
собственное  колебание,  представляющее  собой  сумму двух волноводных
волн.  Из этого примера становится очевидным,  что собственная частота
зависит от размера резонатора.

     Ввиду сказанного выше поучительно начать исследование собственных
колебаний с  волноводных  резонаторов.  Будем  считать,  что  торцевые
стенки  волноводного  (или  цилиндрического)  резонатора расположены в
плоскостях $z=0$ и $z=d$,  на  которых  должны  выполняться  граничные
условия
     $$E_x=E_y=0\;,\eqno(15.1)$$
и пусть объём   резонатора    заполнен    однородным    веществом    с
проницаемостями $\varepsilon,\mu$.

     Система собственных   волн   волновода   распадается   на  два типа волн:
электрические и   магнитные.  Напомним,  что эта система является полной
(любое поле в пространстве,  свободном от источников, может  быть
представлено  в  виде  совокупности таких волн), поэтому никаких  других
собственных  колебаний,  отличных от  стоячих  волн волновода, в
волноводном резонаторе существовать не может.

     Для системы стоячих  волноводных  волн  (как  и  для  бегущих)  в
качестве  потенциальных  функций  удобно  взять  продольные компоненты
электрического  и  магнитного  векторов  Герца:   $\Pi^e_z(x,y,z)$   и
$\Pi^m_z(x,y,z)$.   При   выбранном   расположении  торцов  резонатора
зависимость  этих  функций  от   продольного   волнового   числа   $h$
определяется, соответственно, множителями $\cos{hz}$ и $\sin{hz}$:
     $$\Pi_z^e(x,y,z)=\Pi^e(x,y)\cos{hz}\,,\qquad \Pi^m_z(x,y,z)=\Pi^m
         (x,y)\sin{hz}\,.\eqno(15.2)$$
Выбор функций  обусловлен  граничными  условиями  (15.1)  в  плоскости
$z=0$.  Компоненты  электромагнитного  поля  в  соответствии  с общими
соотношениями (3.26) и (3.34) выражаются через  потенциальные  функции
(15.2) следующим образом:

{\it электрические колебания ---}
     $$\left.\begin{array}{llllll}E_x&=&-h\displaystyle{\frac
         {\partial\Pi^e}{\partial x}}\sin hz\,,\qquad &H_x&=&-ik
         \varepsilon\displaystyle{\frac{\partial\Pi^e}{\partial y}}
         \cos hz\,,\\[.5cm]E_y&=&-h\displaystyle{\frac{\partial\Pi^e}
         {\partial y}}\sin hz\,,\qquad &H_y&=&\phantom{-}ik\varepsilon
         \displaystyle{\frac{\partial\Pi^e}{\partial x}}\cos hz\,,
         \\[.5cm]E_z&=&\phantom{-}g^2\Pi^e\cos hz\,,\qquad &H_z&=&
         \phantom{-}0\;;\end{array}\right \}\eqno(15.3)$$

{\it магнитные колебания ---}
     $$\left.\begin{array}{llllll}E_x&=&\phantom{-}ik\mu\displaystyle
         {\frac{\partial\Pi^m}{\partial y}}\sin hz\,,\qquad &H_x&=&h
         \displaystyle{\frac{\partial\Pi^m}{\partial x}}\cos hz\,,
         \\[.5cm]E_y&=&-ik\mu\displaystyle{\frac{\partial\Pi^m}{
         \partial x}}\sin hz\,,\qquad &H_y&=&h\displaystyle{\frac{
         \partial\Pi^m}{\partial y}}\cos hz\,,\\[.5cm]E_z&=&\phantom{-}
         0\,,\qquad &H_z&=&g^2\Pi^m\sin hz\,.\end{array}\right \}
         \eqno(15.4)$$

Граничные  условия(15.1) в плоскости $z=d$ выполняются для обоих
типов волн при  $\sin{hd}=0$; поэтому  для собственных колебаний волноводного
резонатора всегда
     $$h=\frac{l\pi}d\,,\eqno(15.5)$$
где $l$ --- целое число,  равное числу вариаций поля  вдоль  $z$.  В случае
электрических  волн $l$ может принимать все значения,  начиная с нуля;
для магнитных волн $l=0$ не определяет колебания,  поскольку  согласно
(15.2), (15.4) все компоненты поля в этом случае равны нулю.  Отметим,
что  продольное  волновое  число  $h$  в волноводном резонаторе уже не
является  функцией  частоты,   как   в   волноводе,   а   определяется
соотношением  (15.5),  то есть только продольным размером резонатора.

     Собственные частоты  волноводного резонатора образуют бесконечный
дискретный набор и определяются общим соотношением
     $$g^2=k^2\varepsilon\mu-h^2\eqno(15.6)$$
между частотой  $k$,  продольным  волновым  числом  $h$ и собственными
значениями $g$ соответствующей  двумерной  краевой  задачи.  В  случае
поперечных   сечений  тех  форм,  для  которых  переменные  в  волновом
уравнении  в  соответствующих  координатах  разделяются,   собственным
значениям  $g$ присваиваются два индекса,  определяющих число вариаций
поля по каждой координате. У собственных частот резонатора добавляется
третий  индекс  согласно  формуле  (15.5); в результате спектр частот
волноводного резонатора записывается в виде
     $$k_{pql}=\frac 1{\sqrt{\varepsilon\mu}}\sqrt{g_{pq}^2+\Bigl
         (\frac{\pi l} d \Bigr)^2}\;.\eqno(15.7)$$
     Собственные колебания  резонатора, представляющие   собой  стоячие
магнитные  волны, обозначаются $H_{pql}$, электрические волны --- $E_{pql}$.

     Поскольку проницаемости  $\varepsilon$  и  $\mu$  сами   являются
функциями частоты,  то для резонатора, заполненного веществом,
формула (15.7) представляет собой уравнение,  которое при заданных
значениях $g_{pq}$  и  $l$  может иметь несколько решений.
Соответствующие этим корням колебания  называются свободными:  из
суммы  таких  колебаний состоит  поле  в резонаторе  при
отключении  источника  возбуждения. Отметим,  что в  пустом
резонаторе  различие  между  собственными  и свободными
колебаниями отсутствует,  а значения $k_{pql}$ определяются
исключительно  геометрией  резонатора.  Собственные  колебания
играют важную  роль  в теории возбуждения резонаторов.  Далее
будет показано, что  при  расчёте  возбуждения  поля
гармоническим источником  (то  есть  вынужденных колебаний,
представляющих основной интерес во  всех  приложениях),
собственные  частоты  вычисляются  по формуле  (15.7)  при
значениях $\varepsilon$ и $\mu$,  соответствующих частоте
возбуждения  (то  есть  частоте  внешнего  по  отношению  к
резонатору генератора).

\begin{wrapfigure}[15]{l}{6cm}
\begin{picture}(65,47)
\put(0,50){\special{em:graph fig15-1.bmp}}
\end{picture}
\hbox to 7 cm{\hfil\footnotesize{Рис.~15.1.~Круглый
цилиндрический}
\hfill}
\hbox to 6.5cm{\hfil\footnotesize{ резонатор.}\hfil}
\end{wrapfigure}
     В волноводном резонаторе с круглым  поперечным  сечением  радиуса
$a$ (pис.~15.1) согласно формулам (9.14) и (9.15) собственные значения
краевой  задачи  и  потенциальная  функция  для   магнитных   волн   в
цилиндрической системе координат даются следующими выражениями:
     $$g_{pq}= \mu_{pq}/a\,,\eqno(15.8\mbox{\textit a})$$
     $$\Pi^m_{pql}=AJ_p\Bigl(\frac{\mu_{pq} r} a\Bigr)\;\cos{p(\varphi
         -\varphi_0)}\;\sin{\frac {\pi l} d z}\,,\eqno(15.8\mbox{\textit б})$$
где $\mu_{pq}$ --- корни  производной  функции  Бесселя:  $J_p'  (\mu_
{pq})   =0$.   Таким  образом,  спектр  собственных  частот  колебаний
магнитного  типа  круглого  цилиндрического  резонатора   определяется
формулой
     $$k_{pql}=\frac 1{\sqrt{\varepsilon\mu}}\sqrt{\Bigl(\frac{\mu_
         {pq}} a\Bigr)^2+\Bigr(\frac{\pi l}d\Bigr)^2}\;,\quad p=0,
         1,\dots;\,\,\,q=1, 2,\dots;\,\,\,l=1, 2,\dots\,.
         \eqno(15.9)$$

     Для электрических волн собственные  значения  $g_{pq}=\nu_{pq}/a$
($\nu_{pq}$    ---    корни   функции   Бесселя,   $J_p(\nu_{pq})=0$),
потенциальная функция
     $$\Pi^e_{pql}=AJ_p\Bigl(\frac{\nu_{pq} r}a\Bigr)\,\cos{p
         (\varphi-\varphi_0)}\cos{\frac {\pi l} d z}\,,\eqno(15.10)$$
а спектр частот собственных колебаний определяется формулой
     $$k_{pql}=\frac 1{\sqrt{\varepsilon\mu}}\sqrt{\Bigl(\frac{\nu_
         {pq}} a\Bigr)^2+\Bigl(\frac{\pi l}d\Bigr)^2}\;,\quad p=0,
         1,\dots;\,\,\,q=1, 2,\dots;\,\,\,l=0, 1,\dots\,.
         \eqno(15.11)$$

     Обычно для работы в электронных  приборах  СВЧ  используется  так
называемое  {\it  основное}  колебание  резонатора,  определяемое  как
имеющее самую низкую собственную частоту.  Поскольку наименьший корень
$\mu_{pq}$  есть $\mu_{11}=1,841\dots$,  то среди колебаний магнитного
типа низшей  частотой  обладает  $H_{111}$.  Среди  корней  $\nu_{pq}$
наименьший  $\nu_{01}=2,4048\dots$,  низшую  частоту  среди  колебаний
электрического типа имеет $E_{010}$.  Какое  из  этих  двух  колебаний
является основным --- зависит от соотношения геометрических параметров
резонатора.  Путём сравнения (15.9) и (15.11) нетрудно убедиться,  что
при $d>2,03\,a$ основное колебание --- $H_{111}$,  при меньших $d$ ---
$E_{010}$.

     Компоненты поля  в цилиндрических координатах находятся из (15.8)
с помощью общих формул (7.12),(7.13) и (7.17). Структура силовых линий
поля   двух   основных  типов  колебаний  рассматриваемого  резонатора
показана на рис.~15.2.

\begin{picture}(150,55)
\put(5,50){\special{em:graph  fig15-2a.bmp}}
\put(84,50){\special{em:graph fig15-2b.bmp}}
\end{picture}
\begin{center}\footnotesize{Рис.15.2.~Основные колебания
цилиндрического резонатора: {\it а)} $d<2,03 a$; {\it б)} $d>2,03 a.$
}\end{center}

     Среди колебаний  круглого  цилиндрического  резонатора выделяются
симметричные колебания ($p=0$).  В них  отличны  от  нуля  только  три
составляющие      поля:      у     электрических     колебаний     ---
$E_z,\,E_r,\,H_{\varphi}$, у магнитных --- $H_z,\,H_r,\, E_{\varphi}$.
При  этом  компоненты электрического поля всегда синфазны между собой;
синфазны и компоненты магнитного поля,  а  поля  $\rv  E$  и  $\rv  H$
отличаются   на   множитель   $i$   и,   следовательно,  смещены  друг
относительно друга на  четверть  периода  колебаний.  В  результате  в
некоторые моменты времени электрическое поле исчезает во всем объёме
резонатора,  а  магнитное  принимает  максимальное   значение;   через
четверть  периода  исчезает магнитное поле,  а электрическое достигает
максимума.

     Частоты симметричных   колебаний   свободны  от  поляризационного
вырождения,  присущего всем колебаниям  с  $p\ne  0$.  Это  вырождение
обусловлено   азимутальной   симметрией   резонатора   и   связано   с
неопределённостью    положения    плоскости    поляризации     поля,
определяемого  потенциалом  (15.8)  или (15.10):  угол $\varphi_0$ ---
произволен.  Только  сумма  двух  колебаний  с  различными  значениями
$\varphi_0$   служит  базисом,  разложением  по  которому  может  быть
представлено произвольное поле на собственной частоте.  Таким образом,
все  частоты  с  $p\ne  0$  оказываются  по  крайней  мере  двухкратно
вырожденными.  Наличие двух  опорных  колебаний  на  одной  частоте  с
однотипной потенциальной функцией приводит к возможности возбуждения в
резонаторе колебания,  имеющего вид бегущей  по  азимуту  волны,  поле
которого     с    учетом    временного    множителя    пропорционально
$\cos{(p\varphi-ck_{pql}t)}$. Резонаторы, работа которых рассчитана на
возбуждение  колебания  подобного  типа,  называются {\it резонаторами
бегущей волны}.

     Поляризационное вырождение   является   не   единственным   видом
вырождения,  присущим  резонаторам  рассматриваемого типа.  Одинаковые
собственные частоты имеют колебания $H_{01l}$ и $E_{11l}$ ($l>0$). При
определённых   соотношениях  длины  и  радиуса  резонатора  возможно
совпадение частот и целого ряда других колебаний.

     Одной из    разновидностей   волноводных   резонаторов   является
коаксиальный  резонатор  (рис.~15.3),  представляющий  собой   отрезок
коаксиальной   линии,   закороченный   с   обоих   концов  поперечными
металлическими перегородками $z=0$ и $z=d$.  В спектре его собственных
колебаний  помимо  обычных  $E$- и~$H$-волн присутствуют стоячие $TEM$
волны.  Для них $g=0$ и $h=k$,  а  среди  компонент  поля  только  две
отличны от нуля:
     $$E_r=\frac A r \sin{\frac{\pi l} d z}\;,\quad H_{\varphi}=-i
         \frac A r \cos{\frac{\pi l}d z}\,, \quad l=1,2,\dots\,.
         \eqno(15.12)$$

\begin{wrapfigure}[14]{l}{6.5cm}
\begin{picture}(60,55)
\put(-5,50){\special{em:graph fig15-3.bmp}}
\end{picture}
\hbox to 7 cm{\hfil\footnotesize{Рис.~15.3.~Коаксиальный
резонатор.}\hfil}
\end{wrapfigure}

     Если присваивать  колебанию  индексы по числу вариаций поля вдоль
соответствующей координаты,  то эти колебания следует  обозначить  как
$E_{00l}$.  Соответствующая  длина  волны в свободном пространстве ---
$\lambda=2d/l$,  собственная частота --- $k_{00l}=\pi l/d$.  Колебание
$E_{001}$ является основным только в достаточно длинном резонаторе,  а
в коротких резонаторах основное колебание --- $E_{010}$.  Коаксиальный
резонатор   можно   рассматривать   и  как  тороидальный  резонатор  с
прямоугольной формой поперечного сечения. Именно такие резонаторы чаще
всего используются как резонаторы бегущей волны.

     К волноводным резонаторам относится  и  прямоугольный  резонатор,
если   его   рассматривать   как   отрезок  прямоугольного  волновода,
ограниченный перпендикулярными  к  его  оси  проводящими  плоскостями.
Обозначим  размеры  резонатора и выберем привязку к координатным осям
так, как это показано на рис.~15.4. Поскольку для  волн типов $E_{pq}$
и $H_{pq}$ в прямоугольном волноводе согласно (9.5)
     $$g^2_{pq}=\left(\frac{p\pi}a\right)^2+\left(\frac{q\pi}b\right)
         ^2\;,\eqno(15.13)$$
то собственные частоты резонатора в соответствии с (15.7) равны
     $$k_{pql}=\pi\sqrt{\left(\frac p a\right)^2+\left(\frac q b
         \right)^2+\Bigl(\frac l d\Bigr)^2}\;.\eqno(15.14)$$

\newpage
\vspace{-.3cm}
\begin{wrapfigure}[13]{l}{7.5cm}
\begin{picture}(80,45)
\put(5,45){\special{em:graph fig15-4.bmp}}
\end{picture}
\hbox to 7.5cm{\hfil\footnotesize{Рис.~15.4.~Прямоугольный резонатор.
}\hfil}
\end{wrapfigure}

     Однако волноводная трактовка прямоугольного резонатора с присущим
ей делением на $E$- и $H$-колебания является  искусственной,  так  как
ось  $z$ в этом случае ничем не отличается от осей $x$ и $y$.  Поэтому
поучительно построить  систему  собственных  колебаний  резонатора  не
опираясь  на теорию волноводов,  а исходя непосредственно из уравнений
для электрического поля
     $$\Delta{\rv E}+k^2{\rv E}=0\;,\quad \div {\rv E}=0\eqno(15.15)$$
с граничными условиями
     $$\left.\begin{array}{l}E_y=E_z=0\qquad\mbox{при}\quad x=0\quad\mbox{и}
         \quad x=a\;,\\[.2cm]E_z=E_x=0\qquad\mbox{при}\quad y=0\quad\mbox{и}
         \quad y=b\;,\\[.2cm]E_x=E_y=0\qquad\mbox{при}\quad z=0\quad\mbox{и}
         \quad z=d\;.\end{array}\right \} \eqno(15.16)$$

     Из первого уравнения (15.15) следует,  что составляющие $\;E_x,\;
E_y,\;E_z\;$  удовлетворяют  скалярным  волновым уравнениям.  Решая их
методом  разделения  переменных,  получим  следующие   выражения   для
компонент поля:
     $$\left.\begin{array}{l}E_x=A\displaystyle{\cos\frac{p\pi x}{a}
         \sin\frac{q\pi y}{b}\sin\frac{l\pi z}{d}}\,,\\[.5cm]E_y=B
         \displaystyle{\sin\frac{p\pi x}{a}\cos\frac{q\pi y}{b}\sin
         \frac{l\pi z}{d}}\,,\\[.5cm]E_z=C\displaystyle{\sin\frac
         {p\pi x}{a}\sin\frac{q\pi y}{b}\cos\frac{l\pi z}{d}}\,,\\
         \end{array}\right\}\eqno(15.17)$$
где $A,\,B,\,C$  ---  комплексные постоянные;  $p$,~$q$,~$l$ --- целые
числа.

     В формулах   (15.17)  синусы  обеспечивают  выполнение  граничных
условий (15.16),  а косинусы выбраны  для  того,  чтобы  удовлетворить
второму уравнению (15.15) пут\"ем  подбора  постоянных  $A$,~$B$,~$C$.
Действительно, с помощью выражений (15.17) находим
     $$\div {\rv E}=-\pi\Bigl(\frac{p}{a}A+\frac{q}{b}B+\frac{l}{d}C
         \Bigr)\sin\frac{p\pi x}{a}\sin\frac{q\pi y}{b}\sin\frac{l
         \pi z}{d}\,,\eqno(15.18)$$
и $\div  {\rv  E}=0$   при   всех   $x$,~$y$,~$z$,   если   постоянные
удовлетворяют соотношению
     $$\frac{p}{a}A+\frac{q}{b}B+\frac{l}{d}C=0\;.\eqno(15.19)$$

     Подставляя выражения  (15.17)  в  волновое   уравнение   (15.15),
получаем   формулу  (15.14)  для  собственной  частоты,  а  с  помощью
уравнения ${\rv H}=\rot{\rv E}/ik $ находим  выражения  для  компонент
магнитного поля:
\vspace{.3 cm}
     $$\left.\begin{array}{l}H_x=\displaystyle{\frac{\pi}{ik}\left
         (\frac{q}{b}C-\frac{l}{d}B\right )\sin\frac{p\pi x}{a}\cos
         \frac{q\pi y}{b}\cos\frac{l\pi z}{d}}\,,\\[.5cm]H_y=
         \displaystyle{\frac{\pi}{ik}\left (\frac{l}{d}A-\frac{p}{a}C
         \right)\cos\frac{p\pi x}{a}\sin\frac{q\pi y}{b}\cos\frac{l\pi
         z}{d}}\,,\\[.5cm]H_z=\displaystyle{\frac{\pi}{ik}\left (\frac
         {p}{a}B-\frac{q}{b}A\right)\cos\frac{p\pi x}{a}\cos\frac{q\pi
         y}{b}\sin\frac{l\pi z}{d}}\,.\end{array}\right\}
         \eqno(15.20)$$
\vspace{.2cm}

     Таким образом,  каждая  тройка  чисел   $p$,~$q$,~$l$ определяет
решение   уравнений  поля  согласно формулам  (15.17),  (15.20),
обладающее   собственным волновым  числом  (15.14).  Если  среди  чисел
$p$,~$q$,~$l$ нет нулей, то такой тройке соответствует два собственных
колебания с одной и той же частотой,  но различным распределением поля.
Действительно, в формулах для полей имеется две произвольные
постоянные ---  пусть  это будут $A$ и~$B$,  поскольку третья постоянная
$C$ выражается через $A$ и $B$ из  соотношения  (15.19).  Полагая
$A=1$,~$B=0$,  получим  одно собственное  колебание,  полагая
 $A=0$,~$B=1$  --- колебание с той же частотой, но c другим распределением
 поля. Таким образом, если индексы $p$,~$q$,~$l$  все  отличны  от нуля,
 то определяемая ими собственная частота (15.14) оказывается,  по крайней
 мере, двухкратно вырожденной. Если размеры резонатора $a$,~$b$,~$d$
 соизмеримы, то есть их отношение рациональное число,  то возможно
 вырождение более  высокой  кратности. Так, например, при $a=b$ и~$p\ne q$
 частота (15.14) имеет не менее чем четырёхкратное вырождение,  поскольку
 тройкам чисел $p$,~$q$,~$l$  и $q$,~$p$,~$l$  соответствует  одна  частота,  но
 разные распределения поля.

     Собственные колебания  прямоугольного  резонатора,  у  которых  один
из  индексов  равен  нулю,  занимают особое положение. Так, например,
тройке  чисел    $p$,~$q$,~$0$     соответствует  следующее  электромагнитное
поле:
\vspace*{.3cm}
     $$\left.\begin{array}{l}E_z=C\displaystyle{\sin\frac{p\pi x}{a}
         \sin\frac{q\pi y}{b}}\,,\\[.5cm]H_x=-i\displaystyle{\frac{q
         \pi}{kb}C\sin\frac{p\pi x}{a}\cos\frac{q\pi y}{b}}\,,\\[.5cm]
         H_y=i\displaystyle{\frac{p\pi}{ka}C\cos\frac{p\pi x}{a}\sin
         \frac{q\pi y}{b}}\,,\\[.4cm]\end{array}\right\}
         \eqno(15.21)$$
остальные компоненты  равны  нулю.  Это  колебание  является   простым
(невырожденным),  поскольку  оно  зависит только от одной произвольной
постоянной  $C$.  Как  и  во  всяком  простом  колебании,   компоненты
магнитного  поля  синфазны и сдвинуты на четверть периода относительно
электрического поля.  Силовые линии колебания $E_{110}$  изображены  на
рис.~15.5.  По  своей  конфигурации  они сходны с аналогичными линиями
колебания $E_{010}$ круглого цилиндрического резонатора  (рис.~15.2\textit{a}).
На  примере этого колебания хорошо видно,  что классификация колебаний
прямоугольного резонатора является в значительной степени  условной  и
неоднозначной.   Так,   то  же  самое  колебание  $E_{110}$  с  равным
основанием можно считать либо колебанием $H_{011}$, либо $H_{101}$ ---
для   этого  достаточно  переобозначить  соответствующим  образом  оси
координат.  Если   размеры   резонатора   удовлетворяют   неравенствам
$a>d$,~$b>d$,   то  колебание  $E_{110}$  является  основным  ---  его
собственная частота наименьшая.  Собственных колебаний
прямоугольного    резонатора,  у   которых   два  индекса    равны  нулю,
не  существует,  поскольку в этом случае согласно (15.17) и (15.20) все
компоненты поля тождественно равны нулю.

\begin{wrapfigure}[14]{l}{7.5cm}
\begin{picture}(65,55)
\put(-5,50){\special{em:graph fig15-5.bmp}}
\end{picture}
\hbox to 7 cm{\hfil\footnotesize{Рис.~15.5.~Колебание $E_{110}$.}
\hfil}
\end{wrapfigure}
     Кроме рассмотренных выше резонаторов  строгое  решение  задачи  о
собственных   колебаниях  допускают  ряд  цилиндрических  резонаторов,
например,  эллиптический цилиндр,  сферический  резонатор,  резонатор,
представляющий  собой  эллипсоид,  и  некоторые  другие.  У всех у них
стенки полости совпадают  с  координатными  поверхностями  тех  систем
координат,  в  которых  разделяются  переменные  в волновом уравнении.
Однако эти  резонаторы  не  нашли  достаточно  широкого  практического
применения   ни   раньше  (в  первую  очередь,  из-за  технологических
сложностей изготовления), ни сейчас (когда ряд технологических проблем
уже решён).

     С другой  стороны,  в  технике СВЧ широко используются резонаторы
более сложной формы,  для  которых  задача  о  собственных  колебаниях
изложенным строгим методом разделения переменных неразрешима.  Поэтому
большое значение имеют численные и приближённые аналитические методы
расчёта  таких  резонаторов.  К  прямым  численным методам относятся
различные  модификации  метода  сеток,  используемые  для   численного
решения волнового уравнения при заданных форме граничной поверхности и
граничных условиях на ней.  В настоящее время разработаны  и  отлажены
программы  расчёта  всех  основных  характеристик  резонаторов очень
сложной  геометрии.  При  современных  быстродействии  компьютеров   и
объёме их оперативной памяти расчёт резонаторов даже очень сложной
геометрии стал  операцией,  позволяющей  за  вполне  приемлемое  время
достигнуть точности,  превышающей потребности практики.  Такой подход,
однако,  оставляет в стороне  все  физические  стороны  явления  и  на
начальном этапе разработки не всегда оптимален.

     Промежуточное место  между  прямыми  численными и приближёнными
аналитическими   методами   занимает   метод    частичных    областей,
использованный  в  предыдущих  разделах при рассмотрении периодических
структур.  С одной  стороны,  для  его  реализации  необходим  большой
объём  аналитических  выкладок  (проясняющих при этом в значительной
степени физическую сторону решения),  с другой,  он фактически требует
на конечном этапе большого численного счёта (в стандартных операциях
решения системы линейных алгебраических уравнений), который может быть
реализован только при применении компьютеров.

     К аналитическим приближённым  методам следует отнести квазистационарный
 расчёт собственной  частоты основного колебания и метод возмущений.  Метод
 возмущений будет  рассмотрен  в следующем    разделе    после    ознакомления  с
некоторыми общими  свойствами  резонаторов,  а прежде,  чем  переходить  к
квазистационарным расчётам, сделаем несколько полезных замечаний
относительно  метода   частичных  областей, которые   позволяют   в  значительной
степени сократить   объём   вычислений.   Если  резонатор обладает  плоскостью
симметрии,   то достаточно рассмотреть   только половину   резонатора, ограниченную
этой   плоскостью,   поставив  на ней определённые граничные условия.

\begin{wrapfigure}[13]{l}{7cm}
\begin{picture}(55,45)
\put(0,50){\special{em:graph fig15-6.bmp}}
\end{picture}
\hbox to 7 cm{\hfil\footnotesize{Рис.~15.6.~Резонатор, обладающий}
\hfil}
\hbox to 7 cm{\hfil\footnotesize{плоскостью симметрии.}
\hfil}
\end{wrapfigure}

     Возьмём в   качестве   примера   изображённый   на  рис.~15.6
резонатор,  представляющий собой  закороченную  проводящими  торцевыми
стенками    ячейку    периодического   диафрагмированного   волновода.
Собственные частоты такого  резонатора  являются  верхней  или  нижней
граничной  частотой  полос  пропускания  волновода,  изображённых на
диаграмме 14.2. Плоскость $z=0$ представляет собой плоскость симметрии
резонатора. Все его собственные частоты будут найдены, если решить две
задачи для половинки резонатора  $z>0$,  поставив  в  плоскости  $z=0$
последовательно   следующие   граничные   условия:  $\rv  E_{\tau}=0\;
(H_n=0)$  (стенка  из  идеального  проводника)  и  $\rv   H_{\tau}=0\;
(E_n=0)$ (стенка из идеального магнетика). По терминологии, перешедшей
из электротехники,  они называются соответственно  условием  короткого
замыкания  и  условием холостого хода.  Во всех собственных колебаниях
первой задачи продольное электрическое поле  $E_z$  является  чётной
функцией $z$, а силовые линии электрического поля пересекают плоскость
$z=0$ по нормали;  во второй задаче  (идеальный  магнетик)  $E_z$  ---
нечётная  функция  $z$, и  электрические силовые линии не пересекают
плоскости $z=0$.

\vspace{-.1cm}
\begin{wrapfigure}[21]{l}{7.7cm}
\begin{picture}(55,85)
\put(0,85){\special{em:graph fig15-7.bmp}}
\end{picture}
\hbox to 7 cm{\hfil\footnotesize{Рис.~15.7.~Тороидальные резонаторы.}
\hfil}
\end{wrapfigure}

     Квазистационарный метод  пригоден   для расчёта  основной  собственной
частоты резонаторов,  полость  которых  может  быть  чётко разделена  на  две
области. В одной из  них   явно преобладает  электрическое  поле  (обычно  это
емкостной  зазор между  параллельными  близко расположенными   достаточно
большими металлическими  пластинами),  в  другой  в основном   сосредоточено
магнитное  поле. Вычислив   соответствующие   ёмкость   $C$   и индуктивность
$L$,    собственную   частоту колебания определяют по формуле Томпсона
     $$\omega  =\frac 1{\sqrt{LC}}\,.\eqno(15.22)$$
Для возможности применения этой формулы необходимо соблюдение  условия
квазистационарности:  $kD\ll  1\,,$  где  $D$  --- максимальный размер
резонатора, то есть размеры резонатора должны быть малы по сравнению с
длиной  волны  колебания.  Примерами  таких  квазистационарных  систем
являются тороидальные  резонаторы,  представленные  на  рис.~15.7, где
цифрой 1 отмечена ёмкостная (электрическая) область,  цифрой  2  ---
индуктивная (магнитная) область.

     Рассмотрим тороидальный  резонатор общего вида,  электрическая область
которого представляет собой пространство между двумя расположенными на
расстоянии $d$ друг от друга  параллельными соосными круглыми пластинами
радиуса $a\;(d\ll a)$,  а магнитная образована вращением вокруг оси
резонатора произвольной  замкнутой на края пластин кривой, ограничивающей
область
$\Sigma$ (рис.~15.7{\it {а}}).   Без учёта краевого
эффекта ёмкость электрической области
составляет
     $$C=\frac{a^2}{4d}\,.\eqno(15.23)$$
Поля и  поверхностные  токи  на  внутренних  стенках  тора  в основном
колебании естественно  считать  азимутально  симметричными,  а  вектор
плотности тока лежащим в азимутальной плоскости. Тогда отлична от нуля
только одна составляющая магнитного поля --- $H_{\varphi}$,  а силовые
линии   магнитного  поля  представляют  собой  окружности,  лежащие  в
плоскостях, перпендикулярных оси резонатора.

     Если на  основании  условия  квазистационарности пренебречь током
смещения,  то уравнение Максвелла $(1.12\mbox{\textit б}')$
запишется в данном случае в виде
     $$ \oint H_{\varphi}\,dl_{\varphi}=\frac{4\pi} c J\,,
         \eqno(15.24)$$
где $dl_{\varphi}=rd\varphi$,  $J$  --  полный  ток,  протекающий   по
внутренним стенкам тороидальной камеры. Отсюда следует, что
     $$H_{\varphi}=\frac{2J}{cr}.\eqno(15.25)$$
Индуктивность $L$ связана с энергией магнитного поля
     $$U=\frac 1{8\pi}\int H^2_{\varphi}\,dV=\frac{J^2}{c^2}\int
         \limits_{\Sigma}\frac{dS}r\eqno(15.26)$$
известным соотношением   $U=LJ^2/2$.   В   результате   получаем   для
индуктивности выражение
     $$L=\frac 2{c^2}\int \limits_\Sigma\frac{dS} r\;,\eqno(15.27)$$
подставляя которое  в (15.22) и учитывая (15.23),  находим длину волны
основного колебания:
     $$\lambda=\pi a\sqrt{\frac 2 d \int\limits_\Sigma {\frac{dS}r}}\;.
         \eqno(15.28)$$

     В результате   несложных    вычислений    получаем    для    тора
прямоугольного сечения (рис.~15.7{\it {б}}):
     $$\lambda=\pi a\sqrt{\frac{2h} d \ln{\displaystyle{\frac
            b a }}} \,;\eqno(15.29)$$
для тора круглого сечения (рис.~15.7{\it в}) длина волны основного колебания
равна
     $$\lambda=2\pi a\sqrt{\frac{\pi} d (b-\sqrt{a(2b-a)})}\;.
         \eqno(15.30)$$

     Аналогично может быть вычислена и  основная  собственная  частота
для отдельного резонатора магнетрона. Собственные частоты высших типов
колебаний в рассмотренных резонаторах таким способом вычислены быть не
могут, поскольку для них условие квазистационарности не выполняется.

%\end{document}

\newpage
\oddsidemargin=-0.4mm \evensidemargin=-0.4mm
\topmargin=-0.4mm
\headsep=7mm
\textheight=231.875mm
\textwidth=160mm
\mathsurround=2.5pt
\unitlength=1mm
%\begin{document}
%\input{macr.tex}
\thispagestyle{empty}
%\addtocounter{page}{166}

\begin{center}\subsubsection*{16.~Общая теория резонаторов}\end{center}
\vspace*{0.5cm}

\markboth{Глава 5.~Объёмные резонаторы}{16.~Общая теория резонаторов}

\begin{center}\begin{minipage}[c]{0.75\textwidth}
\footnotesize{\parindent=0.5cm
         Общие свойства  идеальных резонаторов.  Потери в резонаторах.
         Добротность и сдвиг собственной частоты. Добротность основных
         колебаний    прямоугольного    и   круглого   цилиндрического
         резонаторов.  Вычисление собственных  частот  деформированных
         резонаторов методом возмущений.
}\end{minipage}\end{center}\vspace*{0.5cm}

     Для цилиндрических  (волноводных) резонаторов собственные частоты
и поля собственных колебаний вычисляются единым образом  по  известным
собственным  значениям и собственным функциям скалярной краевой задачи
для двух потенциальных функций  соответствующего  волновода.  В  общем
случае   резонатора   произвольной   формы   краевая  задача  является
трёхмерной векторной и составляет предмет  исследования  специальной
главы  математической  физики.  Однако,  не  вдаваясь в сложный вопрос
нахождения  её  решения,  на  основании  только  вида  уравнений   и
граничных   условий   можно  сделать  некоторые  важные  общие  выводы
относительно собственных частот и полей собственных колебаний.

     Краевая задача  для  идеального  пустого  объёмного  резонатора
включает в себя векторные уравнения
     $$\rot{\rv H}_n=-ik_n{\rv E}_n\,,\qquad\rot{\rv E}_n=ik_n{\rv H}
         _n\,\eqno(16.1)$$
и граничное условие
     $$E_t=0 \eqno(16.2)$$
на замкнутой поверхности резонатора $S$.

     Из общих физических соображений ясно, что при отсутствии потерь в
материале   заполнения   и   идеально   проводящих  замкнутых  стенках
собственные колебания в резонаторе,  если они существуют,  должны быть
{\it  незатухающими}.  Поскольку  зависимость от времени в этом случае
определяется множителем $\exp(-ik_nct)$,  то  все  $k_n$  должны  быть
действительными   величинами.  Докажем,  что  этот  факт  есть  прямое
следствие условий  краевой  задачи.  Для  этого  преобразуем  величину
$\div{[{\rv   E_n\rv  H_n^*}]}$  с  помощью  известной  формулы
векторного  анализа  $  \div{[{\rv  { AB}}]}={\rv   B}\rot{\rv   A}-{\rv
A}\rot{\rv B}$ и, используя уравнения (16.1), получим
     $$k_n^2\left |{\rv H}_n\right |^2=\left |\rot{\rv H}_n\right |^2-
         ik_n\div{[{\rv E}_n{\rv H}_n^*]}\,.\eqno(16.3)$$
Проинтегрируем это равенство по объёму резонатора  $V$
и воспользуемся теоремой Гаусса (1.10); в результате имеем:
     $$k_n^2=\frac{\displaystyle{\int\limits_V}{\left |\rot{\rv H}_n
         \right |^2\,dV}-ik_n\displaystyle{\oint\limits_S}{[{\rv E}_n
         {\rv H}^*_n]\,d{\rv S}}}{\displaystyle{\int\limits_V}{\left
         |{\rv H}_n\right |^2\,dV}}\;.\eqno(16.4)$$
При условии  (16.2)  поверхностный  интеграл в (16.4) равен нулю,  что
свидетельствует об отсутствии потока энергии в  стенки  резонатора,  и
для $k_n^2$ получается положительная величина.

     Легко показать,   что   для   идеального  резонатора  выполняется
равенство средних  по  времени  энергий  электрического  и  магнитного
полей:
     $$\overline{U}_E=\frac{1}{16\pi}\int\limits_V{\left |{\rv E}_n
         \right |^2\,dV}\qquad\mbox{и}\qquad\overline{U}_H=\frac{1}{16
         \pi}\int\limits_V{\left |{\rv H}_n\right |^2\,dV}\;.
         \eqno(16.5)$$
Действительно, из (16.1) следует, что
     $$\div{[{\rv E}_n{\rv H}_n^*]}=ik_n|{\rv H}_n|^2-ik_n
         ^*|{\rv E}_n|^2;\eqno(16.6)$$
интегрируя это   равенство   по   объёму   резонатора   и   учитывая
вещественность $k_n$,  убеждаемся в  равенстве  энергий  (16.5).  Если
таким   же  образом  проинтегрировать  величину  $\div{[{\rv  E}_n{\rv
H}_n]}$, то получим
     $$\int\limits_V {{\rv H}_n^2\,dV}+\int\limits_V {{\rv E}_n^2\,dV}
         =0\,.\eqno(16.7)$$
Это равенство имеет совсем другой физический смысл, который будет
разъяснён чуть ниже.

     Поля различных  собственных  колебаний  {\it  ортогональны} между
собой в том смысле, что при $k_n\ne k_m$
     $$\int\limits_V{ {\rv E}_n{\rv E}_m\,dV}=0\,,\qquad\int\limits_V{
         {\rv H}_n{\rv H}_m\,dV}=0\,.\eqno(16.8)$$
Эти равенства   легко   доказать,  интегрируя  выражения  $\div[  {\rv
E}_n{\rv H}_m]$ и $\div[ {\rv E}_m{\rv H}_n]$ по объ\"ему  резонатора,
используя   (1.5),   (16.1)   и   (16.2).   В   результате   очевидных
преобразований получаем
     $$(k^2_m-k^2_n)\int\limits_V\,{\rv E}_n{\rv E}_m\,dV=0\;,\qquad
         (k^2_m-k^2_n)\int\limits_V\,{\rv H}_n{\rv H}_m\,dV=0\;.
         \eqno(16.9)$$
При $k^2_n\ne  k^2_m$  равенства  (16.9)  и  (16.8)  эквивалентны   и,
следовательно,  соответствующие  собственные  колебания  ортогональны.
Однако,  как следует из  предыдущего  раздела,  некоторые  собственные
частоты  являются  вырожденными  ---  на  одной  и  той  же  частоте в
резонаторе  могут  быть  возбуждены   поля   различных   конфигураций.
Кратность   вырождения   определяется   числом   линейно   независимых
колебаний,   суперпозицией    которых    представимо    любое    поле,
удовлетворяющее  краевой  задаче.  Например,  в круглом цилиндрическом
резонаторе   все   азимутально    несимметричные    колебания    имеют
поляризационное  вырождение,  кратность  которого  равна  двум.  Любые
линейно  независимые  колебания  могут   быть   выбраны   в   качестве
собственных  на вырожденной частоте.  Вообще говоря,  эти колебания не
будут ортогональными в смысле (16.8).  Однако в математике  существует
стандартная процедура ортогонализации (ввиду её очевидности здесь не
место  на  ней  останавливаться),  позволяющая  обеспечить  выполнение
равенств (16.8) и при вырождении.

     Построенные таким образом ортогональные колебания обладают тем же
свойством, как и все простые (невырожденные) колебания резонатора: они
определяются с  точностью  до  одной  произвольной  постоянной  и  все
компоненты  электрического  и  все  компоненты  магнитного  поля в них
синфазны во всем объёме резонатора, и, следовательно, каждое колебание
представляет  собой  стоячую волну.  Для них равенство (16.7) обобщает
отмеченную  ранее  для  простых  колебаний   закономерность   смещения
электрического  и  магнитного  полей по времени на четверть периода на
все ортогональные собственные колебания.  Оно позволяет  также  ввести
норму $n$-ого колебания $N_n$, определив её, как это часто делается,
следующими соотношениями:
     $$N_n=\frac 1{4\pi}\int\limits_V \rv E_n^2\,dV=-\frac 1{4\pi}\int
         \limits_V \rv H_n^2\,dV\;.\eqno(16.10)$$
Нормируя колебания   условием   $N_n=1   $,   находим   тем   самым
произвольную постоянную, определяющую амплитуду колебания.

     Таким образом,   каждому     резонатору     можно     сопоставить
последовательность  ортогональных нормированных собственных колебаний.
Соответствующая последовательность собственных частот  может  включать
несколько   одинаковых   значений,   число   которых  равно  кратности
вырождения.  При выбранной норме (16.10) все компоненты электрического
поля  являются  действительными величинами,  все компоненты магнитного
поля  ---  мнимыми  величинами.  Ортогональные  собственные  колебания
всегда  представляют  собой  стоячую  волну.  Резонаторы бегущей волны
могут работать только на вырожденных частотах и  рабочее  колебание  в
них  есть  сумма двух ортогональных колебаний.  Практически это всегда
частоты,  обладающие   поляризационным   вырождением   в   азимутально
симметричных структурах.

     Все формулы   для   резонатора,   полость   которого    заполнена
непоглощающей   диэлектрической   средой  с  вещественными  значениями
$\varepsilon$ и $\mu$,  получаются из формул  для  пустого  резонатора
путем следующей замены:
     $$k,\;\;{\rv E,\;\;\rv H}\quad\Rightarrow\quad k\sqrt{\varepsilon
         \mu},\;\;\sqrt\varepsilon{\rv E},\;\;\sqrt\mu {\rv H}\,.
         \eqno(16.11)$$
Это ясно из  того,  что  при  таком  преобразовании  уравнения
(16.1) переходят в  уравнения Максвелла в среде:
     $$\rot{\rv H}_n=-ik_n\varepsilon{\rv E}_n\,,\qquad\rot{\rv
         E}_n=ik_n\mu{\rv H}_n\,.\eqno(16.12)$$
В частности,  наличие  среды  уменьшает  все  собственные  частоты   в
$\sqrt{\varepsilon\mu}$ раз.

     До сих   пор   речь  шла  об  идеальных  резонаторах,  в  которых
отсутствуют потери и колебания являются незатухающими.  Можно  указать
три  вида  потерь,  которые  в  той  или  иной  степени присутствуют в
реальных резонаторах: 1 --- потери в стенках резонатора, обусловленные
конечностью  их проводимости;  2 --- потери в материале заполнения;  3
--- излучение через отверстия связи  с  подводящими  волноводами.  Два
первых   вида   потерь   почти   всегда  являются  нежелательными  (за
исключением  случаев   использования   селективного   поглощения   для
разрежения спектра собственных колебаний).  Третий вид потерь называть
собственно потерями можно  только  условно,  поскольку  это  излучение
определяет  сам  принцип  работы  устройства,  в  котором используется
резонатор.

     При наличии потерь собственные частоты $\omega_n=ck_n$ становятся
комплексными:  $\omega_n=\omega_n'-i\omega_n''\;  (\omega''_n>0)$. Это
значит, что поля колебаний, определяемые выражениями
     $$\mbox{\boldmath$\gv E$}(t)=\re {({\rv E}e^{-i\omega't})}\,
       e^{-\omega''t},\quad \mbox{\boldmath$\gv H$}(t)=\re {({\rv H}\,
       e^{-i\omega't})}\,e^{-\omega''t}, \eqno(16.13)$$
затухают со   временем   (в   формулах  (16.13)--(16.28)  индекс  $n$,
сопоставляемый  сочетанию  трёх  индексов,  характеризующих   данное
собственное  колебание,  опускается  --- содержание этой части раздела
справедливо не только для собственно резонаторов,  но и для  некоторых
других систем). Величина $\omega''$ называется коэффициентом затухания
и полностью  характеризует  качество  каждого  собственного  колебания
резонатора.  В  радиотехнике  для  этих  целей  пользуются  обычно  не
коэффициентом затухания,  а безразмерной  величиной,  называемой  {\it
добротностью} колебания $Q$ и определяемой равенством
     $$Q=\frac{\omega'}{2\omega''}\;.\eqno(16.14)$$

     При наличии  затухания  поле  собственного колебания в резонаторе
уже не является монохроматическим, а зависит от времени по закону
     $$\mbox{\boldmath$\gv E$}(t)=
     \mbox{\boldmath$\gv E$}_0e^{-i\omega't-\omega't/2Q}\;.\eqno(16.15)$$
Разлагая $E(t)$ в интеграл Фурье
     $$\mbox{\boldmath$\gv E$}(t)=\int\limits_{-\infty}^\infty
       {\rv E(\omega)e^{-i\omega t}\,dt}\,,\eqno(16.16)$$
получим для  спектральной плотности
     $$\rv E(\omega)=\frac{\rv E_0}{\pi}\int\limits_0^\infty{e^{i(\omega-
         \omega')t-\omega' t/2Q}}\,dt\,.\eqno(16.17)$$
Интеграл легко вычисляется и в результате
     $$\left |\rv E(\omega)\right |^2\sim\frac{1}{(\omega-\omega')^2+\left
         (\omega'/{2Q}\right )^2}\,.\eqno(16.18)$$
Очевидно, что    $|E|^2$    принимает    максимальное   значение   при
$\omega=\omega'$, а вдвое меньшее значение --- при $\omega-\omega'=\pm
\;\omega'/2Q$.  Таким  образом,  добротность  может  быть выражена через
ширину $\Delta\omega$ резонансной кривой на уровне 1/2 от максимума:
     $$Q=\frac{\omega'}{\Delta \omega}.\eqno(16.19)$$
Это соотношение    широко    используется,    в     частности,     для
экспериментального измерения добротности.

     Добротность колебания нетрудно выразить через
энергетические величины. Если поле затухает по закону (16.13), то
запасённая   в резонаторе  энергия поля,
усреднённая по периоду колебания,
изменяется во времени как
     $$\overline{U}(t)=\overline{U}(0)e^{-2\omega''t}\,\eqno(16.20)$$
и, следовательно,
     $$\frac{d\overline {U}}{dt}=-2\omega''\overline U\,.
         \eqno(16.21)$$
Усреднение в  (16.20)  по  времени  подразумевает,  что за один период
амплитуда поля  меняется  незначительно  и,  следовательно,  $\omega''
\ll\omega'$, то есть $Q$ --- величина большая. Производя усреднение по
времени в законе сохранения энергии (1.21), получим:
     $$\frac{d\overline U}{dt}+P_{sum}=0\,,\eqno(16.22)$$
где мощность  суммарных  потерь $P_{sum}=\overline
P +\overline \Sigma_w +\overline\Sigma_{rad}$   включает  в  себя
потери  в  среде   заполнения $\overline  P$,  поток  энергии
$\overline\Sigma_w$,  поглощающейся в стенках,  и поток энергии
$\Sigma_{rad}$, излучаемой через отверстия связи в стенках.
Из (16.20), (16.21)  тогда следует соотношение
     $$\omega''=\frac{P_{sum}}{2\overline U}\,\eqno(16.23)$$
и --- согласно (16.14) --- получаем выражение для добротности:
     $$Q=\frac{\omega' \overline U}{P_{sum}}\;.\eqno(16.24)$$

     Разбиение потерь мощности на несколько  частей  позволяет  ввести
соответствующие   добротности.   Добротность   (16.14),
обусловленную полными  потерями,  называют   нагруженной
добротностью   резонатора (колебания) $Q_{\mbox{\footnotesize
н}}$, обусловленную потерями на  излучение  ---  внешней
добротностью $Q_{\mbox{\footnotesize{вн}}}$, потерями в стенках и
заполняющем веществе
--- собственной добротностью $Q_0$.  Разделяя потери в стенках и
веществе, можно ввести добротности $Q_{0\Sigma}$ и $Q_{0 P}$.  Все
эти  величины связаны очевидным соотношением:
     $$\frac 1 Q_{\mbox{\footnotesize н}}
          =\frac 1 Q_{0\Sigma} + \frac1 Q_{0P}+
          \frac 1 Q_{\mbox{\footnotesize{вн}}}
         \;.\eqno(16.25)$$
Вывод формулы (16.24) предполагал при усреднении  по  времени  малость
затухания  и  соответственно  большую  величину  Q.  Только  при  этих
условиях она может быть полезной при вычислении добротности  и  только
при  дополнительном  предположении  о возможности вычисления потерь на
основе распределения полей,  получаемых для собственного  колебания  в
идеальном  резонаторе.  Эти предположения позволяют рассчитывать каждый
вид потерь и,  следовательно,  каждую добротность независимо  друг  от
друга.

     Вычисление добротности   можно  производить  как  основываясь  на
формуле (16.14), определив из (16.4) коэффициент затухания $\omega''$,
так   и   с   помощью   формулы   (16.24),   вычислив   предварительно
энергетические характеристики.  В некотором  отношении  первый  способ
предпочтительнее,   поскольку  позволяет  в  ряде  случаев  попутно  с
затуханием вычислить и смещение максимума резонансной кривой.  В  силу
сказанного   выше   расчёт   каждой   добротности   производится   в
предположении  отсутствия  других  потерь.  Проще  всего   вычисляется
добротность $Q_{0P}$,  обусловленная потерями в среде заполнения.  При
$\mu=1$ мощность потерь в объёме резонатора согласно формуле  (2.32)
составляeт
     $$P=\frac{\omega'\varepsilon''}{8\pi}\int\limits_V|\rv E|^2
         \,dV\,,\eqno(16.26)$$
а запасённая энергия
     $$\overline U=\frac{\varepsilon'}{8\pi}\int\limits_V|\rv E|^2
         \,dV\,,\eqno(16.27)$$
так что
     $$Q_{0P}=\frac{\varepsilon'}{\varepsilon''}=\frac 1{\tg{\theta}}
         \,,\eqno(16.28)$$
где $\theta$--  угол потерь в среде.  Заметим,  что добротность в этом
случае не  зависит  от распределения поля в объёме и различается для
разных колебаний только из-за зависимости $\varepsilon(\omega)$.

     Поскольку в большинстве приложений используется пустой  резонатор
или  заполненный  воздухом,  для  которого $\varepsilon''$ очень малая
величина,  то собственная добротность определяется  в  первую  очередь
потерями в  стенках.  Вычислим  её,   исходя   из  формулы   (16.4),
предполагая  выполненными  условия  сильного  скин-эффекта и пользуясь
вместо (16.2) граничным условием Щукина-Леонтовича  (6.17).  При  этом
будем  искать  поля  и собственные частоты в виде разложения по малому
параметру $W$-- волновому сопротивлению стенок резонатора:
     $${\rv E}_n={\rv E}_{n0}+W{\rv E}_{n1}+\dots,\quad{\rv H}_n=
        {\rv H}_{n0}+W{\rv H}_{n1}+\dots\,,\quad k_n=k_{n0}+Wk_{n1}+
         \dots,\eqno(16.29)$$
где величины  ${\rv  E}_{n0},\,{\rv H}_{n0}\,$ и $k_{n0}$ берутся из решения
для идеального резонатора.  Оставляя в (16.4)  только  члены  не  выше
первой степени $W=W'+iW''$, получим для мнимой части уравнения:
     $$2k_{n0}k_n''=-\frac{k_{n0}W'}{D},\eqno(16.30)$$
где
   $$D=\frac{\displaystyle{\int\limits_V|\rv H_{n0}|^2\,dV}}{
    \displaystyle{\oint\limits_S|\rv H_{n0t}|^2\,dS}}\;\eqno(16.31)$$
($\rv H_{n0t}$ -- тангенциальная составляющая вектора $\rv H_{n0}$ на
внутренней поверхности резонатора). Очевидно, что для резонаторов
простых форм, соразмерных по всем осям, величина $D$ может служить в
качестве характерного размера резонатора. В результате для добротности
имеем
     $$Q_{0\Sigma}=\frac{k_n'}{2|k_n''|}=\frac{2D}\delta\,,
         \eqno(16.32)$$
где $\delta$  --  глубина  скин-слоя,  связанная  с  $W$  соотношением
$W=(1-i)k\delta/2$ (при $\mu=1$), причём $k$ следует положить равным
$k_{n0}$. Действительную и мнимую части $k_n$ можно записать в виде
     $$k_n'=k_{n0}+W'k_{n1}'-W''k_{n1}''+\dots\,,\qquad k_n''=W'k_{n1}
         ''+W''k_{n1}'+\dots.\eqno(16.33)$$
Из сопоставления соотношений (16.30) и (16.33) следует
     $$k_{n1}'=0\,,\qquad k_{n1}''=-\frac 1{2D}\,,\eqno(16.34)$$
так что  из-за проникновения поля в стенки резонатора  вещественная  часть
собственных частот   колебаний   составляет
     $$k_n'=k_{n0}+\frac{W''}{2D}\,.\eqno(16.35)$$
Поскольку для  металлов  $W''=-W'<0$,  то  собственные частоты реального
резонатора меньше собственных частот идеального; при этом сдвиг  частоты
численно равен затуханию ($k_n'-k_{n0}=k_n''$).

     Оставляя в  действительной  части  уравнения (16.4) члены не выше
первого порядка по $W$, приходим к соотношению
   $$\int\limits_V(|\rv H_n|^2-|\rv E_n|^2)\,dV=\frac{W''}{k_{n0}}\oint
         \limits_S|\rv H_{tn0}|^2\,dS=-\frac\delta 2\oint\limits_S
         |\rv H_{tn0}|^2\,dS\,,\eqno(16.36)$$
которое позволяет  связать  разность  средних  энергий  магнитного   и
электрического  полей  в  реальном  резонаторе  с  потоком  мощности в
стенки:
     $$\overline U_H-\overline U_E=-\frac 1{ck_{n0}}\Sigma_w\,.
         \eqno(16.37)$$
В идеальном  резонаторе  эта разность равна нулю;  в реальном резонаторе
при учёте потерь в стенках сумма средних энергий
 согласно  (16.24)  выражается  через  добротность  $Q_{0\Sigma}
=ck_{n0}(\overline U_H+\overline U_E)/\Sigma_w$,  поэтому
     $$\overline U_H=\frac {Q_{0\Sigma}-1}{Q_{0\Sigma}+1}\,\overline
         U_E\,.\eqno(16.38)$$

     Для резонаторов простейших  форм  определяемый  формулой  (16.31)
параметр  $D$,  через который выражается добротность согласно (16.32),
вычисляется  без  каких  либо  затруднений,  хотя   в   общем   случае
соответствующие    выражения    достаточно    громоздки.    Так,   для
$H_{pql}$-колебания круглого цилиндрического резонатора (радиус  $a$,
длина $d$) этот параметр определяется соотношением
     $$2D=\frac {a\Bigl[\mu_{pq}^2+\displaystyle{\Bigl(\frac{\pi la}d
         \Bigr)^2}\Bigr]\Bigl(1-\frac {m^2}{\mu_{pq}^2}\Bigr)}{\mu_{pq}
         ^2+\displaystyle{\Bigl(\frac{\pi la}d\Bigr)^2\Bigl[2\frac a d
         \Bigl(1-\frac{m^2}{\mu_{pq}^2}\Bigr)+\frac{m^2}{\mu_{pq}^2}
         \Bigr]}}\;.\eqno(16.39)$$

     Для $E_{pql}$--колебания формула проще:
     $$2D=\frac d{\displaystyle{2-\delta_{0l}+\frac d a}}\;
         \eqno(16.40)$$
и, в частности,  для колебания типа $E_{010}$ в круглом цилиндрическом
резонаторе   добротность   $Q=ad/\delta    (a+d)$.    Для    колебания
прямоугольного резонатора типа $E_{110}$
     $$2D=\frac{ab(a^2+b^2)d}{2(a^3+b^3)d+ab(a^2+b^2)}\;,
         \eqno(16.41)$$
а для одиночного резонатора магнетрона с учётом (16.31),  (16.32)  и
того обстоятельства, что магнитное поле $\rv H$ в цилиндрической части
резонатора не зависит от координат,  получим $D=a/2$ и $Q=a/\delta$,
где $a$ --- расстояние между <<пластинами конденсатора>>.

     На основании соотношения (16.32) и  приведенных  выражений  для
$D$  можно  сделать  некоторые  общие выводы о зависимости добротности
колебаний от размеров резонатора и их частоты.  Проще всего выявляется
зависимость  $Q$ от размеров подобных резонаторов фиксированной формы.
В  этом  случае  все  размеры  пропорциональны  и  определяются  одной
величиной (обозначим её $R$), с которой $k_n$,~$\delta$,~$D$ связаны
соотношениями:  $k_n\sim  R^{-1},\;\delta\sim  R^{1/2},\;D\sim  R$   и
поэтому в результате получаем,  что $Q\sim R^{1/2}$.  Снижение $Q$ при
уменьшении размеров резонатора  является  одной  из  основных  причин,
затрудняющих их использование в миллиметровом диапазоне.

     В резонаторе  заданных  размеров  имеется   общая   тенденция   к
повышению добротности более высокочастотных колебаний, что обусловлено
в первую очередь уменьшением толщины  скин-слоя,  хотя  для  некоторых
колебаний  фактор  формы резонатора и структура поля оказываются более
важными.  В этом смысле следует отметить колебания  вида  $H_{0nl}$  в
круглом   цилиндрическом  резонаторе,  в  которых  в  боковых  стенках
отсутствуют продольные токи и  добротность  которых  заметно  выше  по
сравнению даже с более высокочастотными колебаниями других типов.

\vspace*{-.1cm}
\begin{wrapfigure}[14]{l}{7.8cm}
\begin{picture}(80,55)
\put(0,50){\special{em:graph fig16-1.bmp}}
\end{picture}
\hbox to 7.8cm{\hfil\footnotesize{Рис.~16.1.~Возмущение объёма
резонатора}\hfil}
%\hbox to 7.5cm{\hfil\footnotesize{резонатора.}\hfil}
\end{wrapfigure}

     Однако добротность  реальных  резонаторов   оказывается   заметно
меньшей, чем расчётное  значение, и почти всегда это связано с плохой
обработкой  поверхности  стенок,   требование   к   качеству   которой
повышается с уменьшением $\delta$.  Поэтому реальная величина потерь в
стенках оказывается больше расчётной.  Последние десятилетия широкое
применение  находят сверхпроводящие резонаторы,  у которых зависимость
потерь в стенках от частоты имеет другой закон, что меняет зависимость
$Q$ от размеров резонатора,  да и сами значения добротности возрастают
на  несколько  порядков.  Часто  добротность   записывается   в   виде
$Q=G/R_s$, где $G=4\pi\omega D/c^2$ называется структурным параметром,
а поверхностное  сопротивление  $R_s=1/\sigma\delta\sim\sqrt{f}$  (для
сверхпроводящих  резонаторов в области частот,  представляющей интерес
для электродинамики СВЧ,  оказывается, что $R_s\sim f^2$). Структурный
параметр  (или геометрический фактор) равен,  как правило,  нескольким
сотням Ом.

     В предыдущем  разделе был упомянут метод возмущений для расчёта
собственных частот резонаторов неправильной формы. Основное применение
его  результаты  находят  для  оценки  сдвига  частоты  при  небольшой
деформации стенок,  в  частности,  с  помощью  настроечных  элементов.
Однако  используемый  при  этом  подход  к  задаче имеет более широкое
применение и может быть  положен  в  основу  при  численном  расчёте
резонаторов сложной формы.

     Суть метода   заключается   в  следующем.  Предположим,  что  для
некоторого  резонатора  объёма  $V_0$,  ограниченного
поверхностью $S_0$, известно решение уравнений (16.1) при
граничном условии (16.2), то есть известна система собственных
значений $k_{n0}$  и  собственных функций  ${\rv  E}_{n0},\,{\rv
H}_{n0}$.  Исказим  форму  резонатора, заменив участок поверхности
$S_0'$ участком $S'$,  который  расположен внутри объёма $V_0$,
как это показано на рис.~16.1.  Получившийся в результате новый
"возмущённый" резонатор (его объ\"ем $V$) образован из  старого
пут\"ем  удаления  объёма  $V'$  (показан  штриховкой),
ограниченного поверхностями $S_0'$ и $S'$.  Необходимо  найти
решение уравнений  (16.1)  с  граничным условием (16.2),  но
теперь на границе "возмущённой" поверхности $S$,  состоящей  из
неискажённой  части $S_0$  и  участка $S'$,  ограничивающей
объём $V$ нового резонатора. Будем обозначать это решение теми
же буквами,  но без нижних индексов: $\rv  E,\,\rv H,\,  k$.
Образуем величины $\div [\rv H \rv E_{n0}]$ и $\div [\rv H_{n0}\rv
E]$,  выразим  их  с  помощью  известной  формулы векторной
алгебры $\div[\rv A \rv B]=\rv B \rot\rv A\rv- A\rot\rv B$,
домножим первую на $ik$,  вторую  на  $ik_{n0}$  и  вычтем  из
первой вторую. В результате получим, что
     $$k\div[\rv H\rv E_{n0}]-k_{n0}\div[\rv H_{n0}\rv E]=k\rv E_{n0}
         \rot \rv H -k_{n0}\rv E\rot \rv H_{n0}\;.\eqno(16.42)$$
Преобразуем правую  часть  этого выражения с помощью уравнений (16.1),
проинтегрируем по  объёму  $V_0$  и  воспользуемся  теоремой  Гаусса
(1.16); тогда
     $$(k^2-k_{n0}^2)\int\limits_{V_0} \rv E_{n0}\rv E\,dV=ik\int
         \limits_{S_0} [\rv H\rv E_{n0}]\,d\rv S-ik_{n0}\int\limits_
         {S_0} [\rv H_{n0}\rv E]\,d\rv S\;.\eqno(16.43)$$

     В силу  граничных условий для функций $\rv E_{n0}$ на поверхности
$S_0$ первый интеграл в правой  части  равен  нулю.  Вклад  во  второй
интеграл по той же причине даёт только участок поверхности $S_0'$, к
которому можно добавить участок $S'$,  не дающий вклада в интеграл.  В
результате в правой части стоит интеграл по замкнутой поверхности:
     $$(k^2-k_{n0}^2)\int\limits_{V_0} \rv E_{n0}\rv E\,dV=-ik_{n0}
         \int\limits_{S'+S_0'}[\rv H_{n0}\rv E]d\rv S\,,
         \eqno(16.44)$$
который опять с помощью теоремы Гаусса сводится к виду
     $$(k^2-k_{n0}^2)\int\limits_{V_0} \rv E_{n0}\rv E\,dV=-ik_{n0}
         \int\limits_{V'}\div[\rv H_{n0}\rv E]\, dV\,.\eqno(16.45)$$

     Разложим теперь собственные функции  краевой  задачи  для  нового
резонатора по известным собственным функциям старого:
     $$\rv E=\sum\limits_{m=1}^{\infty}\,A_m\rv E_{m0}\;,
         \eqno(16.46)$$
что всегда возможно,  поскольку последние составляют  полную  систему.
Ввиду ортогональности функций имеем
     $$(k^2-k_{n0}^2)A_n\int\limits_{V_0} \rv E_{n0}^2\,dV=-k_{n0}^2
         \sum\limits_{m=1}^{\infty}A_m\Bigl(\int\limits_{V'}\rv E_{m0}
         \rv E_{n0}\,dV+\frac{k_{m0}}{k_{n0}}\int\limits_{V'}\rv H_
         {m0}\rv H_{n0}\,dV\Bigr)\;.\eqno(16.47)$$
Индекс $n$   в   этом   уравнении   может   пробегать   все  значения:
$n=1,2,\dots$ и, следовательно, (16.47) представляет собой бесконечную
систему  однородных  линейных  алгебраических  уравнений  относительно
коэффициентов $A_n$.  Эта система имеет ненулевое решение  только  при
дискретных   значениях  $k$,  совпадающих  с  нулями  соответствующего
детерминанта матрицы,  составленной из коэффициентов  при  неизвестных
$A_m$.  По найденным собственным значениям легко находятся собственные
функции.

     Отметим, что до сих пор не было  сделано  никаких  приближений  и
фактически был построен метод решения краевой задачи,  сводящий её к
бесконечной   системе   уравнений.   Выведенная   система   аналогична
получаемой  в  методе  частичных  областей.  Однако  приведенный здесь
подход является более  общим,  поскольку  требует  знания  собственных
функций  только  для  одной  простой области.  Любой резонатор сложной
формы  может  быть  представлен  как  результат  деформации  простого.
Получение  системы  уравнения  в  рассматриваемом  методе  сводится  к
интегрированию известных собственных функций по областям, удаляемым из
исходного резонатора.

     Сам метод  возмущений  заключается  в  предположении  о   малости
деформации и е\"е плавности.  Так как в правой части уравнения (16.47)
уже  есть  параметр  малости  (интегрирование  проводится  по   малому
объёму $V'$),  то  в  подынтегральных выражениях достаточно оставить
собственные функции  невозмущённого  резонатора.  Это  соответствует
учёту  в  сумме  (16.46)  только  одного члена с $m=n$ и $A_m=1$.  В
результате характеристические уравнения для собственных  значений  $k$
становятся независимыми и имеют вид
     $$(k^2-k_{n0}^2)\int\limits_{V_0}\rv E_{n0}^2\,dV=-k_{n0}^2
         \biggl(\int\limits_{V'}\rv E_{n0}^2\,dV+
         \int\limits_{V'}\rv H_{n0}^2\,dV\biggr)\;.\eqno(16.48)$$
Поскольку интегралы $\int\limits_{V'}\rv E_{n0}^2\,dV$, ~$\int\limits_
{V'}    \rv   H_{n0}^2\,dV$,~$\int\limits_{V_0}\rv   E_{n0}^2\,dV$   c
точностью до общего  множителя  и  знака  совпадают  соответственно  с
максимальной   электрической   энергией  объёма  $V'$,  максимальной
магнитной энергией объёма $V'$ и максимальной электрической энергией
всего резонатора, то формулу (16.48) можно записать в виде
     $$k^2=k_{n0}^2\Bigl(1+\frac{\Delta U_E-\Delta U_H}{U_0}\Bigr)\;,
         \eqno(16.49)$$
где $\Delta U_E$ и $\Delta U_H$ ---  изменения  максимальных  значений
электрической  и  магнитной  энергии  резонатора  при  возмущении  его
объёма (они отрицательны при уменьшении объема), а $U_0$ --- энергия
всего  резонатора в невозмущённом состоянии.  Формула (16.49) в ряде
случаев даёт возможность легко сделать качественное суждение о знаке
изменения   собственной  частоты  при  той  или  иной  деформации  его
объёма.

     В связи с этим отметим,  что поле в электродинамической структуре
оказывает  давление  на  стенки,  что,  естественно,  в  общем  случае
приводит  к  изменению собственных частот и тогда при заданной частоте
внешнего   генератора    изменяется    ускоряющее    поле.    Действие
пондеромоторных  сил  на  стенки  структуры  может вызвать,  например,
автоколебания системы и хотя вибрация стенок незначительна,  в  случае
высокой  добротности  возникает  модуляция амплитуды ускоряющего поля.
Характер  воздействия  пондеромоторных  сил  зависит  от   того,   как
соотносятся  между  собой  частота  генератора  и  частота резонансной
системы.  Существует также некоторое пороговое  значение  запасённой
энергии,   при   которой   возникают   механические   колебания;   для
сверхпроводящих структур это значение заметно меньше  по  сравнению  с
аналогичными  структурами,  работающими  при  комнатных  температурах.
Механические  колебания  существенны,  например,   в   сверхпроводящих
спиральных  замедляющих  структурах,  когда  при определённом уровне
мощности  величина  изменения  частоты  может   быть   больше   ширины
резонансной кривой.

%\end{document}

\newpage
\oddsidemargin=-0.4mm \evensidemargin=-0.4mm
\topmargin=-0.4mm
\headsep=7mm
\textheight=231.875mm
\textwidth=160mm
\mathsurround=2.5pt
\unitlength=1mm
%\begin{document}
%\input{macr.tex}
\thispagestyle{empty}
%\addtocounter{page}{177}

\begin{center}
   \subsubsection*{\rm Г\,Л\,А\,В\,А\, 6}
      \vspace{-1.15em}
      \line(6,0){160}\\
      \vspace{-1em}
      \line(6,0){160}
      \vspace{-1.15em}
   \subsubsection*{СФЕРИЧЕСКИЕ ВОЛНЫ И ИХ ВОЗБУЖДЕНИЕ}
      \vspace{31mm}
\subsubsection*{17. Излучение сферических волн в безграничном пустом
               пространстве}
\end{center}
\vspace{.5cm}

\markboth{Глава 6. Сферические волны и их возбуждение}{17. Излучение
         сферических волн в безграничном пространстве}

\begin{center}\begin{minipage}[c]{0.75\textwidth}
\footnotesize{\parindent=0.5cm
         Поле элементарного электрического диполя. Сферическая волна в
         ближней  и  дальней зоне.  Функция Грина скалярного волнового
         уравнения. Векторный потенциал поля излучения системы токов в
         пустоте. Сложение сферических волн. Сопротивление излучения.
}\end{minipage}\end{center}
\vspace{.5cm}

     До сих пор рассматривались электромагнитные поля,  которые  всюду
удовлетворяют  однородным  уравнениям  Максвелла.  В  частности,  были
изучены плоские и цилиндрические волны,  а также собственные колебания
замкнутых  полостей  (объёмных  резонаторов).  Вопрос об источниках,
возбуждающих эти поля,  оставался в стороне. В случае волн упомянутого
типа  такой  подход  оправдан  отсутствием  в  их полях особенностей и
равноправностью всех  участков  оси,  вдоль  которой  распространяется
волна.  Поэтому  всегда  можно  полагать,  что  источник расположен за
границей данного участка.  В случае  резонаторов  возбуждённое  поле
продолжает  существовать  в  полости  ещё  долгое время (по временным
меркам,  характерным для данного процесса) после прекращения  действия
источника. Помимо рассмотренных волн существуют и играют важную роль в
теории и в практических приложениях так  называемые  {\it  сферические
волны},  которые  значительно  ближе связаны с источниками,  поскольку
поля этих волн имеют  {\it  особенности}  в  точках,  где  расположены
источники.  Поэтому  структуру  сферических волн целесообразно изучать
одновременно с задачей их возбуждения.

     Простейшей сферической  волной  является  волна,  возбуждаемая  в
неограниченном  пустом пространстве,  то есть в отсутствии других тел,
{\it элементарным диполем} (элементарным электрическим излучателем). В
электростатике  момент электрического диполя определяется соотношением
$\rv d=q\rv l$,  где $\rv l$ --- вектор, начало которого лежит в месте
расположения  заряда $q$,  конец --- в месте расположения заряда $-q$.
Переменному  монохроматическому  диполю  соответствует  ток  $\rv   J$,
направленный  вдоль  $\rv  l$  и  равный  по  величине  $J=dq/dt$  или
$J=-i\omega q$,  то есть $J\rv l=-i\omega\rv d$.  Предельный переход к
элементарному  диполю  $|\rv  l|\to  0,\quad  q\to\infty$  при условии
сохранения произведения этих величин приводит к  плотности  тока
 $$\rv j^e=-i\omega\rv  d\delta(\rv  r-\rv  r_0)\,,\eqno(17.1)$$
 где  $\delta(\rv r-\rv  r_0)$ --- трехмерная $\delta$-функция,
  $\rv r,\;   \rv r_0$ ---  радиусы-векторы соответственно точки наблюдения
 и точки,  в которой расположен диполь. Будем считать плотность тока диполя
 нормированной таким  образом,  что комплексная   амплитуда   плотности
 стороннего  электрического  тока записывается в виде
     $$\rv j^{e}=\rv a\delta(\rv r-\rv r_0)\,,\eqno(17.1\mbox{\textit а})$$
где $\rv  a$  ---  единичный  вектор,  определяющий   ориентацию
диполя.  Для  вычисления  поля,  возбуждаемого  заданным током (17.1)
в пустоте, достаточно найти   векторный   потенциал   $\rv   A$,
удовлетворяющий волновому уравнению (3.22) при $\varepsilon =\mu=1$:
     $$\Delta \rv A+k^2\rv A=-\frac{4\pi} c \rv j^{e}\,.\eqno(17.2)$$
После этого  поля  находятся  дифференцированием  $\rv   A$   согласно
формулам (3.23).

     Удобство такого способа решения задачи обусловлено тем,  что  для
декартовых  компонент  векторное  уравнение  (17.2) распадается на три
независимых  уравнения,  каждое  из  которых  является   скалярным   и
связывает между собой лишь соответствующие компоненты векторов $\rv A$
и $\rv j^{e}$.  Поместим начало координат в точку $\rv r_0$ и направим
ось $z$ вдоль вектора $\rv a$.  Тогда компонента $A_z$   удовлетворяет
скалярному волновому уравнению (уравнению  Гельмгольца)
     $$\Delta A_z+k^2 A_z=-\frac{4\pi} c \rv \delta(\rv r)\,,
         \eqno(17.3)$$
которое удобно  решать  в  {\it  сферической  системе  координат} $(r,
\theta,\varphi)$.  Оператор Лапласа $\Delta$, действующий на скалярную
функцию $u(r,\theta,\varphi)$, имеет вид
     $$\Delta u=\displaystyle{\frac 1{r^2}}\frac{\partial}{\partial r}
         \Bigl (r^2\frac{\partial u}{\partial r}\Bigr ) +\frac 1{r^2
         \sin\theta}\frac{\partial}{\partial \theta}\Bigl (\sin \theta
         \frac{\partial u}{\partial \theta}\Bigr ) +\frac 1{r^2\sin^2
         \theta}\frac{\partial^2u}{\partial\varphi^2}\,.\eqno(17.4)$$

     Правая часть (17.3) не зависит от углов и,  следовательно,  $A_z$
является функцией одной переменной $r$. Всюду, кроме начала координат,
компонента $A_z(r)$  удовлетворяет  однородному  волновому  уравнению,
которое   в   данном  случае  является  обыкновенным  дифференциальным
уравнением и имеет два линейно независимых регулярных на бесконечности
решения
     $$\frac {e^{ikr}} r\qquad\mbox{и}\qquad \frac {e^{-ikr}} r\,,
         \eqno(17.5)$$
линейной комбинацией  которых  может  быть  представлено  его общее
решение.  Дополнительным физическим соображением, позволяющим получить
однозначное  решение  уравнения  (17.3),  является  выбор  направления
движения  фазы  в  соответствующей  решениям  (17.5)  волне.  Учитывая
временной множитель  $e^{-i\omega  t}$,  убеждаемся,  что  поверхности
постоянных  значений  фаз  $\pm\; kr-\omega t= const$,  где верхний и
нижний знаки соответствуют первому и второму решению (17.5),  в первом
случае  движутся в сторону возрастающих значений $r$,  во втором --- в
сторону убывающих.

     Естественно потребовать,  чтобы волна была расходящейся,  то есть
фаза бежала от диполя --- далее будет показано,  что при этом и  поток
энергии  в возбуждаемой волне распространяется от источника излучения.
Это  требование,  совместно  с  требованием   регулярности   поля   на
бесконечности,   может   быть   сформулировано   для   поля  излучения
произвольных источников в неограниченном пустом  пространстве  в  виде
так   называемых  {\it  условий  излучения}  Зоммерфельда  (см.  также
раздел 3):
     $$u\to f(\theta,\varphi)\,\frac{e^{ikr}}r\,,\qquad \frac 1 u\frac{
         \partial u}{\partial r}\to \Bigl(ik-\frac 1 r\Bigr)\,,
         \eqno(17.6\mbox{\textit а})$$
где $u$ --- любая декартова компонента векторного потенциала.

     Условия (17.6{\it а})  позволяют  выделить класс единственных (физически
разумных и согласующихся с  опытом)  решений  неоднородного  волнового
уравнения   в   неограниченных   областях,   внешних  по  отношению  к
ограниченным.  В теории уравнений математической физики эти условия  в
общем случае формулируются следующим образом:
     $$u(x)=O(r^{-1})\,,\qquad\frac{\partial u(x)}{\partial r}-iku(x)=
        o(r^{-1})\,,\qquad r\to\infty\,.\eqno(17.6\mbox{\textit б})$$
где $x=(x_1,x_2,x_3)$,~$r=\sqrt{x_1^2+x_2^2+x_3^2}$                ---
соответственно   точка   и   норма   в  трёхмерном вещественном евклидовом
пространстве.  При $k>0$  всякое   решение  однородного  уравнения
Гельмгольца,  удовлетворяющее второму из условий (17.6{\itб}), удовлетворяет и
первому условию.

     Таким образом,  подходящим по  физическим  соображениям  решением
уравнения (17.3) является
     $$A_z=B\,\frac {e^{ikr}} r\,.\eqno(17.7)$$
Коэффициент $B$  легко  находится  путём  интегрирования  (17.3)  по
объёму  сферы  малого  радиуса  $b\;(kb\ll1)$  с  центром  в  начале
координат.   Так   как  $\Delta=\div\grad$,  то  интеграл  от  первого
слагаемого  согласно  теореме  Гаусса  преобразуется  в  поверхностный
интеграл  от $dA_z/dr$,  который равен $-4\pi B$.  Интеграл от второго
слагаемого имеет порядок $k^2b^2$ (можно  считать,  что  внутри  и  на
поверхности  сферы  $A_z=B/r$) и исчезает при $kb\to 0$.  В результате
$B=1/c$ и искомое решение (17.3) имеет вид
     $$A_z= \frac{e^{ikr}}{cr}\,.\eqno(17.8)$$

     Отсюда дифференцированием  по   формулам   (3.23)   легко   найти
компоненты  полей.  Поскольку  полученное  решение  представляет собой
частный   вид   азимутально   симметричной   сферической   волны,   то
целесообразно  вычислить  {\it  сферические  компоненты полей},  среди
которых три отличны от нуля:
     $$\left.\begin{array} {lll}H_{\varphi}&=&\displaystyle{\frac 1 c
         \Bigl(\frac 1 r-ik\Bigr)\frac{e^{ikr}}r}\sin{\theta}\,,\\
         [.4cm]E_r&=&\displaystyle{\frac{2i}{kc}\Bigl(\frac 1 {r^2}-
         \frac{ik}r\Bigr)\frac{e^{ikr}}r}\cos{\theta}\,,\\[.4cm]E_{
         \theta}&=&\displaystyle{\frac i{kc}\Bigl(\frac 1 {r^2}-\frac
         {ik} r -k^2\Bigr)\frac{e^{ikr}}r}\sin{\theta}\,.\end{array}
         \right\}\eqno(17.9)$$

     На малых расстояниях от диполя,  определяемых условием $kr\ll 1$,
в  так  называемой  {\it  ближней  зоне},  в поле (17.9) преобладающим
является электрическое поле,  компоненты которого с  учетом  выбранной
нормировки плотности тока (17.1) совпадают со статическим полем диполя
(с моментом $d$)
     $$E_r=2d\,\frac{\cos{\theta}}{r^3}\,,\quad E_{\theta}=d\,\frac{\sin
         {\theta}}{r^3}\,,\eqno(17.10)$$
изменяющимся со временем по гармоническому закону; волнового характера
поле в этой зоне не имеет.

     На больших  расстояниях,  при  $kr\gg  1$,  в так называемой {\it
дальней зоне},  компонента  $E_r$  становится  малой  по  сравнению  с
остальными,  так  как  она  убывает  как  $1/r^2$,  а  $E_{\theta}$  и
$H_{\varphi}$  медленнее,  как  $1/r$.  Две   последние   составляющие
стремятся к одному и тому же значению
     $$E_{\theta}=H_{\varphi}=-i\frac k c \,\sin{\theta}\,\frac{e^{ikr}}r
         \,.\eqno(17.11)$$

     Эти предельные  формулы  позволяют  вычислить  мощность излучения
$\Sigma_{\mbox{\textit{изл}}}$,  то есть {\it полный поток
энергии}, уносимый волной. Очевидно,  что  этот поток одинаков для
сферы любого радиуса,  поэтому его достаточно вычислить для сферы
большого радиуса $R$,  для которого выполняется  условие  $kR\gg
1$. Вектор Умова-Пойнтинга (1.22) в этом случае имеет только одну
компоненту
     $$\overline{\gv S}_r=\frac c{8\pi}\re{E_{\theta}H_{\varphi}^*}=
         \frac {k^2}{8\pi c R^2}\,\sin^2{\theta}\,.
         \eqno(17.12)$$
Интегрируя по поверхности сферы, получим
     $$\Sigma_{\mbox{\footnotesize\textit{изл}}}=
       \oint \overline{\gv S}_r\,dS=\frac {k^2}{3c}\,,\eqno(17.13)$$
где элемент поверхности сферы $dS=R^2\sin{\theta}\,d\theta\, d\varphi$.

     Характерным для  сферической  волны  является  её  поведение на
больших расстояниях от источника.  В  дальней  зоне  поле  сферической
волны   в  каждой  точке  поперечное:  векторы  электрического  и
магнитного  поля  лежат  в   плоскости,   нормальной   к   направлению
распространения, ортогональны между собой и равны по величине, то есть
локально волна является  плоской.  Однако  при  этом  компоненты  поля
убывают как $1/r$ и зависят,  вообще говоря, от обоих углов $\theta$ и
$\varphi$.  Возбуждаемая элементарным электрическим диполем  волна  не
содержит   компоненты  $H_r$  во  всем  пространстве.  По  аналогии  с
цилиндрической волной,  поле которой не содержит $H_z$,  её  принято
называть  сферической волной типа TM.  Поля любой сферической TM волны
могут  быть  выражены  через  одну  скалярную  потенциальную  функцию.
Существуют и сферические TE волны,  не содержащие компоненты $E_r$,  и
их поля также выражаются через одну потенциальную  функцию;  такую
волну представляет собой,  в частности,  поле элементарного магнитного
диполя.

     Отвлекаясь пока  от  физического  смысла понятия магнитного тока,
сопоставим  монохроматическому   диполю   с   комплексной   амплитудой
магнитного момента $\rv m$ плотность {\it магнитного тока}
     $$\rv j^m=-i\omega \rv m\delta(\rv r-\rv r_0)\,.\eqno (17.14)$$
Нормируя эту  плотность  по аналогии с (17.1) на единичный вектор $\rv
a$,  получим согласно (3.32) волновое уравнение для  {\it  магнитного}
векторного потенциала $\rv A^m$ в том же виде (17.3),  как и уравнение
для   обычного   (электрического)   векторного   потенциала  в  случае
электрического диполя.  Поэтому при расположении магнитного  диполя  в
начале  координат  и  ориентации  $\rv  m$  вдоль  оси $z$ решение для
компоненты $A_z^m$ имеет тот же вид (17.8).  Однако теперь  компоненты
поля  следует вычислять по формулам (3.33),  из чего следует,  что они
могут быть получены из (17.9) заменой
     $$\rv H\to-\rv E^m\,,\quad \rv E\to \rv H^m\,.\eqno(17.15)$$
В результате  для  элементарного  магнитного   диполя   (элементарного
магнитного излучателя) получаем
     $$\left.\begin{array} {lll}E^m_{\varphi}&=&\displaystyle{-\frac 1
         c\Bigl(\frac 1 r-ik\Bigr)\frac{e^{ikr}}r}\sin{\theta}\,,\\
         [.4cm]H^m_r&=&\displaystyle{\frac{2i}{kc}\Bigl(\frac 1 {r^2}-
         \frac{ik} r\Bigr)\frac{e^{ikr}}r}\cos{\theta}\,,\\[.4cm]
         H^m_{\theta}&=&\displaystyle{\frac i{kc}\Bigl(\frac 1 {r^2}-
         \frac{ik} r -k^2\Bigr)\frac{e^{ikr}}r}\sin{\theta}\,.\end
         {array}\right\}\eqno(17.16)$$
В дальней зоне преобладающими  являются компоненты
     $$E^m_{\varphi}=-H^m_{\theta}=i\frac k c \sin{\theta}\frac{e^{ikr}
         }r\,,\eqno(17.17)$$
так что полная мощность излучения даётся той же формулой (17.13).

     Уравнение (17.2)  определяет  векторный  потенциал  $\rv  A$,  а,
следовательно,  и  поле,  создаваемое произвольной системой токов $\rv
j^{e}$.  Каждая декартова компонента $\rv A$ удовлетворяет  скалярному
волновому уравнению вида
     $$\Delta u+k^2 u=-f\,,\eqno(17.18)$$
решение которого  в неограниченном пустом пространстве удобно искать с
помощью  {\it  основной  функции  Грина}  $G(\rv  r,\rv  r_0)$   этого
уравнения. Последняя удовлетворяет уравнению
     $$\Delta G+k^2 G=-\delta(\rv r-\rv r_0)\,,\eqno(17.19)$$
где оператор $\Delta$ действует на координаты точки наблюдения $\rv r$
($\rv r_0$ --- радиус-вектор  точки  расположения  источника,  который
играет  роль  параметра),  а  также  удовлетворяет  условиям излучения
(17.6). При этом подразумевается отсутствие во всем пространстве каких
либо   граничных  поверхностей,  на  которых  наложены  дополнительные
условия (например,  металлических тел с требованием обращения  в  нуль
тангенциальной  компоненты  электрического  поля).  Выше  эта  функция
фактически была уже найдена при вычислении поля  элементарного  диполя
(17.1) в пустом пространстве:
     $$G(\rv r,\rv r_0)=\frac 1{4\pi|\rv r -\rv r_0|}\, e^{ik|
         \rv r -\rv r_0|}\,.\eqno(17.20)$$
Отметим, что функция $G(\rv r,\rv r_0)$ симметрична по переменным
$\rv  r$  и $\rv  r_0$;  это свойство  функции Грина сохраняется и тогда,
когда в пространстве находятся проводящие тела. В этом случае
уравнение (17.19) не полностью определяет функцию $G(\rv r,\rv  r_0)$,
необходимо ещё поставить дополнительные граничные условия.

     Для любых  двух функций $u_1$ и $u_2$,  удовлетворяющих уравнению
(17.18), справедливо очевидное равенство
     $$\div(u_1\grad u_2-u_2\grad u_1)=u_2f_1-u_1f_2\,.\eqno(17.21)$$
Его легко получить,  умножив уравнение для $u_2$ на $u_1$ и  уравнение
для  $u_1$  на $-u_2$ и сложив результаты.  Проинтегрировав (17.21) по
произвольному объёму $V$, получим вторую формулу Грина:
     $$\oint\limits_S\left(u_1\frac{\partial u_2}{\partial\rv n}-u_2
         \frac{\partial u_1}{\partial \rv n}\right)\,dS=\int\limits_V
         (u_2f_1-u_1f_2)\,dV\,,\eqno(17.22)$$
где $\rv  n$  ---  внешняя  нормаль к поверхности $S$,  ограничивающей
объем $V$.

     Возьмём в  качестве $u_1$ и $f_1$ функции $u$ и $f$ из (17.18),
а в качестве $u_2$ и  $f_2$  ---  основную  функцию  Грина  (17.20)  и
$\delta(\rv r-\rv r_0)$,   причём    будем   далее   считать   $\rv   r_0$
радиус-вектором точки наблюдения, а $\rv r$ --- радиус-вектором точек,
по   которым   в   (17.22)   производится   интегрирование. Используя
основное свойство $\delta$-функции, получаем для произвольного
 объема  $V$, ограниченного замкнутой поверхностью $S$, интегральное
 уравнение

 $$u(\rv r_0)=\int\limits_V G(\rv r,\rv r_0)f(\rv r)\,dV-\oint
         \limits_S\left(u\frac{\partial G}{\partial{\rv n}}-G\frac{
       \partial u}{\partial{\rv n}}\right)\,dS\,,\eqno(17.23)$$
эквивалентное волновому уравнению (17.18).

Возьмём в качестве  объема  $V$ пространство, ограниченное сферой
и включающее в себя все источники, то есть вне этой сферы $f(\rv r)\equiv 0$.
При стремлении радиуса этой сферы к бесконечности поверхностный интеграл
в (17.23) обращается в  нуль, так как обе функции $u$ и $G$ удовлетворяют
условиям излучения (17.6). В результате для всего пустого пространства
получаем решение  волнового уравнения (17.18) в квадратурах
   $$u(\rv r_0)=\int\limits_V G(\rv r,\rv r_0)f(\rv r)\,dV\,.    \eqno(17.24)$$

         Следовательно, решение уравнения (17.2) для ограниченной
системы  токов  в безграничном пустом пространстве представляется в
виде интеграла по всем токам:
 $$\rv A(\rv r_0)=\frac 1 c\int \rv j^{e}(\rv r)\frac {e^{ik|\rv r
         -\rv r_0|}}{|\rv r-\rv r_0|}\,dV\,.\eqno(17.25)$$
Подчеркнём лишний раз,  что эта формула связывает между собой только
соответствующие {\it декартовы} компоненты $\rv A$ и $\rv j$.

   Рассмотрим теперь свойства входящего в (17.23) интеграла по произвольной
замкнутой поверхности $S$ в случае непрерывной функции $u$,
удовлетворяющей уравнению (17.18).   Пусть в  некоторой  ограниченной
области  пространства   имеются сторонние  токи.  На рис.~17.1 занятая этими
токами часть пространства показана   штриховкой.   Проведём   три   взаимно
непересекающиеся поверхности:  $\Sigma$ -- сферу бесконечно большого радиуса,
$S_1$ -- замкнутую поверхность произвольной формы,  охватывающую  только
часть токов,  занимающих  объём  $V_1$,  и  замкнутую  поверхность  $S_2$,
заключающую внутри себя все  токи.  Выделим  три  точки  пространства:
$r_1$,   $r_2$   и   $r_3$,   расположенные   относительно   указанных
поверхностей так, как показано на рисунке. Во всех точках пространства
$\rv  r_0$,  в  том  числе $r_1\,,r_2\,,r_3$ и на обеих поверхностях $S_1$ и $S_2$,  при
выполнении условия излучения (17.6) точные значения функции $u$ согласно
(17.24) представляются в виде
     $$u(\rv r_0)=\int\limits_{V_1+V_2} G(\rv r,\rv r_0)f(\rv r)\,dV\,.
         \eqno(17.24\textit а)$$
Но  в соответствии с формулой (17.22)
функция  $u(\rv  r_0)$  может   быть   выражена   и   с   привлечением
поверхностных  интегралов.  Поскольку функция  $u(\rv  r_0)$  непрерывная,  то
нет принципиальных затруднений при вычислении на поверхностях $S_1$ и
$S_2$  производных $\partial  u/\partial  \rv  n$.  При  этом  выражения  для
выделенных точек имеют разный вид.

     В точке $\rv r_1$ функция $u$ может  быть  выражена помимо (17.24\textit а)
ещё двумя способами:  либо как интеграл по объёму $V_1$ и по поверхности
$S_1$,  либо как  интеграл  по  суммарному  объёму  $V_1+V_2$  и  по
поверхности  $S_2$ ).  Чтобы все три выражения для $u(\rv r_1)$  давали  одно  и
то  же  значение, необходимо,  чтобы интеграл по объёму $V_2$ совпадал с
интегралом (со знаком минус) по поверхности $S_1$,  а интеграл по поверхности $S_2$
равнялся  нулю.

 В точках $\rv r_2$ функция $u$ может  быть  выражена помимо (17.24\textit а)
также ещё двумя способами: либо как интеграл по объёму $V_2$ и по двум
поверхностям $S_1$ и $S_2$  ,  либо как  интеграл  по  суммарному  объёму
$V_1+V_2$  и  по поверхности  $S_2$ ). Отметим, что для точек $\rv r_2$ интеграл
по  $S_1$ отличается знаком от соответствующего интеграла для точек  $\rv r_1$,
поскольку направление внешней нормали к поверхности меняется на
противоположное. И для точек  $\rv r_2$ все три выражения для $u$ совпадают
при выполнении перечисленных выше условий для точек  $\rv r_1$.

%\vspace{-.1cm}
\begin{wrapfigure}[12]{l}{7.5cm}
\begin{picture}(80,50)
\put(0,50){\special{em:graph fig17-1.bmp}}
\end{picture}
\hbox to 7.5cm{\hfil\footnotesize{Рис.~17.1.~К выяснению физического
}\hfil}
\hbox to 7.5cm{\hfil\footnotesize{смысла поверхностного интеграла.}
\hfil}
\end{wrapfigure}

В точках $\rv r_3$, лежащих между    поверхностями  $S_2$  и  $\Sigma$,
 источников нет ($f(\rv r)\equiv 0$) и
     $$u(\rv r_3)=-\oint\limits_{S_2}\left(u\frac{\partial G}{\partial
         {\rv n}}-G\frac{\partial u}{\partial{\rv n}}\right)\,dS\,
         \eqno(17.26)$$
(интеграл по $\Sigma$ равен нулю из-за одинакового  поведения  функций
$u$ и $G$ на бесконечности). Из требования совпадения значений функции
$u$ по формулам  (17.24\textit а) и (17.26) следует, что интеграл по поверхности
$S_2$  представляет собой разрывную функцию: внутри этой поверхности он
равен нулю,  а вне --- с точностью до знака совпадает с интегралом в (17.24\textit а).

 Необходимо   особо подчеркнуть,  что  результат вычислений по формуле (17.26)
будет правильным  только в том случае, когда значения $u$ и $\partial u/\partial
\rv n$ на поверхности в (17.26) точные,  например,  вычисленные с помощью
 формулы (17.25).  При задании произвольных значений этих величин на
поверхности   $S_2$   (17.26)   определяет   функцию,  удовлетворяющую однородному
волновому уравнению,  но её значения в точках вблизи поверхности  не
стремятся к заданным, а в точках внутри $S_2$ не равны нулю.

     Всё сказанное свидетельствует о том,  что поверхностный интеграл в (17.23)
в любой точке как внутри,  так и  вне замкнутой  поверхности  $S$  представляет
вклад  в  поле  источников, расположенных  по  другую  сторону   поверхности.

     Возможность вычислить   с   помощью   формулы   (17.26)
функцию $u$ в некоторой области без источников по {\it
точным}  значениям  $u$  и  $\partial  u/\partial\rv   n$   на
поверхности   лежит   в   основе   {\it   принципа Гюйгенса-Кирхгофа},
 находящего  широкое применение  в  теории дифракции.  Решение ищется
 путем замены неизвестных точных значений на граничной поверхности
 на   приближённые,   и близость полученного таким образом решения во
 всей области к  точному определяется тем, насколько удачно  были  выбраны
 эти   приближённые  значения.  Более подробно  об  этом  будет  сказано  далее
 при рассмотрении векторного аналога формулы (17.26).

     Существенным моментом,  определяющим важность формулы (17.24) для
практических расчётов     антенных     устройств,     является    то
обстоятельство,  что в  дальней  зоне  результат  сложения  нескольких
сферических  волн,  имеющих  особенности в разных точках пространства,
представляет собой также сферическую волну.  Убедимся в этом и попутно
уточним  понятие  дальней  или,  иначе,  волновой  зоны  для излучения
системы токов.   Введём  сферическую  систему  координат  с  центром
где-нибудь в пределах области токов.  Пусть $(r_0,\theta_0,\varphi_0)$
---  координаты точки наблюдения,  определяемой вектором $\rv r_0$,  а
$(r,\theta,\varphi)$  ---  координаты  точки   $\rv   r$,   являющейся
переменной  интегрирования  в  (17.24).  Тогда  расстояние между этими
точками
     $$|\rv r -\rv r_0|=r_0\Bigl(1-2\frac r{r_0}\cos \gamma +\frac
         {r^2}{r_0^2}\Bigr )^{1/2}\,,\eqno(17.27)$$
где $ \gamma$ --- угол между векторами $ \rv r_0$ и $\rv r$:
     $$\cos{\gamma}=\cos{\theta}\cos{\theta_0}+\sin{\theta}\sin{
         \theta_0}\cos{(\varphi-\varphi_0)}\,.\eqno(17.28)$$

     Определим линейный размер $D$  области,  занятой токами (размер антенны),
условием $r\leqslant D$.  Тогда в дальней зоне во всяком случае должно
быть выполнено условие
     $$r_0\gg D\,,\eqno(17.29)$$
которое позволяет  разложить  (17.27)  по  степеням  малого  отношения
$r/r_0$:
     $$|\rv r -\rv r_0|=r_0\left[1- \frac r{r_0}\cos{\gamma}+O\Bigl(
         \frac{r^2}{r_0^2}\Bigr )\right ]\,.\eqno(17.30)$$
В дальней  зоне  в  знаменателе  подынтегрального  выражения в (17.24)
достаточно ограничиться первым членом этого разложения,  в  показателе
экспоненты  может  оказаться  необходимым  удерживать  и  второй.  Это
зависит  от  соотношения  размера  $D$   и   длины   волны   излучения
$\lambda=2\pi/k$.     При    $D\ll\lambda$    показатель    экспоненты
$kr\cos\gamma$ малая величина и (17.25) можно записать в виде
     $$\rv A(\rv r_0)=\frac{e^{ikr_0}}{cr_0}\int\rv j^e(\rv r)\,dV
         \,.\eqno(17.31)$$

     С помощью интегрирования по частям
     $$\int\rv j^e(\rv r)\,dV=-\int\rv r\div{\rv j^e(\rv r)}\,dV=
         -i\omega\int \rv r \rho(\rv r)\,dV\,,\eqno(17.32)$$
где использовано уравнение непрерывности
     $$i\omega\rho(\rv r)=\div \rv j^e(\rv r)\,,\eqno(17.33)$$
векторный потенциал преобразуется к виду
     $$\rv A(\rv r_0)=-ik\rv d\frac {e^{ikr_0}}{r_0}\,,\eqno(17.34)$$
причём
     $$\rv d=\int\rv r\rho(\rv r)\,dV\,\eqno(17.35)$$
--- электрический дипольный момент системы.

     Таким образом,   при   условии   $D\ll\lambda$    дальняя    зона
определяется  условием  $r_0\gg\lambda$  и преобладающим является {\it
электрическое дипольное излучение}. В случае равенства нулю дипольного
момента   (17.35)  излучение  будет  определяться  следующими  членами
разложения экспоненты в (17.25),  первыми среди которых выступают {\it
магнитное дипольное} и {\it электрическое квадрупольное} поля.

     Отметим, что при расчёте реальных антенных устройств такое {\it
мультипольное} разложение экспоненты в (17.25) не представляет интереса,
поскольку  для  них  практически всегда   выполняется   обратное   условие:
$D\gg\lambda$.   Иначе   не   представляется   возможным    обеспечить
достаточную  мощность  излучения и острую диаграмму направленности.  В
этом случае $kr$ большая величина и разложение экспоненты в  степенной
ряд даёт мало пользы.  Дальняя зона определяется теперь возможностью
пренебрежения  третьим  членом  в  правой  части  (17.30) и сводится к
условию
     $$r_0\gg kD^2\,.\eqno(17.36)$$
Следовательно, граница дальней  зоны  лежит значительно  дальше от
источников излучения,   чем следует из простого условия $r_0\gg D$.

     Итак, при $D\gg\lambda$ векторный потенциал $\rv A$ в дальней зоне
 может быть записан в виде
     $$\rv A= \frac {e^{ikr_0}}{cr_0}\rv P\,,\eqno(17.37)$$
где вектор $\rv P$,  определяющий зависимость поля излучения от углов,
даётся интегралом
     $$\rv P(\theta_0,\varphi_0)=\int\limits_V \rv j(\rv r)\,e^{-ikr
         \cos{\gamma}}\,dV\,,\eqno(17.38)$$
и, следовательно,  как  и  в  длинноволновом  случае,  поле  излучения
системы  токов  определяется  векторным  сложением  токов элементарных
источников.  При этом необходимо учитывать  разность  фаз  сферических
волн,   приходящих  от  разных  элементов  тока.  Прямые,  по  которым
отсчитываются фазы для дальней зоны, параллельны между собой.

     Для вычисления полей $\rv E$ и $\rv H$ в дальней зоне по формулам
(3.23) (в сферической системе координат)  достаточно  сохранять  члены
порядка  $1/r_0$,  отбрасывая  члены  $\sim1/r_0^2$  и  более  высоких
степеней,  а потому дифференцировать только множитель $e^{ikr}$  и  не
учитывать  производных  по  $\theta_0,\,  \varphi_0$.  Компоненты поля
$E_r,\,H_r$ малы ($\sim 1/r_0^2$),  а  остальные  в  этом  приближении
записываются  через сферические компоненты вектора $\rv P$ в следующем
виде:
     $$E_{\theta}=H_{\varphi}=-i\frac k c P_{\theta}\frac{e^{ikr_0}}
         {r_0},\quad E_{\varphi}=-H_{\theta}=-i\frac k c P_{\varphi}
         \frac{e^{ikr_0}}{r_0}\,.\eqno(17.39)$$

     Таким образом,  при  любом  распределении токов их поле в дальней
зоне представляет собой сферическую волну.  Выбор начала координат  не
сказывается  на угловой зависимости полей,  определяемой вектором $\rv
P$ (17.38),  а приводит к появлению дополнительного  экспоненциального
множителя  в  (17.37),  компенсирующего  изменение расстояния до точки
наблюдения.

     Поле элементарного   диполя  (17.8)  представляет  собой  частный
случай   (17.37),   для   которого   $\rv   P=\{0,0,1\}$   и   поэтому
$P_{\theta}=-\sin\theta_0$.  Для  произвольной  антенны $P_{\theta}$ и
$P_{\varphi}$ --- некоторые,  вообще говоря, комплексные функции углов
$\theta_0\,,\varphi_0$,  определяющие  {\it диаграмму направленности},
поляризацию и полную излучаемую мощность
     $$\Sigma_{\mbox{\footnotesize\textit{изл}}}=\frac {k^2}{8\pi c}\int |P(\theta_0,\varphi_0)|^2
         \sin{\theta_0}\,d\theta_0\,d\varphi_0\,.\eqno(17.40)$$
Если диполь  образован  током  $J_0$,  протекающим  по  прямолинейному
отрезку  тонкого провода длиной $l$,  расположенного вдоль оси $z$,  и
при этом $kl\ll 1$,  то $P_z=-J_0 l\sin{\theta_0}$, и  полная  мощность
излучения    согласно    (17.40)    составит   $k^2l^2J_0^2/3c$.   Для
характеристики этой величины введём посредством соотношения
     $$\Sigma_{\mbox{\footnotesize\textit{изл}}}=\frac 1 2 RJ_0^2\eqno(17.41)$$
независящий от   тока   $J_0$   коэффициент   $R$,   называемый   {\it
сопротивлением излучения};  для рассматриваемого диполя он оказывается
равным
     $$R=\frac 2 {3c} (kl)^2\,.\eqno(17.42)$$
Величины, имеющие  размерность  сопротивления,  удобнее представлять в
СИ,  для чего достаточно заменить  величину  $1/c$  множителем  30.  В
результате для диполя
     $$R=20(kl)^2\;[\mbox{Ом}]\,;\eqno(17.43)$$
омическое сопротивление провода на практике,  как правило,  исчезающе
мало по сравнению с сопротивлением излучения $R$.

     Если элемент  тока  свернуть  в  кольцо  радиуса  $a=l/2\pi$,  то
возбуждаемое в дальней зоне поле совпадает с рассмотренным выше  полем
{\it магнитного диполя} с моментом $m=\pi a^2J_0/c$,  расположенного в
центре  кольца  и   направленного   перпендикулярно   его   плоскости.
Действительно, считая кольцо расположенным в плоскости $xy$, получаем:
     $$j_x=-j_{\varphi}\sin{\varphi}\,,\quad j_y=j_{\varphi}\cos{\varphi }\,,
     \eqno(17.44)$$
где
     $$j_{\varphi}=\frac{J_0} r\,\delta(r-a)\,\delta(\theta-\frac \pi 2)\,.
         \eqno(17.45)$$
Подставляя (17.44),  (17.45)  в  (17.38)   и учитывая,  что $ka\ll 1$,
находим в результате интегрирования
     $$P_x=-ik\pi a^2 J_0\sin{\theta_0}\sin{\varphi_0}\,,\quad P_y=ik
         \pi a^2 J_0\sin{\theta_0}\cos{\varphi_0}\,,\eqno(17.46)$$
так что из сферических компонент отлична от нуля только одна:
     $$P_{\varphi}=ik\pi a^2 J_0\sin{\theta_0}\,.\eqno(17.47)$$

     Определяемое этой  компонентой  согласно  формулам  (17.39)  поле
совпадает   с   полем  элементарного  магнитного
диполя с плотностью $j^m= -i\omega m \delta(r)$ магнитного тока. Таким
образом,  возможно  приближённое  моделирование  магнитного диполя с
помощью  токового  кольца.  Полное  излучение  (а,  следовательно,   и
сопротивление  излучения)  токового  кольца  отличается  от  излучения
соответствующего прямолинейного тока дополнительным  малым  множителем
$(ka/2)^2$.  Таков  результат взаимной частичной компенсации излучения
от элементов тока, расположенных на одном диаметре.

%\end{document}

Результат
интегрирования существенным образом зависит от  положения  точки  $\rv
r_0$.  Если  она  расположена  внутри  объема  $V$,  то  на  основании
основного свойства $\delta$-функции получаем
     $$u(\rv r_0)=\int\limits_V G(\rv r,\rv r_0)f(\rv r)\,dV-\oint
         \limits_S\left(u\frac{\partial G}{\partial{\rv n}}-G\frac{
       \partial u}{\partial{\rv n}}\right)\,dS\,.\eqno(17.23)$$
Если же точка $\rv r_0$ находится вне поверхности $S$, то правая часть
этого  равенства  обращается в нуль.  Поверхностный интеграл в (17.23)
представляет собой функцию $\rv r_0$,  которая непрерывна  как  внутри
области  $V$,  так  и  вне  её,  но терпит разрыв при переходе через
поверхность $S$. Если объём $V$ заключает в себе все сторонние токи,
то  интеграл  для внутренних точек области равен нулю,  если же не все
(или не включает вообще), то он определяет вклад в поле от источников,
расположенных вне произвольно выбранной поверхности $S$.  При удалении
поверхности на бесконечность интеграл стремится к нулю,  поскольку обе
функции $u$ и $G$ должны удовлетворять условиям излучения.  Именно это
обстоятельство позволяет получить решение задачи о  поле  ограниченной
системы  токов  в безграничном пустом пространстве в виде интеграла по
всем токам:

Основным в  изложенном  методе  решения уравнения (17.2) является
применение функции Грина  (17.20)  и  использование  формулы  (17.23).
Отметим, что  только  обращение в нуль поверхностного интеграла в
этой  формуле  позволяет  получить  решение  (17.24),   сводящееся   к
квадратурам.   Без   этого  (17.23)  представляет  собой  интегральное
уравнение,  поскольку значения $u$ и  $\partial  u/\partial\rv  n$  на
поверхности  $S$  не могут быть заданы произвольно.  Поясним сказанное
следующим рассуждением.
(вторым способом выражается и $u(\rv r_2)$
 Результат
интегрирования существенным образом зависит от  положения  точки  $\rv
r_0$.  Если  она  расположена  внутри ,  то  на  основании

Если же точка $\rv r_0$ находится вне поверхности $S$, то правая часть
этого  равенства  обращается в нуль.  Поверхностный интеграл в (17.23)
представляет собой функцию $\rv r_0$,  которая непрерывна  как  внутри
области  $V$,  так  и  вне  её,  но терпит разрыв при переходе через
поверхность $S$. Если объём $V$ заключает в себе все сторонние токи,
то  интеграл  для внутренних точек области равен нулю,  если же не все
(или не включает вообще), то он определяет вклад в поле от источников,
расположенных вне произвольно выбранной поверхности $S$.  При удалении
поверхности на бесконечность интеграл стремится к нулю,  поскольку обе
функции $u$ и $G$ должны удовлетворять условиям излучения.  Именно это
обстоятельство позволяет получить решение задачи о  поле  ограниченной
системы  токов  в безграничном пустом пространстве в виде интеграла по
всем токам: 

\newpage
\oddsidemargin=-0.4mm \evensidemargin=-0.4mm \topmargin=-0.4mm
\headsep=7mm \textheight=231.875mm \textwidth=160mm
\mathsurround=2.5pt \unitlength=1mm
%\begin{document}
%\input{macr.tex}
\thispagestyle{empty}
%\addtocounter{page}{188}

\begin{center}\subsubsection*{18.~Лемма  Лоренца  и  её   следствия}
         \end{center}
\vspace*{0.5cm}

\markboth{Глава 6.~Сферические  волны  и   их   возбуждение}{18.~Лемма
         Лоренца и её следствия}

\begin{center}\begin{minipage}[c]{0.75\textwidth}
\footnotesize{\parindent=0.5cm
        Лемма Лоренца.  Теоремы  взаимности.  Магнитные токи.  Принцип
        эквивалентности.  Векторные функции Грина уравнений Максвелла.
        Поле системы токов в безграничном пространстве.
}\end{minipage}\end{center}\vspace*{0.5cm}

     В предыдущем разделе была решена задача о поле заданных  токов  с
помощью  неоднородного  скалярного  волнового уравнения для декартовых
компонент  векторного  потенциала.  Такой   метод   в   ряде   случаев
оказывается неудобным,  хотя  бы  потому,  что для вычисления реальных
физических     полей  приходится   проводить   дополнительное
дифференцирование.    При   наличии   же   вблизи   излучающих   токов
поверхностей,  на которых резко изменяются свойства  среды,  например,
антенных  отражателей,  использование скалярного уравнения оказывается
недостаточным. Приходится работать с векторными функциями, причём во
многих  случаях  удобнее решать сами уравнения Максвелла.  Методически
векторный  способ  решения  {\it  граничных  задач}  очень  близок   к
изложенному  выше  для  скалярного  уравнения  и  в простейших случаях
приводит к замкнутым выражениям в виде квадратур для  самих  компонент
поля.   Развитие   этого   аппарата   позволяет   получить  ряд  общих
результатов, дающих возможность строить приближённые решения сложных
задач дифракции и излучения волн в разнообразных структурах.

     В основе векторного метода решения перечисленных задач лежит {\it
лемма Лоренца},  которая устанавливает связь между двумя произвольными
решениями неоднородных уравнений Максвелла.  Приведём её  вывод  в
наиболее общем виде,  для чего целесообразно с самого начала формально
ввести в уравнения {\it магнитные токи}:
     $$\left.\begin{array}{lll}\rot{\rv H}&=&-ik\varepsilon\rv E+
         \displaystyle{\frac{4\pi}c}\rv j^{e}\,,\\[.3cm]\rot{\rv E}&=
         &\phantom{-}ik\mu\rv H-\displaystyle{\frac{4\pi}c}\rv j^{m}\,
         .\end{array}\right\}\eqno(18.1)$$
Уравнения записаны  для  произвольной  линейной  среды,  в  том  числе
неоднородной и анизотропной (то есть $\varepsilon$ и $\mu$ не  зависят
от значений полей $\rv E$ и $\rv H$, но могут быть функциями координат
и тензорами).

     Тогда между  двумя  произвольными  решениями  этих  уравнений для
одной и той же среды,  то есть полями  $\rv  E_1$,~$\rv  H_1$  и  $\rv
E_2$,~$\rv H_2$, возбуждаемыми соответственно токами $\rv j^e_1$,~$\rv
j^m_1$ и $\rv j^e_2$,~$\rv j^m_2$,  существует соотношение, являющееся
векторным аналогом формулы (17.21):
     $$\div\{[\rv E_1\rv H_2]-[\rv E_2\rv H_1]\}=\frac{4\pi} c(\rv j^e
         _1\rv E_2-\rv j^e_2\rv E_1-\rv j^m_1\rv H_2+\rv j^m_2\rv H_1)
         \,.\eqno(18.2)$$

     Его легко  можно  получить,  если  умножить  первое  из уравнений
Максвелла (18.1) для полей с индексом (1) на $\rv E_2$,  второе --- на
$\rv H_2$, затем умножить первое из уравнений для полей с индексом (2)
на $-\rv E_1$,  второе на $-\rv H_1$ и,  наконец,  сложить все  четыре
равенства,  воспользовавшись  при этом векторной формулой (1.5).  Если
теперь  проинтегрировать  (18.2)  по  произвольному  объему  $V$,   то
согласно теореме Гаусса (1.10) получим:
     $$\oint\limits_S\{[\rv E_1\rv H_2]-[\rv E_2\rv H_1]\}\,d\rv S=
         \frac{4\pi}c\int\limits_V(\rv j^e_1\rv E_2-\rv j^e_2\rv E_1-
         \rv j^m_1\rv H_2+\rv j^m_2\rv H_1)\,dV\,,\eqno(18.3)$$
где интеграл слева взят по поверхности $S$,  ограничивающей объем $V$,
а векторный элемент поверхности $d\rv S$ направлен по внешней нормали.

     Две последние   формулы   и   называются  {\it  леммой  Лоренца},
соответственно в дифференциальной и интегральной форме. Отметим, что с
математической точки зрения (18.3) представляет собой векторный аналог
второй теоремы Грина
     $$\int\limits_V (\rv Q\rot\rot{\rv P}-\rv P\rot\rot\rv Q)\,dV=
         \oint\limits_S\{[\rv P\rot\rv Q]-[\rv Q\rot\rv P]\}\,d\rv S
         \,,\eqno(18.4)$$
где произвольные векторные функции $\rv P,  \rv Q$  имеют  непрерывные
вторые   производные  в  области  $V$,  для  частного  случая  функций,
удовлетворяющих  уравнениям  (18.1).  Использовавшееся  в   предыдущем
разделе  соотношение  (17.5)  также  представляет собой частный случай
второй  теоремы   Грина   для   скалярных   функций,   удовлетворяющих
неоднородному уравнению Гельмгольца.

     Уточним прежде   всего  {\it   условия  применимости}  леммы Лоренца.
Изложенный выше  вывод  леммы подразумевал  взаимное  сокращение
слагаемых   в   выражениях
     $$\rv  E_1\varepsilon\rv E_2-\rv E_2\varepsilon\rv E_1\,,\quad
         \rv H_1\mu\rv H_2-\rv H_2\mu\rv H_1\,.\eqno(18.5)$$
Для этого  необходимо,  во-первых,  чтобы  $\varepsilon$  и  $\mu$  не
зависили от полей,  то есть среды были линейными,  и, во-вторых, чтобы
$\varepsilon$ и $\mu$ были скалярами или симметричными тензорами.  Для
сред,  у  которых тензоры проницаемостей имеют антисимметричную часть,
формулировка  леммы  требует  модификации.  Из  таких   сред   широкое
применение в технике СВЧ находят ферриты.

     Лемма Лоренца имеет многообразные следствия.  Начнем с вывода  на
её  основе  так  называемых {\it теорем взаимности} для элементарных
диполей, позволяющих существенно упростить решение многих практических
задач электродинамики.  Пусть все источники (сторонние токи) находятся
в  конечной  области  пространства,  то   есть   нет   приходящих   из
бесконечности волн. Применим соотношение (18.3) ко всему пространству;
тогда поверхностный интеграл исчезнет,  так как на больших расстояниях
поля  $\rv  E_1,\,\rv  H_1$  и  $\rv E_2,\,\rv H_2$ представляют собой
уходящие  волны   одинаковой   структуры.   Исчезает   он   также   на
металлических   поверхностях,   так   как   нормальная  к  поверхности
компонента $[\rv E\rv H]$ содержит только тангенциальные к поверхности
компоненты  $\rv E$.  Таким образом,  для неограниченного пространства
лемма Лоренца сводится к соотношению
     $$\int(\rv j^e_1\rv E_2-\rv j^e_2\rv E_1-\rv j^m_1\rv H_2+\rv j^m
         _2\rv H_1)\,dV\,=0\,,\eqno(18.6)$$
где интеграл вычисляется по области, содержащей все токи.

     Пусть теперь  имеются  два  электрических диполя с моментами $\rv
d_1$ и $\rv d_2$,  расположенные соответственно в точках  $\rv
r_1$ и $\rv r_2$, так что
     $$\rv j^e_1=-i\omega\rv d_1\delta(\rv r-\rv r_1)\,,\quad\rv j^e
         _2=-i\omega\rv d_2\delta(\rv r-\rv r_2)\,,\quad\rv j^m_1
         =0\,,\quad \rv j^m_2=0\,.\eqno(18.7)$$
Тогда (18.6) сводится к соотношению
     $$\rv E_2(\rv r_1)\rv d_1=\rv E_1(\rv r_2)\rv d_2\,.\eqno(18.8)$$
Это и  есть   {\it   теорема   взаимности}   для   двух   элементарных
электрических диполей. Она сопоставляет результаты двух опытов в одной
среде. В первом опыте диполь помещается в точку $\rv r_1$, а в некоторой
точке  $\rv r_2$ измеряется поле.  Во втором опыте диполь помещается в
точку $\rv r_2$,  поле измеряется в $\rv r_1$. Технические устройства, для
которых выполняется теорема (18.8),  называются {\it взаимными}.  В них
возможна   передача  электромагнитной  энергии  в прямом и обратном
направлении.  Только при использовании невзаимных  устройств,  содержащих
такие элементы,  как  ферриты,  удаётся  создать вентили,  находящие
широкое применение в волноводных трактах.

     Аналогичное (18.8)    соотношение    имеет   место   для   полей,
возбуждаемых двумя магнитными диполями с моментами $\rv  m_1$  и  $\rv
m_2$. Плотности токов в этом случае выражаются в виде
     $$\rv j^m_1=-i\omega\rv m_1\delta(\rv r-\rv r_1)\,,\quad\rv j^m_2
         =-i\omega\rv m_2\delta(\rv r-\rv r_2)\,,\quad\rv j^e_1=0\,,
         \quad \rv j^e_2=0\,,\eqno(18.9)$$
в результате  чего из (18.6) следует:
     $$\rv H_2(\rv r_1)\rv m_1=\rv H_1(\rv r_2)\rv m_2\,.
         \eqno(18.10)$$
Точно также между полями,  возбуждёнными электрическим  и  магнитным
диполями с токами
     $$\rv j^e_1=-i\omega\rv d_1\delta(\rv r-\rv r_1)\,,\quad\rv j^m_2
         =-i\omega\rv m_2\delta(\rv r-\rv r_2)\,,\quad\rv j^e_2=0\,,
         \quad\rv j^m_1=0\,,\eqno(18.11)$$
существует соотношение
     $$\rv E_2(\rv r_1)\rv d_1=-\rv H_1(\rv r_2)\rv m_2\,.
         \eqno(18.12)$$

     Лемму Лоренца  и теоремы взаимности можно переформулировать таким
образом,  что они будут выполняться и в невзаимных средах.  Для  этого
надо сопостовлять результаты двух опытов, проведенных не в одной и той
же среде,  а в двух {\it различных} средах,  в которых тензоры $\mu$ и
$\varepsilon$   не   одинаковы,   а   получены   друг  из  друга  {\it
транспонированием},  то   есть   связаны   условиями   $\mu_{ik}^{(1)}
=\mu_{ki}^{(2)},\;\varepsilon_{ik}^{(1)}= \varepsilon_{ki}^{(2)}$.

     Подобная теорема   взаимности   верна  и  для скалярной задачи ---
согласно  (17.22)  между  двумя решениями уравнения (17.18),
соответствующими двум $\delta$-источникам $f_1=\delta(\rv  r-\rv  r_1)$
и  $f_2=\delta(\rv r- \rv r_2)$,  имеет место соотношение
$u_1(\rv r_2)=u_2(\rv r_1)$.

     Теорема взаимности  находит широкое применение при замене решения
сложной электродинамической задачи  решением  более  простой  или  при
сведении  задачи  к  уже решённой.  Рассмотрим два типичных примера.
Пусть диполь $\rv d_1$ расположен в точке 1 вблизи поверхности земли и
требуется  вычислить возбуждаемое им поле на большой высоте в точке 2,
где  находится  приёмная  антенна  некоторого   летящего   объекта.
Вычисление электромагнитного поля передающей антенны,  расположенной в
непосредственной близости от земли, является довольно сложной задачей.
Однако  поле  в  точке  2  может быть найдено без решения этой задачи.
Поместим в  точке  2  вспомогательный  излучающий  диполь  $\rv  d_2$;
поскольку он находится далеко от точки 1, то излучаемую им сферическую
волну вблизи последней можно считать плоской,  а поле, возникающее при
падении   плоской  волны  на  плоскую  проводящую  поверхность,  хорошо
известно (см.  раздел 5). По известному полю $\rv E_2(1)$ на основании
теоремы  взаимности  легко  находится  искомое  поле  $\rv E_1(2)$.  С
помощью леммы Лоренца теорему взаимности можно обобщить таким образом,
чтобы она относилась не только к элементарным диполям,  но и к сложным
антенным  системам.   Здесь   не   место   останавливаться   на   этих
формулировках, но, например, можно доказать, что так называемые диаграммы
направленности антенны при работе на приём и передачу совпадают.

     Более сложным   является   пример  двух  полубесконечных
волноводов,  связанных между  собой  через  какой-нибудь  нерегулярный
участок  волновода  или свободное пространство.  Пусть решена задача о
поле,  возникающем  при  падении  $n$-й  собственной  волны  единичной
амплитуды первого волновода на сочленение,  и,  в частности, известна
амплитуда  $A$  $m$-й  волны,  уходящей  во  второй  волновод.   Можно
утверждать, что  одновременно  решена  задача и об амплитуде $B$ $n$-й
волны,  уходящей в первый волновод при падении из второго $m$-й  волны
единичной амплитуды.

     При практическом  применении  леммы  Лоренца  для  решения  задач
важную   роль   играет   аппарат  векторных  функций  Грина  уравнений
Максвелла.  Эти функции представляют собой решения уравнений (18.1)  с
плотностью  сторонних токов в правой части в виде $\delta$-функции.  В
зависимости  от  уравнения,  в  котором  присутствует  сторонний  ток,
различают две векторные функции Грина --- электрическую и магнитную.

     Электрическая векторная   функция   Грина   представляет    собой
совокупность   двух   векторов  $\rv  E^e(\rv  r,\rv  r_0)\,,\;\rv
H^e(\rv r,\rv r_0)$, удовлетворяющих системе уравнений
     $$\left.\begin{array}{lll}\rot\rv H^e&=&-ik\rv E^e+\displaystyle
         {\frac{4\pi}c}\delta(\rv r-\rv r_0)\rv a\,,\\[.3cm]\rot\rv E
         ^e&=&ik\rv H^e\,\end{array}\right\}\eqno(18.13)$$
и условиям   излучения,   обеспечивающим   распространение   волны  от
источника. Следовательно,  она представляет собой  поле  элементарного
электрического   диполя,   расположенного   в   точке   $\rv   r_0$  и
направленного вдоль  вектора  $\rv  a$.  Для  неограниченного  пустого
пространства  и  диполя,  расположенного  в начале системы координат и
ориентированного вдоль  оси  $z$,  компоненты  поля  даются  формулами
(17.9).  При произвольном расположении диполя компоненты функции Грина
легко находятся с помощью обычного преобразования координат. Очевидно,
что   функция   Грина   симметрична  относительно  перестановки  своих
аргументов, что является прямым следствием теоремы взаимности.

     Аналогичным образом  определяется  и  магнитная векторная функция
Грина. Компоненты составляющих её векторов  $\rv  E^m(\rv  r,\rv
r_0),\,\rv H^m(\rv r,\rv r_0)$ удовлетворяют уравнениям
     $$\left.\begin{array}{lll}\rot\rv H^m&=&-ik\rv E^m\,,\\[.3cm]\rot
         \rv E^m&=&ik\rv H^m-\displaystyle{\frac{4\pi}c} \delta(\rv r-
         \rv r_0)\rv a\,\end{array}\right\}\eqno(18.14)$$
и условиям излучения на бесконечности.  Для пустого  пространства  они
фактически определены формулами (17.16).

     Рассмотрим теперь   систему   заданных    сторонних    токов    в
неограниченном  пространстве  и попробуем найти возбуждаемые ими
поля. Предположим,  что все токи сосредоточены в конечной области,
так  что вне   некоторой  поверхности  $S_0$  плотности  сторонних
токов  $\rv j_{\mbox{\footnotesize\textit{ст}}}^e$ и $\rv
j_{\mbox{\footnotesize\textit{ст}}}^m$ равны нулю.  Выделим
некоторый  объем  $V$, ограниченный поверхностью  $S$.
Воспользуемся леммой Лоренца (18.3), полагая в ней поля с индексом
1 равными искомым  полям:  $\rv  E_1=\rv E(\rv r)$,  $\rv H_1=\rv
H(\rv r)$,  а токи с тем же индексом равными сторонним токам: $\rv
j^e_1=\rv j_{\mbox{\footnotesize\textit{ст}}}^e$, $\rv j^m_1=\rv
j_{\mbox{\footnotesize\textit{ст}}}^m.$ В  качестве полей с
индексом~2 возьмем последовательно электрическую и магнитную
функции Грина, а токов --- соответствующую $\delta$-функцию.
Результат  интегрирования  в (18.3) зависит от расположения точки
$\rv r_0$  (определяющей положение элементарного  диполя  в
определении функции Грина) относительно поверхности $S$.  Если
$\rv r_0$ находится {\it внутри} $S$, то получаем:
     $$\begin{array}{l}\rv a\rv E(\rv r_0)=\displaystyle{\int\limits_V
         \{\rv j_{\mbox{\footnotesize\textit{ст}}}^e
         (\rv r)\rv E^e(\rv r,\rv r_0)-\rv j_{\mbox{\footnotesize\textit{ст}}}^m
         (\rv r)\rv H^e(\rv r,\rv r_0)\}\,dV}\,-\\\displaystyle{\qquad
         {}-\frac c{4\pi}\oint\limits_S\{[\rv E(\rv r)\rv H^e(\rv r,
         \rv r_0)]\}-[\rv E^e(\rv r,\rv r_0)\rv H(\rv r)]\,d\rv S}\,,
         \end{array}\eqno(18.15)$$
     $$\begin{array}{l}\rv a\rv H(\rv r_0)=\displaystyle{\int\limits_V
         \{\rv j_{\mbox{\footnotesize\textit{ст}}}^m(\rv r)\rv H^m(\rv r,\rv r_0)-\rv j_{\mbox{\footnotesize\textit{ст}}}^e(
         \rv r)\rv E^m(\rv r,\rv r_0)\}\,dV}\,+\\\displaystyle{\qquad
         {}+\frac c{4\pi}\oint\limits_S\{[\rv E(\rv r)\rv H^m(
         \rv r,\rv r_0)]-[\rv E^m(\rv r,\rv r_0)\rv H(\rv r)]\}
         \,d\rv S}\,.\end{array}\eqno(18.16)$$
Если же  точка  $\rv  r_0$ лежит {\it вне} поверхности $S$,  то правые
части  в  (18.15)  и  (18.16)  равны  нулю,  и,  следовательно,  между
интегралами выполняются соотношения
     $$\begin{array}{l}\displaystyle{\frac c{4\pi}\oint\limits_S\{[
         \rv E(\rv r)\rv H^e(\rv r,\rv r_0)]-[\rv E^e(\rv r,\rv r_0)
         \rv H(\rv r)]\}\,d\rv S}\,=\\\displaystyle{\qquad{}
         =\int\limits_V \{\rv j_{\mbox{\footnotesize\textit{ст}}}^e(\rv r)\rv E^e(\rv r,\rv r_0)-
         \rv j_{\mbox{\footnotesize\textit{ст}}}^m(\rv r)\rv H^e(\rv r,\rv r_0)\}\,dV}\,,\end
         {array}\eqno(18.17)$$
     $$\begin{array}{l}\displaystyle{\frac c{4\pi}\oint\limits_S\{[
         \rv E^m(\rv r,\rv r_0)\rv H(\rv r)]-[\rv E(\rv r)
         \rv H^m(\rv r,\rv r_0)]\}\,d\rv S}\,=\\\displaystyle{\qquad{}
         = \int\limits_V \{\rv j_{\mbox{\footnotesize\textit{ст}}}^m(\rv r)\rv H^m(\rv r,\rv r_0
         )-\rv j_{\mbox{\footnotesize\textit{ст}}}^e(\rv r)\rv E^m(\rv r,\rv r_0)\}\,dV}\,.\end
         {array}\eqno(18.18)$$

     Следует отметить,  что формулы (18.15) и (18.16) не дают  решения
задачи  о  поле  заданных  токов в замкнутом виде,  а только связывают
между собой значения полей в некoторой точке $\rv r_0$ c интегралом от
их значений на выделенной поверхности $S$.  Тем не менее,  эти формулы
лежат в  основе  строгой  формулировки  задач  излучения  и  дифракции
путём сведения их к интегральным уравнениям и при решении формальных
граничных задач при различных приближённых подходах. Кроме того, они
позволяют   получить   решение   задачи  о  заданных  токах  в  пустом
пространстве  в  замкнутом  виде.  Получаемое  с  их  помощью  решение
обладает  определёнными  преимуществами по  сравнению  с найденным в
предыдущем   разделе   в   скалярной    формулировке.    Прежде    чем
останавливаться   на  всех  этих  вопросах,  обратим  внимание  на  те
особенности   выражений   (18.15)   и   (18.16),   которые   позволяют
сформулировать   утверждение,   составляющее   предмет  {\it  принципа
эквивалентности}.

     Основу нижеследующих  рассуждений  составляет  то  примечательное
обстоятельство,  что одни и те же функции Грина $\rv E^e$, $\rv
H^e$ в первой  формуле  и $\rv E^m$,  $\rv H^m$ во второй
умножаются в первых строчках на $\rv
j_{\mbox{\footnotesize\textit{ст}}}^e$,  $\rv
j_{\mbox{\footnotesize\textit{ст}}}^m$,  а  во вторых  ---  на
тангенциальные к поверхности составляющие искомых полей $E_t$,
$H_t$. Следовательно, поля в произвольной точке $\rv r_0$ будут
одинаковыми в двух  задачах  ---  в  одной,  где на границе
касательные составляющие полей равны $E_t$, $H_t$, и в другой, где
на границе $E_t=0$, $H_t=0$, но на ней распределены поверхностные
токи с плотностями
     $$\rv I^e=\phantom{-}\frac c{4\pi}[\rv n\rv H]\,,\eqno (18.19)$$
     $$\rv I^m=-\frac c{4\pi}[\rv n\rv E]\,.\eqno (18.20)$$
В этих  выражениях подразумевается,  что нормаль $\rv n$ направлена из
выделенного объема $V$, для точек которого поля определяются формулами
(18.15) и (18.16).

     Формула (18.19)  устанавливает  привычную   связь   поверхностной
плотности тока с магнитным полем на поверхности идеального проводника.
С  её помощью   удобно   разъяснить   физический   смысл  {\it принципа
эквивалентности}. Пусть  поверхность  $S$  или  какая-то   её   часть
совпадает   с   поверхностью   идеального  проводника.  Представляется
очевидным,  что поле в точке $r_0$  не  может  зависеть  от  положения
мысленной  поверхности  $S$,  в частности,  от того сдвинута ли она на
бесконечно малое расстояние вглубь проводника или от  него.  В  первом
случае   в   области,  по  которой  проводится  интегрирование,  будут
присутствовать токи,  а  на  поверхности  интегрирования  $H_t=0$.  Во
втором  случае плотность тока не войдет в объёмный интеграл,  и поле
будет выражено через значение $H_t$ на границе.  Формула  (18.19)  как
раз  и обеспечивает такое соотношение между плотностью тока и полем на
границе, при котором результаты вычислений совпадают.

     Наличие аналогичного  соответствия  между   полем,   определяемым
граничным  значением  $E_t$,  и полем,  возбуждаемым магнитным током с
поверхностной  плотностью  (18.20),  является  основной  побудительной
причиной для введение в теорию дифракции {\it магнитных токов}.

     Соотношения (18.15)  и  (18.16)  позволяют  вычислить  поле  произвольной
системы заданных  сторонних   токов  в  неограниченном  пустом
пространстве  в  замкнутой  форме  в виде квадратур.  Хотя эта задача в
скалярной формулировке уже была решена в  предыдущем разделе,
поучительно  привести решение и с помощью аппарата векторных
функций Грина.

     Будем для   простоты   записи   рассматривать   заданную  систему
электрических токов,  не забывая,  что по  современным
представлениям только  они  и  являются  физической реальностью.
Поэтому в (18.15) и (18.16) положим $\rv
j_{\mbox{\footnotesize\textit{ст}}}^m=0$,  а у $\rv
j_{\mbox{\footnotesize\textit{ст}}}^e$ опустим  верхний индекс.
Простейший  способ состоит в выборе в качестве ограничивающей
поверхности $S$ сферы  бесконечно большого  радиуса.  В  этом
случае поверхностные  интегралы обращаются  в  нуль,  так как и
поле системы токов и функция Грина имеют  на  бесконечности
одинаковую  структуру расходящейся сферической  волны.  Нетрудно
убедиться при этом,  что с увеличением радиуса сферы
подынтегральное выражение  убывает быстрее, чем  растёт площадь
сферы.  Поэтому из (18.15) и (18.16) сразу следуют выражения для
полей в замкнутом виде в произвольной точке пространства $\rv r_0$:
     $$\rv a\rv E(\rv r_0)=\int\limits_V \rv j_{\mbox{\footnotesize\textit{ст}}}(\rv r)\rv E^e(
         \rv r,\rv r_0)\,dV\,,\eqno(18.21)$$
     $$\rv a\rv H(\rv r_0)=-\int\limits_V \rv j_{\mbox{\footnotesize\textit{ст}}}(\rv r)\rv E^m(
         \rv r,\rv r_0)\,dV\,.\eqno(18.22)$$

     Таким образом,   чтобы  найти  в  точке  $\rv  r_0$  какую-нибудь
компоненту вектора $\rv E$,  необходимо в эту точку мысленно поместить
элементарный  электрический  диполь,  вычислить  поле  $\rv E^e$ этого
диполя во всех точках,  где расположены сторонние токи,  и  произвести
интегрирование.  Отметим  при  этом,  что магнитное поле может быть по
известному полю $\rv E$ вычислено путем дифференцирования,  исходя  из
второго   уравнения   Максвелла   (18.1),   а   может   быть   найдено
непосредственно из (18.22) с помощью магнитной векторной функции Грина
$\rv E^m$.  При численном расчёте полей сложной системы токов на ЭВМ
второй способ предпочтительнее. Три формулы (17.24), (18.21) и (18.22)
представляют  собой  разные  формы решения одной и той же задачи.  Они
являются основой теории антенн, точнее, наиболее простой её главы, в
которой рассматриваются поля, возбуждаемые заданными токами.

     На примере   решённой   задачи  целесообразно  посмотреть,  как
видоизменяется ход решения при  другом  выборе  поверхности  $S$,  что
позволяет  установить  ряд  закономерностей в поведении поверхностного
интеграла в (18.15) и (18.16),  которые важны при  использовании  этих
формул   в  более  сложных  граничных  задачах.  Проведём  мысленно
вспомогательную замкнутую поверхность,  которую обозначим как $S_1$, и
которая делит все пространство на объём $V_1$,  расположенный внутри
этой поверхности,  и объём $V_2$ вне её.  Возможны три  случая:  в
первом  из  них  все  заданные токи находятся внутри $S_1$,  во втором
внутри $S_1$ находится только часть токов,  в третьем в объёме $V_1$
токов нет. Во всех трёх случаях формула (18.15) (ниже только о ней и
пойдёт  речь,  поскольку  для  (18.16)  все  рассуждения  проводятся
аналогично)  верна  и  должна  давать одинаковый результат.  Поэтому в
первом случае поверхностный интеграл
     $$I_{S_{1}}=\oint\limits_{S_1}\{[\rv E^e(\rv r,\rv r_0)\rv H
         (\rv r)]-[\rv E(\rv r)\rv H^e(\rv r,\rv r_0)]\}\,d\rv S
         \eqno(18.23)$$
для всех точек $\rv r_0$ из  $V_1$  равен  нулю  и  для  этой  области
получается  тот  же  самый  результат  (18.21).  Для вычисления поля в
объёме $V_2$  опять  достаточно  воспользоваться  формулой  (18.15),
только   теперь   в  ней  поверхность  $S$  состоит  из  двух  частей:
поверхности $S_1$ и сферы бесконечно  большого  радиуса,  интеграл  по
которой  в  силу сказанного выше равен нулю.  Так как токов в пределах
$V_2$ нет,  то объёмный интеграл в (18.15) по  области  $V_2$  равен
нулю. Из формулы (18.17) с учётом изменения направления нормали
следует, что для точек $\rv r_0$ из $V_2$
     $$I_{S_1}=\int\limits_{V_1} \rv j_{\mbox{\footnotesize\textit{ст}}}(\rv r)\rv E^e(\rv r,
         \rv r_0)\,dV\,,\eqno(18.24)$$
и, следовательно,
     $$\rv a\rv E(\rv r_0)=\int\limits_{V_1} \rv j_{\mbox{\footnotesize\textit{ст}}}(\rv r)\rv E^
         e(\rv r,\rv r_0)\,dV\,,\eqno(18.25)$$
что совпадает с (18.21),  поскольку объём $V_1$ заключает в  себе  все
сторонние токи.

     Аналогичным образом  проводятся  рассуждения  для   двух   других
вариантов расположения поверхности $S_1$ относительно сторонних токов.
Опуская  их,  сформулируем  окончательный   результат:   поверхностные
интегралы  в  (18.15)  и  (18.16)  представляют  собой  по обе стороны
замкнутой поверхности $S$ поле сторонних токов,  расположенных по {\it
другую}  сторону  поверхности,  чем  та,  которая  примыкает к области
вычисления полей.  Возможность замены  системы  токов  при  вычислении
полей  на  значения  полей на некоторой замкнутой поверхности вне этих
токов лежит в основе {\it принципа Гюйгенса-Кирхгофа},  который в свою
очередь является фундаментом теории решения граничных задач дифракции,
рассматриваемых в следующих разделах.

%\end{document}

\newpage
\oddsidemargin=-0.4mm \evensidemargin=-0.4mm
\topmargin=-0.4mm
\headsep=7mm
\textheight=231.875mm
\textwidth=160mm
\mathsurround=2.5pt
\unitlength=1mm
%\begin{document}
%\input{macr.tex}
\thispagestyle{empty}
%\addtocounter{page}{196}

\begin{center}
   \subsubsection*{\rm Г\,Л\,А\,В\,А\, 7}
      \vspace{-1.15em}
      \line(6,0){160}
      \vspace{-1em}
      \line(6,0){160}
      \vspace{-1.15em}
   \subsubsection*{ГРАНИЧНЫЕ ЗАДАЧИ И ВОЗБУЖДЕНИЕ ОГРАНИЧЕННЫХ СТРУКТУР}
      \vspace{31mm}
\subsubsection*{19.~Типы граничных задач в электродинамике СВЧ}
\end{center}\vspace{.5cm}

\markboth{Глава~7.~Граничные задачи и возбуждение ограниченных
                структур}{19.~Типы граничных задач в  электродинамике
                СВЧ}

\begin{center}\begin{minipage}[c]{0.75\textwidth}
\footnotesize{\parindent=0.5cm
           Общая формулировка   граничной  задачи  в  электродинамике.
           Разновидности  граничных  (дифракционных)  задач.  Сведение
           граничной задачи к интегральному уравнению для наведённых
           токов на поверхности проводников с помощью  функций  Грина.
           Многообразие функций Грина. Интегральное уравнение для поля
           на отверстии в проводящей поверхности.  Условие на ребре  и
           проблема     выбора    единственного    решения.    Теорема
           двойственности для тонкого  плоского  идеально  проводящего
           экрана. Приближённые методы решения дифракционных задач.
}\end{minipage}\end{center}\vspace{.5cm}

     До сих    пор    рассматривались   два   вида   монохроматических
электромагнитных полей:  поля,  удовлетворяющие однородным
уравнениям Максвелла  без источников,  и поля,  возбуждаемые
заданными сторонними токами в бесконечном {\it однородном}
пространстве.  Поля первого вида исследовались   и   для
неоднородных   сред;  при  этом  в  основном рассматривался
частный  вид   неоднородности   свойств   бесконечного
пространства,  позволяющий  разделить  его  на  несколько областей
с однородными,  но различающимися между собой свойствами по
отношению  к электромагнитному   полю.   Такое
<<кусочно-однородное>>  пространство подразумевает  наличие  {\it
граничных}  поверхностей,   на   которых свойства среды изменяются
скачком.  При этом терпят разрыв и некоторые компоненты поля,
подчиняясь определённым граничным условиям.

     Настоящий раздел  посвящён  постановке  задач и описанию основных
методов их решения при наличии сторонних  токов  в  кусочно-однородном
пространстве. Для упрощения и без того довольно громоздких выражений и
избежания дополнительных, зачастую очевидных оговорок, рассматривается
лишь  частный  случай  двух  сред:  одна  из  них пустота,  другая ---
идеальный  проводник.  На  границе  между  ними  выполняется   условие
равенства  нулю  двухкомпонентного вектора тангенциальной составляющей
электрического поля:  $\rv E_t=0$.  Многообразие топологических  видов
поверхности  проводника  делает  и  этот частный класс граничных задач
электродинамики достаточно обширным, требующим различного подхода к их
решению.

     Общая постановка граничной задачи  состоит  в  задании  сторонних
токов  и  граничной  поверхности  проводника.  Поскольку задача о поле
токов в свободном пространстве уже решена выше (в  виде  квадратур)  в
самом  общем  случае,  то  эквивалентная  постановка  граничной задачи
подразумевает  задание  первичного   или   падающего   поля,   которое
рассеивается на поверхности проводника, приводя к появлению вторичного
(или  {\it   дифрагированного})   поля.   Поэтому   граничные   задачи
рассматриваемого   типа   с   равным  правом  называют  дифракционными
задачами.  Если  максимальный   линейный   размер   области,   занятой
сторонними  токами,  существенно  меньше  минимального  расстояния  от
поверхности проводника,  то первичное поле можно в районе  проводящего
тела  считать  сферической  волной  и говорить о дифракции сферической
волны на данном теле. Если же, кроме того,  расстояние до поверхности
проводника от области источников существенно превышает и размеры тела,
то сферическую  волну  в  его  окрестности  можно  считать  плоской  и
говорить о дифракции плоской волны.

     Можно выделить    несколько    характерных    типов    проводящих
поверхностей.  В  простейшим варианте граничной задачи тело проводника
сплошное с  достаточно  гладкой  односвязной  поверхностью.  Типичными
примерами  являются  задачи  дифракции на шаре,  эллипсоиде и подобных
телах. К этому топологическому типу задач примыкают и такие, в которых
объём тела простирается до бесконечности --- например,  дифракция на
цилиндре или клине.  Среди задач этого класса следует выделить случаи,
когда  проводящие  поверхности совпадают с координатными поверхностями
одной из  систем ортогональных  координат,  в  которых  в  волновом
уравнении  переменные разделяются.

     Все такие   задачи   могут   быть   решены  классическим  методом
разделения   переменных   (методом   Фурье).   При   этом    замкнутое
аналитическое  решение  получается  в  виде двойных или тройных рядов,
которые  зачастую  плохо  сходятся,  во  всяком  случае  в   некотором
диапазоне   значений   параметров.   Для   улучшения   их   сходимости
используются тонкие математические методы и  приёмы.  Такие  решения
хорошо изучены и очень полезны для проверки более общих приближённых
методов  решения,  используемых  для  тел  произвольной  формы.  Здесь
нецелесообразно  на  них  останавливаться  не  только потому,  что эти
решения очень громоздки и изложены в  ряде  монографий,  но  в  первую
очередь потому,  что в них слабо проглядывается физический смысл задач
и они, как правило, не допускают обобщения.

     Наиболее просто  анализируются дифракционные задачи для областей,
ограниченных  замкнутыми  проводящими  поверхностями.  При   сделанных
предположениях  такие области представляют собой идеальные резонаторы.
Тогда задача о  возбуждении  поля  сторонними  токами,  расположенными
внутри полости,  легко решается в общем случае в виде квадратур,  если
известна  система   собственных   частот   и   собственных   колебаний
резонатора.  Соответствующие  формулы  и  примеры  приведены  далее  в
разделе 21.

     При идеальной проводимости стенок их толщина  не  играет  роли  и
дифракционная задача для внешней области никак не связана с внутренней
задачей  для  резонатора.  Сторонние  токи  по   разную   сторону   от
проводящего слоя возбуждают поля только в своих областях, но положение
полностью изменяется,  если в стенке (она может быть нулевой  толщины)
имеется   хотя   бы   маленькое   отверстие.  В  этом  случае  решение
дифракционной  задачи  удобнее  всего   строить   по   тангенциальному
электрическому  полю  в  плоскости отверстия.

     Прежде чем переходить к развёрнутому  анализу  подобных  задач,
необходимо ещё  сказать  несколько  слов  о   специфическом   классе
граничных  задач,  который  возникает  в  волноводах  ---  бесконечных
однородных   вдоль   выделенной   оси   и   ограниченных    проводящей
цилиндрической поверхностью структурах.

     Для них  можно выделить три разновидности дифракционных задач.  К
первой  из  них  можно  отнести  возбуждение  идеального   однородного
волновода системой расположенных внутри него сторонних токов (методика
решения подобных задач излагается  в  следующем  разделе).  Ко  второй
разновидности относятся задачи рассеяния волноводных волн на различных
нерегулярностях в волноводе (диафрагмы,  штыри,  изгибы); их  решение
часто  связано  с понятием {\it матрицы рассеяния},  рассматриваемой в
разделе 22.  К третьему  виду  следует  отнести  задачи  излучения  из
волноводов  через  щели  и  отверстия  в  боковых стенках,  а также из
открытого конца полубесконечного волновода;  подобные задачи  решаются
теми же методами, какие пригодны для отверстий в стенках резонаторов.

     Начнём с физически наиболее ясной  граничной  задачи.  Пусть  в
конечной  области пространства имеется система сторонних электрических
токов $\rv j(\rv r)$ (индексы у токов для упрощения записи  опускаем).
Помимо  заданных  токов  в  этой  же  области пространства и вне токов
расположено идеально проводящее тело.  Будем считать поверхность  тела
(замкнутая  поверхность)  достаточно  гладкой  и обозначим её $S_0$.

\begin{wrapfigure}[16]{l}{7.5cm}
\begin{picture}(80,60)
\put(2,58){\special{em:graph fig19-1.bmp}}
\end{picture}
\hbox to 7.5cm{\hfil\footnotesize{Рис.~19.1.~К дифракционной задаче
}\hfil}
\hbox to 7.5cm{\hfil\footnotesize{для ограниченного проводящего тела.}
\hfil}
\end{wrapfigure}
     Из общих  соображений  очевидно,   что   под   действием   полей,
возбуждаемых   сторонними   токами,   на  поверхности  тела  наведутся
дополнительные токи и полное  поле  в  пространстве  представит  собой
сумму полей от заданных и наведённых токов. В частности, если забыть
о существовании тела и считать наведённые токи также сторонними,  то
полное   поле,   вычисленное   по   формулам   (18.21),   (18.22)  для
неограниченного  пустого  пространства,  в  области,  занятой   телом,
окажется   тождественно   равным  нулю.  Введём  мысленно  еще  одну
вспомогательную замкнутую поверхность $S_1$,  заключающую в  себя  все
токи  и  тело.  Положение  этой  поверхности  не важно для последующих
рассуждений,  в  частности,  она  может   представлять   собой   сферу
бесконечно  большого радиуса;  важно только,  что вне этой поверхности
нет ни сторонних, ни наведённых токов.

     Поле в   любой   точке   объёма   $V$,   заключённого   между
поверхностями  $S_0$  и  $S_1$,  может быть в соответствии с формулами
(18.15) и  (18.16)  выражено  с  помощью  векторных  функций  Грина  и
значений полей на выбранных поверхностях.  Выпишем эти формулы еще раз
для  рассматриваемого  здесь  частного  случая  только   электрических
сторонних токов:
     $$\rv a\rv E(\rv r_0)=\int\limits_V \rv j(\rv r)\rv E^{e}(\rv r,
         \rv r_0)\,dV\,\displaystyle{-\frac c{4\pi}\oint\limits_S\{[
         \rv E(\rv r)\rv H^e (\rv r,\rv r_0)]-[\rv E^e(\rv r,\rv r_0)
         \rv H(\rv r)]\}\,d\rv S}\,,\eqno(19.1)$$
     $$\rv a\rv H(\rv r_0)=-\int\limits_V \rv j(\rv r)\rv E^{m}(\rv r
         ,\rv r_0)dV+\frac c{4\pi}\oint\limits_S\{[\rv E(\rv r)\rv H
         ^{m}(\rv r,\rv r_0)]-[\rv E^{m}(\rv r,\rv r_0)\rv H(\rv r)]
         \}d\rv S.\eqno(19.2)$$
В этих формулах $\rv r_0$ ---  точка  {\it  внутри}  объёма  $V$,  в
которой ищется  поле,  $\rv E^e$,~$\rv H^e$ и $\rv E^m$,~$\rv H^m$ ---
соответственно электрическая и магнитная векторные функции  Грина  для
{\it неограниченного пустого пространства}, удовлетворяющие уравнениям
(18.13) и (18.14) и ---  дополнительно  ---  условиям  излучения.  Эти
функции  были  найдены выше и они представляют собой поле в точке $\rv
r$    элементарного    электрического    или    магнитного     диполя,
ориентированного  вдоль  единичного вектора~$\rv a$ и расположенного в
точке~$\rv  r_0$.  Поверхность  $S$  представляет  собой  совокупность
замкнутых  непересекающихся  поверхностей  $S_0$ и $S_1$,  а векторный
элемент поверхности $d\rv S$ направлен по внешней нормали,  то есть на
поверхности $S_0$ внутрь тела, а на $S_1$ --- во внешнее пространство.

     Проанализируем формулу  (19.1)  подробнее.  Отметим прежде всего,
что она ни в коем случае не представляет собой решение рассматриваемой
граничной  задачи,  так  как  искомое  поле  $\rv E$ в точке $\rv r_0$
выражено не только  через  известные  функции  $\rv  j(\rv  r)$,  $\rv
E^e(\rv r,\rv r_0)$,  $\rv H^e(\rv r,\rv r_0)$, но и через неизвестные
значения тангенциальных составляющих полей  на  поверхностях  $S_0$  и
$S_1$.  На самом деле то, что нам неизвестны значения на $S_1$, не так
важно --- в конце предыдущего раздела было показано,  что интеграл  по
этой  поверхности,  вычисленный  для  всех  точек внутри объёма $V$,
равен нулю,  поскольку по другую сторону от этой  поверхности  никаких
токов нет;  интеграл по поверхности $S_0$,  очевидно,  заведомо даёт
вклад в поле.  Обратим внимание, что формула (19.1) подразумевает, что
поверхность  $S_0$  чуть  смещена  от  поверхности  тела,  так  что  в
объёмный интеграл наведённые на поверхности тела  токи  не  вошли.
Однако вошли    тангенциальные   составляющие   магнитного   поля   на
поверхности $S_0$,  непосредственно связанные с  наведёнными  токами
соотношением (18.19), и по {\it принципу эквивалентности} вклад в поле
в точке $\rv r_0$ от этих  величин  одинаков.  Всё  вышесказанное  в
равной  степени  относится  и  к  формуле  (19.2)  с учётом замены в
рассуждениях величин $\rv  E^e$,~$\rv  H^e$  соответственно  на  $\rv
E^m$,~$\rv H^m$ (с точностью до знака).

     Обозначим через $\rv E^0$,  $\rv H^0$ поля,  которые возбудили бы
сторонние токи $\rv j$ в отсутствии тела (они определяются объёмными
интегралами в (19.1) и (19.2)). Тогда с учётом граничного условия на
$S_0$ и нулевого вклада поверхности $S_1$ эти  формулы  перепишутся  в
виде
     $$\rv a\rv E(\rv r_0)=\rv a\rv E^0(\rv r_0)\displaystyle{+\frac c
         {4\pi}\oint\limits_{S_0}[\rv E^e(\rv r,\rv r_0)\rv H(\rv r)]
         \,d\rv S}\,,\eqno(19.3)$$
     $$\rv a\rv H(\rv r_0)=\rv a \rv H^0(\rv r_0)-\frac c{4\pi}\oint
         \limits_{S_0}[\rv E^{m}(\rv r,\rv r_0)\rv H(\rv r)]d\rv S.
         \eqno(19.4)$$

     Итак, граничная  задача   определения   поля   {\it   во   всём
пространстве}  будет  решена  в  замкнутом  виде,  если  удастся найти
тангенциальное магнитное поле (или,  что  то  же  самое,  наведённые
токи)  на  поверхности  тела  $S_0$.  Для  двухкомпонентной  векторной
функции $\rv  H_t(\rv  r)$,  где  точки  $\rv  r$  принадлежат  $S_0$,
нетрудно  вывести  интегральные  уравнения.  Но прежде чем это делать,
вернёмся немного назад с целью расширить  представления  о  функциях
Грина уравнений Максвелла и о корректной постановке граничной задачи.

     Возьмём некоторую произвольную точку  $\rv  r_0$,  лежащую  вне
тела  и  свободную от сторонних токов $\rv j$.  Окружим её замкнутой
поверхностью $S_2$ таким образом, чтобы все точки внутри обладали теми
же  свойствами.  Тогда  на  основании  леммы Лоренца поле внутри $S_2$
определяется формулами
     $$\rv a\rv E(\rv r_0)=\frac c{4\pi}\oint\limits_{S_2}\{[\rv E^e(
         \rv r,\rv r_0)\rv H(\rv r)]- [\rv E(\rv r)\rv H^e(\rv r,\rv r
         _0)]\}\,d\rv S\,,\eqno(19.5)$$
     $$\rv a\rv H(\rv r_0)=-\frac c{4\pi}\oint\limits_{S_2}\{[\rv E^
         {m}(\rv r,\rv r_0)\rv H(\rv r)]- [\rv E(\rv r)\rv H^m(\rv r,
         \rv r_0)]\}d\rv S\,,\eqno(19.6)$$
где элемент  поверхности  $d\rv  S$  направлен  наружу от  $S_2$,  а $\rv
E^e,\;\rv H^e$ и $\rv  E^m,\;\rv  H^m$  ---  векторные  функции  Грина
свободного пространства.

     Формальная граничная  задача  в   математике   (там   она   часто
называется ещё {\it краевой задачей}) ставится следующим образом:  в
данной  области   найти   функцию,   удовлетворяющую   определённому
дифференциальному  уравнению  в  частных  производных и принимающую на
границах области заданные  значения.  В  рассматриваемом  случае  речь
идёт  об однородных уравнениях Максвелла и о двух векторных функциях
$\rv E$ и $\rv H$.  Если исходить из формул (19.5)  и  (19.6),  то  на
первый  взгляд  можно  прийти к выводу,  что постановка задачи требует
задания на границе двух двумерных векторных функций $\rv E_t$  и  $\rv
H_t$,  но  этот  вывод неверен.  При произвольных функциях $\rv E_t$ и
$\rv  H_t$  поля,  вычисленные  по  (19.5)  и  (19.6),   удовлетворяют
уравнениям Максвелла внутри $S_2$, но они не стремятся при приближении
к поверхности $S_2$ к заданным значениям.  И только в том единственном
случае, когда $\rv E_t$ и $\rv H_t$ являются решениями рассматриваемой
выше граничной задачи для всего пространства,  выполняется и физически
необходимое   условие  непрерывности  $\rv  E$  и  $\rv  H$  в  пустом
пространстве.  Более того, нетрудно убедиться, что для нахождения поля
внутри $S_2$ достаточно задать только одно какое-нибудь тангенциальное
поле на граничной поверхности или только какое-нибудь  одно  на  части
границы и только другое на оставшейся части границы.

     Предположим, что  на $S_2$ задано поле $\rv E_t$,  и введ\"ем две
новые векторные функции Грина $\rv{\tilde  E}^e,\,\rv{\tilde  H}^e$  и
$\rv{\tilde  E}^m,\,\rv{\tilde  H}^m$,  которые определены только {\it
внутри} $S_2$,  удовлетворяют там тем же уравнениям (18.13) и (18.14),
но  в  отличие  от  $\rv  E^e,\,\rv H^e$ и $\rv E^m,\,\rv H^m$ для них
вместо условия излучения на  бесконечности  выполняются  условия  $\rv
{\tilde  E}^e_t=0$  и  $\rv {\tilde E}^m_t=0$ на $S_2$.  Такие функции
заведомо   существуют,   поскольку   они   представляют   собой   поля
элементарного  электрического  и  магнитного  диполя  внутри  области,
ограниченной замкнутой идеально проводящей поверхностью  $S_2$.  Более
того,   они  сравнительно  просто  находятся,  если  известна  система
собственных частот и собственных функций области  (как  это  делается,
рассмотрено далее в разделе 21).  С помощью этих функций вместо (19.5)
и (19.6) для внешней к поверхности $S_2$ нормали имеем
     $$\rv a\rv E(\rv r_0)=- \frac c{4\pi}\oint\limits_{S_2}[\rv E(
         \rv r)\rv{\tilde H}^e(\rv r,\rv r_0)]\,d\rv S\,,\eqno(19.7)$$
     $$\rv a\rv H(\rv r_0)=\frac c{4\pi}\oint\limits_{S_2}[\rv E(\rv r
         )\rv{\tilde H}^m(\rv r,\rv r_0)]\,d\rv S\,.\eqno(19.8)$$
Формула (19.8)   определяет   значение  магнитного  поля  и  на  самой
граничной поверхности $S_2$.  И только в том случае,  если в (19.5)  и
(19.6)  вместе  с  заданным $\rv E_t$ будет подставлена вычисленная по
(19.8)  функция  $\rv  H_t$,  они   дадут   правильное   и   физически
единственное  решение  дифракционной  задачи  для  внутренней области,
ограниченной $S_2$.

     Если на  $S_2$  задано  тангенциальное  магнитное  поле,  то  все
рассуждения остаются в силе, но при введении ещё одних новых функций
Грина,  которые следует подчинить условию $\rv{\tilde H}^{e,m}_t=0$ на
$S_2$,  то есть считать $S_2$ поверхностью идеального магнетика. Таким
образом,  видно,  что многообразие векторных функций  Грина  уравнений
Максвелла   бесконечно   велико,   все  они  представляют  собой  поле
соответствующего элементарного  диполя  при  определённых  граничных
условиях,  но  практическая ценность каждой такой функции определяется
возможностью её нахождения и достаточной аналитической простотой для
вычисления соответствующих интегралов.

     После всего сказанного вернёмся к нашей граничной задаче.  Если
теперь в (19.1) и (19.2) выбрать функцию Грина для внешней области  по
отношению   к   $S_0$   и  удовлетворяющей  на  бесконечности  условию
излучения,  а на $S_0$ условию $\rv  {\tilde  E}^e_t=0$,~$\rv  {\tilde
E}^m_t=0$, то решение задачи получается в замкнутом виде:
     $$\rv a\rv E(\rv r_0)=\int\limits_V \rv j(\rv r)\tilde{\rv E}^e
         (\rv r,\rv r_0)\,dV\,,\eqno(19.9)$$
     $$\rv a\rv H(\rv r_0)=-\int\limits_V \rv j(\rv r)\tilde{\rv E}^m
         (\rv r,\rv r_0)\,dV\,.\eqno(19.10)$$
Существенный недостаток приведенного решения состоит в том,  что функция 
Грина со значком  $\sim$ неизвестна и задача определения этой функции, 
которая зависит от формы тела,  лишь незначительно проще первоначальной 
граничной задачи. Более того,   полученный   результат   представляется   
очевидным  из  общих соображений с  учётом  линейности  уравнений  
Максвелла  и  граничных условий.   Если   нам   известно  поле  элементарного 
 диполя  {\it  в присутствии проводящего тела},  то полное поле системы сторонних 
 токов находится  простым  суммированием  (интегрированием с весом $\rv j(\rv
r)$) вкладов отдельных диполей.  Тем не менее  изложенные  рассуждения
подсказывают  нам о существующей возможности.  И в тех редких случаях,
когда удаётся построить функцию  Грина  в  присутствии  дополнительных
проводящих  поверхностей,  пренебрегать этим ни в коем случае не надо.
Фактически можно указать лишь два простых примера, когда такая функция
легко находится.  К ним относятся полупространство,  ограниченное либо
идеально проводящей плоскостью,  либо плоской поверхностью  идеального
магнетика,  и  идеальный  волновод,  собственные  волны которого можно
считать обобщёнными функциями Грина для источника, расположенного на
бесконечности; оба эти случая будут рассмотрены далее.

     Таким образом,  призрачная  простота   решения   (19.9),   (19.10)
возвращает  нас  к необходимости получения интегрального уравнения для
$\rv H_t$ на $S_0$.  Для этого поместим точку наблюдения $\rv r_0$  на
поверхность   тела  и  будем  считать  орт  $\rv  a$  в  каждой  точке
поверхности $S_0$  лежащим  в  касательной  плоскости.  Тогда  (19.4)
становится  интегральным уравнением второго рода для тангенциальной компоненты
магнитного поля $\rv H_t$ на поверхности проводящего тела (нормальная
компонента  равна  на  ней  нулю),  а  из  (19.3) следует интегральное
уравнение первого рода:
     $$\rv a\rv E^0(\rv r_0)=-\frac c{4\pi}\oint\limits_{S_0}[\rv E^e(
         \rv r,\rv r_0)\rv H(\rv r)]\,d\rv S\,,\eqno(19.11)$$
где, напомним,  функция  Грина  определена для неограниченного пустого
пространства.    Отметим,  что  оба  интегральные  уравнения  являются
сингулярными,  поскольку  функция Грина в точке $\rv r= \rv  r_0$  обращается
в  бесконечность.  Поиск  решения  этих  интегральных уравнений и
составляет  основную  и  очень  серьёзную  математическую  трудность
задачи.  В  подавляющем  большинстве случаев решение приходится искать
численными методами с помощью ЭВМ.  Если же интегральные уравнения для
функции  $\rv  H_t(\rv  r)$  на поверхности тела решены,  то формулами
(19.3) и (19.4) граничная задача сведена к квадратурам.

     Существует очень широкий класс граничных задач, в которых решение
целесообразней    строить,    переходя   к   интегральному   уравнению
относительно   касательного   электрического   поля    на    некоторой
вспомогательной   поверхности.   Типичным  примером  служит  излучение
сторонних токов,  расположенных на некотором расстоянии от отверстия в
сплошной  проводящей  поверхности.  Эта  поверхность  может  быть либо
замкнутой,  как в случае волновода или резонатора,  в  стенке  которых
прорезана  щель,  либо  простирающейся  до бесконечности и разделяющей
всё пространство на две области,  соединяющиеся между  собой  только
посредством  этого отверстия.  Вспомогательной поверхностью для вывода
интегрального уравнения служит произвольная поверхность,  затягивающая
отверстие.

     Простейшим примером является плоский идеально проводящий экран  с
проделанным в нём отверстием.  Будем считать, что экран расположен в
плоскости $z=0$; вспомогательную поверхность $S_0$ естественно выбрать
в  той  же  плоскости.  Идея  вывода  интегрального  уравнения проста:
полагаем,  что  тангенциальное  электрическое  поле   $\rv   E_t$   на
поверхности  $S_0$  известно  и  по  нему  с  помощью  леммы Лоренца и
подходящей функции Грина вычисляется магнитное  поле  по  обе  стороны
этой  поверхности.  Приравнивая  теперь  в  соответствии с требованием
непрерывности  поля  на  отверстии  тангенциальные   магнитные   поля,
вычисленные  в  непосредственной  близости  от  $S_0$  справа и слева,
получаем искомое уравнение.  Действительно,  пусть в правом $(z>0)$  и
левом   $(z<0)$   полупространстве  имеются  сторонние  токи,  которые
обозначим соответственно $\rv j^{+}(\rv  r)$  и  $\rv  j^{-}(\rv  r)$.
Выразим далее $\rv H_t(\rv r)$ в области $z>0$ с помощью леммы Лоренца
(18.3),  в которой в  качестве  поверхности  $S$  возьмём  плоскость
$z=0$, замкнутую полусферой бесконечно большого радиуса. За $\rv E_1$,
$\rv H_1$ примем искомые поля и положим $\rv  j^e_1=\rv  j^{+}$,  $\rv
j^m_1=0$,  а за $\rv E_2$,  $\rv H_2$ --- магнитную функцию Грина $\rv
{\tilde E}^{m}_{z>0}$,  $\rv  {\tilde  H}^{m}_{z>0}$,  удовлетворяющую
условию  излучения  на  замыкающей  полусфере  и  условию $\rv {\tilde
E}^m_t=0$  при  $z=0$.

     Эту функцию  легко  построить  с  помощью соответствующей функции
Грина для свободного пространства:  она представляет собой сумму поля
элементарного   магнитного   диполя, ориентированного  таким  образом,
что  тангенциальная составляющая его электрического  поля  на   плоскости  $z=0$ равна нулю,  и
поля зеркального  отражения  этого диполя в этой плоскости.  В результате
проекция   магнитного   поля   на   направление   магнитного   диполя,
определяемое ортом $\rv a$, имеет вид
     $$\rv a\rv H(\rv r_0)= -\int\limits_{V_{z>0}}\rv j^{+}(\rv r)\rv
         {\tilde  E}^m_{z>0}(\rv r,\rv r_0)\,dV+\frac c{4\pi}\int
         \limits_{S_0}[\rv E(\rv r) \rv{\tilde H}^m_{z>0}(\rv r,\rv r
         _0)]\rv n_1\,dS\,;\eqno(19.12)$$
аналогичным образом  для  левого  полупространства  в  соответствующих
обозначениях получаем:
     $$\rv a\rv H(\rv r_0)= -\int\limits_{V_{z<0}}\rv j^{-}(\rv r)\rv
         {\tilde E}^m_{z<0}(\rv r,\rv r_0)\,dV+\frac c{4\pi}\int
         \limits_{S_0}[\rv E(\rv r)\rv{\tilde H}^m_{z<0}(\rv r,\rv r_0
         )]\rv n_2\,dS\,.\eqno(19.13)$$
Функция Грина  $ \rv {\tilde E}^{m}_{z<0}$,  $\rv{\tilde H}^{m}_{z<0}$
отлична от функции $ \rv{\tilde E}^{m}_{z>0}$,  $\rv  {\tilde  H}^{m}_
{z>0}$  из-за  смены  знака  нормали  на плоскости $z=0$,  $n_1=-n_2$,
причём $n_2$ направлена вдоль оси $z$.

     Объёмные интегралы в  формулах  (19.12)  и  (19.13)  определяют
магнитные  поля  соответствующих сторонних токов при наличии сплошного
экрана;  обозначим их $\rv a\rv H^{+}_0$ и $\rv a\rv  H^{-}_0$.  Выбирая  теперь
точку  наблюдения  $\rv  r_0$  в  обоих  выражениях в непосредственной
близости от $S_0$,  располагая  вектор  $\rv  a$  в  плоскости  $z=0$,
получаем  из условия непрерывности поля интегральное уравнение первого
рода:
     $$ \rv a\{\rv H^+_0(\rv r_0)-\rv H^-_0(\rv r_0)\}=\frac c{4
         \pi}\int\limits_{S_0}\{[\rv{\tilde H}^{m}_{z>0}(\rv r,\rv r_
         0)\rv E(\rv r)]\rv n_1-[\rv{\tilde H}^{m}_{z<0}(\rv r,\rv r_0
         )\rv E(\rv r)]\rv n_2\}\,dS\,,\eqno(19.14)$$
где точка $\rv r_0$ принадлежит $S_0$.

     Решение полученного   интегрального   уравнения   встречает   ряд
специфических    математических    трудностей,    на   которых   здесь
останавливаться не будем --- за исключением одной, которая обусловлена
физически  нереализуемой  постановкой  задачи:  не  совсем  корректным
является предположение о бесконечно тонком проводящем экране.  С одной
стороны,  нулевая  толщина $d$ экрана позволяет убрать из формулировки
задачи  несущественный  параметр.  Физически  ясно,  что  при  условии
$d\ll\lambda$  и  $d\ll  D$  ($D$  ---  минимальный характерный размер
отверстия) толщина экрана не может  сказаться  на  значениях  полей  в
точках,  отстоящих  от  отверстия  на  расстояние  порядка  нескольких
$\lambda$ или $D$.  С другой стороны,  поверхность  экрана  необходимо
считать двухсторонней, поскольку наводимые на ней с разных сторон токи
сильно различаются. При нулевой толщине экрана кривизна поверхности на
кромке  отверстия  оказывается бесконечно большой,  что в свою очередь
приводит к бесконечно большим полям вблизи кромки,  в частности, имеет
особенность  нормальная  к линии кромки тангенциальная компонента $\rv
E$.  Поэтому  интеграл  в  уравнении  (19.9)  несобственный,  а  такие
интегральные   уравнения   первого   рода  имеют  несколько  или  даже
бесконечно  много  решений.  Естественно,  что   для   рассматриваемой
физической задачи встаёт проблема выбора единственного решения.

     Такой выбор   осуществляется   путём   наложения   на   решение
интегрального уравнения дополнительного условия,  которое в литературе
принято называть  {\it  условием  на  ребре}.  Физически  это  условие
заключается   в  требовании,  чтобы  энергия  электромагнитного  поля,
сосредоточенная в окрестности кромки  отверстия  или  любого  элемента
поверхности  тела,  где  радиус  кривизны равен нулю (например,  ребро
клина или вершина конуса),  оставалась конечной,  несмотря на  наличие
особенности у   самого   поля.   Таким   образом,   необходимо,  чтобы
выполнялось условие
     $$ \int\limits_{V_{\varepsilon}}\{|\rv E|^2+|\rv H|^2\}\,dV\;<\;
         \infty\;,\eqno(19.15) $$
где $V_{\varepsilon}$ --- область вблизи ребра.

     Этого условия (условия Мейкснера) оказывается  достаточно,  чтобы
из  решений  интегрального  уравнения  (19.14)  выделить единственное,
соответствующее физическому смыслу граничной задачи.  Следует сказать,
что  условие  (19.15)  ---  краткий  вывод которого будет приведён в
разделе 24 --- может нарушаться только в окрестности  сторонних  токов
(например,  элементарного  диполя)  и  позволяет различать сторонние и
наведённые токи в граничной задаче. С (19.15) тесно связано и другое
утверждение,  а  именно  --- равенство нулю среднего по времени потока
мощности через поверхность,  охватывающую область с особенностью поля,
причём  в  случае  выполнения  условия  Мейкснера поток всегда равен
нулю.  Обратное может и  не  иметь  места,  и  поэтому  требование равенства
нулю среднего  потока  мощности  не может заменить условия на
ребре,  обеспечивающее единственность решения.  В свою очередь отличие
среднего  потока  мощности  через замкнутую поверхность от нуля всегда
свидетельствует о  наличие  внутри  неё  сторонних  токов.  Обратное
утверждение также неверно:  например, в идеальном замкнутом резонаторе
средний поток через поверхность, окружающую элементарный диполь, равен
нулю.

     На рассмотренном примере плоского тонкого  проводящего  экрана  с
отверстием  хорошо  видны  преимущества определения поля за экраном по
тангенциальной составляющей электрического поля. Именно они показывают
целесообразность   введения   понятия   {\it   магнитных   токов}  при
одновременной  металлизации   отверстия.   Посмотрим   кратко,   какие
неудобства  возникают,  если  в  этой  граничной  задаче  пользоваться
интегральным уравнением для тангенциального магнитного поля. На первый
взгляд,  наоборот, просматриваются несомненные выгоды. Дело в том, что
наводимые на плоском  бесконечно  тонком  проводящем  экране  токи  не
создают   тангенциальной  компоненты  магнитного  поля  на  отверстии.
Наведённые   поверхностные   электрические   токи    всегда    можно
рассматривать  как  совокупность  элементарных  электрических диполей,
вектор дипольного момента которых лежит в плоскости  $z=0$.  Магнитное
поле  каждого  такого  диполя имеет в этой плоскости только нормальную
компоненту,  поэтому тангенциальное магнитное поле в отверстии  такого
экрана  {\it  точно равно падающему полю}.

     Однако возникает  проблема,   обусловленная   появлением   токов,
затекающих за кромку отверстия.  Тангенциальное магнитное поле отлично
от нуля на экране,  даже если экран рассматривать  двухсторонним,  при
этом  интегральное  уравнение  оказывается  сформулированным  для всей
бесконечной проводящей области экрана.  Интуитивно это  представляется
большим неудобством даже в том случае, когда в качестве падающего поля
выбирается поле,  возбуждаемое  при  наличии  сплошного  экрана.  Хотя
решение  соответствующего  интегрального уравнения будет в этом случае
быстро спадать при удалении от отверстия,  всё равно область,  которая
фигурирует в уравнении,  бесконечна. Такие же неудобства возникают при
вычислении  дифрагированного  поля  по  его  значениям  на   граничной
поверхности.

     Плоский проводящий экран с отверстием представляет особый интерес
ещё  и потому,  что для него выполняется так называемая {\it теорема
двойственности},    которая    формулируется    следующим     образом.
Сопоставляются  две граничные задачи,  причём предполагается,  что в
обеих задачах сторонние токи расположены в полупространстве  $z<0$.  В
плоскости $z=0$ в первой задаче расположен идеально проводящий экран с
отверстием;  удобно  обозначить  поверхность  экрана  через   $S$,   а
отверстие   ---   через  $s$.  Во  второй  задаче  в  плоскости  $z=0$
расположена тонкая проводящая пластинка,  по своей форме и положению в
пространстве  совпадающая  с  отверстием  в первой задаче (рис.~19.2);
поверхность пластинки также обозначим через  $s$,  а  остальную  часть
плоскости  ---  через  $S$.  Первичным  назовём  поле,  возбуждаемое
сторонними токами в безграничном пустом пространстве; для первой задачи
обозначим его $\rv  E^0_1,\;\rv  H^0_1$,  для  второй  --- $\rv
E^0_2,\;\rv H^0_2$.

\begin{picture}(160,52)
\put(5,50){\special{em:graph  fig19-2a.bmp}}
\put(84,50){\special{em:graph fig19-2b.bmp}}
\end{picture}
\begin{center}\begin{minipage}[c]{0.9\textwidth}
\footnotesize{
Рис.19.2.~К  теореме двойственности: {\it{ а)}} -- первая задача;
{\it {б)}} -- вторая задача.
}\end{minipage}\end{center}\vspace*{0.25cm}

     Теорема утверждает, что если первичные поля удовлетворяют
соотношениям
     $$ \rv E^0_2=\rv H^0_1\,,\quad \rv H^0_2=-\rv E^0_1\,,
         \eqno(19.16)$$
то в полупространстве $z>0$ (где нет  сторонних  токов)  {\it  полное}
поле  в  первой задаче ($\rv E_1,\;\rv H_1$) и {\it вторичное} поле во
второй задаче ($\rv E_2-\rv E^0_2,\;\rv H_2-\rv H^0_2$) связаны  между
собой  следующим  образом:
     $$\rv  E_2-\rv  E^0_2=-\rv H_1\,,\quad \rv H_2-\rv H^0_2=\rv E_1
         \,.\eqno(19.17)$$

     Поясним, что  условие  (19.16)  требует,  чтобы  сторонние токи в
первой задаче ($\rv j^e_1,\;\rv j^m_1$) и во второй ($\rv  j^e_2,\;\rv
j^m_2$) удовлетворяли соотношениям:
     $$\rv j^e_1=-\rv j^m_2\,\quad \rv j^m_1=\rv j^e_2.\eqno(19.18)$$
Эквивалентность условий   (19.16)  и  (19.18)  есть  прямое  следствие
симметрии  уравнений  Максвелла  (18.1)  относительно  соответствующей
замены.  Эта  симметрия уже была использована выше при вычислении поля
элементарного магнитного диполя.  В случае,  если сторонние  источники
расположены  очень  далеко  (на  бесконечности) и в качестве первичных
полей могут быть взяты плоские волны,  то поляризация этих волн в двух
задачах должна быть взаимно перпендикулярной.

     Доказательство теоремы основывается  на  сопоставлении  граничных
условий  на  плоскости  $z=0$  в  первой  и  второй задачах.  При этом
используется то обстоятельство, что поскольку в обеих задачах источники
при    $z>0$    отсутствуют,   то   электромагнитное   поле   в   этом
полупространстве  однозначно  определяется  значениями  тангенциальных
полей $\rv E_t$ и $\rv H_t$ на границе.  Точнее,  как следует из леммы
Лоренца,  при оптимальном выборе  функции  Грина  для  каждой  области
граничной  плоскости достаточно знания только касательной составляющей
одного поля. Для первой задачи граничные условия имеют следующий вид:
     $$\rv E_{1,t}=0\quad\mbox{на S}\,,\qquad \rv H_{1,t}=\rv H^0_{1,
         t}\quad\mbox{на s}\,;\eqno(19.19)$$
для второй ---
     $$ \rv E_{2,t}=0\quad\mbox{на s}\,,\qquad \rv H_{2,t}=\rv H^0_{2,
         t}\quad\mbox{на S}\,.\eqno(19.20)$$

     Пояснения требуют только граничные условия для  магнитных  полей.
Фактически утверждается,  что в части плоскости, свободной от металла,
тангенциальные составляющие вторичного магнитного поля, обусловленного
токами,  наведёнными  на проводящих участках плоскости,  равны нулю.
Выше уже было разъяснено, почему это действительно так.

     Предположим теперь,   что   нам   удалось  найти  решение  первой
граничной задачи $\rv  E_1,\;\rv  H_1$,  и  определим  решение  второй
задачи с помощью формул (19.17):
     $$\rv E_2=\rv E^0_2-\rv H_1\,,\quad \rv H_2=\rv H^0_2+\rv E_1\,.
         \eqno(19.21)$$
Это решение удовлетворяет однородным  уравнениям  Максвелла  из-за  их
симметрии  относительно  замены  $\rv H_1\to \rv E_2,\;\rv E_1\to -\rv
H_2$. Остаётся проверить, удовлетворяет ли решение (19.21) требуемым
граничным   условиям   (19.20).  Граничные  условия  (19.19),  которым
удовлетворяет решение $\rv  E_1,\;\rv  H_1$,  переписываются  согласно
(19.21) в виде
     $$\rv H_{2,t}-\rv H^0_{2,t}=0\quad\mbox{на S}\,,\qquad-\rv E_{2,
         t}+\rv E^0_{2,t}=\rv H^0_{1,t}\quad\mbox{на s}\,,
         \eqno(19.22)$$
что в  силу условия теоремы (19.16) и есть требуемые граничные условия
(19.20) для второй задачи.  Следовательно, (19.21) есть решение второй
задачи, что и доказывает теорему двойственности.

     Теорема двойственности    устанавливает    соответствие     между
излучением  из  узкой щели в проводящей плоскости и излучением тонкого
вибратора.  Она в значительной степени послужила  исходным  пунктом  в
развитии  теории  щелевых  антенн  и  в  обосновании введения в теорию
магнитных  токов.  Из  теоремы  вытекает,   в   частности,   очевидное
следствие:  электрическое  поле в узкой щели направлено поперек неё,
поскольку в дополнительной узкой проводящей ленте вторичное  магнитное
поле не имеет компоненты вдоль ленты.

     До сих пор речь шла о фундаментальных подходах к точному  решению
граничных  задач.  На  этом  пути исследователя встречают значительные
трудности, хотя при современном развитии теории дифракции они носят не
столько  принципиальный,  сколько вычислительный характер.  Однако
для многих практических задач  высокочастотной  электродинамики  вполне
удовлетворительные  результаты дают приближённые  методы.  Для  двух
выделенных   типов   граничных   задач   такие  приближённые   решения
строятся  разными способами.

     При излучении   системы   сторонних   токов,   расположенных    в
окрестности   достаточно   большого  (по  сравнению  с  длиной  волны)
проводящего тела,  наведённые на его поверхности  токи  определяются
теми  соотношениями,  которые возникают в случае падения плоской волны
на   проводящую   плоскость:   тангенциальная   компонента   падающего
магнитного поля удваивается,  чем и определяется плотность тока.  Если
размеры тела велики,  а поверхность достаточно гладкая, так что радиус
кривизны в каждой её точке велик по сравнению с длиной волны,  и при
этом падающее поле не сильно отличается от поля плоской  волны,  то  с
хорошей точностью можно считать,  что на освещённой поверхности тела
сохраняется указанное соотношение между магнитным полем на  проводнике
и  в  падающем поле,  а на теневой стороне токи равны нулю.  Тем самым
задача сводится к нахождению полей известных токов.

     Значительно лучшие  результаты  получаются  при учёте затекания
наведённых токов в область геометрической тени.  Такой учёт  легче
всего  провести  и  он  наиболее  эффективен  в  случае  металлических
поверхностей с большим радиусом  кривизны.  В  этом  случае  отношение
тангенциального   магнитного   поля   на   металле  к  соответствующей
компоненте  в  падающей  волне  описывается  универсальной   функцией,
обеспечивающей  плавный  переход  от  освещённой  области к тени.  В
освещённой области вдали  от  границы  геометрической  тени  функция
равна  двум,  а  в  глубокой  тени  стремится к нулю.  Аргументом этой
функции является отношение расстояния  от  границы  тени  к  параметру
$(R^2\lambda)^{1/3}$, определяющему размер области {\it полутени} ($R$
-- радиус кривизны поверхности).

     К другому   классу   можно  отнести  задачи,  где  приближённым
способом определяется поле на отверстии --- щели, прорезанной в стенке
волновода или резонатора,  открытом конце волновода или,  например, на
раскрыве  рупорной  антенны.  Речь  идёт  именно  о   приближённом
способе,  поскольку  строгий  подход на основе интегрального уравнения
зачастую мало перспективен.

     Наиболее универсальным можно считать приближение,  используемое в
случае дифракции известного падающего поля по одну сторону  отверстия.
Оно  состоит  в  прямой  замене  истинного  поля на отверстии падающим
полем,  то есть в  самой  плоскости  отверстия  не  учитывается  поле,
возбуждаемое  наведёнными  токами  на  металле.  Такой подход даёт
очень  хорошие  результаты  при  расчете   дифракционной   картины   в
оптических  задачах  и вообще во всех случаях,  когда {\it длина волны
мала по сравнению со всеми размерами отверстия}. Например, при падении
плоской   волны   на   металлический   экран   с   большим  отверстием
электромагнитное  поле  отлично  от  падающего  только  вблизи  кромки
отверстия  на  расстояниях порядка нескольких длин волн,  а поле по ту
сторону экрана определяется интегрированием по поверхности отверстия и
вклад  искажений вблизи кромки оказывается несущественным,  даже в тех
случаях,  когда из-за нулевой  толщины  экрана  эти  искажения  велики
(поскольку  истинное поле на кромке,  как было сказано выше,  обладает
особенностью).

     Точно также  поле  на открытом конце волновода принимается таким,
каким оно было бы в данном сечении бесконечного волновода,  а поле  на
раскрыве рупора принимается равным полю в сечении бесконечного конуса.
К сожалению,  в задачах,  с которыми имеет дело  электродинамика  СВЧ,
критерий,  определяющий применимость такого приближения, в большинстве
случаев не выполняется ---  длина  волны  излучения  близка  или  даже
несколько  больше  размеров  отверстия.  И тем не менее в ряде случаев
приходится  удивляться  хорошему  совпадению  результатов,  получаемых
таким  способом,  с точным решением или экспериментом.  Надо добавить,
что в  большинстве  практических  задач  такая  аппроксимация  поля  в
отверстии  далеко  не  единственное  приближение,  которое  приходится
использовать при расчёте поля во внешнем пространстве.

     Так, для  строгого  решения  задачи  по известному электрическому
полю на отверстии  наиболее  удобным  способом  является  металлизация
отверстия  и  размещение на нём эквивалентного магнитного тока.  При
этом для вычисления поля  во  внешнем  пространстве  необходимо  знать
электрическую   векторную   функцию   Грина   при   наличии   наружной
металлической   поверхности,    например,    волновода,    резонатора,
параболического  отражателя.  Зачастую определение такой функции Грина
само по себе является практически неразрешимой  задачей  и  приходится
дополнять  реальную  поверхность  до бесконечной плоскости,  например,
фланцем у волновода или  рупорной  антенны,  на  поверхности  которого
тангенциальную компоненту поля можно считать равной нулю. В результате
не так просто понять,  что привносит большее  искажение  ---  неточное
задание  поля  на отверстии или дополнительные металлические элементы.

     Аналогичные, если не большие,  трудности возникают при вычислении
дифрагированного  поля  по приближённому значению магнитного поля на
отверстии.  Ведь  в  этом  случае   даже   на   поверхности   сплошной
металлической   плоскости   нет   оснований   полагать  тангенциальную
компоненту равной нулю вблизи отверстия, поскольку реально имеет место
затекание  токов  за  его  кромку.  Опять  же  требуемая в этом случае
магнитная векторная функция Грина достаточно  легко  находится  только
для бесконечной проводящей плоскости. Не удаётся избежать трудностей
и  при  одновременном   приближённом   задании   и   магнитного,   и
электрического  поля  в плоскости отверстия.  Хотя в этом случае можно
пользоваться известной  функций  Грина  для  свободного  пространства,
однако  к  проблеме  учёта  затекающих токов и доопределения поля на
некоторой    воображаемой    поверхности     добавляется     трудность
непротиворечивого задания обеих компонент поля.

     Существует определённый вид  отверстий,  для  которых   использование
приближённого  значения тангенциального электрического поля особенно
эффективно.  Это  узкие  по  сравнению  с  длиной  волны  щели  в стенках
волноводов или  резонаторов,  протяженность  которых  значительно  больше
их  ширины.  Нормальная  к  кромке  щели  компонента  электрического  поля
на  ней  значительно  больше  продольной,  её  зависимость   от   координаты,
перпендикулярной   щели,   одинакова   для   всех  щелей.  Это  делает
целесообразным введение напряжения $V$ на щели как интеграла от  поля,
взятого  от  одной кромки до другой.

     В результате определение  поля  на  щели  сводится  к  одномерной
скалярной  задаче  вычисления  напряжения на щели в функции продольной
координаты $s$.  Если щель заметаллизировать  и  ввести  эквивалентную
скачку поля плотность магнитного тока, то задача о напряжении сводится
к задаче о распределении линейного магнитного  тока  вдоль  щели.  Для
щелей в плоском бесконечном экране согласно теореме двойственности эта
задача  эквивалентна  задаче  о  распределении  тока   вдоль   тонкого
вибратора.

     Наибольшее напряжение на щели возбуждается в  том  случае,  когда
она перерезает поперёк токи,  которые протекали бы в сплошной стенке
волновода  или  резонатора,  в  месте   их   максимальной   плотности.
Практическое  применение  в  качестве  антенных  устройств  в основном
находят щели,  длина которых близка к половине  длины  волны.  В  этом
случае напряжение вдоль щели распределяется по закону
     $$ V(s)=V_0\cos{ks}\,.\eqno(19.23)$$
Амплитуда $V_0$  пропорциональна  интегралу $\int \Delta H(s) \cos{ks}
\,ds$, где $\Delta H$ --- разность магнитных полей, параллельных щели,
которые  возникли  бы  при  её  металлизации  на  кромках  (под
действием возбуждающих полей).

     Ещё один способ приближённого  расчёта  поля  на  отверстии
применим в том случае, когда все размеры отверстия малы по сравнению с
длиной волны.  При этом всегда есть примыкающая к  отверстию  область,
которая   значительно   больше  размеров  отверстия  и  в  то же  время
расстояние до точек которой от середины отверстия  значительно  меньше
длины волны.  Распределение полей в этой области определяется законами
электростатики и магнитостатики,  поскольку в волновом уравнении можно
пренебречь членом, пропорциональным $k^2$, и оно переходит в уравнение
Лапласа;  таким  образом,  поле  вблизи   отверстия   определяется   в
статическом  приближении,  а  затем  продлевается  в  остальную  часть
пространства по волновым законам (для двумерных задач  в  этом  случае
очень   эффективен   метод   конформных   преобразований).   Поскольку
эквивалентные  токи  сосредоточены  в  малой  области,  то   отверстие
равносильно    действию    электрического    и   магнитного   диполей.
Электродинамической   характеристикой    малого    отверстия    служат
коэффициенты электрической и магнитной поляризуемости в падающем поле,
пропорциональные для круглого отверстия кубу его радиуса (в частности,
для плоского экрана поле  малого  круглого  отверстия  радиуса $a$ есть поле
элементарного магнитного диполя,  а поток энергии через это отверстие равен
половине излучения  диполя  и  $\sim\omega^4a^6|\rv H_0|^2$,  где $\rv H_0$ ---
падающее на  отверстие  поле).  Подчеркнём,  что  в  равной  степени
сказанное   относится  и  к  дифракции  возбуждаемых  полей  на  малых
проводящих телах.

%\end{document}

\newpage
\oddsidemargin=-0.4mm \evensidemargin=-0.4mm \topmargin=-0.4mm
\headsep=7mm \textheight=231.875mm \textwidth=160mm
\mathsurround=2.5pt \unitlength=1mm
%\begin{document}
%\input{macr.tex}
\thispagestyle{empty}
%\addtocounter{page}{210}

\begin{center}
\subsubsection*{20.~Возбуждение волноводов}
\end{center}
\vspace*{0.5cm}

\markboth{Глава 7.~Граничные   задачи   и   возбуждение   ограниченных
                структур} {20.~Возбуждение волноводов}

\begin{center}\begin{minipage}[c]{0.75\textwidth}
\footnotesize{\parindent=0.5cm
         Поперечные компоненты  поля   системы монохроматических
         сторонних токов в бесконечном  волноводе  и волноводе с  торцевой
         проводящей   стенкой. Продольные компоненты полей. Сопротивление
         излучения диполя в волноводе.   Возбуждение  прямоугольного
         волновода  идеально сгруппированным  пучком  зарядов,
         движущихся  по  нормали  к  стенке.   Поле   заряда,   движущегося
         вдоль  оси  круглого  волновода. Излучение Вавилова-Черенкова
         в круглом волноводе.
}\end{minipage}\end{center}\vspace*{0.5cm}

     Теория возбуждения  волноводов   сторонними   токами   фактически
представляет   собой   пример  специального  выбора  функции  Грина  и
существенным образом опирается на {\it  полноту}  системы  собственных
волноводных волн, рассмотренных в разделе 8. Напомним, что эта система
включает в себя всю совокупность $E$- и $H$-волн обоих направлений,  в
том  числе  и затухающих.  В силу полноты системы волноводных волн при
любом возбуждении волновода в нём ничего,  кроме цилиндрических волн
вполне   определённой   структуры,   возникнуть  не  может.  Поэтому
поперечные      компоненты      произвольного       монохроматического
электромагнитного   поля,   удовлетворяющего  уравнениям  Максвелла  и
граничным условиям на стенках волновода,  всегда можно разложить в ряд
по полям волноводных волн.

\begin{wrapfigure}[14]{l}{7.5cm}
\begin{picture}(80,50)
\put(-4,50){\special{em:graph fig20-1.bmp}}
\end{picture}
\hbox to 7.5cm{\hfil\footnotesize{Рис.~20.1.~Возбуждение волновода
}\hfil}
\hbox to 7.5cm{\footnotesize{ограниченной системой
сторонних токов.}\hfil}
\end{wrapfigure}

     Проще всего  рассматривается
возбуждение  волновода  ограниченной  системой
монохроматических токов. Пусть плотность токов
$\rv   j^e$   и   $\rv   j^m$   отлична  от  нуля  только  в
области,  ограниченной поперечными сечениями $z=z_1$ и  $z=z_2$,
которые  обозначим как $S_1$ и $S_2$,  так что все источники излучения
расположены в области $z_1<z<z_2$  (рис.~20.1).  Тогда  из  физических
соображений представляется очевидным,  что справа от плоскости $z=z_2$
поле  состоит  из  волн,   распространяющихся   (или   затухающих)   в
положительном направлении оси $z$,  а слева от плоскости $z=z_1$ --- в
отрицательном  направлении,  то  есть  волны  расходятся  от  области,
занятой  источниками.  Эти соображения эквивалентны условиям излучения
Зоммерфельда (17.6),  позволяющим  выбрать  единственное  решение  при
возбуждении    неограниченного   пустого   пространства   ограниченной
совокупностью сторонних токов.  Таким образом,  искомые поля $\rv E$ и
$\rv  H$  (во  всяком  случае  их  поперечные  компоненты) в областях,
свободных от токов, могут быть представлены в виде
     $$\rv E=\sum\limits_n\,A_n\rv E_n\,,\qquad \rv H=\sum\limits_n\,
         A_n\rv H_n\quad\mbox{при}\quad z>z_2\eqno(20.1\mbox{\textit a})$$
и
     $$\rv E=\sum\limits_n\,A_{-n}\rv E_{-n}\,,\qquad\rv H=\sum
         \limits_n\,A_{-n}\rv H_{-n}\quad\mbox{при}\quad z<z_1\,,
         \eqno(20.1\mbox{\textit б})$$
где $\rv  E_n$,  $\rv  H_n$  и  $\rv  E_{-n}$,  $\rv  H_{-n}$ --- поля
собственных   волноводных   волн   соответственно   положительного   и
отрицательного  направлений  (для  упрощения  записи  в  эти  величины
включены соответственно множители $e^{ih_nz}$ и $e^{-ih_n z}$). Индекс
у   полей  будем  считать  совпадающим  с  номером  члена  неубывающей
последовательности собственных значений краевых  задач  для  $E\,$-  и
$H\,$-волн, так что в суммы (20.1) входят и магнитные, и электрические
волны.   Волны,   соответствующие   разным   собственным    значениям,
ортогональны  в  смысле  (8.17),  а в случае вырождения предполагаются
ортогонализированными. Кроме того, будем считать поля собственных волн
нормированными  условиями  (8.23)  --  (8.24);  можно  выбрать и любую
другую нормировку,  при этом изменятся величины  коэффициентов  $A_n$,
$A_{-n}$,  однако  значения  полей  $\rv  E$  и $\rv H$,  естественно,
останутся неизменными.

     Коэффициенты $A_n$ и $A_{-n}$ легко  находятся  с  помощью  леммы
Лоренца  (18.3),  в  которой за $V$ следует принять объ\"ем волновода,
заключённый между  сечениями  $S_1$  и  $S_2$.  Ограничивающая  этот
объ\"ем  поверхность  $S$  включает  в себя помимо $S_1$ и $S_2$ также
поверхность стенок волновода $S_0$ (рис.  20.1).  В качестве  полей  и
токов  $\rv  E_1$,  $\rv  H_1$,  $\rv  j^e_1$,  $\rv  j^m_1$  в (18.3)
возьм\"ем искомые поля $\rv E$, $\rv H$ и заданные сторонние токи $\rv
j^e$,  $\rv j^m$,  а в качестве $\rv E_2$,  $\rv H_2$ --- поочерёдно
поля  собственной  волноводной   волны   с   номером   $m$   и   $-m$.
Соответствующие  им  токи  $\rv  j^e_2$,  $\rv  j^m_2$ следует считать
расположенными достаточно далеко от области $V$, так что они во всяком
случае не войдут в объёмный интеграл.  Если считать стенки волновода
идеально проводящими, и, следовательно, тангенциальные компоненты  электрического
поля  на  них  равны  нулю  и  поверхность $S_0$ вклада в интегралы не
даёт,  то  в  результате  получим  два   следующих   уравнения   для
коэффициентов:
     $$-\sum\limits_n A_{-n}\displaystyle{\int\limits_{S_1}\{[\rv E_
         {-n}\rv H_m]_z-[\rv E_m\rv H_{-n}]_z\}\,dS}+\sum\limits_n A_n
         \displaystyle{\int\limits_{S_2}\{[\rv E_n\rv H_m]_z-[\rv E_m
         \rv H_n]_z\}\,dS}=$$
     $$\qquad\qquad=\displaystyle{\frac{4\pi}c\int\limits_V(\rv j^e
         \rv E_m-\rv j^m\rv H_m)\,dV},\eqno(20.2)$$
     $$-\sum\limits_n A_{-n}\displaystyle{\int\limits_{S_1}\{[\rv E_
         {-n}\rv H_{-m}]_z-[\rv E_{-m}\rv H_{-n}]_z\}\,dS}+\sum
         \limits_n A_n\displaystyle{\int\limits_{S_2}\{[\rv E_n\rv H_
         {-m}]_z-[\rv E_{-m}\rv H_n]_z\}\,dS}=$$
     $$\qquad\qquad=\displaystyle{\frac{4\pi}c\int\limits_V(\rv j^e
         \rv E_{-m}-\rv j^m\rv H_{-m})\,dV}\,.\eqno(20.3)$$
Используя условия  ортогональности  (8.17)  и  нормировки  (8.23)  или
(8.24),  получаем отсюда выражения для коэффициентов в замкнутом виде.
Если $n$-ый член в сумме (20.1) представляет $E\,$-волну, то
     $$A_n=-\frac{2\pi}{\omega h_n \varepsilon}\int \limits_V(\rv j^e
         \rv E_{-n}-\rv j^m\rv H_{-n})\,dV\,,\eqno(20.4)$$
     $$A_{-n}=-\frac{2\pi}{\omega h_n \varepsilon}\int \limits_V(\rv j
         ^e\rv E_n-\rv j^m\rv H_n)\,dV\,,\eqno(20.5)$$
а если $H$-волну, то в этих выражениях следует заменить $\varepsilon$ на
$-\mu$.

     Найденные коэффициенты $A_n$ и $A_{-n}$ полностью решают задачу о
поле  излучения  системы  токов в дальней зоне,  где согласно основным
свойствам  волноводов  может  распространяться  лишь  конечное   число
волноводных волн.  Полный поток мощности излучения $\overline{\Sigma}$
в каждом направлении равен сумме потоков  отдельных  волн,  которые  в
свою  очередь  равны  потоку  мощности  в  нормированной волне (8.26),
умноженному на квадрат модуля соответствующего коэффициента $A_n$  или
$A_{-n}$. Таким образом,  для электрических волн полный поток мощности
излучения составляет
     $$\overline\Sigma=\displaystyle{\frac{\pi}{2\omega\varepsilon}}
         \displaystyle{\sum\limits_n\frac 1{h_n}}\times\displaystyle
         {\left\{\left |\int\limits_V(\rv j^e\rv E_{-n}-\rv j^m\rv H_
         {-n})\,dV\right |^2+\left |\int\limits_V(\rv j^e\rv E_n-
         \rv j^m\rv H_n)\,dV\right|^2\right\}}\,,\eqno(20.6)$$
где суммирование  производится лишь по распространяющимся волнам.  Для
магнитных волн в этой формуле следует заменить $\varepsilon$ на $\mu$.

     Полученное выражение позволяет сделать ряд определённых выводов
о свойствах излучения в волноводе. Наличие в знаменателе $h_n$ говорит
о  существенном  возрастании  мощности  излучения  вблизи  критической
частоты  соответствующей волны.  В этом случае сопротивление излучения
стремится к бесконечности и, следовательно, трудно говорить о заданных
сторонних  токах --- для их поддержания на постоянном уровне требуется
бесконечная мощность генератора. Необходимо учитывать обратное влияние
поля   излучения   на  сторонние  токи,  генератор  оказывается  плохо
согласованным  с  волноводом.  Поэтому   вблизи   критических   частот
стараются  не  работать;  к  тому  же  в этой области резко возрастают
потери в стенках волновода из-за конечной проводимости.

     Согласно (20.6) для эффективного возбуждения волновода  сторонние
электрические токи следует помещать вдоль силовых линий электрического
поля встречной волны,  а сторонние магнитные токи,  которые фактически
представляют собой щели в боковых стенках волновода, должны пересекать
поверхностные токи тех волн,  которые требуется  возбудить.  Очевидно,
что  продольные  электрические токи не возбуждают $H$-волн,  так как у
них $E_z=0$;  по аналогичной причине  продольные  щели  не  возбуждают
$E$-волн.

     Нетрудно обобщить  полученные  выше  формулы  на  полубесконечный
волновод,  ограниченный с одной стороны поперечной идеально проводящей
перегородкой  ---  возбуждаемое  в этом случае одностороннее излучение
как раз и находит основное применение на практике.  Пусть эта торцевая
стенка   расположена   при  $z=-L$;  тогда  естественно  считать,  что
$z_1>-L$.  Поле в области $z>z_2$ по-прежнему следует  искать  в  виде
(20.1а),  а  в  области $-L<z<z_1$ необходимо учитывать и отражённую
волну,  обеспечивающую выполнение граничного условия на перегородке.
В результате получаем:
     $$\rv E=\sum\limits_n\,A_{-n}(\rv E_{-n}\pm\rv E_ne^{2ih_nL})\,,
         \qquad\rv H=\sum\limits_n\,A_{-n}(\rv H_{-n}\pm\rv H_ne^{2i
         h_nL}),\eqno(20.7)$$
где верхний  знак  соответствует электрическим,  а нижний -- магнитным
волнам.  В качестве вспомогательного поля $\rv E_2$,~$\rv H_2$ в лемме
(18.3) также необходимо взять сумму волн обоих направлений, обращающую
в нуль поперечные компоненты электрического поля на перегородке:
     $$\rv E_2=\rv E_{-m}\pm\rv E_me^{2ih_mL}\,,\qquad \rv H_2
         =\rv H_{-m}\pm \rv H_m e^{2ih_mL}.\eqno(20.8)$$
После несложных преобразований,  аналогичных  приведённым  выше  для
бесконечного  волновода,  получаем из леммы (18.3) следующие выражения
для коэффициентов $A_n$: в случае электрических волн
     $$A_n=-\frac{2\pi}{\omega h_n \varepsilon}\int\limits_V\{\rv j^e
         (\rv E_{-n}+\rv E_n e^{2ih_nL})-\rv j^m(\rv H_{-n}+\rv H_n e^
         {2ih_nL})\}\,dV\,;\eqno(20.9)$$
в случае магнитных волн
     $$A_n=\frac{2\pi}{\omega h_n \mu}\int\limits_V\{\rv j^e(\rv E_
         {-n}-\rv E_n e^{2ih_nL})-\rv j^m(\rv H_{-n}-\rv H_n e^{2ih_nL
         })\}\,dV\,.\eqno(20.10)$$
К этим же результатам можно прийти,  рассматривая зеркальные отражения
источников возбуждения в идеально проводящей  поперечной  перегородке.
Отметим, что  выбор  знака  в формулах (20.7) и (20.8) непосредственно
следует из выражений для полей (7.16), (7.17):  поперечные  компоненты
электрического поля $E\,$-волны  при отражении меняют знак,  а
$H\,$-волны --- нет.

     Следует сказать  дополнительно  несколько  слов   о   возбуждении
волноводов магнитными токами. В практических расчётах магнитные токи
используются в основном при рассмотрении возбуждения  волновода  через
щели  или отверстия в его стенках.  Считая поле на отверстии известным
(приближённо или полагая его  неизвестной  функцией  координат,  для
которой  предстоит  получить интегральное уравнение),  в силу принципа
эквивалентности на отверстии  вводятся  поверхностные  магнитные  токи
согласно формуле (18.20),  после чего стенки волновода рассматриваются
как металлизированные.  Поле, возбуждаемое в таком волноводе заданными
поверхностными  токами,  вне  области  щели или отверстия определяется
теми же выражениями (20.1), (20.4), (20.5). Поскольку в двух последних
формулах   объёмные   интегралы   сводятся   к   поверхностным,   то
коэффициенты $A_n$ и $A_{-n}$ можно представить в виде
     $$A_n=\frac{2\pi}{\omega h_n}\int\limits_S \rv I^m \rv H_{-n}\,dS
         \,;\qquad A_{-n}=\frac{2\pi}{\omega h_n}\int\limits_S \rv I
         ^m\rv H_n\,dS\,,\eqno(20.11)$$
где $\rv   I^m$   ---   поверхностная  плотность  магнитного  тока,  а
интегрирование производится по поверхности отверстия или щели в стенке
волновода. На основании (18.20) эти выражения можно записать иначе:
     $$A_n=-\frac 1 {2kh_n}\int\limits_S[\rv n\rv E]\rv H_{-n}\,dS\,,
         \qquad A_{-n}=-\frac 1 {2kh_n}\int\limits_S[\rv n\rv E]\rv H
         _n\,dS\,,\eqno(20.12)$$
или
   $$A_n=\frac{2\pi}{\omega h_n}\int\limits_S \rv E\rv I^e_{-n}\,dS\,,
   \qquad A_{-n}=\frac{2\pi}{\omega h_n}\int\limits_ S \rv E\rv I^e_n
         \,dS\,,\eqno(20.13)$$
где
     $$\rv I^e_n=\frac c{4\pi} [\rv n\rv H_n]\,,\qquad\rv I^e_{-n}=
         \frac c{4\pi} [\rv n\rv H_{-n}]\eqno(20.14)$$
--- обычные   электрические    поверхностные    токи,
сопутствующие $n$-й  и  $-n$-й  волнам  на  металлизированной   стенке
волновода  (напомним,  что в (20.14) нормаль $\rv n$ направлена внутрь
металла).

     Отметим, что  если  возбуждение   электродинамической   структуры
осуществляется щелью   длиною   $l_0<\lambda$,   к  которой  приложено
постоянное (вдоль  щели)  напряжение  $V_0$,  то  оказывается  удобным
приближённо описывать её магнитным током,  направленным вдоль щели
и равным  $cV_0l_0\delta(\rv   r-\rv   r_0)/4\pi$,   где   $\rv   r_0$
соответствует середине  щели.  В  случае  возбуждения  структуры малой
петлёй с током $J_0$ её можно заменить на круговой  ток,  которому
--- как  это следует,  например,  из леммы Лоренца --- можно приписать
магнитный момент,  равный  $\mu  J_0S/c$,  где  $S$ --- площадь петли;
последнее даёт возможность ввести согласно формуле (17.14) магнитный
ток  $j^m=-i\omega\rv  m\delta(\rv  r-\rv r_0)$,  где вектор $\rv r_0$
направлен к центру петли.

     Все полученные до сих пор выражения  для  коэффициентов  $A_n$  и
$A_{-n}$   в   разложении   полей  (20.1)  обладают  тем  существенным
недостатком,  что  они  неприменимы   в   области   $z_1<z<z_2$,   где
расположены   источники   возбуждения.  Однако  в  теории  электронных
приборов волноводного типа, в которых электронный пучок движется вдоль
оси   волновода   и   взаимодействует   с  возбуждаемым  им  полем  на
протяжённом  отрезке,  именно  знание  поля  в  этом  интервале  $z$
позволяет  определить  его  обратное  влияние  на  пучок  и понять сам
принцип работы прибора.  Поскольку для любого $z$ из  этого  интервала
источники есть и справа,  и слева,  то представляется очевидным, что в
нём разложение  поперечных  компонент  поля  по  волноводным  волнам
должно  содержать  волноводные волны обоих направлений и,  более того,
коэффициенты разложения должны зависеть от продольной координаты:
     $$\rv E=\sum\limits_n[A_n(z)\rv E_n+A_{-n}(z)\rv E_{-n}]\,,\qquad
         \rv H=\sum\limits_n[A_n(z)\rv H_n+A_{-n}(z)\rv H_{-n}]\,.
         \eqno(20.1\mbox{\textit{в}})$$
Выражения для коэффициентов  $A_n(z)$  и  $A_{-n}(z)$  можно  найти  с
помощью  следующего  не  очень  строгого  (во  всяком  случае требующего
дополнительного обоснования) приёма:  удалим из узкого плоского слоя
$(z-\delta$,~$z+\delta)$   источники   и   тогда   в   соответствии  с
приведенными выше    результатами   для   областей,   свободных   от
источников, в пределе $\delta\to 0$ получим
     $$A_n(z)=-\frac{2\pi}{\omega h_n}\int \limits_{V_1}(\rv j^e\rv E_
         {-n}-\rv j^m\rv H_{-n})\,dV\,,\eqno(20.15)$$
     $$A_{-n}(z)=-\frac{2\pi}{\omega h_n}\int \limits_{V_2}(\rv j^e\rv
         E_n-\rv j^m\rv H_n)\,dV\,,\eqno(20.16)$$
где объём $V_1$ ограничен плоскостями $z_1$ и $z$,  а объём  $V_2$
--- плоскостями $z$ и $z_2$.  Таким образом, зависимость коэффициентов
от $z$  определяется  зависимостью от этой координаты объёма, по
которому проводится  интегрирование  в  (20.15)  и (20.16).

     Более строгий вывод этих формул  состоит  в  разбиении  уравнений
Максвелла  и компонент поля на продольные и поперечные и представлении
последних в виде рядов (20.1\mbox{\textit{в}}). В результате громоздких преобразований
получаются  обыкновенные  дифференциальные уравнения для коэффициентов
$A_n$ и $A_{-n}$, решение которых методом вариации постоянных приводит
к тем же самым выражениям (20.15) и (20.16).  Отметим, что эти формулы
представляют собой общее решение для  поперечных  компонент  поля  ---
(20.4) и (20.5) являются их частным случаем.

     Продольные компоненты поля,  а именно они особенно важны в теории
электронных  приборов,  в  общем  случае  могут быть найдены с
помощью известных  поперечных  компонент  только   из   уравнений
Максвелла; например, для $E_z$ из (18.1) следует, что
     $$E_z=\frac i{\varepsilon k}\rot_z\rv H-\frac{4\pi i}{\varepsilon
         \omega}j^e_z\,.\eqno(20.17)$$
Первое слагаемое справа содержит  только  производные  по  $x,  y$  от
поперечных компонент $\rv H$.  Следовательно,  его можно представить в
виде  суммы  $E_{n,z}$  и  $E_{-n,z}$  с  теми же коэффициентами $A_n,
A_{-n}$,  что и для поперечных компонент. Зависимость коэффициентов от
$z$ в данном случае не важна,  поскольку $z$-компонента $\rot\rv H$ не
содержит производных по $z$.  Второе слагаемое содержит в себе  только
заданные величины и поэтому
     $$E_z=\sum\limits_n(A_n E_{n,z}+A_{-n}E_{-n,z})-\frac{4\pi i}{
         \omega\varepsilon} j^e_z\,.\eqno(20.18)$$
Аналогичное выражение   получается   и   для   продольной   компоненты
магнитного поля:
     $$H_z=\sum\limits_n(A_n H_{n,z}+A_{-n}H_{-n,z})-\frac{4\pi i}{
         \omega\mu}  j^m_z\,.\eqno(20.19)$$

     Изложенная методика  расчёта  полей сторонних токов в волноводе
очень эффективна при вычислении потока излучения в дальней  зоне,  где
существенными  являются  лишь  несколько коэффициентов $A_n$,  которые
определяют амплитуды  распространяющихся  на  данной  частоте  волн  в
волноводе.  Вблизи же от области, занятой источниками излучения, вклад
в поля дают и затухающие волны. В ряде случаев ряды (20.1) оказываются
медленно  сходящимися  и  требуется учёт очень большого числа членов
ряда.  Физически  это  обусловлено  вкладом   квазистатического   поля
пространственного  заряда,  которое  не  носит  волнового  характера и
поэтому его разложение по решениям  волнового  уравнения  не  является
оптимальным.  Строго разделить поля в ближней зоне на волновую часть и
квазистатическую  не  представляется   возможным,   однако   удаётся
приближённо  найти  квазистатическое  поле,  отбросив  в  уравнениях
Максвелла  индукционные  члены,  но  оставив  токи   смещения.   Такая
возможность  связана  с  тем,  что  движение зарядов,  обусловливающее
переменные  токи,  практически  всегда  является  нерелятивистским.  В
результате  все  выражения  для  волновых полей становятся значительно
более громоздкими,  но ряды для компонент поля  сходятся  быстрее.  Не
будем останавливаться на этом вопросе здесь подробнее,  скажем только,
что целесообразность выделения поля пространственного заряда в  каждом
случае  определяется  спецификой  конкретной задачи и далеко не всегда
оправдана.

     Рассмотрим теперь   некоторые   частные   примеры   использования
изложенной  теории  возбуждения  волноводов.  Начнём   с   излучения
элементарных  диполей  и для упрощения записей будем полагать волновод
пустым  ($\varepsilon=  \mu=1$),  а  диполь  ---  электрическим.   При
плотности тока диполя
     $$\rv j^e=J\rv l\delta (x-x_0)\delta (y-y_0)\delta (z-z_0)\,,
         \eqno(20.20)$$
где $J$ --- линейный ток, $\rv l$ --- векторная длина диполя, из общей
формулы (20.6) следует, что суммарный поток мощности в оба направления
в волноводе составляет
     $$\overline \Sigma=\frac{\pi J^2l^2}{\omega}\sum\limits_m\frac 1
         h_m|E_{m,l}(x_0,y_0)|^2\,,\eqno(20.21)$$
где $E_{m,l}$  ---  проекция  вектора  $\rv E_m$ на направление диполя
$\rv l$,  а суммирование  производится  только  по  распространяющимся
волнам. Таким образом, сопротивление излучения диполя в волноводе есть
     $$R_s=\frac{2\pi l^2}{\omega}\sum\limits_m\frac 1 h_m|E_{m,l}
         (x_0,y_0)|^2\,,\eqno(20.22)$$
а его  отношение  к  сопротивлению излучения в пустоте $R$,
определяемому формулой (17.42), выражается в виде
     $$\frac {R_s}R=\frac{3\pi}{k^3}\sum\limits_m\frac 1 h_m|E_
         {m,l}(x_0,y_0)|^2\,.\eqno(20.23)$$

     В частности,  для прямоугольного волновода $(a>b)$, работающего в
одномодовом  режиме $(\pi/a<k<\pi/b)$,  при возбуждении волны $H_{10}$
диполем, ориентированным вдоль оси $y$, в соответствии с (9.8)
     $$\frac {R_s}R=\frac{6\pi}{kb\sqrt{k^2a^2-\pi^2}}
        \sin^2\frac{\pi x_0}a\,.\eqno(20.24)$$
Излучение максимально при помещении диполя в  пучность  электрического
поля волны.  При продольной ориентации диполя первой излучаемой волной
является $E_{11}$ (см.~(9.11)) и для неё
     $$\frac {R_s}R=\frac{12\pi g^2_{11}}{abk^3 h_{11}}\sin^2\frac
         {\pi x_0}a\sin^2\frac{\pi y_0} b\,,\qquad g^2_{11}=\frac{\pi
         ^2}{a^2}+\frac{\pi^2}{b^2},\quad h_{11}=\sqrt{k^2-g^2_{11}}
         \,.\eqno(20.25)$$

     Таким образом, соотношение излучения диполя в волноводе и пустоте
зависит от геометрических параметров волновода.  Волновод существенным
образом  изменяет  диаграмму  направленности излучения,  поскольку все
волны распространяются только в направлении оси  волновода.  При  этом
продольный  диполь  в волноводе излучает,  а в пустоте излучение вдоль
оси диполя отсутствует.

     Перейдём теперь к анализу излучения в  прямоугольном  волноводе
идеально сгруппированного заряженного пучка, пролетающего по нормали к
широкой  стенке  в  её  середине.  Плотность  тока  в  этом   случае
составляет
     $$\gv j_y=Qv\delta (z)\delta (x- a/ 2)\sum\limits_
         {n=-\infty}^{\infty}\delta(y-nL-vt)\,,\eqno(20.26)$$
где $Q,\;v$  --  заряд и скорость сгустков,  а $L$ -- расстояние между
ними.  Разлагая плотность тока в  ряд  Фурье  по  гармоникам  основной
частоты  $\omega_0=  2\pi  v/L$  (или  воспользовавшись соотношениями,
приведенными в Приложении П-6) , получаем:
     $$\gv j_y=\bigl(j_{0,y}+\re\sum\limits_{n=1}^\infty j_{n,y}e^
         {-in\omega_0t}\bigr)\,\delta(z)\delta(x- a/ 2)\,,
        \eqno(20.27)$$
где
     $$j_{0,y}=\frac{Qv} L\,,\qquad j_
         {n,y}=\frac{2Qv} L e^{in\frac{2\pi} L y}\,.\eqno(20.28)$$
Для вычисления  потока  мощности излучения на $n$-й гармонике (частота
$\omega_n=n\omega_0$)   по    формуле    (20.6)    необходимо    знать
$y$-компоненты электрического поля нормированных волноводных волн. Для
магнитных и электрических волн в  прямоугольном  волноводе  они  имеют
одинаковое   пространственное   распределение,  поскольку  собственные
значения $g_{pq}$ и продольные волновые числа $h_{pq}$ у них одни и те
же (отметим, что при этом поля волн $E_{pq}$ и $H_{pq}$ ортогональны в
смысле (8.17)):
     $$E^H_{pq,y}=ik_nN_{pq}\frac{\pi p} a e^{ih_{pq}z}
         \sin{\frac{\pi p}a x} \cos{\frac{\pi q} b}y,\quad
         q=0,1,\dots;\;p=0,1,\dots, \eqno(20.29)$$
     $$E^E_{pq,y}=ih_{pq}N_{pq}\frac{\pi q} b e^{ih_{pq}z}
         \sin{\frac{\pi p}a x} \cos{\frac{\pi q} b}y,\quad
         q=1,2,\dots;\;p=1,2,\dots, \eqno(20.30)$$
где
     $$k_n=\frac{\omega_n} c,\qquad h^2_{pq}=k_n^2-g_{pq}^2,\qquad
          g_{pq}^2=\pi^2\Bigl(\frac{p^2}{a^2}+\frac{q^2}{b^2}\Bigr),
         \eqno(20.31)$$
а
     $$N_{pq}^2=\frac 4{g_{pq}^2\, ab\,(1+\delta_{0p})(1+\delta_{0q})}\,.
          \eqno(20.32)$$
При такой записи составляющей поля собственных  волн  суммирование  по
$n$   в   формулах   (20.1)  и  (20.6)  следует  заменить  на  двойное
суммирование по $p$ и $q$,  в котором каждой  паре  значений  индексов
суммирования соответствуют два члена ряда, помечаемые верхним индексом
$H$ и $E$.  Очевидная симметрия излучения в обе  стороны  по  оси  $z$
позволяет учитывать в (20.6) только одно слагаемое в фигурных скобках,
удвоив при этом результат.  Таким образом, поток мощности излучения на
$n$-той гармонике оставляет
     $$\left.\begin{array}{l}\overline{\Sigma}_n=\displaystyle{\frac
         {\pi}{\omega_n}}\sum\limits_{p,q}\frac 1 h_{pq}\left\{\left|
         \int\limits_Vj_{n,y}E_{pq,y}^H\,dV\right|^2+\left|\int
         \limits_Vj_{n,y}E_{pq,y}^E\,dV\right|^2\right\}=\\[.8cm]
         =\displaystyle{\frac{8Q^2v}{abLn}}\sum\limits_{p,q}\frac
         {\displaystyle{\sin^2{\frac{\pi p}2}}\displaystyle{\left[
         \Bigl(\frac{2\pi\beta np}{La}\Bigr)^2+\Bigl(\frac{h_{pq}q}b
         \Bigr)^2\right]}}{\displaystyle{h_{pq}\Bigl(\frac{p^2}{a^2}+
         \frac{q^2}{b^2}\Bigr)(1+\delta_{0q})}}\left|\int\limits_0^b
         \cos{\frac{\pi q}b y}\,e^{in\frac{2\pi}L y}\,dy\right|^2\,,
         \end{array}\right.\eqno(20.33)$$
где, как и в (20.32), $\delta_{q0}$ --- символ Кронекера, а $\beta=v/c$
--- безразмерная скорость.

     В результате вычисления интеграла получаем окончательно:
     $$\overline{\Sigma}_n=\frac{8Q^2v}{\pi abn^2}\sum\limits_{p=1}
         \sum\limits_{q=0}\frac{\Bigl[\beta^2-\displaystyle{\Bigl(
         \frac{qL}{2bn}\Bigr)^2}\Bigr]\Bigl[1-(-1)^q\cos{\frac{2
         \pi nb}L}\Bigr]}{(1+\delta_{q0})\displaystyle{\Bigl[1-
         \Bigl(\frac{qL}{2bn}\Bigr)^2\Bigr]^2\sqrt{\beta^2-\Bigl(
         \frac{pL}{2na}\Bigr)^2-\Bigl(\frac{qL}{2nb}\Bigr)^2}}}\,,
         \eqno(20.34)$$
где суммирование по $p$ производится  лишь  по  нечётным  значениям.
Верхние  пределы  суммирования  по  $p$  и  $q$  определяются условием
излучения для соответствующей гармоники:
     $$\beta^2>\Bigl(\frac{pL}{2na}\Bigr)^2 +\Bigl(\frac{qL}{2nb}
         \Bigr)^2\,.\eqno(20.35)$$
Для основной  гармоники  $(n=1)$   существует   минимальное   значение
скорости   сгустков   $\beta=L/2a$,  начиная  с  которой  возбуждается
основная волноводная волна $H_{10}$.  Однако при любой скорости всегда
найдется  достаточно  большой  номер гармоники,  для которой излучение
имеет место. Более того, нетрудно видеть, что суммарное излучение всех
гармоник расходится.

     Физическая причина этого результата очевидна и состоит в том, что
для  идеально  сгруппированного,  то  есть  точечного,  сгустка работа
выхода из идеально проводящего металла равна бесконечности. Поэтому во
всех задачах,  где излучающий заряд вылетает из или влетает в идеально
проводящую среду (например,  в  случае  переходного  излучения)  имеет
место  расходимость излучаемой мощности на высоких частотах.  Отметим,
что  физически  расходимость  обусловлена  не  идеализацией  точечного
сгустка (пучок может,  в частности, состоять из отдельных электронов),
а предположением, что и   при очень высоких  частотах   стенки волновода
обладают свойствами идеального  проводника  (напомним,  что  при
$\omega\to\infty$ в любой среде  комплексная  диэлектрическая  проницаемость
$\varepsilon(\omega)\to  1$).  Но и при идеальных стенках расходимость
излучаемой мощности отсутствует,  если группировка пучка  неидеальная.
Так, если считать, что заряд сгустка равномерно распределён на длине
$d<L$,  то в разложении плотности тока (20.27) появляется  характерный
множитель
     $$\frac{\sin{({2\pi n d}/L})} {{2\pi nd}/L}\,,\eqno(20.36)$$
который и обеспечивает сходимость на высоких гармониках.

     Рассмотрим теперь типичный пример,  когда в результате разложения
плотности  стороннего  тока в интеграл Фурье весь волновод оказывается
заполненным  источниками  возбуждения.  Для  этого  решим  задачу   об
излучении  точечного  заряда,  пролетающего  параллельно  оси круглого
волновода,  заполненного диэлектриком с проницаемостью  $\varepsilon$,
не зависящей от частоты,  и посмотрим,  к каким результатам приводит в
такой сравнительно простой и ясной  задаче  пренебрежение  дисперсией.
Для  удобства  предположим,  что  заряд  $Q$  движется  равномерно  со
скоростью $v$ вдоль оси волновода;  тогда из-за азимутальной симметрии
задачи  в  волноводе  будут возбуждены только электрические волны типа
$E_{0q}$.  Если считать,  что ниже у всех величин (в  том  числе  и  у
корней  $\nu_{0q}$ функции Бесселя $J_0(x)$) индексу $n$ соответствует
совокупность азимутального (равного нулю) и радиального (равного  $q$)
индексов,  то  при  выбранной  ранее  нормировке  волноводных  волн их
потенциальная функция Герца равна
     $$\Pi^e(r)=\frac 1{\sqrt{\pi}\nu_nJ_1(\nu_n)}\cdot J_0\Bigl(
         \frac{\nu_n} a r\Bigr)\,,\eqno(20.37)$$
а поля представляются в виде
     $$\left .\begin{array}{lcl}E_{\pm n,r}&=&\displaystyle{\mp i
         \frac{h_n}{\sqrt{\pi}a J_1(\nu_n)}J_1\Bigl(\frac{\nu_n}a r
         \Bigr)e^{\pm ih_n z}},\\[.8cm]E_{\pm n,z}&=&\displaystyle{
         \frac{\nu_n}{\sqrt{\pi}a^2 J_1(\nu_n)}J_0\Bigl(\frac{\nu_n}a
         r\Bigr)e^{\pm i h_n z}},\\[.8cm]H_{\pm n,\varphi}&=&
         \displaystyle{-i\frac{k\varepsilon}{\sqrt{\pi} a J_1(\nu_n)}
         J_1\Bigl(\frac{\nu_n}a r\Bigr)e^{\pm i h_n z}}\,,
         \end{array}\right\}\eqno(20.38)$$
где $h^2_n=\varepsilon  k^2-\nu_n^2/a^2,$.

     Плотность тока   движущегося   заряда  в  цилиндрической  системе
координат имеет вид
     $$\gv j_z(\rv r,t)=\frac {Qv}{2\pi}\frac{\delta(r)}r \delta(z-vt)\,;\eqno(20.39)$$
разложим её и искомые   компоненты поля   $\gv  E_r(\rv  r,t)$,~$\gv  E_z(\rv  r,t)$,
~$\gv H_{\varphi}(\rv r,t)$ в интегралы Фурье
      $$\left .\begin{array}{lcl}\gv j_z(\rv r,t)&=&\int\limits_{-\infty}^\infty j_z
         (\rv r,\omega)e^{-i\omega t} \,d\omega\,,\\[.8cm]
      \gv E_r(\rv r,t)&=&\int\limits_{-\infty}^\infty
        E_r(\rv r,\omega)e^{-i\omega t} \,d\omega\,,\\[.8cm]\gv E_z(\rv r,t)&=&
         \int\limits_{-\infty}^\infty E_z
         (\rv r,\omega)e^{-i\omega t} \,d\omega\,,\\[.8cm]\gv H_{\varphi}(\rv r,t)&=&
         \int\limits_{-\infty}^\infty H_{\varphi}
         (\rv r,\omega)e^{-i\omega t} \,d\omega\,.\end{array}\right\}\eqno(20.40)$$
  Тогда   образы  Фурье  $j_z(\rv  r,\omega)$,~$E_r(\rv
r,\omega)$,~$E_z(\rv r,  \omega)$ и~$H_\varphi(\rv r,\omega)$ являются
комплексными     амплитудами,   используемыми   при   рассмотрении
монохроматических полей.

     В соответствии с известной формулой
      $$\delta(x)=\frac 1{2\pi}\int\limits_{-\infty}^{\infty}
         e^{-i\alpha x}\,d\alpha\eqno(20.41)$$
комплексная амплитуда плотности тока равна
     $$j_z(\rv r,\omega)=\frac{Q}{4\pi^2}\frac{\delta(r)}{r}\,
         e^{i\frac {\omega}{v}z}\,,\eqno(20.42)\vspace{.5cm}$$
а комплексные   амплитуды   полей   согласно   вышеизложенной   теории
возбуждения волноводов представляются в виде разложения по волноводным
волнам (20.38):
     $$\left.\begin{array}{l}E_ r(\rv r,\omega)=\sum\limits_n(A_nE_
         {n, r}+A_{-n}E_{-n, r})\,,\\[.5cm]E_z(\rv r,\omega)=\sum
         \limits_n(A_nE_{n,z}+A_{-n}E_{-n,z})-i\displaystyle{\frac{4
         \pi}{\omega\varepsilon}} j_z(\rv r,\omega)\,,\\[.5cm]H_
         {\varphi}(\rv r,\omega)=\sum\limits_n(A_nH_{n,\varphi}+A_{-n}
         H_{-n,\varphi})\,.\end{array}\right\}\eqno(20.43)$$

     Коэффициенты разложения  $A_n$ и $A_{-n}$ вычисляются по формулам
(20.4), (20.5) и, например, для $A_n$ получается следующее выражение:
     $$A_n=-\frac{2\pi}{\omega\varepsilon h_n}\int\limits_Vj_z E_{-n,z
         }\,dV=-\frac{Q\nu_n}{a^2\sqrt{\pi}\omega\varepsilon h_n J_1
         (\nu_n)}\int\limits_{-\infty}^z e^{i(\frac {\omega} v -h_n)z}
         \,dz.\eqno(20.44)$$
При вычислении последнего  интеграла  необходимо  учитывать  небольшую
мнимую  добавку  у  $h_n$,  обусловленную  поглощением  в  стенках ---
предположение об их идеальной  проводимости  является  некорректным  в
подобных  задачах  и  не  позволяет  выделить  единственное физически
разумное решение.  Тогда первообразная на нижнем пределе обращается в
нуль и в результате
     $$A_n=i\frac{Q\nu_n}{\sqrt{\pi}a^2\omega\varepsilon h_n J_1(
         \nu_n)}\;\frac 1{\displaystyle{\frac {\omega} v} -h_n}\;
         e^{i(\frac{\omega} v  -h_n)z}\,.\eqno(20.45)$$
Аналогичное выражение,  отличающееся лишь заменой  знака  перед  всеми
$h_n$, имеет место и для $A_{-n}$.

     Преобразуем выражение  для   плотности   тока
(20.42).  Как известно, для $\delta$-функции имеет место
разложение
     $$\delta(x-x_0)=\sum\limits_{n=1}^{\infty}\varphi_n(x)\varphi_n
         (x_0)\,,\eqno(20.46)$$
где $\{\varphi_n(x)\}$  ---  любая  полная  ортонормированная  система
функций;  в частности,  такую систему  в  интервале  $(0,a)$  образуют
функции
     $$\varphi_n( r)=\frac {\sqrt{2 r}}a\displaystyle{\frac{J_0
         (\nu_n r/a)}{J_1(\nu_n)}}\,.\eqno(20.47)$$
Поэтому
     $$\frac{\delta( r)} r=\frac 2{a^2}\sum\limits_{n=1}\displaystyle
         {\frac{J_0(\nu_n r/a)}{J_1^2(\nu_n)}}\,\eqno(20.48)$$
и, следовательно, плотность тока (20.42) может быть записана в виде
     $$j_z(\rv r,\omega)=\frac{2Q}{\pi^2a^2}\sum\limits_{n=1}
         \displaystyle{\frac{J_0(\nu_n r/a)}{J_1^2(\nu_n)}}
         \, e^{i\frac {\omega} v z}\,.\eqno(20.49)$$

     Теперь после  несложных  преобразований  из  (20.43)   с   учетом
(20.38), (20.45)   и   (20.49)   следует   выражение   для  продольной
составляющей электрического поля
     $$E_z(\rv r,\omega)=-i\frac{2Q\omega(1-\varepsilon \beta^2)}{\pi
         a^2v^2\varepsilon}\, e^{i\frac {\omega} v z}\sum\limits_{n=1}
         \displaystyle{\frac{J_0(\nu_n r/a)}{J_1^2(\nu_n)}}\frac 1{
         \displaystyle{\frac{\omega^2}{v^2}(1-\varepsilon\beta^2)+
         \Bigl(\frac{\nu_n}a\Bigr)^2}}\,;\eqno(20.50)$$
аналогичный вид имеют выражения и для  других  компонент  поля.  Таким
образом,  во  временном  представлении  выражения  для компонент полей
сводятся к интегралам Фурье вида
     $${\gv E}_z(\rv r,t)=-i\frac{2Q}{\pi a^2 v^2}
         \sum\limits_{n=1}\displaystyle{\frac{J_0(\nu_n r/a)}{J_1^2(
         \nu_n)}}\int\limits_{-\infty}^{\infty}\frac{1-\varepsilon
         \beta^2}{\varepsilon\Bigl[\displaystyle{\frac{\omega^2}{v^2}
         (1-\varepsilon \beta^2)+\Bigl(\frac{\nu_n} a\Bigr)^2}\Bigr]}\,
         e^{i\frac{\omega}v(z-vt)} \omega\,d\omega\,,\eqno(20.51)$$
вычисление которых в случае $\varepsilon=const$ не представляет труда.

     Наиболее простые результаты получаются при  условии  $\varepsilon
\beta^2<1$,  которое,  в  частности,  всегда  имеет место при движении
заряда  в  пустом  волноводе.  Вычисление  интеграла  (20.51)   удобно
проводить на комплексной плоскости $\omega$. Подынтегральное выражение
имеет два полюса на мнимой оси:
     $$\omega=\pm\, i \frac{\nu_n v}{a\sqrt{1-\varepsilon \beta^2}}\,,
         \eqno(20.52)$$
а контур   интегрирования,   проходящий   по   действительной  оси,  в
зависимости  от  знака  выражения  $z-vt$   приходится   замыкать   по
окружности   бесконечно   большого   радиуса   в  верхней  или  нижней
полуплоскости.  В каждом случае вклад в интеграл  даёт  только  один
полюс и в результате получается
     $${\gv E}_z(\rv r,t)=\mbox{sign}(z-vt)\frac{2Q}
         {a^2\varepsilon}\sum\limits_{n=1}^{\infty}\displaystyle{\frac
         {J_0(\nu_n r/a)}{J_1^2(\nu_n)}}\,e^{-\nu_n|z-vt|/a\sqrt{1-
         \varepsilon\beta^2}}\,.\eqno(20.53)$$
При $v\to   0$   (20.53)   переходит   в   компоненту   поля   заряда,
расположенного в точке $z=0$  на  оси  проводящей  трубы,  заполненной
диэлектриком:
     $${\gv E}_z(\rv r)=\mbox{sign}(z)\frac {2Q}{a^2
         \varepsilon}\sum\limits_{n=1}^{\infty}\frac {J_0(\nu_n r/a)}
         {J_1^2(\nu_n)}\,e^{-\nu_n|z|/a}.\eqno(20.54)$$
Это выражение   может   быть   получено    как    результат    решения
электростатической  задачи,  а  формула  (20.53)  из  него  с  помощью
известных формул преобразования полей в движущемся диэлектрике.

     Более интересные  результаты  получаются  при  выполнении условия
$\varepsilon\beta^2>1$. В этом случае в среде возникает так называемое
излучение   Вавилова-Черенкова,   имеющее   в   волноводе  характерные
особенности. В подынтегральном выражении (20.50) на действительной оси
$\omega$ для каждого значения индекса $n$ появляются два полюса
     $$\omega_n=\pm\,\frac{\nu_n v}{a\sqrt{\varepsilon\beta^2-1}}\,,
         \eqno(20.55)$$
в результате чего интеграл становится несобственным и неопределённым
до тех пор, пока не будет установлено правило обхода полюсов. Нетрудно
видеть,  что  по  физическому  смыслу  полюсы  соответствуют   условию
синхронизма  волны  и  частицы:  фазовая скорость волноводной волны на
этих  частотах  совпадает  со  скоростью  движения  заряда.

     Введение небольшого   затухания   в   среде,   которое  неизбежно
присутствует в силу дисперсионных  соотношений,  показывает,  что  оба
полюса  смещены  вниз  относительно  действительной  оси.  Поэтому при
замыкании  контура  интегрирования  в   верхней   полуплоскости   (что
приходится  делать  при  $z>vt$ для обнуления вклада от полуокружности
большого радиуса) интеграл  оказывается  равным  нулю.  Наоборот,  при
замыкании  контура  в нижней полуплоскости вычеты в обоих полюсах дают
вклад в интеграл и в результате
     $$\gv E_z(\rv r,t)=\left\{\begin{array}{l}0
         \hspace{8.3cm}\mbox{     при}\quad z>vt\,,\\[.5cm]-\displaystyle
         {\frac{4Q}{a^2\varepsilon}}\sum\limits_{n=1}^{\infty}
         \displaystyle{\frac{J_0(\nu_n r/a)}{J_1^2(\nu_n)}}\cos{
         \displaystyle{\frac{\nu_n}{a\sqrt{\varepsilon \beta^2-1}}
         (z-vt)}}\qquad\mbox{при }\quad z<vt\,.\end{array}\right.
         \eqno(20.56)$$

     Таким образом,  при  сделанных предположениях о свойствах среды в
области перед  зарядом  поле  отсутствует,  а  в  области,  пройденной
зарядом,  возбуждена  система незатухающих волноводных волн дискретных
частот,  определяемых условием  синхронизма.  При  выполнении  условия
излучения Вавилова-Черенкова $\varepsilon\beta^2>1$ заряженная частица
тратит  энергию  на  возбуждение  поля  излучения,  которое  по   мере
продвижения  частицы  заполняет  всё  большую  часть  пространства в
волноводе.  В  случае  невыполнения  условия  излучения  поле  (20.53)
обладает  свойствами  трансляционной симметрии и перемещается вместе с
зарядом, не увеличивая занимаемый им объём и сохраняя запасённую в
нём энергию.

     Потери энергии частицей в единицу времени  могут  быть  вычислены
как   работа   поля   над   током   $P=-\int\mbox{\boldmath  $\gv  j$}
\mbox{\boldmath $\gv E$}\,dV$,  однако непосредственно основываясь  на
формулах  (20.39)  и  (20.56)  этого  сделать  нельзя,  поскольку поле
$\mbox{$\gv E$}_z$ в точке $z=vt$  и  обращается  в  бесконечность,  и
терпит разрыв.  Дело в том, что (20.56) представляет собой полное поле
заряда,  включающее в себя  и  кулоновское  поле,  и  поле  излучения.
Кулоновское  поле  в  точке  нахождения  заряда  всегда  обращается  в
бесконечность и терпит  разрыв.  Но  продольное  поле  антисимметрично
относительно  плоскости  $z=vt$  и вклада в реакцию излучения не дает.
Для выделения антисимметричной части поля следует провести  вычисления
аккуратнее,  начиная  с  вычисления  интеграла  по $dV$;  в результате
получается интеграл по частоте $\omega$ вида
     $$P=-i\frac{Q^2}{\pi a^2v}\sum\limits_{n=1}^{\infty}\frac 1{ J_1
         ^2(\nu_n)}\cdot\int\limits_{-\infty}^{\infty}\frac{1-
         \varepsilon\beta^2}{\varepsilon\Bigl[\displaystyle{\frac{
         \omega^2}{v^2}(1-\varepsilon\beta^2)+\Bigl(\frac{\nu_n} a
         \Bigr)^2}\Bigr]}\omega\,d\omega\,,\eqno(20.57)$$
который в отличие от (20.51) не содержит экспоненты и потому не  может
быть  вычислен  путем  замыкания  контура  по  дуге  большого радиуса.
Имеющиеся на пути интегрирования полюсы в точках (20.55)  должны  быть
смещены вниз  и  в  результате интеграл (20.57) сводится к интегралу в
смысле главного значения и сумме полувычетов  в  полюсах.  Интеграл  в
смысле  главного  значения описывает вклад в реакцию антисимметричного
кулоновского  поля  и,  как  нетрудно  видеть,   равен   нулю.   Сумма
полувычетов представляет работу поля излучения на единицу длины
     $$P=\frac{2Q^2 v}{a^2\varepsilon}\sum\limits_{n=1}^{\infty}\frac 1
         {J_1^2(\nu_n)}\,.\eqno(20.58)$$

     Полученное выражение,  несмотря  на  свою  простоту  и изящество,
обладает  тем  существенным  недостатком,   что   представляет   собой
расходящийся   ряд.   Таким  образом,  реакция  излучения  оказывается
бесконечно  большой,  что  не  соответствует  физической   реальности.
Причина  этого  очевидна  и  весьма  поучительна  ---  она  состоит  в
пренебрежении дисперсией диэлектрической проницаемости и поэтому вклад
высоких   частот   оказывается  существенно  завышенным.  При  учёте
дисперсии всё явление сильно  усложняется:  пропадает  излучение  на
высоких частотах,  зато появляются дополнительные полюсы,  связанные с
функцией $\varepsilon(\omega)$,  в частности,  с нулями этой  функции.
Соответствующую  часть  реакции излучения связывают с поляризационными
потерями в  веществе.  Большое  своеобразие  излучение  приобретает  в
области аномальной дисперсии,  где всегда велико поглощение в среде,
но здесь не место останавливаться на всех этих деталях.

%\end{document}

\newpage
\oddsidemargin=-0.4mm \evensidemargin=-0.4mm \topmargin=-0.4mm
\headsep=7mm \textheight=231.875mm \textwidth=160mm
\mathsurround=2.5pt \unitlength=1mm
%\begin{document}
%\input{macr.tex}
\thispagestyle{empty}
%\addtocounter{page}{224}

\begin{center}\subsubsection*{21.~Возбуждение резонаторов}\end{center}
\vspace*{0.5cm}

\markboth{Глава 7.~Граничные   задачи   и   возбуждение   ограниченных
         структур}{21.~Возбуждение резонаторов}

\begin{center}\begin{minipage}[c]{0.76\textwidth}
\footnotesize{\parindent=0.5cm
         Постановка задачи   о   возбуждении   резонатора    заданными
         монохроматическими   токами. Продольная  и  поперечная  части
         поля  вынужденных  колебаний. Разложение поперечного  поля по
         собственным  колебаниям  резонатора.  Резонансный    характер
         поперечного  поля,  нерезонансный  фон.   Резонанс  в  случае
         комплексных собственных частот. Альтернативный способ решения
         задачи о возбуждении  волноводных резонаторов. Поле точечного
         заряда,  пролетающего  по  оси  цилиндрического резонатора.
}\end{minipage}\end{center}
\vspace*{0.5cm}

     Изложение теории  возбуждения  резонаторов  начнём с вычисления
электромагнитного поля  системы  заданных  монохроматических  токов  в
пустом   (диэлектрическая и   магнитная  проницаемости  равны единице)
замкнутом резонаторе с идеально  проводящими  стенками.  Для  общности
рассмотрим  совокупность сторонних электрических $\rv j^e$ и магнитных
$\rv j^m$ токов.  Таким образом,  задача сводится  к  решению  системы
уравнений Максвелла

     $$\rot\rv  E-ik \rv H=-\frac{4\pi} c\rv j^m,\quad\rot\rv  H+ik
         \rv  E=\frac{4\pi} c\rv  j^e\eqno(21.1)$$
с граничным условием $E_t=0$ на стенках резонатора.  В  данном  случае
волновое  число  $k$  задано  ---  оно определяется частотой сторонних
токов.

     Возбуждаемое в резонаторе электромагнитное поле можно  однозначно
разделить на две части --- {\it продольное} $\rv E^l,\;\rv H^l$ и {\it
поперечное} $\rv E^t,\;\rv H^t$, так что

     $$\rv E=\rv E^l+\rv E^t\,,\quad \rv H=\rv H^l+\rv H^t\,.
         \eqno(21.2)$$
Поперечная часть поля удовлетворяет уравнениям $\div \rv E^t=0$, $\div
\rv  H^t=0$,  поэтому  её  называют  ещё   {\it   соленоидальной}.
Продольное  поле может быть выражено через скалярные потенциалы:  $\rv
E^l=-\grad\Phi^e,  \;\rv H^l=-\grad\Phi^m$ и поэтому часто  называется
{\it   потенциальным}.   Обратим  внимание,  что  здесь  продольная  и
поперечная части поля имеют другой смысл, чем в теории волноводов, где
существует  выделенное  направление  ---  ось волновода,  относительно
которой  и  производится  разбиение  поля   на   части.   Терминология
{\it продольное}  и  {\it поперечное}  поле в резонаторе связана с разложением
поля  в  трёхмерный  пространственный  интеграл  Фурье:  $\rv  E(\rv
r)=\int  \rv  E(\rv  k)e^{i\rv  k  \rv r}\,d\rv k$.  В продольном поле
вектор $\rv E^l$ параллелен вектору $\rv k$,  в поперечном поле вектор
$\rv E^t$ ортогонален $\rv k$.

     Рассмотренная выше   (раздел  16) ортонормированная  система
векторных собственных  функций  резонатора  $\rv  E_n$,~$\rv
H_n$  в  определённом  смысле  является  полной: соленоидальная часть
произвольного  поля, удовлетворяющего неоднородным
уравнениям (21.1),  может быть разложена в ряд по этим
функциям.  Напомним, что каждое собственное колебание резонатора
представляет собой соленоидальное поле,  поскольку уравнения $\div \rv
E_n=0$   и  $\div\rv  H_n=0$  являются  прямым  следствием  однородных
уравнений (16.1).  Это позволяет  представить  искомые
полные поля $\rv E,\;\rv H$ в виде:

     $$\rv E=\sum\limits_nA_n\rv E_n-\grad\Phi^e,\qquad \rv H=\sum
         \limits_nB_n\rv H_n-\grad\Phi^m\,.\eqno(21.3)$$
Отметим, что коэффициенты в разложении для поперечных электрического и
магнитного  полей  ($A_n$  и  $B_n$) различны,  тогда как в разложении
искомых полей  по  собственным  волноводным  волнам  (20.1)  в  теории
возбуждения  волноводов  они одинаковы.

     Подстановка разложений  (21.3)  для  полей  $\rv  E$,~$\rv  H$  в
уравнения Максвелла (21.1) приводит к следующим соотношениям:

     $$\left.\begin{array}{l}i\displaystyle{\sum\limits_n}(k_nA_n-k
         B_n)\rv H_n=-\displaystyle{\frac{4\pi} c}\rv j^m-ik\grad\Phi
         ^m,\\[.4cm]i\displaystyle{\sum\limits_n}(kA_n-k_nB_n)\rv E_n=
         \phantom{-}\displaystyle{\frac{4\pi} c}\rv j^e+ik\grad\Phi^e.
         \end{array}\right\}\eqno(21.4)$$
Вычислив дивергенцию от обеих частей этих уравнений и воспользовавшись
уравнениями непрерывности

     $$\div\rv j^e=i\omega\rho^e\,,\quad\div\rv j^m=i\omega\rho^m,
         \eqno(21.5)$$
получим уравнения для определения потенциалов продольного поля:

     $$\left.\begin{array}{l}\triangle\Phi^e=-4\pi\rho^e\,,\\[.3cm]
         \triangle\Phi^m=-4\pi\rho^m.\end{array}\right\}\eqno(21.6)$$

     Это ---  типичные  уравнения  электростатики  и   магнитостатики,
которые   решаются   соответствующими  методами  (пример  решения  для
круглого  цилиндрического  резонатора  будет  приведён  ниже).   Вся
зависимость  от  частоты  источников возбуждения сводится в продольном
поле к множителю $e^{-i\omega t}$ (при  заполнении  резонатора веществом
  может  возникнуть  дополнительная   зависимость,    обусловленная
дисперсией проницаемостей $\varepsilon$ и $\mu$) и,  следовательно,
резонансными  свойствами оно не обладает.

     Поэтому основной интерес,  и  именно  из-за  своего  резонансного
характера,  в  практических  приложениях представляет поперечное поле.
Уравнения для определяющих  его  коэффициентов  разложения  получаются
путём скалярного умножения первого соотношения (21.4) на $\rv  H_m$,
второго   соотношения  на  $\rv  E_m$  и  интегрирования  по  объёму
резонатора.  Важным  моментом  при   этом   является   ортогональность
потенциального   поля   и   поля   собственных   функций   резонатора.
Действительно,  интегрируя выражения $\div[\rv H_n  \grad  \Phi^e]$,~$
\div[\rv  E_n\grad\Phi^m]$ по объёму резонатора,  легко убеждаемся в
том, что

     $$\int\rv E_n\grad\Phi^e\,dV=0,\qquad\int\rv  H_n\grad\Phi^m
         \,dV=0.\eqno(21.7)$$
Учитывая  ортогональность  собственных  колебаний  в  смысле  (16.8) и
нормировку (16.10), получаем

     $$\left.\begin{array}{l}i(kA_n-k_nB_n)=\displaystyle{\frac 1 c}
         \int\rv j^e\rv E_n\,dV\,,\\[.4cm]i(k_nA_n-kB_n)=\displaystyle
         {\frac 1 c}\int\rv j^m\rv H_n\,dV\,,\end{array}\right\}
         \eqno(21.8)$$
откуда легко находятся и сами коэффициенты:

     $$\left.\begin{array}{l}A_n=\displaystyle{\frac i{\omega^2-\omega
         _n^2}}\Bigl(\omega_n\int\rv j^m\rv H_n\,dV-\omega\int\rv j^e
         \rv E_n\,dV\Bigr)\,,\\[.4cm]B_n=\displaystyle{\frac i{\omega
         ^2-\omega_n^2}\Bigl(\omega\int\rv j^m\rv H_n\,dV-\omega_n
         \int\rv j^e\rv E_n\,dV\Bigr)}\,.\end{array}\right\}
         \eqno(21.9)$$

     При $\omega\to \omega_n$ коэффициенты $A_n$ и $B_n$ неограниченно
возрастают и всё меньше  отличаются  друг  от  друга.  В  результате
вблизи   резонанса   структура  поля  практически  совпадает  с  полем
выделенного  собственного  колебания.  Небольшое  отличие  обусловлено
вкладом колебаний,  собственные частоты которых далеки от резонанса, а
также продольным нерезонансным полем.  Все  вместе  они  образуют  так
называемый {\it нерезонансный фон}.

     Из формул  (21.9)  следует,  что сторонний электрический ток $\rv
j^e$ наиболее эффективно возбуждает колебание в том случае,  когда  он
находится  в  пучности электрического поля $\rv E_n$ и параллелен ему.
При возбуждении магнитными токами $\rv  j^m$  их  следует  помещать  в
пучности  магнитного поля параллельно $\rv H_n$.  В случае петли связи
её следует помещать перпендикулярно силовым линиям магнитного  поля,
в области, где они гуще всего. Щель нужно прорезать в области наиболее
сильного  поверхностного  тока,  при  этом  узкие  щели  должны   быть
прорезаны поперёк линий тока.  В этом смысле имеется полная аналогия
с возбуждением волноводов.

     Следует особо  отметить,  что до тех пор,  пока стенки резонатора
считаются идеально проводящими,  а другие виды  потерь  (поглощение  в
материале  заполнения  или излучение через отверстие) отсутствуют,  то
есть пока собственные частоты $\omega_n$ величины действительные, даже
при  очень  малой  отстройке  от  резонансной  частоты  возбуждённое
электромагнитное поле и сторонние токи сдвинуты по фазе на $\pi/2$ и в
среднем  по  времени  не  обмениваются энергией.  Положение изменяется
только строго на собственной частоте,  когда $\omega=\omega_n$ и  сами
формулы  (21.9)  теряют  смысл.  Вблизи  собственной частоты колебания
необходимо учитывать потери в резонаторе,  какими  бы  малыми  они  не
были.

     Учёт потерь приводит к тому, что собственные частоты становятся
комплексными: ${\omega}_n=\omega_n'-i\omega_n''=\omega_n(1-i
/2Q_n)$,  где $Q_n$ ---  добротность  $n$-го  колебания.  Если  потери
обусловлены    неидеальной    проводимостью    стенок,    то   условия
ортогональности (16.8) могут быть обобщены на резонатор с неоднородным
заполнением    $\varepsilon(\rv   r)$   путём   увеличения   области
интегрирования на приповерхностный слой металла,  в пределах  которого
протекает  ток.  Таким  образом,  можно  строго показать,  что формулы
(21.9) справедливы и для резонатора с малыми  потерями.  В  результате
резонансный   знаменатель  ни  при  каком  значении  частоты  $\omega$
возбуждающего  тока  не  обращается  в   нуль   и   $|A_n(\omega)|^2$,
определяющий энергию поля в резонаторе,  остаётся конечным, достигая
максимума на резонансной частоте

     $${\omega}_{\mbox{\textit{рез}}}=\omega_n'\sqrt{1-\Bigl(\frac{\omega_n''}
           {\omega_n'}\Bigr)^2}=\omega_n'\sqrt{1-\frac 1{4Q^2_n}}\,.
           \eqno(21.10)$$
Отметим, что  в  реальных  резонаторах   сама   величина   $\omega_n'$
отличается  от собственной частоты $\omega_n$ идеального резонатора на
величину  того  же  порядка,   что   и   $\omega_n''$.   Комплексность
собственной  частоты  резонатора  приводит к появлению дополнительного
сдвига фаз между  током  $\rv  j^e$  и  параллельной  ему  компонентой
электрического поля,  что обеспечивает равенство работы тока над полем
и потоком энергии  в  стенки  резонатора.  Если  потери  в  резонаторе
обусловлены потоком мощности в подводящий волновод,  то простой учёт
комплексности собственной частоты даёт  мало  пользы  для  расчёта
взаимодействия резонатора с волноводом. В этом случае необходимо более
тщательное   исследование,   при   котором   волновод   и    резонатор
рассматриваются как единая структура.  Для простейшей модели оно будет
приведено в следующем разделе.

     Обычно стараются     добиться     того,    чтобы резонатор работал на
одном  определённом  типе колебания,  а  другие  типы колебаний не
возбуждались.  В случае разреженного спектра и высокой добротности
рабочего  колебания  это получается  само  собой,  но  чаще  всего для
подавления нежелательных колебаний  приходится  применять  специальные
меры   ---    собственные частоты таких колебаний смещаются  из диапазона
рабочих частот или же искусственно снижается их добротность за счёт
избирательного поглощения.

     Таким образом,  теория  возбуждения резонаторов строится по общей
схеме теории возбуждения системы с конечным числом  степеней  свободы.
Система  собственных  колебаний  резонатора  в некотором смысле
полнее,  чем система волноводных волн:  в волноводе произвольное поле
не  является  суммой  волноводных  волн  с постоянными амплитудами ---
такое разложение существует только в сечениях,  где нет источников.  В
резонаторах  же  полнота системы функций  обеспечивается во всех
трёх измерениях.

     Особое место среди резонаторов занимают  {\it волноводные}      резонаторы,
которые представляют    собой   отрезок  волновода, ограниченный   с  двух
сторон  торцевыми плоскостями,  нормальными  к   его  оси.   Теория
возбуждения  волноводных  резонаторов   может   быть построена    по
аналогии    с   теорией  возбуждения  волноводов.  В ряде практически
важных  случаев   такой   подход   обладает  определёнными преимуществами
и    поэтому  целесообразно кратко рассмотреть его.

    Не уменьшая   общности,   будем  считать,  что  торцы  резонатора
находятся  при $z=0$ и $z=d$, а область, где расположены возбуждающие
сторонние  токи,  ограничена  сечениями  $z=z_1$  и  $z=z_2$,  так что
$0<z_1<z_2<d$. Представим поле в резонаторе в виде

     $$\left.\begin{array}{l}\rv E=\sum\limits_n\bigl(C_n\rv E_n^R+C_
         {-n}\rv E_{-n}^R\bigr)-i\displaystyle{\frac{4\pi}{\omega}}j_
         z^e\rv e_z\,,\\[.4cm]\rv H=\sum\limits_n\bigl(C_n\rv H_n^R+C_
         {-n}\rv H_{-n}^R\bigr)-i\displaystyle{\frac{4\pi}{\omega}} j_
         z^m\rv e_z\,,\end{array}\right\}\eqno(21.11)$$
где $\rv  e_z$  ---  единичный орт в направлении оси $z$,  а векторные
функции  $\rv  E_{\pm  n}^R,\;\rv   H_{\pm   n}^R$   есть   комбинация
нормированных  в  смысле  (8.23),  (8.24)  полей  прямых  (волна  $\rv
E^R_n$,~$\rv H^R_n$ распространяется в направлении положительных  $z$)
и обратных (волна $\rv E^R_{-n}$,~$\rv H^R_{-n}$ набегает на плоскость
$z=0$) волноводных волн:

     $$\rv E_{\pm n}^R=\rv E_{\pm n}+R_{\pm n}\rv E_{\mp n}\,,\quad
         \rv H_{\pm n}^R=\rv H_{\pm n}+R_{\pm n}\rv H_{\mp n}\,.
         \eqno(21.12)$$
Коэффициенты $R_{\pm n}$ определяются  из  условия  обращения  в  нуль
поперечных  (относительно оси $z$) компонент функций $\rv E^R_{\pm n}$
на торцевых стенках,  от которых они полностью отражаются:  $R_n$  ---
функции  $\rv E^R_n$ на правом торце $z=d$,  $R_{-n}$ --- функции $\rv
E^R_{-n}$ на левом торце $z=0$. Легко видеть, что
     $$  R_n=\pm\, e^{2ih_n d}\,,\qquad R_{-n}=\pm\, 1\,,\eqno(21.13)$$
где $h_n=\sqrt{k^2-g^2_n}$ --- продольное волновое число ($  g_n$  ---
собственные значения  краевой задачи для мембранных функций),  знак
$+$ в правой части формул (21.13)  соответствует  электрическим
волнам, знак $-$  соответствует магнитным волнам.

     Выражения для  $C_n$  и  $C_{-n}$  находятся  с  помощью  условия
ортогональности функций $\rv E^R_n$ и $\rv H^R_n$
     $$\int\limits_S\bigl([\rv E^R_m\rv H^R_n]_z-[\rv E^R_n\rv H^R_m
         ]_z\bigr)\,dS=0\qquad \mbox{при}\quad m\ne -n\,,
         \eqno(21.14)$$
где $S$  ---  поперечное  сечение  волновода,  точно
также,   как   это   было   сделано    выше  в    теории
возбуждения  волноводов,  то есть на основе леммы Лоренца.
При $n=-m$ интеграл в (21.14) равен $2kh_n(R_nR_{-n}  -  1))$,
так что в результате получаем:

     $$C_{\pm n}=-\frac{ i\pi e^{-ih_nd}}{\omega h_n\sin{h_n d}}\int
         \limits_V(\rv j^e\rv E^R_{\mp n}-\rv j^m\rv H^R_{\mp n})\,
         dV\,,\eqno(21.15)$$
где объём интегрирования $V$ и определяет  всю  зависимость  величин
$C_n$  и $C_{-n}$ от $z$.  Для $C_n$ объём $V$ ограничен плоскостями
$z=z_1$ и $z$;  если $z<z_1$,  то $C_n=0$;  если  $z>z_2$,  то  правая
граничная  плоскость  объёма $V$ есть $z=z_2$ и $C_n$ --- постоянные
величины.  Соответственно,  для $C_{-n}$ объём ограничен плоскостями
$z$  и  $z=z_2$,  причём  при $z<z_1$ левая плоскость есть $z=z_1$ и
$C_{-n}$ не зависят от $z$,  а при $z_2<z<d$ коэффициенты  $C_{-n}=0$.
Все  резонансные  свойства возбуждаемого поля определяются близостью к
нулю $\sin{h_n d}$;  напомним,  что обращение в нуль этой  величины  и
определяет собственные частоты волноводного резонатора.  Следует ещё
сказать, что в формулах (21.11)--(21.15) под индексом суммирования $n$
подразумевается   совокупность   двух   индексов,  определяющих  число
вариаций поля в волноводной  волне  по  соответствующей  координате  в
поперечной  плоскости,  а  в  формуле  (21.3)  --- совокупность трёх
индексов.  Поскольку  (21.3)  и  (21.11)   описывают   одно   и   тоже
электромагнитное поле,  то можно сказать,  что при волноводном способе
описания  поля  автоматически  выполняется  суммирование  по  индексу,
определяющему    продольные   вариации.   Всегда   ли   это   является
достоинством, будет обсуждено ниже.

     Для этого   решим  задачу  о  возбуждении  идеального  резонатора
пролетающим точечным зарядом двумя способами и сопоставим  результаты.
Рассмотрим  круглый  цилиндрический резонатор радиуса $a$ и длины $d$.
Будем считать,  что заряд $Q$ движется по оси резонатора с  постоянной
скоростью  $v$ (обозначение заряда таким способом не должно привести к
недоразумениям,  так как добротность  в  данном  разделе  обозначается
символом  $Q$ с индексом).  Плотность стороннего электрического тока в
резонаторе может быть тогда записана в виде

     $$\hbox{$\gv j_z(t)=$}\hbox{$\left\{\begin{array}{ll}0&\mbox
         {при}\qquad t<0,\\[0.25cm]\displaystyle{\frac{Qv}{2\pi}\frac
         {\delta(r)}r}\delta(z-vt)\qquad &\mbox{при}\qquad0<t<
         \displaystyle\frac{d}{v},\\[0.25cm]0&\mbox{при}\qquad t>
         \displaystyle\frac{d}{v},\end{array}\right.$}
         \eqno(21.16)$$
а его  комплексная  амплитуда  на частоте $\omega$,  вычисляемая путем
разложения в интеграл Фурье функции $\delta(z-vt)$, составляет
     $$j^e_z(\omega)=\frac Q{4\pi^2}\frac{\delta(r)} r \,e^{i\frac
         \omega v  z}\,.\eqno(21.17)$$

     Рассмотрим ожидаемый  процесс формирования поля по мере пролёта
заряда через резонатор.  До влёта заряда через торцевую стенку $z=0$
в  момент  $t=0$  поле  в  резонаторе  отсутствует.  В течение времени
пролёта поле носит сложный характер,  обладает непрерывным  спектром
частот  (его,  впрочем,  можно  считать состоящим из суммы колебаний с
дискретными частотами,  амплитуды которых меняются со  временем).  При
разложении  в  интеграл  Фурье и при таком подходе спектр непрерывный.
Это поле должно включать в себя и собственное  поле  заряда,  которое,
как  известно,  расходится  в  точке  нахождения заряда.  После вылета
заряда через  торцевую  стенку  $z=d$  в  момент  $t=d/v$  поле  может
представлять  собой  только  сумму  незатухающих (резонатор идеальный)
собственных колебаний.  На возбуждение этого  поля  заряд  должен  был
затратить энергию,  которая может быть вычислена либо как запасённая
в резонаторе к моменту вылета заряда,  либо как работа поля  излучения
над зарядом по мере его продвижения по резонатору.  Посмотрим, как эта
картина  реализуется  в  результате  расчётов  по  изложенной   выше
методике.

     Начнём с вычисления поперечной части поля, поскольку именно она
несёт в себе основную физическую нагрузку.  Магнитные токи в  задаче
отсутствуют  и имеет место азимутальная симметрия.  Поэтому поперечная
часть возбуждаемого в резонаторе поля может быть представлена  в  виде
совокупности $E$-колебаний без вариаций по азимуту:

     $$\rv E^t(\rv r,\omega)=\sum\limits_{q=1}^\infty\sum\limits_{l=
         0}^\infty A_{ql}(\omega)\rv E_{ql}(\rv r)\,,\qquad\rv H^t(
         \rv r,\omega)=\sum\limits_{q=1}^\infty\sum\limits_{l=0}^
         \infty B_{ql}(\omega)\rv H_{ql}(\rv r)\,,\eqno(21.18)$$
где отличные от нуля компоненты поля

     $$\left.\begin{array}{l}E_{z,ql}=E_0J_0\displaystyle{\Bigl(
         \frac {\nu_q} a r\Bigr)}\cos{\Bigl(\frac{l\pi} d z\Bigr)}\,,
         \\[.4cm] E_{r,ql}=\displaystyle{\frac{l\pi a}{d\nu_q}} E_0J_1
         \Bigl(\frac{\nu_q} a r\Bigr)\sin{\Bigl(\frac{l\pi} d z\Bigr)}
         \,,\\[.4cm]H_{\varphi,ql}=-\displaystyle{\frac{i\omega_{ql}a}
         {c\nu_q}}E_0J_1\Bigl(\frac{\nu_q} a r\Bigr)
         \cos{\Bigl(\frac{l\pi} d z\Bigr)}\,,\end{array}\right\}
         \eqno(21.19)$$
а $\omega_{ql}=c\displaystyle{ \sqrt{\Bigl(\frac{\nu_q} a\Bigr)^2+
\Bigl(\frac{l\pi} d\Bigr)^2}}$  --- собственные частоты резонатора,
$J_0(\nu_q)=0$,  и у всех величин для упрощения записи  опущен  равный
нулю  индекс,  соответствующий  вариациям поля по азимуту.  Постоянный
коэффициент $E_0$ в  (21.19)  определён  условием  нормировки  $\int
H_{\varphi,ql}^2\,dV=-4\pi$  и  составляет

     $$E_0=\frac{2\sqrt{2}\nu_q c}{a^2\omega_{ql}
     J_1(\nu_q)\sqrt{d(1+\delta_{0l})}}\,,\eqno(21.20)$$
 где $\delta_{0l}$  --- символ Кронекера.

     Коэффициенты $A_{ql}$  и  $B_{ql}$  легко  находятся  по формулам (21.9):

     $$\left.\begin{array}{l}
         A_{ql}(\omega)=-\displaystyle{\frac{i\omega}{\omega^2-\omega_
         {ql}^2}\frac{QE_0}{4\pi^2}}\int\limits_0^{2\pi}d\varphi\int
         \limits_0^a\frac{\delta(r)} rJ_0\Bigl(\frac{\nu_q} a r\Bigr)
         r\,dr\int\limits_0^d e^{i\frac{\omega}  v z}\,\cos\Bigl(
         \frac{l\pi} d z\Bigr)\,dz\,,\\[.4cm] B_{ql}(\omega)=
         \displaystyle{\frac {\omega_{ql}} {\omega}}A_{ql}(\omega)
         \,.\end{array}\right\}  \eqno(21.21) $$
Определяя функцию $F(z,u)$ (она потребуется  нам  и  в  дальнейшем)  с
помощью интеграла
     $$F(z,u)=\int\limits_0^ze^{i\frac{\omega} v x}\cdot\cos{ux}
         \,dx=\frac {e^{i\frac{\omega} v z}\Bigl(i\displaystyle{\frac
         \omega v}\cos uz+u\sin uz\Bigr)-i\displaystyle{\frac
         {\omega} v}}{ u^2-\displaystyle{\frac{\omega^2}{v^2}}},
         \eqno(21.22)$$
получаем:

     $$A_{ql}(\omega)=-\frac{i\omega}{\omega^2-\omega^2_{ql}}
         \frac{QE_0}{2\pi}F(d,\frac{l\pi} d)=
         \frac{QE_0v}{2\pi}\cdot\frac{\omega^2}{\omega^2-
         \omega_{ql}^2}\cdot\frac{1-(-1)^le^{i\frac{\omega} v d}}
         {\omega^2-l^2\pi^2v^2/d^2}\,.\eqno(21.23)$$
В результате   комплексная   амплитуда   $z$-компоненты    поперечного
электрического  поля,  которая  нам только и понадобится в дальнейшем,
представляется в виде:
     $$E_z^t(\omega)=\frac{4Qvc^2\omega^2}{\pi a^4d}\sum\limits_{q=1}
         ^\infty\frac{\nu_q^2\displaystyle{J_0\Bigl(\frac{\nu_q} a r
         \Bigr)}}{J_1^2(\nu_q)}\sum\limits_{l=0}^\infty\frac {[1-(-1)
         ^le^{i\frac \omega v d}]\;
         \cos\Bigl(\displaystyle{\frac{l\pi}
          d}z\Bigr)
}{(1+\delta_{0l})\omega_{ql}^2(\omega^2-
         \omega_{ql}^2)(\omega^2- l^2\pi^2 v^2/d^2)}\,.\eqno(21.24)$$

     Для нахождения зависимости компоненты поля от времени  необходимо
вычислить интеграл Фурье:
     $${\gv E}_z(r,z,t)=\int\limits_{-\infty}^{\infty} E_z(r,z,\omega)
         \,e^{-i\omega t}\,d\omega\,.\eqno(21.25)$$
Согласно (21.24) вся зависимость поля от времени определяется  простой
комбинацией интегралов по $\omega$:

     $${\gv E}_z(r,z,t)=\frac{4Qvc^2}{\pi a^4d}\sum\limits_{q=1}^\infty
         \frac{\nu_q^2J_0\Bigl(\displaystyle{\frac{\nu_q}a}r\Bigr)}
         {J_1^2(\nu_q)}\sum\limits_{l=0}^\infty\frac{
         \cos\Bigl(\displaystyle{\frac{l\pi} d}z\Bigr)}
         {(1+\delta_{0l})\,\omega_{ql}^2}\,\Bigl[I_{ql}(t)-(-1)^lI_{ql}
         (t-\frac d v)\Bigr]\,,\eqno(21.26)$$
где
     $$I_{ql}(x)=\int\limits_{-\infty}^{\infty}\frac{\omega^2}{(\omega
         ^2-\omega^2_{ql})(\omega^2-l^2\pi^2v^2/d^2)}\,e^{-i\omega x}
         \,d\omega\,.\eqno(21.27)$$
Этот интеграл   удобно  вычислять  в  комплексной  плоскости  $\omega$,
учитывая, что  из  соображений причинности (или при учёте небольшого
затухания) все полюсы  подынтегрального  выражения  смещены  в  нижнюю
полуплоскость.  Замыкая контур интегрирования полуокружностью большого
радиуса в зависимости от знака $x$ в верхней или нижней  полуплоскости
и вычисляя вычеты в полюсах, получаем:
     $$I_{ql}(x)=\left\{\begin{array}{l} 0\hspace{6.6cm}\mbox{при}
         \quad x<0\,,\\[.2cm]-\displaystyle{\frac{2\pi a^2}{c^2\nu_q
         ^2}}\,\displaystyle{\frac{\omega_{ql}\sin{\omega_{ql}x}
         -\displaystyle{\frac{l\pi v}d}\sin{\displaystyle{\frac{l\pi v
         }d} x}}{1+\Bigl(\displaystyle{\frac{l\pi a}{\gamma d\nu_q}}
         \Bigr)^2}}\qquad \mbox{при}\quad x>0\,,\end{array}\right.
         \eqno(21.28)$$
где $\gamma=(1-\beta^2)^{-1/2}$  ---  релятивистский  фактор,  $\beta=
v/c$.  Комбинируя в (21.26) интегралы (21.28) в зависимости  от  знака
величин $t$ и $t-d/v$,  убеждаемся,  что при $t<0$ поле равно нулю,  а
при $0<t<d/v$

     $${\gv E}^t_z(t)=-\frac{8Qv}{a^2d}\sum\limits_{q=1}^\infty\frac{J_0
         \Bigl(\displaystyle{\frac{\nu_q}a}r\Bigr)}{J_1^2(\nu_q)}
         \sum\limits_{l=0}^\infty\frac{\Bigl(\omega_{ql}\sin{\omega_
         {ql}t}-\displaystyle{\frac{l\pi v}d}\sin{\displaystyle{\frac
         {l\pi v}d t}}\Bigr)\,\cos{\displaystyle{\frac{l\pi}d z}}}{(1+
         \delta_{0l})\omega_{ql}^2[1+(l\pi a/\gamma d \nu_q)^2]}\,.
         \eqno(21.29)$$

     Теперь можно вычислить потери энергии зарядом на возбуждение поля
в резонаторе как работу поля излучения вдоль траектории,  то есть  как
работу   силы   торможения,   действующей   на   частицу   со  стороны
возбуждаемого ею поля:

     $$U= - \int\!\!\int\mbox{\boldmath$\gv j$}\mbox{\boldmath$\gv E$}
          \,dVdt= - Q\int\limits_0^d {\gv E}_z(r=0,z,
         t= z/v)\,dz\,.\eqno(21.30)$$
Подставляя сюда  (21.29)  при  соответствующих  значениях  аргументов,
получаем в результате вычисления интеграла:
     $$U=\frac{8Q^2\beta^2}d\sum\limits_{q=1}^\infty\sum\limits_{l=0}
         ^\infty\frac 1 {(1+\delta_{0l})\nu_q^2J_1^2(\nu_q)}\cdot\frac
        {1-(-1)^l\cos\displaystyle{\frac{\omega_{ql}} v d}}{[1+(l\pi a/
         \gamma d\nu_q)^2]^2}\,.\eqno(21.31)$$

     Потери энергии зарядом могут быть найдены и другим способом:  как
энергия  электромагнитного  поля,  запасённая  в  резонаторе   после
пролёта   заряда.   Эту   энергию  можно  выразить  через  амплитуды
возбуждённых собственных колебаний.  При $t>d/v$ поле  в  резонаторе
может быть представлено в виде

     $$\mbox{\boldmath$\gv E$}^t(\rv r,t)=\sum\limits_{q=1}^
       \infty\sum\limits_{l=0}
         ^\infty C_{ql}\rv E_{ql}\cos{(\omega_{ql}t-\varphi_{ql})}\,,
         \eqno(21.32)$$
где $C_{ql}$ --- амплитуда,  $\varphi_{ql}$ --- несущественная для нас
фаза  колебаний.  С  помощью  формул (21.26) и (21.28) компонента поля
$E^t_z$ для этих моментов времени записывается как

     $${\gv E}^t_z(t)=-\frac{8Qv}{a^2d}\sum\limits_{q=1}^\infty\frac{J_0
         \Bigl(\displaystyle{\frac{\nu_q}a}r\Bigr)}{J_1^2(\nu_q)}
         \sum\limits_{l=0}^\infty\frac{[\sin{\omega_{ql}t}-(-1)^l\sin{\displaystyle{
         \omega_{ql}(t-\frac d v)}}]\,\cos{\displaystyle{\frac{l\pi}d z} }}
         {(1+\delta_{0l})\,\omega_{ql}[1+(l\pi a/\gamma d \nu_q)^2]}\,.
         \eqno(21.33)$$
Сопоставляя формулы (21.32) и (21.33) с учётом (21.19) и (21.20)
убеждаемся, что амплитуды собственных колебаний равны
     $$C_{ql}=\frac{4Q\beta}{\nu_q J_1(\nu_q)[1+(l\pi a/\gamma d
         \nu_q)^2]}\sqrt{\frac{1-(-1)^l\cos\displaystyle{{\frac{\omega_{ql}} v  d}}}
         {d(1+\delta_{0l})}}\,.\eqno(21.34)$$
При выбранной  выше нормировке  энергия каждого колебания
$U_{ql}=C_{ql}^2/2$; как уже говорилось, в  силу  ортогональности  собственных
колебаний  их энергии складываются  аддитивно и поэтому для запасённой
энергии получается та же  формула (21.31), что и для работы поля над зарядом.

     Остановимся на этой формуле более подробно.  Нетрудно видеть, что
при   каждом   значении   индекса  $l$  сумма  по  $q$  логарифмически
расходится.  Такой результат физически обусловлен тем обстоятельством,
что  большие  значения $q$ (как,  впрочем и $l$) соответствуют высоким
частотам,  для которых модель идеальной проводимости  торцевых  стенок
резонатора  неприменима.  Однако  потери на возбуждение низкочастотных
колебаний  описываются  формулой  (21.31)  правильно.  В  этом   можно
убедиться,  решив  существенно более сложную задачу о пролёте заряда
через структуру,  состоящую из цилиндрического круглого  резонатора  и
подводящих круглых волноводов. Дискретный спектр собственных колебаний
такой структуры ограничен критической частотой подводящих волноводов и
никаких  расходимостей  потерь  энергии  в задаче не возникает.  А вот
численные значения потерь на  низких  частотах  при  достаточно  малом
радиусе волноводов совпадают с вычисленными по формуле (21.31).

     Вычислим теперь продольное поле; для этого представим комплексную
амплитуду потенциала $\Phi^e(\rv r,\omega)$ в виде разложения

     $$\Phi^e(\rv r,\omega)=\sum\limits_q\sum\limits_lD_{ql}(\omega)
         f_{ql}(\rv r)\eqno(21.35)$$
по полной ортогональной системе собственных функций  $f_{ql}$  краевой
задачи
     $$\triangle f+\alpha^2 f=0\,,\qquad f=0\quad\mbox{при}\;r=a,\;z=0,
         \;z=d.\eqno(21.36)$$
Ввиду азимутальной симметрии задачи оператор $\triangle$  не  содержит
производных  по  углу $\varphi$,  а собственные значения и собственные
функции легко находятся методом разделения переменных:
     $$\alpha^2_{ql}=\Bigl(\frac{\nu_q} a\Bigr)^2+\Bigl(\frac{l\pi} d
         \Bigr)^2=\frac{\omega_{ql}^2}{c^2}\,,\quad f_{ql}=J_0\Bigl(
         \frac{\nu_q} a r\Bigr)\sin{\frac{l\pi}d z}\,,\quad q,l=
         1,2,\dots\,. \eqno(21.37)$$
Подставляя разложение (21.35) в (21.6), где, как это видно из (21.17),

     $$\rho^e=\frac{Q}{4\pi^2v}\frac{\delta (r)}r\,e^{i\frac
         \omega v z}\,,\eqno(21.38)$$
домножая на  $f_{mn}$ и интегрируя по объему резонатора $V$,  учитывая
при этом уравнение (21.36) и соотношение ортогональности

     $$\int\limits_V f_{ql}f_{mn}\,dV=\left\{\begin{array}{l}0\,,\hspace
         {3cm}q\ne m\quad\mbox{или}\quad l\ne n\;,\\[.3cm]
         \displaystyle{\frac{\pi a^2} 2} J_1^2(\nu_q)\,d\,,\quad
         \qquad q=m
         \quad\mbox{и}\quad l=n\;,\end{array}\right.\eqno(21.39)$$
получаем следующее выражение для коэффициентов:
\vspace{.1cm}
     $$D_{ql}=\frac{2Q}{\pi^2 a^2 vd J_1^2(\nu_q)}\int \limits_V\frac
         {\delta(r)} r e^{i\frac{\omega} v z} f_{ql}\,dV.
         \eqno(21.40)$$

     Вычисляя интегралы  и  подставляя  $D_{ql}$  в  (21.35),  находим
потенциал
     $$\Phi^e(\omega)=\frac{4Q}{a^2d^2v}\sum\limits_{q=1}^\infty\frac
         {J_0\Bigl(\displaystyle{\frac{\nu_q}a r}\Bigr)}{J_1^2(\nu
         _q)}\sum\limits_{l=1}^\infty l\frac{1-(-1)^l e^{i\frac \omega
         v d}}{\alpha_{ql}^2\Bigl[\displaystyle{\Bigl(\frac{l
         \pi}d\Bigr)^2-\Bigl(\frac \omega v\Bigr)^2}\Bigr]}\sin\frac
         {l\pi}d z\,,\eqno(21.41)$$
после чего компоненты  продольного поля определяются
дифференцированием.  Приведём  выражение  для  комплексной амплитуды
$E_z^l$:

     $$E_z^l(\omega)=-\frac{4Q\pi}{a^2d^3v}\sum\limits_{q=1}^\infty
         \frac{J_0\Bigl(\displaystyle{\frac{\nu_q}ar}\Bigr)}{J_1^2(
         \nu_q)}\sum\limits_{l=1}^\infty l^2\frac{1-(-1)^l e^{i\frac
         \omega v d}}{\alpha_{ql}^2\Bigl[\displaystyle{\Bigl(
         \frac{l\pi}d\Bigr)^2-\Bigl(\frac \omega v\Bigr)^2}\Bigr]}
         \cos\frac{l\pi}d z\,.\eqno(21.42)$$

     Нетрудно видеть,  что продольное поле  не  обладает  резонансными
свойствами  (на  частотах  $\omega=\pm  l\pi v/d$ выражение (21.42) не
имеет особенностей,  поскольку на них обращается в  нуль  и  числитель
дроби).  Полная работа продольного поля над зарядом за всё время его
нахождения в резонаторе равна нулю,  так что вклада в  потери  энергии
продольное  поле не даёт.  Естественно,  что после пролёта зарядом
резонатора продольная часть поля  в  нём  отсутствует.  Полное  поле
находится  суммированием выражений (21.24) и (21.42);  после несложных
(но громоздких) преобразований имеем в результате:
     $$\begin{array}{l}E_z(\omega)=-\displaystyle{\frac{4Q}{\pi a^2dv}
         \sum\limits_{q=1}^\infty}\displaystyle{\frac{J_0\Bigl(
         \displaystyle{\frac{\nu_q}a r}\Bigr)}{J_1^2(\nu_q)\Bigl[
         \Bigl(\displaystyle{\frac{\omega}{v\gamma}}\Bigr)^2+\Bigl(
         \displaystyle{\frac{\nu_q}a}\Bigr)^2\Bigr]}}
         \sum\limits_{l=0}
         ^\infty\frac{1-(-1)^l e^{i\frac \omega v d}}{1+\delta_{0l
         }}\times \\[.9cm]\phantom{E_z(\omega)=}\times\Biggl[\frac{
         \Bigl(\displaystyle{\frac{\omega}{v\gamma}}\Bigr)^2}{\Bigl(
         \displaystyle{\frac{l\pi}d}\Bigr)^2-\Bigl(\displaystyle{\frac
         \omega v}\Bigr)^2}+\frac{\Bigl(\displaystyle{\frac{\nu_q}{a}}
         \Bigr)^2}{\Bigl(\displaystyle{\frac{l\pi}d}\Bigr)^2-h_q^2}
         \Biggr]\,\cos{\displaystyle{\frac{l\pi} d z}}\,,\end{array}
         \eqno(21.43)$$
где $h_q$ --- продольное волновое число симметричной волны  в  круглом
волноводе радиуса $a$.

     Найдём теперь  поле в резонаторе методом,  основанным на теории
возбуждения  волноводов.  Это  возможно,   поскольку   рассматриваемый
резонатор  представляет собой типичный пример волноводного резонатора.
В  соответствующих  формулах  (21.11)  --  (21.15)  под  индексом  $n$
подразумевается   совокупность   двух   индексов,  определяющих  число
вариаций  поля  в  поперечной  к  оси  $z$  плоскости  по  координатам
$\varphi$  и  $r$.  В  данной  задаче  первый  индекс равен нулю (поле
обладает осевой симметрией) и его можно опустить,  а второй индекс
обозначим через $q$.  Ограничимся вычислением только
компоненты поля  $E_z$;  тогда  по  сложившейся  в  теории  волноводов
терминологии  эта  компонента  и  есть продольное электрическое поле в
волноводе.  Во  избежание  путаницы  напомним,  что  в  этом   разделе
используется другая терминология и $E_z$ в (21.11) включает в себя как
продольную,  так  и  поперечную  части  поля.   Искомая   составляющая
векторных  функций  $\rv  E_{\pm  q}^R$,  определённых соотношениями
(21.12) и  (21.13),  для  круглого  цилиндрического  резонатора  имеют
вид
     $$\left.\begin{array}{l}E_{z,q}^R=\displaystyle{\frac{2\nu_q\,e^
         {ih_q d}}{\sqrt{\pi}a^2J_1(\nu_q)}}\,J_0\Bigl(\displaystyle{
         \frac{\nu_q}a}r\Bigr)\cos{h_q(z-d)}\,,\\[.5cm]E_{z,-q}
         ^R=\displaystyle{\frac{2\nu_q}{\sqrt{\pi}a^2J_1(\nu_q)}}\,J_0
         \Bigl(\displaystyle{\frac{\nu_q}a}r\Bigr)\cos{h_q z}\,;
         \end{array}\right\}\eqno(21.44)$$
естественно, что  в  этих   выражениях   используется   принятое   для
волноводов (а не для резонаторов) условие нормировки (8.23).

     Входящие в  (21.11)  коэффициенты  $C_{\pm  q}$  вычисляются   по
формулам   (21.15).   Используя   функцию   $F(z,u)$,   определённую
соотношением (21.22),  легко видеть,  что $C_q\sim F(z,h_q)$,~ $C_{-q}
\sim   F(z-d,h_q)$.  В  результате  искомые  коэффициенты  могут  быть
записаны в следующем виде:
     $$  C_q=2Q\nu_q \frac{e^{i\frac{\omega}v z}\Bigl[i\displaystyle
         {\frac{\omega} v}\,\cos{h_q z}+h_q\,\sin{h_q z}\Bigr]-i
         \displaystyle{\frac{\omega} v}}{\sqrt{\pi}a^2
         \omega h_q J_1(\nu_q)(1-e^{2ih_qd})\Bigl[\Bigl(
         \displaystyle{\frac{\omega}{v\gamma}}\Bigr)^2+\Bigl(
         \displaystyle{\frac{\nu_q}a}\Bigr)^2\Bigr]}\,,\eqno(21.45)$$
     $$  C_{-q}=-2Q\nu_q\,e^{ih_q d} \frac{e^{i\frac{\omega}v z}\Bigl
         [i\displaystyle{\frac{\omega} v}\,\cos{h_q(z-d)}+h_q\,\sin{
         h_q (z-d)}\Bigr]-i\displaystyle{\frac{\omega} v}e^{i\frac{
         \omega} v d}}{\sqrt{\pi}a^2\omega h_q J_1(\nu_q)(1-e^{2ih_qd
         })\Bigl[\Bigl(\displaystyle{\frac{\omega}{v\gamma}}\Bigr)^2+
         \Bigl(\displaystyle{\frac{\nu_q}a}\Bigr)^2\Bigr]}\,.
         \eqno(21.46)$$

     После подстановки найденных коэффициентов $C_{\pm q}$ в (21.11) и
несложных преобразований с учётом разложения функции $\delta(r)/r$ в
ряд по функциям Бесселя
    $$\frac{\delta(r)}r=\frac 2{a^2}\sum\limits_{n=1}\frac {J_0
         \Bigl(\displaystyle{\frac {\nu_n} a}r\Bigr)}{J_1^2(\nu_n)}
         \,\eqno(21.47)$$
находим  комплексную  амплитуду искомой компоненты поля:
     $$E_z(\omega)=-\frac{2iQ}{\pi a^2v}\sum\limits_{q=1}^\infty
         J_0\Bigl(\displaystyle{\frac{\nu_q}ar}\Bigr)\frac{
         \Bigl[\displaystyle{\frac{\omega}{v\gamma^2}\,e^{i\frac{
         \omega}v z}-i\Bigl(\frac{\nu_q}a\Bigr)^2\;\frac{e^{i\frac{
         \omega}v d}\,\cos{h_q z}-\cos{h_q(z-d)}}{h_q\sin{h_qd}}\Bigr]
         }}{J_1^2(\nu_q)\Bigl[\Bigl(\displaystyle{\frac{\omega}{v
         \gamma}}\Bigr)^2+\Bigl(\displaystyle{\frac{\nu_q}a}\Bigr)^2
         \Bigr]}
         \,.\eqno(21.48)$$
Обратим внимание,  что резонансный  характер  поля  в  этом  выражении
обусловлен приближением к нулю стоящего в знаменателе $\sin{h_q d}$, а
то есть  условием  обращения  в  нуль  этого  синуса  и  определяются
собственные частоты волноводного резонатора.

     Выражения (21.43)  и  (21.48)  описывают  одно  и  то  же поле и,
естественно,  должны давать один и тот же результат. Так оно и есть, в
чём  нетрудно  убедиться,  если  в  (21.43) провести суммирование по
индексу $l$, воспользовавшись известными формулами:
     $$\sum\limits_{l=0}^\infty \frac{\cos{\displaystyle{\frac{l\pi}
         d z}}}{\Bigl[u^2-\Bigl(\displaystyle{\frac{l\pi}d}\Bigr)^2
         \Bigr](1+\delta_{0l})}=\frac d{2u}\,\frac{\cos{u(d-z)}}{
         \sin{ud}}\,,\eqno(21.49)$$\vspace{.1cm}
     $$\sum\limits_{l=0}^\infty \frac{(-1)^l\,\cos{\displaystyle{
         \frac{l\pi}d z}}}{\Bigl[u^2-\Bigl(\displaystyle{\frac{l\pi}d}
         \Bigr)^2\Bigr](1+\delta_{0l})}=\frac d{2u}\,\frac{\cos
         {uz}}{\sin{ud}}\,,\eqno(21.50)$$
где $u$ --- произвольный параметр.

     Таким образом,  на первый взгляд расчёт возбуждения  резонатора
по  методу,  используемому в теории волноводов,  обладает несомненными
преимуществами:  во-первых,  он сразу даёт полное поле,  и, во-вторых,
все  вычисления  продвинуты  на одно суммирование.  При этом,  однако,
необходимо  помнить,  что  метод  применим  лишь   к   частному   виду
волноводных  резонаторов,  да  и  сами  преимущества во многих случаях
оборачиваются  недостатками.  В  большинстве  задач  основной  интерес
представляет   лишь   поперечное  поле,  поскольку  только  оно  имеет
резонансный  характер  и   вблизи   собственной   частоты   резонатора
фактически  совпадает с полем собственного колебания.  Суммирование же
по индексу,  определяющему число вариаций поля вдоль  оси  резонатора,
также  не  всегда  даёт  физически  осмысленный  результат,  хотя  и
безупречный с точки зрения математики.  В частности, в формуле (21.48)
учтён вклад самых высоких собственных частот, для которых вся модель
резонатора с  идеально  проводящими  торцевыми  стенками  неприменима.
Использование  того  или  иного  метода расчёта существенным образом
зависит как от частотного спектра стороннего возбуждающего тока, так и
от  спектра  собственных  значений возбуждаемой структуры,  которая во
всех  практических  приложениях  представляет  собой  {\it  комбинацию
резонатора   и   волновода},  по  которому  отводится  или  подводится
высокочастотная мощность.

     Можно сказать,  что вынужденные колебания объёмных  резонаторов
допускают различные математические трактовки,  также как,
например, и колебания струны, закреплённой на концах, решение
задачи для которой может  быть  получено  и  по методу Бернулли
(разложение  по  собственным функциям) и  по  методу Даламбера  (сумма
бегущих  и  отражённых  волн). Отметим  также,  что  теория
возбуждения резонаторов в изложенном выше виде позволяет
рассмотреть,  например, и такие вопросы, как возмущение
собственной  частоты  резонатора  вследствие конечной проводимости
или деформации его стенок --- фактически  задачи  этого  типа
сводятся  к правильному заданию стороннего тока; они изложены в
других разделах на основе иного математического аппарата.

%\end{document}

\newpage
\oddsidemargin=-0.4mm \evensidemargin=-0.4mm
\topmargin=-0.4mm\headsep=7mm \textheight=231.875mm
\textwidth=160mm \mathsurround=2.5pt \unitlength=1mm
%\begin{document}
%\input{macr.tex}
\thispagestyle{empty}
%\addtocounter{page}{236}
%\baselineskip=\normalbaselineskip
\baselineskip=0.99\normalbaselineskip

\begin{center}
   \subsubsection*{\rm Г\,Л\,А\,В\,А\, 8}
      \vspace{-1.15em}
      \line(6,0){160}\\
      \vspace{-1em}
      \line(6,0){160}
      \vspace{-1.15em}
   \subsubsection*{НЕРЕГУЛЯРНОСТИ В ЭЛЕКТРОДИНАМИЧЕСКИХ СТРУКТУРАХ}
      \vspace{31mm}
\subsubsection*{22. Матрица рассеяния}
\end{center}
\vspace{.5cm}

\markboth{Глава 8. Нерегулярности в электродинамических структурах}
         {22. Матрица рассеяния}

\begin{center}\begin{minipage}[c]{0.75\textwidth}
\footnotesize{\parindent=0.5cm
         Альтернативный способ  описания   поля   основной   волны   в
         регулярном волноводе и волноводе,  нагруженном
         нерегулярностью.  Одиночная   нерегулярность   в   волноводе.
         Матрица   рассеяния  и  её  основные  свойства.  Нахождение
         матрицы рассеяния для нерегулярных структур, обладающих осями
         и  (или)  плоскостями  симметрии.  Матрица рассеяния элемента
         связи для волновода, соединённого с резонатором.  Разбиение сложного
         волноводного    тракта   на  отдельные   нерегулярности.
         Взаимодействующие   и   невзаимодействующие   нерегулярности.
         Теория СВЧ цепей для описания  волноводного  тракта.  Матрицы
         сопротивлений  и  проводимостей.    Эквивалентные   схемы
         нерегулярностей.
}\end{minipage}\end{center}\vspace{.5cm}

     Однородный (или  регулярный)  бесконечный   волновод   несомненно
является одним из важнейших элементов, рассматриваемых в теории СВЧ. И
в  то  же  время  очевидно,  что  он  представляет  собой   физическую
абстракцию: в реальных устройствах встречаются лишь отрезки регулярных
волноводов  вполне  определённой  длины.  Между   ними   расположены
различные   элементы   СВЧ   тракта,   которые   будем  называть
 {\it нерегулярностями}.  Конструкция  тракта  может  быть  очень   сложной,
включать  в  себя волноводные разветвления и петли.  Расчёт такого
устройства строгим методом решения граничной задачи электродинамики  в
большинстве  случаев  практически  не реализуем  и,  во  всяком случае,
нецелесообразен.  Для подавляющего числа практических задач  детальное
распределение  электромагнитного  поля  во всех элементах волноводного
тракта не представляет интереса  ---  необходимо  и  достаточно  уметь
рассчитывать прохождение СВЧ мощности по тракту.

     В результате расчёт СВЧ тракта распадается на несколько этапов.
Прежде   всего   необходимо   разбить   все  устройство  на
отдельные нерегулярности,  соединённые отрезками регулярных
волноводов.  После этого    находятся    параметры,   определяющие
свойства   отдельных нерегулярностей по их воздействию  на
собственную  волну  регулярного волновода.  Дальнейший  расчёт
удаётся  провести  по  той  хорошо разработанной методике,
которая используется для  электротехнического анализа сложных
низкочастотных цепей переменного тока.  В связи с этим для
характеристики свойств нерегулярности и поля в волноводе  зачастую
используется   терминология,   не  имеющая  для  волновых  полей
того физического смысла,  который она заключает в себе  для
низкочастотных процессов.   К   таким   терминам   относятся
{\it напряжение},   {\it ток}, {\it сопротивление}, {\it многополюсник} и ряд
других,  что требует известной осторожности при анализе
волноводных трактов СВЧ.

     Прежде чем    рассматривать    способы     описания     отдельных
нерегулярностей, целесообразно несколько изменить использованное ранее
(раздел 8) представление  собственных  волн  в  волноводе.
Упомянутая выше методика расчёта подразумевает,
что в каждом отрезке регулярного волновода  на  данной  частоте  может
распространяться  только  одна  основная  собственная волна (остальные
являются затухающими).  Если таких волн несколько,  то в эквивалентной
схеме волноводного тракта реальному волноводу сопоставляется несколько
волноводов (по числу распространяющихся волн), хотя все они фактически
расположены  в  одном  и  том  же месте пространства.  Физически такая
возможность  обусловлена  независимостью  величины   среднего   потока
мощности данной собственной волны от амплитуды любой другой.

     Использованный ранее    способ    описания    собственных    волн
подразумевает различную нормировку для магнитных и электрических волн.
Это  приводит  к тому,  что поток мощности согласно формуле (8.26) при
одной и  той  же  амплитуде  зависит  от  рабочей  частоты,  материала
заполнения,  номера  и  типа волны.  Изменим нормировку таким образом,
чтобы  поток  мощности  для  любой  волны  определялся   только   её
амплитудой.  Вспомним,  что средний поток мощности $\overline{\Sigma}$
вдоль  оси  волновода   согласно   формуле
     $$\overline{\Sigma}=\frac c{8\pi}\re\int\limits_S[\rv E(\rv r)
         \rv H^*(\rv r)]_z\, dS\eqno(22.1)$$
определяется только   поперечными   компонентами   поля.   Для   любой
цилиндрической  волны  вида  (7.1),  распространяющейся вдоль оси $z$,
согласно (7.16) двумерные векторы поперечного поля могут быть записаны
в виде

     $$\rv E_t^+(\rv r)=a \rv F(x,y)\,e^{ihz}\,,\qquad \rv H_t^+
         (\rv r)=a \rv G(x,y)\,e^{ihz}\,,\eqno(22.2)$$
где комплексная  постоянная  $a$  ---  амплитуда  волны.   Для   волны
амплитуды   $b$,  распространяющейся  в  противоположном  направлении,
аналогично имеем     $$\rv E_t^-(\rv r)=b \rv F(x,y)\,e^{-ihz}\,,\qquad \rv H_t^-
         (\rv r)=- b \rv G(x,y)\,e^{-ihz}\,,\eqno(22.3)$$
причём необходимость   смены  знака  в  выражении  для  $\rv  H^-_t$
непосредственно следует из (7.16).

     Поскольку поперечные  поля  $\rv  E_t$  и  $\rv  H_t$  синфазны и
ортогональны,  то векторные функции $\rv F(x,y)$ и $\rv G(x,y)$  также
ортогональны   и  могут  быть  выбраны  действительными.  Кроме  того,
отношение  их  соответствующих  компонент  в  каждой   точке   $x,\,y$
поперечного сечения волновода одно и то же:
     $$\frac {F_x}{G_y}=-\frac {F_y}{G_x}=\frac c{4\pi} Z_0\,,
         \eqno (22.4)$$
где действительная  положительная  постоянная  $Z_0$  называется  {\it
характеристическим сопротивлением волновода} (для  данной  собственной
волны). Из формул (7.16) и (10.3) следует:
     $$Z_0=\frac{4\pi} c W\times\left\{\begin{array}{ll}
         K/h &\quad\mbox{--- для магнитных волн,}\\
         h/K &\quad\mbox{--- для электрических волн,}\\
         1   &\quad\mbox{--- для TEM волны,}\end {array}\right.
         \eqno(22.5)$$
где, как обычно, $K=k\sqrt{\varepsilon\mu}$, $W=\sqrt{\mu/\varepsilon}$.

    Функции $\rv  F(x,y)$  и  $\rv  G(x,y)$  определены с точностью до
общего множителя,  который удобно выбрать таким образом,  чтобы  поток
(22.1)  для  волн  с  поперечными  компонентами  поля  (22.2) и (22.3)
составил соответственно
     $$\overline{\Sigma}^+=\frac { a\, a^*} 2\qquad\mbox{и}\qquad
         \overline{\Sigma}^-= - \frac { b\, b^*} 2\,.\eqno(22.6)$$
С учётом  предыдущих  формул и (8.26) для этого необходимо выполнение
следующих соотношений:
     $$\int\limits_S[\rv F\rv G]\,d\rv S\!=\!\frac{4\pi}{c},\qquad
         \int\limits_S|\rv F|^2\,dS\!=\!Z_0,\qquad\int\limits_S
         |\rv G|^2\,dS\!=\!\frac{1}{Z_0}\Bigl(\frac{4\pi}{c}\Bigr)^2.
         \eqno(22.6\textit{а})$$

Именно такая  нормировка  собственных  волн  и  будет использоваться в
дальнейшем.  Необходимо отметить,  что амплитуды $a$ и $b$  однозначно
определены  только  в  том  случае,  если  в  волноводе оговорена {\it
плоскость отсчёта} фаз.  В качестве таковой может быть  выбрана  как
плоскость  $z=0$  (как  в (22.2) и (22.3)),  так и любая другая.  Eсли
плоскость  отсчёта  сдвигается  влево  на  отрезок  $l$,  то   новые
амплитуды волн становятся равными $a'=a\,e^{-ihl}$ и $b'=b\,e^{ihl}$;  при
этом  потоки  мощности  (22.6)  сохраняются   неизменными,   поскольку
$|a'|=|a|$ и $|b'|=|b|$.

     Рассмотрим теперь,  что  представляет   собой   поперечное   поле
основной  волны  в  том  случае,  когда в волноводе присутствуют волны
обоих направлений.  Структура поля существенно зависит от  соотношения
амплитуд $a$ и $b$.  Поскольку появление волны обратного направления в
большинстве  случаев  обусловлено   отражением   падающей   волны   от
какой-нибудь  нерегулярности  в  волноводе,  то  отношение  поперечных
компонент электрического поля в плоскости отсчёта  (примем  здесь  в
качестве   таковой   плоскость   $z=0$),  одинаковое  для  всех  точек
поперечного сечения  и  равное  отношению  амплитуд,  называется  {\it
коэффициентом отражения} $\Gamma_0$:
    $$ \Gamma_0= \left.\frac{\rv E^-_t}{\rv E^+_t}\right |_{z=0}=
         \frac  b a\,,\eqno(22.7)$$
при этом отношение поперечных компонент магнитного поля этих двух волн
в той же плоскости равно  $-\Gamma_0$.  В  других  сечениях  волновода
коэффициент  отражения  зависит от расстояния до плоскости отсчёта и
равен
     $$\Gamma=\Gamma_0e^{-2ihz}.\eqno(22.8)$$

     Суммарные поперечные поля
     $$\left.\begin{array}{llll}\rv E_t(\rv r)&=\rv E^+_t(\rv r)+
         \rv E^-_t(\rv r)&=\rv E^+_t(\rv r)(1+\Gamma)&=a(e^{ihz}+
         \Gamma_0 e^{-ihz})\rv F(x,y)\,,\\[.3cm]\rv H_t(\rv r)&=
         \rv H^+_t(\rv r)+\rv H^-_t(\rv r)&=\rv H^+_t(\rv r)(1-
         \Gamma)&=a(e^{ihz}-\Gamma_0 e^{-ihz})\rv G(x,y)\,\end
         {array}\right\}\eqno(22.9)$$
образуют {\it  стоячую  волну},  огибающие  амплитуд   полей   которой
являются  периодическими  функциями  $z$ с периодом $\Lambda/2=\pi/h$.
Минимумы и  максимумы  огибающих  сдвинуты  на  $\Lambda/4$,  причём
минимуму  электрического  поля  соответствует  максимум  магнитного  и
наоборот,  а огибающая магнитного поля может быть получена сдвигом  на
$\Lambda/4$ огибающей электрического. Отношение максимального значения
огибающей к минимальному  обозначается  как  KCBH  и  называется  {\it
коэффициентом  стоячей  волны  напряжения}.  Слово {\it напряжение} в этом
определении является  той  исторически  сложившейся  терминологической
издержкой,  которой на данном этапе трудно дать обоснование. Из (22.9)
следует, что
     $$ \mbox{KCBH}=\frac {1+|\Gamma|}{1-|\Gamma|}\,,\eqno(22.10)$$
а средний поток мощности в стоячей волне составляет
     $$\overline{\Sigma}= \frac 1 2 |a|^2-\frac 1 2 |b|^2=\overline
         {\Sigma}^+ (1-|\Gamma|^2)\,.\eqno(22.11)$$

     Все нерегулярности,  расположенные правее отсчётной  плоскости,
по  своему  совокупному  действию  могут рассматриваться как некоторая
{\it нагрузка},  помещённая в этой плоскости. Если отражения от нагрузки
нет ($\Gamma_0=0$),  то KCBH=1 и говорят, что нагрузка и волновод {\it
согласованы};  в частности,  такой  согласованной  нагрузкой  является
поглощающая насадка. Введение понятия согласованной нагрузки позволяет
рассматривать часто используемый далее  полубесконечный  волновод  как
элемент конструкции,  близкий к реальности. Если же волновод ограничен (закорочен)
нормальной к его оси идеально проводящей  плоскостью  $z=0$,
 то  $\Gamma_0=-1$;  если  эта плоскость  идеальный магнетик или ограничивающая
 идеально проводящая плоскость отстоит от плоскости отсчета на четверть длины волны, то $\Gamma_0=1$.
 В обоих этих случаях $\mbox{КСВН}=\infty$ --- в волноводе чисто стоячая волна
 с узлами поля и нулевым потоком энергии.

     Рассмотрим теперь возможные способы  описания  элементарной  (или
одиночной)  нерегулярности  волноводного  тракта.  Под  элементарной в
данном случае понимается такая нерегулярность,  которая не может  быть
представлена   совокупностью   нескольких   более   простых  по  своей
конструкции  нерегулярностей,   соединённых   отрезками   регулярных
волноводов.  Конструкция  элементарной  нерегулярности  при этом может
быть очень сложной и поэтому начнём с наиболее простой из них.

\begin{wrapfigure}[14]{l}{7cm}
\begin{picture}(80,52)
\put(-10,50){\special{em:graph fig22-1.bmp}}
\end{picture}
\hbox to 7cm{\hfil\footnotesize{Рис.~22.1.~Нерегулярность в
волноводе.
}\hfil}
\end{wrapfigure}

     Пусть имеется бесконечный регулярный  волновод  и  где-то  внутри
него  расположена  одиночная  нерегулярность.  Она  может  быть самого
общего  вида,  например,  плавное  сужение  и  расширение  поперечного
сечения волновода или диафрагма произвольной формы и толщины.  В такой
структуре  (рис.  22.1)  всегда  можно  выделить  три   области:   два
полубесконечных    регулярных   волновода   и   саму   нерегулярность,
ограниченную плоскостями $z=z_L$ и $z=z_R$  (занумеруем  их  так,  как
показано  на рисунке).  Будем считать,  что частота монохроматического
поля лежит в одноволновом диапазоне частот  волновода;  тогда  поле  в
обоих  волноводах  при  достаточном  удалении  от нерегулярности может
представлять собой только сумму основных волн обоих направлений.

     Возможная постановка  граничной (дифракционной) задачи излагается
ниже.  Пусть слева на нерегулярность падает основная  волна  некоторой
амплитуды  $a_1$;  из  самых  общих  соображений  ясно,  что дифракция
(другими  словами,  рассеяние)  волны  на  нерегулярности  приводит   к
возможному  появлению  слева  от нерегулярности отражённой волны,  а
справа --- прошедшей волны.  Конечно, в области самой нерегулярности и
вблизи неё  в выделенных волноводах электромагнитное поле может иметь
очень сложное строение и,  следовательно,  определяться большим числом
параметров.   Однако   вдали   от   нерегулярности  (в  дальней  зоне)
существенными  становятся  только  две  величины  (обе   комплексные):
амплитуда отражённой и амплитуда прошедшей волны.  Подчеркнём, что
обе эти величины могут  быть  найдены  только  в  результате  строгого
решения электродинамической задачи.

     В силу линейности уравнений  Максвелла  очевидно,  что  изменение
амплитуды  падающей  волны приводит лишь к пропорциональному изменению
амплитуд прошедшей и  отражённой  волн.  Поэтому  достаточно  решить
задачу  для единичной амплитуды падающей волны,  но вот имеет ли смысл
решать  ещё  и  задачу  о  падении  волны  справа?  В  случае,  если
какая-нибудь   поперечная   плоскость  является  плоскостью  симметрии
нерегулярности,  в этом нет необходимости:  достаточно  переобозначить
волноводы.  Однако в общем случае произвольной нерегулярности ответ на
этот вопрос может быть получен  в  рамках  её описания    с
помощью  {\it  матрицы  рассеяния}  --- понятия,  исторически и идейно
восходящего  к  квантовой  теории  поля.

     В соответствии   с   принципом   суперпозиции   амплитуды   волн,
распространяющихся  от  нерегулярности,  линейно связаны с амплитудами
падающих волн, что может быть записано в виде равенства
     $$ \rv b=\rv S \rv a\,,\eqno(22.12)  $$
где $\rv a$,  $\rv b$ --- векторы-столбцы соответственно  из  амплитуд
падающих   и   рассеянных   волн,   $\rv   S$  ---  матрица  рассеяния
рассматриваемой нерегулярности:
     $$ \rv a=\left(\begin{array}{c}a_1\\a_2\end{array}\right)\,,
         \qquad \rv b =\left(\begin{array}{c}b_1\\b_2\end{array}
         \right)\,,\qquad \rv S =\left(\begin{array}{l r}s_{11}&s_{12}
         \\s_{21}&s_{22}\end{array}\right)\,.\eqno(22.13)$$

     Физический смысл  элементов  матрицы  рассеяния  в  данном случае
очевиден:  $s_{11}$ ---  амплитуда  отражённой  волны,  когда  волна
единичной  амплитуды  падает  слева,  а  справа  падающей  волны  нет,
$s_{21}$ -- амплитуда прошедшей волны при тех же условиях. Аналогичным
образом  понимаются  и  элементы матрицы $s_{22}$ и $s_{12}$.  Матрица
рассеяния однозначно определена только в том случае,  если для каждого
волновода  выбрана плоскость отсчета.  Эти плоскости $z=z_1$ и $z=z_2$
могут  не   совпадать   с   весьма   условно   выделенными   границами
полубесконечных  волноводов  $z=z_L$ и $z=z_R$ --- какие-то их отрезки
могут  быть  отнесены  к  нерегулярности.  Так  как   зависимость   от
продольной  координаты  $z$ в волноводных волнах выражается множителем
$e^{ihz}$,  то   перенос   отсчётных   плоскостей   вглубь   каждого
полубесконечного  волновода на расстояния $l_1$ и $l_2$ соответственно
приводит к преобразованию векторов $\rv a$ и $\rv b$:
     $$\rv  a'=\rv L^{-1}\rv  a\,,\qquad  \rv  b'=\rv  L\rv b\,,
         \eqno(22.14)$$
где
     $$ \rv L=\left(\begin{array}{cc} e^{ihl_1}&0\\0&e^{ihl_2}
         \end{array}\right)\,.\eqno(22.15)$$
Новые векторы  $\rv  a'$  и  $\rv  b'$  связаны  между  собой  тем  же
соотношением вида (22.12),  в котором новая матрица рассеяния $\rv S'$
есть
     $$\rv S'=\rv L\rv S\rv L\,.\eqno(22.16)$$
Таким образом,  дифракция основной волны  в  регулярном  волноводе  на
произвольном препятствии в дальней зоне полностью описывается четырьмя
комплексными величинами,  которые  определяются  совокупным  действием
бесконечного  числа затухающих волн высших видов.  Однако эти величины
связаны между  собой  дополнительными  соотношениями,  вытекающими  из
общих свойств матрицы рассеяния.

     Прежде чем  говорить  об  этих  общих  свойствах,   распространим
понятие   матрицы  рассеяния  на  структуру  более  общего  вида,  чем
единичное препятствие в регулярном  волноводе.  Во-первых,  регулярные
полубесконечные  волноводы  справа и слева от нерегулярности могут
различаться   по форме и даже не  иметь общей   оси   $z$.   Во-вторых,
число   волноводов,   примыкающих  к нерегулярности  (будем  далее
называть   её   многополюсником   --- происхождение термина выяснится
чуть позже) может быть произвольным, в частности,  имеет смысл рассматривать
и  случай  одного  волновода.  В третьих,  на  данной частоте распространяющихся
волн может быть больше одной и --- как уже отмечалось ранее --- каждой такой
волне необходимо  сопоставить   свой  волновод.  Условно  такая  обобщённая
структура  представлена на рис.~22.2.

\begin{wrapfigure}[14]{l}{7.25cm}
\begin{picture}(80,55)
\put(-3,50){\special{em:graph fig22-2.bmp}}
\end{picture}
\hbox to 7.25cm{\hfil\footnotesize{Рис.~22.2.~К определению матрицы
рассеяния.}\hfil}
\end{wrapfigure}

     Перечислим теперь   (без   доказательства)   основные    свойства
квадратной  матрицы  рассеяния $\rv S$ размерности $n$,  совпадающей с
числом примыкающих к $2\times n$-полюснику волноводов.

     1. Если многополюсник включает в себя лишь  изотропные  материалы
($\varepsilon,  \,\mu$  ---  скалярные величины),  то матрица рассеяния
$\rv S$ симметрична,  то есть $s_{pq}=s_{qp}$.  Это  свойство  матрицы
рассеяния  эквивалентно  теореме  взаимности  для всего пространства и
доказывается во многом аналогичным способом.  Особо подчеркнём,  что
симметрия   $\rv  S$  не  связана  с  реальной  физической  симметрией
нерегулярности.

     2. Если  в  объёме  нерегулярности  нет  потерь,   то   матрица
рассеяния унитарна:
     $$ \rv S^H\rv S = \rv 1\,,\eqno(22.17)$$
где $\rv S^H$ --- эрмитово-сопряжённая матрица  ($\rv  S^H=\bigl(\rv
S^T\bigr)^*$  ,  $\rv S^T $ --- транспонированная матрица,  * --- знак
комплексного сопряжения),  $\rv 1$ --- единичная матрица. Это свойство
является  прямым  следствием  теоремы Умова-Пойнтинга и выражает собой
закон сохранения энергии. В случае изотропного заполнения $\rv S^T=\rv
S$ и, следовательно, $\rv S^H=\rv S^*$, так что
     $$ \rv S^* \rv S= \rv 1\,.\eqno (22.18) $$
Поскольку между  матрицей  $\rv  S$  и  обратной  к  ней матрицей $\rv
S^{-1}$ имеет место такое же соотношение, то
     $$\rv S^*=\rv S^{-1}\,.\eqno(22.19)$$

     3. Для нерегулярностей,  матрица рассеяния  которых  симметрична,
выполняется   соотношение,  представляющее  собой  один  из  вариантов
формулировки {\it теоремы Фостера}:
     $$-\frac i 2 \rv a^T\rv S^*\frac {d\rv S}{d\omega}\rv a=
         \overline{U}_E+\overline{U}_H\geq 0\,,\eqno(22.20)$$
где $\rv  a^T$  ---   вектор-строка   из   амплитуд   $a_p$,   матрица
$\displaystyle{\frac   {d\rv   S}{d\omega}}$   состоит   из  элементов
$\displaystyle{\frac     {ds_{pq}}{d\omega}}$,      $\overline{U}_E,\;
\overline{U}_H$  ---  запасённая  в объёме нерегулярности энергия,
соответственно, электрического и магнитного поля.

     Матрица рассеяния   выявляет  на  данной  частоте  характеристики
собственно самого многополюсника как  участка  волноводного  тракта  и
определяется  только  его  конструкцией.  Следует  особо сказать,  что
элементы матрицы рассеяния могут быть не только вычислены в результате
решения  электродинамической  задачи,  но  и  измерены в эксперименте.
Фактически измерение параметров  многополюсника  производится  путём
измерения характеристик поля в подводящих волноводах.  На практике ---
помимо   частоты   ---   стандартными   экспериментальными   способами
определяются поток мощности через сечение волновода,  КСВН и положение
минимума и максимума поля в волноводе.  Из изложенного выше ясно,  что
знание  этих величин для всех волноводов многополюсника достаточно для
вычисления всех элементов матриц $\rv S$.

     Если СВЧ устройство обладает реальной физической симметрией, то в
ряде  случаев  структура  матрицы  рассеяния  может  быть  во   многом
определена до решения электродинамической задачи или измерений.  Более
того,  такое  предварительное   исследование   может   выявить   число
параметров,  определяющих  свойства  нерегулярности,  и  те  величины,
которые должны быть рассчитаны или  измерены.  При  наличии  симметрии
относительно    некоторой    оси   возможна   операция   симметричного
преобразования --- поворот относительно оси  симметрии,  в  результате
которого  отдельные  элементы устройства меняются местами,  в то время
как геометрическая форма устройства в целом остаётся без  изменений.
Другой  важной для нас операцией симметричного преобразования является
зеркальное  отражение  относительно  плоскости   симметрии.   Операция
симметричного преобразования описывается матрицей {\it симметрического
оператора} $\gv G$, который преобразует электрические поля $\rv a +\rv
b$, или $\rv a$, или $\rv b$ в поля $\widetilde{\rv a}'+\widetilde{\rv
b}'$,   или   $\widetilde{\rv   a}'$,   или    $\widetilde{\rv    b}'$
соответственно:
     $$ \widetilde{\rv a}'=\gv G \rv a\,,\qquad\widetilde{\rv b}'=
         \gv G\rv b\,,\eqno(22.21)$$
где $\widetilde{\rv a}'$ и  $\widetilde{\rv  b}'$  ---  также  решения
уравнений Максвелла, удовлетворяющие граничным условиям.

     Симметрический оператор   $\gv   G$   для   каждого   конкретного
устройства  следует  искать  на  основе  общих  рассуждений.  Так  как
вектор-столбец $\widetilde{\rv a}'$  представляет  собой  совокупность
тех  же  элементов,  из  которых состоит вектор $\rv a$ ($a_p$ и $a_q$
поменялись местами или  $a_r$  заменён  на  $-  a_r$),  то  элементы
матрицы $\gv G$ могут быть равны либо $\pm1$,  либо 0, прич\"ем только
один элемент в каждой строке и в каждом столбце отличен от 0;  поэтому
матрица $\gv G$ ортогональна, то есть
     $$\gv G^T\gv G=\rv 1\,.\eqno(22.22)$$
Из определения обратной матрицы $\gv G^{-1}\gv G=\rv 1$ следует
     $$\gv G^{-1}=\gv G^T\,,\eqno(22.23)$$
а так как элементы матрицы $\gv G$ --- действительные величины, то она
также и унитарна.  Отметим,  что матрица  $\gv  G$  симметрична,  если
определяет  операцию  отражения;  она может быть несимметричной,  если
определяет операцию поворота около  оси  симметрии.  Для  симметричных
$\gv  G$  и~$\rv  S$  собственные векторы,  связанные с невырожденными
собственными значениями, ортогональны друг другу.

     Важнейшим свойством  матриц  $\gv  G$  и  $\rv  S$  является   их
коммутативность:
     $$ \gv G \rv S=\rv S\gv G\,,\eqno(22.24) $$
что непосредственно следует из  (22.21)  и  того  факта,  что  векторы
$\widetilde{\rv  a}'$  и  $\widetilde{\rv  b}'$  связаны  соотношением
(22.12) с той же самой матрицей $\rv S$.  Коммутирующие матрицы  могут
иметь  разные  собственные  значения,  но  при условии невырожденности
собственных значений их  собственные  векторы  совпадают.  Собственные
значения  $g,\;s$ матриц $\gv G,\;\rv S$ и соответствующий собственный
вектор $\rv a$ определяются уравнениями
     $$\gv G \rv a= g\rv a\,\qquad \rv S \rv a= s\rv a\,.
         \eqno(22.25)$$

     Число собственных значений и собственных векторов  матрицы  равно
её  размерности   $n$.   Будем   помечать  собственные  значения  и
собственные      векторы       верхним       индексом:$g^i,\;s^i,\;\rv
a^i\;(i=1,\dots,n)$.   Симметрической   матрице  $\gv  G$  ставится  в
соответствие {\it матрица преобразования}  $\gv  F$,  столбцы  которой
составлены  из  собственных векторов $\gv G$ и нормированы к единичной
амплитуде.  Если собственные значения невырождены (все различны), то с
помощью  матрицы  преобразования  $\gv  F$  матрицы  $\gv G$ и $\rv S$
приводятся к диагональной форме:
     $$\gv F^{-1}\gv G \gv F =\gv G_d\,,\qquad \gv F^{-1}\rv S \gv F =
         \rv S_d\,,\eqno(22.26)$$
где $\gv  G_d$ и $\rv S_d$ --- диагональные матрицы,  элементы которых
являются собственными значениями соответствующих  матриц.  С  учётом
ортогональности матрицы $\gv F$ отсюда находим,  что матрица рассеяния
выражается через свои собственные значения следующим образом:
     $$\rv S=\gv F\rv S_d\gv F^T\,.\eqno(22.27)$$

     Важно отметить, что матрица преобразования $\gv F$ должна не быть
особой,  для чего необходимо,  чтобы все собственные значения  матрицы
$\gv G$ были невырождены. Вырождение всегда имеет место, когда порядок
матрицы $n$ больше порядка симметрии $m$,  определяемого условием $\gv
G^m=\rv 1$ ($m$ --- число последовательных  преобразований  симметрии,
возращающих  структуру  в  исходное  состояние).  Напомним,  что  сама
матрица рассеяния  и  её  собственные  значения  зависят  от  выбора
плоскостей отсчёта, а её собственные векторы --- нет. Кроме этого,
важным свойством эрмитово-сопряжённых матриц является то, что модули
всех  их собственных значений равны единице и если для матрицы $\rv S$
справедлива формула (22.17), то $|s^i|=1\;(i=1,\dots,n)$.

     В качестве  простейшего примера применения симметрической матрицы
$\gv  G$  для  выяснения  структуры   матрицы   рассеяния   рассмотрим
симметричную неоднородность в волноводе,  изображённую на рис.~22.3.
Будем считать,  что плоскости отсчёта в левом  и  правом  волноводах
выбраны  на  равном расстоянии от плоскости симметрии.  Тогда операция
отражения относительно плоскости симметрии меняет  поля  в  подводящих
волноводах местами, что позволяет записать симметрическую матрицу $\gv
G$ в виде
     $$\gv G=\hbox{$\left(\begin{array}{cc}0&1\\1&0\\\end{array}
         \right)$}.\eqno(22.28)$$
Собственные значения этой матрицы равны $1$ и $-1$,  а соответствующие
собственные векторы, которые являются и собственными векторами матрицы
рассеяния $\rv S$, равны
     $$\rv a^1=\hbox{$\left(\begin{array}{c}1\\1\\\end{array}\right)$
         },\qquad \rv a^2=\hbox{$\left(\begin{array}{c}\phantom{-}1
         \\-1\\\end{array}\right)$}.\eqno(22.29)$$
Они схематически показаны на  рис.~22.3,  причём  вектор  $\rv  a^1$
симметричный   и   соответствующее  ему  поле  принимает  максимальное
значение   в   плоскости   симметрии,   а   вектор   $\rv   a^2$   ---
антисимметричный и поле в этой же плоскости обращается в нуль.

     Матрица преобразования $\gv F$ симметрична и имеет вид
     $$  \gv F =\frac1{\sqrt{2}}\left(\begin{array}{lr}1&1\\1&-1
         \end{array}\right)\,.\eqno(22.30)$$
Используя далее преобразование (22.27), выражаем матрицу $\rv S$ через
её  собственные значения:
     $$   \rv S=\left(\begin{array}{lr}s_{11}&s_{12}\\s_{12}&s_{22}
         \end{array}\right)=\frac 1 2 \left(\begin{array}{lr}
         s^1+s^2&s^1-s^2\\s^1-s^2&s^1+s^2\end{array}\right)\,.
         \eqno(22.31)$$
Отсюда следует,  что нерегулярность будет  согласована  с  волноводом,
если  $s^1=-s^2$,  то  есть   отражённой  волны  нет,  но фаза
прошедшей волны отлична от фазы падающей.

\begin{wrapfigure}[13]{l}{7.25cm}
\begin{picture}(70,45)
\put(-7,50){\special{em:graph fig22-3.bmp}}
\end{picture}
\hbox to 7.25cm{\hfil\footnotesize{Рис.~22.3.~Симметричная
нерегулярность.
}\hfil}
\end{wrapfigure}

     Если нерегулярность  не  имеет  потерь,  то  $|s^1|=|s^2|=1$,   и
комплексные  числа  $s^1+s^2$  и  $s^1-s^2$  находятся  в  квадратуре;
поэтому
     $$ |s_{11}|^2+|s_{12}|^2=1\,.\eqno(22.32)$$
Это соотношение  легко  может  быть  выведено  и  из условия эрмитовой
сопряжённости матрицы $\rv S$.  Оно автоматически выполняется,  если
элементы матрицы представить в виде
     $$ s_{11}=r\,e^{i\theta}\,,\qquad s_{12}=i\sqrt{1-r^2}\,e^{i\theta}\,,
         \eqno(22.33)$$
где $r$ --- вещественное положительное число,  не превышающее единицы.
За счёт выбора плоскости отсчёта можно  добиться,  чтобы  $s_{11}$
равнялся  действительному  числу  $r$  и тогда $s_{12}=i\sqrt{1-r^2}$.
Таким образом,  симметричная нерегулярность в волноводе по отношению к
основной  волне определяется двумя вещественными параметрами,  которые
либо  должны  быть  измерены,  либо  найдены  в   результате   решения
дифракционной    задачи.    В   частном   случае   бесконечно   тонкой
нерегулярности (диафрагма,  толщиной которой можно пренебречь) в  её
плоскости  должно  выполняться  условие  непрерывности  тангенциальных
составляющих поля,  откуда следует  равенство  $s_{11}  +1=s_{12}$,  и
поэтому  нерегулярность определяется одним параметром $\theta$,  через
который элементы матрицы рассеяния выражаются так:
     $$s_{11}=i\sin{\theta}e^{i\theta} ,\qquad s_{12}=\cos{\theta}
         e^{i\theta}\,.\eqno(22.34)$$
Заметим, что   в  случае  несимметричной  нерегулярности  в  волноводе
(рис.~22.1) число независимых параметров равно трём и все они  могут
быть  вычислены,  если  решена  задача  о  падении  основной  волны  с
какой-нибудь одной стороны.

\begin{wrapfigure}[13]{l}{7.25cm}
\begin{picture}(80,47)
\put(-5,45){\special{em:graph fig22-4.bmp}}
\end{picture}
\hbox to 7.25cm{\hfil\footnotesize{Рис.~22.4.~Направленный ответвитель.
}\hfil}
\end{wrapfigure}

     Другим примером  нерегулярности  с  высокой  степенью   симметрии
служит  направленный  ответвитель,  представленный  на  рис.~22.4.  Он
состоит  из  двух  прямоугольных   волноводов   одинакового   сечения,
примыкающих  друг  к  другу  широкими  стенками,  в  которых проделано
отверстие,  симметричное по отношению к узким стенкам  и  относительно
некоторой плоскости,  нормальной к оси волноводов.  Структура обладает
двумя  плоскостями  симметрии  $P_1$  и  $P_2$,   представляет   собой
симметричный  восьмиполюсник,  матрица  рассеяния  которого состоит из
$4\times 4$ элементов.  Если плоскости отсчёта в волноводах  выбраны
симметрично,  то  из  общих соображений ясно,  что несовпадающих между
собой  элементов  в  матрице  рассеяния  четыре,  и  она  может   быть
представлена в виде
    $$\rv S=\left(\begin{array}{cccc}\alpha&\beta&\gamma&\delta\\
        \beta&\alpha&\delta&\gamma\\\gamma&\delta&\alpha&\beta\\
        \delta&\gamma&\beta&\alpha\end{array}\right)\,.\eqno(22.35)$$

     Пусть в структуре  отсутствуют  потери;  тогда  матрица  $\rv  S$
унитарна,  и  все  её  собственные  значения  $s^i$  по модулю равны
единице.  Направленный ответвитель называется идеальным в том  случае,
когда  падающая  по  волноводу волна не испытывает отражения,  то есть
$\alpha=0$.  Это может быть достигнуто за  счёт  подбора  размера  и
формы  отверстия  связи  между  волноводами  (ответвитель может быть и
многодырочным).  Из унитарности  матрицы  $\rv  S$  следует,  что  при
$\alpha=0$ один из оставшихся элементов матрицы $\beta$,  $\gamma$ или
$\delta$ также равен нулю,  то есть соответствующий волновод развязан,
и,  кроме того,  фазы двух других элементов различаются на $\pi/2$. Не
будем приводить соответствующих  выкладок,  поскольку  рассматриваемое
ниже   использование   матрицы   преобразования  $\gv  F$  ---  помимо
установления этих свойств --- позволяет ещ\"е больше  конкретизировать
вид матрицы рассеяния рассматриваемого ответвителя.

     При показанной   на  рис.~22.4  нумерации  волноводов  отражениям
относительно плоскостей $P_1$  и  $P_2$  соответствуют  симметрические
матрицы
     $$\gv G_1=\hbox{$\left(\begin{array}{cccc}0&0&1&0\\0&0&0&1\\1&0
         &0&0\\0&1&0&0\end{array}\right)$}\qquad\mbox{и}\qquad\gv G_2=
         \hbox{$\left(\begin{array}{cccc}0&1&0&0\\1&0&0&0\\0&0&0&1\\
         0&0&1&0\end{array}\right)$}\,.\eqno(22.36)$$
Обе эти  матрицы  имеют  вырожденные  собственные значения,  поскольку
каждая из них возвращает структуру в  первоначальное  состояние  после
двух преобразований,  а размерность матриц равна четыр\"ем.  Требуется
найти такую матрицу $\rv M$,  представляющую собой линейную комбинацию
матриц $\gv G_1$ и $\gv G_2$,  которая имеет невырожденные собственные
значения.  Для этого  необходимо,  чтобы  выполнялось  равенство  $\rv
M^4=\rv  1$,  так  что  собственные значения $\rv M$ будут равны $+1$,
$i$,  $-1$  и  $-i$;  собственные  векторы  этой   матрицы   и   будут
собственными векторами матрицы $\rv S$.

     Опуская несложные   промежуточные   выкладки,  выпишем    матрицу
преобразования $\gv F$, столбцы которой состоят из этих векторов:
     $$\gv F=\frac1 2\left(\begin{array}{rrrr}1&1&1&1\\1&-1&1&-1\\
         1&1&-1&-1\\1&-1&-1&1\end{array}\right)\,.\eqno(22.37)$$
Матрица $\gv F$ ортогональна и симметрична,  так что элементы  матрицы
рассеяния (22.35) находятся с помощью формулы (22.27):
     $$ \left.\begin{array}{lcr}\alpha&=&\displaystyle
         (s^1+s^2+s^3+s^4)/4,\\[.3cm]\beta&=&
         (s^1-s^2+s^3-s^4)/4,\\[.3cm]\gamma&=&
         (s^1+s^2-s^3-s^4)/4,\\[.3cm]\delta&=&
         (s^1-s^2-s^3+s^4)/4.\end{array}\right\}\eqno(22.38)$$
Нетрудно видеть, что в отношении собственных векторов $\rv a^3$ и $\rv
a ^4$ верхний и нижний  волноводы  развязаны,  поскольку  наличие  или
отсутствие  в  плоскости  $P_1$  проводящей  стенки  не сказывается на
структуре  поля.  Так  как  $\rv  a^3$  относительно  плоскости  $P_2$
симметричен, а $\rv a^4$ --- антисимметричен, то $s^3=+1$, а $s^4=-1$.
В результате
     $$\left.\begin{array}{lcl}\alpha&=&
         (s^1+s^2)/4,\\[.3cm]\beta&=&
         (s^1-s^2+2)/4,\\[.3cm]\gamma&=&
         (s^1+s^2)/4,\\[.3cm]\delta&=&
         ( s^1-s^2-2)/4;\end{array}\right\}\eqno(22.39)$$
таким образом, для рассматриваемого ответвителя
     $$\alpha=\gamma\qquad\mbox{и}\qquad\beta-\delta=1.\eqno(22.40)$$

     Направленный ответвитель становится идеальным при $s^1=-s^2$; при
этом  волновод  3  развязан  с  волноводом  1 ($\gamma=0$),  $\beta$ и
$\delta$ сдвинуты по фазе на $\pi/2$  и  $\delta$  опережает  $\beta$.
Выбором  плоскости  отсчёта матрица рассеяния ответвителя может быть
приведена к виду
     $$\rv S=\left(\begin{array}{cccc}0&\sqrt{1-r^2}&0&ir\\
         \sqrt{1-r^2}&0&ir&0\\0&ir&0&\sqrt{1-r^2}\\ir&0&\sqrt{1-r^2}&
         0\end{array}\right)\,,\eqno(22.41)$$
где $r$  ---  действительное  положительное  число.   Подбором   формы
отверстия можно добиться того, что мощность, поступающая в волноводы 2
и 4, одинакова. В этом случае $r$ в (22.41) равно $1/\sqrt{2}$ и такой
ответвитель   называется   трёхдецибельным.   Происхождение  термина
связано с понятием {\it переходного ослабления} $C$,  определяемого  в
децибелах соотношением
     $$C=-20 \lg{|\delta|}\,;\eqno(22.42)$$
при $r=1/\sqrt{2}$ ослабление $C$ близко к трём децибелам.

     Матрица рассеяния  с  успехом   может   быть   использована   при
исследовании  такой  важной  для  практики структуры,  как резонатор с
подводящим волноводом.  Рассмотрим волноводный резонатор,  в одной  из
торцевых стенок которого проделано отверстие для стыковки с подводящим
волноводом.  Величина  связи  может  регулироваться  выбором   размера
диафрагмы,  помещённой  в  плоскости  стыка.  Конкретная конструкция
элемента  связи  (рис.~22.5)  для  нас   не   важна   и   её   можно
характеризовать матрицей рассеяния
     $$\rv S=\left(\begin{array}{ll}s_{11}&s_{12}\\s_{21}&s_{22}
         \end{array}\right)\,.\eqno(22.43)$$
Плоскости отсчёта удобно выбрать  так,  чтобы  элементы  $s_{11}$  и
$s_{22}$   были   действительными   отрицательными   величинами.  Если
устройство  связи  не  имеет  потерь  и   не   содержит   анизотропных
материалов,  то  матрица  $\rv  S$ унитарна и симметрична и может быть
записана в виде
     $$\rv S=\left(\begin{array}{cc}-\sqrt{1-r^2}&ir\\ir&
         -\sqrt{1-r^2}\end{array}\right)\,,\eqno(22.44)$$
где коэффициент связи $r$ --- действительное число (положительное  или
отрицательное).

     Пусть расстояние  от  плоскости  отсчёта 2 до сплошной торцевой
стенки равно $L$.  При слабой связи  (малые  $r$,  чему  соответствует
малое   отверстие   диафрагмы)   это   расстояние  лишь  незначительно
отличается от расстояния $l$ между торцами резонатора, которое и будет
использоваться   во   всех  последующих  формулах.  Тогда  набег  фазы
волноводной волны в резонаторе  (электрическая  длина)  на  длине  $l$
составляет $\varphi=h_2 l$,  где $h_2$ --- продольное волновое число в
волноводе,  образующем резонатор.  Если затухание  волны  на  пути  до
стенки  и  обратно равно $\alpha$,  то в случае идеальной проводимости
торца между $a_2$ и $b_2$ имеет место соотношение
     $$a_2=-b_2\,e^{-\alpha+2i\varphi}\,,\eqno(22.45)$$
позволяющее с помощью (22.12) и (22.44)  выразить  амплитуды  $b_1$  и
$b_2$ через $a_1$:
     $$\begin{array}{lcl}b_1&=&-\left(\sqrt{1-r^2}-\displaystyle{\frac
         {r^2e^{-(\alpha-2i\varphi)}}{1-\sqrt{1-r^2}\,e^
         {-(\alpha-2i\varphi)}}}\right) a_1\,,\\
         b_2&=&\displaystyle{\frac{ir}{1-\sqrt{1-r^2}\,
         e^{-(\alpha-2i\varphi)}}} \,a_1\,.\end{array}\eqno(22.46)$$

\begin{wrapfigure}[13]{l}{7.25cm}
\begin{picture}(80,50)
\put(-5,45){\special{em:graph fig22-5.bmp}}
\end{picture}
\hbox to 7.25cm{\hfil\footnotesize{Рис.~22.5.~Волноводный резонатор
}\hfil}
\hbox to 7.25cm{\hfil\footnotesize{и подводящий волновод.
}\hfil}
\end{wrapfigure}

     Для заданных $\alpha$ и $r$ модуль амплитуды $b_2$ максимален,  а
амплитуды $b_1$ --- минимален, при
     $$\varphi=n\pi\,,\qquad n=1, 2\dots\,,\eqno(22.47)$$
что соответствует {\it условию резонанса}; на резонансных частотах
     $$ \begin{array}{lcl}b_1&=&-\left(\sqrt{1-r^2}-\displaystyle{\frac
         {r^2\,e^{-\alpha}}{1-\sqrt{1-r^2}\;e^
         {-\alpha}}}\right) a_1\,,\\
         b_2&=&\displaystyle{\frac{ir}{1-\sqrt{1-r^2}\;
         e^{-\alpha}}}\, a_1\,.\end{array}\eqno(22.48)$$

Таким образом,   резонансные   частоты
$\omega_{\footnotesize\textit{рез}}$ волноводного резонатора, связанного
с внешним волноводом, определяются условием, что на электрической
длине от выделенной плоскости отсчёта  элемента связи до
сплошного торца резонатора укладывается целое число полуволн.
Следовательно, они сдвинуты относительно собственных частот
идеального резонатора как за счёт потерь в стенках, так и из-за
наличия связи с волноводом.

     Если регулировать  коэффициент  связи  $r$,  то  при   постоянном
$\alpha$ максимум величины $b_2$ имеет место при условии
     $$ \frac{d|b_2|}{dr}=\frac{\sqrt{1-r^2}-e^{-\alpha}}{\sqrt{1-r^2}
         (1-\sqrt{1-r^2}\;e^{-\alpha})^2}|a_1|=0\,,\eqno(22.49)$$
то есть при
     $$r=\pm \sqrt{1-e^{-2\alpha}}\,.\eqno(22.50)$$
При критическом коэффициенте связи $r$, определяемом этим выражением,
     $$b_1=0\,,\qquad b_2=\pm \, i\frac 1 {\sqrt{1-e^{-2\alpha}}}\,a_1\,,
           \eqno(22.51)$$
и, следовательно,  вся падающая энергия поглощается  в  резонаторе.  В
случае  малых  потерь  экспонента  $e^{-2\alpha}$ близка к единице и в
резонансе при критической связи $|b_2|$ может  на  несколько  порядков
превышать  $|a_1|$,  так  что в резонаторе запасено большое количество
энергии.  В этом случае $r$ согласно  (22.50)  очень  мало  (небольшое
отверстие диафрагмы).

     С физической  точки  зрения  эффект  согласования  резонатора   и
волновода  обусловлен  тем,  что  <<просачивающаяся>>  из  резонатора  в
волновод небольшая часть волны большой амплитуды гасит отражённую от
диафрагмы     большую     часть     падающей     волны.     Если    же
$|r|\ne\sqrt{1-e^{-2\alpha}}$, то амплитуда $b_1$ при резонансе уже не
равна  $0$:  действительно,  при $|r|<\sqrt{1-e^{-2\alpha}}$ резонатор
{\it недосвязан}, $b_1<0$, то есть плоскость отсчёта $1$ расположена
в минимуме стоячей волны,  а при $|r|>\sqrt{1-e^{-2\alpha}}$ резонатор
{\it  пересвязан},  $b_1>0$  и  плоскость  отсчёта  1  находится   в
максимуме   стоячей   волны.   В   первом  случае  отражённая  волна
создаётся в основном за счёт отражения от диафрагмы, во втором ---
за счёт волны, просачивающейся из резонатора.

     Так называемые  {\it  резонаторы  бегущей   волны}   представляют
значительный  интерес  с точки зрения рекуперации СВЧ энергии.  В
этом случае мощность от генератора поступает через  линию  к
направленному ответвителю,  после  которого часть её попадает в
резонансное кольцо (в которое может быть встроен,  например,
диафрагмированный волновод), а  часть идёт в согласованную
нагрузку.  Пройдя по кольцу,  мощность снова   делится   на
направленном   ответвителе   и   если    волна, распространяющаяся
в кольце,  синфазна с волной,  вновь поступающей в кольцо от
генератора,  то амплитуда ускоряющего поля может стать  очень
большой по сравнению с полем волны от генератора.  Расчёт
резонатора бегущей волны производится аналогично случаю
волноводного  резонатора, однако  накопление  энергии  в нём
обусловлено увеличением амплитуды бегущей по кольцу волны, а не
переотражением волн.

     Для практики   большой  интерес  представляет  величина  энергии,
запасаемой в резонаторе. Сразу отметим, что говорить о запасённой
энергии  и  резонансных  свойствах  структуры имеет смысл только в том
случае,  когда за  один  период  высокочастотного  колебания  энергия,
поглощаемая внутри резонатора,  и энергия,  просачивающаяся в
волновод  (при  отсутствии   падающей   волны),   существенно   меньше
запасённой.  В  рассматриваемой  модели  волноводного резонатора эти
условия сводятся к виду
     $$\alpha\ll 1\qquad\mbox{и}\qquad r^2\ll 1\,.\eqno(22.52)$$
Только при этих условиях  можно  ввести  рассмотренные  в  разделе
16 понятия  собственной,  внешней  и  нагруженной добротности
резонатора. Оказывается,  что если  заданы  мощность  внешнего
генератора  $P_G$, определяющая  амплитуду  $a_1$  падающей волны,
и отстройка $\Delta \omega$ его частоты $\omega$  от резонансной
частоты  $\omega_{\footnotesize\textit{рез}}$ ($\Delta
\omega=\omega-\omega_{\footnotesize\textit{рез}}$), то только через эти
добротности и выражается запасённая энергия.

     Поскольку при малых $\alpha$ поле  в резонаторе незначительно
отличается от стоячей  волны  с  амплитудой  $2b_2$,  то
запасённая энергия легко выражается через средний поток мощности
$\overline{\Sigma}$  бегущей  волны  с амплитудой $b_2$, плотность
энергии которой на единицу длины составляет
$\overline{\Sigma}/v_{\footnotesize\textit{гр}}$:
     $$U_{\footnotesize\textit{зап}}=\frac {2l\overline{\Sigma}}
     {v_{\footnotesize\textit{гр}}}=
     \frac{|b_2|^2 \omega \,l}{c^2h_2}\,.\eqno(22.53)$$
С учётом  условий (22.52) из (22.46) получаем,  что вблизи резонанса
($\sin\varphi\sim r^2\sim\alpha$), определяемого условием (22.47),
     $$|b_2|^2=\frac{4 r^2}{(r^2+2\alpha)^2}\cdot
         \frac {|a_1|^2} {1+\displaystyle{\frac {16\sin^2{\varphi}}
         {(r^2+2\alpha)^2}}}\,.\eqno(22.54)$$
Разлагая функцию $\varphi (\omega)$ по степеням
$\omega-\omega_{\footnotesize\textit{рез}}$, убеждаемся, что при малой
отстройке
     $$\sin{\varphi}=(-1)^n\frac{l\,\omega_{\footnotesize\textit{рез}}}
     {c^2 h_2}(\omega-\omega_
         {\footnotesize\textit{рез}})\,.\eqno(22.55)$$

     Мощность потерь в резонаторе составляет $P_{\footnotesize\textit{пот}}
     =\alpha |b_2|^2$, так что его собственная добротность
     $$Q_0=\frac{\omega_{\footnotesize\textit{рез}}\,U_{\footnotesize\textit{зап}}}
     {P_{\footnotesize\textit{пот}}}=\frac{\omega_{\footnotesize\textit{рез}}^2 \,l}
         {\alpha h_2 c^2}\,.\eqno(22.56)$$
Мощность излучения из резонатора при  выключенном  внешнем
генераторе $P_{\footnotesize\textit{изл}}=|b_1|^2/2=|b_2|^2 r^2/2$;
поэтому внешняя добротность
     $$Q_{\footnotesize\textit{вн}}=
     \frac{2\omega_{\footnotesize\textit{рез}}^2\, l}{r^2 h_2 c^2}\,\eqno(22.57)$$
и, следовательно, нагруженная добротность равна
     $$Q_{\footnotesize\textit{н}}=\frac {Q_0 Q_{\footnotesize\textit{вн}}}
     {Q_0+Q_{\footnotesize\textit{вн}}}=\frac 2{r^2+2\alpha}\cdot
         \frac{\omega_{\footnotesize\textit{рез}}^2 \,l}{h_2 c^2}\,.\eqno(22.58)$$

     Полученные выражения  для добротностей позволяют записать формулу
(22.53) в виде
     $$  U_{\footnotesize\textit{зап}}=
          \frac {4 Q_{\footnotesize\textit{н}}^2 P_{G}}
           {Q_{\footnotesize\textit{вн}}\,\omega_{\footnotesize\textit{рез}}}\cdot
           \frac 1 {1 +\left(2Q_{\footnotesize\textit{н}}
           \displaystyle{\frac{\Delta \omega}
           {\omega}}\right)^2}\,,\eqno(22.59)$$
откуда непосредственно  следует,   что   ширина   резонансной
кривой определяется нагруженной добротностью
$Q_{\footnotesize\textit{н}}$.

     Из формулы (22.59) следует, что  пассивный резонатор
может служить эффективным усилителем мощности.  Действительно,
если от генератора  мощностью  $P_{G}$  в  резонатор  на
резонансной  частоте подводится энергия в течение времени,
достаточного  для  установления стационарной   амплитуды,  то  при
резком  отключении  генератора  из резонатора в первый момент
излучается  мощность,  заметно  превышающая мощность  $P_G$.  При
$\alpha\ll  r^2$,  то  есть при $Q_{\footnotesize\textit{вн}}\ll Q_0$
($Q_{\footnotesize\textit{н}}\approx Q_{\footnotesize\textit{вн}}$),
коэффициент усиления по мощности вблизи резонанса приближается  к
четырем. Этот эффект в своё время нашёл практическое применение
в линейных ускорителях электронов.

     Как видно  даже  из  нескольких рассмотренных примеров,  описание
отдельной нерегулярности с помощью  матрицы  рассеяния  очень  удобно,
эффективно и для большинства устройств СВЧ соответствует таким реалиям
практики,  как  постоянство  падающей  мощности.  Более  того,   такое
описание  пригодно  и используется для описания свойств устройства или
волноводного тракта целиком. Неудобства возникают только в том случае,
когда  устройство  естественным  образом  можно  разбить  на несколько
отдельных  нерегулярностей,  соединённых  друг  с  другом  отрезками
регулярного  волновода.  Пусть  в этом случае матрицы рассеяния каждой
отдельной нерегулярности  известны;  тогда  задача  выражения  матрицы
рассеяния  целого устройства через матрицы отдельных нерегулярностей и
параметры  соединяющих  волноводов  существенным  образом  усложняется
из-за   необходимости   учёта   многократных  отражений  волн  между
соседними нерегулярностями.  Однако существует  альтернативный  способ
представления  нерегулярности,  который легко позволяет преодолеть эти
трудности, и, помимо того, позволяет сопоставить каждой нерегулярности
эквивалентную схему из используемых в низкочастотных цепях переменного
тока сосредоточенных элементов,  таких как  ёмкость,  индуктивность,
трансформатор и прочее.  Такие схемы оказываются полезными при синтезе
устройств СВЧ с требуемыми свойствами, однако их вид зависит от выбора
плоскости отсчёта.

     Следует предварительно сказать несколько слов о тех условиях, при
которых  какой-то  нерегулярный  участок  тракта  допустимо разбить на
несколько отдельных нерегулярностей.  Возьмём простейший пример  ---
толстую диафрагму в прямоугольном волноводе.  Её можно рассматривать
как два последовательных скачкообразных изменения размеров  волновода,
соединённых  отрезком  волновода  меньшего  размера.  Каждый  скачок
представляет собой бесконечно тонкую нерегулярность. Возникает вопрос,
при  каком расстоянии между ними (толщине диафрагмы) матрица рассеяния
диафрагмы определяется только элементами матриц уединённых скачков и
параметрами  одномодового  волновода  между  ними.  Из вышеизложенного
способа введения матрицы рассеяния  ясно,  что  для  этого  необходимо
отсутствие  перекрытия  полей  высших типов волн,  возбуждаемых вблизи
каждого скачка. Очевидно, что требуемое расстояние существенно больше,
если  скачки  обращены  друг  к  другу  со стороны волноводов большего
размера.  Нерегулярность в целом  в  этом  случае  представляет  собой
проходной  резонатор  с  сильной  связью  с  подводящими  волноводами.
Увеличение области влияния волн высших типов ясно из  картины  силовых
линий  электрического  поля,  которые  должны  подходить  по нормали к
торцевой поверхности резонатора.

     Формально альтернативный способ описания нерегулярности состоит в
замене  одномодовых  волноводов  на  эквивалентные  длинные  линии,  а
фактически представляет собой переход от бегущих волн двух направлений
к   стоячим   волнам  электрического  и  магнитного  поля.  Рассмотрим
волновод, в отсчётной плоскости $z=0$ которого расположена нагрузка,
обусловливающая   коэффициент  отражения  $\Gamma_0$.  Запишем  полные
поперечные поля (22.9) в виде
     $$ \rv E_t(\rv r)=\rv F(x,y) V(z)\,,\qquad \rv H_t(\rv r)=
         \rv G(x,y) I(z)\,;\eqno(22.60)$$
тем самым введены {\it напряжение}     $$ V(z)=ae^{ihz}+be^{-ihz}\eqno(22.61)$$
и {\it ток}     $$ I(z)=ae^{ihz}-be^{-ihz}\eqno(22.62)$$
эквивалентной линии передачи.  Их отношение называется
{\it нормированным сопротивлением};   оно   также  зависит  от  $z$  и
выражается  через коэффициент отражения $\Gamma$ следующим
образом:
     $$Z(z)=\frac{V(z)}{I(z)}=\frac{ae^{ihz}+be^{-ihz}}{ae^{ihz}-be^
         {-ihz}}=\frac{1+\Gamma}{1-\Gamma}\,.\eqno(22.63)$$
Средний поток мощности,  который в теории длинных  линий  обозначается
как $P$, выражается через $V(z)$ и $I(z)$ по известной формуле
     $$P=\overline{\Sigma}=\frac 1 2 V(z)I^*(z)\,.\eqno(22.64)$$

     Все введённые    величины    становятся    значительно    менее
формальными,  когда после подстановки (22.60) в уравнения Максвелла  и
учёта  соотношений  (22.4) между $\rv F(x,y)$ и $\rv G(x,y)$ (или из
соотношений  (22.61),  (22.62))  убеждаемся,  что  $V(z)$   и   $I(z)$
удовлетворяют тем же дифференциальным уравнениям
     $$ \frac{dV(z)}{dz}=ih I(z)\,,\qquad \frac {dI(z)}{dz}=ih
         V(z)\,,\eqno(22.65)$$
что и соответствующие величины в теории длинных линий.

     С помощью  очевидного решения уравнений (22.65) нетрудно выразить
$V(z)$ и $I(z)$ через их значения $V_0$ и $I_0$ в плоскости отсчёта:
     $$\left.\begin{array}{lclcl}V(z)&\!\!=\!\!&V_0\cos{hz}&\!\!+
         \!\!&iI_0\sin{hz}\,,\\[.3cm]I(z)&\!\!=\!\!&I_0\cos{hz}&\!\!+
         \!\!&i V_0\sin{hz}\,.\end{array}\right\}\eqno(22.66)$$

     Простейшей нерегулярности в волноводе  (рис.~22.1),  для  которой
формулами  (22.12)  и  (22.13)  была определена матрица рассеяния 2-го
порядка   $\rv   S$,   можно    теперь    сопоставить    эквивалентный
четырёхполюсник,   используемый  для  расчета  низкочастотных  цепей
(рис.~22.6).  Свойства четырёхполюсника определяются  {\it  матрицей
сопротивлений} $\rv Z$, связывающей два вектора-столбца $\rv V$ и $\rv
I$:
     $$ \rv V=\rv Z \rv I\,,\quad\rv V=\left(\begin{array}{c}V_1\\V_2
         \end{array}\right)\,,\quad\rv I=\left(\begin{array}{c}I_1
         \\I_2\end{array}\right)\,,\quad\rv Z=\left(\begin{array}{ll}
         Z_{11}&Z_{12}\\Z_{21}&Z_{22}\end{array}\right)\,,
         \eqno(22.67)$$
где $V_1,\,V_2,\,I_1,\,I_2$ --- напряжения и токи на  входе  и  выходе
четырёхполюсника,  которые в тоже время можно считать напряжениями и
токами в отсчётных плоскостях примыкающих  слева  и  справа  длинных
линий  (в рассматриваемом примере эти линии имеют одинаковые параметры
$h$ и $Z_0$,  поскольку являются частями одного и того же  волновода).
Аналогичным образом вводится и матрица проводимостей $\rv Y$:
     $$ \rv I=\rv Y \rv V\,\qquad \rv Y=\rv Z^{-1}=\left(\begin{array}
         {ll}Y_{11}&Y_{12}\\Y_{21}&Y_{22}\end{array}\right)\,.
         \eqno(22.68)$$

     Поскольку матрица  рассеяния $\rv S$ и матрица сопротивлений $\rv
Z$ полностью и адекватно описывают одну и  ту  же  нерегулярность,  то
между  ними  должна  быть  вполне определённая связь,  которую легко
установить,  сопоставляя компоненты поперечных полей (22.9) и (22.60);
в результате получаем
     $$ \rv S=(\rv Z - \rv 1)(\rv Z+\rv 1)^{-1}\,, \qquad\rv Z=(\rv 1
         +\rv S)(\rv 1-\rv S)^{-1}\,.\eqno(22.69)$$

     Матрицы сопротивлений    и    проводимостей   очевидным   образом
обобщаются  на  произвольный  многополюсник,  связывая   между   собой
напряжения  и  токи  на входе каждого плеча.  Однако соотношения между
матрицей сопротивления и матрицей рассеяния усложняются по сравнению с
формулой  (22.69),  поскольку  параметры  подводящих  волноводов $h$ и
$Z_0$ в разных плечах в общем случае различны.  Следует сказать, что в
тех случаях,  когда требуется перенести отсчётные плоскости, матрица
сопротивлений приводит к дополнительным усложнениям,  поскольку в этом
случае  изменяются  и модули,  и фазы как её элементов,  так и самих
переменных $\rv V$ и $\rv I$.  Элементы же матрицы $\rv S$ и амплитуды
волн $a_p$, $b_p$ при этом изменяют только свою фазу.

     Связь между матрицами $\rv S$ и $\rv Z$ позволяет установить  ряд
важных  общих  свойств матрицы $\rv Z$,  исходя из перечисленных ранее
свойств матрицы $\rv S$.  Из симметрии матрицы $\rv S$  (то  есть  при
изотропии  материалов  заполнения  нерегулярности)  следует  симметрия
матрицы сопротивлений:  $Z_{pq}=Z_{qp}$.  Кроме этого,  из унитарности
$\rv S$ (при отсутствии потерь в объёме нерегулярности) следует, что
все элементы матрицы $\rv Z$ являются чисто мнимыми величинами,  и  их
принято записывать в виде
     $$ Z_{pq}=-iX_{pq}\,;\eqno(22.70)$$
элементы $X_{pq}$ (действительные числа)  образуют  матрицу  $\rv  X$,
называемую матрицей реактивных сопротивлений. Матрица $\rv X$ является
положительно или отрицательно определённой в зависимости от разности
электрической   и   магнитной   энергии,  накопленной  высшими  видами
колебаний вблизи нерегулярности. Это свойство следует из равенства
     $$2\omega(\overline{U}_H-\overline{U}_E)=\frac1 2(\rv I^*,\rv X
         \rv I)\eqno(22.71)$$
(доказательство его  с  помощью  первого  тождества  Грина  и  формулы
(22.64) здесь опускаем), поскольку $\overline{U}_E$ и $\overline{U}_H$
положительно определены.

     Из теоремы Фостера (22.20) следует, что производная по частоте от
матрицы  реактивных сопротивлений всегда положительно определена,  что
выражается соотношением
     $$ \frac 1 2 \Bigl(\rv I^*,\frac {d\rv X}{d\omega} \rv I\Bigr)=
         2(\overline{U}_H+\overline{U}_E)\geqslant 0\,.\eqno(22.72)$$

\begin{picture}(150,55)
\put(0,50){\special{em:graph  fig22-7a.bmp}}
\put(80,50){\special{em:graph fig22-7b.bmp}}
\end{picture}
\begin{center}\begin{minipage}[c]{0.9\textwidth}
\footnotesize{\parindent=0.5cm
Рис.~22.6. Эквивалентные схемы четырёхполюсника:  {\it {а)}}
---  Т-схема, {\it { б)}} ---  П-схема.
}\end{minipage}\end{center}\vspace*{0.25cm}

     С помощью  матриц  сопротивлений  $\rv  Z$ и проводимости $\rv Y$
четырёхполюснику можно сопоставить одну из  двух  представленных  на
рис.~22.6   эквивалентных  схем:  Т-схему  или  П-схему,  которые  уже
учитывают симметрию матриц $\rv  Z$  и  $\rv  Y$.  Это  позволяет  при
анализе  или  синтезе  сложного  СВЧ устройства представлять отдельные
нерегулярности элементами эквивалентной схемы,  что зачастую  упрощает
процесс и делает его более наглядным.

%\end{document}

\newpage
\oddsidemargin=-0.4mm \evensidemargin=-0.4mm
\topmargin=-0.4mm
\headsep=7mm
\textheight=231.875mm
\textwidth=160mm
\mathsurround=2.5pt
\unitlength=1mm
%\begin{document}
%\input{macr.tex}
\thispagestyle{empty}
%\addtocounter{page}{254}
\baselineskip=0.99\normalbaselineskip

\begin{center}\subsubsection*{23. Нерегулярности   в
         волноводах}\end{center}
\vspace*{.5cm}

\markboth{Глава 8.~Нерегулярности в электродинамических структурах}
         {23.~Нерегулярности в волноводах}

\begin{center}\begin{minipage}[c]{0.75\textwidth}
\footnotesize{\parindent=0.5cm
         Индуктивная диафрагма в прямоугольном волноводе. Интегральное
         уравнение  для  поля в окне диафрагмы.  Эквивалентная схема и
         стационарный функционал для проводимости шунта.  Интегральное
         уравнение   для  плотности  тока  на  поверхности  диафрагмы.
         Ступенчатое  изменение   высоты   волновода.   Приближённое
         решение.    Толстая   диафрагма   как   двойная   ступенчатая
         нерегулярность.
}\end{minipage}\end{center}
\vspace*{.5cm}

     Под   нерегулярностью  в  волноводе   принято
понимать  нарушение его однородности вдоль направления
распространения собственных  волн  ---  оси  $z$.   Всё
многообразие   волноводных нерегулярностей  можно  разделить  на
<<тонкие>> и  <<протяжённые>> в зависимости от того, насколько
успевает измениться фаза волны на длине нерегулярности.   Среди
тонких   нерегулярностей   большой   интерес представляют плоские
нерегулярности,   при    которых    нарушение однородности
примыкающих  волноводов происходит в одной плоскости.  К ним
относятся бесконечно тонкие диафрагмы (как математическая  модель
реальной диафрагмы,  толщина которой существенно меньше длины
волны) и скачкообразное изменение поперечных размеров  волновода,
при  котором обязательно возникает  плоский  проводящий  торец
---  примеры таких нерегулярностей и рассматриваются  ниже.  Что
касается протяжённых нерегулярностей,   то   всегда   следует
проанализировать,   нет  ли возможности  разбить  их  на несколько
отдельных,   более   тонких, нерегулярностей  и проводить
дальнейшие  расчёты  на основе теории цепей.  Общий признак
одиночной  нерегулярности  состоит  в  том,  что краевые поля
затухающих  высших  мод  не  простираются  до  соседней
нерегулярности.

     В качестве   простейшего   примера   одиночной  нерегулярности  в
однородном  волноводе  рассмотрим  бесконечно   тонкую   диафрагму   в
прямоугольном волноводе,  представляющую собой две идеально проводящие
полоски,  примыкающие к узким стенкам.  Пусть диафрагма расположена  в
плоскости  $z=0$;  система  координат  и  соответствующие  обозначения
приведены на рис.  23.1.  Будем считать,  что рабочая частота лежит  в
диапазоне  одноволновости,  то есть распространяющейся является только
волна  типа  $H_{10}$.  Нерегулярность  в  этом   случае   описывается
двумерной  матрицей рассеяния $S$;  при вакуумном заполнении волновода
матрица $S$ симметрична ($s_{12}=s_{21}$), а из-за симметрии структуры
$s_{11}=s_{22}$.   Для упрощения  записи у   продольного   волнового  числа
основной  волны $h_{10}=\sqrt{k^2-\pi^2/a^2}$   опустим  индексы,
что   позволит  представить  единственную  отличную  от  нуля  комноненту
электрического поля в падающей на нерегулярность волне в виде
     $$E^{in}_y=\sin{\frac{\pi x} a}\; e^{ihz}\,;\eqno(23.1)$$
в этой формуле опущен  нормировочный  коэффициент  $ik\mu\sqrt{2/ab}$,
присутствующий,  например,  в  (9.8).

     Из структуры падающего поля и цилиндрической симметрии  структуры
относительно  оси  $y$  следует,  что установившееся поле представляет
собой совокупность волн  типа  $H_{n0}$;  в  одноволновом  режиме  все
$h_{n0}$   при  $n\geqslant  2$  ---  чисто  мнимые  величины  и  их  удобно
представить в виде $h_{n0}=ip_n$,  где $p_n=\sqrt{(\pi n/a)^2-k^2}$. С
учётом  всего сказанного полное электрическое поле в структуре может
быть записано в виде:
     $$\left.\begin{array}{lclll}E_y&=&\sin{\displaystyle
         {\frac{\pi x} a}}\; \Bigl(e^{ihz}+s_{11} e^{-ihz}\Bigr)
         +\sum\limits_{n=2}^\infty A_n\sin{\displaystyle
         {\frac{n\pi x} a}}\; e^{p_n z}&\mbox{при}& z<0;\\[.5cm]
     E_y&=&\sin{\displaystyle{\frac{\pi x} a}}\; s_{21}\; e^{ihz}+
         \sum\limits_{n=2}^\infty B_n\;\sin{\displaystyle{\frac
         {n\pi x} a}}\; e^{-p_nz},&\mbox{при}&z>0.\end{array}\right\}
         \eqno(23.2)$$

\begin{wrapfigure}[12]{l}{7.5cm}
\begin{picture}(80,45)
\put(0,45){\special{em:graph fig23-1.bmp}}
\end{picture}
\hbox to 7.5cm{\hfil\footnotesize{Рис.~23.1.~Индуктивная диафрагма.
}\hfil}
\end{wrapfigure}

\vspace{.1cm}
     Требование непрерывности поля в плоскости  диафрагмы  приводит  к
равенствам:
     $$1+s_{11}=s_{21}\;,\qquad A_n=B_n\,.\eqno(23.3)$$
Эти коэффициенты  легко могут быть выражены через электрическое поле в
окне диафрагмы,  которое обозначим как ${\cal E}(x)$.  Домножая (23.2)
на  $\sin{(m\pi x/a)}$ и интегрируя по $x$ от 0 до $a$,  получаем,  с
учётом обращения в нуль поля на самой диафрагме,  что
     $$1+s_{11}=\frac 2 a\;\int\limits_{d_1}^{d_2} {\cal E}(x)\sin
         {\frac{\pi x} a}\,dx,\qquad A_n=\frac 2 a\;\int\limits_{d_1}
         ^{d_2}{\cal E}(x)\sin{\frac{n\pi x} a}\,dx.\eqno(23.4)$$

     Поперечная компонента магнитного поля  в  структуре  $H_x$  легко
находится  из  (23.2)  с  помощью уравнения Максвелла $ikH_x=-\partial
{E_y}/\partial z$.  Из непрерывности этой компоненты в окне  диафрагмы
следует уравнение
     $$h(1-s_{11}-s_{21})\sin{\frac {\pi x} a}=2i\sum\limits_{n=2}
         ^\infty p_n A_n\sin{\frac{\pi n x} a}\qquad\mbox{при}
         \qquad d_1<x<d_2\;,\eqno(23.5)$$
которое с помощью соотношений (23.3) и (23.4) сводится к интегральному
уравнению относительно функции ${\cal E}(x)$:
     $$ s_{11}\sin{\frac{\pi x} a}=-i \frac 2{ha}\sum\limits_{n=2}
          ^\infty p_n\sin{\frac{\pi n x}a}\int\limits_{d_1}^{d_2}
          {\cal E}(x')\sin{\frac{\pi n x'}a}\,dx'\,.\eqno(23.6)$$
С помощью первого соотношения (23.4) это уравнение приводится  к  виду
     $$ i\frac{s_{11}}{1+s_{11}}\sin{\frac {\pi x} a}\int\limits_
         {d_1}^{d_2} {\cal E}(x')\sin{\frac{\pi x'} a}\,dx'=
         \frac 1{h}\sum\limits_{n=2}^\infty p_n\sin{\frac{\pi n x}a}
         \int\limits_{d_1}^{d_2}{\cal E}(x')\sin{\frac{\pi n x'}a}
         \,dx'\,.\eqno(23.7)$$
Домножая правую и левую части на ${\cal E^*}(x)$  (${\cal  E}(x)$  ---
комплексная  функция)  и  интегрируя по $x$ в пределах окна диафрагмы,
получаем, что
     $$ i\frac{ s_{11}}{1+s_{11}}=\frac 1 h\,\frac{\displaystyle
         {\sum\limits_{n=2}^\infty p_n \left |\int\limits_{d_1}^{d_2}
         {\cal E}(x)\sin{\frac{\pi n x} a}\,dx\right|^2}}{\left |
         \displaystyle{\int\limits_{d_1}^{d_2}}{\cal E}(x)\sin{\frac
         {\pi x} a}\,dx\right|^2}\,.\eqno(23.8)$$

     Из анализа этого выражения  ясно,  что  стоящая  справа  величина
является    действительной    и   положительной,   а,   следовательно,
$s_{11}/(1+s_{11})$ --- чисто мнимая и лежащая на отрицательной полуоси
величина,  которую обозначим как $-i{\cal B}$, где ${\cal B}>0$. Кроме
того,  из (23.7)  следует,  что  комплексная  функция  действительного
переменного  ${\cal E}(x)$ может быть представлена в виде произведения
комплексной постоянной на действительную функцию.  Имея это в виду,  в
дальнейшем в выражениях типа (23.8) будем опускать знаки модуля.

     Для выяснения физического  смысла  ${\cal  B}$  найдём  матрицу
сопротивлений  $\rv  Z$  и  построим  эквивалентную  схему.  С помощью
формулы (22.69), учитывая симметрию матрицы рассеяния $\rv S$ и первое
соотношение (23.3), получим, что
     $$\rv Z=-\frac {1+s_{11}}{2s_{11}}\left(\begin {array}{ll}1&1
         \\1&1\end{array}\right)=-\frac i {\cal B} \left(\begin{array}
         {ll}1&1\\1&1\end{array}\right)\,,\eqno(23.9)$$
откуда на   основании   рис.~22.6  следует,  что  эквивалентная  схема
нерегулярности  представляет  собой  одиночный  шунт  с  чисто  мнимой
проводимостью.   Положительный   знак   ${\cal   B}$   (при  временной
зависимости  $e^{-i\omega  t}$)  свидетельствует,  что  шунт  является
индуктивностью, чем и обусловлено название представленной на рис.~23.1
диафрагмы.

     Преобразуем уравнение (23.7), выполнив интегрирование по частям и
учитывая,  что на краях диафрагмы ${\cal  E}(x)$  обращается  в  нуль.
Вводя   вспомогательные   переменные
     $${\cal   F}(x)=\frac   {d{\cal E}(x)}{d\,x},\quad \delta_n=
         1-\sqrt{1-\frac{k^2a^2}{n^2\pi^2}},\quad\theta=\frac
         {\pi x} a,\quad \theta_i=\frac{\pi d_i} a\quad(i=1,2)
         \,,\eqno(23.10)$$
приведём интегральное уравнение к виду
     $$ \frac {a{\cal B}}\Lambda \sin{\theta}\int\limits_{\theta_1}
         ^{\theta_2}{\cal F}(\theta\,')\cos{\theta\,'}\,d\theta\,'
         =\sum\limits_{n=2}^\infty (1-\delta_n)\sin{n\theta}\int
         \limits_{\theta_1}^{\theta_2}{\cal F}(\theta\,')\cos{n\theta\,'}
         \,d\theta\,'\;,\eqno(23.11)$$
где $\Lambda$ --- длина основной волны в волноводе.

     Из этого   однородного   интегрального    уравнения    необходимо
определить  функцию  ${\cal  F}(\theta)$  (а, значит, и ${\cal
E}(x)$, причём из (23.7) следует,  что $\re {\cal E}(x)$ и $\im
{\cal E}(x)$ различаются  только  постоянным  множителем)  и
постоянную ${\cal B}$. Найти решение (23.11) в замкнутом
аналитическом виде не представляется возможным,     однако
удаётся    построить    последовательность приближённых
решений,  используя  быстрое  убывание  с  $n$  величин
$\delta_n$. Самым   грубым   является   {\it   квазистатическое
приближение},  в котором для $n\geqslant 2$ все $\delta_n=0$.
Отметим, что в основе   приближённых   решений   уравнения
(23.11) лежит  замена переменных  $\theta,\;\theta'$  на
$\varphi,\;\varphi'$,   связанных соотношениями
     $$\cos{\theta}=c+s\,\cos{\varphi}\;,\qquad
         \cos{\theta\,'}=c+s\,\cos{\varphi'}\;,\eqno(23.12)$$
где
     $$c=\cos\frac{\theta_1+\theta_2}{2}\,\cos\frac{\theta_1-
         \theta_2}{2}\,,\qquad s=\sin\frac{\theta_1+\theta_2}{2}\,
         \sin\frac{\theta_2-\theta_1}{2}\,.\eqno(23.13)$$
В результате этой замены область интегрирования по $\theta'$ в (23.11)
переходит при интегрировании по $\varphi'$ в  интервал  $(0,\pi),$  на
котором функции $\cos{n\varphi}$ ортогональны.

     Переход в   (23.11)   от   сумм,   содержащих  $\sin{n\theta}$  и
$\cos{n\theta\,'}$,  к суммам,  содержащим  $\sin{n\varphi}$  и
 $\cos{n\varphi'}$, осуществляется на основе соотношения
     $$\sum\limits_{n=1}^{\infty}\sin{n\theta}\,
         \cos{n\theta\,'}=\frac{\sin{\theta}}{s\,\sin{\varphi}}
         \sum\limits_{n=1}^{\infty}
         \sin{n\varphi}\,\cos{n\varphi'}\,,\eqno(23.14)$$
вывод которого производится с помощью известного ряда
     $$\sum\limits_{n=1}^{\infty}\frac 1 n\cos{n\theta}\cos{n\theta\,'}=
         -\frac 1 2 \ln(2\,|\cos{\theta}-\cos{\theta'}|)\,.
         \eqno(23.15)$$
Этот ряд представляет собой  разложение  в  ряд  Тейлора
функции  $\ln{(1-x)}$,  в которое в качестве $x$ подставляются функции
$e^{i(\theta\pm\theta')}$.  Осуществляя в  (23.15)  замену  переменных
(23.12), приходим к соотношению
     $$\sum\limits_{n=1}^{\infty} \frac 1 n \cos{n\theta}\cos{n
         \theta'}=-\frac 1 2 \ln{s}+\sum\limits_{n=1}^{\infty}
         \frac 1 n\cos{n\varphi} \cos{n\varphi'}\,,\eqno(23.16)$$
дифференцирование которого  по  $\varphi$  приводит  к (23.14).  Здесь
нецелесообразно останавливаться на вопросе сходимости рядов в  (23.14)
и допустимости почленного дифференцирование рядов в (23.16), поскольку
получаемые далее с их помощью решения  уравнения  (23.11)  могут  быть
проверены непосредственной подстановкой.

     Замена переменных  (23.12)  и  использование  соотношения  (23.14)
преобразуют уравнение (23.11) к виду
     $$\begin{array}{l}\displaystyle{\Bigl[\Bigl(\frac {a{\cal B}}
         {\Lambda}+1\Bigr)\,s^2-1\Bigr]\,\sin{\varphi}
         \int\limits_0^\pi {\cal F}(\theta\,')\frac{d\theta'}
         {d\varphi'}\cos{\varphi'}\,d\varphi'}=\\[.4cm]
         =\displaystyle{\sum\limits_{n=2}^\infty \sin{n\varphi}
         \int\limits_0^\pi {\cal F}(\theta\,')\frac{d\theta'}
         {d\varphi'}\cos{n \varphi'}\,d\varphi'}-\\[.4cm]
         -\displaystyle{s\,\frac{\sin{\varphi}}{\sin{\theta}}
         \sum\limits_{n=2}^\infty \delta_n\,\sin{n\theta}
         \int\limits_0^\pi {\cal F}(\theta\,')\frac{d\theta\,'}
         {d\varphi'}\cos{n\theta\,'}\,d\varphi'}
         \end{array} \eqno(23.17)$$
для всех точек интервала  $(0,\,\pi)$  переменной  $\varphi$.  Функция
${\cal F}(\theta) $ и постоянная ${\cal B}$ --- помимо этого уравнения
--- должны удовлетворять ещё и уравнению, следующему из формул (23.4):
     $$ \frac 1 {1+i {\cal B}/2}=\frac {2 s}{\pi}
         \int\limits_0^\pi {\cal F}(\theta\,')\frac{d\theta'}
         {d\varphi'}\cos{\varphi'}\,d\varphi'\,.\eqno(23.18)$$

     Решение подобных  интегральных  уравнений  целесообразно искать в
виде ряда
     $${\cal F}(\theta)\frac {d\theta}{d\varphi}=\sum\limits_{m=1}
         C_m\cos{m\varphi},\eqno(23.19)$$
причём электрическое   поле   в   окне   диафрагмы   в  этом  случае
представляется как
     $${\cal E}(\theta)=\sum\limits_{m=1}\frac1 m C_m\sin{m\varphi}\,,
         \eqno(23.20)$$
что обеcпечивает  требуемое  обращение поля в нуль на краях диафрагмы.
Уравнение  (23.18)  сводится  тогда  к  следующему  соотношению  между
постоянными $C_1$ и~${\cal B}$:
     $$\frac 1 {1+i{\cal B}/2}=sC_1\,.\eqno(23.21)$$

     Точного решения уравнения (23.17) в виде ряда (23.19),  в котором
все  коэффициенты  $C_m$  имеют  замкнутое  аналитическое   выражение,
получить  не  удаётся,  и  приходится  ограничиваться приближениями,
учитывающими конечное число быстро убывающих с номером  малых  величин
$\delta_n$. Возможность аналитического выражения коэффициентов $C_m$ в
каждом таком приближении в замкнутом виде связана с тем, что в рядах
     $$ \cos{n\theta}=\sum\limits_{m=1}\alpha_{mn}\cos{m\varphi}
         \quad\mbox{и}\quad \frac {\sin{n\theta}}{\sin{\theta}}
         =\frac1{\sin{\varphi}}\sum\limits_{m=1}\beta_{mn}\sin{m
         \varphi}\eqno(23.22)$$
коэффициенты $\alpha_{mn}$  и  $\beta_{mn}$ отличны от нуля только при
$m\leqslant n$.  В результате максимальный номер отличного от нуля  $C_m$  в
(23.19) совпадает с номером последнего учтённого $\delta_n$.

     В простейшем квазистатическом приближении все $\delta_n$, начиная
с $n=2$, полагаются нулями, так что
     $${\cal F}(\theta)\frac{d\theta}{d\varphi}=C_1\,
                 \cos{\varphi}\eqno(23.23)$$
и
     $${\cal B}=\frac{\Lambda}a \Bigl(\frac 1{s^2}-1\Bigr)\,.
         \eqno(23.24)$$
C учётом  соотношения  (23.21) находим распределение поля в раскрыве
диафрагмы:
     $$ {\cal E}(x)=\frac i {i-{\cal B}/2}\sqrt{\sin^2{\frac
         {\pi x_0}a}\sin^2{\frac{\pi d}{2 a}}-\Bigl(\cos{\frac{\pi x}
         a}-\cos{\frac{\pi x_0}a}\,\cos{\frac{\pi d}{2a}}\Bigr)^2}\,,
         \eqno(23.25)$$
где $x_0=(d_1+d_2)/2,\quad d=d_2-d_1$ (см. рис.~23.1).

     Последующие приближения  строятся  аналогичным   способом.   Так,
оставляя  в  последней сумме уравнения (23.17) слагаемое с $\delta_2$,
используя  получаемые   при   помощи   простейших   тригонометрических
преобразований равенства
     $$ \cos{2\theta'}=s^2\,\cos{2\varphi'}+4cs\,\cos{\varphi'}+
         2c^2+s^2-1\eqno(23.26)$$
и
     $$ \frac{\sin{2\theta}}{\sin{\theta}}=\frac{s\,\sin{2\varphi}+2c
         \,\sin{\varphi}}{\sin{\varphi}}\eqno(23.27)$$
и приравнивая нулю множители перед $\sin{\varphi}$ и $\sin{2\varphi}$,
приходим к системе уравнений для коэффициентов $C_1,\,C_2$ и $B$:
     $$\left.\begin{array}{l}
         \displaystyle{\Bigl[\Bigl(\frac{aB} \Lambda+1\Bigr]s^2-
         1\Bigr]C_1+ 2cs\delta_2(s^2C_2+4csC_1)=0,}\\[.4cm]
         \displaystyle{C_2-s^2\delta_2(s^2C_2+4csC_1)=0}\,.\end{array}
         \right\}\eqno(23.28)$$
В результате получаем уточнённое значение для ${\cal B}$:
     $$ {\cal B}=\frac \Lambda a \Bigl(\frac 1{s^2}-1-\frac{8c^2
         \delta_2}{1-s^4\delta_2}\Bigr)\,,\eqno(23.29)$$
причём в  случае  симметричного  расположения образующих препятствие
полосок, когда $c=0$, формулы (23.29) и (23.24) совпадают.

     Последующие приближения  быстро  становятся  очень  громоздкими и
практического применения не находят.  Однако важно помнить,  что таким
способом  могут  быть учтены все распространяющиеся волны при работе в
многоволновом режиме,  когда размерность матрицы рассеяния  необходимо
увеличить.

     Следует сказать,  что  прямое  решение  интегрального  уравнения,
описывающего нерегулярность в волноводе, даже путём последовательных
приближений удаётся провести только в  простейших  случаях.  Широкое
применение   находят   поэтому   {\it  вариационные  методы}  решения.
Воспользуемся   таким   подходом   для   рассматриваемой   индуктивной
диафрагмы.   Докажем,  что  следующее  из  (23.4)  с  учётом  замены
переменных (23.10), (23.12) выражение для сопротивления эквивалентного
шунта
     $$ {\cal B}=\frac{\Lambda} a \frac{\displaystyle
         {\sum\limits_{n=2}^\infty (1-\delta_n) \left|\int
         \limits_0^{\pi}{\cal F}(\theta)\frac{d\theta}{d\varphi}\,
         \cos{n \theta} \,d\varphi\right|^2}} {\displaystyle{s^2
         \left|\int\limits_0^{\pi}{\cal F}(\theta)\frac{d\theta}{d
         \varphi}\,\cos{\varphi} \,d\varphi\right|^2}}\eqno(23.30)$$
представляет собой стационарный  функционал,  принимающий  минимальное
значение  для  функции  ${\cal  F}(\theta)$,  являющейся  {\it точным}
решением уравнения (23.17).  Следовательно,  при любой пробной функции
${\cal   F}(\theta)$   выражение   (23.30)   даёт   верхнюю   оценку
проводимости шунта.

     Чтобы сделать доказательство наиболее общим,  рассмотрим типичное
интегральное уравнение
     $$ \lambda \psi_0(x)\int F(x') \psi_0(x')\,dx'=
         \sum\limits_{n=1}^\infty\psi_n(x)\int F(x')\psi_n(x')\,dx'\,,
         \eqno(23.31)$$
к которому часто  сводятся  задачи  о  ступенчатых  нерегулярностях  в
волноводах.  Под  $x$  в  нём  понимается  совокупность  независимых
переменных,  представляющих  поперечные  координаты  в  сечении,   где
расположена    нерегулярность,   $\psi_n$   ---   известные   функции,
определяющие  собственные  волны  волновода,  $F(x)$  ---  неизвестная
функция,   соответствующая   полю  в  окне  или  току  на  поверхности
диафрагмы, $\lambda$ --- собственное значение, однозначно определяемое
самим  интегральным уравнением.  Величина $\lambda$ представляет собой
либо проводимость,  либо сопротивление элемента эквивалентной схемы. В
большинстве   случаев  именно  эта  величина,  а  не  функция  $F(x)$,
представляет основной интерес.  Отметим,  что  $F(x)$  определяется  с
точностью   до   произвольного   множителя,  который  находится  из
дополнительного    соотношения     между     $\lambda$     и     $\int
F(x)\psi_0(x)\,dx$.

     С помощью  несложной  процедуры,   описанной   выше,   представим
$\lambda$ через функционал:
     $$\lambda=\frac{\displaystyle{\sum\limits_{n=1}^\infty \biggl
         [\int F(x)\psi_n(x)\,dx\biggr]^2}}{\displaystyle{\biggl
         [\int F(x)\psi_0(x)\,dx\biggr]^2}}\,.\eqno(23.32)$$
Докажем, что   $\lambda$  не  только  стационарно  относительно  малых
вариаций $F(x)$ вблизи истинного решения  $F_0$,  но  и  достигает  на
нём   абсолютного   минимума.   Действительно,   пусть   $F(x)$  ---
произвольная пробная функция; рассмотрим тогда неравенство
     $$\sum_{n=1}^\infty\left\{\int \psi_n(x)\left[\frac {F(x)}
         {\displaystyle{\int F(x)\psi_0(x)\,dx}}-\frac{F_0(x)}
         {\displaystyle{\int F_0(x)\psi_0(x)\,dx}}\right]\,dx\right\}
         ^2\geqslant 0\,,\eqno(23.33)$$
которое после возведения в квадрат и опускания аргументов  у  функций,
входящих в подинтегральные выражения, принимает вид
     $$\sum\limits_{n=1}^\infty\left[\frac{\displaystyle{\int F
         \psi_n\,dx}}{\displaystyle{\int F\psi_0\,dx}}\right]^2+\sum
         \limits_{n=1}^\infty\left[\frac{\displaystyle{\int F_0\psi_n
         \,dx}}{\displaystyle{\int F_0\psi_0\,dx}}\right]^2-
         \frac{\displaystyle{2\sum\limits_{n=1}^\infty\int F
         \psi_n\,dx\int F_0\psi_n\,dx}}{\displaystyle{\int F\psi_0
         \,dx\int F_0\psi_0\,dx}}\geqslant 0.\eqno(23.34)$$

     Изменив в последнем слагаемом в левой части порядок  суммирования
и интегрорования, с учётом (23.32) получим:
     $$  \lambda +\lambda_0 -\frac{\displaystyle{2\int F(x)\,dx
         \sum\limits_{n=1}^\infty \psi_n(x)\int F_0(x')\psi_n(x')
         \,dx'}}{\displaystyle{\int F(x)\psi_0\,dx\int F_0(x)\psi_0(x)
         \,dx}}\geqslant 0\,\eqno(23.35)$$
и поскольку $F_0$ представляет собой решение уравнения (23.31), то
      $$ \lambda-\lambda_0 \geqslant 0\,.\eqno(23.36)$$
Знак равенства  имеет  место  только  в  том  случае,   когда   $F(x)$
отличается  от  $F_0(x)$ лишь на постоянный множитель.  Таким образом,
искомое  поле  в  окне   диафрагмы   может   быть   найдено   методами
вариационного исчисления как минимизирующее функционал (23.30).

     Ценность стационарных  функционалов  для   практики   существенно
повышается   в   том   случае,   когда   удаётся  получить  ещё  и
альтернативное  представление   параметров   нерегулярности   в   виде
функционала, принимающего для истинного решения максимальное значение,
что  позволяет  легко  получить  верхнюю  и  нижнюю  оценку   значения
параметра.  Для  индуктивной диафрагмы в прямоугольном волноводе такой
функционал  находится  без  особого  труда  на  основе   интегрального
уравнения  для  плотности тока на поверхности диафрагмы.  С этой целью
выпишем с учётом равенств (23.3)  поперечную  компоненту  магнитного
поля, соответствующую электрическому полю (23.2):
     $$\left.\begin{array}{lclll}H_x&\!\!=\!\!&-\displaystyle
         {\frac h k}\sin{\displaystyle {\frac{\pi x} a}}\Bigl(e^{ihz}
         -s_{11}e^{-ihz}\Bigr)+\displaystyle{\frac i k}\sum\limits_
         {n=2}^\infty A_n\,p_n\sin{\displaystyle{\frac{n\pi x} a}}e^
         {p_n z}&\mbox{при}&z<0;\\[.5cm]H_x&\!\!=\!\!&-\displaystyle
         {\frac h k}\sin{\displaystyle{\frac{\pi x}a}}s_{21}e^{ihz}-
         \displaystyle{\frac i k}\sum\limits_{n=2}^\infty A_n\,p_n\sin
         {\displaystyle{\frac{n\pi x}a}}e^{-p_nz}&\mbox{при}&z>0.
         \end{array}\right\}\eqno(23.37)$$

     Плотность тока  на   диафрагме   $ I_y(x)$   определяется
граничным условием на скачок магнитного поля:
     $$ \frac {4\pi} c I_y(x)=\left.H_x(x,z)\right|_{z=+0}-
         \left.H_x(x,z)\right|_{z=-0}\,,\eqno(23.38)$$
что позволяет записать е\"е в виде
     $$ I_y(x)=-\frac{h c}{2\pi k}s_{11}\,\sin{\frac{\pi x} a}-
         \frac {i c}{2\pi k}\sum_{n=2} p_n A_n\sin{\frac{\pi n x} a}\,.
         \eqno(23.39)$$
Используя далее ортогональность синусов, находим, что
     $$  s_{11}=-\frac{4\pi k}{c h a}\int\limits_{{\cal D}}
          I_y(x)\sin{\frac {\pi x} a}\,dx \,,\qquad
         A_n=i\frac{4\pi k}{ac p_n}\int\limits_{{\cal D}}
         I_y(x)\sin{\frac {\pi n x} a}\,dx\,.\eqno(23.40)$$

     Из равенства нулю компоненты $E_y$ на диафрагме следует,  что для
всех  $x$  в  интервалах  $(0,d_1)$  и $(d_2,a)$ плотность тока $I_y(x)$
удовлетворяет интегральному уравнению
     $$ \sin{\frac{\pi x} a}\,(1+s_{11}) + i\frac{4\pi k}{ac}\sum
         \limits_{n=2}\frac 1 {p_n}\sin{\frac{\pi n x}a}\int
         \limits_{\cal D}I_y(x')\sin{\frac{\pi n x'} a}\,dx'
         =0\,,\eqno(23.41)$$
где ${\cal D}=(0,d_1)\bigcup(d_2,a)$. Домножая это уравнение на $I_y(x)$
и интегрируя по  поверхности  диафрагмы,  получаем  с  учетом
первого соотношения (23.40) выражение для стационарного функционала
     $$ \frac 1 {{\cal B}} = \frac h 2 \displaystyle{\frac
         {\displaystyle{\sum\limits_{n=2}}\displaystyle{\frac 1{g_n}}
         \left[\int\limits_{\cal D} I_y(x')\displaystyle{\sin
         {\frac{\pi n x'} a}}\,dx'\right]^2}{\left[\displaystyle{\int
         \limits_{\cal D} I_y(x')\sin{\displaystyle{\frac
         {\pi x'} a}}\,dx'}\right]^2}}\,,\eqno(23.42)$$
позволяющее получить  нижнюю  оценку  проводимости   ${\cal   B}$.   В
заключение  рассмотрения  индуктивной  диафрагмы следует сказать,  что
помимо знака ${\cal B}$ имеется ещё одно дополнительное соображение,
в пользу используемой терминологии: в запасённой энергии поля высших
типов волн вблизи диафрагмы преобладает энергия магнитного поля.

В качестве примера  нерегулярности   {\it   ёмкостного}   типа
рассмотрим   скачкообразное    изменение    размера    узкой    стенки
прямоугольного  волновода.  Оказывается,  что  при  дифракции основной
волны   $H_{10}$   на   нерегулярностях,   обладающих   цилиндрической
симметрией  вдоль оси $x$,  достаточно решить задачу о дифракции волны
типа $TEM$ на стыке  двух  плоских  волноводов.  Действительно,  ввиду
указанной   симметрии  полное  поле  в  ней  может  быть  представлено
совокупностью продольных (или смешанных) волн двух типов ---  волн,  у
которых  $E_x\equiv  0$  и  волн,  у которых $H_x\equiv 0$.  Поскольку
падающая волна $H_{10}$ (классификация относительно оси $z$) относится
к  первому  типу,  то  можно  ограничиться  только им.  Поля этих волн
определяются  магнитным  вектором  Герца  с  одной  отличной  от  нуля
компонентой --- $\Pi_x^{m}$, которая может быть представлена в виде
$$ \Pi^{m}_x=C\Psi(y,z)\,\sin{\frac {\pi x}a}\,,\eqno(23.43)$$
где функция $\Psi(y,z)$ удовлетворяет волновому уравнению
$$(\Delta_2+h^2)\Psi(y,z)=0\,,\qquad h=\sqrt{k^2-\Bigl(\frac
     {\pi}a\Bigr)^2}\,,\eqno(23.44)$$
а $C$ --- нормировочный множитель, который опускается в двух последующих
формулах.
Тогда компоненты поля согласно (3.34) выражаются следующим образом:
$$\left.\begin{array}{lcl}
      E_y&=&ik\displaystyle{\frac{\partial
     \Psi}{\partial z} \sin{\frac{\pi x}a}}\,,\\[.4cm]
      E_z&=&-ik\displaystyle{\frac{\partial
     \Psi}{\partial y} \sin{\frac{\pi x}a}}\,,\\[.4cm]
      H_x&=&h^2\displaystyle{\Psi\sin{\frac{\pi x}a}},\\[.4cm]
      H_y&=&\displaystyle{\frac {\pi}a\,\frac{\partial
     \Psi}{\partial y} \cos{\frac{\pi x}a}}\,,\\[.4cm]
      H_z&=&\displaystyle{\frac {\pi}a\frac{\partial \Psi}
     {\partial z} \cos{\frac{\pi x}a}}\,. \end{array}\right\}
     \eqno(23.45)$$

   Граничное условие $E_y=0$ на поверхности стыка волноводов в
виду цилиндрической симметрии будет выполнено при всех $x$ в
интервале $(0,a)$, если оно выполнено при каком-нибудь одном.
В качестве такого удобно выбрать $x=a/2$; в этой плоскости отличны
от нуля компоненты поля
$$\left.\begin{array}{lcl}
      E_y&=&ik\displaystyle{\frac{\partial
     \Psi}{\partial z}}, \\[.4cm]
      E_z&=&-ik\displaystyle{\frac{\partial
     \Psi}{\partial y}}, \\[.4cm]
      H_x&=&h^2\Psi. \end{array}\right\}
     \eqno(23.46)$$

     Эти  выражения для  полей  совпадают  с соответствующими
выражениями для  плоского волновода, неограниченного вдоль оси $x$,
если в них заменить $h$ на $k$  (это  замечание справедливо и в случае
уравнения Гельмгольца  (23.44)). Поэтому для решения любой задачи
о дифракции волны $H_{10}$ на металлической неоднородности в
прямоугольном волноводе, сохраняющей однородность вдоль $x$,
достаточно рассмотреть дифракцию $TEM$-волны в  волновод из
параллельных пластин с тем же поперечным сечением и в полученных
формулах заменить $k$ на $h$.

     Введём обозначения размеров и нумерацию однородных областей, на
которую разбивается рассматриваемая структура, согласно рис.~23.2. Все
величины  будем  помечать верхним индексом,  соответствующим нумерации
областей. Произвольные решения уравнения (23.44) могут быть записаны в
виде
     $$ \Psi^{(i)}= A_0^{(i)}e^{ih z}+B_0^{(i)}e^{-ih z}+
         \sum\limits_{m=1}^{\infty}[A_m^{(i)}e^{-p_m^{(i)} z}+
         B_m^{(i)}e^{p_m^{(i)} z}]\cos\frac{m\pi y}{b_i}\,,
         \qquad i=1,2\;,\eqno(23.47)$$
где $p_m^{(i)}=\sqrt{(\pi  m/b_i)^2-h^2}$.  В  одноволновом  режиме,
который  далее  только  и  рассматривается,  $A_m^{(1)}$ и $B_m^{(2)}$
равны нулю ($m\geqslant 1$), а все $p_m^{(i)}$-- действительные положительные
величины.

\begin{wrapfigure}[14]{l}{7.5cm}
\begin{picture}(80,55)
\put(0,50){\special{em:graph fig23-2.bmp}}
\end{picture}
\hbox to 7.5cm{\hfil\footnotesize{Рис.~23.2.~Ступенчатая
нерегулярность.
}\hfil}
\end{wrapfigure}

     Множитель $C$  выбирается  таким  образом,  чтобы  средний  поток
мощности  основной  волны  с  амплитудой  $A_0^{(i)}$ в соответствии с
(22.11) составлял
     $$\overline{\Sigma}_0^{(i)}=\frac 1 2 |A_0^{(i)}|^2\,,
         \eqno(23.48)$$
что достигается при
     $$C_i=\frac 1 h\sqrt{\frac {8\pi}{cab_ih k}}\,.
         \eqno(23.49)$$
С учётом  этого  множителя условия непрерывности тангенциальных
составляющих полей в плоскости $z=0$ и обращения в нуль  электрической
составляющей на металле имеют вид
     $$\frac1{\sqrt{b_1}}\frac{\partial\Psi^{(1)}}{\partial z}=\left
         \{\begin{array}{l}0\qquad\qquad\quad\quad\mbox{при}\qquad
         b_2<y<b_1,\\[0.25cm]\displaystyle{\frac1{\sqrt{b_2}}\frac
         {\partial\Psi^{(2)}}{\partial z}}\qquad\mbox{при}\qquad0<y
         <b_2\end{array}\right.\,\eqno(23.50)$$
и
     $$  \frac 1 {\sqrt{b_1}}\Psi^{(1)}=\frac 1{\sqrt{b_2}}\Psi^{(2)}
         \qquad\mbox{при}\qquad 0<y<b_2\,.\eqno(23.51)$$
\vspace{.3cm}

     Домножая условия  (23.50)  и  (23.51) на $\cos{(\pi n y/ b_1)}$ и
$\cos{(\pi n y/b_2)}$ и интегрируя по $y$ от 0  до  $b_1$  или  $b_2$,
получим  после  несложных  преобразований  следующие соотношения между
коэффициентами:
     $$ A_0^{(1)}-B_0^{(1)}=\sqrt{\alpha}(A_0^{(2)}-B_0^{(2)})\,,
         \eqno(23.52)$$
     $$ B_n^{(1)}=\frac 2{\pi p_n^{(1)}\sqrt{\alpha}}\sin{\pi n\alpha}
         \Bigl[\frac{ih} n (A_0^{(2)}-B_0^{(2)})-n\sum\limits_{m=1}^
         {\infty} \frac {(-1)^m p_m^{(2)} A_m^{(2)}}{n^2 -
         \displaystyle{\frac{m^2}{\alpha^2}}}\Bigr],\quad n\geqslant 1,
         \eqno(23.53)$$
     $$ A_0^{(1)}+B_0^{(1)}+\frac 1 {\pi\alpha}\sum\limits_{m=1}^
         {\infty}\frac{B_m^{(1)}}m \sin{\pi m \alpha}=
         \frac 1 {\sqrt{\alpha}}(A_0^{(2)}+B_0^{(2)})\,,\eqno(23.54)$$
     $$  A_n^{(2)}=\frac{2(-1)^n}{\pi \sqrt{\alpha}}\sum\limits_{m=1}
         ^{\infty}B_m^{(1)}\frac{m\,\sin{\pi m \alpha}}{m^2-
         \displaystyle{\frac{n^2}{\alpha^2}}},\quad n\geqslant 1,
         \eqno(23.55)$$
в которых введено обозначение $\alpha= {b_2}/{b_1}$.

     Исключим теперь  коэффициенты  $B_m^{(1)}$; для упрощения записи
удобно  при  этом  ввести   вспомогательные   переменные   $\sigma_0$,
$\sigma_n$ и $\tilde A_n$ следующим образом:
     $$\sigma_0=\frac{4}{\pi^2\alpha}\sum\limits_{s=1}
         \frac{\sin^2{\pi s \alpha}}{p_s^{(1)}\,s^2}\,,\qquad
         \sigma_n=\frac{4 \sqrt{p_n^{(2)}}}{\pi^2\alpha}\sum\limits_
         {s=1}\frac{\sin^2{\pi s \alpha}}{p_s^{(1)}(s^2-\displaystyle{
         \frac{n^2}{\alpha^2})}}\,,\qquad (n=1,2\dots)\eqno(23.56)$$
     $$\sqrt{p_n^{(2)}}A_n^{(2)}=i(-1)^n h (A_0^{(2)}-B_0^{(2)})
         \tilde A_n\,.\eqno(23.57)$$
В результате для коэффициентов  $\tilde  A_n$  получается  бесконечная
система линейных алгебраических уравнений
     $$ \sum\limits_{m=1}^{\infty}\tilde A_m M_{nm}=\sigma_n\,,
         \eqno(23.58)$$
где
     $$M_{nm}=\left\{\begin{array}{l}\displaystyle{\frac{ n^2\sigma_n
         \sqrt{p_m^{(2)}}-m^2\sigma_m\sqrt{p_n^{(2)}}}{n^2-m^2}}\qquad
         \mbox{при}\qquad m\ne n,\\[.4cm]\displaystyle{1+\sqrt
         {p_n^{(2)}}\bigl(\sigma_n+\frac{n^2}{\alpha^2}\Sigma_n\bigr)}
         \qquad\quad\mbox{при}\qquad m=n,\end{array}\right.\,
         \eqno(23.59)$$
а
     $$\Sigma_n=\frac{4\sqrt{p_n^{(2)}}}{\pi^2\alpha}\sum\limits_{s=1}
         \frac{\sin^2{\pi s \alpha}}{p_s^{(1)}(s^2-\displaystyle{
         \frac{n^2}{\alpha^2})^2}}\,.\eqno(23.60)$$

     Формальное решение системы (23.58)
     $$\tilde A_n=\sum\limits_{m=1}^{\infty} M_{nm}^{-1}\,\sigma_m
         \eqno(23.61)$$
позволяет записать  соотношение (23.54) после исключения коэффициентов
$B_m$ в виде
     $$\sqrt{\alpha}(A_0^{(1)}+B_0^{(1)})=A_0^{(2)}+B_0^{(2)}+i
         {\cal B}(A_0^{(2)}-B_0^{(2)})\,,\eqno(23.62)$$
где
     $${\cal B}=-\frac{h} 2 (\sigma_0-\sum_{m=1}^{\infty}\sigma_m
         \sum\limits_{n=1}^{\infty}M_{mn}^{-1}\sigma_n)\,.
         \eqno(23.63)$$

     Таким образом, в  одноволновом  режиме  все   электродинамические
свойства  рассматриваемой  структуры  определяются одной величиной ---
$\cal B$,  через которую нетрудно выразить матрицу рассеяния $\rv  S$.  Для
этого  вспомним,  что  при  использованной нормировке (22.11) элементы
матрицы рассеяния и коэффициенты  $A_0^{(i)}$  и  $B_0^{(i)}$  связаны
следующими соотношениями:

если
     $$ A_0^{(1)}=1\quad\mbox{и}\quad B_0^{(2)}=0\,,\qquad\mbox{то}
         \qquad s_{11}=-B_0^{(1)}\quad\mbox{и}\quad s_{12}=A_0^{(2)}
         \,;\eqno (23.64)$$

если
     $$ A_0^{(1)}=0\quad\mbox{и}\quad B_0^{(2)}=1\,,\qquad\mbox{то}
         \qquad s_{22}=-A_0^{(2)}\quad\mbox{и}\quad s_{21}=B_0
         ^{(1)}\,.\eqno (23.65)$$
В результате с учётом (23.52) и~(23.62) получаем следующее выражение
для  матрицы  рассеяния:
     $$\rv S=\frac1{1+\alpha+i{\cal B}}\left(\begin{array}{cc}
         \alpha-1-i{\cal B}&2\sqrt{\alpha}\\[.2cm]2\sqrt{\alpha}&1-
         \alpha-i{\cal B}\end{array}\right) \,.\eqno(23.66)$$
Как и должно быть в структурах с изотропным заполнением,  матрица $\rv
S$ симметрична.  Поскольку потерь нет (идеальная проводимость стенок),
то матрица и унитарна,  то есть в данном случае $\rv S^{-1}= \rv S^*$,
в чём легко убедиться непосредственным вычислением матрицы, обратной
матрице (23.66).

     Вычислим теперь  матрицу  сопротивлений  $\rv Z$,  основываясь на
соотношении   (22.69).   Опуская   несложные   выкладки,    приведём
окончательный результат:
     $$\rv Z=-\frac i{{\cal B}}\left(\begin {array}{cc}\alpha& \sqrt
         {\alpha}\\[.2 cm]\sqrt{\alpha}&1\end{array}\right)\,.
         \eqno(23.67)$$
Полученной матрице соответствует  Т-образная  эквивалентная  схема
(рис.~23.3,  сплошные  линии),  параметры  которой  находятся  с  помощью
обозначений на рис~22.6a:
     $$Z_1=-\frac i {\cal B} \sqrt{\alpha}\,;\qquad Z_2=-\frac i{{
         \cal B}}(\alpha-\sqrt{\alpha}),\qquad Z_3=-\frac i{{\cal B}}
         (1-\sqrt{\alpha})\,.\eqno (23.68)$$

\vspace{.2cm}
\begin{wrapfigure}[14]{l}{7.3cm}
\begin{picture}(80,50)
\put(-5,50){\special{em:graph fig23-3.bmp}}
\end{picture}
\hbox to 7.3cm{\hfil\footnotesize{Рис.~23.3.~Эквивалентная схема.
}\hfil}
\end{wrapfigure}

     Теперь нетрудно по обычным электротехническим правилам рассчитать
эквивалентную  нагрузку  левой  линии  в  плоскости  $z=0$.  Для этого
необходимо  включить  в  схему  дополнительное  сопротивление   $Z_4$,
показанное  на рис.~23.3 штриховыми линиями,  которое учитывает правый
волновод.  В используемой здесь нормировке,  при которой сопротивления
всех  элементов  эквивалентной  схемы  отнесены  к характеристическому
сопротивлению волновода,  $Z_4=1$.  Отметим, что во всех рассмотренных
до  сих пор нерегулярностях характеристическое сопротивление волновода
справа и слева от нерегулярности одно и  то  же;  для  этого  частного
случая выписаны и формулы (22.69). Обобщение их на четырёхполюсник с
разными сопротивлениями правой и левой линий не вызывает сколько-нибудь
серьёзных   принципиальных   осложнений,   но   делает  все  формулы
значительно более громоздкими,  и  останавливаться  на  них  не  имеет
смысла.

     Используя правило  сложения  сопротивлений,   находим   некоторое
эквивалентное  (нормированное)  сопротивление  нагрузки,  которая  для
левого волновода заменяет собой нерегулярность и правый волновод:
     $$Z_{1\mbox{\footnotesize\textit{экв}}}=
        Z_2+\frac{ Z_1(Z_3+1)}{Z_1+Z_3 +1}=\frac {\alpha}
         {1+i{{\cal B}}}\,.\eqno(23.69)$$
Аналогичное выражение  для  сопротивления  нагрузки  правого волновода
имеет вид
     $$Z_{2\mbox{\footnotesize\textit{экв}}}=
       Z_3+\frac{ Z_1(Z_2+1)}{Z_1+Z_2 +1}=\frac 1
         {\alpha+i{\cal B}}\,.\eqno(23.70)$$
Вычисляя коэффициент  отражения  $\Gamma_0$  для  обоих  волноводов  по
формуле двухполюсника (см.,  например,  (22.63)  в  плоскости  $z=0$),
имеющей вид
     $$\Gamma_0=\frac {Z_{i\mbox{\footnotesize\textit{экв}}}-1}
       {Z_{i\mbox{\footnotesize\textit{экв}}}+1}\,,\eqno(23.71)$$
убеждаемся, что,   как   и   должно   быть,   $\Gamma_0$  совпадает  с
соответствующим элементом $s_{ii}$ матрицы рассеяния (23.66).

    Из сказанного  может  сложиться  впечатление,  что  само  введение
матрицы  сопротивлений и эквивалентной схемы не даёт ничего нового и
весь этот аппарат несколько устарел.  Однако  чуть
дальше  можно будет убедиться,  что это не так.  В случае двух и более
последовательных нерегулярностей в волноводном тракте  он  оказывается
очень   удобным   при   расчёте  коэффициента  отражения,  если  эти
нерегулярности можно считать невзаимодействующими  между  собой  через
поля   нераспространяющихся   высших   типов  волн.  Получить  тот  же
результат,  используя   отдельные   матрицы   рассеяния   для   каждой
нерегулярности, значительно сложнее.

     Прежде чем  переходить к двойным нерегулярностям,  сопоставим два
рассмотренных выше метода решения  задачи  о  дифракции  на  одиночной
нерегулярности.  Выведем  для  этой  цели  интегральное  уравнение для
электрического поля в  плоскости  стыка  волноводов  разной  высоты  и
найдём   стационарный   функционал   для   величины   $\cal   B$.  В
непосредственной  близости  от  плоскости  $z=0$  справа  и  слева   в
соответствии  с  (23.45)  и  (23.47)  электрическое  поле  может  быть
записано в виде
     $$E^{(1)}_y=ikC_1\Bigl[ih(A_0^{(1)}-B^{(1)}_0)+\sum
         \limits_{m=1}^{\infty}p_m^{(1)}B_m^{(1)}\cos{\frac{\pi m y}
         {b_1}}\Bigr ]=\left \{ \begin{array}{l} 0\;\;\mbox{при $b_2
         <y<b_1$},\\{\cal E}(y)\;\;\mbox{при $y<b_2$},\end{array}
         \right.\eqno(23.72)$$
     $$E^{(2)}_y=ikC_2\Bigl[ih(A_0^{(2)}-B^{(2)}_0)-\sum\limits_
         {m=1}^{\infty}p_m^{(1)}A_m^{(2)}\cos{\frac{\pi m y}{b_2}}
         \Bigr ]={\cal E}(y)\;\;\mbox{при $y<b_2$}\,,\eqno(23.73)$$
где ${\cal  E}(y)$  --- распределение поля в плоскости,  общей для двух
волноводов.

     Используя ортогональность  косинусов,   нетрудно   выразить   все
коэффициенты через интегралы от функции ${\cal E}(y)$:
     $$\int\limits_0^{b_2}{\cal E}(y)\,dy=-kC_1h b_1 (A_0^{(1)}
         -B_0^{(1)})=-kC_2h b_2 (A_0^{(2)}-B_0^{(2)})\,,
         \eqno(23.74)$$
     $$\int\limits_0^{b_2} {\cal E}(y)\cos{\frac{\pi m y} {b_1}}\,dy=
         ikC_1 b_1 p_m^{(1)} \frac{B_m^{(1)}}2\,,\eqno(23.75)$$
     $$\int\limits_0^{b_2} {\cal E}(y)\cos{\frac{\pi m y} {b_2}}\,dy=
         - ikC_2 b_2 p_m^{(2)} \frac{A_m^{(2)}}2\,.\eqno(23.76)$$

     Условие непрерывности  (23.51)  магнитного поля в плоскости $z=0$
при $0<y<b_2$ с учётом (23.62) и (23.74)  сводится  к  интегральному
уравнению
     $$ \begin{array}{c} \displaystyle{\frac {{\cal B}} {2h}\int
         \limits_0^{b_2}{\cal E}(y')\,dy'+\alpha
         \sum\limits_{m=1}^{\infty}\frac 1{p_m^{(1)}}\cos{\frac
         {\pi m y}{b_1}} \,\int\limits_0^{b_2}{\cal E}(y')\cos{\frac
         {\pi m y'}{b_1}}\,dy'}+\\[.4 cm]+\displaystyle{\sum\limits_
         {m=1}^{\infty}\frac 1{p_m^{(2)}}\cos{\frac{\pi m y}{b_2}}
         \int \limits_0^{b_2}{\cal E}(y')\cos{\frac{\pi m y'}{b_2}}
         \,dy'=0}\,.\end{array}\eqno(23.77)$$
Умножая это уравнение на ${\cal E}^*(y)$ и интегрируя по $y$ от  0  до
$b_2$,    находим    стационарный   функционал   основного   параметра
нерегулярности
     $$ {\cal B}= - \frac{\displaystyle{2h\sum\limits_{m=1}^{\infty}
         \left\{\frac \alpha{p_m^{(1)}}\left|\int\limits_0^{b_2}
         {\cal E}(y)\cos{\frac{\pi m y}{b_1}}\,dy\right|^2+ \frac 1
         {p_m^{(2)}}\left|\int\limits_0^{b_2}{\cal E}(y)\cos{\frac
         {\pi m y}{b_2}}\,dy\right|^2\right\}}}{\displaystyle{\left|
         \int\limits_0^{b_2}{\cal E}(y)\,dy\right|^2}}\,.
         \eqno(23.78)$$

     Уравнение (23.77)   может   быть   решено   в    квазистатическом
приближении  с  использованием  преобразования типа (23.12) и уточнено
последовательными приближениями во многом аналогично  тому,  как  выше
было решено уравнение (23.7). Точно так же --- на основе стационарного
функционала (23.78) --- вариационными методами может быть  определён
параметр   ${\cal   B}$.  Результаты  такого  решения  имеют  довольно
громоздкий вид,  не очень высокую точность и здесь не  приводятся.  Во
всяком случае, они уступают по точности тому приближённому решению в
замкнутом виде,  которое легко получить из  точного  решения  (23.63).
Само  по себе это решение считать аналитическим нельзя,  поскольку оно
включает в себя  операцию  обращения  матрицы  $\rv  M$,  определяемой
формулой  (23.58).  Но  если отбросить в (23.63) сумму по $m$,  то для
${\cal B}$ получаем простое  выражение  ${\cal  B}=-h\sigma_0/2$,  в
которое  входит сумма $\sigma_0$,  определённая формулой (23.56).  В
квазистатическом приближении она может быть представлена  в  замкнутом
аналитическом виде,  а в общем случае легко вычисляется, поскольку ряд
быстро сходится.  Ещё более  точный  результат  получается,  если  в
(23.63) оставить член, соответствующий $m=1$ в сумме, стоящей в правой
части. В этом случае обратная матрица легко вычисляется и в результате
     $$ {\cal B}=-\frac{h } 2\left (\sigma_0-\frac{\sigma_1^2}{1+
         \sqrt{g_1^{(2)}}\bigl (\sigma_1+\displaystyle{\frac{n^2}
         {\alpha^2}\Sigma_1}\bigr )}\right )\,,\eqno(23.79)$$
где $\sigma_1$ и $\Sigma_1$ определены выше формулами (23.56) и (23.60).

     Решение (23.63)  удобно  и  при  его  реализации  на  компьютере,
поскольку оно включает в себя лишь  простейшие  стандартные  процедуры
обращения  и  умножения  матриц.  Переход  к более точному приближению
состоит в увеличении порядка матрицы $\rv M$,  которую  при  численном
решении,   естественно,   приходится  редуцировать.  При  сравнительно
невысоком порядке матрицы ($N>50$)  достигаемая  точность  значительно
превышает   потребности   практики   во  всём  диапазоне  параметров
структуры, соответствующем режиму одноволновости.

\begin{picture}(160,55)
\put(-5,50){\special{em:graph  fig23-4a.bmp}}
\put(75,50){\special{em:graph fig23-4b.bmp}}
\end{picture}

\begin{center}\begin{minipage}[c]{0.9\textwidth}
\footnotesize{\parindent=0.5cm
Рис.~23.4.~Толстая диафрагма в прямоугольном волноводе: {\it а)}
---  продольное сечение структуры; {\it б)} --- эквивалентная схема
нерегулярности и её преобразование.
}\end{minipage}\end{center}\vspace*{0.25cm}

     Из сказанного  можно  сделать поспешный вывод,  что прямой способ
решения задачи путём разложения поля по обе стороны от плоскости,  в
которой    расположена    нерегулярность,    по   собственным   волнам
соответствующего волновода и последующего переразложения одних функций
по другим обладает преимуществом перед методом интегрального уравнения
в общем случае.  Однако это не так.  Высокая  точность  приближённой
формулы   (23.79)   и  быстрая  сходимость  результата  при  численном
обращении матрицы  в  (23.63)  с  увеличением  порядка  редуцированной
матрицы   обусловлены   физическими   свойствами,   присущими   данной
нерегулярности.  Картина силовых линий  электрического  поля,  которые
должны  везде  подходить по нормали к идеально проводящей поверхности,
говорит о том, что в правом волноводе (с меньшей высотой узкой стенки)
отличие  полного  поля  от  поля  основной волны $H_{10}$ невелико,
поэтому  коэффициенты  $A_m^{(2)}$  в  разложении  (23.47)   малы   по
сравнению  с  $A_0^{(2)}$  ---  формула же (23.79) получена фактически
отбрасыванием этих коэффициентов.  Кроме того, угол излома поверхности
в  плоскости  стыка  равен $\pi/2$,  поэтому запасённая энергия поля
высших типов волн стремится к конечному  пределу  по  мере  увеличения
порядка редуцированной матрицы, хотя само поле и расходится.

     В случае бесконечно  тонкой  диафрагмы  также  нетрудно  получить
бесконечную систему линейных алгебраических коэффициентов. Более того,
фактически получаются две такие системы: одна из условия непрерывности
поля   в   окне,  другая  из  требования  равенства  нулю  касательной
составляющей электрического поля на поверхности диафрагмы  ---  но  ни
одна  из  них  не  даёт  разумного  результата  при  решении методом
редукции соответствующей матрицы.  Напомним, что физически осмысленным
можно   считать   только  решение,  обеспечивающее  конечное  значение
запасённой  вблизи  нерегулярности  энергии  поля   и   именно   это
требование являет собой предмет обсуждавшегося ранее условия на ребре.
Такое  единственное  решение  может   быть   получено   в   результате
совместного   решения   обеих   систем   при   строго  определённом
соотношении между числом уравнений,  учитываемых в каждой из  них  по
мере увеличения порядка матрицы. Это соотношение зависит от параметров
структуры и его определение само по себе непростая задача.  Физическая
причина такого результата очевидна,  она обусловлена большей величиной
угла излома поверхности на краю диафрагмы (он  составляет  $2\pi$)  и,
соответственно,   более  быстрым  ростом  поля  вблизи  точки  излома.
Отметим, что использованный выше метод решения интегрального уравнения
в  квазистатическом  приближении  автоматически обеспечивал конечность
энергии поля.

    Рассмотрим теперь  кратко  толстую   диафрагму   в   прямоугольном
волноводе, представленную на  (рис.~23.4\textit а).  При  достаточной
толщине  диафрагмы $d$ её можно   рассматривать как последовательность
двух  одиночных нерегулярностей,  соединённых  отрезком  регулярного
волновода.  Обе нерегулярности в данном случае представляют собой один и
тот  же  стык двух волноводов с разным размером узкой стенки. Матрица
рассеяния этой одиночной нерегулярности определена формулой (23.66),
полученной выше в  результате решения электродинамической задачи.
Наша цель состоит в нахождении матрицы рассеяния толстой диафрагмы
методом  теории  цепей, основываясь на матрице (23.66).

     При построении матрицы рассеяния диафрагмы в качестве  плоскостей
отсчёта  обоих  примыкающих  волноводов  удобно  выбрать  одну и
ту же плоскость --- плоскость симметрии $z=0$.  Тогда матрица $\rv
S$  будет не  только  симметрична  (устройство взаимно),  но и
её  диагональные элементы  $s_{11}$   и   $s_{22}$   будут равны.
Построим   теперь эквивалентную схему диафрагмы. Узкий волновод
(область 2 на рис.~23.4) нагружен  справа  на эквивалентное
сопротивление   правого   стыка, определяемое формулой  (23.70),
которое  дополнительно  шунтируется единичным сопротивлением ---
характеристическим сопротивлением правого полубесконечного
волновода    (рис.~23.5a).    Пересчитаем теперь сопротивление
нагрузки $Z_4$,  расположенной в  плоскости $z=d/2$,  в
эквивалентное сопротивление в плоскости $z=-d/2$.  Для этого
выберем в волноводе  две  произвольные  плоскости  $z=z_1$ и
$z=z_2$,  которым соответствуют  сопротивления  нагрузки
$Z_{\mbox{\footnotesize\textit{н1}}}$ и
$Z_{\mbox{\footnotesize\textit{н2}}}$.  Очевидно, что коэффициенты
отражения $\Gamma(z_1)$ и $\Gamma(z_2)$ связаны между собой
следующим из (22.8) соотношением:
     $$\Gamma (z_2)=\Gamma (z_1)\,e^{2ih(z_1-z_2)}\,,\eqno(23.80)$$
а каждый   из   коэффициентов   отражения   выражается   через  своё
сопротивление нагрузки как
     $$\Gamma(z_i)=\frac {Z_{\mbox{\footnotesize\textit{н}}i}-1}
       {Z_{\mbox{\footnotesize\textit{н}}i}+1}\,,\eqno(23.81)$$
причём отсюда получается, что
     $$Z_{\mbox{\footnotesize\textit{н2}}}=
        \frac{Z_{\mbox{\footnotesize\textit{н1}}}
         +i\tg{h(z_2-z_1)}}{1+iZ_{\mbox{\footnotesize\textit{н1}}}
         \tg{h(z_2-z_1)}}\,.\eqno(23.82)$$
На рис.~23.4б представлена соответствующая эквивалентная схема, где
     $$Z_4= \frac{1-(i\alpha-{\cal B})\tg{h d}}{\alpha+i{\cal B}-i
         \tg{h d}}\,.\eqno(23.83)$$

     Теперь нетрудно  выразить  эквивалентное  сопротивление  нагрузки
левого волновода в плоскости $z=-d/2$:
     $$ Z_{\mbox{\footnotesize\textit{н}}}(-d/2)=
         Z_2+\frac {Z_1(Z_3+Z_4)}{Z_1+Z_3+Z_4}\,,
         \eqno(23.84)$$
где $Z_1$, $Z_2$ и $Z_3$ определены выше формулой (23.68). После этого
с помощью  (23.81)  находится  коэффициент  отражения  $\Gamma(-d/2)$,
который  согласно  (22.16) связан с элементом $s_{11}$ искомой матрицы
$\rv S$ соотношением
     $$  s_{11}=\Gamma(-d/2)\,e^{-ih d}\,.\eqno(23.85)$$

     Опуская несложные,    но    громоздкие    выкладки,    приведём
окончательный результат:
     $$s_{11}=s_{22}=\frac {e^{-ih d}} 2 \left [ \frac
         {\displaystyle{(\alpha-i{\cal B})\tg{h \frac d 2} -i}}
         {\displaystyle{(\alpha+i{\cal B})\tg{h\frac d 2}+i}}+\frac
         {\displaystyle{\alpha-i{\cal B}+i\tg{h \frac d 2}}}
         {\displaystyle{\alpha+i{\cal B}-i\tg{h \frac d 2}}}
         \right ]\,.\eqno(23.86)$$
Остальные элементы матрицы $\rv S$ легко  находятся  из  условия  её
унитарности:
     $$  s_{12}=s_{21}=\frac {e^{-ih d}} 2 \left [ \frac
         {\displaystyle{\alpha-i({\cal B}-\tg{h \frac d 2})}}
         {\displaystyle{\alpha+i({\cal B}-\tg{h \frac d 2})}}+\frac
         {\displaystyle{i-\tg{h \frac d 2}(\alpha-i{\cal B})}}
         {\displaystyle{i+\tg{h \frac d 2}(\alpha+i{\cal B})}}
         \right ]\,.\eqno(23.87)$$

     Теория цепей является хорошо разработанным и мощным  инструментом
расчёта сложных устройств СВЧ.  В частности,  для вычисления матрицы
каскадно расположенных четырёхполюсников вводится еще  одна  матрица
--- {\it матрица передачи}.  Но  останавливаться здесь на таких деталях
нет возможности.

%\end{document}

\newpage
\oddsidemargin=-0.4mm \evensidemargin=-0.4mm
\topmargin=-0.4mm
\headsep=7mm
\textheight=231.875mm
\textwidth=160mm
\mathsurround=2.5pt
\unitlength=1mm
%\begin{document}
%\input{macr.tex}
\thispagestyle{empty}
%\addtocounter{page}{271}
\baselineskip=0.99\normalbaselineskip

\begin{center}
   \subsubsection*{\rm Г\,Л\,А\,В\,А\, 9}
      \vspace{-1.15em}
      \line(6,0){160}\\
      \vspace{-1em}
      \line(6,0){160}
      \vspace{-1.15em}
   \subsubsection*{МЕТОДЫ ТФКП В ЭЛЕКТРОДИНАМИКЕ}
      \vspace{31mm}
   \subsubsection*{24. Метод Винера-Хопфа-Фока }
\end{center}
\vspace{0.5cm}

\markboth{Глава 9. Методы ТФКП в электродинамике}{24. Метод
         Винера-Хопфа-Фока }

\begin{center}\begin{minipage}[c]{0.75\textwidth}
\footnotesize{\parindent=0.5cm
         Дифракционная задача для плоского разветвлённого волновода.
         Прямое  решение  бесконечной  системы  линейных  уравнений по
         правилу  Крамера.  Поведение  компонент  поля  вблизи   ребра
         структуры.   Выбор   единственного  решения.  Решение  задачи
         методом  Винера-Хопфа-Фока.  Метод  Джонса.  Метод   вычетов.
         Сравнение аналитических методов теории волноводов.
}\end{minipage}\end{center}\vspace{0.5cm}

     Метод Винера-Хопфа-Фока  представляет  собой хорошо разработанный
математический аппарат  решения  дифракционных  задач  электродинамики
СВЧ.  Он применяется в тех случаях,  когда задача может быть сведена к
линейному интегральному уравнению определённого вида, ядром которого
служит  функция Грина.  Решение этого уравнения производится на основе
теории функций комплексного переменного.  Важнейшими моментами в  ходе
решения  являются {\it факторизация} комплексного преобразования Фурье
функции Грина,  то есть представление его  в  виде  произведения  двух
функций,  и последующее {\it разложение} на сумму функций,  обладающих
определёнными  аналитическими  свойствами   в   верхней   и   нижней
полуплоскостях комплексного переменного.

     Ознакомление с  методом Винера-Хопфа-Фока поучительно провести на
примере волноводной задачи,  которая (едва ли не  единственная)  имеет
решение  в  замкнутом  аналитическом виде,  причём это решение может
быть получено и другими способами,  в частности,  прямым  решением  по
правилу  Крамера бесконечной системы линейных алгебраических уравнений
для коэффициентов разложения по собственным волнам отдельных  областей
структуры.  Такой  задачей,  часто  называемой ключевой задачей теории
дифракции в волноводах,  является  рассеяние  волны  типа  $H_{10}$  в
разветвлённом   плоском   волноводе,  изображённом  на  рис.~24.1.
Структура   однородно   заполнена    веществом    с    проницаемостями
$\varepsilon$   и   $\mu$;   стенки   волновода  и  бесконечно  тонкая
разветвляющая пластина имеют идеальную проводимость.  В падающей волне
отличны от нуля компоненты поля
     $$\left.\begin{array}{l}E_y^{(in)}=\sin{\displaystyle{\frac
         {\pi x}a}\,e^{ih_{1a}z}}= \varphi^{(in)}(x,z)\,,\\[.5cm] H_x^
         {(in)}=\displaystyle{\frac i{k\mu}\,\frac{\partial\varphi^
         {(in)}}{\partial z}}\,,\\[.5cm] H_z^{(in)}=-\displaystyle
         {\frac i{k\mu}\,\frac{\partial\varphi^{(in)}}{\partial x}}
         \,;\end{array}\right\}\eqno(24.1)$$
здесь и везде далее $h_{nr}=\sqrt{K^2-  (\pi  n/r)^2}$, где
$r=a$,~$b$~или  $d$,  $K=k\sqrt{\varepsilon \mu}=K_1+iK_2\;  (K_1>0,
\,K_2>0)$. Требуется найти рассеянное поле в областях 1, 2 и 3.

\begin{wrapfigure}[12]{l}{7.5cm}
\begin{picture}(70,45)
\put(-1,50){\special{em:graph fig24-1.bmp}}
\end{picture}
\hbox to 7.5cm{\hfil\footnotesize{Рис.~24.1.~Плоский разветвлённый
волновод.}\hfil}
\end{wrapfigure}
     Ввиду однородности   структуры  вдоль  оси  $y$  рассеянное  поле
состоит  только  из  волн  типа  $H_{n0}$,  все  компоненты   которого
выражаются    формулами    (24.1)    через    функцию   $\varphi=E_y$,
удовлетворяющую скалярному волновому уравнению
     $$\left(\frac{\partial^2}{\partial x^2}+\frac{\partial^2}
         {\partial z^2}+K^2\right)\varphi(x,z)=0\eqno(24.2)$$
и дополнительно:  1) граничным условиям $\varphi=0$ при $x=0,\,a$  для
всех  $z$  и при $x=b$ для $z>0$;  2) условиям излучения при $z\to \pm
\infty$;   3)   непрерывности    касательных    составляющих    полных
электрического и магнитного поля в плоскости $z=0$;  4) и, кроме того,
условию на ребре $x=b,\, z=0$, конкретный вид которого уточним далее.

     Приведём кратко  решение  задачи  прямым  методом разложения по
волноводным волнам. Опустим при этом громоздкие промежуточные выкладки
и остановимся на основных принципиальных моментах решения.  Рассеянное
поле,  удовлетворяющее уравнению (24.2) и условиям 1) и 2), может быть
записано в виде
     $$E_y=\left\{\begin{array}{ll}\displaystyle{\sum_{n=1}^\infty
         A_n\sin{\frac{n\pi x}a}\,e^{-ih_{na}z}}&\qquad\mbox{при}\quad
         z<0;\\[.4cm] \displaystyle{\sum_{n=1}^\infty B_n\sin{\frac
         {n\pi x}b}\,e^{ih_{nb}z}}&\qquad\mbox{при}\quad 0<x<b,z>0\,;
         \\[.4cm]\displaystyle{\sum_{n=1}^\infty C_n\sin{\frac{n\pi
         (x-b)} d}\,e^{ih_{nd}z}}&\qquad\mbox{при}\quad b<x<a, z>0\,.
         \end{array}\right.\eqno(24.3)$$
     Сшивание полей  в  плоскости  $z=0$ приводит к следующим системам
функциональных уравнений:
     $$\left.\begin{array}{l}\displaystyle{\sin{\frac{\pi x}a}+
         \sum_{n=1}^\infty A_n\sin{\frac{\pi n x}a}=\sum_{n=1}^\infty
         B_n\sin{\frac{\pi n x}b}}\,,\\[.5cm]\displaystyle{h_{1a}
         \sin{\frac{\pi x}a}-\sum_{n=1}^\infty h_{na}A_n\sin{\frac
         {\pi n x}a}=\sum_{n=1}^\infty B_n h_{nb}\sin{\frac{\pi n x}
         b}}\,,\end{array}\right\}\quad\mbox{при}\; 0<x<b\,;
         \eqno(24.4)$$
     $$\left.\begin{array}{l}\displaystyle{\sin{\frac{\pi x}a}+
         \sum_{n=1}^\infty A_n\sin{\frac{\pi n x}a}=\sum_{n=1}^
         \infty C_n\sin{\frac{\pi n (x-b)}d}}\,,\\[.5cm]\displaystyle
         {h_{1a}\sin{\frac{\pi x}a}-\sum_{n=1}^\infty h_{na}A_n\sin
         {\frac{\pi n x}a}=\sum_{n=1}^\infty C_n h_{nd}\sin{\frac
         {\pi n (x-b)}d}}\,,\end{array}\right\}\quad\mbox{при}\;
         b<x<a\,.\eqno(24.5)$$

     Используя ортогональность  синусов,  перейдём   к   бесконечной
системе  линейных  алгебраических уравнений относительно коэффициентов
$A_m$,  $B_m$ и $C_m$.  Исключая из полученной системы $B_m$ и  $C_m$,
получаем {\it парную} бесконечную систему уравнений для $A_n$:
     $$\left.\begin{array}{lcl}\displaystyle{\sum_{n=1}^\infty\frac 1
         {h_{na}-h_{mb}}\,A'_n}&=&\displaystyle{\frac{A'}{h_{1a}+h_
         {mb}}}\,,\\[.5cm]\displaystyle{\sum_{n=1}^\infty\frac 1
         {h_{na}-h_{md}}\,A'_n}&=&\displaystyle{\frac {A'}{h_{1a}+h_
         {md}}}\,,\end{array}\right\}\quad m=1,2,\dots ,\eqno(24.6)$$
где введены   обозначения  $A'_n=A_n\sin{(n\pi  b/a)}$,  $A'=\sin{(\pi
b/a)}$.

     Дальнейшие этапы   решения   задачи   включают  в  себя  редукцию
полученной бесконечной парной системы,  решение редуцированной системы
по правилу Крамера и предельный переход к решению бесконечной системы.
Редуцированная система включает в себя $P$ уравнений из первой системы
(24.6),   $Q$   уравнений   из  второй  системы,  $N=P+Q$  неизвестных
коэффициентов $A'_m$ и может быть записана в  матричном  виде.  Каждый
элемент  определителя  матрицы  представляет  собой частное от деления
единицы на разность двух величин,  например $h_{na}$  и  $h_{pb}$  или
$h_{qd}$,  прич\"ем  одна  из этих величин --- $h_{na}$ --- изменяется
при движении вдоль строк,  а  $h_{pb}$  и  $h_{qd}$  изменяются  вдоль
столбцов.  Такой  специфический  вид определителя подсказывает искомое
решение для $A_m$:
     $$A_m=-\frac{\sin{(\pi b/a)}}{\sin{(m\pi b/a)}}\;\prod^P_{p=1}
         \frac{h_{pb}-h_{ma}}{h_{pb}+h_{1a}}\;\prod^Q_{q=1}\frac
         {h_{qd}-h_{ma}}{h_{qd}+h_{1a}}\;{\prod_{n=1}^N}'\frac{h_{na}
         +h_{1a}}{h_{na}-h_{ma}}\,,\eqno(24.7)$$
где знаком $'$  в последнем произведении обозначен пропуск множителя  с
$n=m$.

     Полученное выражение представляет собой {\it точное} решение {\it
редуцированной} системы.  Следующий этап решения состоит в  предельном
переходе $P\to \infty,\;  Q\to \infty$. Для этого необходимы некоторые
преобразования,  поскольку входящие в (24.7)  произведения  расходятся
при  стремлении  верхнего  предела  к  бесконечности.  Для преодоления
возникшей  трудности  учтём,  что  произведение  $  \prod_{p=1}^P(1-
iw/h_{pb})\exp{(wb/p\pi)}$  сходится  равномерно  при $P\to \infty$ на
всей комплексной плоскости $w$.  Пут\"ем тождественных  преобразований
представим (24.7) в виде
     $$ \begin{array} {l}A_m=-\displaystyle{\frac{\sin{(\pi b/a)}}
         {\sin{(m\pi b/a)}}\exp{\bigl[\bigl(-i\frac{h_{ma}+h_{1a}}
         {\pi}\bigr)\bigl(\chi-\frac a m\bigr)\bigr]}\,\prod^P_{p=1}
         \frac{1-h_{ma}/h_{pb}}{1+h_{1a}/h_{pb}}}\times\\[.3cm]
         \times\displaystyle{\exp{\bigl[-\frac{i b}{p\pi}\bigl
         (h_{ma}+h_{1a}\bigr)\bigr]}\,\prod^Q_{q=1}\frac{1-h_{ma}
         /h_{qd}}{1+h_{1a}/h_{qd}}\exp{\bigl[-\frac{i d}{q\pi}\bigl
         (h_{ma}+h_{1a}\bigr)\bigr]}}\times\\[.3cm]\times
         \displaystyle{{\prod_{n=1}^N}'\frac{1+h_{1a}/h_{na}}
         {1-h_{ma}/h_{na}}\,\exp{\bigl[\frac{i a}{n\pi}\bigl(
         h_{ma}+h_{1a}\bigr)\bigr]}}\,,\end{array}\eqno(24.8)$$
где
     $$\chi=a\sum_{n=1}^N\frac 1 n-b\sum_{p=1}^P\frac 1 p-d\sum_
         {q=1}^Q\frac 1 q\,,\eqno(24.9)$$
а знаком  $'$  в  последнем произведении обозначен пропуск множителя с
$m$, равным $n$. Такая запись позволяет вычислить пределы сомножителей
в  (24.8)  при  неограниченном  возрастании $P$ и $Q$.

     Действительно, предельное  значение  $\chi$  находится при помощи
формулы, определяющей постоянную Эйлера,
     $$ C=\lim_{M\to \infty}\Bigl(\sum_{m=1}^M\frac 1 m-\ln M\Bigr)=
         0,5772\dots\;, \eqno(24.10)$$
и легко можно убедиться, что
     $$\lim _{\begin {array}{l}\scriptstyle{\; P\to\infty}\\[-.1cm]
         \scriptstyle{\; Q\to\infty}\end{array}}\chi=\lim _{\begin
         {array}{l}\scriptstyle{\; P\to\infty}\\[-.1cm]\scriptstyle
         {\; Q\to\infty}\end{array}}\Bigl[b\ln{\Bigl(1+\frac Q P
         \Bigr)}+d\ln{\Bigl(1+\frac P Q\Bigr)} \Bigr].\eqno(24.11)$$
Таким образом,   предельное   значение  $\chi$  определяется  пределом
отношения $P/Q$ при $P\to\infty$ и $Q\to\infty$.  Последнее  означает,
что   решение  (24.7)  системы  уравнений  (24.6)  при  неограниченном
увеличении порядка редуцированной матрицы сходится к различным наборам
коэффициентов   $A_m$   в   зависимости   от  отношения  $P/Q$.  Выбор
единственного  физически  разумного  решения   возможен   только   при
использовании  условия  на  ребре,  но  прежде  чем  переходить  к его
конкретизации,  приведём без вывода  следующую  из  анализа  формулы
(24.8) асимптотику коэффициентов $A_m$ для больших $m$:
     $$ A_m=O(m^{-3/2}e^{m\chi'/a})\qquad\mbox{при}\quad m\to\infty\,,
         \eqno(24.12)$$
где
     $$ \chi'=   \lim _{\begin {array}{l}\scriptstyle{\; P\to\infty}
         \\[-.1cm]\scriptstyle{\; Q\to\infty}\end{array}}\Bigl[b\ln
         \Bigl(\frac{1+Q/P}{1+d/b}\Bigr)+d\ln\Bigl(\frac{1+P/Q}{1+
         b/d}\Bigr)\Bigr]\,.\eqno(24.13)$$

     Условие на ребре было сформулировано в разделе 19 в виде
     $$\int\limits_V(\varepsilon|\rv E|^2+\mu|\rv H|^2)\,dV\to 0
         \eqno(24.14)$$
при $V\to 0$ в  окрестности  ребра.  Физический  смысл  этого  условия
состоит в конечности энергии рассеянного поля,  сосредоточенной вблизи
ребра --- без этого условия не может  быть  достигнуто  установившееся
решение.  При этом составляющие полей $\rv E$ и $\rv H$ могут на ребре
обращаться в бесконечность,  что обусловлено используемой идеализацией
структуры  поверхности  реального тела.  Если ребро представляет собой
гладкую кривую (в рассматриваемой задаче прямая),  то  в  каждой  её
точке   можно   ввести  локальные  цилиндрические  координаты  $(\rho,
\varphi,z)$; тогда из условия (24.14) следует, что в окрестности ребра
{\it  ни одна составляющая электромагнитного поля} не может возрастать
быстрее,  чем $\rho^{-1+\tau}\;(\tau>0)$ при $\rho\to  0$.

     Рассмотрим теперь  на  примере  двумерного  идеально  проводящего
клина,  изображённого на рис.~24.2,  как,  основываясь на уравнениях
Максвелла,  можно  вычислить  значение  $\tau$.  При этом для простоты
будем считать, что (как и в анализируемой  задаче) отличны от нуля только
составляющие   $E_z,\;H_{\rho}$   и   $H_{\varphi}$,   удовлетворяющие
уравнениям
     $$ \frac 1{\rho}\frac{\partial E_z}{\partial \varphi}=ik\mu H_
         {\rho}\,,\qquad \frac{\partial E_z}{\partial\rho}=-ik\mu H_
         {\varphi}\,,\qquad\frac 1 \rho \frac{\partial(\rho H_{
         \varphi})}{\partial\rho}-\frac 1 \rho\frac {\partial H_
         {\rho}}{\partial \varphi}=-ik\varepsilon E_z\,.
         \eqno(24.15)$$

     Все составляющие  поля  вблизи ребра $\rho=0$ можно представить в
виде разложений по степеням $\rho$.  Поскольку они не могут возрастать
быстрее, чем $\rho^{-1+\tau}$, где $\tau>0$, то их можно записать как
     $$ \begin{array}{lcl}
         E_z&=&\rho^{-1+\tau}(a_0+a_1\,\rho+\dots)\,,\\[.2cm]
         H_{\rho}&=&\rho^{-1+\tau}(b_0+b_1\,\rho+\dots)\,,\\[.2cm]
         H_{\varphi}&=&\rho^{-1+\tau}(c_0+c_1\,\rho+\dots)\,.
         \end{array}\eqno(24.16)$$
\begin{wrapfigure}[10]{l}{7.0cm}
\begin{picture}(60,40)
\put(-5,45){\special{em:graph fig24-2.bmp}}
\end{picture}
\hbox to 7.0cm{\hfil\footnotesize{Рис.~24.2.~Идеально проводящий
клин.}\hfil}
\end{wrapfigure}
Коэффициенты в  этих разложениях зависят лишь от координаты $\varphi$.
После подстановки  разложений  в  уравнения  (24.15)  и  приравнивания
коэффициентов  при одинаковых степенях $\rho$ получим ряд соотношений,
из которых, в частности, следует,  что $a_0\equiv 0$,  $b_0\sim d  a_1/d
\varphi$, $c_0\sim\tau\,a_1$, а $a_1$ удовлетворяет уравнению
     $$ \frac{d^2 a_1}{d\varphi^2}+\tau^2a_1=0\,.\eqno(24.17)$$
Из общего решения этого уравнения
     $$ a_1=\alpha\cos{\tau\varphi}+\beta\sin{\tau\varphi}\,
         \eqno(24.18)$$
вытекает, что граничные условия на поверхности клина  будут  выполнены
при
     $$  \alpha=0\,,\qquad  \tau=\frac \pi \psi\,.\eqno(24.19)$$
Подставляя эти   результаты   в   (24.17),  находим,  что  зависимость
составляющих полей от координат вблизи ребра имеет вид
     $$  \begin{array}{lcl} E_z&\sim&\rho^{\tau}\sin{\tau\varphi}+
         \dots\,,\\[.1cm]H_{\rho}&\sim&\rho^{-1+\tau}\cos{\tau
         \varphi}+\dots\,,\\[.1cm]H_{\varphi}&\sim&\rho^{-1+\tau}
         \sin{\tau\varphi}+\dots\,;\end{array}\eqno(24.20)$$
в случае   бесконечно   тонкой  полуплоскости  ($\psi=2\pi$)  согласно
(24.19) параметр $\tau=1/2$ и, следовательно,
     $$ E_z\sim \sqrt{\rho}\,,\qquad H_{\varphi}\sim\frac 1 {\sqrt
         {\rho}}\,.\eqno(24.21)$$

     Таким образом,  конкретизирован  вид  условия  4)  для  уравнения
(24.2).  Асимптотические  оценки  рядов  (24.3)  позволяют  установить
(соответствующие выкладки здесь опущены),  что поля вблизи ребра будут
иметь вид (24.21) только при убывании всех коэффициентов с  ростом  их
номера по закону
     $$ A_n,\,B_n,\,C_n\,=O(n^{-3/2})\qquad\mbox{при}\quad
         n\to\infty\,. \eqno(24.22)$$
Сопоставляя полученный  результат  с  формулами  (24.12)  и   (24.13),
убеждаемся,  что  требуемая  асимптотика  для коэффициентов разложения
полей будет верна,  если при решении  парной  системы  уравнений
(24.6)  выполняется  следующее  соотношение  между  числом  уравнений,
учитываемых в каждой из систем пары:
     $$ \frac P Q = \frac b d\,.\eqno(24.23)$$

     В результате   коэффициенты   $A_m$  могут  быть  представлены  в
следующем замкнутом виде:
     $$ \begin{array} {l}A_m=-\displaystyle{\frac{\sin{(\pi b/a)}}
         {\sin{(m\pi b/a)}}\exp{\Bigl(-i\frac{h_{ma}+h_{1a}}{\pi}
         \Bigl\{\Bigl[b\ln{\Bigl(\frac a b\Bigr)}+d\ln{\Bigl
         (\frac a d\Bigr)}\Bigr]-\frac a m \Bigr\}\Bigr)}}\times
         \\[.4cm]\qquad\times\displaystyle{\prod\limits^\infty_{n=1}}
         \displaystyle{\frac{1-h_{ma}/h_{nb}}{1+h_{1a}/h_{nb}}}\,e^
         {-i(h_{ma}+h_{1a}) b/n\pi}\,\prod\limits^\infty_{n=1}
         \displaystyle{\frac{1-h_{ma}/h_{nd}}{1+h_{1a}/h_{nd}}}\,e^
         {-i (h_{ma}+h_{1a})d/n\pi}\times\\[.4cm]\qquad\times
         \displaystyle{{\prod_{n=1}^N}'}\displaystyle{\frac{1+h_{1a}
         /h_{na}}{1-h_{ma}/h_{na}}}\,e^{i (h_{ma}+h_{1a})a/n\pi}\,,
         \qquad m=1,2,3,\ldots\;.\end{array}\eqno(24.24)$$
Аналогичным способом могут быть найдены коэффициенты  $B_m$  и  $C_m$.
Таким   образом,   аналитическое  решение  задачи  получается  в  виде
бесконечных рядов (24.3),  коэффициенты в которых содержат бесконечные
произведения.

     Перейдём теперь    к    решению    этой   же   задачи   методом
Винера-Хопфа-Фока.  Поскольку   метод   самым   существенным   образом
базируется   на   теории  функций  комплексного  переменного,  то  для
избежания  разночтений  уточним  используемую  далее  терминологию   и
привед\"ем без вывода важнейшие для последующего изложения сведения из
ТФКП и преобразования Фурье для комплексных переменных.  Функция  {\it
аналитическая  в точке},  если она однозначна и дифференцируема в этой
точке; функция {\it аналитическая в области}, если она аналитическая в
каждой точке этой области, за исключением, быть может, конечного числа
{\it особых точек}. Если аналитическая функция не имеет особых точек в
некоторой области,  то она в ней {\it регулярная}. Функция, регулярная
во всей конечной  области  комплексного  переменного  называется  {\it
целой}. Функция, у которой все особенности в конечной области являются
полюсами,  называется  {\it  мероморфной}.  Произвольная   мероморфная
функция представима в виде отношения двух целых функций.

     Преобразование Фурье  функции  $f(z)$  в  комплексной   плоскости
$\alpha=\sigma+i\tau$ определяется интегралом
     $$F(\alpha)=\frac 1 {\sqrt{2\pi}}\int\limits_{-\infty}^{\infty}
         f(z)\,e^{i\alpha z}\,dz\,,\eqno(24.25)$$
который часто удобно записать в виде
     $$F(\alpha)=F_{+}(\alpha)+F_{-}(\alpha)\,,\eqno(24.26)$$
где
     $$F_{+}(\alpha)=\frac 1 {\sqrt{2\pi}}\int\limits_0^{\infty}
         f(z)e^{i\alpha z}\,dz\,,\qquad F_{-}(\alpha)=\frac 1
         {\sqrt{2\pi}}\int\limits_{-\infty}^0f(z)e^{i\alpha z}\,dz
         \,.\qquad\eqno(24.27)$$

     Для дальнейшего   важными   являются   следующие   свойства  этих
интегралов.  {\it Если $f(z)\sim Ae^{\tau_- z}$ при $z\to +\infty$, то
функция $F_+(\alpha)$     регулярна    в    полуплоскости    $\tau=\im
\alpha>\tau_-$.} И аналогично,  {\it если $f(z)\sim Be^{\tau_+ z}$ при
$z\to  -\infty$,  то  функция  $F_-(\alpha)$ регулярна в полуплоскости
$\tau<\tau_+$.}  Если  выполнены  оба  перечисленных   выше   условия,
прич\"ем   $\tau_{+}>   \tau_{-}$,   {\it   то   преобразование  Фурье
$F(\alpha)$  регулярно  в  полосе  $\tau_-<\tau<\tau_+$,  а   обратное
преобразование даётся интегралом}
     $$ f(z)=\frac 1 {\sqrt{2\pi}}\int\limits_{-\infty+i\tau}^
         {\infty+i\tau}\;F(\alpha)\,e^{-i\alpha z}\,d\alpha,\qquad
         \tau_-<\tau<\tau_+\,.\eqno(24.28)$$
Необходимо ещё упомянуть {\it теорему о свёртке}, которая в этих
обозначениях записывается в виде
     $$\int\limits_{-\infty}^{\infty} f(z')g(z-z')\,dz'=\int\limits_
         {-\infty+i\tau}^{\infty+i\tau}F(\alpha) G(\alpha)\,e^{-i
         \alpha z}\,d\alpha\,.\eqno(24.29)$$

     Интегральное уравнение  Винера-Хопфа-Фока  выводится  на   основе
второй теоремы Грина (в данном случае двумерной):
     $$\int\limits_S(\varphi\Delta_2 g- g\Delta_2 \varphi)\,dx\,dz=
         \oint\limits_C \Bigl(\varphi\frac{\partial g}{\partial n}-
         g\frac{\partial \varphi}{\partial n}\Bigr)\,ds\,,
         \eqno(24.30)$$
где в данном случае область $S$ ограничена стенками широкого волновода
и полубесконечной пластиной,  разветвляющей волновод (рис.~24.1),  $C$
-- граница области $S$ (показана штриховой линией), $\partial/\partial
n$  --  оператор  дифференцирования  в направлении внешней нормали.  В
качестве функции $\varphi$  берётся  составляющая  рассеянного  поля
$E_y$,  удовлетворяющая  уравнению (24.2) и нулевым граничным условиям
на поверхностях $x=0$ и $x=a$, а также в сумме с падающим полем (24.1)
обращающаяся  в  нуль при $x=b$ и $z>0$ (предполагается,  что падающее
поле имеет вид (24.1) при всех значениях $z$).

     В качестве функции  $g$  в  (24.30)  используется  функция  Грина
$g(x,z;x_0z_0)$, определяемая уравнением
     $$\left(\frac {\partial^2}{\partial x^2}+\frac{\partial^2}
         {\partial z^2}+K^2\right)g(x,z;x_0,z_0)=-\sqrt{2\pi}\delta(x
         -x_0)\delta(z-z_0)\,,\eqno(24.31)$$
(множитель $\sqrt{2\pi}$  введён  здесь  в  отличие,  например,   от
(17.19)  для  упрощения  записи  ряда  последующих формул),  граничным
условиям $g=0$ при $x=0$ и $x=a$ для всех $z$ и условию излучения  при
$z\to\pm  \infty$.  Отметим  сразу,  что  явный  вид  функции  Грина в
дальнейшем не понадобится (хотя  в  данном  случае  он  находится  без
особого  труда),  но  принципиально  важно,  что  $g$  зависит лишь от
разности $z-z_0$. В дальнейшем используется только Фурье-образ функции
Грина, а его во многих случаях найти проще, чем саму функцию.

     Используя уравнения (24.2) и (24.31), убеждаемся, что интеграл по
области  $S$  в  левой  части  (24.30)  равен  $-\sqrt{2\pi}   \varphi
(x_0,z_0)$. От интеграла по контуру остаётся только
     $$\int\limits_0^\infty g(b,x_0;z-z_0)j(z)\,dz,\eqno(24.32)$$
где величина
     $$j(z)=\left.\frac{\partial \varphi}{\partial x}\right|_{b+
         \varepsilon}-\left.\frac{\partial \varphi}{\partial x}
         \right|_{b-\varepsilon}\qquad\mbox{при}\quad\varepsilon
         \to 0\,\eqno(24.33)$$
и пропорциональна  плотности  наведённого на пластине поверхностного
тока.  В результате получаем выражение для поля в  произвольной  точке
$(x_0,z_0)$ структуры:
     $$ \varphi(x_0,z_0)=-\frac 1{\sqrt{2\pi}}\int\limits_0^\infty g
         (b,x_0;z-z_0)j(z)\,dz.\eqno(24.34)$$
Граничное условие  $\varphi^i+\varphi=0$  при  $x=0,\,z>0$  приводит к
интегральному уравнению относительно неизвестной функции $j(z)$:
     $$\frac 1{\sqrt{2\pi}}\int\limits_0^\infty j(z')g(z-z')\,dz'=
         a(z)\,\qquad \mbox{при}\quad z>0\,,\eqno(24.35)$$
где введены обозначения $g(z-z')=g(b,b;z-z')$,
     $$   a(z)=\sin\frac{\pi b} a\,e^{ih_{1a}z}\,.\eqno(24.36)$$
Уравнение (24.35)  ---  типичная  форма  записи   {\it   интегрального
уравнения   Винера-Хопфа-Фока},   в  котором  $a(z)$  и  $g(z-z')$  --
известные функции, а $j(z)$ -- искомая функция.

     Стандартный подход   к   решению   уравнения  (24.35)  состоит  в
применении  теоремы  о  свёртке  (24.29)  с  целью  сведения  его  к
алгебраическому  уравнению,  включающему  в себя Фурье-образы функций.
Для того,  чтобы (24.35) имело вид свёртки,  функции $j(z)$ и $a(z)$
необходимо  доопределить на всю действительную ось,  для чего введём
новые функции
     $$\hbox{$j_+(z)=$}\hbox{$\cases{j(z)&при $z>0$,
         \cr\noalign{\vskip4pt}0&при $z<0$;\cr}$}\qquad
         \hbox{$a_+(z)=$}\hbox{$\cases{a(z)&при $z>0$,\cr
         \noalign{\vskip4pt}0&при $z<0$.\cr}$}\eqno(24.37)$$
На первый взгляд теперь уравнение (24.35) можно переписать в виде
     $$\frac 1{\sqrt{2\pi}}\int\limits_{-\infty}^\infty j_+(z_0)g(z-
         z_0)\,dz_0=a_+(z),\qquad -\infty<z<\infty.\eqno(24.38)$$
Однако такой  переход  является некорректным,  поскольку нет оснований
считать интеграл в левой части равным нулю для $z<0$.

     Для расширения  области  определения  исходного уравнения (24.34)
приходится вводить новую неизвестную функцию  $b_-(z)$,  в  результате
чего уравнение принимает вид
     $$\frac 1{\sqrt{2\pi}}\int\limits_{-\infty}^\infty j_+(z_0)g(z-
         z_0)\,dz_0=a_+(z)-b_-(z),\qquad-\infty<z<\infty,
         \eqno(24.39)$$
где
     $$\hbox{$b_-(z)=\cases{0,&$z>0;$\cr\noalign{\vskip4pt}-
         \displaystyle{\frac 1{\sqrt{2\pi}}}\displaystyle{\int
         \limits_0^\infty}j_+(z_0)g(z-z_0)\,dz_0,&$z<0.$}$}
         \eqno(24.40)$$
Теперь к уравнению (24.39) можно применить теорему о свёртке, но для
этого   предварительно   следует   определить   области   регулярности
Фурье-образов     $A_+(\alpha)$,    $J_+(\alpha)$,    $G(\alpha)$    и
$B_-(\alpha)$ функций $a_+(z)$,  $j_+(z)$,  $g(z)$ и  $b_-(z)$.  Проще
всего   это  сделать  для  Фурье-образа  функции  $a_+(z)$.  Поскольку
$a_+(z)\sim \exp{(-\tau_0 z)}$ при  $z\to  +\infty$,  где  $\tau_0=\im
h_{1a}>0$,   то  согласно  сказанному  после  формул  (24.27)  функция
$A_+(\alpha)$ регулярна в  верхней  полуплоскости  $(\tau>  -\tau_0)$;
е\"е явное выражение имеет вид
     $$   A_+(\alpha)=\frac 1{\sqrt{2\pi}}\,\frac i{\alpha+h_{1a}}
         \sin{\frac {\pi b} a}\,,\qquad \tau>-\tau_0\,.\eqno(24.41)$$

     Здесь необходимо обсудить вопрос о том,  что  представляет  собой
величина  $\tau_0$.  При отсутствии поглощения в заполняющей
структуру среде,  например,  в пустой структуре  (что  не
противоречит  никаким физическим принципам электродинамики),
сделанное предположение об {\it идеальной} проводимости всех
стенок волновода приводит к тому, что для распространяющейся волны
типа $H_{10}$ $\im h_{1a}=\tau_0=0$ и поэтому пропадает полоса на
плоскости  $\alpha$,  внутри  которой  справедливо уравнение
(24.35).  Как  будет видно дальше,  помимо этого становится
невозможным провести обратное преобразование Фурье,  поскольку
полюсы подынтегрального  выражения  оказываются  точно  на
действительной оси $\sigma$,~$\alpha=\sigma+i\tau$.  Оставаясь  в
рамках   эффективного математического  аппарата,  физически
некорректное  предположение  об идеальной  проводимости   стенок
на   высоких   частотах   удаётся <<обезвредить>>  только с помощью
дополнительного предположения о наличии поглощения в заполняющей
среде,  в результате чего  волновое  число  в среде
$K=k\sqrt{\varepsilon\mu}$  становится комплексным,  у $h_{1a}$
появляется небольшая мнимая часть и возникает полоса регулярности.

     Ценность получаемого  в  последнем  случае решения состоит в том,
что его предел при $\im K\to 0$ совпадает с пределом точного
решения в пустой структуре,  если устремить к бесконечности
проводимость стенок. В этом нетрудно усмотреть прямую аналогию  с
условием  на  ребре  --- необходимость его использования (для
получения единственного физически разумного решения) обусловлена
физически некорректным предположением о {\it  нулевой}  толщине
разветвляющей пластины.  Отметим,  что из двух этих неправомерных
идеализаций вторая более  <<вредная>>  --- так,  задача о   рассеянии
при  конечной  толщине  пластины  не  требует  никаких
дополнительных условий и при  идеальной  проводимости  стенок,
однако получить  решение в замкнутом аналитическом виде уже не
представляется возможным.

     Функцию $j_+(z)$  можно  представить  в  виде линейной комбинации
собственных волн областей 2 и 3 рис.~24.1,  а поскольку для  всех  $n$
$\im  h_{nb}\geq\tau_0$  и  $\im  h_{nd}\geq\tau_0$ (в разветвлённом
волноводе собственные волны затухают не медленнее,  чем основная волна
в  широком  волноводе),  то  $J_+(\alpha)$ регулярно по крайней мере в
полуплоскости $\tau>-\tau_0$.

     Преобразование Фурье   функции   $g(x,x_0;z)$   можно   найти  из
уравнения  (24.29),  не  определяя  саму  функцию;  её  образ  Фурье
$G(x,x_0;\alpha)$ удовлетворяет уравнению
     $$ \Bigl(\frac{\partial^2}{\partial x^2}- \gamma^2\Bigr)
         G(x,x_0;\alpha)=-\delta(x-x_0)\,,\eqno(24.42)$$
где $\gamma^2=\alpha^2-K^2$. Так как функция $g(x,x_0;z)$ представляет
собой линейную комбинацию собственных волн типа $H_{n0}$ области 1, то
$G(x,x_0;\alpha)$ регулярна в полосе $|\tau|<\tau_0$. Непосредственной
подстановкой   в  уравнение  нетрудно  убедиться,  что  его  решением,
удовлетворяющим граничным условиям при $x=0$ и $x=a$, является функция
     $$ G(x,x_0;\alpha)=\left\{\begin{array}{l}\displaystyle{\frac{
         \sh{\gamma(a-x_0)}\,\sh{\gamma x}}{\gamma\sh{\gamma a}}}\,,
         \quad x<x_0\,,\\[.4cm]\displaystyle{\frac{\sh{\gamma x_0}\,
         \sh{\gamma (a-x)}}{\gamma\sh{\gamma a}}}\,,\quad x>x_0\,;
         \\[.2cm]\end{array}\right.\eqno(24.43)$$
отметим здесь,  что  $G(x,x_0;\alpha)$  ---  {\it  чётная}   функция
$\alpha$.  Полагая теперь в (24.42) $x=x_0=b$, находим требуемый образ
Фурье функции $g(z)$:
     $$G(\alpha)=\frac{\sh\gamma b\,\sh\gamma d}{\gamma\,\sh
         \gamma a},\qquad\gamma=\sqrt{\alpha^2-K^2}.\eqno(24.44)$$

     Используя асимптотику  функции  $g(z-z_0)$  при  $z\to  -\infty$,
убеждаемся,  что преобразование Фурье $B_-(\alpha)$  неизвестной  пока
функции  $b_-(z)$,  определённой  интегралом  (24.40),  регулярно  в
нижней полуплоскости $\tau<\tau_0$. Поэтому образы Фурье всех функций,
входящих в уравнение (24.39),  регулярны в полосе $|\tau|<\tau_0$, и к
этому  уравнению  может  быть  применена  теорема  о  свёртке.   В
результате   получаем  алгебраическое  {\it  функциональное  уравнение
Винера-Хопфа-Фока}
     $$J_+(\alpha)G(\alpha)=A_+(\alpha)-B_-(\alpha),\qquad
         |\tau|<\tau_0\,,\eqno(24.45)$$
где известные функции $A_+(\alpha)$ и $G(\alpha)$ определены формулами
(24.41) и (24.44),  а {\it две} неизвестные  функции  $J_+(\alpha)$  и
$B_-(\alpha)$ необходимо найти из {\it одного} уравнения.

     Важнейшим моментом   решения   уравнения  (24.45)  является  {\it
факторизация}  входящей  в   него   функции   $G(\alpha)$,   то   есть
представление её в виде произведения двух функций:
     $$G(\alpha)=G_+(\alpha)G_-(\alpha),\eqno(24.46)$$
одна из  которых   ---   $G_+(\alpha)$   ---   регулярна   в   верхней
полуплоскости ($\tau>-\tau_0$),  другая --- $G_-(\alpha)$ --- в нижней
полуплоскости  ($\tau<\tau_0$)  комплексной  переменной  $\alpha$.  Не
вдаваясь  во все тонкости ТФКП,  укажем достаточно общие условия,  при
которых факторизация функции всегда может быть  выполнена,  а  именно:
{\it   если   $G(\alpha)$   регулярна   и  не  имеет  нулей  в  полосе
$\tau_-<\tau<\tau_+,\;  -\infty<\sigma<\infty$  и   $G(\alpha)\to   1$
равномерно  при  $\sigma\to\pm\infty$  в  этой  полосе,  то существует
представление}  (24.46),  причём   обе   функции   $G_+(\alpha)$   и
$G_-(\alpha)$ не имеют нулей в областях своей регулярности.  Очевидно,
что функция (24.44) удовлетворяет этим условиям  и,  более  того,  эта
функция  мероморфная,  то  есть представима в виде частного двух целых
функций;  всякая  же  целая  функция  $G(\alpha)$,  в  свою   очередь,
представима в виде произведения сомножителей $1-\alpha/\alpha_n$,  где
$\alpha_n$ --- простые нули $G(\alpha)$.

В рассматриваемом случае факторизация упрощается из-за того,  что
целые функции {\it чётные} по $\alpha$ и поэтому могут быть записаны
в виде
     $$G(\alpha)=G(0)\prod\limits_{n=1}^\infty\Bigl[1-\Bigl(\frac
         {\alpha}{\alpha_n}\Bigr)^2\Bigr]\,,\eqno(24.47)$$
откуда сразу находим, что
     $$G_{\pm}(\alpha)=\sqrt{G(0)}e^{\pm\chi(\alpha)}\prod\limits_
         {n=1}^\infty\Bigl(1\pm\frac{\alpha}{\alpha_n}\Bigr)
         e^{\pm\alpha/\alpha_n}\,, \eqno(24.48)$$
где везде надо брать одновременно либо верхние,  либо нижние знаки,  а
$\chi(\alpha)$ --- произвольная  целая  функция.  Выбор  этой  функции
производится  с учётом условия на ребре,  которое требует степенного
характера  роста  функций  $G_{+}(\alpha)$   и   $G_{-}(\alpha)$   при
$|\alpha|\to\infty$.   Опуская   подробности,  приведем  окончательный
результат факторизации функции $G(\alpha)$:

     $$\qquad G_-(\alpha)=G_+(-\alpha);$$
     $$ G_+(\alpha)=\sqrt{\frac{\sin kb\,\sin kd}{ k\sin ka}}\,\exp
         \left\{i{\alpha\over\pi}\left[b\ln\Bigl({a\over b}\Bigr)+d\ln
         \Bigl({a\over d}\Bigr)\right]\right\}\times\eqno(24.49)$$
     $$\times\frac{\prod\limits_{n=1}^\infty(1+\alpha/h_{nb})\,e^{i
         \alpha b/n\pi}\;\prod\limits_{n=1}^\infty(1+\alpha/
         h_{nd})\,e^{i\alpha d/n\pi}}{\prod\limits_{n=1}^\infty
         (1+\alpha/h_{na})\,e^{i\alpha a/n\pi}}.$$

     Теперь уравнение (24.45) может быть переписано в виде

     $$J_+(\alpha)G_+(\alpha)=\frac{A_+(\alpha)}{G_-(\alpha)}-\frac
         {B_-(\alpha)}{G_-(\alpha)},\qquad|\tau|<\tau_0;\eqno(24.50)$$
левая часть этого  равенства  регулярна  в  верхней  полуплоскости,  а
последний  член  правой  части  ---  в нижней.  Следующий этап решения
задачи методом Винера-Хопфа-Фока заключается в том,  чтобы осуществить
{\it   разложение}   функции   ${A_+(\alpha)/   G_-(\alpha)}$  на  два
слагаемых,  одно из которых регулярно в верхней, а другое --- в нижней
полуплоскости:

     $$\frac{A_+(\alpha)}{G_-(\alpha)}=S_-(\alpha)+S_+(\alpha).
          \eqno(24.51)$$
Единственной особенностью этой функции в нижней полуплоскости является
полюс первого порядка в точке $\alpha=-h_{1a}$,  что  позволяет  после
несложного преобразования

     $$\frac{A_+(\alpha)}{G_-(\alpha)}=\frac{i\sin(\pi b/a)}{\sqrt
         {2\pi}(\alpha+h_{1a})}\left[\frac 1{ G_-(\alpha)}-\frac1{G_-
         (-h_{1a})}\right]+\frac{i\sin(\pi b/a)}{\sqrt{2\pi}(
         \alpha+h_{1a})}\frac 1{G_-(-h_{1a})}\eqno(24.52)$$
определить $S_{-}(\alpha)$ и $S_{+}(\alpha)$. Записав теперь уравнение
(24.51) в виде
     $$J_+(\alpha)G_+(\alpha)-S_+(\alpha)=S_-(\alpha)-B_-(\alpha)/G_-
         (\alpha),\qquad|\tau|<\tau_0,\eqno(24.53)$$
убеждаемся, что его левая часть регулярна при $\tau>-\tau_0$, а правая
---  при  $\tau<\tau_0$,  и   при   этом   имеется   полоса   $-\tau_0
<\tau<\tau_0$,  в  которой  обе  части  регулярны  и равны.  Последнее
позволяет на основании свойств аналитического продолжения  утверждать,
что  уравнение определено во всей плоскости и его правая и левая части
представляют собой одну и ту же целую функцию $P(\alpha)$.  Из условия
на ребре следует,  что функция $P(\alpha)$ ограничена и равна нулю при
$|\alpha|\to\infty$;   следовательно,   согласно   теореме   Луивилля,
$P(\alpha)$  тождественно  равно  нулю  во  всей  плоскости  и,  таким
образом,  определяется решение для  обеих  искомых  функций  уравнения
(24.45):
     $$ J_+(\alpha)=\frac{S_+(\alpha)}{G_+(\alpha)}\,,\qquad
         B_-(\alpha)=S_-(\alpha)\,G_-(\alpha)\,.\eqno(24.54)$$

\begin{wrapfigure}[10]{l}{7.cm}
\begin{picture}(60,45)
\put(0,45){\special{em:graph fig24-3.bmp}}
\end{picture}
\hbox to 7.0cm{\hfil\footnotesize{Рис.~24.3.~Контуры интегрирования.
}\hfil}
\end{wrapfigure}
     Для нахождения  плотности  наведённого тока $j_+(z)$ необходимо
вычислить обратное преобразование Фурье:
     $$j_+(z)=\frac 1{\sqrt{2\pi}}\int\limits_{-\infty}^\infty\frac
         {S_+(\alpha)}{G_+(\alpha)}e^{-i\alpha z}\,d\alpha,
         \eqno(24.55)$$
где функции $S_+(\alpha)$  и~$G_+(\alpha)$  известны.  Подынтегральная
функция  мероморфна  и  имеет  в нижней полуплоскости простые полюсы в
точках
     $$\alpha=-h_{nb},\quad\alpha=-h_{nd},\quad n=1,2,3\ldots\,
         \,;\eqno(24.56)$$
полюс функции  $S_+(\alpha)$  в  точке $\alpha=-h_{1a}$ компенсируется
нулём функции $G_+^{-1}( \alpha)$ в этой же точке.  Выполним  теперь
интегрирование  в  плоскости комплексной переменной $\alpha$,  выбирая
контуры интегрирования при $z<0$ и $z>0$  так,  как  это  показано  на
рис.~24.3.   При   $z<0$   функция  $j_+(z)=0$,  так  как  в  области,
ограниченной контуром $C_{z<o}$,  подынтегральная  функция  регулярна.
При $z>0$ интеграл (24.55) удобно преобразовать к виду
     $$j_+(z)=\frac i{2\pi}\,\sin{\Bigl(\frac{\pi b}a\Bigr)}\,\frac 1
         {G_-(-h_{1a})}\,\int\limits_{C_{z>0}}\frac{G_-(\alpha)}
         {\alpha+h_{1a}}\,\frac{\gamma\sh{\gamma a}}{\sh{\gamma b}
         \,\sh{\gamma d}}\, e^{-i\alpha z}\,d\alpha\eqno(24.57)$$
и нахождение   функции  $j_+(z)$  сводится  к  вычислению  вычетов.  В
результате получаем, что
     $$j_+(z)=\sum\limits_{m=1}^\infty(\widetilde{B}_m e^{ih_{mb}
         z}+\widetilde{C}_me^{i h_{md}z}),\qquad z>0,\eqno(24.57)$$
где
     $$\widetilde{B}_m=\frac {(m\pi)^2\sin{(\pi b/a)}}{b^3 h_{mb}
         (h_{1a}-h_{mb})}\,\frac{G_-(-h_{mb})}{G_-(-h_{1a})}\,,
         \eqno(24.59)$$
а коэффициенты $\widetilde C_m$  определяются  такой  же  формулой,  в
которой все символы $b$ (в том числе и в индексах) заменены на $d$.

     При вычислении рассеянного поля  удобно  вместо  формулы  (24.34)
воспользоваться теоремой о свёртке
     $$\varphi(x,z)=-\frac 1{\sqrt{2\pi}}\int\limits_{-\infty+i\tau}
         ^{\infty+i\tau} G(b,x;\alpha)J_+(\alpha)e^{-i\alpha z}\,d
         \alpha,\eqno(24.60)$$
где образ Фурье $G(b,x;\alpha)$ определен формулой (24.43). Вычисление
этого интеграла производится аналогично предыдущему с помощью вычетов.
В  результате  получаются  те же выражения (24.3) и (24.24),  что были
получены ранее прямым методом  переразложения  и  решения  системы  по
правилу Крамера.

     Перед тем как провести сравнение двух методов решения одной и той
же задачи о разветвлённом волноводе,  остановимся кратко ещё на их
двух разновидностях. Широко используемый {\it метод Джонса} отличается
от метода Винера-Хопфа-Фока тем,  что в нём функциональное уравнение
выводится    не   на   основе   интегрального   уравнения,   а   путем
непосредственного  применения  преобразования  Фурье  к  уравнению   в
частных  производных,  определяющему  рассеянное  поле  (как  это было
сделано выше для функции Грина).

     Преобразование Фурье  уравнения (24.2) приводит к уравнению
     $$\Bigl(\frac{\partial^2}{\partial x^2}-\gamma^2\Bigr)\Phi(x,
         \alpha)=0\,,\qquad \gamma^2=\alpha^2-K^2\,,\eqno(24.61)$$
которое справедливо  в  полосе  $|\tau|<\tau_0$,   где   его   решение
$\Phi(x,\alpha)$  ---  регулярная  функция.  Решение,  удовлетворяющее
граничным условиям при $x=0$ и $x=a$, может быть записано в виде
     $$  \Phi(x,\alpha)=\Phi_+(x,\alpha)+\Phi_-(x,\alpha)=
         \left\{\begin{array}{ll}C(\alpha)\sh{\gamma(a-x)}\quad&\mbox
         {при} \quad a>x>b\,,\\[.2cm]D(\alpha)\sh{\gamma x}\quad&\mbox
         {при} \quad b>x>0\,.\\[.2cm]\end{array}\right.\eqno(24.62)$$

     Полное электрическое поле на разветвляющей пластине  равно  нулю,
поэтому
     $$  \varphi(b,z)=-\sin{\frac{\pi b} a}\,e^{i h_{1a}z},\qquad
         z>0\,,\eqno(24.63)$$
так что образ Фурье
     $$\Phi_+(b,\alpha)=-\frac 1{\sqrt{2\pi}}\,\sin{\Bigl(\frac
         {\pi b}a\Bigr)}\frac i{\alpha+h_{1a}}=-A_+(\alpha).
         \eqno(24.64)$$
Из условия непрерывности поля  при  $x=b$  следует  соотношение  между
функциями $C(\alpha)$ и $D(\alpha)$:
     $$  C(\alpha)\, \sh{\gamma d}=D(\alpha)\,\sh{\gamma b}\,.
         \eqno(24.65)$$

     Магнитная составляющая поля $H_z$ в  плоскости  $x=b$  при  $z>0$
терпит  разрыв,  в  результате  чего  между образами Фурье имеет место
соотношение
     $$\left.\frac{\partial \Phi(x,\alpha)}{\partial x}\right|_{x=b+0}
         -\left.\frac{\partial \Phi(x,\alpha)}{\partial x}\right|_
         {x=b-0}=J_+(\alpha);\eqno(24.66)$$
подстановка в  эту  формулу  выражения  (24.62) для $\Phi(x,  \alpha)$
позволяет представить $J_+(\alpha)$ в виде
     $$J_+(\alpha)=-\gamma\, C(\alpha)\,\ch\gamma d-\gamma\,
         \ch\gamma b\, D(\alpha)\,.\eqno(24.67)$$
Отсюда с учётом (24.65) находим следующие соотношения:
     $$C(\alpha)=-\frac{\sh{\gamma b}}{\gamma\sh{\gamma a}}\,
          J_+(\alpha),\qquad D(\alpha)=-\frac{\sh{\gamma d}}{\gamma
         \sh{\gamma a}}\, J_+(\alpha).\eqno(24.68)$$
Подставляя теперь (24.68) в (24.62) и полагая $x=b$, получаем уравнение
     $$\hbox{$-A_+(\alpha)+\Phi_-(b,\alpha)=\cases{-\displaystyle
         {\frac{\sh{\gamma b}}{\gamma\sh{\gamma a}}}\,J_+(\alpha)
         \sh{\gamma d}&\cr\noalign{\vskip4pt}-\displaystyle{\frac
         {\sh\gamma d}{\gamma\sh\gamma a}}\,J_+(\alpha)\sh\gamma b&
         \cr}=-G(\alpha)J_+(\alpha)\,,$}\eqno(24.69)$$
которое, как нетрудно видеть,  тождественно совпадает с функциональным
уравнением  Винера-Хопфа-Фока  (24.45),  поскольку   $\Phi_-(b,\alpha)
\equiv B_-(\alpha)$.

     Несомненно, описанный  вывод  функционального  уравнения  методом
Джонса проще, чем при использовании интегрального уравнения, так что в
подавляющем большинстве случаев именно его и  используют  при  решении
конкретных  задач.  Существует  и широко используется ещё один метод
решения дифракционных задач в области СВЧ --- {\it метод  вычетов},  в
основе   которого   также   лежит  ТФКП.  Исходная  система  уравнений
получается  прямым  методом  сшивания  разложений  полей  в  отдельных
областях   структуры.   В  рассматриваемой  задаче  о  разветвлённом
волноводе это --- парная  система  (24.6);  её  решение  строится  с
помощью интегралов
     $$\frac 1{2\pi i}\oint\limits_{C}\frac{f(w)}{w+i h_{mb}}\,dw,
         \qquad\frac 1{2\pi i}\oint\limits_{C}\frac{f(w)}{w+i h_{md}}
         \,dw\eqno(24.70)$$
по окружности $C$ бесконечно большого радиуса в плоскости комплексного
переменного $w$, проходимой против часовой стрелки.

     Предположим, что функция $f(w)$ удовлетворяет следующим условиям:

     1. $f(w)$  ---  аналитическая  функция  во всей плоскости,  кроме
полюсов первого порядка в точках $w=-i h_{ma}\;(m=1,2,\dots)$ и  $w=+i
h_{1a}$.

     2. $f(w)$  имеет  простые  нули  в  точках  $w=-i h_{mb}$ и $w=-i
h_{md}$.

     3. Вычет $Res\,f(ih_{1a})=A'$.

     4. При $|w|\to\infty$ функция $f(w)=O(w^{-3/2})$.

     Выражая интегралы (24.70)  в  виде  суммы  вычетов  в  полюсах  и
приравнивая  результаты  нулю  (вычет в бесконечно удалённой точке),
получаем парную сиcтему
     $$\left.\begin{array}{lclcc}\displaystyle{ \sum_{n=1}^\infty
         \frac 1{h_{na}-h_{mb}}\,Res\,f(-i h_{na})}&-&\displaystyle
         {\frac{A'}{h_{1a}+h_{mb}}}&=&0 \,,\\[.5cm]\displaystyle
         {\sum_{n=1}^\infty\frac 1{h_{na}-h_{md}}\,Res\,f(-i h_{na}
         )}&-&\displaystyle{\frac{A'}{h_{1a}+h_{md}}}&=&0\,,\end
         {array}\right\}\quad m=1,2,\dots ,\eqno(24.71)$$
которая совпадает с системой (24.6) при условии
     $$A_n\sin{\frac{\pi b} a}=Res\;f(-i h_{na}).\eqno(24.72)$$

     Таким образом,   метод  вычетов  сводится  к  нахождению  функции
$f(w)$,  обладающей перечисленными выше свойствами.  В рассматриваемой
задаче такой функцией является
     $$\begin{array}{l} f(w)=\displaystyle{\frac{\sin{(\pi b/a)}}
         {-i h_{1a}+w}}\,e^{\frac{w-i h_{1a}} {\pi}[b\ln{
        ( a/b)}+d\ln{( a/d)}]
         }\times\\[.5cm]\qquad\times\prod\limits_{n=1}^\infty
         \displaystyle{\frac{1-i w/h_{nb}}{1+h_{1a}/h_{nb}}\,e^{(w-
         ih_{1a})b/n\pi}}\prod\limits_{n=1}^\infty\displaystyle{\frac
         {1-i w/h_{nd}}{1+h_{1a}/h_{nd}}\,e^{(w-ih_{1a})d/n\pi}
         }\;\times\\[.6cm]\qquad\times\prod\limits_{n=1}^\infty
         \displaystyle{\frac{1+h_{1a}/h_{na}}{1-iw/h_{na}}\,e^{-(w
         -ih_{1a})a/n\pi}}\,.\end{array}\eqno(24.73)$$
С помощью этой функции из равенства (24.72) находим коэффициенты
$A_n$, которые совпадают с полученным ранее решением (24.24).

         Изложенное выше  решение  задачи  о   разветвлённом   волноводе
фактически   четырьмя  разными  методами  позволяет  в  какой-то  мере
провести их сравнительный анализ.  Эти методы попарно объединяются  по
совпадающей  части  процесса решения:  метод прямого обращения и метод
вычетов  используют  одну  и  ту  же  бесконечную   систему   линейных
алгебраических   уравнений   для   коэффициентов  разложения  поля  по
собственным волнам частичных областей структуры и  различаются  только
по  способу  решения  этой системы,  а метод Винера-Хопфа-Фока и метод
Джонса,  наоборот,  разными способами приводят  к  одному  и  тому  же
алгебраическому  функциональному  уравнению  для  преобразования Фурье
искомой величины,  которое решается путем факторизации  и  разложения.
Для    получения   аналитических   решений   задач   дифракции   метод
Винера-Хопфа-Фока следует считать наиболее универсальным.  Он пригоден
для  решения  краевых  задач в структурах,  которые представляют собой
сочленение двух  или  более  полубесконечных  подобластей,  являющихся
частью   области,  представляющей  собой  координатную  поверхность  в
системе координат, где переменные в волновом уравнении разделяются.

     Метод Джонса,   несмотря   на   свою   сравнительную  простоту  и
привлекательность,  не является столь общим,  поскольку для
некоторых структур    получить    функциональное    уравнение без
применения модифицированного интегрального уравнения не удаётся,
например,  в случае,  когда  одно  из  разветвлений рассмотренного
волновода имеет диэлектрическое заполнение, отличное от остальных
областей структуры. Большим  достоинством этого метода перед
другой парой является хорошо разработанный математический  аппарат
факторизации  в   ТФКП.   Если сравнивать между  собой  метод
прямого обращения и метод вычетов,  то второй способ несомненно
обладает  большими  возможностями  в  смысле модификации  и
получении приближённых аналитических решений. Метод прямого
обращения (или решения системы уравнений по  правилу Крамера)
весьма ограничен случаями простого <<угадывания>>  решения для
небольшого числа определителей весьма специального вида. Однако,
как это видно на  ряде  примеров  из предыдущих разделов, этот
метод сравнительно прост и  удобен  для  получения формального
решения  через  обратную матрицу и дальнейшего численного
обращения матрицы.  Для структур,  не обладающих острыми
рёбрами,  при современном развитии вычислительной техники на
этом пути быстро и легко могут быть получены решения
--- по своей точности намного превышающие запросы  практики  ---
для  весьма сложных по своей геометрии структур.

%\end{document}

\newpage
\oddsidemargin=-0.4mm \evensidemargin=-0.4mm
\topmargin=-0.4mm
\headsep=7mm
\textheight=231.875mm
\textwidth=160mm
\mathsurround=2.5pt
\unitlength=1mm
%\begin{document}
%\input{macr.tex}
\thispagestyle{empty}
%\addtocounter{page}{286}

\begin{center}\subsubsection*{25. Метод задачи Римана-Гильберта.
         Расчёт дифракционного излучения сторонних источников
         методом Винера-Хопфа-Фока   }
\end{center}

\markboth{Глава 9.  Методы  ТФКП  в  электродинамике}{25. Метод
         задачи Римана-Гильберта}
\vspace*{0.5cm}

\begin{center}\begin{minipage}[c]{0.75\textwidth}
\footnotesize{\parindent=0.5cm
         Нормальное падение  плоской  волны  на периодическую решётку.
         Метод задачи Римана-Гильберта.  Решение Карлемана.  Излучение
         заряженной  нити  при  пролёте  с постоянной скоростью мимо
         края  идеально  проводящей  полуплоскости;  сопоставление   с
         задачей      Зоммерфельда.    О   дифракции   излучения
         Вавилова-Черенкова   при   движении   источника   вблизи
         полубесконечного   экрана.  Вывод  интегральных  уравнений  в
         случае пролёта токовой нити через конец плоского волновода.
}\end{minipage}\end{center}
\vspace*{0.5cm}

     Рассмотренный в  предыдущем   разделе   для решения задач дифракции
математический метод   Винера-Хопфа-Фока является частным случаем
более общего метода, который принято называть
{\it методом задачи Римана-Гильберта}. Прежде чем излагать
его суть и приводить решение частного случая этой задачи, предложенное
Карлеманом  и  находящее  широкое  применение  в   теории   дифракции,
остановимся   на   следующем   классическом  примере.  Речь  идёт  о
нормальном  падении   плоской   волны   на   периодическую   решётку
(постановка задачи),  образованную системой бесконечно тонких идеально
проводящих параллельных полос.  Обозначения геометрических  параметров
решётки и используемая система координат приведены на рис.~25.1.

     Пусть на решётку сверху $(z>0)$ падает плоская электромагнитная
волна
     $$\rv E^{in}=\rv E_0e^{-ikz}\,,\quad \rv H^{in}=\rv H_0
         e^{-ikz}\,.\eqno(25.1)$$
Рассмотрим далее  лишь  одну поляризацию волны,  когда отличны от нуля
только составляющие $E_{0x}$ и $H_{0y}$, ограничившись замечанием, что
в  случае  ортогональной  поляризации решение находится аналогично,  а
произвольная поляризация может быть представлена линейной  комбинацией
этих двух.  Для упрощения выкладок будем считать,  что амплитуда волны
равна  единице,  а  распространение  волны  происходит   в   свободном
пространстве.  Отметим  при этом,  что метод задачи Римана-Гильберта в
отличие от метода Винера-Хопфа-Фока в  принципе  не  требует  введения
затухания   в   среде,   поскольку   для  него  достаточно  выполнения
соответствующего  уравнения  только  на   некоторой   линии,   и   нет
необходимости в существовании конечной полосы регулярности в плоскости
комплексного переменного.

\begin{wrapfigure}[10]{l}{7.5cm}
\begin{picture}(80,40)
\put(1,40){\special{em:graph fig25-1.bmp}}
\end{picture}
\hbox to 7.5cm{\hfil\footnotesize{Рис.~25.1.~Периодическая решётка.
}\hfil}
\end{wrapfigure}

     Из-за симметрии структуры в рассеянном поле отличны от нуля те же
составляющие,  что и в падающем, отсутствует зависимость от координаты
$x$  и  должна иметь место периодичность по $y$ с периодом $l$.  Решая
волновое уравнение методом разделения переменных,  нетрудно убедиться,
что  электрическое поле в верхнем и нижнем полупространстве может быть
записано в виде
     $$E_x=e^{-ikz}+\sum\limits_{n=-\infty}^\infty a_ne^{i\sqrt{k^2-
         \bigl(\frac{2\pi n} l\bigr)^2}\,z}\,e^{i\frac{2\pi n y}l}\,,
         \qquad z>0\,;\eqno(25.2)$$
     $$E_x=\sum\limits_{n=-\infty}^\infty b_ne^{-i\sqrt{k^2-\bigl(
         \frac{2\pi n} l\bigr)^2}\,z}\,e^{i\frac{2\pi n y}l}\,,
         \qquad z<0\,,\eqno(25.3)$$
где постоянные  коэффициенты  $a_n$  и  $b_n$  являются  неизвестными.
Уравнения  для их определения следуют из граничных условий в плоскости
$z=0$,  то есть непрерывности  полей  на  щелях  и  обращения  в  нуль
электрического  поля  на полосах.  В результате получаем из условий на
электрическое поле равенства
     $$ b_0=1+a_0\,,\qquad b_n=a_n\eqno(25.4)$$
и уравнение
     $$\sum\limits_{n=-\infty}^\infty b_n e^{i\frac{2\pi n y} l} = 0
         ,\eqno(25.5)$$
справедливое на  полосах,  а  из  непрерывности  магнитного  поля  ---
уравнение
     $$-k+\sum\limits_{n=-\infty}^\infty b_n\sqrt{k^2-\bigl(\frac
         {2\pi n}l\bigr)^2}\;e^{i\frac{2\pi n} l y}=0,
         \eqno(25.6)$$
справедливое на щелях.

     Вводя обозначения
     $$ \psi=\frac {\pi d} l\,,\quad \varphi=\frac{2\pi y}l\,,\quad
         \mbox{\ae}=\frac{kl}{2\pi}=\frac l{\lambda}\,,\quad \chi_n=1+i\sqrt
         {\frac{\mbox{\ae}^2}{n^2}-1}\,,\eqno(25.7)$$
где $\lambda$ --- длина волны падающего поля, можно переписать
парные функциональные уравнения (25.5) и (25.6) в виде
     $$ \left.\begin{array}{lr}
         \sum\limits_{n=-\infty}^\infty b_n e^{in\varphi}=0\,,
         & \psi< |\varphi|\leqslant\pi\,;\\
         [.4cm]\sum\limits_{n=-\infty}^\infty b_n
         |n|(1-\chi_n)e^{i n \varphi}=i\mbox{\ae}(b_0-1)\,,&
         |\varphi|< \psi \,.\end{array}\right\}\eqno(25.8)$$
Системы функциональных  уравнений  такого  вида  часто  встречаются  в
задачах дифракции. Их отличительным и важным для дальнейшего свойством
является  стремление  $\chi_n\to  0$  при  $|n|\to \infty$.  Посмотрим
теперь,  каким  образом   система   (25.8)   соотносится   с   задачей
Римана-Гильберта,   которая  ---  без  излишней  для  нашего  краткого
изложения  математической  строгости   ---   формулируется   следующим
образом.  Пусть  в  плоскости  комплексного переменного задана гладкая
линия $L$,  целиком лежащая в ограниченной области и по разные стороны
которой  заданы  предельные  значения некоторой аналитической функции;
требуется восстановить по сумме этих значений  саму  функцию  во
всей плоскости.

     С помощью конформного преобразования линию $L$ можно отобразить в
дугу $L_1$ окружности единичного радиуса (рис.~25.2)  так,  что  концы
дуги  будут  расположены  в комплексно-сопряжённых точках $\alpha$ и
$\alpha^*$.  Введём две  функции  $w^+(z)$  и  $w^-(z)$,  регулярные
соответственно  внутри  и  вне окружности $|z|=1$,  где они могут быть
представлены степенными рядами
     $$w^+(z)=\sum\limits_{n>0}w_n z^n\,;\qquad w^-(z)=-\sum\limits_
         {n<0}w_n z^n\,.\eqno(25.9)$$

\begin{wrapfigure}[12]{l}{7.5cm}
\begin{picture}(80,45)
\put(0,45){\special{em:graph fig25-2.bmp}}
\end{picture}
\hbox to 7.2cm{\hfil\footnotesize{Рис.~25.2.~Единичная окружность
}\hfil}
\hbox to 6.0cm{\hfil\footnotesize{задачи Римана-Гильберта.}\hfil}
\end{wrapfigure}

     Разность этих    функций на окружности единичного  радиуса
представляет  собой  ряд  Фурье по аргументу $\varphi$:
     $$ w^+(e^{i\varphi})-w^-(e^{i\varphi})=\sum\limits_{n\ne 0}w_n
         e^{i n\varphi}\,.\eqno(25.10)$$
Пусть на дуге окружности $L_2$ эти функции совпадают, что записывается
в виде
     $$\sum_{n\ne 0} w_ne^{in\varphi}=0,\qquad |\varphi|>\psi\,,
         \eqno(25.11)$$
где $\psi$  определяется  равенством   $\alpha=e^{i\psi}$.   Тогда   в
соответствии  с  теорией  аналитического  продолжения они представляют
собой одну и ту же аналитическую функцию
     $$w(z)=\left\{\begin {array}{l} w^+(z)\quad\mbox{при}\quad |z|<1;
         \\[.4cm]w^-(z)\quad\mbox{при}\quad|z|>1.\end{array}\right.
         \eqno(25.12)$$

     Обозначим сумму  предельных значений функции $w(z)$ на дуге $L_1$
как
     $$u(e^{i\varphi})=w^+(e^{i\varphi})+w^-(e^{i\varphi})\,,
         \eqno(25.13)$$
что можно записать в виде
     $$\sum\limits_{n\ne 0} w_n\frac {|n|} n e^{i n \varphi}=
         u(e^{i\varphi}), \qquad|\varphi|<\psi\,.\eqno(25.14)$$
Совокупность функциональных  уравнений  (25.11)  и (25.14),  в которых
функция
     $$u(e^{i\varphi})=\sum\limits_{n=-\infty}^\infty u_ne^
         {i n\varphi}\eqno(25.15)$$
задана, а $w(z)$ --- искомая функция,  и есть задача Римана-Гильберта,
простое решение которой предложил Карлеман;  оно  излагается  ниже  по
возможности кратко.

     Проведём в плоскости комплексного переменного разрез вдоль дуги
$L_1$  и  тогда  в  этой  плоскости с разрезом искомая функция $w(z)$,
вспомогательная   функция   $\sqrt{(z-\alpha)    (z-\alpha^*)}$,    а,
следовательно,  и их произведение  однозначны и регулярны.  Поскольку
при $|z|\to \infty$
     $$ w(z)=w_-(z)=z^{-1}(-w_{-1}+O(1)),\eqno(25.16)$$
то остаётся ограниченным и произведение,  которое согласно свойствам
интеграла Коши может быть записано в виде
     $$w(z)\sqrt{(z-\alpha)(z-\alpha^*)}=\frac 1{2\pi i}\oint
         \limits_\Gamma \frac{w(\zeta)\sqrt{(\zeta-\alpha)(\zeta-
         \alpha^*)}}{\zeta-z}\,d\zeta+C,\eqno(25.17)$$
где постоянная  $C$  ---  результат  вычисления  вычета  в  бесконечно
удалённой точке, а контур $\Gamma$ охватывает разрез вдоль $L_1$.

     Стянем контур  $\Gamma$ к дважды пробегаемой дуге $L_1$;  тогда с
учётом знака корня справа и слева от дуги формула (25.17)  переходит
в формулу
     $$w(z)\sqrt{(z-\alpha)(z-\alpha^*)}=\frac 1{2\pi i}\int
         \limits_{L_1} \frac{[w^+(\zeta)+w^-(\zeta)]\sqrt{(\zeta-
         \alpha)(\zeta-\alpha^*)}}{\zeta-z}\,d\zeta+C\,,\eqno(25.18)$$
которая и  представляет  собой  решение задачи Римана-Гильберта в виде
интеграла типа Коши.

     Основной интерес  для  дальнейшего  представляет значение функции
$w(z)$ на дуге единичного  радиуса,  точнее,  коэффициенты  разложения
$w_n$  разности  (25.10) в ряд Фурье.  Для их нахождения воспользуемся
формулами Сохоцкого,  определяющими предельные значения интеграла типа
Коши
%     $$ F(z)=\frac 1 {2\pi i}\int\limits_L \frac{f(\zeta)}{\zeta-z}
%         \,d\zeta\eqno(25.19)$$
в точке $\zeta_0$ линии $L$:
     $$F^+(\zeta_0)=F(\zeta_0)+\frac 1 2 f(\zeta_0),\qquad F^-
         (\zeta_0)=F(\zeta_0)-\frac 1 2 f(\zeta_0),\eqno(25.19)$$
где интеграл  $F(\zeta_0)$  понимается  в смысле главного значения.  В
результате из (25.18) получаем
     $$\begin{array}{c}[w^+(\zeta_0)-w^-(\zeta_0)]\sqrt{(\zeta_0-
         \alpha)(\zeta_0-\alpha^*)}=\\[.4cm]=\displaystyle{\frac 1
         {\pi i}\int\limits_{L_1}\frac {[w^+(\zeta)+w^-(\zeta)]\sqrt{
         (\zeta-\alpha)(\zeta-\alpha^*)}}{\zeta-\zeta_0}\,d\zeta}
         +2C\,,\end{array} \eqno(25.20)$$
где точка $\zeta_0$ лежит на дуге $L_1$. Вводя вспомогательную функцию
     $$R(\zeta)=\left\{\begin{array}{cl}\displaystyle{\frac 1
         {\sqrt{(\zeta-\alpha)(\zeta-\alpha^*)}}}&\qquad
         \mbox{на дуге}\quad L_1,\\[.4cm]  0&\qquad\mbox{на дуге}
         \quad L_2\end{array}\right.\eqno(25.21)$$
и используя обозначение (25.13), преобразуем (25.20) к виду
     $$w^+(\zeta_0)-w^-(\zeta_0)=\frac{R(\zeta_0)}{\pi i}\int\limits_
         {L_1}\frac{u(\zeta)\sqrt{(\zeta-\alpha)(\zeta-\alpha^*)}
         }{\zeta-\zeta_0}\,d\,\zeta+2CR(\zeta_0).\eqno(25.22)$$
Разложим обе части этого равенства в ряд Фурье по $\varphi$ и введ\"ем
следующие обозначения:
     $$ \left.\begin{array}{c}V_n(e^{i\varphi})=\displaystyle{\frac1{\pi}
         \int\limits_{-\psi}^{\psi}\frac{e^{i n\varphi'+i\varphi'}}
         {e^{i\varphi'}-e^{i\varphi}}\sqrt{
         (e^{i\varphi'}-e^{i\psi})(e^{i\varphi'}-e^{-i\psi})}}\,
         d\varphi'\,,\\[.4cm]V_m^n=\displaystyle\frac 1{2\pi}\int
         \limits_{-\pi}^{\pi}V_n(e^{i\varphi})\,R(e^{i\varphi})\,e^
         {-i m \varphi}\,d\varphi\,,\\[.4cm]R_m=\displaystyle\frac 1
         {2\pi}\int\limits_{-\pi}^{\pi}R(e^{i\varphi})\,e^{-i m
         \varphi}\,d\varphi\,.\end{array}\right \}\eqno(25.23)$$
Интегралы, определяющие  величины  $V_m^n$  и   $R_m$   и   являющиеся
функциями только параметра $\psi$,  выражаются через полиномы Лежандра
от  аргумента  $\cos{\psi}$  (соответствующие  выражения  приведены  в
Приложении П-8).

     После исключения  постоянной  $C$    с  учётом  (25.10) и (25.14) получаем
для коэффициентов $w_m$ следующее  выражение:
     $$w_m=\sum\limits_{n=-\infty}^{\infty}\,u_nV_m^n-\frac{R_m}{R_0}
         \sum\limits_{n=-\infty}^{\infty}\,u_nV_0^n\,,\eqno(25.24)$$
что и есть искомое решение задачи.

     Система уравнений (25.8),  описывающая дифракцию плоской волны на
решётке,  совпадает  с парой уравнений (25.11),  (25.14),  к которым
сводится задача Римана-Гильберта, если сопоставить
     $$ w_n\Longleftrightarrow nb_n\,\qquad u_0\Longleftrightarrow
         i\mbox{\ae}(b_0-1)\,\qquad u_n\Longleftrightarrow |n|b_n\chi_n
         \eqno(25.25)$$
и продифференцировать первое уравнение системы  (25.8)  по  $\varphi$.
Поскольку  величины $u_n$ при такой замене включают в себя неизвестные
$b_n$,   то   замкнутое   аналитическое   решение    (25.24)    задачи
Римана-Гильберта    переходит    в    бесконечную   систему   линейных
алгебраических уравнений относительно $b_n$:
     $$mb_m=i\mbox{\ae}(b_0-1)\Bigl(V_m^0-\frac{R_m}{R_0}V_0^0\Bigr)+
         \sum\limits_{n\ne 0}|n|b_n\chi_n\Bigl(V_m^n-\frac{R_m}{R_o}
         V_0^n\Bigr),\qquad m\ne 0,\eqno(25.26)$$
которое необходимо дополнить соотношением
     $$b_0=-\sum\limits_{n\ne 0}(-1)^nb_n\,,\eqno(25.27)$$
вытекающим из первого уравнения (25.8) при $\varphi=\pi$.

     Бесконечная система (25.26) удобна для численного решения методом
редукции.  При определённых значениях  параметров  $\mbox{\ae}$
и  $\psi$, соответствующих  малому  отношению  периода  решётки
к длине волны, методом последовательных приближений может быть
получено аналитическое решение,  хорошо  описывающее основные
характеристики дифрагированного поля.   Подробнее  останавливаться
на  этих  вопросах  здесь  нецелесообразно.

     Рассмотрим теперь,   используя   метод   Винера-Хопфа-Фока,   две
классические задачи  {\it  дифракционного  излучения}.  Этим  термином
принято обозначать поле,  возбуждаемое равномерно движущимся сторонним
заряженным или токовым источником при пролёте его вблизи  оптической
неоднородности. Начнём с анализа излучения, испускаемого бесконечной
однородно  заряженной  нитью,   пролетающей   вблизи   края   идеально
проводящей бесконечно тонкой полуплоскости. Нить с линейной плотностью
заряда $q$ ориентирована вдоль оси $y$ и  движется  со  скоростью  $v$
вдоль  оси  $z$  на  расстоянии  $a$ от края полуплоскости $z=0\;,x>0$
(рис.~25.3). Плотности заряда и тока записываются в этом случае в виде
     $$ \rho(\rv r,t)=q\delta(x+a)\delta(z-vt)\,,\qquad \gv j_z(\rv r,t)
         = vq\delta(x+a)\delta(z-vt)\,.\eqno(25.28)$$
При решении задач такого класса  естественно  использовать  разложение
всех физических величин в интеграл Фурье по времени, например,
     $$\mbox{\boldmath$\gv E$}(\rv r,t)=\int\limits_
        {-\infty}^\infty
         \rv E(\rv r,\omega)e^{-i\omega t}\,d\omega\,,\qquad \mbox{\boldmath
         $\gv j$}
         (\rv r,t)=\int\limits_{-\infty}^\infty \rv j
         (\rv r,\omega)e^{-i\omega t}\,d\omega.\eqno(25.29)$$
%\vspace{-.2cm}
\begin{wrapfigure}[12]{l}{7.5cm}
\begin{picture}(80,40)
\put(-2,45){\special{em:graph fig25-3.bmp}}
\end{picture}
\hbox to 7.5cm{\hfil\footnotesize{Рис.~25.3.~Пролёт нити вблизи }\hfil}
\hbox to 5.8cm{\hfil\footnotesize{ края полуплоскости. }\hfil}
\end{wrapfigure}

     Поскольку физические величины действительные,  то их Фурье-образы
удовлетворяют  соотношениям  вида  $\rv  E(\rv r,-\omega)=\rv E^* (\rv r,\omega)$,
так что достаточно найти их  для  $\omega>0$.  Фактически
Фурье-образы  представляют собой комплексные амплитуды соответствующих
величин монохроматического поля.  В дальнейшем везде, где это не может
привести  к  недоразумениям,  у  них  опущен аргумент $\omega$,  а все
выражения приведены для $\omega>0$.

     Дополнительное и   весьма   существенное   осложнение  в  задачах
дифракционного излучения по сравнению с задачами дифракции  состоит  в
том,  что  основной  интерес  в  этом  случае представляют зависимости
возбуждаемого поля и его обратного действия на источник  от  реального
времени.  Это  приводит к необходимости интегрирования по частоте всех
компонент возбуждаемого поля.  В остальном подход к задачам фактически
одинаков:  исходя  из  граничных  условий  на  поверхности  проводника
выводится интегральное уравнение (или  система  таких  уравнений)  для
наводимого  на  ней тока.  Основная трудность обычно состоит в решении
этих уравнений, и именно для этого используется метод Винера-Хопфа-Фока
(часто  в  несколько  модифицированном  по  сравнению  с  изложенным в
предыдущем  разделе  виде  и  называемом  в  литературе  {\it  методом
факторизации}).

     Начнём решение   задачи   с  нахождения  комплексной  амплитуды
падающего поля, то есть поля движущейся нити в свободном пространстве.
Скалярный  потенциал нити $\Phi'$ в её собственной системе координат
$x',\,y',\,z'$ (где она покоится) удовлетворяет уравнению Лапласа
     $$\Delta \Phi'=-4\pi q\delta(x'+a)\delta(z')\,.\eqno(25.30)$$
Имея в виду известное интегральное представление $\delta$-функции
     $$\delta(x)=\frac 1{2\pi}\int\limits_{-\infty}^\infty e^{i
         \alpha x}\,d\alpha\,,\eqno(25.31)$$
будем искать $\Phi'$ в виде двухкратного интеграла Фурье
     $$\Phi'(x',z')=\int\limits_{-\infty}^\infty\int\limits_{-\infty}
         ^\infty\varphi(\xi,\eta)e^{i[\xi(x'+a)+\eta z']}\,d\xi d
         \eta\,,\eqno(25.32)$$
то есть  фактически  в  виде  разложения  по плоским волнам с волновым
вектором $\rv k=\{\xi,0,\eta\}$. Подставляя это выражение в уравнение
(25.30), получаем
     $$\varphi(\xi,\eta)=\frac q \pi \;\frac 1 {\xi^2+\eta^2}\,.
         \eqno(25.33)$$

     При таком $\varphi(\xi,\eta)$ интеграл по $\xi$ в  (25.32)  легко
вычисляется   с  помощью  вычетов.  Замыкая  контур  интегрирования  в
комплексной  плоскости  $\xi$  полуокружностью  большого   радиуса   в
зависимости  от  знака  величины  $\eta  (x'+a)$  в верхней или нижней
полуплоскости, находим $\Phi(x',z')$ в виде однократного интеграла:
     $$\Phi'(x',z')=q\int\limits_{-\infty}^\infty\frac{e^{-|\eta(x'+a
         )|}}{|\eta|}\;e^{i\eta z'}\,d\eta\,.\eqno(25.34)$$
Векторный потенциал $\mbox{\boldmath$\gv A',$}$ очевидно, можно считать
равным нулю. При использовании  лоренцовской  калибровки  (3.15)
скалярный   потенциал  $\Phi$ и векторный потенциал $\mbox{\boldmath$\gv A$}$
в свободном пространстве образуют  4-вектор.  Переходя к лабораторной системе
координат  $x\,,y\,,z$,  в  которой  покоится  проводящая полуплоскость,
с помощью преобразования  Лоренца
     $$ x'\longrightarrow x\,,\qquad z'\longrightarrow\gamma(z-vt)\,,
         \eqno(25.35)$$
где релятивистский фактор  $\gamma=1/\sqrt{1-\beta^2},\;\beta=v/c$,  и
используя формулы преобразования компонент 4-вектора
     $$ \Phi=\gamma(\Phi'+\beta\gv A_z')\,,\qquad \gv A_x=\gv A_x'
         \,,\qquad \gv A_y=\gv A_y'\,,\qquad \gv A_z=(\gv A_z'+\beta
         \Phi')\gamma\,,\eqno(25.36)$$
находим единственную отличную от нуля компоненту векторного потенциала:
     $$ \gv A_z(x,z,t)=\gamma\beta q\int\limits_{-\infty}^\infty
         \frac{e^{-|\eta(x+a)|}}{|\eta|}\;e^{i\eta(z-vt)\gamma}\,
         d\eta\,.\eqno(25.37)$$
     Для того, чтобы это интегральное представление имело вид (25.29),
необходимо   заменить   переменную   интегрирования  $\eta  v\gamma\to
\omega$.  В результате  комплексная  амплитуда  компоненты  векторного
потенциала выражается в виде
     $$A_z(x,z,\omega)= q\beta\gamma\frac{e^{-\frac{\omega|x+a|}
         {v\gamma}}}{\omega}\,e^{i\frac{\omega}v z}\,.\eqno(25.38)$$

     При лоренцовской  калибровке  электромагнитное  поле   выражается
через $\rv A$ по формулам
     $$ \rv E=ik\rv A+ \frac i k \grad\div\rv A\,,\qquad \rv H=\rot
         \rv A\,,\eqno(25.39)$$
так что отличны от нуля компоненты поля
     $$\left.\begin{array}{lcl}E^0_x(x,z)=\displaystyle{\frac{ q}
         v \,\sign (x+a)\,e^{-\frac{\omega|x+a|}{v\gamma}}\,e^{i
         \frac {\omega} v z}}\,,\\ [.6cm]E^0_z(x,z)=-i\displaystyle{
         \frac{ q} {v\gamma} e^{-\frac{\omega |x+a|}{v\gamma}}\,e^
         {i\frac \omega v z}}\,,\\ [.6cm]H^0_y(x,z)=\beta E^0_x(x,z),
         \\[.1cm]\end{array}\right\}\;\eqno(25.40)$$
которые как компоненты падающего поля отмечены верхним индексом 0.

     Итак, спектральное  разложение  поля равномерно движущейся нити в
пустоте представляет собой совокупность неоднородных  плоских  волн  с
комплексным  волновым  вектором  $\rv  k=\rv  k'+i\rv  k''$,  где $\rv
k'=\{0,0,k'_z\},\;  \rv k''=\{\omega/v\gamma,0,0\}$.  При этом,  как и
для  любого другого источника,  равномерно движущегося в пустоте вдоль
оси  $z$,  $k'_z=\omega/v$.  Плоскости  постоянной   амплитуды   волны
параллельны плоскости $yz$, в которой перемещается нить.

     Под действием  падающего  поля  источника  на идеально проводящей
полуплоскости наводится  поверхностный  ток  $\mbox{\boldmath$\gv  I$}^1
(\rv  r,t)$.  Из  соображений  симметрии  очевидно,  что отлична от нуля лишь
компонента $\gv  I^1_x(x,t)$.  Именно  этот  ток  возбуждает  то  рассеянное
(дифрагированное)  поле,  которое  принято называть полем дифракционного
излучения.  Векторный потенциал $\rv A^1(x,z)$ этого поля также  имеет
только   одну,   отличную  от  нуля  компоненту  $A^1_x$,  связанную  с
комплексной амплитудой тока $I^1_x(x)$ соотношением
     $$A^1_x(x,z)=\frac 1 c \int\limits_0^\infty\int\limits_{-\infty}
         ^\infty \frac{e^{ikR}} R  I_x(x')\,dx'dy'\,,\eqno(25.41)$$
где $R=\sqrt{(x-x')^2+(y-y')^2+z^2}$.  Из этой  формулы  следует,  что
векторный  потенциал  дифрагированного поля является чётной функцией
координаты $z$  и  поэтому  оно  симметрично  относительно  бесконечно
тонкой  проводящей  полуплоскости.  Это  свойство  рассеянного поля не
зависит от источника возбуждения поверхностного тока и имеет место как
в   задачах  дифракции,  так  и  при  дифракционном  излучении  любыми
движущимися заряженными или токовыми источниками.

     Интегрирование по  переменной  $y'$  в   (25.41)   проводится   с
использованием формулы
     $$H_0^{(1)}(k|D|)=\frac 1 {\pi i}\int\limits_{-\infty}^\infty
         \frac{e^{ik\sqrt{D^2+\xi^2}}}{\sqrt{D^2+\xi^2}}\,d\xi,
         \eqno(25.42)$$
так что
     $$A^1_x(x,z)=\frac{i\pi}c \int\limits_0^\infty H_0^{(1)}(k\sqrt
         {(x-x')^2+z^2})\,I^1_x(x')\,dx'\,.\eqno(25.43)$$
Представим неизвестную плотность поверхностного тока $I_x^1(x)$ в виде
интеграла Фурье
     $$I^1_x(x)=\int \limits_{-\infty}^\infty F(\alpha)e^{i\alpha x}\,
          d\alpha \eqno(25.44)$$
и, учитывая,  что $I^1_x(x)\equiv 0$ при $x<0$, произведём в (25.43)
интегрирование по $x'$, воспользовавшись интегралом
     $$\int\limits_{-\infty}^\infty e^{i\alpha \xi}H_0^{(1)}(k\sqrt
         {D^2+\xi^2})\,d\xi=\frac{2e^{iu|D|}} u,\eqno(25.45)$$
где $u=\sqrt{k^2-\alpha^2},\;\im  u\geqslant  0$;  в результате получается,
что
     $$ A^1_x(x,z)=\frac{ i}c\int\limits_{-\infty}^\infty \frac
         {e^{iu|z|}} u F(\alpha)e^{i\alpha x}\,d\alpha\,.
         \eqno(25.46)$$

     Согласно формулам    (25.39)    этому    векторному    потенциалу
соответствует электрическое поле,  поляризованное в плоскости $xz$,  и
магнитное поле, направленное вдоль оси $y$:
     $$\left.\begin{array}{lcl}E_x^1(x,z)&=&-\displaystyle{\frac 1
         \omega}  \int\limits_{-\infty}^\infty uF(\alpha)e^{i(u|z|+
         \alpha x)}\,d\alpha\,,\\[.6cm]E_z^1(x,z)&=&\displaystyle
         {\frac 1 \omega} \sign z \int\limits_{-\infty}^\infty
         \alpha F(\alpha)e^{i(u|z|+\alpha x)}\,d\alpha\,,\\[.6cm]
         H_y^1(x,z)&=&-\displaystyle{\frac 1 c} \sign z\int
         \limits_{-\infty}^\infty F(\alpha)e^{i(u|z|+\alpha x)}\,d
         \alpha\,.\end{array}\right\}\eqno(25.47)$$
Таким образом,  поле   излучения   полностью   определяется   функцией
$F(\alpha)$.

     Выведем теперь  уравнения,  которым удовлетворяет эта неизвестная
функция $F(\alpha)$.  Для этого воспользуемся очевидными  условиями  в
плоскости   $z=0$:   при   $x>0$   касательная   составляющая  полного
электрического поля $E_x=E^0_x+E^1_x$ равна нулю,  а при  $x<0$  равна
нулю  плотность  тока.  Эти  два  условия  записываются в виде системы
парных интегральных уравнений, типичных для метода Винера-Хопфа-Фока:
     $$\left.\begin{array}{lcll}\displaystyle{\int\limits_{-\infty}^
         \infty uF(\alpha)e^{i \alpha x}}\,d\alpha&=&\displaystyle
         {\frac {2\pi q k}{ \beta}\,e^{-\frac{k(x+a)}{\beta\gamma}}}\qquad
         &\mbox{при}\quad x>0\,,\\[.5cm]\displaystyle{\int\limits_
         {-\infty}^\infty F(\alpha)e^{i \alpha x}\,d\alpha}&=&0
         &\mbox{при}\quad x<0\,.\end{array}\right\}\eqno(25.48)$$

\begin{wrapfigure}[13]{l}{7.5cm}
\begin{picture}(80,45)
\put(-5,45){\special{em:graph fig25-4.bmp}}
\end{picture}
\hbox to 7.5cm{\hfil\footnotesize{Рис.~25.4.~Разрезы на комплексной
}\hfil}
\hbox to 7.5cm{\hfil\footnotesize{плоскости $\alpha$.}\hfil}
\end{wrapfigure}

     Прежде, чем  решать  эту  систему,  необходимо   сделать   важное
отступление.  Поскольку  в дальнейшем при вычислении полей по формулам
(25.47) придётся деформировать контур интегрирования  в  комплексной
плоскости $\alpha$,  то необходимо, чтобы подынтегральные функции были
однозначными и аналитическими. Входящая в них функция $u$ двухзначная и
имеет  точки  ветвления $\alpha=\pm k$.  Поэтому на плоскости $\alpha$
необходимо  провести  разрезы,   позволяющие   образовать   двулистную
риманову  поверхность,  на  которой  подынтегральные  функции обладают
указанными выше свойствами.  При этом разрезы  целесообразно  провести
таким образом,  чтобы на всём верхнем листе выполнялось условие $\im
u>0$,  в результате чего решение (25.47) будет  удовлетворять  условию
излучения.  Так как метод Винера-Хопфа-Фока предполагает, что среда, в
которой происходит дифракция, обладает хотя бы небольшими потерями, то
волновое   число   в   ней   комплексное:   $K=K'+iK'',\,K''>0$.   Для
математической корректности всех преобразований  и  в  рассматриваемом
случае  свободного  пространства волновое число $k$ необходимо считать
комплексным, а к пределу $k''\to 0$ переходить на завершающем этапе. В
результате точки ветвления смещаются от действительной оси комплексной
плоскости $\alpha=\sigma+i\tau$,  а разрезы следует проводить от  этих
точек  вдоль  линий  $\im  u^2=0$,  как  это  показано  на  рис.~25.4.
Очевидно,  что уравнение  этих  линий  $\sigma\tau=k'k''$,  и  что  они
разграничивают области $\alpha$, где $\re u<0$ и $\re u>0$.

     Представим теперь  ядро первого из интегральных уравнений системы
(25.47) в виде произведения двух множителей:
     $$u=\frac{L_+(\alpha)L_-(\alpha)}{ \alpha^2+\Bigl(\displaystyle
         {\frac{\omega}{v\gamma}}\Bigr)^2}\,,\eqno(25.49)$$
где
     $$L_{\pm}(\alpha)=\sqrt{k\pm  \alpha}(\alpha\pm i\frac{\omega}
         {v\gamma})\,.\eqno(25.50) $$
Функция $L_+(\alpha)$  регулярна  и   не   имеет   нулей   в   верхней
полуплоскости  комплексной  переменной  $\alpha$  ($\im \alpha>0$),  а
функция  $L_-(\alpha)$  обладает  аналогичными  свойствами  в   нижней
полуплоскости; при этом подразумевается, что $\im k>0$.

     Из второго   интегрального   уравнения   следует,   что   функция
$F(\alpha)$  должна быть регулярной в нижней полуплоскости комплексной
переменной  $\alpha$;  поэтому  решение  системы  парных  интегральных
уравнений можно искать в виде
     $$F(\alpha)=\frac {C_0} {L_-(\alpha)},\eqno(25.51)$$
где $C_0$  ---  постоянная,  не  зависящая  от  $\alpha$.  Подстановка
выражения  для  $F(\alpha)$  в  первое  интегральное  уравнение даёт
возможность определить $C_0$:

     $$C_0=-i\frac{qk}{\beta}\,\frac {e^{-ka/\beta\gamma}}
         {\sqrt{k+i\omega/v\gamma}}.\eqno(25.52)$$
В результате   получаем   решение   для   $F(\alpha)$   в    замкнутом
аналитическом виде:

     $$F(\alpha)=-i\frac{qk}{\beta}\,\frac{e^{-ka/\beta\gamma}}
         {\sqrt{k-\alpha}\,(\alpha-i\omega/v\gamma)\sqrt{k+i\omega/v
         \gamma}},\eqno(25.53)$$
причём в этой формуле берётся та  ветвь  корня  $\sqrt{k-\alpha}$,
действительная часть которой положительна при $\alpha\to-\infty$.

     Следует отметить,  что использованное разбиение на множители ядра
интегрального  уравнения  (25.49)  не  является единственно возможным.
Сомножители $L_+(\alpha)$ и $L_-(\alpha)$ сохранят  свои  свойства  на
плоскости  комплексного  переменного  $\alpha$,  если  первый  из  них
домножить на произвольную целую функцию,  а второй на неё разделить.
Выбранное   разбиение   было   сделано  исходя  из  условия  на  ребре
полуплоскости:  нормальная к ребру составляющая плотности тока $I^1_x$
как  функция  расстояния  от  ребра  должна вести себя пропорционально
$\sqrt{x}$.  Такая зависимость согласно интегралу (25.44) имеет  место
только при условии
     $$F(\alpha)\sim\frac1{\alpha^{3/2}}\quad\hbox{ при $\alpha\to
         \infty$},\eqno(25.54)$$
что и соответствует выбранному разбиению (25.49).

     Таким образом,   рассеянное   поле   (точнее,   Фурье-образы  его
компонент по переменной $t$),  возникающее  при  равномерном  движении
заряженной  нити  мимо  проводящей  полуплоскости,  определено  в виде
интегралов Фурье по переменной $\alpha$ совокупностью формул (25.47) и
(25.53).  Прежде чем приводить оценки интегралов, преобразуем (25.53),
введя формально угол $\theta_0$ по  формуле  $\cos{\theta_0}=1/\beta$,
так что $\sin{\theta_0}=i/\gamma\beta$. В результате
     $$F(\alpha)=-i q\sqrt{2k}\,e^{-ka/\gamma\beta}\cos
         \Bigl(\frac{\theta+\theta_0}2\Bigr)\,[\sqrt{k-\alpha}(\alpha-
         k\sin\theta_0)]^{-1},\eqno(25.55)$$
где $\theta=\pi/2$   --   угол  между  направлением  движения  нити  и
полуплоскостью.  Такое представление решения позволяет сопоставить его
с  решением  классической  задачи  о  дифракции  плоской волны на краю
плоского идеально проводящего экрана (задача  Зоммерфельда).  Нетрудно
убедиться,  что  взяв  в  качестве падающего поля вместо (25.40) поле
плоской волны
     $$\rv E^0(x,z)=E_0 e^{ik(x\cos{\theta}+z\sin{\theta}-ct)}
         \,,\eqno(25.56)$$
поляризованной в плоскости $xz$,  получим для  $F(\alpha)$  выражение,
отличающееся  от  (25.55)  только  множителем  и  соответствующее углу
$\theta_0=0$.  Такое сопоставление в  общем  случае  имеет  формальный
характер,   поскольку   угол   $\theta_0$  в  (25.55)  комплексный.

     С физической  точки  зрения  различие  обусловлено тем,  что поле
заряженной  нити  разлагается  по  неоднородным   плоским   волнам   с
комплексным  волновым  вектором.  Только в предельном случае $\beta=1$
волны  становятся  однородными  и  тогда  всё  различие  сводится  к
постоянному множителю, несущественному для последующего интегрирования
по $\alpha$.  Угол $\theta_0$  может  стать  действительной  величиной
ещё  и  в  том случае,  когда всё  явление имеет место в однородной
среде с диэлектрической проницаемостью  $\varepsilon$  и  в  некотором
диапазоне     частот     $\omega$    выполнено    условие    излучения
Вавилова-Черенкова:  $\varepsilon \beta^2>1$.  Как нетрудно убедиться,
рассеянное  поле,  возбуждаемое наведёнными на полуплоскости токами,
описывается тем же векторным потенциалом (25.46),  в  котором  функция
$u$ имеет вид
     $$u=\sqrt{K^2-\alpha^2}=\sqrt{\varepsilon k^2-\alpha^2}\,,
         \qquad \im u>0\,,\eqno(25.57)$$
так что для $F(\alpha)$ аналогично предыдущему получаем
     $$F(\alpha)=-i q\sqrt{2K}\,e^{i\omega a/v\,\sqrt
         {\varepsilon\beta^2-1}}\cos\Bigl(\frac{\theta+\theta_0}2
         \Bigr)[\sqrt{K-\alpha}(\alpha-K\sin\theta_0)]^{-1},
         \eqno(25.58)$$
где угол   $\theta_0$  есть  реальный  угол  между  волновым  вектором
излучения  Вавилова-Черенкова  и  скоростью  источника,   определяемый
соотношением
     $$\cos{\theta_0}=\frac1{\beta\sqrt{\varepsilon}}\,.\eqno(25.59)$$
Отметим, что  в этом случае падающее поле само уже является регулярным
излучением  нити  при  движении  её  в  однородной  среде,  и  только
рассеянное поле представляет собой поле дифракционного излучения.

     Вычисление интегралов по $\alpha$ в формулах (25.47) представляет
собой  самостоятельную  и непростую задачу.  Рассмотрим её кратко на
примере магнитного  поля,  отличная  от  нуля  компонента  которого  в
принятых обозначениях имеет вид
     $$H_y^1(x,z)= i\,\frac {q\sqrt{2k}} {2\pi c}\, e^{-ka/\gamma\beta}\,\cos
         \Bigl(\frac{\theta +\theta_0}2\Bigr)\sign{y}\int\limits_{-
         \infty}^\infty\frac{e^{i(u|x|+\alpha z)}}{(\alpha-k\sin
         \theta_0)\,\sqrt{k-\alpha}}\,d\alpha,\eqno(25.60)$$
когда (в   общем   случае)    целесообразно    деформировать    контур
интегрирования в комплексной плоскости.  Однако на больших расстояниях
от края полуплоскости рассеянное  поле  может  быть  вычислено  хорошо
известным  методом  перевала.  Вводя  обозначения  $x=r\cos {\varphi},
\;z=r \sin{\varphi}, \;0\leqslant \varphi \leq \pi$, получим в результате
     $$H_y^1(x,z)=\frac{2 q}c\,\frac{e^{i(kr-\pi/4)}}{\sqrt{2\pi kr
         }}\,e^{-ka/\beta\gamma}\,\frac{\displaystyle{\cos{\frac{
         \varphi}2}\,\cos{\frac{\theta+\theta_0}2}}}{\cos{\varphi}+
         \cos{(\theta+\theta_0)}}\,.\eqno(25.61)$$
     Область применимости этой формулы определяется  ограничениями  на
метод перевала: полюс подынтегрального выражения в (25.60) должен быть
достаточно удален от  точки  перевала  $\alpha=k\,\cos{\varphi}$,  что
сводится к условию
     $$\sqrt{2\pi k r}[\cos{\varphi}+\cos{(\theta+\theta_0)}]
         \gg 1\,.\eqno(25.62)$$

     Полученное выражение  (25.61)  позволяет   оценить   длительность
вспышки  дифракционного  излучения.  Для  этого  найдём  зависимость
магнитного поля от времени:
     $$\gv H_y(r,t)=2\,\re\int\limits_0^\infty H^1_y(x,z)\,
         e^{-i\omega t}\,d\omega\sim\frac 1  {[(r-ct)^2+a^2/\gamma^2
         \beta^2]^{1/4}}.\eqno(25.63)$$
Отсюда следует,  что пространственная протяжённость волнового пакета
дифракционного   излучения  равна  $a/\beta  \gamma$,  а  длительность
вспышки составляет $a/v\gamma$. В свободном пространстве этот волновой
пакет распространяется без расплывания.

     Точное выражение   (25.60)  для  поля  дифракционного  излучения,
пригодное и для тех областей пространства,  где не выполняется условие
(25.62),  может  быть  сведено  к  интегралу  Френеля  от комплексного
аргумента,  аналогично тому,  как это делается в задаче  дифракции  на
полуплоскости     плоской     волны    соответствующей    поляризации.
Нецелесообразно приводить здесь эти  довольно  громоздкие  формулы,  а
можно ограничиться  описанием  различий  при  вычислении  интегралов в
плоскости  комплексного  переменного  $\alpha$  для  разных   областей
пространства.  Для  $x>0$  контур  интегрирования замыкается в верхней
полуплоскости, и интеграл (25.60) сводится к  интегралу  по  разрезу  в
верхней  полуплоскости,  показанному  на рис.~25.4,  и вычету в полюсе
$\alpha=-k \cos( \theta+ \theta_0)$.  Интеграл по  разрезу  определяет
рассеянное  поле,  а  вычет  в  полюсе  даёт  так  называемое  {\it поле
изображения},  то  есть  поле  воображаемого  источника   с   зарядом,
противоположным  по  знаку  реальному заряду,  и движущегося зеркально
относительно плоскости $z=0$.  Это  поле  компенсирует  падающее  поле
нити,  так  что  в результате создается область геометрической тени по
другую сторону проводящей полуплоскости, чем та, которую <<видит>> нить.
При  $x<0$ контур интегрирования замыкается вниз и интеграл сводится к
интегралу по берегам разреза в нижней  полуплоскости.  Полное  поле  в
этой  области  слагается  из поля нити в свободном пространстве и поля
рассеяния. Отметим, что при вычислении интеграла методом перевала поля
изображения  отсутствуют,  поскольку  они,  как и исходное поле нити в
свободном пространстве,  экспоненциально затухают по мере удаления  от
траектории.

     Рассмотрим теперь  ещё  одну  {\it  ключевую}   задачу   теории
дифракционного  излучения,  решаемую  методом  Винера-Хопфа-Фока,  ---
возбуждение движущейся нитью плоского  волновода  с  открытым  концом.
Пусть  волновод  образован двумя полубесконечными идеально проводящими
пластинами нулевой толщины $(x=\pm\,a,\,z>0)$.  В  качестве  источника
возбуждения возьмём для разнообразия токовую нить,  параллельную оси
$y$,  по которой протекает постоянный ток  $\gv  j_0$.  Нить  движется
равномерно  вдоль  оси $z$ на расстоянии $b\,\,(b<a)$ от плоскости $x=0$
со скоростью $v$ и в момент $t=0$ влетает в волновод (Рис.~25.5).  Эта
задача  интересна  ещё  и тем,  что указывает путь получения точного
решения для простейшей модели открытого резонатора.

\begin{wrapfigure}[13]{l}{7.0cm}
\begin{picture}(80,45)
\put(-5,45){\special{em:graph fig25-5.bmp}}
\end{picture}
\hbox to 7.0cm{\hfil\footnotesize{Рис.~25.5.~Плоский волновод с }\hfil}
\hbox to 5.3cm{\hfil\footnotesize{открытым концом.}\hfil}
\end{wrapfigure}
     Очевидно, что в рассматриваемом случае
электромагнитное поле может быть описано одним векторным
потенциалом  $\gv  A$  с единственной отличной от нуля
$y$-составляющей.  Только такую же составляющую
имеют и все поверхностные токи,  наводимые
на пластинах волновода. Будем искать
Фурье-образ   (по $t$) составляющей
векторного потенциала полного поля
(опуская, как и выше, у всех величин аргумент
$\omega$) в виде
     $$A_y(x,z)=A_y^0(x,z)+A_y^1(x,z)\,,\eqno(25.64)$$
где $A_y^0$ определяет падающее, а $A_y^1$ --- рассеянное поле.

     Возьмём в качестве падающего поля поле токовой нити, движущейся
в однородном плоском волноводе.  Такой выбор вовсе не обязателен, но в
данном  случае  он  несколько  удобнее  при  вычислении  интегралов на
заключительном  этапе  решения,  чем,  например,  выбор  поля  нити  в
свободном пространстве,  и, кроме того, позволяет отчётливее выявить
тот  класс  граничных  задач,  которые  могут  быть   решены   методом
Винера-Хопфа-Фока.

     Векторный потенциал однородной нити,  расположенной  между  двумя
идеально   проводящими  параллельными  плоскостями  и  несущей  ток  с
объёмной плотностью
     $$\gv j_y=J\delta(x'-b)\delta(z')\,,\eqno(25.65)$$
в собственной системе координат  $x',\,y',\,z'$  (где  нить  покоится)
удовлетворяет уравнению
     $$\Delta \gv A'_y=-\frac{4\pi} c \gv j_y\eqno(25.66)$$
с граничными  условиями  $\partial \gv A'_y /\partial z'=0$ при $x=\pm
a$. Разложив $\gv A'_y(x',z')$ в интеграл Фурье
     $$ \gv A'_y(x',z')=\int\limits_{-\infty}^\infty A'_y(x',\eta)
         e^{i\eta z'}\,d \eta\,,\eqno(25.67)$$
убеждаемся (при  помощи  выкладок,  аналогичных  проделанным  выше при
вычислении потенциала заряженной нити),  что $A_y(x',\eta)$ может быть
представлено в виде
     $$ A'_y(x',\eta)=\frac {J}{c}\,\frac{e^{-|\eta||x'-b|}}{|
         \eta|}+C_1e^{\eta x'}+C_2 e^{-\eta x'}\,,\eqno(25.68)$$
где постоянные  $C_1$  и  $C_2$  определяются  из  граничного  условия
$A'_y(-a,\eta)=A'_y(a,\eta)=0$;    тогда    в   результате   несложных
преобразований получаем, что
     $$\gv A'_y(x',z')=\frac{J}{c}\int\limits_{-\infty}
         ^\infty\frac {e^{i\eta z'}}{|\eta|}\Bigl\{e^{-|\eta||x'-
         b|}-e^{-|\eta| a}\,\Bigl[\frac{\ch{\eta b}}{\ch{\eta a}}
         \ch{\eta x'}+\frac{\sh{\eta b}}{\sh{\eta a}}\sh{\eta x'}
         \Bigr]\Big\}d\eta\,.\eqno(25.69)$$

     Перейдя с помощью преобразования Лоренца (25.35)  в  лабораторную
систему   координат,   учитывая   правило   преобразования  векторного
потенциала $\gv  A_y=\gv  A'_y$  и заменяя переменные в соответствии с
соотношением $\eta  v\gamma=\omega$,  находим  Фурье-образ  векторного
потенциала  падающего  поля  (для  получения искомого результата можно
было воспользоваться и методом изображений):
     $$ A^0_y(x,z)=\frac{ J\gamma}c\frac {e^{i\frac{\omega}v z}}
         {\omega}\Biggl\{e^{-\frac{\omega}{v\gamma}|x-b|}-e^{-\frac
         {\omega}{v\gamma} a}\,\Biggl[\,\frac{\ch{\displaystyle{\frac{\omega}{v\gamma}}
        \, b}}{\ch{\displaystyle{\frac{\omega}{v\gamma}}\, a}}\,\ch{\frac{\omega}{v\gamma}\,
         x}+\frac{\sh\displaystyle{{\frac{\omega}{v\gamma}}\,b}}{\sh{\displaystyle{\frac{\omega}
         {v\gamma}}\, a}}\,\sh{\frac{\omega}{v\gamma}\, x}\Biggr]\Bigg\}\,.
         \eqno(25.70)$$
Если падающее  поле  задаётся   в   таком   виде,   то   тем   самым
предполагается,  что  в плоскостях $x=\pm a$ имеют место поверхностные
токи
     $$\left. I^0_{\pm a}(z)=\frac c{4\pi}H_z(\pm a,z)=-\frac c {4\pi}
         \frac{\partial A_y^0(x,z)}{\partial x}\right |_{x=\pm a}\,.
         \eqno(25.71)$$
Полный поверхностный  ток  складывается из этого тока и тока $I^1_{\pm
a}(z)$,  возбуждающего  рассеянное  поле.  Условие   на   полный   ток
определяет  одно  из  двух интегральных уравнений,  к которым сводится
задача.

     Разложим обе    части   поверхностного   тока   на   симметричную
(относительно плоскости $x=0$) и антисимметричную  части,  представив,
например, ток $I^1$ в виде
     $$I^1_{\pm a}(z)=I_+(z)\pm I_-(z)\,,\qquad I_{\pm}(z)=\int
         \limits_{-\infty}^\infty F_{\pm}(\alpha)\,e^{i\alpha z}\,d
         \alpha\,.\eqno(25.72)$$
Из условия,  что  полный  поверхностный ток при всех $z<0$ равен нулю,
получаем первое интегральное уравнение для определяющей симметричную и
антисимметричную части тока функции $F_{\pm}(\alpha)$:
     $$ \int\limits_{-\infty}^\infty F_{\pm}(\alpha)\,e^{i\alpha z}\,d
         \alpha=B_{\pm}e^{i\frac \omega v z}\qquad\mbox{при}\quad z<0
         \,,\eqno(25.73)$$
где
     $$B_+=\frac J {2v}\,\displaystyle{\frac{\ch{\displaystyle{\frac{\omega b}{v\gamma}}}}
        { \ch{\displaystyle{\frac{\omega a}{v\gamma}}}}}\,,\qquad
         B_-=\frac J {2v}\,\displaystyle{\frac{\sh{\displaystyle{\frac{\omega b}{v\gamma}}}}
        { \sh{\displaystyle{\frac{\omega a}{v\gamma}}}}}\,.\eqno(25.74)$$

     Второе интегральное   уравнение  выводится  из  условия  $E_y(\pm
a,z)=0$  при  $z>0$,  то  есть  на  проводящих  пластинах.  Для  этого
необходимо  выразить  векторный потенциал рассеянного поля $A^1_y(x,z)$
через наведённые на обеих пластинах поверхностные токи $I_a^1(z)$  и
$I_{-a}^1(z)$. Общее соотношение между этими величинами
     $$A^1_y(x,z)=\frac 1 c\int\limits_0^\infty\int\limits_{-\infty}
         ^\infty \Bigl[\frac {e^{ikR_a}}{R_a}I^1_a(z')+\frac {e^{ikR_
         {-a}}}{R_{-a}}I^1_{-a}(z')\Bigr]\,dz'dy'\,,\eqno(25.75)$$
где
     $$R_{\pm a}=\sqrt {(x\mp a)^2+(y-y')^2+(z-z')^2}\,,\eqno(25.76)$$
после интегрирования  по  $y'$  с  помощью формулы (25.42) и по $z'$ с
помощью формулы (25.45) принимает вид
     $$A_y^1(\pm a,z)=\frac i c \int\limits_{-\infty}^\infty
         L_{\pm}(\alpha)F_{\pm}(\alpha)e^{i\alpha z}\,d\alpha\,,\qquad
         L_{\pm}(\alpha)=\frac{1\pm e^{2iua}} u\,,\eqno(25.77)$$
где $u=\sqrt{k^2-\alpha^2},\;\im{u}>0$.

     Поскольку составляющая  $E_y^1$  пропорциональна  $A_y^1$,  то  в
качестве второго интегрального уравнения системы имеем:
     $$\int\limits_{-\infty}^\infty L_{\pm}(\alpha)F_{\pm}(\alpha)e^
         {i\alpha z}\,d\alpha=0\,\qquad \mbox{при}\quad z>0.
         \eqno(25.78)$$
Таким образом,  интегральные  уравнения (25.73) и (25.78) представляют
собой  две   независимых   системы   парных   интегральных   уравнений
относительно функций $F_+(\alpha)$ и $F_-(\alpha)$,  которые допускают
решение методом Винера-Хопфа-Фока.

     Решение этой системы производится  по  той  же  схеме,  что  и  в
предыдущей задаче. Ядро интегрального уравнения (25.78) разлагается на
множители:
     $$L_{\pm}(\alpha)=\frac{1\pm e^{2i u a}} u=\frac{\psi_{\pm}
         (\alpha)\varphi_{\pm}(\alpha)} u\,,\eqno(25.79)$$
где функции $\psi_+$ и $\varphi_+$ аналитические  в  верхней,  а  функции
$\psi_-$  и  $\varphi_-$  ---  в  нижней полуплоскости комплексного
переменного  $\alpha$.  Явные  выражения  для  этих  функций  довольно
громоздки и здесь не приводятся, но именно через функции $\varphi_+$ и
$\varphi_-$  в  замкнутом  аналитическом   виде   выражаются   искомые
$F_{\pm}(\alpha)$:
     $$F_{\pm}(\alpha)= i B_{\pm}\frac{\varphi_{\pm}
         (\omega/v)}{\varphi_{\pm}(\alpha)}\frac{\sqrt{k-\alpha}}
         {\sqrt{k-\omega/v}}\frac 1 {\alpha-\omega/v}\,.\eqno(25.80)$$

     По известным    $F_{\pm}$   легко   выразить   все   составляющие
рассеянного поля  в  виде  интегралов  по  $\alpha$.  Так,  компонента
электрического поля $E_y^1(x,z)$,  с которой остальные две отличные от
нуля компоненты магнитного поля связаны очевидными соотношениями
     $$ H_x^1=-\frac 1{ik}\frac{\partial E^1_y}{\partial z}\,,
         \qquad H^1_z=\frac 1{ik}\frac{\partial E^1_y}{\partial x}
         \,,\eqno(25.81)$$
представляется  при $|x|<a$ в виде
     $$E_y^1=-\frac{2 k}c\int\limits_{-\infty}^\infty e^{i(
         \alpha z+ua)}[F_+(\alpha)\cos(ux)-iF_-(\alpha)\sin{(ux)}]
         \frac {d\alpha}u\,.\eqno(25.82)$$
При $|x|>a$ в подынтегральном выражении этой формулы следует $a$ и $x$
поменять местами.

     Обсуждать здесь  детали  вычисления  интегралов  путем деформации
контура в   комплексной   плоскости  нет  возможности.  Скажем  только
несколько слов о характере поля в разных  областях  структуры.  Внутри
волновода  рассеянное поле представляет собой совокупность собственных
магнитных волн волновода,  верхняя граница спектра частот которых  при
малых   скоростях  нити  пропорциональна  $v$,  а  при  релятивистских
скоростях пропорциональна $\gamma$.  В последнем  случае  имеет  место
резкая асимметрия излучения в зависимости от знака $v$:  при влёте в
волновод нить излучает значительно больше энергии, чем при вылете.

     В свободном  пространстве вне волновода рассеянное поле имеет вид
расходящейся цилиндрической волны с продольным волновым вектором вдоль
оси  $y$,  равным  нулю.  В  полярной  системе координат $r,\varphi$ в
плоскости $y=const$ в  дальней  зоне  существенные  составляющие  поля
зависят от координат следующим образом:
     $$E^1_y=H^1_\varphi\sim \frac {e^{ikr}}{\sqrt{kr}}\,f(\varphi)\,,
        \eqno(25.83)$$
где функция    $f(\varphi)$,    определяющая   угловое   распределение
излучения,  очень сложным образом зависит от соотношения между  длиной
волны и единственным геометрическим параметром задачи $a$.

     На примере  рассмотренной  задачи  хорошо  выявляется  тот  класс
граничных  задач электродинамики,  которые эффективно решаются методом
Винера-Хопфа-Фока.  В этих задачах  граничные  условия  для  волнового
уравнения  задаются на координатной поверхности (системы координат,  в
которой переменные разделяются)   таким образом, что на части её они
являются условиями Дирихле, а на другой части --- условиями Неймана.

     Рассмотренные в этом разделе задачи занимают особое  положение  в
теории  дифракционного излучения,  так как они являются {\it ключевыми}
--- таковыми принято называть задачи,  которые могут служить  примером
применения   к   простейшему   случаю   метода,   позволяющего  решать
аналогичные задачи и для других,  в том числе более сложных, структур.
Как  правило,  подробный анализ ключевой задачи даёт представление о
возможностях применённого  метода  и  о  классе  электродинамических
задач, для решения которых он может быть использован.

%\end{document}
%\end{document} 

\newpage
\oddsidemargin=-0.4mm \evensidemargin=-0.4mm
\topmargin=-0.4mm
\headsep=7mm
\textheight=231.875mm
\textwidth=160mm
\mathsurround=2.5pt
\unitlength=1mm
%\begin{document}
%\input{macr.tex}
\thispagestyle{empty}
%\addtocounter{page}{303}

\begin{center}\subsubsection*{26. Дифракция на телах, имеющих форму
         кругового цилиндра}\end{center}
\vspace*{0.5cm}

\markboth{Глава 9. Методы ТФКП в электродинамике}{26. Дифракция на
          телах, имеющих форму кругового цилиндра}

\begin{center}\begin{minipage}[c]{0.75\textwidth}
\footnotesize{\parindent=0.5cm
         Общая задача   теории   дифракции.  Простейшие  дифракционные
         задачи.  Дифракция цилиндрической волны на круговом цилиндре.
         Разложение  плоской  и  сферической  волны  на цилиндрические
         волны.  Дифракция поля диполя на стенках круглого  волновода.
         Нахождение   полного   поля   интегрированием  по  контуру  в
         комплексной плоскости.  Анализ этой же задачи с точки  зрения
         возбуждения волновода сторонним током.
}\end{minipage}\end{center}
\vspace*{0.5cm}

     {\it Общую   задачу   теории   дифракции}   можно  сформулировать
следующим образом.  В некоторой структуре (в частности, это может быть
всё  свободное  пространство) задано электромагнитное поле сторонних
источников,   которое    удовлетворяет    уравнениям    Максвелла    и
соответствующим  граничным условиям на всех поверхностях структуры.  В
большинстве дифракционных задач  сторонние  источники  расположены  на
бесконечности  и  поэтому  их  поле  имеет  вид  приходящей  волны  и,
следовательно,  не  удовлетворяет  условиям  излучения   Зоммерфельда.
Где-то  внутри  структуры находится тело,  которое для самой структуры
следует считать посторонним.  На  этом  теле  и  происходит  дифракция
заданного  поля  (которое  принято  называть падающим полем),  то есть
возникновение  дифрагированного  (рассеянного)  поля.  Дифрагированное
поле должно удовлетворять уравнениям Максвелла,  граничным условиям на
поверхностях  основной  структуры,  не   иметь   особенностей   внутри
структуры  и  соответствовать  условиям излучения на бесконечности,  а
{\it в сумме со сторонним полем} (то есть полное поле)  ---  граничным
условиям  на  поверхности рассеивающего тела.  Задача теории дифракции
заключается в нахождении рассеянного поля.

     В простейших  задачах  дифракции падающее поле имеет вид какой-то
одной  приходящей  (по  отношению   к   рассеивающему   телу)   волны.
Практический   интерес   для   свободного   пространства  представляют
сферическая волна (на достаточно большом расстоянии  от  антенны  поле
всегда имеет такой вид) и плоская (вблизи рассеивающего тела,  размеры
которого  в  большинстве  случаев  меньше  радиуса  волнового  фронта,
сферическую  волну  можно  считать плоской).  Рассеяние цилиндрических
волн для свободного пространства имеет скорее  теоретический  интерес,
который  становится  практическим  в  ограниченных структурах,  таких,
например, как цилиндрический волновод.

     В рассматриваемом  ниже  варианте геометрия фронта падающей волны
совпадает с геометрией поверхности тела и  в  этом  случае  рассеянное
поле  представляет  собой  {\it  одну}  волну той же геометрии,  что и
падающая,  но только уходящую.  Этих двух волн вне тела достаточно для
того,  чтобы  обеспечить  выполнение  граничных условий на поверхности
тела. Примером служит падение плоской волны на плоскую границу раздела
двух  сред,  когда отражённое поле представляет собой одну волну той
же структуры, что и падающая.

     Аналогичное явление  имеет место при падении цилиндрической волны
на поверхность кругового цилиндра радиуса  $a$ в том  случае,  когда  ось
цилиндра $z$ совпадает с осью цилиндрической системы координат, в которой
представлена падающая волна.  Напомним,  что  цилиндрические  волны  в
однородном пространстве распадаются на два типа: магнитные (отлична от
нуля продольная компонента $H_z\!$) и  электрические  (отлична  от  нуля
продольная  компонента  $E_z\!$).  Эти  компоненты  могут быть выбраны в
качестве потенциальных функций, полностью описывающих электромагнитное
поле  соответствующей  волны,  и обе они удовлетворяют в свободном пространстве
без источников одному и тому же волновому уравнению. Поэтому
обозначим  потенциальную  функцию  как $u$, и тогда всё
различие в описании дифракции  обоих  типов  волн  сведётся  к  виду
граничных условий на поверхности цилиндра.

     Основной характеристикой  цилиндрической  волны   является   её
продольное волновое число $h$,  определяющее зависимость потенциальной
функции  от координаты $z$:  $u(\rv r)=u(r,\varphi) e^{ihz}$. В последующих
формулах  множитель   $e^{ihz}$ опускается. Функция $u(r,\varphi)$ в
свободном пространстве без источников удовлетворяет волновому уравнению
$\Delta_2 u(r,\varphi)+(k^2-h^2)u(r,\varphi)=0\,,$  где оператор $\Delta_2$
определён формулой (7.9). Разложим функцию $u(r,\varphi)$ в ряд по
азимутальному углу $\varphi$:

$$u(r,\varphi)=\sum\limits_{m=-\infty}^\infty u_m(r)\,e^{im\varphi} \eqno(26.1)$$
Общее решение волнового уравнения определяется функцией

     $$u_m(r)=B\, H_m^{(1)}(\sqrt{k^2-h^2}\,r)+
         C\, H_m^{(2)}(\sqrt{k^2-h^2}\,r)\,. \eqno(26.2)$$
Первое слагаемое  описывает  волну, расходящуюся  от  оси $z$,
второе   ---  сходящуюся к ней,  что  следует  из  асимптотического
поведения функций Ханкеля при больших значениях аргумента:
     $$H_n^{(1)}(x)\sim\frac{e^{ix}}{\sqrt{x}},\qquad H_n^{(2)}(x)\sim
         \frac{e^{-ix}}{\sqrt{x}}.\eqno(26.3)$$

     Возьмём в   качестве   падающей   сходящуюся   волну  единичной
амплитуды:

     $$u_m^{\mbox{\footnotesize\textit{{пад}}}}(r)=H_m^{(2)}(\sqrt{k^2-h^2}\,r)\,.\eqno(26.4)$$
Амплитуда  $B$  расходящейся (дифрагированной) волны

 $$ u_m^{\mbox{\footnotesize\textit{{диф}}}}(r,\varphi)=
 B\,H_m^{(1)}(\sqrt{k^2-h^2}\,r),\eqno(26.5)$$
находится с помощью граничного условия для полного поля
 $u_m=u_m^{\footnotesize\mbox{\textit{пад}}}+u_m^{\mbox{\footnotesize\textit{{диф}}}}$
 на поверхности  цилиндра.

  Если цилиндр идеально проводящий, а падающая  волна  ---
  электрическая,  то  это условие   записывается в виде
     $$  u_m(a)=0\,,\eqno(26.6)$$
и для его выполнения достаточно выбрать амплитуду $B$ равной
     $$B=-\frac{H_m^{(2)}(\sqrt{k^2-h^2}\,a)}{H_m^{(1)}(\sqrt{k^2
         -h^2}\,a)}\,.\eqno(26.7)$$

В результате полное поле вне цилиндра будет имеет вид
     $$ u_m(r)=H_m^{(2)}(\sqrt{k^2-h^2}\,r) -\frac{H_m^
         {(2)}(\sqrt{k^2-h^2}\,a)}{H_m^{(1)}(\sqrt{k^2-h^2}\,a)}\,
         H_m^{(1)}(\sqrt{k^2-h^2}\,r)\,.\eqno(26.8)$$

     Если падающая волна является магнитной ($u=H_z$), то единственной
тангенциальной компонентой электрического поля на поверхности цилиндра
является $E_{\varphi}=\displaystyle{\frac  1{ik}\,\frac{\partial  H_z}
{\partial r}}$, и граничное условие принимает вид
     $$\left.\frac{d u_m(r)}{d r}\right|_{r=a}= 0\,,\eqno(26.9)$$
что даёт для полного поля вместо (26.8) следующее выражение:

     $$ u_m(r)=\left [H_m^{(2)}(\sqrt{k^2-h^2}\,r) -
         \frac{{H_m^{(2)}}'(\sqrt{k^2-h^2}\,a)}{{H_m^{(1)}}'(\sqrt
         {k^2-h^2}\,a)}\,H_m^{(1)}(\sqrt{k^2-h^2}\,r)\right ] \,.\eqno(26.10)$$

     Если цилиндр,  на  который  падает  волна,
диэлектрический,  то вне  цилиндра  дифрагированное  поле  по-прежнему
следует искать  в виде (26.4).  Внутри же цилиндра подходящим решением
волнового уравнения является
     $$   u_m(r)=D\,J_m(\sqrt{\varepsilon k^2-h^2}\,r)\,,\eqno(26.11)$$
поскольку из  всех  цилиндрических  функций  только  $J_m(x)$ не имеет
особенности в нуле. Для определения двух констант $B$ и $D$ достаточно
граничных    условий   на   непрерывность   тангенциальных   компонент
электрического и магнитного полей на поверхности цилиндра.

     Простота решения  задачи при падении на поверхность цилиндра соосной
цилиндрической волны  подсказывает  способ  решения  в  случае
падения   волн  другой  геометрии.  Несомненный  практический  интерес
представляет  дифракция  на   цилиндре   плоской   волны.   Рассмотрим
простейший вариант, когда волновой вектор падающей на цилиндр плоской
волны направлен вдоль оси $x$,  то есть по нормали к
оси цилиндра $z$, и её электрическое поле поляризовано вдоль этой же оси:
     $$u^{\mbox{\footnotesize\textit{{пад}}}}=e^{ikx}=e^{ikr\cos{\varphi}}\,.\eqno(26.12)$$

     Представим падающее поле в виде  ряда Фурье по углу $\varphi$,
воспользовавшись для этого известным интегральным представлением
функции Бесселя
     $$ J_n(x)=\frac{i^{-n}}{2\pi}\int\limits_0^{2\pi} \,e^{i(x\cos
         \varphi+n\varphi)}\,d\varphi\,.\eqno(26.13)$$
В результате получаем для падающей волны следующее выражение:

     $$u^{\mbox{\footnotesize\textit{пад}}}=\sum\limits_{m=-\infty}^\infty i^m J_m
         (kr)\,e^{im\varphi}.\eqno(26.14)$$

     Будем искать дифрагированное поле в виде ряда
     $$u^{\mbox{\footnotesize\textit{диф}}}=\sum\limits_{m=
     -\infty}^\infty B_m H_m^{(1)}(kr)e^{im\varphi}\,,\eqno(26.15)$$
который почленно  удовлетворяет  волновому  уравнению  и  представляет
собой совокупность уходящих на  бесконечность  цилиндрических  волн  с
равным  нулю  продольным волновым числом $h$.  Тогда граничное условие
для полного поля $u(a,\varphi)=0$ записывается в виде
     $$\sum\limits_{m=-\infty}^\infty\left[B_mH_m^{(1)}(ka)+(-i)^m
         J_m(ka)\right]e^{im\varphi}=0.\eqno(26.16)$$
Исходя из полноты системы функций $e^{im\varphi}$, находим коэффициенты
     $$B_m=-(-i)^m\frac{J_m(ka)}{H_m^{(1)}(ka)}\,,\eqno(26.17)$$
так что полное поле равно
     $$u(r,\varphi)=-\sum\limits_{m=-\infty}^\infty (-i)^m\Bigl[J_m(kr)-\frac{J_m(ka)}
         {H_m^{(1)}(ka)}\, H_m^{(1)}(kr)\Bigr]\,e^{im\varphi}\,.
         \eqno(26.18)$$
Полученный  ряд достаточно быстро сходится   при   произвольном   соотношении
между   параметрами задачи.  Не представляет большого труда решить таким же
способом и более общую задачу при  произвольной  поляризации плоской волны
и произвольном угле между её волновым вектором и осью цилиндра.

     Перейдём теперь  к  рассмотрению   дифракции  поля  элементарного
электрического  диполя  на внутренней  поверхности  идеально  проводящей
цилиндрической оболочки. Эта задача по существу ничем  не  отличается  от
рассмотренной  ранее задачи   о  возбуждении  круглого  цилиндрического
волновода  заданным распределением стороннего тока,  в качестве которого в
данном  случае рассматривается  расположенный в начале координат
 элементарный диполь, направленный вдоль оси $z$, с плотностью тока
         $$j_z=j_0\delta(x)\delta(y)\delta(z)\,.\eqno(26.19)$$
Как показано в разделе 17,  поле такого диполя полностью  определяется
векторным  потенциалом $\rv A$  с  единственной  отличной  от нуля компонентой
$A_z$, которая в сферической системе координат $(\rho,\theta,\varphi)$
зависит только от $\rho$ и имеет вид
     $$   A_z= \frac 1 c\frac {e^{ik\rho}}{\rho}\,.\eqno(26.20)$$
Электромагнитное поле   диполя   в  цилиндрической  системе  координат
$(r,\varphi,z)$,  линейные  координаты  которой  связаны  с   линейной
координатой  сферической системы соотношением $\rho^2=r^2+z^2$,  имеет
отличными от нуля только три компоненты поля, которые выражаются через
$A_z$ следующими формулами:
     $$  E_z=ikA_z+\frac i k\frac{\partial^2 A_z}{\partial z^2}\,,
         \qquad E_r=\frac i k \frac{\partial^2 A_z}{\partial r
         \partial z}\,,\qquad H_{\varphi}=-\frac{\partial A_z}
         {\partial r}\,.\eqno(26.21)$$

     Обозначим для краткости записи $A_z=u$  и  нормируем  ток  диполя
таким образом, чтобы падающее поле имело вид
     $$ u^{\mbox{\footnotesize\textit{пад}}}=\frac {e^{ik\sqrt{r^2+z^2}}}{\sqrt{r^2+z^2}}\,.
         \eqno(26.22)$$
Разложим это поле сферической волны в интеграл Фурье по $z$
     $$  \frac {e^{ik\sqrt{r^2+z^2}}}{\sqrt{r^2+z^2}}=\int\limits
         _{-\infty}^\infty f(r,h)\,e^{ihz}\,dh\,\eqno(26.23)$$
и преобразуем Фурье-образ
     $$f(r,h)=\frac 1{2\pi}\int\limits_{-\infty}^\infty\frac {e^{ik
         \sqrt{r^2+z^2}}}{\sqrt{r^2+z^2}}\,e^{-ihz}\,dz\,
         \eqno(26.24)$$
с помощью   подстановок   $z=r\sh{\gamma}$,   $h=g\,\sh{\delta}$,  где
величина $g$ определена формулой  $g^2=k^2-h^2$.  Перейдём  далее  к
интегрированию по переменной $\gamma$ и тогда,  учитывая,  что в новых
переменных $k=g\,\ch{\delta}$, интеграл (26.24) сводится к виду
     $$f(r,h)=\frac 1{2\pi}\int\limits_{-\infty}^\infty
         e^{ig r\ch{(\delta-\gamma)}}\,d\gamma\,.\eqno(26.25)$$
Если теперь   воспользоваться  известным  интегральным  представлением
функции Ханкеля
     $$ H_0^{(1)}(x)=\frac 1{\pi i}\int\limits_{-\infty}^{\infty}
         e^{ix\ch{t}}\,dt\,,\eqno(26.26)$$
то получится    интегральное    разложение    сферической   волны   по
цилиндрическим волнам:
     $$\frac {e^{ik\rho}}{\rho}=\frac i 2\int\limits_{-\infty}^\infty
         H_0^{(1)}(g r)\,e^{ihz}\,dh\,,\qquad g^2=k^2-h^2\,.
         \eqno(26.27)$$

     Имея в виду последующее интегрирование в  комплексной  плоскости,
проанализируем  эту формулу и определим нужную ветвь корня в аргументе
функции Ханкеля.  При заданном $k$ в пределах  области  интегрирования
(то  есть  на  вещественной  оси  $h$)  функция  $g^2$  принимает  как
положительные, так и отрицательные значения, обращаясь в нуль в точках
$h=\pm \,  k$.  Асимптотическое  поведение  $H_0^{(1)}(x)$  при  больших
значениях аргумента (26.2)  подсказывает  необходимость  выбора  знака
корня таким образом, чтобы выполнялись следующие условия:
     $$\re g>0\qquad\mbox{при}\quad g^2>0,\eqno(26.28a)$$
     $$\im g>0\qquad\mbox{при}\quad g^2<0.\eqno(26.28\mbox{\textit{б}})$$
Тогда подынтегральная функция в (26.27) при $h$ в  интервале  $-k<h<k$
будет представлять собой уходящую волну, а при $h$ в интервалах $h<-k$
и  $h>k$  ---  затухающую.  В  результате  весь  интеграл  в   (26.27)
представляет собой уходящую волну, а условие (26.28{\it б)} обеспечивает его
сходимость.

     При переходе  в  комплексную  плоскость  переменной   $h=h'+ih''$
следует  учесть,  что  точки,  в которых $g=0$,  то есть точки $h=k$ и
$h=-k$,    являются    точками    ветвления    не    только    функции
$g=\sqrt{k^2-h^2}$,  но и самой функции $H_0^{(1)}(x)$,  которая,  как
известно, может быть представлена в виде
     $$H_0^{(1)}(x)=\frac{2i}{\pi}J_0(x)\ln{x}+
         F(x)\,,\eqno(26.29)$$
где $F(x)$  ---  однозначная  функция  своего  аргумента.  При   малых
значениях аргумента функция $H_0^{(1)}(x)$ ведёт себя как $\ln{x}$ и
чтобы сделать  функцию  $H_0^{(1)}(g  r)$  однозначной  функцией  $h$,
необходимо  на  комплексной  плоскости  $h$  провести разрезы из точек
$h=\pm \,k$  и  выбрать  подходящий  лист  на  многолистной  поверхности
Римана.   Выберем   лист,   на  котором  $\ln{g  r}$  вещественен  при
вещественном $g$,  и рассмотрим  окрестность  точки  $h=k$,  для  чего
введём   в   этой   точке   локальную   полярную  систему  координат
$(R,\psi)$,$R\ll k$.  В ближайшей  окрестности  рассматриваемой  точки
функция $g^2$ представляется в виде
     $$g^2=-2kRe^{i\psi}+O(R^2)\,,\eqno(26.30)$$
так что
     $$ g=|g|e^{i(\psi+\pi)/2}\qquad \mbox{или}\qquad g=|g|e^{i(\psi-
         \pi)/2}\,.\eqno(26.31)$$

\begin{wrapfigure}[12]{l}{7.5cm}
\begin{picture}(80,40)
\put(-5,40){\special{em:graph fig26-1.bmp}}
\end{picture}
\hbox to 7.5cm{\hfil\footnotesize{Рис.~26.1.~Разрезы в комплексной
}\hfil}
\hbox to 6.8cm{\hfil\footnotesize{ плоскости $h$.
}\hfil}
\end{wrapfigure}
     Из этих двух ветвей  необходимо  выбрать  первую,  поскольку  при
$\psi=0$,  согласно (26.28б),  должно быть $\im g>0$.  Тогда для того,
чтобы при действительных $g$ выполнялось условие (26.28а),  им  должен
соответствовать угол $\psi=-\pi$,  и,  следовательно,  для перехода на
выбранном листе поверхности Римана  от  мнимых  $g$  к  действительным
обход  точки  $h=k$  должен происходить в нижней полуплоскости.  Таким
образом, оба условия (26.28) будут выполнены, если разрезы в плоскости
комплексного переменного $h$ провести согласно рис.~26.1.

     Вернёмся к    рассматриваемой    задаче    и    будем    искать
дифрагированное поле в виде интеграла Фурье
     $$ u_{\footnotesize{\textit{диф}}}(r,z)=\int\limits_{-\infty}^\infty f(h)J_0
         (g r)e^{ihz}\,dh.\eqno(26.32)$$
В таком  виде  оно удовлетворяет волновому уравнению при любой функции
$f(h)$,  не имеет в интересующей  нас  области  $r<a$  особенностей  и
позволяет обеспечить выполнение граничного условия $u(a,z)=0$ {\it при
всех $z$} полагая равным  нулю  множитель  при  $e^{ihz}$  в
разложении Фурье полного поля, то есть
     $$  \frac i 2  H_0^{(1)}(g a)+f(h)\, J_0(g a)=0\,.\eqno(26.33)$$
Найдя отсюда функцию $f(h)$, получим решение задачи для полного поля в
виде интеграла
     $$u(r,z)=\frac i 2\int\limits_{-\infty}^\infty\Bigl[H_0^{(1)}
         (g r)-\displaystyle\frac{H_0^{(1)}(g a)}{J_0(
         g a)}\,J_0(g r)\Bigr]\, e^{ihz}\,dh.
         \eqno(26.34)$$

     Для того,  чтобы  этот  интеграл  представлял  собой  однозначное
решение и удовлетворял всем вышеперечисленным требованиям,  необходимо
сделать ряд важных уточнений.  Прежде  всего  заметим,  что  если  под
областью интегрирования понимать вещественную ось переменной $h$,  как
это предполагается при разложении  в  интеграл  Фурье,  то  полученное
выражение является неопределённым,  поскольку на пути интегрирования
встречаются  полюсы,  обусловленные  нулями  функции  $J_0(ga)$.  Нули
функции   Бесселя   $J_0(x)$   образуют  счётную  последовательность
вещественных корней $\nu_{0n}$.  При заданном действительном  значении
волнового  числа $k$ в интервале на вещественной оси $h$ между точками
$h=-k$  и  $h=k$  может  находиться  только  конечное  число   полюсов
подынтегрального   выражения   в   (26.34),   расположенных  в  точках
$h=h_n=\pm\,\sqrt{k^2-(\nu_0n/a)^2}$.  Полюсов нет только в том  случае,
когда  $ka<\nu_{01}$,  что одновременно является условием отсутствия в
круглом  волноводе  распространяющихся  волн  типа  $E_{0n}$,  которые
только   и  могут  возбуждаться  рассматриваемым  источником.  Попутно
заметим,  что тем корням функции $J_0(ga)$,  которым не  соответствует
полюс  на  действительной оси $h$,  всегда сопоставлен полюс на мнимой
оси.

     Физическая причина возникшей неопределённости состоит,  как уже
отмечалось в предыдущих  разделах,  в  неправомочности  одновременного
использования   двух  идеализаций:  идеальной  проводимости  стенок  и
отсутствия  потерь  в   среде,   заполняющей   структуру.   Достаточно
отказаться  от  одного  из  этих предположений и тогда соответствующий
интеграл становится однозначно определённым (отметим,  что физически
необоснованной является только первая идеализация). Однако цена такого
отказа слишком велика --- получающиеся в результате выражения для поля
непомерно сложны для их аналитического исследования.

%\vspace{-.2cm}
\begin{wrapfigure}[14]{l}{7.5cm}
\begin{picture}(80,50)
\put(5,50){\special{em:graph fig26-2.bmp}}
\end{picture}
\hbox to 7.5cm{\hfil\footnotesize{Рис.~26.2.~Замкнутый контур в
}\hfil}
\hbox to 7.8cm{\hfil\footnotesize{комплексной плоскости $h$ при $z>0.$
}\hfil}
\end{wrapfigure}
     Существует, однако, eщё  один способ устранения    неопределённости,    которым
чаще всего и пользуются в задачах дифракции.  Он состоит в изменении
области  интегрирования  путем  перехода в комплексную плоскость $h$ и
частичного смещения контура интегрирования с действительной оси  таким
образом,  чтобы  на  контуре  не оставалось полюсов.  С математической
точки зрения  этой  процедуре  нет  оправдания:    математика
объективно  фиксирует  некорректность постановки задачи.  С физической
точки зрения оправданием (и весьма убедительным) служит тот факт,  что
получающийся результат является пределом, к которому стремится решение
правильно поставленной задачи.

     Само смещение   контура   приходится   производить,   исходя   из
физических соображений.  В данном случае достаточно  утверждения,  что
полное поле не может содержать волн,  приходящих из $z=\pm\,\infty$.  Из
этого  в  соответствии   с   (26.34)   сразу   следует,   что   контур
интегрирования  слева  от  мнимой  оси  $h$  должен обходить полюсы на
действительной оси сверху,  а справа от мнимой оси --- снизу,  как это
показано на рис.~26.2. Отметим при этом, что строгий учет поглощения в
среде или в металлических стенках приводит к соответствующему смещению
самих полюсов в верхнюю или нижнюю полуплоскость.

     Второе важное   замечание  состоит  в  том,  что  подынтегральная
функция в  выражении  для  полного  поля  (26.34)  ---  в  отличие  от
соответствующего выражения (26.27) для поля диполя (сферической волны)
в свободном пространстве (26.22)  ---  имеет  одинаковое  значение  на
обоих берегах разрезов в комплексной плоскости $h$. Это связано с тем,
что скачок второго слагаемого  в  (26.34)  точно  компенсирует  скачок
первого,  в  чём  легко  убедиться,  используя представление функции
Ханкеля (26.29).  В результате на всей плоскости  $h$  подынтегральное
выражение   не   имеет  других  особых  точек,  кроме  {\it  полюсов},
соответствующих нулям $J_0(ga)$.

     Всё вышесказанное  позволяет провести интегрирование в (26.34),
используя теорию вычетов.  При  $z>0$  контур  интегрирования  следует
замкнуть  дугой  окружности  большого радиуса в верхней полуплоскости.
Интеграл по дуге исчезает согласно лемме Жордана,   интеграл по разрезу
равен нулю,  так что интеграл по контуру сводится к сумме вычетов
в полюсах, лежащих в верхней полуплоскости:
     $$u(r,z)=\pi\sum\limits_{n=1}^\infty\frac{H_0^{(1)}(g_n a)}{
         \displaystyle{\left.\frac{dJ_0(g a)}{dh}\right|_{h=h_n}}}
         J_0(g_n r)e^{ih_nz}.\eqno(26.35)$$
При $z<0$    контур   интегрирования   следует   замыкать   в   нижней
полуплоскости,  интеграл  опять  сводится   к   вычетам   в   полюсах,
расположенных  теперь  ниже  первоначального  контура.  Выражение  для
$u(r,z)$ отличается в результате только общим знаком (замкнутый контур
обходится теперь по часовой стрелке) и знаком перед всеми $h_n$.

     Таким образом,  полное поле в волноводе удалось выразить  в  виде
ряда, который можно записать в виде
     $$u(r,z)=\sum\limits_{n=1}^\infty B_n J_0(g_n r)e^{ih_nz},
         \qquad z>0, \eqno(26.36)$$
то есть представить в виде суммы волн типа $E_{0n}$, причём полюсам,
лежащим  на  действительной  оси  при  заданном   $k$,   соответствуют
распространяющиеся  волны,  а  лежащим  на  мнимой оси --- затухающие.
Амплитуды  волн  $B_n$  находятся  из  (26.35)  с  помощью  вронскиана
цилиндрических функций
     $$ J_0(x){H_0^{(1)}}'(x) - {J_0'(x)}H_0^{(1)}(x)=\frac {2i}
         {\pi x}\eqno(26.37)$$
и равны
     $$B_n=\frac{2i}{h_n a^2 J_1^2(g_n a)}\,.\eqno(26.38)$$

     Полученный результат  является   прямым   следствием   известного
свойства   полноты   системы   волноводных  волн:  любое  поле  внутри
волновода, в области, свободной от источников, может быть разложено по
этой  системе.  Обращение  в  нуль  интеграла по разрезу является,  по
существу,  прямым следствием полноты системы функций. Если бы решалась
задача  о  дифракции  поля  электрического  диполя  на  цилиндрической
поверхности,  отделяющей  диэлектрический   цилиндр   от   окружающего
пространства,  то  поскольку  в  такой  структуре собственные волны не
образуют  полную  систему  функций,  то  помимо  вычетов   в   точках,
соответствующих собственным волнам,  существенный вклад в интеграл дал
бы интеграл по разрезу.

     Решим теперь  эту  же  задачу,  с самого начала используя полноту
системы собственных волноводных волн кругового цилиндра. При источнике
вида  (26.19)  из  этой  системы достаточно взять только электрические
волны  типа  $E_{0n}$.  Ввиду   выбранной   ориентации   источника   и
азимутальной  симметрии  его  поля  задача  относительно потенциальной
функции $u=A_z$ является  скалярной.  Поэтому  при  её  решении  нет
необходимости пользоваться общей теорией возбуждения векторных полей в
волноводе,  изложенной  в  разделе   20,   а   можно   непосредственно
рассмотреть  скалярный  вариант  задачи.  Для  этого необходимо решить
волновое уравнение
     $$\Delta u+k^2u=-4\pi\delta(r)\delta(z)\eqno(26.39)$$
в цилиндрической системе координат с граничным условием
     $$ \left.u(r,z)\right|_{r=a}=0.\eqno(26.40)$$
Будем искать  потенциал  полного  поля  при  $z>0$ в виде совокупности
волноводных волн, распространяющихся (или затухающих) вправо:
     $$u(r,z)=\sum\limits_{n=1}^{\infty} B_n u_n=\sum  \limits_{n=1}^\infty
        B_nJ_0(g_n r)e^{ih_nz},\qquad z>0.\eqno(26.41)$$
При $z<0$ поле ищется  в  таком  же  виде,  но  только  как  сумма
встречных волн $u_{-n}$; поэтому у всех $h_n$ следует сменить знак.

     Воспользуемся теперь вместо леммы Лоренца её скалярным аналогом
--- формулой Грина
     $$\int\limits_S\Bigl[v_1\frac{\partial v_2}{\partial n}-v_2
         \frac{\partial v_1}{\partial n}\Bigr]\,dS=\int
         \limits_V (v_1f_2-v_2f_1)\,dV,
         \eqno(26.42)$$
где $n$ --- внешняя нормаль к поверхности $S$, окружающей объём $V$,
$v_{1,2}$ --- решения, а $f_{1,2}$ --- правые части волновых уравнений
     $$\Delta  v_{1,2}+k^2 v_{1,2}=f_{1,2}\,.\eqno(26.43)$$
Выберем в  качестве  $v_1$  искомое поле (26.41),  а в качестве $v_2$ --- встречную
волну   $u_{-m}= J_0(g_m r)e^{-ih_mz}$.
 При этом  $f_1=-4\pi\delta(\rv  r)$, а  $f_2=0$ .
Проведём   в  (26.42)  интегрирование  по  объёму  волновода  $V$,
ограниченному сечениями $z=z_1<0$ и $z=z_2>0$.  Интеграл по  объёму,
очевидно,  равен  $4\pi$,  интеграл по боковой поверхности $r=a$ равен
нулю из-за граничного условия $u(a,z)=0$;  интеграл по сечению $z=z_1$
также  равен  нулю,  поскольку функции $J_0(g_n r)$ и $J_0(g_m r)$ при
$m\ne  n$  ортогональны  на  сечении,  а  $m$-тый  член   равен   нулю
тождественно.  В  интеграл по правому сечению $z=z_2$  вклад даёт только член
ряда (26.41) с $n=m$, который составляет
     $$  \begin{array}{l}\displaystyle{\int\limits_0^a}[B_m J_0(g_m r)
         e^{ih_mz_2}\,J_0(g_m r)e^{-ih_mz_2}(-ih_m)-\\[.5cm]-J_0(g_m r
         )e^{-ih_mz_2}\cdot B_m J_0(g_m r)e^{ih_mz_2}(ih_m)]2\pi r\,dr=
         \\[.2cm]=-4\pi ih_m B_m\displaystyle{\int\limits_0^a}J_0^2(g_
         m r)r\,dr=-2\pi ih_m B_m a^2 J_1^2(g_ma)\,.\end{array}
         \eqno(26.44)$$
Приравнивая полученную  величину  $4\pi$,  получаем  для коэффициентов
$B_m$ то же самое выражение (26.38), как и в дифракционной задаче.

     Сравнивая два способа  решения  одной  и  той  же  задачи,  можно
сделать  вывод  о  преимуществе  (в смысле простоты выкладок) способа,
использующего  теорию  возбуждения  волноводов.  При   этом,   однако,
необходимо  помнить,  что  он  применим  только в том случае,  когда у
структуры существует полная система ортогональных собственных  функций
и она уже известна.

%\end{document}

\newpage
\oddsidemargin=-0.4mm \evensidemargin=-0.4mm
\topmargin=-0.4mm
\headsep=7mm
\textheight=231.875mm
\textwidth=160mm
\mathsurround=2.5pt
\unitlength=1mm
%\begin{document}
%\input{macr.tex}
\thispagestyle{empty}
%\addtocounter{page}{313}

\begin{center}
   \subsubsection*{\rm Г\,Л\,А\,В\,А\, 10}
      \vspace{-1.15em}
      \line(6,0){160}\\
      \vspace{-1em}
      \line(6,0){160}
      \vspace{-1.15em}
   \subsubsection*{КВАЗИОПТИЧЕСКИЕ СТРУКТУРЫ}
      \vspace{31mm}
   \subsubsection*{27. Квазиоптические линии передачи}
\end{center}
\vspace{0.5cm}

\markboth{Глава 10.   Квазиоптические  структуры}{27.  Квазиоптические
         линии передачи}

\begin{center}\begin{minipage}[c]{0.75\textwidth}
\footnotesize{\parindent=0.5cm
         Непригодность обычных  волноводов  для  передачи  волн короче
         нескольких  миллиметров.  Открытые  волноводные  линии  и  их
         исследование методами квазиоптики. Параболическое уравнение и
         его решение в виде пучков  Гаусса-Эрмита.  Собственные  волны
         линзовой   линии.  Метод  фазовой  коррекции  и  интегральное
         уравнение.  Радиационные потери в линзовой линии. Особенности
         диафрагменной линии.
}\end{minipage}\end{center}
\vspace{0.5cm}

     Обычные волноводы    круглого    или    прямоугольного   сечения,
рассмотренные в разделе 9, начиная с волн длиною несколько миллиметров
и короче,  практически непригодны для передачи энергии или информации.
Уже в самом начале этого диапазона затухание достигает нескольких дБ/м
и   при   работе  в  одномодовом  режиме  растёт  с  частотой  $\sim
\omega^{3/2}$,  если с её увеличением соответственно  уменьшается  и
размер  поперечного  сечения.  Помимо этого,  из-за уменьшения площади
сечения  волновода  $\sim  \omega^2$  быстро  падает  его   пропускная
мощность,  а  само  изготовление  миниатюрных  волноводов представляет
собой сложную технологическую задачу.  Не спасает положения и  попытка
работать с волноводами привычных для сантиметрового диапазона размеров
путём перехода в многомодовый режим.  Хотя в этом  случае  затухание
растёт  существенно медленнее ($\sim \omega^{1/2}$),  но в волноводе
возбуждается много волн  с  различными  скоростями  распространения  и
распределениями полей,  что приводит к существенному искажению сигнала
на  выходе.  Число  новых   распространяющихся   в   волноводе   волн,
появляющихся   в   частотном   интервале  $\Delta\omega$  при  высоких
частотах,  определяется  асимптотической  формулой,  которая  является
следствием  теоремы  Куранта  и которая в двумерном случае (поперечное
сечение волновода) без  учёта  поправок  на  длину  контура  сечения
имеет вид
     $$ \Delta N=\frac S{\pi c^2}\omega \Delta \omega\,,\eqno(27.1)$$
где $S$ --- площадь поперечного сечения.

     Для волн субмиллиметрового диапазона единственный выход состоит в
переходе  к  линиям передачи,  работающим на иных принципах --- такими
линиями являются открытые  волноводы.  Во  всяком  открытом  волноводе
имеет  место излучение через боковые поверхности и,  на первый взгляд,
это излучение может привести только  к  увеличению  затухания.  Однако
 действуют  два  фактора,  делающие  рассматриваемые структуры
пригодными для  каналирования  очень  коротких  волн,  а  именно:  1)
почти  полное отсутствие металлических стенок,  что избавляет от омических
потерь в волноводах;  2) резкая  неравномерность радиационных   потерь   для
различных   типов   волн,  приводящая  к  существенному разрежению спектра
собственных  волн  линии.  При  этом  оказывается   возможным   использовать
в   качестве  линии  передачи  структуры,  поперечные размеры  которых
значительно  превышают  длину  волны.

     Исторически самой первой открытой линией для волн  миллиметрового
и  субмиллиметрового диапазона явилась {\it линзовая линия} --- прямой
аналог оптических систем,  используемых  для  передачи  изображений  в
существенно  более  короткой  части  спектра  электромагнитных волн (в
области видимого излучения).  Линзовая линия  (рис.~27.1)  состоит  из
последовательности расположенных на общей оси на одинаковом расстоянии
$2D$ друг от друга  {\it  длиннофокусных}  диэлектрических  линз.  Все
геометрические параметры линии существенно больше длины волны.  Обычно
используются тонкие  линзы,  толщина  которых  $2d$  много  меньше  их
диаметра $2a$, и при этом $a\ll D$.

\begin{wrapfigure}[11]{l}{7.0cm}
\begin{picture}(80,40)
\put(0,40){\special{em:graph fig27-1.bmp}}
\end{picture}
\hbox to 7.0cm{\hfil\footnotesize{Рис.~27.1.~Линзовая линия.
}\hfil}
\end{wrapfigure}

     Для видимого излучения (света),  то есть при длинах волн  порядка
$(4,0-7,6)\cdot10^{-5}$~см, расчёт таких систем производится методами
{\it геометрической оптики}. Полагается, что распространение света
происходит  вдоль  лучей,  направление   которых   в
каждой  точке  пространства  совпадает  с  нормалью  к
волновой  поверхности  (поверхности   равной
фазы электромагнитной волны). Геометрическая
оптика   фактически  не  учитывает    волновой
природы    света,   и   получаемые   с   её   помощью
результаты   точны  в  пределе    $\lambda\to     0$.
Дифракционные эффекты,  описываемые волновым
уравнением,  приходится  учитывать  лишь   в   таких
случаях,     когда,     например,  интересуются
интенсивностью   светового   пятна:  по  законам
геометрической   оптики в фокусе  линзы  или
сферического зеркала лучи сходятся в точку,  а
волновая     оптика    показывает,  что  пятно
размывается  по  объёму,   линейные   размеры
которого составляют несколько длин волн.

     При распространении  волновых  пучков  миллиметрового   диапазона
волновые  эффекты  играют  определяющую  роль.  Однако строгое решение
волнового уравнения аналитическими методами даже в  случае  простейшей
линзовой  линии получить невозможно --- как будет видно из дальнейшего
изложения,  в  этом  нет  и  необходимости.  Возникла  новая   область
теоретических  исследований,  названная {\it квазиоптикой},  в которой
удачно  сочетаются  методы  геометрической  оптики  и  методы  решения
дифракционных  задач,  опирающиеся на волновое уравнение.  Квазиоптика
оказалась очень эффективной при расчёте не только линий передачи, но
и   таких   важнейших  устройств  субмиллиметрового,  инфракрасного  и
светового диапазонов,  как {\it открытые  резонаторы},  которые  будут
рассмотрены в следующем разделе.

     Качественное описание работы линзовой линии состоит в  следующем.
Каждая  линза фокусирует падающий на неё пучок электромагнитных волн
и направляет его  сходящимся  пучком  к  следующей  линзе.  Поперечный
размер области, занятой полем, вблизи линзы не превышает её апертуры
$2a$,  а по мере удаления от линзы он  сначала  уменьшается,  а  потом
возрастает  в  соответствии  с законами волновой оптики.  На следующую
линзу попадает уже расходящийся пучок и необходимо  обеспечить,  чтобы
его  размер  не  превысил  апертуру линзы;  далее картина повторяется.
Затухание волны обусловлено тремя основными причинами:  часть  энергии
всё-таки  не  попадает  на  линзу и уходит в окружающее пространство
(радиационные потери),  часть отражается от поверхности линзы  (потери
на  отражение),  часть  поглощается в материале линзы (диэлектрические
потери).  В дальнейшем,  когда говорится о потерях,  всегда имеются  в
виду    радиационные   потери,   а   двумя   другими   видами   потерь
пренебрегается.

     Линзовая линия,  как  и  другие открытые квазиоптические системы,
формирует  {\it  длинные  параксиальные  пучки},  поперечные   размеры
которых  много  меньше  длины  пучка  и  много  больше длины волны,  а
волновой  вектор  в  каждой  точке  пучка  составляет  малый  угол   с
направлением   оптической   оси.   Для   квазиоптических   систем  два
безразмерных параметра всегда являются большими числами:
     $$  ka\gg1,\qquad  \frac  L  a\gg 1,\eqno(27.2)$$
где $a$  ---   поперечный   размер   системы,   $L$   ---   продольная
протяжённость  пучка  (в  линзовой  линии $2a$ --- поперечный размер
линзы,  а $L$ определяется расстоянием между  линзами). Отношение же
этих двух параметров
     $$ \delta =\frac {ka^2} L\eqno(27.3)$$
может принимать любые значения и является существенной характеристикой
конкретной системы при её квазиоптическом описании.

     Собственная волна  линзовой  линии не может быть цилиндрической в
смысле (7.1),  поскольку структура неоднородна вдоль оси  $z$.  Однако
она  несомненно  периодическая с периодом $2D$,  навязанным геометрией
структуры.  Так как этот период очень  велик  по  сравнению  с  длиной
волны, то характерные свойства собственных волн периодических структур
сантиметрового диапазона,  такие,  как наличие полос непропускания,  в
ней не проявляются. Следует отметить, что и математический аппарат для
описания волн в линзовых линиях заметно отличается  от  изложенного  в
разделах, например, 13 и 14.

     Приближения, позволяющие   построить    удовлетворительную    для
практических  целей  аналитическую  теорию,  приводят  к независимости
решений,  в которых отличны от нуля только  компоненты  поля  $H_x$  и
$E_y$,  и решений,  в которых не равны нулю только $E_x$ и $H_y$.  Эти
решения оказываются  независимыми  при  любой  форме  контура  линз  и
поэтому  все  расчёты  можно  проводить со скалярными величинами,  в
качестве которых для линзовой линии удобно взять компоненту магнитного
поля.   В   дальнейшем   изложении   везде   использовано  обозначение
$u(x,y,z)=H_x$.

     Квазиоптический подход  к  нахождению  собственных  волн линзовой
линии состоит в разбиении периода структуры на две области, в одной из
которых  решение  строится методами геометрической оптики,  а в другой
--- находится из волнового уравнения.  Такими областями  на  рис.~27.1
для  периода  $-D+d<z<D+d$  являются  область,  ограниченная сечениями
$z=-D+d$ и $z=D-d$, в которой решается волновое уравнение
     $$ \Delta u+k^2 u=0\,,\eqno(27.4)$$
и прилегающая область,  ограниченная справа  плоскостью  $z=D+d$,  где
расположена  сама  линза  и  где  рассматриваются  только свойства
фазовой поверхности волны, что характерно для геометрической оптики.

     Исходя из   предположения   о   параксиальном  волновом  пучке  и
связанных с этим условий (27.2),  волновое уравнение (27.4) может быть
существенно упрощено.  Действительно,  будем искать его решение в виде
произведения двух множителей
     $$ u(x,y,z)=w(x,y,z)\,e^{ikz}\,,\eqno(27.5)$$
из которых второй множитель быстро осциллирует вдоль оси $z$, а первый
является   сравнительно   медленно   меняющейся   функцией  поперечных
координат  $x,\,y$  и  ещё  более  медленной  функцией  $z$.   Чтобы
убедиться в этом, заметим, что функция $w(x,y,z)$ должна удовлетворять
уравнению
     $$ \frac {\partial^2 w}{\partial x^2}+\frac {\partial^2 w}
         {\partial y^2}+\frac {\partial^2 w}{\partial z^2}+
         2ik\frac{\partial w}{\partial z}=0\,,\eqno(27.6)$$
и введём безразмерные координаты
     $$ \xi=x\sqrt{\frac k L}\,,\qquad  \eta=y\sqrt
         {\frac k L}\,,\qquad \zeta=\frac z L\,.\eqno(27.7)$$

     В этих переменных уравнение (27.6) имеет вид
     $$ \frac {\partial^2 w}{\partial \xi^2}+\frac {\partial^2 w}
         {\partial \eta^2}+\frac 1{kL}\frac {\partial^2 w}{\partial
         \zeta ^2}+2i\frac{\partial w}{\partial \zeta}=0\,;
         \eqno(27.8)$$
из  дальнейшего  станет  ясно,  что при таком   выборе безразмерных
координат их максимальные значения  в области,  занятой пучком,
имеют один и тот же порядок.  Ввиду малости  величины  $1/kL$
(согласно  условиям  27.2)  имеются  все  основания  отбросить член со
второй производной по  $z$  и  тем  самым  перейти  к  параболическому
уравнению
     $$ \frac {\partial^2 w}{\partial \xi^2}+\frac {\partial^2 w}
         {\partial \eta^2}+2i\frac{\partial w}{\partial \zeta}=0\,,
         \eqno(27.9)$$
поиск решения  которого  практически  во  всех  случаях   проще,   чем
эллиптического уравнения, каковым является волновое уравнение (27.4).

     Начнём рассмотрение с двумерных решений этого уравнения, имея в
виду,   что   оно   разделяется   по   переменным   $\xi$   и  $\eta$.
Непосредственной подстановкой  легко  убедиться,  что  независящая  от
$\eta$ и определяемая только $L$ функция
     $$w_m(\xi,\zeta)=\frac {A^{2m+1}}{{(AA^*)}^m}\,H_m(AA^*\xi)\,
         e^{-\frac{\xi^2}2 A^2},\qquad  m=0,1,2,\dots\,,
         \eqno(27.10)$$
где $H_m(x)$ --- полиномы Эрмита, удовлетворяет уравнению (27.9), если
функция $A(\zeta)$ удовлетворяет уравнению
     $$\frac{dA}{d\zeta}=-\frac i 2\,A^3\,.\eqno(27.11)$$
Решение последнего уравнения при дополнительном условии $A(0)=1$ есть
     $$A(\zeta)=\frac1{\sqrt{1+i\zeta}}\,,\quad\hbox{причём}\quad
         \frac{1}{A^{*2}}+\frac{1}{A^2}=2\,.\eqno(27.12)$$

     Напомним, что  полиномы  Эрмита  удовлетворяют  дифференциальному
уравнению
     $$ \frac{d^2H_m}{dx^2}-2x\frac{dH_m}{dx}+2mH_m=0\,\eqno(27.13)$$
и обычно  нормируются  таким  образом,  что  могут  быть  вычислены по
формуле
     $$H_m(x)=(-1)^me^{x^2}\frac{d^m}{dx^m}\,e^{-x^2}\,;\eqno(27.14)$$
поэтому $H_0=1,\,H_1=2x,\,H_2=4x^2-2x$ и так далее.

     Из всех  решений  (27.10)  наибольшее значение для практики имеет
простейшая функция
     $$w_0(\xi,\zeta)=A(\zeta)\,e^{-\frac{\xi^2} 2 A^2(\zeta)}\,,
         \eqno(27.15)$$
определяющая так   называемый   {\it  гауссов  пучок}.  Однако  и  вся
совокупность функций $w_m$,  представляющих собой {\it волновые  пучки
Гаусса-Эрмита},  имеет  фундаментальное значение в квазиоптике и на их
свойствах целесообразно остановиться подробнее.

     Функцию $w_0$ (27.15) при $ A(\zeta)$, определяемой формулой (27.12),
можно записать в виде
     $$ w_0(\xi,\zeta)=A(\zeta)\;e^{i\frac{\xi^2}{2{\cal R}(\zeta)}}
         \;e^{-\frac{\xi^2}{2\beta^2(\zeta)}}\,,\eqno(27.16) $$
где
     $${\cal R}(\zeta)= \frac 1 \zeta +\zeta,\qquad \beta^2=1+\zeta^2\,.
         \eqno(27.17)$$
В этом  выражении  безразмерные  величины   $\beta$   и   ${\cal
R}$ представляют собой так называемую {\it полуширину} пучка и радиус
кривизны его  волновой  поверхности;  в  переменных  $x,\,z$  им
соответствуют $b=\sqrt{L/k}\beta$    и    $R={\cal   R}L$.
Происхождение   термина {\it полуширина} пучка очевидно из вида
формулы  (27.16),  а  в  том,  что ${\cal  R}$  есть нормированный
радиус кривизны поверхности постоянной фазы вблизи оси пучка,
нетрудно убедиться,  выписав  уравнение этой поверхности,
проходящей через точку $\xi=0,\;\zeta=\zeta_0$:
     $$ kL(\zeta-\zeta_0) -\frac 1 2 \arctg{\zeta}+\frac 1 2
         \arctg{\zeta_0} +\frac {\xi^2}2\,\frac {\zeta}{1+\zeta^2}
           =0\,.\eqno(27.18)$$

     Отметим, что  функция  $w_m$  симметрична  относительно плоскости
$\zeta=0$,  радиус кривизны при переходе через  эту  плоскость
меняет знак,  проходя  через  бесконечность.  Ширина  волнового
пучка в этой плоскости минимальна и говорят,  что  в  ней
расположена  {\it перетяжка} пучка.  Величина  $\beta$  в  равной
степени относится ко всем пучкам Гаусса-Эрмита,  но реальная
ширина пучка, очевидно, растёт с номером $m$,  поскольку
степенной  рост полиномов при отклонении от оси пучка может быть
подавлен  экспоненциальным  зарезанием  лишь  при  большем
значении поперечной координаты.

     Так как параболическое уравнение,  в частности, описывает явление
диффузии,  то  поведение  поля  в пространстве можно рассматривать как
диффузию амплитуды поля,  происходящую, разумеется, не во времени, как
в  обычных  задачах  диффузии  и  теплопроводности,  а  с координатой,
отсчитываемой вдоль направления основного  луча  пучка.  При  этом  из
решения   уравнения   следует,  что  скорость  диффузии  в  поперечной
плоскости к этому направлению существенно выше,  чем  вдоль  основного
луча.

     Важным свойством   пучков    Гаусса-Эрмита    является    наличие
поверхностей,  называемых {\it каустиками},  по обе стороны от которых
структура  поля  пучка  существенно   разная.   Модуль   экспонент   в
соотношении   (27.10)   определяется   переменной  $t=\xi  A(\zeta)A^*
(\zeta)$; образуя функцию
     $$\psi_m(t)=H_m(t)\,e^{-\frac {t^2} 2}\,,\eqno(27.19)$$
легко можно  убедиться,  что  с  учётом  (27.13)  она  удовлетворяет
уравнению
     $$\frac{d^2 \psi_m}{d t^2}+(2m+1-t^2)\psi_m=0\,.\eqno(27.20)$$
Согласно теории дифференциальных уравнений решения уравнения
     $$\frac{d^2 f}{d t^2}+q(t)f=0\,\eqno(27.21)$$
в интервале,  где $q(t)>0$,  имеют осциллирующий характер,  а там, где
$q(t)<0$  ---  экспоненциально  затухают  (или  возрастают).   Поэтому
уравнение каустики имеет вид
     $$ A(\zeta)A^*(\zeta)\xi=\pm\sqrt{2m+1}\,\eqno(27.22)$$
или
     $$ \xi^2 -\zeta^2(2m+1)=2m+1\,,\eqno (27.23)$$
то есть  в  двумерной  задаче  каустика  определяется  гиперболой;   в
плоскости   $\zeta=0$   каустики   проходят   через   точки   $\xi=\pm
\sqrt{2m+1}$.

     Вернёмся теперь   к   изучению   линзовой  линии,  когда  ввиду
предполагаемой тонкости линзы её локальное взаимодействие с волновым
пучком  достаточно  хорошо  может быть описано методами геометрической
оптики.  В параксиальном приближении линза полностью определена  своим
фокусным   расстоянием   $F$,  а  для  описания  линии  удобно  ввести
безразмерный параметр $\nu$, равный отношению периода структуры $2D$ к
удвоенному фокусному расстоянию линз:
     $$ \nu=\frac {D}{F}\,.\eqno(27.24)$$
Собственная волна   линзовой   линии   должна   удовлетворять  условию
повторяемости поля на периоде структуры.  В частности,  на выходе двух
соседних  линз  поля  могут  различаться только на постоянный,  вообще
говоря, комплексный множитель:
     $$ u(x,y,-D+d)=\chi \;u(x,y,D+d)\,.\eqno(27.25)$$

     Воспользуемся хорошо  известным свойством линзы --- если на неё
падает  вдоль  оси  параксиальный  пучок  с  расходящейся  сферической
волновой  поверхностью,  радиус  кривизны  которой  равен $2F$,  то на
выходе линзы имеет место сходящийся пучок с тем же радиусом  кривизны.
В результате изображение светящейся точки,  расположенной на удвоенном
фокусном расстоянии от срединной плоскости симметричной линзы, отстоит
от  этой  плоскости  также  на  $2F$.  Фактически  фокусирующая  линза
изменяет знак радиуса кривизны волнового фронта такого пучка, оставляя
величину радиуса неизменной.

     Рассмотрим теперь двумерную линию,  состоящую из однородных вдоль
$y$ цилиндрических линз и попробуем  выбрать  в  качестве  собственной
волны структуры пучок Гаусса-Эрмита,  перетяжка которого расположена в
плоскости $z=0$.  Ввиду симметрии пучка радиус кривизны  его  волновой
поверхности  на  выходе  первой линзы и входе второй будет различаться
только  знаком.  Поэтому  условие  (27.25)  будет  выполнено,  если  в
качестве   решения  параболического  уравнения  в  пространстве  между
линзами выбрать любую функцию (27.10),  подобрав единственный параметр
$L$,  определяющий эти функции,  таким образом,  чтобы радиус кривизны
волнового фронта в  плоскости  линзы  (он  одинаков  для  всех  $w_m$)
оказался равным её удвоенному фокусному расстоянию. Согласно (27.17)
искомый радиус кривизны равен
     $$ R(z)=L\Bigl(\frac L z +\frac z L\Bigl)\,;\eqno(27.26)$$
полагая далее $z=D$ (пренебрегая тем  самым  толщиной  линзы  $2d$  по
сравнению  с  расстоянием  между  линзами)  и  приравнивая  $R(D)=2F$,
находим
     $$ L=D\sqrt{\frac {2-\nu} \nu}\,.\eqno(27.27)$$
В результате  определены  радиус  кривизны  волновой   поверхности   и
полуширина пучка в любой плоскости между линзами:
     $$ R(z)=z+\frac {(2-\nu)D^2}{z\nu}\,,\eqno(27.28)$$
     $$ b^2(z)=\frac D k \sqrt{\frac{2-\nu}\nu}\,\Bigl(
         1+\frac {z^2\nu}{D^2(2-\nu)}\Bigr)\,.\eqno(27.29)$$

     Из приведённого  решения  ясно,  что  собственные  волны  могут
существовать  в  линзовой  линии только при значении параметра $\nu$ в
интервале
     $$ 0<\nu<2\,,\eqno(27.30)$$
вне которого $b^2(z)$ становится мнимой  величиной  и  поле  пучка  не
спадает  в  поперечном  направлении.  В  случае  $\nu=0$  и отсутствия
диафрагм линзовая линия  вырождается  в  свободное  пространство,  для
которого   таких   решений   нет.  Отрицательные  $\nu$  соответствуют
рассеивающим линзам и  отсутствие  требуемых  решений  в  этом  случае
представляется  очевидным.  Значения $\nu>2$ свидетельствуют о слишком
редкой расстановке  линз,  когда  $D>2F$,  а  такие  собирающие  линзы
действуют уже как рассеивающие.

     Полуширина пучка в плоскости линз согласно (27.29) определяется  формулой
     $$ b^2(D)=\frac {2D} k \sqrt{\frac 1 {2\nu-\nu^2}}\,.
         \eqno(27.31)$$
Полуширина минимальна и равна $b(D)=\sqrt{2D/k}$ при $\nu=1$.
Соответствующая  система линз называется \textit {конфокальной}. В такой системе
последовательные линзы отстоят друг от друга на удвоенное фокусное расстояние.
В геометрооптическом приближении каждая линза конфокальной системы
собирает лучи, выходящие из центра предшествующей линзы, в центре последующей.

Отметим также, что равную ширину пучка в  средней  плоскости
симметричных  линз имеют собственные волны двух  линий, у которых
$\nu_1+\nu_2=2$ и, следовательно, фокусные расстояния  связаны соотношением
     $$ \frac 1 {F_1}+\frac 1{F_2}=\frac 2 D\,.\eqno(27.32)$$

     Полученное выше   решение   описывает   идеализированную   линию,
поскольку оно  фактически  предполагает  неограниченное  в  поперечном
направлении  действие  линз.  В  частности,  это приводит к тому,  что
постоянная $\chi$ в (27.25)  для  всех  собственных  волн  оказывается
комплексным  числом с модулем,  равным единице.  По смыслу введённой
таким образом величины $\chi$ она непосредственно связана  с  потерями
энергии  на  каждом  периоде  структуры  ---  относительное уменьшение
потока вектора Умова-Пойнтинга на каждом периоде структуры  составляет
$1-|\chi|^2$. Из физических соображений следует ожидать, что в случае,
когда полуширина пучка $b$ значительно меньше размера  апертуры  линзы
$a$,  собственные  функции не должны заметно отличаться от приведенных
выше.  При промежуточных значениях $a$ и особенно в  случае  $a\ll  b$
отклонения   могут   быть  очень  большие;  тогда  более  удобным  для
расчётов  оказывается  иной  подход  к  задаче,  сводящий   её   к
интегральному уравнению.

     Так как  функция $w_m$,  определяемая соотношением (27.10),  есть
решение параболического уравнения (27.9), то значения $w_m(\xi,\zeta)$
связаны   со  значениями  этой  функции  в  плоскости  $\zeta=\zeta_0$
интегральным соотношением
     $$ w_m(\xi,\zeta)=\int\limits_{-\infty}^\infty G(\xi-\xi_0,
         \zeta-\zeta_0)w_m(\xi_0,\zeta_0)\,d\xi_0\,,\eqno(27.33)$$
где $G(\xi-\xi_0,\zeta-\zeta_0)$  ---  функция  Грина  параболического
уравнения, которая, как известно, имеет вид
     $$G(\xi-\xi_0,\zeta-\zeta_0)=\sqrt{\frac 1{2\pi i (\zeta-\zeta_0
         )}}\;e^{i\frac{(\xi-\xi_0)^2}{2(\zeta-\zeta_0)}}\,.
         \eqno(27.34)$$
Поэтому решение двумерной граничной задачи  нахождения  поля  $u(x,z)$
при  $z>z_0$  по заданному значению $u(x,z_0)$ для волнового уравнения
(27.4) в параксиальном приближении в переменных (27.7) можно  записать
в виде
     $$  u(\xi,\zeta)=\sqrt{\frac 1{2\pi i (\zeta-\zeta_0)}}\;
         e^{ikL(\zeta-\zeta_0)} \int\limits_{-\infty}^\infty
         u(\xi',\zeta_0)\,e^{i\frac {(\xi-\xi')^2}
         {2(\zeta-\zeta_0)}}\,d\xi'\,.\eqno(27.35)$$

     Подставляя в  эту  формулу   функцию   $w_m$,   после   несложных
преобразований находим аналитическое выражение интеграла,  которое нам
понадобится в дальнейшем:
     $$  \begin{array} {l} \int\limits_{-\infty}^\infty H_m(\sqrt
         {\cos{\varphi}} x')\;e^{-[\frac {{x'}^2} 2 (\cos{\varphi}+
          i\sin{\varphi})+ixx']}\,dx'=\\[.2cm]\displaystyle{
          = \sqrt{\frac{2\pi}{\cos{\varphi}+i\sin{\varphi}}}}\,
          (\cos{2\varphi}-i\sin{2\varphi})^{\frac m2}\,e^{-i\frac
          \pi 2 m}\,H_m(\sqrt{\cos{\varphi}} x)\;e^{-\frac {{x}^2} 2
          (\cos{\varphi}-i\sin{\varphi})}\,.\end{array}\eqno(27.36)$$
В последней формуле угол $\varphi$ --- если  иметь  в  виду  уравнение
(27.35)  --- определяется следующим выражением:  $\sqrt{(1+ \zeta^2_0)
(1+\zeta^2)}\cdot\cos\varphi=\zeta-\zeta_0$.

     Решение (27.35)   граничной   задачи   может   быть   получено  и
непосредственно из  волнового  уравнения  в  предположении,  что  поле
представляет   собой   совокупность  плоских  волн,  направленных  под
небольшими углами к оси пучка. Разложим граничную функцию $u(x,z_0)$ в
интеграл Фурье:
     $$u(x,z_0)=\int\limits_{-\infty}^\infty f(\alpha)\,e^{i\alpha x}\,
         d\alpha\,.\eqno(27.37) $$
Умножив подынтегральное  выражение  на   $\exp{(i\sqrt{k^2-\alpha^2}\,
(z-z_0))}$, получим функцию
     $$ u(x,z)= \int\limits_{-\infty}^\infty f(\alpha)\,e^{i[\alpha x+
         \sqrt{k^2-\alpha^2}\,(z-z_0)]}\,d\alpha\,,\eqno(27.38) $$
которая, как легко убедиться, для всех $z>z_0$ удовлетворяет волновому
уравнению,  не  содержит  приходящих  волн,  а при $z=z_0$ переходит в
заданную функцию (27.37).  При $\alpha<k$ можно  ввести  угол  $\beta$
посредством соотношения $\sin{\beta}=\alpha /k$; тогда подынтегральное
выражение будет представлять собой плоскую волну
     $$e^{ik(x\sin{\beta}+z\cos{\beta})} \eqno(27.39)$$
с амплитудой $f(\alpha)$,  распространяющуюся под углом $\beta$ к  оси
$z$.

     Предположим теперь,  что функция $f(\alpha)$ заметно  отлична  от
нуля  только  при  $\alpha\ll  k$,  а  при  б\'ольших  $\alpha$ быстро
стремится к нулю;  тогда для всех $\alpha$,  дающих вклад  в  интеграл
(27.38), имеет место разложение
     $$\sqrt{k^2-\alpha^2}=k-\frac {\alpha^2}{2k}\,,\eqno(27.40)$$
и поле $u(x,z)$ может быть представлено в виде
     $$u(x,z)=e^{ik(z-z_0)}\int\limits_{-\infty}^\infty f(\alpha)\,
         e^{i\alpha x-i\frac{\alpha^2}{2k}(z-z_0)}\,d\alpha\,.
         \eqno(27.41)$$
Подставив сюда Фурье-образ
     $$  f(\alpha)=\frac 1{2\pi}\int\limits_{-\infty}^\infty
         u(x,0)\,e^{-i\alpha x}\,dx\eqno(27.42)$$
и изменив порядок интегрирования, получим выражение
     $$u(x,z)=e^{ik(z-z_0)}\int \limits_{-\infty}^\infty u(x',0)
         \Bigl[\frac 1{2\pi}\int\limits_{-\infty}^\infty e^{i\alpha
         (x-x')-i\frac{\alpha^2}{2k}(z-z_0)}\,d\alpha\Bigr]\,dx'\,,
         \eqno(27.43)$$
в котором внутренний интеграл легко вычисляется.  В  результате  после
замены переменных (27.7) приходим к формуле (27.35).

     Этот же  результат  может  быть получен и с помощью функции Грина
волнового уравнения.  Поучительно привести  соответствующие  выкладки,
поскольку они наиболее отчётливо выявляют те ограничения,  при которых
справедлив окончательный результат;  для этого рассмотрим трёхмерный
случай.  Функция  Грина $G(\rv r,\rv r_0)$ для полупространства $z>0$,
удовлетворяющая уравнению (27.4) с правой  частью  $-\delta(\rv  r-\rv
r_0)$  и  граничному условию $G=0$ при $z=0$,  легко находится методом
зеркального изображения  с  помощью  основной  функции  Грина  (17.20)
скалярного волнового уравнения:
     $$G(x,y,z,x_0,y_0,z_0)=\frac 1{4\pi}\left(\frac{e^{ikR_+}}{R_+}
         -\frac{e^{ikR_-}}{R_-}\right),\eqno(27.44)$$
где $R_{\pm}=\sqrt{(x-x_0)^2+(y-y_0)^2+(z\mp   z_0)^2}$;   $R_+$   ---
расстояние от произвольной точки $(x,y,z)$ до точки $(x_0,y_0,z_0)$, в
которой ищется поле,  $R_-$ --- расстояние от точки $(x,y,z)$ до точки
$(x_0,y_0,-z_0)$,    то    есть   до   зеркального   отражения   точки
$(x_0,y_0,z_0)$ в плоскости $z=0$.

     С помощью второй формулы Грина решение граничной задачи,  то есть
поле в точке наблюдения $(x_0,y_0,z_0)$, выражается в виде
     $$u(x_0,y_0,z_0)=\int\limits_{z=0}u(x,y,0)\frac{\partial G}{
         \partial z}\,dx\,dy\,,\eqno(27.45)$$
где производная  $-\partial/\partial  z$  есть  производная по внешней
нормали к границе  области  $z>0$.  На  плоскости  $z=0$,  по  которой
производится интегрирование,
     $$ R_+=R_-=R_0\,,\qquad \frac{\partial R_+}{\partial z}=-\frac
         {\partial R_-}{\partial z}=-\frac {z_0}{R_0}\,,\eqno(27.46)$$
причём $R_0=\sqrt{(x-x_0)^2+(y-y_0)^2+z_0^2}$.

     Поскольку нас в первую  очередь  интересует  поле  вблизи  второй
линзы, то можно считать условия (27.2) выполненными, и поэтому формула
(27.45) допускает существенные упрощения.  Прежде всего  учтём,  что
при   вычислении   производной   $\partial  G/\partial  z$  достаточно
дифференцировать только экспоненту,  поскольку  вклад  от  знаменателя
даёт слагаемое в $kR_0$ раз меньшее.  В результате получаем формулу,
справедливую лишь  на  расстояниях  от  плоскости  $z=0$,  больших  по
сравнению с длиной волны:
     $$ u(x_0,y_0,z_0)=-\frac{ikz_0}{2\pi}\int\limits_{z=0}\frac
         {e^{ikR_0}}{R_0^2}\,u(x,y,0)\,dx\,dy\,.\eqno(27.47)$$

     Сделаем ещё два упрощения,  допустимые при условии $z_0\gg |x|,
|x_0|,|y|,|y_0|$  ---  когда  расстояния  от   первой   линзы   больше
поперечного  размера  пучка.  Заменим  в  знаменателе (27.47) $R_0$ на
$z_0$,  а в экспоненте разложим  $R_0$  по  обратным  степеням  $z_0$,
оставив в этом разложении только два первых члена:
     $$R_0=z_0+\frac{(x-x_0)^2+(y-y_0)^2}{2z_0}\,.\eqno(27.48)$$
Пренебрежение последующими  членами  разложения  допустимо  лишь   при
условии
     $$ \frac{k|x|^3}{z^2}\ll 1,\qquad \frac{k|y|^3}{z^2}
         \ll 1 \eqno(27.49)$$
для всех точек пучка,  кроме  относительно  небольшой  области  вблизи
первой   линзы.   Отметим,   что  второй  член  в  разложении  (27.48)
существенен для выявления  дифракционных  явлений  в  линзовой  линии.
Расстояние  между  линзами  не настолько велико,  чтобы считать вторую
линзу  расположенной  в  дальней  зоне  первой.  Волновые  эффекты   в
пространстве  между  линзами  обусловлены {\it дифракцией Френеля},  в
которой играют роль лишь небольшие участки волновой поверхности.

     Будем считать,  что  линзы  помещены  в  отверстиях  непрозрачных
экранов,  так  что интегрирование в (27.47) достаточно провести лишь в
пределах апертуры линзы $S$.  Меняя  обозначения  таким  образом,  что
$(x,y,z)$ --- координаты точки наблюдения,  а $x',y,'$ --- переменные,
по  которым  производится  интегрирование,  и  считая,  что  граничное
значение  функции  $u$  задано в плоскости $z=z_0$,  получаем формулу,
лежащую в основе {\it дифракционной теории оптического изображения}  и
используемую в оптике для расчета интенсивности изображения:
     $$u(x,y,z)=-\frac{ik}{2\pi (z-z_0)}\,e^{ik(z-z_0)}\int\limits_S
         u(x',y',z_0)\;e^{\frac{ik}{2 (z-z_0)}[(x-x')^2+(y-y')^2]}
         \,dx'dy'. \eqno(27.50)$$
Несколько способов  решения  одной  и той же граничной задачи были так
подробно   рассмотрены   для   того,   чтобы   представить    основные
математические методы, используемые в задачах квазиоптики.

    В геометрооптическом приближении линзу можно заменить {\it фазовым
корректором},  действие  которого  полностью   определяется   функцией
$\psi(x,y)$, связывающей между собой поле на входе и выходе линзы:
     $$ u(x,y,D+d)=u(x,y,D-d)\,e^{i\psi(x,y)}\,.\eqno(27.51)$$
Функция $\psi(x,y)$  является  единственной  характеристикой  линзы  и
представляет  собой  {\it  оптическую  длину   пути}   вдоль   прямой,
параллельной оси $z$, между плоскостями $z=D-d$ и $z=D+d$:
     $$ \psi(x,y)=k\int\limits_{D-d}^{D+d}\sqrt{\varepsilon(x,y,z)}\;dz
       \,.\eqno(27.52)$$
Такое описание  линзы  как   {\it   фазового   корректора}   полностью
соответствует   законам   геометрической  оптики  в  их  применении  к
прохождению параксиального пучка через тонкую  линзу.  Для  однородной
($\varepsilon=   const   $)   двояковыпуклой   линзы  со  сферическими
поверхностями из (27.52) следует, что
     $$   \psi(x,y) = -\frac {k(x^2+y^2)}{2F}\,,\eqno(27.53)$$
где $F$ --- фокусное расстояние линзы.  Легко видеть, что квадратичный
фазовый корректор (27.53) действует как тонкая линза: так, при падении
пучка параллельных лучей на фазовый корректор фазовая  поверхность  на
выходе  является  частью  сферы,  и  нормальные к ней лучи собираются в
одной точке,  лежащей  в  фокальной  плоскости.  Очевидно,  что  метод
фазового   корректора  позволяет  в  геометрооптическом  представлении
описывать значительно более широкий класс фокусирующих элементов,  чем
тонкая линза со сферическими поверхностями.

     Формул (27.50),  (27.51) и условия (27.25) достаточно для  вывода
интегрального  уравнения линзовой линии.  Введение фазового корректора
позволяет считать линзу бесконечно тонкой, при этом координату входной
плоскости  линзы  будем  помечать нижним индексом $-$,  а выходной ---
индексом $+$. Выразим поле на выходе второй линзы через поле на выходе
первой:
     $$u(x,y,D_+)=-\frac{ik}{4\pi D}e^{i[2kD+\psi(x,y)]}\int
         \limits_S u(x',y',-D_+)\;e^{\frac{ik}{4D}[(x-x')^2+(y-y')^2]}\,
         dx'dy'\,.\eqno(27.54)$$
Потребовав повторяемости  поля  на  каждом  периоде  линии,   получаем
интегральное уравнение
     $$ \int \limits_S u(x',y',-D_+)\;e^{\frac{ik}{4D}
         [(x-x')^2+(y-y')^2]}\,dx'dy'=\chi\frac{4\pi D} k ie^{-i[2kD
         +\psi(x,y)]}u(x,y,-D_+) \,,\eqno(27.55)$$
иногда называемого в литературе уравнением Мандельштама.

     Это уравнение   представляет   собой   однородное    интегральное
уравнение Фредгольма второго рода и из него следует,  что при заданном
$k$ линия  полностью  определяется  расстоянием  между  линзами  $2D$,
апертурой  $S$ и функцией $\psi(x,y)$,  описывающей фазовую коррекцию.
Для широкого спектра этих функций уравнение  (27.55)  имеет  счётную
бесконечную последовательность собственных значений $\chi_n$,  которым
соответствует  совокупность   собственных   функций   $u_n(x,y,-D_+)$,
определяющих  поле  на выходе каждой линзы.  Поле в произвольной точке
$(x_0,y_0,z_0)$ в области между линзами находится по формуле  (27.50).
Модуль   собственного  значения  $\chi_n$  определяет  потери  энергии
собственной  волны  на  каждом  периоде  структуры.  Поскольку   поток
мощности   через   поперечное  сечение  $z=z_0$  пропорционален  $\int
|u_n(x,y,z_0)|^2\,dS$,  то  относительные  потери  на  периоде   равны
$1-|\chi_n|^2$.  Фаза  $\chi_n$  определяет добавку к основному набегу
фазы на периоде линии $2kD$   и позволяет ввести усреднённую фазовую
скорость собственной волны в линзовой линии,  которая в силу сделанных
выше предположений всегда близка к скорости света $c$.

     Уравнение (27.55) удобно  несколько  видоизменить,  введя  вместо
$u(x,y,D_+)$ новую функцию
     $$v(x,y) = u(x,y,D_+)\,e^{-\frac i 2 \psi(x,y)}\,.\eqno(27.56)$$
Для симметричных  линз $v(x,y)$ имеет простой физический смысл --- это
поле  в  средней  плоскости  линзы.   Тогда   интегральное   уравнение
записывается в виде
     $$\int\limits_S v(x',y')\,e^{\frac{ik}{4D}[(x-x')^2+(y-y')^2]+
         \frac i 2 [\psi (x,y)+\psi(x',y')]}\,dx'dy'= \lambda v(x,y)\frac{4\pi D} k
        \,,\eqno(27.57)$$
где постоянная  $\lambda$   отличается   от   постоянной   $\chi$   на
комплексный множитель с модулем, равным единице. Поэтому относительные
потери мощности,  переносимой волной,  на периоде структуры составляют
$1-|\lambda|^2$.

     Из теории  интегральных  уравнений  следует,  что  поскольку ядро
уравнения (27.57) симметричное,  то собственные функции $v_n$ и  $v_m$
ортогональны (при $\lambda_n\ne\lambda_m$), то есть
     $$ \int\limits_S v_n(x,y)v_m(x,y)\,dxdy=0\,,\eqno(27.58) $$
и образуют  полную систему.  Поэтому любое поле $w(x,y)$,  заданное на
некоторой линзе, может быть представлено в виде разложения
     $$v(x,y)=\sum\limits_nC_n v_n(x,y)\eqno(27.59)$$
с коэффициентами
     $$C_n=\frac{\int \limits_S v(x,y)\,v_n(x,y)\,dx\,dy}
         {\int\limits_S v_n^2(x,y)\,dx\,dy}\,.\eqno(27.60)$$
С помощью этих коэффициентов легко находится поле на любой последующей
линзе --- оно представляется тем же рядом (27.59),  в  котором  каждый
коэффициент  $C_n$  домножается  на  $\lambda_n^p$,  где $p$ --- число
пройденных линз.

     Отметим, что ядро  уравнения  (27.57)  не  является  эрмитовым  и
поэтому,  вообще  говоря,  $\int  v_n(x,y)\,v_m^*(x,y)\,dx\,dy\ne 0$ при
$m\ne  n$.  Поэтому  суммарный  поток   энергии   вдоль   линии   {\it
неаддитивно}  складывается из потоков отдельных собственных волн,  так
как
     $$\int|v(x,y)|^2\,dxdy\ne\sum\limits_n|C_n|^2
         \int|v_n(x,y)|^2\,dx\,dy.\eqno(27.61)$$
Впрочем, и в обычных волноводах с потерями поток энергии,  переносимый
различными волнами, также неаддитивен.

     Интегральное уравнение  для  двумерной  линзовой линии может быть
получено из (27.55) интегрированием по $y'$  в  бесконечных  пределах.
Перейдя  предварительно  к  безразмерным  переменным  (27.7),  где $L$
следует положить равным $2D$,  и представляя с помощью (27.24) функцию
коррекции,  зависящей в этом случае только от одной переменной $x$,  в
виде $\psi(\xi)=-\nu\xi^2$, получаем интегральное уравнение для поля в
средней плоскости линзы:

          $$ \int\limits_{-\sqrt \delta}^{\sqrt \delta} v(\xi')e^{\frac i 2[(\xi-
         \xi')^2- \nu(\xi^2+{\xi'}^2)]}\,d\xi'=\sqrt{2\pi}
         \lambda\,v(\xi)\,,\eqno(27.62)$$
где параметр $\delta$ совпадает с параметром $\ae$, определённым
формулой (27.3), если в последней заменить   $L$   на $2D$.

     В приближении    неограниченного    корректора     в     пределах
интегрирования  следует  положить  $\delta=\infty$.  В  этом  случае с
помощью интеграла  (27.36),  в  котором  выбраны  $\cos{\varphi}=\sqrt
{2\nu-\nu^2},\;\sin{\varphi}=1-\nu$, находим собственные функции
     $$ v_m(\xi)=\exp{(-\frac{\xi^2} 2 \sqrt{2\nu-\nu^2})}
         \,H_m({\sqrt[4]{2\nu-\nu^2}}\xi),\qquad m=0,1,2\dots\,;
         \eqno(27.63)$$
при этом модули всех собственных значений равны  единице.  Разумеется,
что  если  по  найденному  $v(\xi)$  вычислить  поле на выходе линзы и
подставить его в  (27.35),  то  после  вычисления  интеграла  опять  с
помощью  формулы  (27.36) получим то же самое выражение для поля между
линзами, что даёт и решение (27.10). При конечных значениях $\delta$
найти  собственные  значения и собственные функции уравнения (27.63) в
замкнутой  аналитической  форме  удаётся  только  для   конфокальных
корректоров  $(\nu=1)$,  но  и  в этом случае решение содержит сложные
трансцендентные  функции,  так  что  приводить  здесь  эти   выражения
нецелесообразно.

     Для всех остальных значений $\nu$  и  для  других  видов  фазовых
корректоров,  определяющих функцию $\psi(\xi)$, вычисление собственных
значений производится либо непосредственным прямым численным  решением
уравнения  (27.62),  либо  путём  разложения неизвестной собственной
функции уравнения с конечными пределами по найденным выше  собственным
функциям для бесконечного корректора, которые образуют полную систему.
Тогда для  коэффициентов  разложения  получается  бесконечная  система
линейных  однородных  алгебраических  уравнений и собственные значения
$\lambda_m$,  определяющие радиационные потери,  находятся из  условия
обращения в нуль детерминанта этой системы.

\begin{wrapfigure}[12]{l}{7.0cm}
\begin{picture}(80,45)
\put(8,45){\special{em:graph  fig27-2.bmp}}
\end{picture}
\hbox to 7.0cm{\hfil\footnotesize
{Рис.~27.2.~Потери в  линзовой  линии.}\hfil}
\end{wrapfigure}
     Вычисленные таким  образом  потери   в   конфокальной   линии   с
квадратичными  корректорами  представлены  на рис.~27.2 для нескольких
первых собственных волн в функции параметра $\delta$. Результаты
расчётов остаются практически неизменными и для неконфокальных линий
при изменении  $\nu$  почти  во  всём  диапазоне  (27.30),  если  по
горизонтальной оси на рис.~27.2 откладывать не $\ae$, а отношение
$a^2/b^2$,  где $b$  ---  полуширина  пучка  в  линии  с  бесконечными
корректорами  при  соответствующем  значении  $\nu$  (при  $\nu=1$ эти
величины совпадают).  Очень близкие результаты получаются и  для  ряда
неквадратичных корректоров.

     Из приведённых  данных следует общий вывод,  что при $\delta\gg
2\pi$ потери незначительны,  а при $\delta \ll  2\pi$,  когда  линзы
расположены друг от друга в дальней зоне,  потери очень велики, а само
понятие собственной волны утрачивает физический смысл.  Нетрудно  дать
геометрооптическое    обоснование   величине   пограничного   значения
параметра $\delta$:  при $\delta=2\pi$ ширина отверстия в  диафрагме
позволяет  попасть на линзу всей первой зоне Френеля в волновом фронте
падающей волны.

     Естественно считать,    что    собственные    функции   линии   с
ограниченными  корректорами  отличны  от   собственных   функций   при
неограниченных  корректорах из-за дифракции волнового пучка на кромках
отверстий  поглощающих   диафрагм.   Эта   дифракция,   приводящая   к
дополнительному размыванию волнового пучка, может оказаться полезной в
линиях,  в которых неограниченный корректор не  формирует  собственных
волн.  В  этом  смысле особый интерес представляет так называемая {\it
диафрагменная линия}, состоящая из параллельных, расположенных на
одном и том же расстоянии друг от друга экранов с одинаковыми отверстиями,
края которых лежат на общей цилиндрической поверхности. В такой линии
полностью отсутствует фазовая коррекция, то есть   $\psi=0$.   В   двумерной
линии  с  отверстием  шириной  $2a$ соответствующее интегральное уравнение
имеет вид
     $$ \int\limits_{-\sqrt \delta}^{+\sqrt\delta} v(\xi')e^{i\frac {(\xi-\xi'
         )^2} 2}\,d\xi'=\sqrt{2\pi}\lambda\,v(\xi)\,.\eqno(27.64)$$

     Это уравнение не может  быть  решено  в  замкнутом  аналитическом
виде;  оно  ещё  встретится нам в следующем разделе при рассмотрении
открытых резонаторов.  Здесь отметим только,  что найденные численными
методами  относительные  потери  на  периоде  диафрагменной линии (они
определяются $|\lambda|$ и  показаны  на  рис.~27.2  кривой  $\psi=0$)
существенно   больше,  чем  в  конфокальной  линии  с  соответствующим
значением параметра $\delta$.  Но для нас важен сам  факт  возможности
формирования    периодического    волнового    пучка,   обусловленного
неравномерностью  распределения  поля  по  площади   отверстия   из-за
дифракционного  размывания  пучка  кромками  диафрагм.  Для прояснения
физической  картины  явления  целесообразно   рассмотреть   в   рамках
параболического  уравнения  дифракцию  плоской волны на полуплоскости,
что и будет сделано  в  следующем  разделе.  Там  же  будут  приведены
приближённые   аналитические   оценки  для  собственных  значений  и
собственных функций уравнения (27.64).

     Все рассмотренные    свойства    двумерных   линзовых   линий   в
значительной степени  присущи  и  реальным  трёхмерным  системам,  в
которых   функции   $w$  и  $\psi$  зависят  от  $x$  и  $y$.  Решение
параболического   уравнения   (27.9)   в   прямоугольных   координатах
представляются в виде
     $$ w_{mn}(\xi,\eta,\zeta)=w_m(\xi,\zeta)\,w_n(\eta,\zeta)\,,
         \eqno(27.65)$$
где функции  $w_m$,$w_n$  определены  формулой  (27.10).   Аналогичным
образом представляется в виде произведения функция Грина трёхмерного
уравнения:
     $$G(\xi-\xi_0,\eta-\eta_0,\zeta-\zeta_0)=G(\xi-\xi_0,\zeta-
         \zeta_0)\,G(\eta-\eta_0,\zeta-\zeta_0)\,.\eqno(27.66)$$
Для квадратичных  корректоров   вида   (27.53)   собственные   функции
интегрального   уравнения  (27.57)  являются  произведениями  $v(x,y)=
X(x)Y(y)$,  где  каждая  из  функций  $X(x)$  и  $Y(y)$  удовлетворяет
двумерному  интегральному уравнению (27.62) и в случае неограниченного
корректора  совпадают.  Если  же   линза   ограничена   диафрагмой   с
прямоугольным  отверстием  со сторонами $a$ и $d$ вдоль осей $x$ и $y$
соответственно,  то собственные значения и функции двумерных уравнений
разные  (уравнения различаются пределами интегрирования).  Собственное
значение трёхмерной  линии  $\lambda_{mn}=\lambda_m(a)\lambda_n(d)$;
при  малом  уровне  потерь  они практически равны сумме потерь в обеих
двумерных линиях.

     Прямоугольные корректоры находят основное применение в зеркальных
линиях,  которые  представляют  собой периодическую последовательность
зеркал,  расположенных таким  образом,  что  после  каждого  отражения
центральный  луч  волнового  пучка попадает в центр следующего зеркала
(рис.~27.3).  Система  из  плоских  зеркал  по  своим  характеристикам
эквивалентна  диафрагменной линии,  из вогнутых сферических зеркал ---
линзовой линии.  Согласно законам геометрической  оптики  действие  на
световой  пучок  линзы  и  сферического  зеркала  с  равными фокусными
расстояниями  одинаково  (если  отвлечься  от  изменения   направления
распространения   пучка   при   отражении  от  зеркала).

\begin{wrapfigure}[12]{l}{7.5cm}
\begin{picture}(80,45)
\put(-3,40){\special{em:graph fig27-3.bmp}}
\end{picture}
\hbox to 7.5cm{\hfil\footnotesize{Рис.~27.3.~Зеркальная линия.
}\hfil}
\end{wrapfigure}

     В двумерной зеркальной линии в том же диапазоне (27.30) параметра
$\nu$,  представляющего  собой  отношение  периода  структуры  $2D$  к
удвоенному  фокусному  расстоянию зеркал $F$ (напомним,  что $F$ равно
половине радиуса  кривизны  поверхности  зеркала),  формируются  пучки
Гаусса-Эрмита, все характеристики которых определяются по существу тем
же интегральным  уравнением,  что  и  для  линзовой  линии.  Некоторые
различия возникают лишь в трёхмерных структурах из-за того,  что оси
$x$ и $y$ неравноправны по  отношению  к  направлению  распространения
основного   луча.   В   результате   для  создания  зеркальной  линии,
конфокальной в обеих  плоскостях,  зеркала  должны  иметь  поверхность
эллипсоида,  что  приводит  к  серьёзным  технологическим проблемам.
Зеркальные линии  подробно рассмотрены в следующем разделе в том
варианте,  когда они состоят всего из двух  расположенных друг против
друга зеркал и образуют  открытый резонатор.

     Для линзовой  линии  наибольший  интерес представляют корректоры,
ограниченные   окружностью.    При    осевой    симметрии    структуры
параболическое  уравнение  естественно решать в цилиндрической системе
координат,  в которой переменные $r$ и $\varphi$ также разделяются.  В
случае  неограниченного  квадратичного корректора с функцией $\psi(r)=
-\nu kr^2/2D$ основная  азимутально  симметричная  волна  определяется
функцией
     $$w_{00}=\exp{\Bigl(-\frac{\sqrt{2\nu-\nu^2}}2\,\frac k{2D}
         \Bigr)}=\exp{\Bigl(-\frac{r^2}{b^2}\Bigr)}\,,\eqno(27.67)$$
так что радиус пучка $b$ совпадает  с  полушириной  двумерного  пучка.
Наиболее узкий пучок образуется в конфокальной линии ($\nu=1$): радиус
пучка в плоскости линзы $b=\sqrt{2D/k}$,  в  средней  плоскости  между
линзами он в $\sqrt{2}$ раз меньше.

     В заключение отметим,  что наименьшие потери имеют  место  в  тех
открытых   линиях,   где  собственные  функции  волнового  пучка  мало
отличаются  от  функций  линии   с   неограниченным   корректором   и,
следовательно,  поле  пучка  имеет  каустики.  Разрежение  же  спектра
собственных  волн  достигается  за  счёт   высокой   неравномерности
радиационных потерь волн различных номеров.

%\end{document}

\newpage
\oddsidemargin=-0.4mm \evensidemargin=-0.4mm
\topmargin=-0.4mm
\headsep=7mm
\textheight=231.875mm
\textwidth=160mm
\mathsurround=2.5pt
\unitlength=1mm
%\begin{document}
%\input{macr.tex}
\thispagestyle{empty}
%\addtocounter{page}{329}
\baselineskip=.99\normalbaselineskip

\begin{center}\subsubsection*{28. Открытые резонаторы}\end{center}
\vspace*{.5cm}

\markboth{Глава 10.    Квазиоптические     структуры}{28.     Открытые
         резонаторы}

\begin{center}\begin{minipage}[c]{0.75\textwidth}
\footnotesize{\parindent=0.5cm
         Необходимость перехода  от замкнутых объёмных резонаторов к
         открытым.  Открытый  резонатор  из  плоских  зеркал   и   его
         интегральное   уравнение.   Вычисление   собственных  частот;
         приближённый расчёт на  основе  анализа  полубесконечного
         плоского   волновода   методом   параболического   уравнения.
         Открытые резонаторы из  фокусирующих  зеркал.  Роль  каустик.
         Геометрооптическое     описание     открытых     резонаторов.
         Цилиндрические и бочкообразные открытые резонаторы.  Открытые
         резонаторы на основе полного  отражения.
}\end{minipage}\end{center}
\vspace*{.5cm}

     Необходимость перехода   на   высоких   частотах   от   замкнутых
объёмных  резонаторов  к  открытым  резонансным системам обусловлена
теми же причинами, что и переход от обычных металлических волноводов к
открытым  волноводам.  С  уменьшением  длины  волны работа на основной
частоте становится неэффективной из-за уменьшения размера  резонатора,
что    приводит    к   снижению   добротности   колебания   $(Q\approx
\omega^{-1/2})$ и запасённой в резонаторе  энергии  поля;  при  этом
само  изготовление  миниатюрных резонаторов представляет собой сложную
технологическую  проблему.  Переход  на  высшие  типы  колебаний   при
сохранении  привычного размера резонатора (когда $Q\sim\sqrt{\omega}$)
также неэффективен из-за сгущения спектра.  По теореме  Куранта  число
колебаний $\Delta N$ в интервале частот $\Delta \omega$ в трёхмерной
области   с   объёмом   $V$   при   учёте   векторного   характера
электромагнитного поля составляет
     $$ \Delta N= \frac V{2\pi^2c^2} \omega^2\Delta \omega\,.
         \eqno(28.1)$$
Так как  коэффициент  затухания  из-за  джоулевых  потерь  в   стенках
пропорционален  $\omega^{1/2}$,  то  резонансные  кривые  высших типов
начинают перекрываться и резонансные свойства исчезают.

     Эффективность открытых     резонаторов     на    первый    взгляд
представляется  парадоксальной,  поскольку  в  этом  случае  неизбежно
возникают дополнительные радиационные потери. Однако именно эти потери
объясняют основное свойство открытых резонаторов  ---  разреженность
спектра собственных частот. Уровень радиационных потерь сильно зависит
от вида колебания и для большинства колебаний настолько велик,  что их
добротность  резко  снижается  и  они практически выпадают  из  спектра.  Для
некоторых же колебаний дополнительные радиационные потери  сравнимы  с
уменьшением  джоулевых  потерь  в  металлических элементах резонатора,
которое обусловлено существенным сокращением их поверхности при том же
объёме,  занятом полем. В результате добротность отдельных колебаний
открытого резонатора может превышать  добротность  основных  колебаний
замкнутого.  Путём дополнительных ухищрений спектр колебаний удаётся
сделать почти эквидистантным, что согласно теореме Куранта свойственно
только одномерным замкнутым структурам.

     Сама возможность удержания поля в открытой структуре  может  быть
объяснена  следующими  физическими  явлениями:  1) {\it дифракционными
эффектами},  приводящими к резкому снижению плотности  тока  на  краях
параллельных плоских зеркал, из-за чего происходит, в частности, почти
полное  отражение  собственной  волны,  падающей  на  открытый   конец
плоского волновода на частотах,  лишь немного превышающих критические;
2) {\it  образованием  каустических   поверхностей},   разграничивающих
области пространства, где волновое поле носит колебательный характер и
где   оно   экспоненциально   спадает   по   амплитуде (с    позиций
электродинамики это явление также носит дифракционный характер, но оно
может  быть  достаточно  просто  объяснено   и   рассчитано   методами
геометрической оптики, что существенно упрощает оценку возможности
использования ряда структур в качестве открытых резонансных систем); 3)
{\it полным внутренним отражением} волны от поверхности оптически менее
плотной окружающей среды.  Во многих используемых на практике открытых
резонаторах одновременно имеют место два,  а то и все три явления,  но
почти всегда какое-то одно из них основное  и его выделение позволяет
существенно   упростить  расчёт  прибора.  Ниже  с  разной  степенью
подробности рассматриваются все три физические причины удержания  поля
в открытой структуре.

\begin{wrapfigure}[14]{l}{7.5cm}
\begin{picture}(80,45)
\put(-1,45){\special{em:graph   fig28-1.bmp}}
\end{picture}
\hbox  to  7.5cm{\hfil\footnotesize
{Рис.~28.1.~Двумерный открытый резонатор}\hfil}
\hbox  to  7.5cm{\footnotesize{\hfilиз
параллельных плоских зеркал.}\hfil}
\end{wrapfigure}

     Начнём рассмотрение  с  открытого  резонатора,  представляющего
собой два параллельных плоских зеркала,  расположенных одно под другим
симметрично  относительно  плоскости  $z=0$   (см.   рис.~28.1).   Для
упрощения  последующих  записей и более чёткого выявления физической
картины  собственных  колебаний  остановимся  подробно  на   двумерном
варианте  резонатора,  когда  зеркала  представляют  собой бесконечные
ленты,  однородные вдоль оси $y$,  а  все  поля  не  зависят  от  этой
координаты.  Исходя  из представлений геометрической оптики можно было
бы ожидать,  что собственные колебания в  такой  структуре  образуются
встречными пучками нормальных к поверхности зеркала лучей, размываемых
вблизи краев из-за дифракции,  что приводит к радиационному затуханию;
однако  оказывается,  что  это  не  так.  Добротные  колебания в таком
резонаторе реализуются благодаря отражению от краев,  подобному  тому,
которое  имеет  место  для  собственной  волны  в плоском волноводе на
частоте,  близкой к критической. Волноводная волна в этом случае может
быть  представлена  совокупностью двух плоских волн,  волновые вектора
которых составляют {\it  небольшой}  угол  с  нормалью  к  поверхности
зеркала.

     Будем искать собственные колебания резонатора в предположении
     $$   kD\gg 1\,,\eqno(28.2)$$
то есть расстояние между  зеркалами  много  больше  длины  волны.  Для
определённости рассмотрим волны,  описываемые электрическим вектором
Герца  с  единственной  компонентой  $\Pi^e_y(x,z)=\Pi(x,z)$,  которая
удовлетворяет двумерному волновому уравнению
     $$\frac{\partial^2\Pi}{\partial x^2}+\frac{\partial^2\Pi}
         {\partial z^2}+k^2\Pi=0\,.\eqno(28.3)$$
Поскольку в этом  случае  единственная  отличная  от  нуля  компонента
электрического поля $E_y=k^2\Pi$,  то на зеркалах $z=\pm D,  \,-a<x<a$
должны выполняться нулевые граничные условия.

     Введём безразмерные величины
     $$\xi=\sqrt{\frac k{2D}}\,x\,, \qquad \zeta=\frac z{2D}\,,
         \qquad \delta=\sqrt{\frac k{2D}}\,a\,.\eqno(28.4)$$
Условие (28.2) позволяет искать решение  в  виде  произведения  быстро
осциллирующей  функции $e^{i2kD\zeta}$ на медленно меняющуюся функцию.
Учитывая,  что из-за  симметрии  структуры  решение  может  быть  либо
чётной, либо нечётной функцией $\zeta$, будем искать его в виде
     $$\Pi(\xi,\zeta)=F(\xi,\zeta)e^{i2kD\zeta}-(-1)^nF(\xi,-\zeta)
         e^{-i2kD\zeta}\,,\eqno(28.5)$$
где $n$ --- произвольное целое число.  Как было показано в  предыдущем
разделе,   из-за  условия  (28.2)  функция  $F(\xi,\zeta)$  с  хорошей
точностью может быть найдена как решение параболического уравнения
     $$\frac{\partial^2 F}{\partial \xi^2}+2i\frac{\partial F}
         {\partial \zeta}=0\,;\eqno(28.6)$$
при этом нулевое граничное условие на зеркалах  записывается в виде
     $$F(\xi,-1/2)=F(\xi,1/2)\,e^{i(2kD-\pi n)}\,.\eqno(28.7)$$
Плотность поверхностного  тока  на  нижнем  зеркале,  вычисляемая  как
обычно,   равна   $I_y(\xi)  =ck^2/2\pi\cdot  F(\xi,-1/2)\,e^{-i2kD}$.
Поэтому   функцию   $F(\xi,\zeta)$,   которая   является   комплексной
амплитудой волны,  распространяющейся от нижнего зеркала к верхнему, в
области $\zeta>-1/2$ можно искать как  решение  граничной  задачи  для
параболического  уравнения  с  граничным  условием $F(\xi,-1/2)=0$ при
$\xi<-\delta$ и $\xi>\delta$.

     С помощью функции Грина (27.34) находим:
     $$F(\xi,\zeta)=\sqrt{\frac 1{2\pi i (\zeta+1/2)}}
         \int\limits_{-\delta}^\delta F(\xi',-1/2)\,\exp
         {\Bigl\{i\frac{(\xi-\xi')^2}
         {(\zeta+1/2}}\Bigr\}\,d\xi'.\eqno(28.8)$$
Вычисляя теперь $F(\xi,1/2)$ и  используя  граничное  условие  (28.7),
приходим к интегральному уравнению
     $$\int\limits_{-\delta}^\delta f(\xi')\,\exp{\Bigl\{i\frac
         {(\xi-\xi')^2}2\Bigr\}}d\xi'=\sqrt{2\pi}\lambda f(\xi)\,,
         \eqno(28.9)$$
где
     $$\lambda=\exp{\{-i[2kD-\pi(n+1/4)]\}}\,\eqno(28.10)$$
и введено   обозначение    $f(\xi)=F(\xi,-1/2)$.    Уравнение    имеет
бесчисленное    множество    собственных    значений   $\lambda_m$   и
соответствующих  им  собственных   функций   $f_m(\xi)$.   Собственные
значения,   являющиеся   функцией  единственного  параметра  уравнения
(28.9), представим в виде
     $$ \lambda_m(\delta)=e^{-i2\pi p_m(\delta)}\,,\eqno(28.11)$$
где $p_m$ --- комплексная функция: $p_m=p_m'+ip_m''$.

     Действительная часть  $p_m$  определяет  поправку  к  собственной
частоте резонатора:
     $$ k_{mn}=\frac \pi{2D}(n+\frac 1 4 +2p'_m)\,.\eqno(28.12)$$
Поскольку $p_m$,  как это видно из (28.4), само является функцией $k$,
то формула (28.12) не даёт явного выражения для собственных частот,  а
представляет собой лишь  уравнение  для  их  определения.  Однако  для
практических   расчётов,   которые  имеют  смысл  лишь  при  больших
значениях $\delta$, когда величина $p_m$ мала, для вычисления $\delta$
достаточно  взять  нулевое приближение для $k$,  получаемое из (28.12)
отбрасыванием слагаемого с $p'_m$.  Мнимая  часть  $p''_m$  определяет
затухание  собственного  колебания.  Поскольку $|p_m|$ мал,  а интерес
представляют  большие  значения  $n$,  то  добротность   колебаний   с
достаточной точностью равна
     $$ Q=\frac{k'_{mn}}{2k''_{mn}}\approx\frac n{4 p''_m}\,.
         \eqno(28.13)$$

     Обратим внимание,  что уравнение (28.9)  совпадает  с  уравнением
(27.65),   определяющим   распределение  поля  в  отверстии  диафрагмы
двумерной диафрагменной линии.  Его решение может быть получено только
численными  методами.  Результаты такого решения частично представлены
на  рис.~27.2  кривой  $\psi=0$,  поскольку  $1-|\lambda|^2\approx4\pi
p''_1$.    Ввиду    сказанного    несомненный   интерес   представляет
приближённая аналитическая оценка собственных значений и собственных
функций  уравнения (28.9).  Такая оценка будет получена ниже на основе
решения задачи об  отражении  собственной  волны  от  открытого  конца
плоского  волновода  методом  параболического  уравнения.  Однако  для
лучшего понимания физической  картины  явления  целесообразно  сначала
решить  таким  способом более простую задачу о дифракции плоской волны
на полуплоскости.

\begin{wrapfigure}[12]{l}{7cm}
\begin{picture}(80,40)
\put(2,40){\special{em:graph fig28-2.bmp}}
\end{picture}
\hbox to 7.0cm{\hfil\footnotesize{Рис.~28.2.~Дифракция на
полуплоскости.}\hfil}
\end{wrapfigure}

     Пусть идеально      отражающая      полуплоскость      определена
полубесконечным   отрезком,   выходящим   из   начала   координат    и
наклонённым  под некоторым углом $\varphi$ к оси $x$ (рис.~28.2),  а
падающая  волна  описывается  функцией  $e^{ikz}$.  Задавая  граничные
значения   в  плоскости  $z=0$  в  виде  $F(\xi,0)=1$  при  $\xi<0$  и
$F(\xi,0)=0$ при $\xi>0$,  находим решение в верхней полуплоскости  по
формуле (27.33) с помощью функции Грина (27.34) в виде
     $$ F(\xi,\zeta)=\Phi(\tau)\,,\qquad \tau=\frac{\xi}
         {\sqrt{\zeta}}\,,\eqno(28.14)$$
где $\Phi(\tau)$ --- интеграл Френеля:
     $$\Phi(\tau)=\frac 1{\sqrt{2\pi i}}\int\limits_{-\infty}^\tau
         e^{i\frac{u^2} 2}\,du\,.\eqno(28.15)$$

     Поскольку $F$ зависит только от $\tau$,  то  линии  $F=const$  на
плоскости $x,z$ совпадают с линиями $z=kx^2/C$, где $C$ --- константа,
и,   следовательно,   являются   параболами.   Выбирая   $C$    равным
положительному   числу   порядка  нескольких  единиц,  такому,  что  с
достаточной точностью можно считать
     $$\Phi(\tau)\approx 1,\quad\mbox{и}\quad \Phi(-\tau)\approx 0
         \qquad \mbox{при}  \quad \tau>C\,,\eqno(28.16)$$
и строя соответствующую параболу (её внутренняя область на рис.~28.2
заштрихована),  можно тем самым разделить верхнюю полуплоскость на три
области.   Слева   от   параболы   поле  практически  не  возмущено  и
представляет  собой  падающую  плоскую  волну.  Справа   от   параболы
расположена  зона тени,  где поле почти полностью отсутствует.  Внутри
параболы находится переходная область --- область  полутени,  прич\"ем
на  оси  $z$  амплитуда  равна  половинному значению в падающей волне.
Аналогичным образом делится на три  области  и  нижняя  полуплоскость,
только  теперь  направление центральной линии параболы зависит от угла
$\varphi$  и  совпадает  с  крайним  лучом,   отразившимся   от   края
полуплоскости.  Очевидно, что в рассматриваемом приближении (напомним,
что  оно  справедливо  при  выполнении  условий  $kz\ll  1,\,   |x|\ll
z,\,kx^3/z^2  \ll  1$) дифрагированное поле в верхней полуплоскости не
зависит от формы всего экрана,  а определяется только  положением  его
края  и  направлением  падающей  волны.  То  же  самое можно сказать о
невлиянии формы задней стенки зеркала на отражённое поле.

\begin{wrapfigure}[12]{l}{7.0cm}
\begin{picture}(80,45)
\put(-1,45){\special{em:graph fig28-3.bmp}}
\end{picture}
\hbox to 7.0cm{\hfil\footnotesize{Рис.~28.3.~Плоский волновод.
}\hfil}
\end{wrapfigure}

     Рассмотрим теперь  отражение  собственной  волны полубесконечного
плоского  волновода  (рис.~28.3)  от  его  открытого  конца  $x=0$  на
частоте,  лишь немного превышающей критическую (обратим внимание,  что
для удобства перехода к резонатору  привязка  к  системе  координат  и
обозначения  отличаются  от  принятых в разделе~10).  Исследуем волны,
определяемые тем же однокомпонентным вектором Герца $\Pi(x,z)$,  как в
резонаторе из плоских зеркал,  будем искать его в том же виде (28.5) и
воспользуемся теми же безразмерными величинами (28.4).  Только  теперь
граничное  значение функции $f(\xi)=F(\xi,-1/2)=0$ имеет место на всей
полуплоскости $\xi<0$,  а граничное  условие  для  $\Pi(\xi,\pm  1/2)$
по-прежнему  определяется  формулой (28.7) на полуплоскости $\xi>0$.  В
результате для функции $f(\xi)$ получаем интегральное уравнение
     $$ \int\limits_0^\infty e^{i\frac{(\xi-\xi')^2} 2} \,f(\xi')\,
         d\xi'= \sqrt{2\pi i}\,{e^{-i2\pi p}} f(\xi)\,,\eqno(28.17)$$
где
     $$ 2\pi p= 2kD - \pi n\,.\eqno(28.18)$$
Поскольку для  наших целей интерес представляют частоты,  лишь немного
превышающие  критические  для  бесконечного  волновода,  то  $p$   ---
величина малая и, во всяком случае, $p<1/2$.

     Если таким  же  способом  искать  собственные  волны бесконечного
плоского волновода,  то в интегральном уравнении нижний предел следует
положить равным $-\infty$.  В этом случае интегральное уравнение имеет
очевидное аналитическое решение
     $$f(\xi)=A\,e^{\pm i h_s\xi}\,,\eqno(28.19)$$
где $h_s=2\sqrt{\pi(s+p)},\;s=0,\pm 1,\,\pm 2,\dots$. Нетрудно видеть,
что $h_s$ приближённо совпадают с волновыми числами собственных волн
бесконечного плоского волновода (распространяющихся или затухающих).

     Поэтому собственные   функции    уравнения    (28.17),    которые
естественно  искать  в  виде  совокупности  падающей и отражённых от
открытого  конца  волноводных  волн  бесконечного   волновода,   можно
представить в виде
     $$f(\xi)=A(e^{-ih_0 \xi}+ \sum\limits_{s=0}^\infty
         R_{0s}e^{ih_s \xi})\,,\eqno(28.20)$$
где слагаемое $ e^{-ih_0\xi}$ --- падающая на открытый конец волна,  а
сумма --- отражённые волны с коэффициентами $R_{0s}$. Подставляя это
выражение для $f(\xi)$  в  интегральное  уравнение  (28.17),  получаем
функциональное уравнение для неизвестных $R_{0s}$:
     $$ e^{-ih_0\xi}\Phi(-\xi-h_0)+\sum\limits_{s=0}^\infty
         R_{0s}e^{ih_s\xi}\Phi(-\xi+h_s)=0\,,\qquad \xi\geqslant 0
         \,,\eqno(28.21)$$
где $\Phi(x)$ --- интеграл Френеля (28.15).

     Наиболее последовательный путь нахождения коэффициентов  $R_{0s}$
состоит  в переходе от функционального уравнения к бесконечной системе
линейных алгебраических уравнений путём умножения уравнения (28.21)
на  $e^{-ih\xi}$ и интегрирования по $\xi$ от 0 до $\infty$ (интегралы
выражаются  в  замкнутом  виде  опять  через  интегралы   Френеля)   и
численного решения этой системы методом редукции. Однако грубую оценку
для интересующего нас в первую  очередь  коэффициента  $R_{00}$  можно
получить,  если  отбросить  в  уравнении  (28.21) все $R_{0s}$,  кроме
$R_{00}$, и положить $\xi=0$. В результате имеем
     $$ R_{00}=-\frac{\Phi(-h_0)}{\Phi(h_0)}\,,\eqno(28.22)$$
что при разложении в ряд по малым $h_0$ сводится к  выражению
     $$ R_{00}\approx -[1+(i-1)\beta h_0+\dots]\,\eqno(28.23)$$
с параметром    $\beta=1,13$.    Строгое    решение   задачи   методом
Винера-Хопфа-Фока даёт для коэффициента  $R_{00}$  при  малых  $h_0$
(малые $p$) следующую формулу:
     $$R_{00}=-e^{i\beta(1+i)h_0}\,,\eqno(28.24)$$
первые члены  разложения которой по малому параметру $h_0$ совпадает с
(28.23), но уточнённое значение $\beta=0,824$.

     Зная коэффициент отражения  $R_{00}$,  нетрудно  получить  оценку
собственных  частот  резонатора с плоскими ленточными зеркалами,  но в
аналитическом  виде  это  удаётся   сделать   лишь   при   не учёте
трансформации  падающей  волны  в  волны другого типа при отражении от
краёв  резонатора  $x=\pm  a$  и  в  предположении,  что   отражение
происходит  так  же,  как  в  полубесконечном волноводе,  то есть края
резонатора не влияют друг на друга.

     Поскольку структура симметрична относительно плоскости $x=0$,  то
собственные   функции   являются   либо  чётными,  либо  нечётными
функциями $\xi$ и их следует искать в виде
     $$f(\xi)=e^{ih_0\xi}-(-1)^m\, e^{-ih_0\xi}\,,\eqno(28.25)$$
где $m$ -- целое число.  Для сопоставления с формулой (28.20), которая
определяет  коэффициент  отражения  $R_{00}$  на  левом  краю  зеркал,
перепишем (28.25) в виде
     $$ f(\xi)=\frac 1 2 e^{ih_0\delta}[e^{-ih_0(\xi+\delta)}
         -(-1)^m\, e^{-2ih_0\delta}\,e^{ih_0(\xi+\delta)}]\,.
         \eqno(28.26)$$
Сравнивая (28.26)  с  формулой  (28.20),  в  которой  в сумме оставлен
только член с $s=0$, получаем характеристическое уравнение
     $$R_{00}=-(-1)^m\,e^{-i2h_0\delta}\,,\eqno(28.27)$$
откуда с учётом (28.24) находим
     $$h_0=\frac {\pi m}{2\delta+(1+i)\beta},\qquad m=1,2,\dots
         ,\eqno(28.28)$$
и, следовательно,
     $$ p_m=\frac{\pi m^2}{4[2\delta+(1+i)\beta]^2}\,.\eqno(28.29)$$

     При больших  значениях  $\delta$  величина  $p'_m$,  определяющая
превышение собственной частоты над критической,  и  величина  $p''_m$,
определяющая добротность колебания, соответственно равны
     $$ p'_m\approx \frac {\pi m^2}{16\delta^2}\,,\qquad p''_m=
          -\frac{\pi\beta m^2}{16\delta^3}\,.\eqno(28.30)$$
Высокую добротность  собственных  колебаний  при   больших   значениях
параметра  $\delta$ легко понять,  если проанализировать распределение
поверхностной плотности  тока  $I_y$,  которая  определяется  функцией
$f_m(x)$.  При  больших  $\delta$  на краях пластин $f_m(\pm a)\sim 1/
\delta$ --- в результате  все  компоненты  поля  малы  и  радиационные
потери минимальны.

     Надо сказать,  что  формула  (28.28)  даёт  решение   уравнения
(28.27)  и  при  $m=0$.  Это  решение  представляет собой незатухающее
колебание на критической  частоте  плоского  волновода.  Однако  такое
решение  отсутствует у интегрального уравнения (28.9) и,  естественно,
не реализуется на опыте.  Отметим также,  что  интегральное  уравнение
(28.17),  определяющее  собственные  колебания в резонаторе с плоскими
зеркалами,  совпадает с интегральным уравнением  (27.65),  описывающим
собственные волны в диафрагменной линии.  Это не случайно, а связано с
тем,  что в обоих  случаях  рассматривается  квазиоптический  пучок  в
системе без фазовой коррекции.

     При переходе к трёхмерным структурам степень усложнения  задачи
существенно  зависит от формы плоских зеркал (при этом предполагается,
что зеркала одинаковы,  параллельны  и  расположены  строго  одно  под
другим).  Если  зеркала  прямоугольные  со  сторонами $2a$ и $2b$,  то
структура пучка в обеих плоскостях различается только параметром  $p$,
собственные   значения  трёхмерного  интегрального  уравнения  равны
произведению  собственных  значений  двух  двумерных, в  результате
собственные частоты определяются формулой, схожей с (28.12):
     $$  k_{mln}=\frac {\pi} {2D}\Bigl(n +\frac 1 2 +p_{l,a}+p_{m,b}
         \Bigr)\,,\eqno(28.31)$$
где
     $$ p_{l,a}=\frac{\pi l^2}{4[2\delta_a+(1+i)\beta]^2}\,,\qquad
         p_{m,b}=\frac{\pi m^2}{4[2\delta_b+(1+i)\beta]^2},
         \qquad l,m=1,2,\dots, \eqno(28.32)$$
и
     $$ \delta_a=\sqrt{\frac{k_{lmn}}{2D}}\,a\,,\qquad
         \delta_b=\sqrt{\frac{k_{lmn}}{2D}}\,b\,.\eqno(28.33)$$
Отметим, что  формула  (28.31),  как  и  (28.12),  явно  не определяет
частоту собственных колебаний,  а является уравнением, из которого эта
частота может быть найдена.

     Для резонатора с круглыми зеркалами радиуса $a$
     $$ k_{mln}=\frac{\pi n}{2D}+\frac{4D\nu^2_{ml}\delta^2}
         {\pi n a^2 [2\delta+(1+i)\beta]^2}\,,\eqno(28.34)$$
где для  колебаний  $E_{mln}^{(z)}$  под  $\nu_{ml}$  следует понимать
корни уравнения $J_m(x)=0$,  а для колебаний $H^{(z)}_{mln}$ --- корни
уравнения $J'_m(x)=0$.

\begin{wrapfigure}[14]{l}{7.5cm}
\begin{picture}(80,45)
\put(2,45){\special{em:graph fig28-4.bmp}}
\end{picture}
\hbox to 7.5cm{\hfil\footnotesize{Рис.~28.4.~Двумерный открытый резонатор из
}\hfil}
\hbox to 7.5cm{\footnotesize{цилиндрических
зеркал постоянного радиуса.
}\hfil}
\end{wrapfigure}

     Существенного уменьшения    радиационных    потерь   в   открытых
резонаторах,  так же,  как и в открытых волноводах, удаётся добиться
за  счёт  фазовой  коррекции  пучка  на  зеркалах,  что достигается,
например,  применением  вогнутых  зеркал,  обеспечивающих  фокусировку
параксиального  пучка.  Двумерный открытый резонатор с цилиндрическими
зеркалами  одинакового  радиуса  кривизны   $R_{з}$   изображён   на
рис.~28.4.   В   геометрооптическом  приближении  фокусное  расстояние
зеркала   $F=R_{з}/2$.   Волновые   свойства    структуры    полностью
определяются    безразмерным    параметром    $\nu=D/F$.    Резонатор,
соответствующий параметру $\nu=1$,  называется конфокальным, поскольку
фокусы  обоих  зеркал в этом случае лежат в одной плоскости.  Значение
$\nu=0$  соответствует  рассмотренному  выше  резонатору  с   плоскими
зеркалами, а резонатор с $\nu=2$ называется концентрическим, так как в
нём  совпадают  центры  кривизны  обоих  зеркал.  Забегая  несколько
вперёд,  скажем,  что  добротные колебания в резонаторах такого типа
возможны лишь в диапазоне  значений  параметра  $0<\nu<2$,  чему  есть
достаточно убедительные физические и математические объяснения.

     Собственные колебания в  открытом  резонаторе  при  использовании
фокусировки  пучка  имеют  много  общего  с  процессом распространения
волнового пучка в соответствующей открытой линии.  В линии пучок после
каждого  корректора  попадает  на  следующий,  а в открытом резонаторе
возвращается  обратно,  потом  опять  попадает  на  второй   корректор
(зеркало)  и так далее.  Установившийся колебательный процесс возможен
лишь на дискретных частотах,  для которых  выполняются  определённые
условия по набегу фаз в волновом пучке между корректорами. Собственные
колебания открытого резонатора состоят из  двух  встречных  пучков.  В
этом   смысле  прослеживается  аналогия  между  обычным  металлическим
волноводом и замкнутым волноводным резонатором.

     Поле в  резонаторе  будем  искать  в  виде  (28.5),  а в качестве
функции  $F(\xi,\zeta)$  выберем  функцию   $w_m$   (см.   соотношение
(27.10)),  определяющую  пучок  Гаусса-Эрмита.  Поле этих пучков резко
ограничено  своей  каустикой,  поэтому  следует  ожидать   значительно
большей концентрации поля вблизи оси, чем это имеет место в резонаторе
с  плоскими  зеркалами.  Параметр  пучка  $L$  определим  из   условия
равенства радиуса кривизны волновой поверхности пучка $R$, одинакового
для пучка любого номера  $m$  и  определяемого  формулой  (27.26)  при
$z=D$, и радиуса кривизны зеркала $R_{з}=2F$:
     $$L\Bigl(\frac L D+\frac D L\Bigr)=2F=\frac {2D} \nu\,.
         \eqno(28.35)$$
Отсюда находим
     $$L=D\sqrt\frac{2-\nu}{\nu}\,\eqno(28.36)$$
и тем  самым  полностью  определена  структура  собственных  колебаний
резонатора.

     Собственные частоты  определяются  из  условия,  что  набег  фазы
волнового  пучка  от  одного зеркала до другого $\Delta\varphi$ должен
быть равен  $n\pi$,  где  $n$  ---  целое  число.  Тогда  пучок  после
возвращения к  исходному  зеркалу оказывается в той же фазе.  Основной
набег фазы обусловлен множителем $e^{i k z}$ и  равен  $2kD$;  к  нему
следует  добавить  изменение  фазы  функции $w_m$.  Из формулы (27.10)
следует,   что   дополнительное   слагаемое   полностью   определяется
множителем   $A^{2m+1}$,   так   как  набег,  связанный  с  множителем
$e^{-\xi^2 A^2/2}$,  равен $\pi$ (он приводит к  требуемому  изменению
знака  радиуса  кривизны  волновой  поверхности).  Поскольку,  согласно
(27.12), $A=\sqrt{1/(1+i z/L)}$, то дополнительный, отличный от целого
числа $\pi$,  набег составляет $-(2m+1)\arctg{\sqrt{\nu/(2-\nu)}}$.  В
результате  для  собственных  частот  двумерного  резонатора  получаем
выражение
     $$ k_{nm}=\frac 1 {2D}\,\Bigl[\pi n+(2m+1)\arctg{\sqrt{
         \frac \nu{2-\nu}}}\;\Bigr]\,.\eqno(28.37)$$

     Отметим два существенных момента.  Во-первых,  частоты получились
действительными,  а  значит  колебания  ---  незатухающими;  последнее
связано с тем,  что  область  действия  зеркал  (фазовых  корректоров)
молчаливо  подразумевалась  неограниченной и не учитывалось неизбежное
<<расплескивание>>  поля пучка,  имеющее место даже в том случае, когда
размер  каустики значительно меньше апертуры зеркала.  Во-вторых,  для
конфокального резонатора  $(\nu=1)$,  привлекательного  тем,  что  его
каустики ближе к оси,  чем у любого другого,  {\it собственные частоты
вырождены},  поскольку  величина  $n+(2m+1)/4$  имеет  одно  и  то  же
значение при разных наборах $n$ и $m$.

     Учёт радиационных потерь в  резонаторах  с  фазовой  коррекцией
удобнее  проводить  в  рамках  интегрального уравнения,  которое,  как
нетрудно видеть,  тождественно  совпадает  с  интегральным  уравнением
линзовой   линии   (27.63).   Собственные   значения  этого  уравнения
определяют собственные частоты резонатора,  которые в этом случае  при
конечных   значениях   предела   интегрирования   $\delta$  получаются
комплексными, а  при  $\delta=\infty$  они,  разумеется,  совпадают  с
(28.37).  Поэтому  все  результаты для линзовой линии,  полученные,  в
частности, для радиационных потерь,  в равной  степени  относятся  и  к
резонатору.  Потери  становятся  очень  большими  для  тех  колебаний,
каустика которых не помещается на апертуре зеркала,  что  имеет  место
при условии
     $$a<\sqrt{\frac {2D}{k} (2m+1)}\,.\eqno(28.38)$$
Это приводит к существенному разрежению спектра собственных частот, из
которого выпадают все частоты с большими $m$.

     Однако это же обстоятельство вместе с отмеченным выше вырождением
собственных   частот   оказывается   губительным   для    конфокальных
резонаторов.  Из-за  совпадения  частот  основное добротное колебание,
соответствующее $m=0$,  легко  перерождается  в  колебание  с  большим
значением $m$, и поле высвечивается из резонатора. Поэтому конфокальные
резонаторы не используются на  практике  и  значение  параметра  $\nu$
выбирают  ближе  к  двум.  Хотя в этом случае основное колебание имеет
пониженную в сравнении с конфокальным резонатором добротность,  однако
высшие  по  $m$  колебания  затухают существенно быстрее и отсутствует
вырождение.

     Рассмотренный открытый   резонатор   с   фокусирующими  зеркалами
основными своими достоинствами обязан наличию  в  поле  его  колебаний
каустик.  Вопрос  о  существовании  у собственных волн квазиоптической
системы каустик во многих случаях значительно проще и нагляднее  может
быть  выяснен,  если основываться на методах геометрической оптики.  В
этой теории каустикой называется поверхность, разделяющая области, где
есть  световые  лучи  и  где  их  нет,  то есть каустика --- огибающая
системы  лучей.  В  двумерном  резонаторе  с  зеркалами  в  виде   дуг
окружности  установить  этим  методом аналитический вид каустики также
весьма затруднительно,  хотя последовательными построениями лучей  при
их  отражении  то  от  одного,  то  от  другого  зеркала  легко  можно
убедиться,  что получаемая таким образом совокупность лучей не выходит
из замкнутой области,  граница которой и есть каустика.  Для некоторых
форм зеркал вид  каустики  нетрудно  установить  путём  элементарных
построений   и  несложных  доказательств.  Примером  такой  оптической
системы служит внутренность зеркального цилиндра,  поперечное  сечение
которого есть эллипс (рис.~28.5 {\it а}).

     Система плоских лучей в такой  структуре  обладает  замечательным
свойством.  Если  взять  за исходный некоторый луч $AB$,  пересекающий
отрезок,  соединяющий между собой фокусы  $O_1$  и  $O_2$  эллипса,  и
построить софокусную данному эллипсу гиперболу, касающуюся этого луча,
то  оказывается,  что  отражённый  от  эллипса  луч  сразу  или  после
нескольких   последовательных   отражений   коснётся   второй  ветви
гиперболы и никакой из лучей не пересечёт  её  (на  доказательстве
этого положения здесь останавливаться нецелесообразно, заметим только,
что  оно  проводится   методами   элементарной   геометрии).   Поэтому
построенная гипербола и есть каустика системы лучей и её естественно
считать  внешней.  Другую  совокупность  лучей,   имеющую   внутреннюю
каустику  в  виде эллипса,  софокусного к исходному,  легко построить,
если выбрать в качестве исходного луч,  не пересекающий отрезок  между
фокусами  (рис.~28.5 {\it б}).  Очевидно,  что  при  деформации эллипса путем
сближения его фокусов в конечном итоге получается структура  (круговой
цилиндр),  в  которой  имеется  только внутренняя каустика,  а внешняя
каустика пропадает.  Это  хорошо  согласуется  с  тем  фактом,  что  у
рассмотренного  выше  двумерного  резонатора  при  значении  параметра
$\nu=2$, которому соответствуют концентрические зеркала, не существует
собственных волн в виде гауссового пучка с каустикой.

\begin{picture}(160,55)
\put(3,50){\special{em:graph  fig28-5a.bmp}}
\put(82,50){\special{em:graph fig28-5b.bmp}}
\end{picture}

\begin{center}\begin{minipage}[c]{0.9\textwidth}
\footnotesize{\parindent=0.5cm
Рис.~28.5.~Каустики и ход лучей в резонаторе эллиптического сечения:  {\it а})
внешняя каустика --- ветви гиперболы, {\it б}) внутренняя каустика --- эллипс.
}\end{minipage}\end{center}\vspace*{0.25cm}

     Методами геометрической  оптики  можно   с   достаточно   хорошей
точностью найти и спектр частот собственных колебаний.  Очевидно,  что
не  всякая  совокупность  лучей  с  соответствующей  каустикой   может
реализоваться в качестве собственного колебания,  а только такая,  для
которой выполнены определённые  фазовые  соотношения.  Разные  лучи,
приходящие в одну и ту же точку поверхности, должны быть синфазны. Так
как все лучи касаются каустик и,  следовательно,  вдоль  каустик  фаза
меняется  со  скоростью  света,  то  длина каждой замкнутой внутренней
каустики должна быть равна целому числу длин волн,  а длина  каустики,
опирающейся  на  зеркала,  целому  числу  полуволн.  Второе  квантовое
условие менее очевидно (таких условий в  двумерном  резонаторе  должно
быть   два   и   это   следует  из  того,  что  собственные  колебания
характеризуются двумя индексами).  Второй инвариантной  величиной  для
любых  пар  лучей,  последовательно  касающихся двух ветвей гиперболы,
является длина линии  $P_1S_1BS_2P_2$  на  рис.~28.5а.  Доказательство
этого  факта  также  проводится  чисто геометрическими методами и этот
инвариант также должен быть равен целому числу полуволн.

     Такой метод определения собственных частот сравнительно  прост  и
приводит к тем более точным результатам, чем ширина пучка больше длины
волны.  Отметим,  что в нём  снято  ограничение  на  параксиальность
волнового пучка. Его обоснование может быть получено и путём решения
волнового уравнения методом ВКБ,  при  этом, однако, оказывается,  что
длина  каустик и линии $P_1S_1BS_2P_2$ должны быть равны не целому,  а
полуцелому  числу  длин  волн.  Соотношение  между  геометрооптическим
решением и асимптотическим решением волнового уравнения примерно такое
же, как между квантованием уровней водородного атома по первоначальной
теории Бора и решением уравнения Шредингера.

\begin{wrapfigure}[14]{l}{7.5cm}
\begin{picture}(80,45)
\put(-2,45){\special{em:graph fig28-6.bmp}}
\end{picture}
\hbox to 7.5cm{\hfil\footnotesize{Рис.~28.6.~Двумерный открытый резонатор
}\hfil}
\hbox to 7.5cm{\hfil\footnotesize{из дуги эллипса и дуг гиперболы.
}\hfil}
\end{wrapfigure}

     Простые геометрические     построения     каустик    подсказывают
существование ряда открытых резонаторов,  в которых  возможно  наличие
добротных  собственных  колебаний.  К  таким,  несколько  экзотическим
устройствам,  можно отнести двумерный резонатор,  у  которого  зеркало
представляет  собой  дугу  эллипса с примыкающими отрезками софокусной
гиперболы (рис.~28.6). Эти отрезки целесообразно несколько продлить за
каустику,  поскольку  за  счет  дифракционных  эффектов  волновое поле
выплёскивается  за  неё.   Как   строгие   расчёты   на   основе
параболического   уравнения,  так  и  практическая  реализация  такого
устройства в трёхмерном варианте показывают, что радиационные потери
в  нём  очень малы.  Правда,  его реальная добротность не так велика
из-за большой площади зеркала на единицу объёма,  занимаемого полем,
и, следовательно, повышенного уровня джоулевых потерь.

     Среди реальных трёхмерных резонаторов  наиболее  распространены
конструкции,  обладающие  осевой симметрией.  Квазиоптическое волновое
поле в таких структурах  естественно  рассматривать  в  цилиндрической
системе  координат.  Важную  роль  среди  этих полей играют обладающие
каустиками волновые пучки  Гаусса-Лагерра,  поле  которых  описывается
потенциалом вида
     $$\Pi(r,\varphi,z)=e^{i k z}\,w_{ml}(r,\varphi)\,\eqno(28.39)$$
и характеризуется  двумя  индексами  $l$ и $m$,  представляющими собой
число  вариаций  поля  соответственно  по  $r$  и  $\varphi$.  Функция
$w_{ml}$  по  своей  структуре  во многом аналогична функции $w_m$ для
двумерных пучков Гаусса-Эрмита:
     $$w_{ml}(r,\varphi)=\Bigl(\frac r b\Bigr)^m\,e^{-\frac {r^2}
         {2b^2}}\,L^m_l\Bigl(\frac {r^2}{b^2}\Bigr)\,
         \cos{m\varphi}\,,\eqno(28.40)$$
где $b$ --- условный радиус пучка, эквивалентный полуширине двумерного
пучка,  $L^m_l(x)$ --- присоединённые полиномы Лагерра, заменяющие в
цилиндрических системах полиномы Эрмита.  Поле  основной  симметричной
моды  по своей структуре не отличается от двумерного гауссового пучка,
если в нём заменить $x$ на $r$.

     В резонаторе  с  круглыми  сферическими  зеркалами при правильном
соотношении  фокусного  расстояния  и   расстояния   между   зеркалами
($0<\nu<2$) все виды колебаний имеют внешнюю каустику,  представляющую
собой однополостный гиперболоид.  Радиус внешней каустики  на  зеркале
минимален  у  основного  колебания  в конфокальном резонаторе,  однако
из-за присущего конфокальным резонаторам вырождения собственных частот
они  не  находят  широкого  применения.  Значительно  больший  интерес
представляют колебания с большими значениями индекса  $m$,  у  которых
помимо внешней каустики появляется и внутренняя.

     Рассмотрим такие колебания на примере  открытого  цилиндрического
резонатора,  представляющего собой тонкий зеркальный обруч радиуса $a$
и высотой лент  $2D$.  Как  и  во  всякой  квазиоптической  структуре,
полагаем,  что $kD\gg 1$,  и будем также считать,  что $a\gg D$. Вдоль
резонатора устанавливается стоячая волна,  поле которой в  зависимости
от  типа  колебаний  описывается  электрическим или магнитным вектором
Герца с компонентой
     $$ \Pi_z^{e,m}(r,\varphi,z)=J_m(gr)e^{\pm i m \varphi}\;f(z)\,,
       \eqno(28.41)$$
где поперечное  волновое  число  $g=\nu_{ml}/a$,  $\nu_{ml}$ --- корни
функции $J_m(x)$ или $J'_m(x)$ в зависимости от граничных  условий  на
потенциал, определяемых типом колебаний.

     Функция $J_m(gr)$ удовлетворяет уравнению
     $$ \frac 1 r\,\frac {d\phantom{r}}{dr}\Bigl(r\frac {dJ_m}{dr}
        \Bigr)+\Bigl( g^2-\frac{m^2}{r^2}\Bigr)\,J_m=0\,,
        \eqno(28.42)$$
которое заменой переменной $gr=e^x$ сводится к уравнению
     $$ \frac {d^2J_m}{dx^2}+(g^2x^2-m^2)\,J_m=0\,.\eqno(28.43)$$
Это уравнение  имеет  экспоненциально затухающее (нарастающее) решение
при $x<m/g$ и осциллирующее решение при  $x>m/g$.  Двумерное  волновое
поле
     $$ \Pi_z^{e,m}(r,\varphi)= J_m(gr)\,e^{\pm im \varphi}
         \eqno(28.44)$$
можно представить в виде системы  лучей,  отражающихся  от  окружности
$r=a$ и касающихся окружности радиуса $r=m/g$,  которая и представляет
собой внутреннюю каустику.  Такое  представление  находится  в  полном
соответствии  с  рассмотренным  выше  в геометрооптическом приближении
зеркальном эллипсе,  у которого фокусы сливаются друг с  другом.  Поле
(28.44) при больших значениях $m$ $(m>ga\gg1)$ называется {\it волнами
шепчущей галлереи};  оно в основном локализовано в кольце $m/g<r<a$  и
при  удалении  от  окружности  радиуса  $m/g$ по направлению к оси $z$
экспоненциально убывает.

     В трёхмерном   зеркальном   обруче   волнам  шепчущей  галлереи
соответствуют колебания,  удерживаемые краями цилиндра.  Они возникают
на  частотах,  лишь  немного  превышающих  критические  для  волн типа
$E_{pq}$ или $H_{pq}$ обычного круглого  волновода  и  по  механизму
удержания  сходны  с  рассмотренными  выше  колебаниями между плоскими
зеркалами:  образование  стоячей  волны   обусловлено   почти   полным
отражением волноводной волны на открытом конце волновода.

     Значительно более  добротные колебания существуют в бочкообразном
резонаторе (рис.~28.7),  представляющем  собой  цилиндрический  обруч,
лента  которого  изогнута  внутрь  по сферической поверхности.  Поле в
таком резонаторе ограничено тремя каустическими  поверхностями:  двумя
внешними  и одной внутренней.  Внешние каустики локализуют поле вблизи
средней  плоскости  резонатора,  а   внутренняя   оттесняет   поле   к
поверхности  обруча.  Бочкообразный  резонатор можно рассматривать как
цилиндрический, сужающийся к концам по закону
     $$a(z)=a-\frac{z^2}{2R_0}\qquad \Bigl(\frac{D^2}{2R_0}\ll a
         \Bigr)\,,\eqno(28.45)$$
где $a(z)$  ---  радиус  резонатора в сечении $z$,  $a$ --- радиус при
$z=0$,  $R_0$ --- радиус кривизны меридионального сечения  при  $z=0$.
Наименьшим   затуханием  обладают  волны  шепчущей  галлереи,  которым
соответствуют большие значения азимутального индекса $m$ и  наименьший
радиальный,  а  между  радиусами выполняется соотношение $R_0=2a$,  то
есть когда диаметрально противоположные участки отражающей поверхности
являются конфокальными зеркалами.

     Заканчивая рассмотрение  открытых   резонаторов,   поле   которых
ограничено  каустиками,  отметим  одно важное обстоятельство.  Наличие
каустик не только существенно  уменьшает  радиационные  потери,  но  и
делает  резонаторы  малочувствительными к технологическим погрешностям
изготовления. Допуски на параметры резонаторов без каустик на коротких
волнах  настолько  малы,  что,  например,  резонатор  с  параллельными
плоскими зеркалами в оптическом диапазоне не может быть реализован.

\vspace{-.2cm}
\begin{wrapfigure}[12]{l}{7.5cm}
\begin{picture}(80,45)
\put(4,45){\special{em:graph fig28-7.bmp}}
\end{picture}
\hbox to 7.5cm{\hfil\footnotesize{Рис.~28.7.~Бочкообразный резонатор.
}\hfil}
\end{wrapfigure}

     Эффекты полного  отражения проявляются в системах, где
прозрачная и оптически  более  плотная  среда  окружена  средой  менее
плотной  или  пустотой.  В разделе 12 были рассмотрены диэлектрическая
пластина и диэлектрический  стержень,  которые  являют  собой  примеры
открытых   волноводов,   основанных  на  явлении  полного
отражения,  и  в  которых  распространяется  без  радиационных  потерь
конечное число собственных волн. Число таких волн может быть сокращено
за счёт выбора диэлектрической проницаемости окружающей  среды  лишь
немного меньшей,  чем самого стержня.  Такие волноводы находят широкое
применение в волоконной оптике.  Очевидно,  что если на торцах стержня
поместить   плоские  зеркала,  то  структура  превратится  в  открытый
резонатор.  Собственные частоты такого резонатора легко  находятся  из
условия набега фаз, кратного $\pi$.

     Использование зеркал вовсе не является обязательным условием  для
создания диэлектрических резонаторов --- например, диэлектрический тор
и диэлектрический шар сами по себе  обладают  добротными  собственными
колебаниями.   Однако   с   помощью   зеркал   их  спектр  может  быть
скорректирован в нужную сторону.  Следует отметить,  что теоретический
расчёт  систем,  включающих  в  себя диэлектрические тела и зеркала,
представляет собой в большинстве  случаев  значительно  более  сложную
задачу, чем рассмотренные до сих пор для открытых резонаторов из одних
зеркал.  Многие из таких  задач,  имеющих  практическое  значение,  не
решены до сих пор даже численными методами.

%\end{document}

\newpage
\oddsidemargin=-0.4mm \evensidemargin=-0.4mm
\topmargin=-0.4mm
\headsep=7mm
\textheight=231.875mm
\textwidth=160mm
\mathsurround=2.5pt
\unitlength=1mm
%\begin{document}
%\input{macr.tex}
\thispagestyle{empty}
%\addtocounter{page}{343}

\begin{center}\subsubsection*{Д\,О\,П\,О\,Л\,Н\,Е\,Н\,И\,Я}
\vspace{10mm}\subsubsection*{Д1. Интегрирование неоднородных  волновых
уравнений} \end{center}\vspace{.5cm}

\markboth{Дополнения}{Д1. Интегрирование    неоднородных    волновых
          уравнений}

\begin{center}\begin{minipage}[c]{0.75\textwidth}
\footnotesize{\parindent=0.5cm
         Функция Грина волнового  уравнения.  Различные  представления
         функции  Грина  в  декартовых  и  цилиндрических координатах.
         Вычисление поля плоского поверхностного  распределения  тока.
         Поле  бесконечно протяжённой трубки тока (линейного тока) и
         его связь с полем собственных волн круглого волновода.
}\end{minipage}\end{center}\vspace{.5cm}

     Решение скалярного  неоднородного волнового уравнения,  которое в
математике принято называть уравнением Гельмгольца,
     $$\Delta u+k^2 u=-f(\rv r),\eqno(\mbox{\rm Д1}.1)$$
где $u$ --- компонента поля или одного из  его  потенциалов,  $k$  ---
волновое  число,  $f(\rv  r)$  --- плотность источников,  в однородном
неограниченном пространстве может быть выражено через основную
функцию  Грина $G(\rv r,\rv r')$, удовлетворяющую уравнению
     $$\Delta G(\rv r,\rv r')+k^2G(\rv r,\rv r')=-\delta(\rv r-\rv r')
         \,,\eqno(\mbox{\rm Д1}.2)$$
в виде интеграла
     $$u(\rv r)=\int\!f(\rv r')G(\rv r,\rv r')\,dV_{r'}
         \eqno(\mbox{\rm Д1}.3)$$
по всему  бесконечному координатному пространству радиуса-вектора
$\rv r'$.  Это решение находится с помощью {\it  второй  формулы
Грина}  в предположении,  что  обе  функции  $u(\rv  r)$  и $G(\rv
r,\rv  r')$ удовлетворяют на бесконечности условию излучения.
Формулы (\mbox{\rm Д1}.1) и (\mbox{\rm Д1}.2) приведены для
свободного пространства $(k=\omega/c)$; их обобщение на однородную
среду с $\varepsilon,\,\mu$ тривиально и здесь не приводится,
чтобы не загромождать ряд последующих выражений.

     Функция Грина $G(\rv r,\rv  r')$,  зависящая  от  радиус-векторов
точки наблюдения $\rv r$ и точки расположения источника $\rv r'$, была
найдена в разделе 17:
     $$ G(\rv r,\rv r')=\frac{e^{ik R}} {4\pi  R}\,,
         \eqno(\mbox{\rm Д1}.4)$$
где $R=|\rv r-\rv r'|$.  Отметим,  что функция $G$ фактически  зависит
лишь  от  разности  своих  аргументов;  при  этом  они  входят в неё
симметричным образом.  Напомним ещё,  что знак у  экспоненты  выбран
исходя   из   условия   излучения   и  выбранного  представления  всех
гармонических величин пропорциональными $e^{-i\omega t}$.

     В ряде   случаев   для  вычисления  интеграла  (\mbox{\rm Д1}.3)  более
удобными оказываются разнообразные интегральные представления  функции
$G(\rv  r,\rv r')$,  которые получаются в результате её разложения в
трёхмерный интеграл Фурье:
     $$G(\rv r,\rv r')=\displaystyle{\frac{1}{8\pi^3}}\!\int\!G(\mbox
         {\bf\ae}) e^{i\mbox{\scriptsize\bf\ae}(\rv r-\rv r')}\,
         dV_{\ae}\,.\eqno(\mbox{\rm Д1}.5)$$
Подынтегральное выражение  имеет  здесь  вид  плоской волны с
волновым вектором  $\mbox{\bf\ae}$,  поэтому  этот  интеграл
Фурье   называют разложением по плоским волнам. Интегрирование в
(\mbox{\rm Д1}.5) проводится по   всему    бесконечному
координатному    пространству    вектора $\mbox{\bf\ae}$.

     Фурье-образ функции Грина $G(\mbox{\bf\ae})$  проще  всего  найти
пут\"ем  подстановки  формулы  (\mbox{\rm Д1}.5)  в  уравнение
(\mbox{\rm Д1}.2). Записывая    правую    часть    уравнения    в
виде    произведения $-\delta(x-x')\delta(y-y')\delta(z-z')$    и
используя    известное представление $\delta$-- функции в виде
интеграла Фурье
     $$ \delta(x)=\frac 1{2\pi} \int\limits_{-\infty}^\infty
            e^{i\alpha x}\,d\alpha\,,\eqno(\mbox{\rm Д1}.6)$$
получаем
     $$  G(\mbox{\bf\ae})=\frac 1{\mbox{\bf\ae}^2-k^2}\,.
         \eqno(\mbox{\rm Д1}.7)$$

     Все приведенные выше выражения записаны через радиусы-векторы и
поэтому являются общими для любой системы координат.  В
зависимости от конкретного  выбора  этой системы и степени
завершённости вычисления трёхкратного  интеграла  в
(\mbox{\rm Д1}.5)  существуют   многочисленные различные
представления функции $G(\rv r,\rv r')$.

     Так, в  прямоугольных координатах функция Грина (\mbox{\rm Д1}.5) имеет
вид
     $$ G(x,y,z;x',y',z')\!=
         \displaystyle{\frac{1}{8
         \pi^3}}\!\!\int\!\!\int\!\!\int\limits_{\hspace{-0.7cm}-
         \infty}^{\hspace{-0.5cm}\infty}\!\!\frac{e^{i\ae_x (x-x')+i
         \ae_y (y-y')+i\ae_z (z-z')}}{{\ae_x^2+\ae_y^2+\ae_z^2-k^2}}
         \,d\ae_x\,d\ae_y\,d\ae_z\,.\eqno(\mbox{\rm Д1}.8)$$
Проведём в  этом  тройном  интеграле  интегрирование по какой-нибудь
переменной, например, по $d\ae_z$, представив его как
     $$ G(x,y,z;x',y',z')\!=
         \displaystyle{\frac{1}{8
         \pi^3}}\!\!\int\!\!\int\limits_{\hspace{-0.4cm}-
         \infty}^{\hspace{-0.2cm}\infty}\!e^{i\ae_x (x-x')+i
         \ae_y (y-y')} L(\gamma,z-z')
         \,d\ae_x\,d\ae_y\,,\eqno(\mbox{\rm Д1}.9)$$
где
     $$ L(\gamma,z-z')=\int\limits_{-\infty}^{\infty}
         \frac{e^{i\ae_z(z-z')}}{(\ae_z+\gamma)(\ae_z-\gamma)}
         \,d\ae_z\,.\eqno(\mbox{\rm Д1}.10)$$
Здесь введено  обозначение
     $$\gamma=\left\{\begin{array}{l}\phantom{i}\sqrt{k^2-\ae_x^2-
         \ae_y^2}\qquad\mbox{при}\quad k^2>\ae_x^2+\ae_y^2, \\[.2cm]
         i\sqrt{\ae_x^2+\ae_y^2-k^2}\qquad\mbox{при}\quad
         k^2<\ae_x^2+\ae_y^2; \end{array}\;\right.\eqno(\mbox{\rm Д1}.11)$$
в обоих случаях подразумевается арифметическое значение корня.

     Интеграл $L$  легко вычисляется с помощью теории вычетов.  Полюсы
подынтегрального  выражения  расположены  либо  на  мнимой  оси   (при
${\ae}_x^2+{\ae}_y^2  >  k^2$),  либо  на действительной (при обратном
неравенстве).  При $z-z'>0$ в соответствии  с  леммой  Жордана  контур
интегрирования следует  замыкать  дугой  бесконечно большого радиуса в
верхней полуплоскости комплексной переменной $\ae_z$,  а при  $z-z'<0$
---   в   нижней   (см.   рис.~Д1.1).   Если   полюсы  расположены  на
действительной оси,  то следует учитывать, что при введении бесконечно
малого  поглощения  в  среде  (или  исходя из условия излучения) левый
полюс смещается чуть ниже действительной оси,  а правый --- чуть выше;
поэтому  при  любом  замыкании  контура вклад в интеграл даёт только
один из полюсов.

\begin{picture}(160,50)
\put(-2,45){\special{em:graph  figA1-1a.bmp}}
\put(78,45){\special{em:graph figA1-1b.bmp}}
\end{picture}

\begin{center}\begin{minipage}[c]{0.9\textwidth}
\footnotesize{\parindent=0.5cm
Рис.~Д1.1.~Расположение полюсов на комплексной плоскости $\ae_z$
и контуры интегрирования при вычислении интеграла $L$:
{\it а)} $\ae_x^2+\ae_y^2>k^2$, {\it б)} $\ae_x^2+\ae_y^2<k^2$\,.
}\end{minipage}\end{center}\vspace*{0.25cm}

     В результате вычисления вычетов получаем
     $$ L=\frac {\pi i} \gamma \,e^{i\gamma |z-z'|}\,,
         \eqno(\mbox{\rm Д1}.12)$$
так что одно из представлений функции Грина записывается в виде
     $$ G(x,y,z;x',y',z')\!= \frac i{8\pi^2}\!\!\int\!\!\int
         \limits_{\hspace{-0.4cm}-\infty}^{\hspace{-0.2cm}\infty}
         \!\frac 1 \gamma\; e^{i\ae_x (x-x')+i\ae_y (y-y')+i
         \gamma|z-z'|}\,d\ae_x\,d\ae_y\,.\eqno(\mbox{\rm Д1}.13)$$
Отметим, что на плоскости  интегрирования  $\ae_x,\,\ae_y$  вне
круга радиуса  $k$ подынтегральное выражение представляет собой
неоднородную плоскую волну,  волновой вектор которой лежит в
плоскости  $z=z'$,  а амплитуда  экспоненциально спадает по мере
удаления от этой плоскости. Внутри этого круга --- локально
однородную  плоскую  волну,  проекция волнового  вектора  которой
на  ось  $z$  в  любой точке пространства направлена  {\it  от}
плоскости  $z=z'$.  Необходимо   указать,   что дальнейшее
вычисление  интеграла  по $d\ae_x$ и $d\ae_y$ представляет собой
более сложную задачу,  но это не должно нас беспокоить,  так как
окончательный результат (\mbox{\rm Д1}.4) известен.

     Рассмотрим теперь   некоторые   представления   функции  Грина  в
цилиндрической системе координат $r,\,\varphi,\,z$.  Произведя
замену переменных    $\ae_x=g\cos{\psi}$,   $\ae_y=g\sin{\psi}$,
$\ae_z=h$, перепишем (\mbox{\rm Д1}.8) в виде
     $$G=\frac 1 {8\pi^3}\int\limits_0^\infty\!\!gdg\!
         \int\limits_0^{2\pi}\!\!d\psi\!\int\limits_{-\infty}^\infty
         \!\! \frac{e^{i g r\cos(\psi-\varphi)-i
         g r'\cos(\psi-\varphi')+ih(z-z')}}{g^2+h^2-k^2}\;dh.
         \eqno(\mbox{\rm Д1}.14)$$
Воспользовавшись далее   известным   из   теории   бесселевых  функций
разложением
     $$e^{i g r\cos{(\psi-\varphi)}}=
         {\sum_{-\infty}^\infty i^n}e^{-in(\psi-
         \varphi)}J_n(g r)\,\eqno(\mbox{\rm Д1}.15)$$
и интегралом
     $$ \frac{1}{2\pi} \int\limits_0^{2\pi}e^{i(m-n)\alpha}\,d\alpha
         = \delta_{mn}\,,\eqno(\mbox{\rm Д1}.16)$$
где $\delta_{mn}$  ---  символ  Кронеккера,  выполним  в
(\mbox{\rm Д1}.14) интегрирование по $\psi$:
     $$G=\displaystyle{\frac{1}{4\pi^2}\sum_{n=-\infty}^
         \infty e^{in(\varphi-\varphi')}\int\limits_0^\infty\int
         \limits_{-\infty}^\infty\frac{e^{ih(z-z')}J_n(g r)J_n(g
         r')}{g^2+h^2-k^2}\,g\,dg\,dh.}\eqno(\mbox{\rm Д1}.17)$$

     Дальнейшее интегрирование  по  $dg$  также  удобно  проводить   в
комплексной   плоскости   с   помощью   теории   вычетов.   Для  этого
предварительно   распространим   область   интегрирования    на    всю
действительную  ось  $g$,  используя  известные формулы для бесселевых
функций:
     $$J_n(-x)=(-1)^nJ_n(x),\qquad 2J_n(x)=H_n^{(1)}(x)-(-1)^nH_n^{(1)}
         (-x).\eqno(\mbox{\rm Д1}.18)$$
В результате преобразований получаем, что
     $$\displaystyle{G=\frac{1}{8\pi^2}\!\sum_{-\infty}^\infty e^{in(
         \varphi-\varphi')}\!\!\int\!\!\int\limits_{\hspace{-0.35cm}-
         \infty}^{\hspace{-0.2cm}\infty}}g\,dgdh\frac{e^{ih(z-
         z')}}{g^2+h^2-k^2}\left\{\begin{array}{rcl}\rule[-0.14cm]
         {0pt}{0.34cm}H_n^{(1)}(gr')J_n(gr)&\!\!&\hbox{при $r
         <r'$},\\[0.25cm]\rule[0.14cm]{0pt}{0.34cm}J_n(gr')H_n^
         {(1)}(gr)&\!\!&\hbox{при  $r>r'$}.\end{array}\;\right.
         \eqno(\mbox{\rm Д1}.19)$$
Контур интегрирования  в  комплексной  плоскости замыкается с учётом
асимптотики функции Ханкеля при больших значениях аргумента
     $$H_n^{(1)}(x)\approx\sqrt{\frac{2}{\pi x}}\;e^{i(x-n
         \frac{\pi}{2}-\frac{\pi}{4})}\,.\eqno(\mbox{\rm Д1}.20)$$

     Вклад в  интеграл  дают  только  полюсы,  соответствующие   нулям
знаменателя.   Вычисляя  вычеты  в  этих  полюсах,  приходим  к  часто
используемому  представлению  функции   Грина,   которое   естественно
рассматривать как её разложение по цилиндрическим волнам:
     $$\displaystyle{G=\frac{i}{8\pi}\sum_{-\infty}^\infty e^{in(
         \varphi-\varphi')}\!\int\limits_{-\infty}^\infty\!dh e^{ih
         (z-z')}}\left\{\begin{array}{rcl}\rule[-0.14cm]{0pt}{0.34cm}
         H_n^{(1)}(\nu r')J_n(\nu r)&\!\!&\hbox{при $r<r'$},
         \\[0.25cm]\rule[0.14cm]{0pt}{0.34cm}J_n(\nu r')H_n^{(1)}(\nu r)
         &\!\!&\hbox{при $r>r'$},\end{array}\right.\;
         \eqno(\mbox{\rm Д1}.21)$$
где
     $$ \nu=\left \{\begin{array}{ll}\phantom{i}\sqrt{k^2-h^2}&
         \quad\mbox{при}\quad k^2>h^2,\\[.2cm] i\sqrt{h^2-k^2}&
         \quad\mbox{при}\quad h^2>k^2\end{array}\right.\;
         \eqno(\mbox{\rm Д1}.22)$$
и в  обоих случаях подразумевается арифметическое значение корня.
При $k^2<h^2$ подынтегральное выражение в (\mbox{\rm Д1}.21)
представляет  собой {\it   цилиндрическую}   волну,  бегущую
вдоль  оси  и  затухающую  в поперечном направлении;  если же
$k^2>h^2$, то волновой вектор имеет и радиальную компоненту и
такие волны часто называют {\it коническими}.

     Нетрудно получить  ещё  одно  представление  функции  Грина   в
цилиндрических координатах,  вычислив интеграл по $dh$ в
(\mbox{\rm Д1}.17). Не будем здесь на этом останавливаться,  так
же  как  и  на  различных широко   используемых   представлениях
функции  Грина  в  сферических координатах,  однако  рассмотрим
кратко  интегральное   представление функции Грина {\it
двумерного} волнового уравнения.

     В том случае, когда правая часть уравнения (\mbox{\rm Д1}.1) не зависит
от   какой-нибудь   из  прямоугольных  координат,  например,  от  $y$,
естественно предполагать,  что и  решение  этого  уравнения  не  будет
зависеть  от  этой координаты,  и имеет смысл ввести двумерную функцию
Грина, удовлетворяющую уравнению
     $$  \frac {\partial^2 G}{\partial x^2}+
         \frac {\partial^2 G}{\partial z^2}+k^2 G=-\delta(x-x')
         \delta(z-z')\,,\eqno(\mbox{\rm Д1}.23)$$
с помощью которой решение уравнения записывается в виде
     $$ u(x,z)=\int\!\!\int\limits_{\hspace {-0.35cm}-\infty}^
         {\hspace{-0.2cm}\infty} f(x',z')\;G(x,z;x',z')
         \,dx'\,dz'\,.\eqno(\mbox{\rm Д1}.24)$$

     Решение уравнения     (\mbox{\rm Д1}.23),    удовлетворяющее    условию
излучения,  легко может быть получено  так  же,  как  была  найдена  в
разделе   17   трёхмерная  функция  Грина,  только  все  выкладки  и
рассуждения  следует  проводить  не  в  сферических,  а   в   полярных
координатах.  В  результате  получаем  выражение для двумерной функции
Грина
     $$ G(x,z;x',z')=\frac i 4\,H_0^{(1)}\Bigl(k\sqrt{(x-x')^2+(z-z')^2}
         \,\Bigr)\,,\eqno(\mbox{\rm Д1}.25)$$
разложение которой в двукратный интеграл Фурье имеет вид
     $$G(x,z;x',z')=\frac{1}{4\pi^2}\int\!\!\int\limits_{\hspace
         {-0.35cm}-\infty}^{\hspace{-0.2cm}\infty}\frac{e^{i\ae_x (x-x')
         +i\ae_z(z-z')}}{\ae_x^2+\ae_z^2-k^2}\,d\ae_x\,d\ae_z.
         \eqno(\mbox{\rm Д1}.26)$$

     Вычисляя интеграл по $d\ae_x$ в комплексной  плоскости,  как  это
было  проделано  выше  для  трёхмерной функции Грина в прямоугольных
координатах,  и вводя обозначение  $h=\ae_z$,  приходим  к  следующему
интегральному представлению двумерной функции Грина:
     $$  G(x,z;x'z')= \frac{i}{4\pi}\int\limits_{-\infty}^\infty
         \frac {e^{ih(z-z')+i\nu |x-x'|}}{\nu}\,dh\, ,
         \eqno(\mbox{\rm Д1}.27)$$
где $\nu$ определено формулой (\mbox{\rm Д1}.22).

     С различными  интегральными  представлениями функции Грина тесным
образом связаны различные представления $\delta$--функции Дирака.
Выше интегральное  представление  функции Грина в прямоугольных
координатах было  получено  с  помощью   основного   интегрального
представления $\delta$-функции   (\mbox{\rm Д1}.6).   Получим
теперь  два  представления $\delta$-функции в  цилиндрических
координатах.  Записывая  уравнение (\mbox{\rm Д1}.2) в виде
     $$ \frac 1 r\,\frac{\partial\phantom{r}}{\partial r}\,\Bigl(
         r\,\frac{\partial G}{\partial r}\Bigr)+\frac 1 {r^2}\,
         \frac{\partial^2 G}{\partial\varphi^2}+\frac
         {\partial^2 G}{\partial z^2}+k^2G=-\frac{1}{r'}\,\delta(r-r')\,
         \delta(\varphi-\varphi')\,\delta(z-z')\,,\eqno(\mbox{\rm Д1}.28)$$
подставляя в  него  функцию Грина (\mbox{\rm Д1}.17) и имея в виду
уравнение Бесселя
     $$ \frac 1 r\,\frac{\partial\phantom{r}}{\partial r}\,\Bigl(
         r\,\frac{\partial J_n(gr)}{\partial r}\Bigr)+
         \Bigl(g^2-\frac{n^2}{r^2}\Bigr)\,J_n(gr)=0
         \,,\eqno(\mbox{\rm Д1}.29)$$
получим   два следующие выражения для $\delta$--функции:
     $$\delta(r-r')=r'\int\limits_0^\infty J_n(g r)
         J_n(g r')\,g\,dg\,,\eqno(\mbox{\rm Д1}.30)$$
и
     $$ \delta(\varphi-\varphi')=\frac 1{2\pi}\sum\limits_
         {-\infty}^\infty e^{-i n(\varphi-\varphi')}\,.
         \eqno(\mbox{\rm Д1}.31)$$

     На первый  взгляд  кажется,  что   от   приведенных   различных
представлений   функции   Грина  мало  пользы  для  решения
уравнения (\mbox{\rm Д1}.1),  поскольку при использовании  функции
Грина  (\mbox{\rm Д1}.4) решение  (\mbox{\rm Д1}.3) выражается
через трёхкратный интеграл Фурье,  а при  использовании  любого
другого   её   представления   кратность интеграла повышается;
однако в ряде случаев это совсем не так.

     Рассмотрим для примера решение  двумерного  волнового  уравнения,
когда в плоскости $x=a$ задан сторонний ток с плотностью
     $$ j_z(x,z)=I_0\,e^{ihz}\,\delta(x-a)\,.\eqno(\mbox{\rm Д1}.32)$$
Поле этого  тока  может  быть  выражено через векторный потенциал $\rv
A=\{0,0,A_z\}$, компонента которого удовлетворяет уравнению
     $$ \Delta A_z+k^2 A_z=-\frac {4\pi} c j_z\,.\eqno(\mbox{\rm Д1}.33)$$
Записывая решение этого уравнения в виде двухкратного  интеграла
типа (\mbox{\rm Д1}.24)   с   функцией  Грина  (\mbox{\rm Д1}.25),
после  тривиального интегрирования по $dx'$ приходим к интегралу
     $$ A_z(x,z)=\frac{\pi i}c I_0\int\limits_{-\infty}^\infty
         e^{i hz'}\,H_0^{(1)}\Bigl(k\sqrt{(x-a)^2+(z-z')^2}
         \,\Bigr)\,dz'\,.\eqno(\mbox{\rm Д1}.34)$$

     Конечно, этот интеграл может быть найден в  толстых  справочниках
или  путём  весьма  кропотливых  выкладок  вычислен.  Но
посмотрим, насколько проще находится решение,  если  для  функции
Грина  выбрать представление   (\mbox{\rm Д1}.27).   В   этом
случае  после  тривиального интегрирования по $dx'$
     $$ A_z= \frac {iI_0}c\int\limits_{-\infty}^\infty dz'\,e^{ihz'}
         \int\limits_{-\infty}^\infty \frac {e^{ih'(z-z')+i
         \nu'|x-a|}}{\nu\,'}\,dh'\,,\eqno(\mbox{\rm Д1}.35)$$
где $\nu'$  определено  формулой  (\mbox{\rm Д1}.22),  в  которой
$h$ следует заменить на  $h'$.  Вычисляя  интеграл  по  $dz'$  с
помощью  формулы (\mbox{\rm Д1}.6),  получаем  в  подынтегральном
выражении  дополнительный множитель $2\pi\delta(h-h')$, после чего
интегрирование по $dh'$ легко проводится, и в результате имеем,
что
     $$ A_z(x,z)=\frac{2\pi i}c\, I_0\,\frac{e^{ihz+i\nu|x-a|}}\nu\,.
         \eqno(\mbox{\rm Д1}.36)$$
В соответствии  с  общими  формулами  (3.23),  связывающими  векторный
потенциал и поля, находим отсюда отличные от нуля компоненты полей:
     $$\left.\begin{array}{lcl} H_y &=&-\displaystyle{\frac {
         \partial A_z}{\partial x}=\frac{2\pi}c\,I_0\sign{(x-a)}\,e^
         {ihz+i\nu|x-a|}}\,,\\[.5cm]E_x&=&\displaystyle{
         \frac i k\frac{\partial^2A_z}{\partial x\partial z}=
         \frac{2\pi h}{ck} \,I_0\sign{(x-a)}\,e^{ihz+i\nu|x-a|}}\,,
         \\[.5cm]E_z&=&ikA_z+\displaystyle{\frac i k\frac{\partial^
         2A_z}{\partial z^2}=-\frac {2\pi\nu}{ck}\,I_0\,e^{ihz+i\nu
         |x-a|}}\,.\end{array}\right\} \eqno(\mbox{\rm Д1}.37)$$

     Согласно уравнению  непрерывности выражению для тока (\mbox{\rm Д1}.32)
необходимо     сопоставить     поверхностную     плотность      заряда
$\rho=I_0h/\omega \;e^{ihz}$.  Нетрудно видеть, что при переходе через
плоскость $x=a$ требуемый скачок нормальной компоненты  электрического
поля   $E_x$   составляет   $4\pi\rho$,  а  тангенциальной  компоненты
магнитного поля $H_y$ --- $4\pi k\rho/h$.

     Структура поля,  определяемого  (\mbox{\rm Д1}.37),  и  соответствующие
этому полю потоки энергии существенным образом зависят от
соотношения между  $k$  и $h$,  то есть между длиной волны поля
$\lambda=2\pi/k$ и длиной волны модуляции тока $\Lambda=2\pi/h$.
Если $k>h$, то параметр $\nu$  (см.  (\mbox{\rm Д1}.22))  величина
действительная,  и  поле в обоих полупространствах $x>a$ и $x<a$
представляет собой однородную  плоскую волну  с  волновым
вектором $\rv k=\{\sign(x-a)\sqrt{k^2-h^2},0,h\}$, направленным к
оси $z$ под углом $\varphi$,  определяемым соотношением
$h=k\cos{\varphi}$.  Обе  отличные  от  нуля  компоненты
комплексного вектора Умова-Пойнтинга --- величины действительные:
     $$ \left.\begin{array}{lclcl} \overline{\gv S}_x & = & -
         \displaystyle{\frac{c}{8\pi}} \re{E_z H_y^*} & = &
         \displaystyle{\frac{\pi I_0^2\sqrt{k^2-h^2}}{2\omega}
         \sign(x-a)}\,,\\ [.4cm] \overline{\gv S}_z & = &
         \phantom{-}\displaystyle{\frac{c}
         {8\pi}} \re {E_x H_y^*} & = & \displaystyle{\frac{\pi h I_
         0^2}{2\omega}}\,, \end{array} \right\}\eqno(\mbox{\rm Д1}.38)$$
и поток  мощности направлен вдоль волнового вектора $\rv k$.  При этом
существует постоянный поток мощности  на  бесконечность,  определяемый
компонентой  ${\gv  S}_x$,  который  компенсируется работой стороннего
тока над полем, равной (на единицу поверхности)
     $$\frac 1 2 \int j_z E^*_z\,dV= -\frac{\pi I_0^2\sqrt{k^2-h^2}}
         {\omega}\,.\eqno(\mbox{\rm Д1}.39)$$
В рассмотренном поле проекция  фазовой  скорости  волны  на  ось  $z$,
равная  $v_{фz}=\omega/\sqrt{k^2-h^2}$,  превышает скорость света $c$,
так что --- согласно сложившейся терминологии --- эта  волна  является
{\it быстрой} волной.

     При $k^2<h^2$ поле (\mbox{\rm Д1}.37) по обе стороны от плоскости $x=a$
представляет  собой  неоднородную  {\it  медленную}   плоскую   волну,
волновой  вектор  которой  направлен  вдоль  оси  $z$.  Волна является
поверхностной,  её поле прижато к плоскости  $x=a$  и  экспоненциально
спадает  при  удалении  от  неё.  Комплексный вектор Умова-Пойнтинга
имеет только одну отличную от нуля действительную компоненту
     $$ \overline{\gv S}_z=\frac{\pi I_0^2h}{2\omega}\,
         e^{-2\sqrt{h^2-k^2}\,|x-a|}\,.\eqno(\mbox{\rm Д1}.40)$$
При этом  отсутствует  поток  мощности,  направленный  от  тока,  и  в
подтверждение этого равна нулю работа поля над током.

     На первый  взгляд   задача   о   стороннем   поверхностном   токе
(\mbox{\rm Д1}.32)    представляется    искусственной, поскольку
трудно представить,  каким  образом  на  практике такой   ток
может   быть возбуждён   и  поддержан.  Однако более
внимательное  рассмотрение указывает  на   тесную   связь этой
задачи   с   двумя   задачами, представляющими несомненный интерес
для электродинамики СВЧ. Так, поле в верхнем полупространстве
$x>a$ по своей структуре совпадает с полем, которое  возбуждается
в свободном полупространстве однородной плоской волны при её
падении на границу раздела со стороны оптически  более плотной
среды.  Быстрая  однородная  плоская волна возбуждается в том
случае,  когда угол падения меньше угла полного внутреннего
отражения, медленная  неоднородная плоская волна --- когда угол
падения превышает этот угол.  Напомним,  что эта  задача  подробно
была  рассмотрена  в разделе  5,  и отметим,  что  в ней поле в
нижнем полупространстве не определяется формулами  (\mbox{\rm
Д1}.37),  а  на  поверхности  раздела не возбуждается ток
(\mbox{\rm Д1}.32).

     Больший интерес  представляет сопоставление полученного решения с
собственными волнами,  возбуждаемыми в плоском волноводе (раздел
10). Рассмотрим   совокупность   двух   плоских  поверхностных
токов  вида (\mbox{\rm Д1}.32) с одинаковой амплитудой $I_0$,
расположенных в плоскостях $x=0$ и $x=b$. В силу линейности
уравнений электродинамики имеет место принцип суперпозиции полей.
Начнём с простого частного случая, когда $h=k$,  а поверхностные
токи направлены в разные стороны.  Сложением полей вида
(\mbox{\rm Д1}.37)  убеждаемся,   что   вне   плоскостей   поле
тождественно  равно  нулю,  а  между  плоскостями  представляет
собой однородную плоскую волну, совпадающую с точностью до
нормировки с {\it собственной  ТЕМ-волной}  плоского  волновода. В
общем случае быстрых волн $k^2>h^2$ поле вне плоскостей равно нулю
при выполнении условия
     $$ e^{-i\sqrt{k^2-h^2}\,a}=\pm\, 1=e^{i\pi p}\,\eqno(\mbox{\rm Д1}.41)$$
(выбор знака необходимо  согласовывать  с  относительным
направлением токов в двух плоскостях). При этом распределение
полей (с точностью до нормировочного множителя) совпадает с полями
собственных электрических волн   плоского   волновода   (10.12)
--  (10.14),  а  само  условие (\mbox{\rm Д1}.41) определяет
продольное волновое число этих волн (10.10).

     Таким образом,  поля собственных волн неоднородной  структуры  (в
данном  случае  состоящей  из  двух  идеально  проводящих параллельных
плоскостей)   совпадают   с   полями,   возбуждаемыми   в   однородном
неограниченном  пространстве  сторонним  поверхностным  током,  равным
току,  который наводится  на  стенках  волновода  при  возбуждении  в   нём
собственной  волны.  Этот  результат  имеет  место в общем случае,  но
убедимся в его справедливости ещё раз на примере круглого волновода.

     Рассмотрим поле   линейного   стороннего   тока,   заданного    в
цилиндрической системе координат,
     $$j_z=\frac{J_0}{r}\delta(r-a)\,\delta(\varphi-\varphi_0)\,
         e^{ihz}=\sum \limits_{n=-\infty}^\infty j_{nz},
         \eqno(\mbox{\rm Д1}.42)$$
где
     $$j_{nz}=\frac{J_0}{2\pi r}\delta(r-a)\,
         e^{ihz+in(\varphi-\varphi_0)}\,,\eqno(\mbox{\rm Д1}.43)$$
и в дальнейшем,  не уменьшая этим общности  выражений,  будем
считать $\varphi_0=0$. Отличная от нуля компонента векторного
потенциала $A_z$ удовлетворяет  уравнению  (\mbox{\rm Д1}.33),
решение  которого  с  помощью функции Грина выражается в виде
тройного интеграла
     $$ A_z(r,\varphi,z)=\frac{4\pi}c\int\limits_0^{2\pi}d\varphi'
         \int\limits_0^\infty r'\,dr'\int\limits_{-\infty}^\infty
         \sum\limits_{n=-\infty}^\infty j_{nz}(r',\varphi',z')\,
         G(r,\varphi,z; r'\varphi',z')\,dz'\,.\eqno(\mbox{\rm Д1}.44)$$
Подставляя сюда функцию Грина в виде  (\mbox{\rm Д1}.21)  и
последовательно вычисляя интегралы по $d\varphi'$,  $dr'$,  $dz'$
и $dh'$,  приходим к следующему выражению для $A_z$:
     $$ A_z=\frac{i\pi J_0}ce^{ihz}\sum\limits_{n=-\infty}^\infty
         e^{in\varphi}\left\{\begin{array}
         {rcl}H_n^{(1)}(\nu a)J_n(\nu r)&\!\!&\hbox{при $r<a$,}
         \\[0.30cm]J_n(\nu a)H_n^{(1)}(\nu r)&\!\!&\hbox{при $r>a$,}
         \end{array}\right.\eqno(\mbox{\rm Д1}.45)$$
где $\nu$   определено   формулой   (\mbox{\rm Д1}.22).   Отметим,
что  при вычислении   интегралов   не   встречается сколько-нибудь
заметных трудностей:    при    интегрировании по   $d\varphi$
используется ортогональность функций $e^{in\varphi}$ и
$e^{im\varphi}$ при $n\ne m$ и  в результате остаётся
однократная сумма,  интегрирование по $dr'$ тривиально из-за
наличия $\delta$-функции,  а интегрирование по $z'$ в соответствии
с (\mbox{\rm Д1}.6) даёт в результате $\delta(h-h')$, после чего
интегрирование по $h'$ становится тривиальным. Если бы в качестве
функции  Грина  в  (\mbox{\rm Д1}.44)  была подставлена функция
(\mbox{\rm Д1}.4), аргумент которой $R$ выражен через
цилиндрические координаты источника и  точки  наблюдения,  то
вычислить интегралы было бы очень и очень непросто.

     Одиночный линейный  ток,  определяемый соотношением (\mbox{\rm Д1}.42),
естественно рассматривать  в  цилиндрической  системе  координат,
ось которой  $z$  совпадает  с  током.  Для  этого  достаточно  в
формуле (\mbox{\rm Д1}.45) положить $a=0$,  а в сумме по $n$
оставить  только  один член, соответствующий $n=0$. В результате
получается, что
     $$A_z=\frac{i\pi J_0}c H^{(1)}_0(\nu r)e^{ihz}\,
         \eqno(\mbox{\rm Д1}.46)$$
и тогда отличны от нуля следующие компоненты поля:
     $$   \left.\begin{array}{lcl}H_{\varphi}&=&-
         \displaystyle{\frac{\partial A_z}
         {\partial r}=\frac{i\pi J_0\nu}c\,H_1^{(1)}(\nu r)\,e^{ihz}}
         \,,\\[.4cm] E_z&=& \displaystyle{ikA_z+\frac i k\frac
         {\partial^2 A_z}{\partial z^2}=
         -\frac{\pi J_0\nu^2}{\omega}}\,H_0^{(1)}(\nu r)\,e^{ihz}\,,
         \\[.4cm] E_r&=& \displaystyle{\frac i k \frac{\partial^2 A_z}
         {\partial r \partial z}= \frac {i\pi J_0 h\nu}{\omega}}
         \,H_1^{(1)}(\nu r)\, e^{ihz}\,.\end{array}\right\}
         \eqno(\mbox{\rm Д1}.47)$$

     Структура этого поля,  как и поля токовой плоскости (\mbox{\rm Д1}.37),
существенным образом зависит от соотношения величин  $k$  и  $h$.  При
$k^2>h^2$  проекция  фазовой скорости волны на ось $z$ больше скорости
света  $c$  (волна  быстрая),  волновой  вектор  $\rv  k$   и   вектор
Умова-Пойнтинга направлены к оси $z$ под углом $\psi$, определяемым из
соотношения   $\cos{\psi}=h/k$,   имеет   место   поток   энергии   на
бесконечность   в   радиальном  направлении  и,  соответственно,  поле
работает   над   током.   При   $k^2<h^2$   волна   медленная,    поле
экспоненциально  спадает при удалении от токовой нити,  действительной
является только одна компонента  комплексного  вектора  Умова-Пойнтинга
--- $\overline{\gv S_z}$,  потока энергии, направленного от нити, нет,
и поле над током не работает.

     Для приложений  значительно  больший  интерес  представляет  поле
трубки тока (считаем его направленным вдоль оси $z$).  Плотность
тока, имеющего $n$ вариаций по азимуту,  в этом случае
определяется формулой (\mbox{\rm Д1}.43),  а  векторный  потенциал
---  формулой (\mbox{\rm Д1}.45),  в которой следует опустить
суммирование по  $n$.  Продольная  компонента электрического поля
легко вычисляется:
     $$E_z=\frac{ic\nu^2}{\omega}\,A_z=-\displaystyle{
         \frac{\pi J_0\nu^2}{\omega}}\, e^{in\varphi+ihz}
         \left\{\begin{array}{rcl}H_n^{(1)}(\nu a)\,J_n(\nu r)&\!\!&
         \hbox {при $r<a$,}\\[0.3cm] H_n^{(1)}(\nu r)\,J_n(\nu a)&\!
         \!&\hbox {при $r>a$.}\end{array}\right.\eqno(\mbox{\rm Д1}.48)$$
Отличны от нуля ещё четыре компоненты поля (только $H_z=0$), которые
легко могут быть найдены по известному $A_z$.  Для нас важным моментом
является то,  что при выполнении условия $J_n(\nu a)=0$ поле {\it вне}
токовой  трубки  тождественно  равно  нулю,  а  внутри  совпадает   (с
точностью  до  нормировочного  множителя) с полем собственных $E$-волн
круглого волновода.  При  этом  из  указанного  выше  условия  следуют
правильные  значения  для критических частот этих волн и их продольных
волновых чисел.  Тем самым ещё раз подтверждена глубокая связь между
полем   собственных   волн   в   неоднородных  структурах  и  полем  в
неограниченном   пространстве   стороннего   тока,    соответствующего
поверхностному току, наведенному на стенках структуры .

     Эту же  закономерность  легко  установить на основании полученных
формул для токовой трубки и в случае  коаксиальной  линии.  Так,  если
взять  две  соосных  токовых  трубки  с  одинаковым  и  противоположно
направленным полным током,  то в частном случае $h=k$  сразу  находим,
что  поле  отлично  от нуля только в пространстве между трубками и там
тождественно совпадает с полем  $TEM$-волны.  Для  быстрых  волн  (при
$k^2>h^2$)  для  бесконечной  последовательности  дискретных значениях
$\nu$,  определяющих критические  частоты  электрических  волн  высших
видов, поле также отлично от нуля лишь между токовыми трубками.

%\end{document}

\newpage
\oddsidemargin=-0.4mm \evensidemargin=-0.4mm
\topmargin=-0.4mm
\headsep=7mm
\textheight=231.875mm
\textwidth=160mm
\mathsurround=2.5pt
\unitlength=1mm
%\begin{document}
%\input{macr.tex}
\thispagestyle{empty}
%\addtocounter{page}{353}
\baselineskip=\normalbaselineskip
%\baselineskip=1.085\normalbaselineskip

\begin{center} \subsubsection*{\bf    Д2.    Кильватерные    поля    в
теории ускорителей заряженных частиц} \end{center}
\vspace{.5cm}

\markboth{Дополнения}{Д2. Кильватерные поля в теории ускорителей}

\begin{center}\begin{minipage}[c]{0.75\textwidth}
\footnotesize{\parindent=0.5cm
         Понятие кильватерных   полей.   Определение   продольного   и
         поперечного   кильватерного   потенциала   в   однородных   и
         неоднородных  структурах.  Кильватерные потенциалы в идеально
         проводящей вакуумной камере кругового сечения.  Разложение по
         мультиполям. Учёт конечной проводимости стенок. Возможность
         ускорения  кильватерными  полями.   Кильватерные   потенциалы
         ступенчатых нерегулярностей камеры кругового сечения. Теорема
         Пановского-Венцеля.
}\end{minipage}\end{center}\vspace{.5cm}

      Одним из основных приложений электродинамики СВЧ является теория
ускорителей заряженных частиц.  Для описания работы ускорителя,
вообще говоря,   нет   необходимости   в   детальном   знании
распределения электромагнитного поля во всём  объёме  его
структуры  ---  чтобы рассчитать  динамику  пучка,  достаточно
иметь некоторые усреднённые характеристики поля вдоль траекторий
основной массы ускоряемых частиц. Электромагнитное поле в
вакуумной камере ускорителя можно разделить на две части:
стороннее поле,  возбуждаемое внешними генераторами СВЧ  с целью
ускорения   заряженных   частиц,  и  собственное  поле  пучка,
включающее в себя и поле,  наводимое в структуре  при  пролёте
мимо металлических  элементов  камеры  заряженных  частиц.
Сторонние  поля локализованы  в  основном  только  в  пределах
ускоряющих   структур (резонаторов, диафрагмированных волноводов и
других), а наводимые поля присутствуют во всём тракте
ускорителя.  В теории ускорителей широко используется  понятие
{\it кильватерного поля} (кильватерной функции, кильватерного
потенциала),  которое  представляет  собой  {\it  часть}
наводимого  пучком  поля  (в  некоторых  работах отечественных
авторов кильватерный потенциал называется  {\it наведённым
потенциалом}).  Само происхождение   термина  {\it кильватерное поле}
подразумевает  поле, остающееся в структуре {\it после} пролёта
заряженного  сгустка  и (или) сопровождающее его.

     Исходным понятием  в  концепции  кильватерных полей является {\it
$\delta$-образный}  (или  точечный)  {\it   кильватерный   потенциал},
который  вводится следующим образом.  Пусть вдоль некоторой траектории
(здесь и в дальнейшем она предполагается прямолинейной и  параллельной
оси  $z$  используемой цилиндрической системы координат) внутри камеры
ускорителя движется с постоянной  скоростью  $v$  точечный  заряд  $Q$
(будем   называть   его   первичным),   который  наводит  в  структуре
электромагнитное поле.  По параллельной траектории с той же  скоростью
$v$  с  некоторым  отставанием  $s$  по $z$ движется пробный заряд $q$
(того же знака), испытывающий действие поля, возбуждённого первичным
зарядом.  Кильватерный  потенциал  описывает  интегральное (вдоль всей
траектории) действие поля,  наведённого единичным первичным зарядом,
на  единичный  пробный заряд.  {\it Продольный кильватерный потенциал}
определяется как
     $${\gv W}_{\parallel}(s,b,\rv r_{\perp})=-\frac 1 Q
         \int\limits_{-\infty}^\infty
         \gv E_z(r,\varphi,z,t=(z+s)/v)\,dz\,,\eqno(\mbox{\rm Д2}.1)$$
где $\mbox{\boldmath$\gv E$}$ -- наведённое поле, $\rv r_{\perp}$ --
проекция радиуса-вектора пробного заряда $q$ на  поперечную  плоскость
$z=const$,  $b$  -- отклонение от оси первичного заряда,  для которого
$z=vt$.  Поскольку далее рассматриваются только аксиально-симметричные
структуры,  то  без  потери  общности  можно  считать,  что  положение
первичного заряда определяется углом $\varphi =0$. Обратим внимание на
знак в определении ${\gv W}_{\parallel}$: положительное значение ${\gv
W}_{\parallel}$ соответствует торможению пробного заряда.

     {\it Поперечный   кильватерный   потенциал}   (двумерный  вектор)
вводится формулой
     $$\mbox{\boldmath$\gv W$}_{\perp}(s,b,\rv r_{\perp})=\frac 1 Q
         \int\limits_{-\infty}^\infty \rv F_{\perp}(r,\varphi,z,t=
         (z+s)/v)\,dz\,,\eqno(\mbox{\rm Д2}.2)$$
где
     $$\rv F_{\perp}= \bigl(\mbox{\boldmath$\gv E$}+\beta[\rv e_z
         \mbox{\boldmath $\gv H$}]\bigr)_{\perp}  \eqno(\mbox{\rm Д2}.3)$$
--- сила   Лоренца,   действующая  на  единичный  заряд  в  поперечном
направлении ($\rv e_z$ --- единичный вектор вдоль оси $z$).

     Скорость зарядов   $v$  полагается  очень  мало  отличающейся  от
скорости света $c$,  так что во всех последующих формулах безразмерную
скорость  $\beta=v/c$  можно  положить равной единице  и единственной
динамической  характеристикой  частиц  считать  релятивистский  фактор
$\gamma=1/\sqrt{1-\beta^2}$.  Без этого предположения, которое с очень
высокой точностью выполняется в большинстве  современных  ускорителей,
нельзя   считать,   что  расстояние  $s$  между  сгустками  остаётся
неизменным вдоль всей траектории,  без  чего  кильватерные  потенциалы
становятся громоздкими и не находят практического применения в теории.

     В однородной (вдоль оси $z$) структуре  поле  движущегося  заряда
обладает  трансляционной  симметрией,  и его действие на пробный заряд
пропорционально  пройденному  пути.  Кильватерные  потенциалы  в  этом
случае  определяются  на  единицу  длины  и  обозначаются  той же,  но
строчной буквой:
     $${\gv w}_{\parallel}(s,b,\rv r_{\perp})=-\frac 1 Q
         \gv E_z(r,\varphi,z,t=(z+s)/v)\,,\eqno(\mbox{\rm Д2}.4)$$
     $$\mbox{\boldmath$\gv w$}_{\perp}(s,b,\rv r_{\perp})=\frac 1 Q
         \rv F_{\perp}(r,\varphi,z,t=(z+s)/v)\,.\eqno(\mbox{\rm Д2}.5)$$

     Отметим, что кильватерные потенциалы при таком определении ни  по
своей физической сути, ни даже по размерности не являются потенциалами
в том  смысле,  как  это  принято  в  электродинамике  ---  но  такова
установившаяся  терминология.  В дальнейшем у кильватерных потенциалов
опускаются все аргументы,  равные нулю (например,  при движении  обоих
зарядов  по  оси  $z$  остаётся  один  аргумент  $s$).  Кильватерные
потенциалы определяются и для первичного  сгустка  с  распределённым
зарядом.  Поскольку точечный кильватерный потенциал является функцией
Грина для сгустка, то такое  обобщение  является  тривиальным и далее не
рассматривается.

     Кильватерные потенциалы  описывают действие поля на пробный заряд
в реальной {\it временной области}.  Если к  кильватерным  потенциалам
применить  преобразование  Фурье  по $s$,  то получим описание того же
взаимодействия  {\it  в  частотной  области}. Величины,   которые
используются в этом случае,  называются соответственно {\it продольным
и поперечным импедансами связи}. Они будут рассмотрены далее в Дополнении
3.

     Простейшим примером,  когда  введённые  выше  потенциалы  имеют
смысл,   является   прямолинейное   движение   зарядов   в   свободном
пространстве.  Поле заряда,  движущего по оси $z$, определяется в этом
случае хорошо известными выражениями:
     $$\left. \begin{array}{lcl} \gv E_z&=&\displaystyle{\frac
         {Q(z-vt)\gamma}{[(z-vt)^2\gamma^2+r^2]^{3/2}}}\,,\\[.5cm]
         \gv E_r&=&\displaystyle{\frac{Q\gamma r}{[(z-vt)^2
         \gamma^2+r^2]^{3/2}}}\,,\\[.5cm] \gv H_{\varphi}&=&
         \beta\gv E_r\,,\end{array}\right\}   \eqno(\mbox{\rm Д2}.6)$$
что непосредственно  приводит  к  следующим  формулам для кильватерных
потенциалов:

     $$ \gv w_{\parallel}(s,r)=\frac{s\gamma}{[s^2\gamma^2+r^2]
         ^{3/2}} \,,\eqno(\mbox{\rm Д2}.7)$$
     $$ \gv w_r(s,r)=\frac r {\gamma[s^2\gamma^2+r^2]^{3/2}}
         \,.\eqno(\mbox{\rm Д2}.8)$$
При релятивистском  движении  (большие  $\gamma$)  потенциалы  малы  и
практически  не  дают  сколько-нибудь  существенного вклада в динамику
частиц.  Отметим  два  момента:  отстающий  заряд  всегда   испытывает
торможение,  а  на  движущийся  впереди  заряд ($s<0$) действует точно
такая же по величине ускоряющая сила.

\begin{wrapfigure}[13]{l}{7.0cm}
\begin{picture}(75,44)
\put(0,45){\special{em:graph FigA2-1.bmp}}
\end{picture}
\hbox to 7 cm{\hfil\footnotesize{Рис.~Д2.1.~К расчёту кильватерного
}\hfil}
\hbox to 7 cm{\hfil\footnotesize{потенциала в круглой трубе.}\hfil}
\end{wrapfigure}

     Ещё меньшее значение имеют кильватерные потенциалы в однородной
камере  круглого сечения при идеальной проводимости стенок.  Физически
этот результат представляется очевидным,  хотя конкретные расчёты  в
этом    случае    значительно   более   громоздкие.   Проведём   их,
воспользовавшись  изложенной  в   разделе   20   теорией   возбуждения
волноводов.  Будем  искать  поле,  возбуждаемое  в  круглом  волноводе
радиуса $a$ точечным  зарядом  $Q$,  пролетающим  параллельно  оси  на
расстоянии $b$ от неё (рис.~Д2.1), чему соответствует плотность тока
     $$ \mbox{$\boldmath{\gv j}$}_z(\rv r,t)=Qv\frac{\delta(r-b)} r\,
         \delta(\varphi)\,\delta(z-vt)=\int\limits_{-\infty}^\infty j
         _z(\rv r,\omega)\,e^{-i\omega t}\,d\omega\,.
         \eqno(\mbox{\rm Д2}.9)$$
Используя сведения,  приведённые в предыдущем
Дополнении 1,  для спектральной  плотности  тока  $j_z(\rv  r,\omega)$
(комплексной амплитуды) получаем выражение
     $$ j_z(\rv r,\omega)=\frac Q{\pi^2 a^2}\,e^{i\frac \omega v z}
         \sum\limits_\lambda
         \frac {J_m(\nu_\lambda r/a)\,J_m(\nu_\lambda b/a)}
         {\epsilon_m J_{m-1}^2(\nu_\lambda)}\cos{m\varphi}\,,\eqno(\mbox{\rm Д2}.10)$$
где $\lambda$  совокупность  двух индексов $m$ и $n$,  $\sum\nolimits_
{\lambda}=\sum\nolimits_{m=0}^  \infty\sum  \nolimits_{n=1}^\infty\,,\epsilon_m=1+\delta_{0m} $,
$\delta_  {0m}$  ---  символ  Кронеккера,  а  $\nu_\lambda$  --- корни
функции Бесселя ($J_m(\nu_\lambda)=0$).

     Поля собственных  электрических  волн  (а  только  эти  волны   и
возбуждаются  током  с  единственной  компонентой  $j_z$) определяются
вектором  Герца  с  одной  компонентой  $\Pi_{\pm   \lambda,z}^{(e)}$,
которая при используемой нами нормировке (8.22) записывается в виде
     $$\Pi_{\pm\lambda,z}^{(e)}=\sqrt{\frac 2 {\pi \epsilon_m}}\frac{J_m(\nu_
         \lambda r/a)}{\nu_\lambda J_{m-1}(\nu_\lambda)}\,
        e^{\pm ih_\lambda z} \,\cos{m\varphi}\,,\eqno(\mbox{\rm Д2}.11)$$
где  $h_\lambda= \sqrt{\omega^2  /c^2-\nu_\lambda^2/a^2}$.  Тогда  в
соответствии с формулами (3.25) и (3.26) получаем следующие  выражения
для  отличных  от  нуля компонент поля собственных волноводных волн:
     $$ \left.\begin{array}{lcl}E_{\pm \lambda,z}&=&\displaystyle
        \sqrt{\frac 2 {\pi \epsilon_m}} {\frac {\nu_\lambda J_m(\nu_
         \lambda r/a)}{a^2 J_{m-1}(\nu_\lambda)}}\,
        e^{\pm ih_\lambda z}\,\cos{m\varphi}\,,\\[.5cm] E_{\pm
         \lambda,r}&=&\pm\, i\,\displaystyle{\sqrt{\frac 2 {\pi \epsilon_m}}\,
         \frac {h_\lambda J'_m(\nu_\lambda r/a )}
          {a J_{m-1}(\nu_\lambda)}}\;e^{\pm ih_\lambda z}\,\cos{m\varphi}\,,
         \\[.5cm]E_{\pm \lambda,\varphi}&=&\mp\, i\,\displaystyle{
         \sqrt{\frac 2 {\pi \epsilon_m}}\,\frac{ h_\lambda m
        J_m(\nu_\lambda r/a)}{r\nu_\lambda J_{m-1}(\nu_\lambda)}}\,
        e^{\pm ih_\lambda z} \,\sin{m\varphi}\,,  \\[.5cm]
         H_{\pm \lambda,r}&=&i\,\displaystyle{\sqrt{\frac 2 {\pi \epsilon_m}}
         \,\frac {k m J_m(\nu_\lambda  r/a)}
         { r\nu_\lambda J_{m-1}(\nu_\lambda)}}\,
         e^{\pm ih_\lambda z}\,\sin{m\varphi}\,,\\[.5cm]
         H_{\pm \lambda,\varphi}&=&i\,\displaystyle{\sqrt{\frac 2 {\pi \epsilon_m}}
         \,\frac {k J'_m(\nu_\lambda r/a)}
         { a J_{m-1}(\nu_\lambda)}}\,e^{\pm ih_\lambda z}\,
         \cos{m\varphi}\,.\\[.5cm]\end{array}\right\}\eqno(\mbox{\rm Д2}.12)$$

     Комплексную амплитуду продольного поля в волноводе ищем  согласно
формуле (20.18) в виде
     $$ E_z(\rv r,\omega)=\sum\limits_\lambda (A_\lambda E_{\lambda,
         z}+ A_{-\lambda}E_{-\lambda,z})-i\frac{4\pi} \omega\,
         j_z(\rv r,\omega)\,,\eqno(\mbox{\rm Д2}.13)$$
причём коэффициенты $A_{\pm \lambda}$, определяемые формулами (20.4)
и (20.5), легко вычисляются:
     $$ A_{\pm \lambda,z}=-\frac{2\pi}{\omega h_\lambda}\int
         \limits_V j_z E_{\mp \lambda,z}\,dV=\pm\, i\,\sqrt{\frac 2 {\pi \epsilon_m}}
         \frac{Q v \nu_\lambda J_m(\nu_\lambda b/a)}{\omega a^2 h_\lambda J_
         {m-1}(\nu_\lambda)(\omega  \mp\, vh_\lambda)}\, e^{i(\frac
         \omega v\mp h_\lambda)z}\,.\eqno(\mbox{\rm Д2}.14)$$
Подставляя теперь   (\mbox{\rm Д2}.10),   (\mbox{\rm Д2}.12)   и
(\mbox{\rm Д2}.14)  в (\mbox{\rm Д2}.13),  после  несложных
выкладок находим выражение   для продольной составляющей
электрического поля:
     $$ E_z(\rv r,\omega)=-i\,\frac{4Q\omega}{\pi}\,e^{i\frac \omega v z}
        \sum\limits_\lambda
       \frac {J_m(\nu_\lambda b/a) J_m(\nu_\lambda r/a)}
        {\epsilon_m  J^2_{m-1}(\nu_\lambda)[(\omega a)^2+
       (\nu_\lambda v\gamma)^2]}\,\cos{m\varphi}.
         \eqno(\mbox{\rm Д2}.15)$$

     Компонента поля как функция времени находится путём  вычисления
интеграла
     $$\gv E_z(\rv r,t)=\int\limits_{-\infty}^\infty E_z(\rv r,
         \omega)\,e^{-i\omega t}\,d\omega\eqno(\mbox{\rm Д2}.16)$$
с помощью теории вычетов:
     $$\gv E_z(\rv r,t)=\sign(z-vt)\frac{4Q}{a^2}\sum
         \limits_\lambda
         \frac {J_m(\nu_\lambda b/a) J_m(\nu_\lambda r/a)}
         {\epsilon_m J^2_{m-1}(\nu_\lambda)}\,e^{-\displaystyle{\frac{\nu_
         \lambda\gamma |z-vt|}a}}\cos{m\varphi}\,.\eqno(\mbox{\rm Д2}.17)$$
Аналогично вычисляются и остальные компоненты поля:
     $$\gv E_r(\rv r,t)=-\frac{4Q\gamma}{a^2}\sum
         \limits_\lambda \frac {J_m(\nu_\lambda b/a) J'_m(\nu_\lambda r/a)}
         {\epsilon_m J^2_{m-1}(\nu_\lambda)}\,e^{-\displaystyle{\frac{\nu_
         \lambda\gamma |z-vt|}a}}\cos{m\varphi}\,,\eqno(\mbox{\rm Д2}.18)$$
     $$\gv E_\varphi(\rv r,t)=\frac{4Q\gamma}{ar}\sum
         \limits_\lambda \frac {J_m(\nu_\lambda b/a) J_m(\nu_\lambda r/a)}
         {\nu_\lambda J^2_{m-1}(\nu_\lambda)}\,e^{-\displaystyle {\frac{\nu_
         \lambda\gamma |z-vt|}a}}m\sin{m\varphi}\,,\eqno(\mbox{\rm Д2}.19)$$
     $$\gv H_r(\rv r,t)=-\frac{4Q\beta\gamma}{ar}\sum
         \limits_\lambda \,\frac {J_m(\nu_\lambda b/a)
         J_m(\nu_\lambda r/a)}{\nu_\lambda J^2_{m-1}(\nu_\lambda)}
         \,e^{-\displaystyle{\frac{\nu_\lambda\gamma
         |z-vt|}a}}m\sin{m\varphi}\,,\eqno(\mbox{\rm Д2}.20)$$
     $$\gv H_\varphi(\rv r,t)=-\frac{4Q\beta\gamma}{a^2}\sum
         \limits_\lambda \frac {J_m(\nu_\lambda b/a) J'_m(\nu_\lambda r/a)}
         {\epsilon_m J^2_{m-1}(\nu_\lambda)}\,e^{-\displaystyle{\frac{\nu_
         \lambda\gamma |z-vt|}a}}\cos{m\varphi}\,.\eqno(\mbox{\rm Д2}.21)$$

     Полученные выражения для полей представляют собой экспоненциально
быстро сходящиеся ряды во всей области внутри волновода за
исключением ближайшей окрестности точки $z=vt$,  $r=b$, где
сходимость существенно более медленная, а в самой этой точке ряды
расходятся, что находится в полном  соответствии  с  формулами
(\mbox{\rm Д2}.6),  если   в   последних перенести  ось
цилиндрической системы координат.  Для нас важно,  что поле на
расстояниях порядка $a/\gamma$ от заряда заметно меньше,  чем в
свободном  пространстве.  Физически  этот  результат очевиден:
поле в волноводе представляет собой совокупность поля заряда в
пустоте и поля поверхностных зарядов противоположного знака,
наведённых на стенках волновода.  Суммарный наведённый заряд
равен  $Q$,  он  локализован вблизи    плоскости   $z=vt$,
причём   распределён   симметрично относительно неё.

     В результате  кильватерный  потенциал  существенно  меньше своего
значения для свободного пространства, и при этом продольный потенциал,
как и в свободном пространстве,  является нечётной, а поперечный ---
чётной функцией $s$.  Приведём выражения для всех трёх компонент
потенциала (на единицу длины):
     $$\hspace*{0.8cm}\gv w_{\parallel}(s,b,r,\varphi)=\frac 4{a^2}
         \sign(s)\sum \limits_\lambda
         \frac {J_m(\nu_\lambda b/a) J_m(\nu_\lambda r/a)}
         {\epsilon_m J^2_{m-1}(\nu_\lambda)}\,e^{-\displaystyle{\frac{\nu_
         \lambda\gamma |s|}a}}\cos{m\varphi}\,,\eqno(\mbox{\rm Д2}.22)$$
     $$\gv w_r(s,b,r,\varphi)=-\frac 4{\gamma a^2}\sum
         \limits_\lambda \frac {J_m(\nu_\lambda b/a) J'_m(\nu_\lambda r/a)}
         {\epsilon_m J^2_{m-1}(\nu_\lambda)}\,e^{-\displaystyle{\frac{\nu_
         \lambda\gamma |s|}a}}\cos{m\varphi}\,,\eqno(\mbox{\rm Д2}.23)$$
     $$\gv w_\varphi(s,b,r,\varphi)=\frac 4{\gamma ar}\sum
         \limits_\lambda  \frac {J_m(\nu_\lambda b/a) J_m(\nu_\lambda r/a)}
         {\nu_\lambda J^2_{m-1}(\nu_\lambda)}\,e^{-\displaystyle
         {\frac{\nu_\lambda\gamma |s|}a}}m\sin{m\varphi}\,.\eqno(\mbox{\rm Д2}.24)$$

     В случае движения обоих зарядов  вдоль  оси  волновода  $(r=b=0)$
поперечный  кильватерный потенциал обращается в нуль,  а выражение для
продольного сильно упрощается:
     $$ \gv w_{\parallel}(s)=\frac 2{a^2}\sign(s)\sum\limits_{n=1}
         ^\infty \frac 1{J_1^2(\nu_{0n})}\,e^{-\displaystyle
         {\frac{\nu_{0n}\gamma|s|}a}}\,.\eqno(\mbox{\rm Д2}.25)$$
При малых значениях параметра  $\alpha=s\gamma/a$  сумма
стремится  к $\alpha^{-2}/2$,  что согласуется с формулой
(\mbox{\rm Д2}.7),  поскольку в этом случае стенки трубы не могут
оказать заметного  влияния  на  поле заряда.

     Антисимметрия продольного кильватерного  потенциала  как  функции
$s$,  имеющая место при любых значениях $\gamma$ в однородной камере с
идеально   проводящими   стенками,   нарушается   при   их    конечной
проводимости.  Убедимся в этом, решив задачу о поле движущегося заряда
в  трубе  кругового  сечения,  проводимость  стенок  $\sigma$  которой
достаточно  велика,  прямым  способом,  не  прибегая  к  разложению по
собственным волнам  волновода  (система  таких  волн  в  волноводе  со
стенками конечной проводимости слишком сложна, а вопрос о её полноте
недостаточно ясен).  Первоначально продемонстрируем этот метод на  уже
решённой задаче для волновода с идеальной проводимостью стенок.

     В данном  случае,  как  и  в  любом другом при наличии аксиальной
симметрии  структуры,  целесообразно  разложить  все  токи,   поля   и
потенциалы  по  мультиполям,  то  есть  в  ряд  по  азимутальному углу
$\varphi$   (существенным   моментом   здесь    является    отсутствие
перемешивания  гармоник  из-за  граничных условий на боковых стенках).
Представим  плотность  тока  точечного  заряда,  траектория   которого
отклонена  от  оси  структуры на расстояние $b$ вдоль оси $x$,  в виде
суммы мультиполей
     $$ j_z(\rv r,\omega)=\sum\limits_{m=0}^\infty j_z^{(m)}(\rv r,
         \omega)\,,\qquad j_z^{(m)}= \frac Q {2\pi^2\epsilon_m}\,\frac {\delta
         (r-b)} r\, e^{i\frac \omega v z}\,\cos{m\varphi}\,.
         \eqno(\mbox{\rm Д2}.26)$$

     Поле этого   тока   однозначно   определяется  вектором  Герца  с
единственной   отличной   от   нуля   компонентой   $\Pi_z$;
каждый мультипольный  момент  удобно  искать  в  виде  суммы
соответствующего момента  в   свободном   пространстве
$\Pi^{(m)}_{z0}$   и   решения однородного  волнового  уравнения
$\Pi^{(m)}_{z1}$,  с зависимостью от $z$ в виде множителя
$\exp{(iz\omega/v)}$.  Первое слагаемое находится путем
разложения  поля  (\mbox{\rm Д2}.6)  в  интеграл Фурье по времени,
и заряду, движущемуся по оси, соответствует компонента
     $$  \Pi_{z}(\rv r,\omega)=i\frac Q{\pi\omega}\,
         K_0(\Gamma r)\,e^{i\frac \omega v z}\,,\eqno(\mbox{\rm Д2}.27)$$
где $\Gamma=\omega/(v\gamma)$.    Используя   теорему   сложения   для
бесселевых функций,  находим $m-$ый мультипольный момент вектора Герца
для заряда, движущегося с отклонением от оси:
     $$ \Pi^{(m)}_{z0}(\rv r,\omega)=i\,\frac Q{\pi\omega\epsilon_m}\left [
         \begin{array}{l} K_m(\Gamma r)I_m(\Gamma b)\\[.3cm]
         I_m(\Gamma r) K_m(\Gamma b)\end{array}\right]
         \,e^{i\frac \omega v z}
         \cos{m\varphi}\,,\eqno(\mbox{\rm Д2}.28)$$
где здесь  и  в  ряде  последующих  формул  верхняя  строчка столбца в
квадратных скобках соответствует $r>b$, а нижняя --- $r<b$.

     Решение однородного волнового уравнения ищем в виде
     $$\Pi^{(m)}_{z1}(\rv r,\omega)=\frac{f(r)}{\epsilon_m}\, e^{i\frac \omega v z}\,
        \cos{m\varphi}\,,\eqno(\mbox{\rm Д2}.29)$$
получая в результате для $f(r)$ уравнение
     $$\frac 1 r\,\frac d{dr}(r\frac {df}{dr})- (\frac {m^2}{r^2}+
         \Gamma^2) f=0\,.\eqno(\mbox{\rm Д2}.30)$$
Выбирая решение  этого  уравнения,  не  имеющее особенности при $r=0$,
находим  для  мультипольного  момента  вектора  Герца  полного   поля
следующее выражение:
     $$\Pi^{(m)}_z=\Pi^{(m)}_{z0}+\Pi^{(m)}_{z1}=\frac {iQ}{\pi \omega \epsilon_m}
         \left\{\left [\begin{array}{l} K_m(\Gamma r)I_m(\Gamma b)
         \\[.3cm]I_m(\Gamma r) K_m(\Gamma b)\end{array}\right ]
         +C_m I_m(\Gamma r)\right\}\,e^{i\frac \omega v z}\,
       \cos{m\varphi}\,.\eqno(\mbox{\rm Д2}.31)$$
Выразив  с помощью формулы (3.26)  продольную компоненту поля
     $$E^{(m)}_z(\rv r,\omega)=-\Gamma^2\Pi^{(m)}_z(\rv r,\omega)
         \,,\eqno(\mbox{\rm Д2}.32)$$
определим неизвестный коэффициент $C_m$ из  условия  $E^{(m)}_z=0$  на
стенке трубы (при $r=a$); в конечном итоге получаем, что
     $$ E^{(m)}_z=-\frac{i Q\omega}{\pi v^2\gamma^2 \epsilon_m}\left\{
         \left[\begin{array}{l} K_m(\Gamma r)I_m(\Gamma b)\\[.3cm]
         I_m(\Gamma r) K_m(\Gamma b)\end{array}\right]
       \!-I_m(\Gamma r) \frac{I_m(\Gamma b) K_m(\Gamma a)}{I_m
         (\Gamma a)}\right\}e^{i\frac \omega v z}
         \cos{m\varphi}.\eqno(\mbox{\rm Д2}.33)$$
Сравнив это   выражение   с   формулой   (\mbox{\rm Д2}.15),
приходим    к математическому   соотношению,   которое   широко
используется   при вычислении  полей  в  структурах,   обладающих
симметрией   цилиндра кругового сечения:
     $$ 2\sum\limits_{n=1}^\infty\frac {J_m(\nu_{mn}\displaystyle{
         \frac b a})J_m(\nu_{mn}\displaystyle{\frac r a})}
         {J_{m-1}^2(\nu_{mn})[\Gamma^2a^2+\nu_{mn}^2]}=
         \left[\begin{array}{l}K_m(\Gamma r)I_m(\Gamma b)\\[.2cm]
         I_m(\Gamma r) K_m(\Gamma b)\end{array}\right]-
         I_m(\Gamma r)\,\frac{I_m(\Gamma b) K_m(\Gamma a)}
         {I_m(\Gamma a)}\,.\eqno(\mbox{\rm Д2}.34)$$

     При конечной  проводимости  стенок  камеры  $\sigma$  решение для
$\Pi^{(m)}_z$ внутри неё имеет тот же вид (\mbox{\rm Д2}.31),
что  и  при идеальной  проводимости,  однако коэффициент $C_m$
определяется теперь условием непрерывности тангенциальных
компонент  поля  на  стенке.  В случае  большой  проводимости
материала стенок можно не рассматривать структуру поля внутри
металла,  а воспользоваться граничными условиями Щукина-Леонтовича
     $$ E^{(m)}_z(a,\omega)=-W H^{(m)}_{\varphi}(a,\omega)
         \,,\eqno(\mbox{\rm Д2}.35)$$
сводящими задачу  к  нахождению  поля  только  внутри  камеры
($W=(1-i)\sqrt{\omega/(8\pi\sigma)}$    ---    волновое
сопротивление материала стенок,  $\mu=1$).  Помимо малости  $|W|$
для  применимости условия   (\mbox{\rm Д2}.35)   необходимо
ещё,  чтобы  глубина  скин-слоя
$\delta=c/\sqrt{2\pi\sigma\omega}$  была  существенно  меньше
радиуса кривизны поверхности стенки $a$ и толщины $d$ стенки
трубы. Если имеет место релятивисткое движение заряда,  то
основной вклад в спектральное разложение  поля  дают высокие
частоты,  и эти условия выполняются для всех реальных конструкций
камер. Тогда в результате простых вычислений получаем:
     $$C_m=-i\frac Q{\pi\omega}\,I_m(\Gamma b)\,\frac {K_m(\Gamma a)+
         iW\beta\gamma K'_m(\Gamma a)} {I_m(\Gamma a)-iW\beta\gamma
         I'_m(\Gamma a)}\,.\eqno(\mbox{\rm Д2}.36)$$

     Ограничимся рассмотрением    значений   $\gamma$,   при   которых
$|W|\gamma\beta\ll  1$.  Поскольку  максимальное   значение   $\omega$
определяется  из соотношения $\Gamma a$ порядка нескольких единиц,  то
для $\gamma$ должно выполняться условие
     $$ \gamma^3\ll \frac{2\pi \sigma a}c\,,\eqno(\mbox{\rm Д2}.37)$$
что для медных стенок  ($\sigma=5\cdot  10^{17}  c^{-1}$)  сводится  к
требованию  $\gamma\ll  5\cdot10^2  a^{1/3}$,  где  $a$ выражено в см.

     Учитывая далее известное соотношение для функций Бесселя
     $$ K_m(\Gamma a)I'_m(\Gamma a) -  K'_m(\Gamma a)
         I_m(\Gamma a)=\frac 1 {\Gamma a}\,,\eqno(\mbox{\rm Д2}.38)$$
можно преобразовать соотношение (\mbox{\rm Д2}.36) к следующему
виду:
     $$C_m=-\frac{iQ}{\pi\omega}\left[\frac {I_m(\Gamma b)K_m(
         \Gamma a)}{I_m(\Gamma a)}+i\frac{W c \beta^2\gamma^2 I_m(
         \Gamma b)}{\omega a I^2_m(\Gamma a)}\right] \,.
         \eqno(\mbox{\rm Д2}.39)$$
Первое слагаемое  в  этом  выражении определяет уже вычисленное поле в
идеально проводящей трубе,  а второе --- даёт добавки к кильватерным
потенциалам,  обусловленные  конечной  проводимостью.  После несложных
преобразований получаем для этих  добавок  (отмеченных  дополнительным
нижним индексом $\sigma$) следующие выражения:
     $$\gv w^{(m)}_{\sigma\parallel}(s,b,r,\varphi)\!=\!
         \frac{(\beta \gamma)^{3/2}}{\pi a^{5/2}\epsilon_m}\sqrt{\frac {2c}
         {\pi \sigma}}\cos{m\varphi}
         \int\limits_0^\infty \frac{\sqrt{x}I_m(x b/a)
         I_m(x r/a)}{I_m^2(x)}(\cos{\alpha x}-\sin{\alpha x})\,
         dx,\eqno(\mbox{\rm Д2}.40)$$
     $$\gv w^{(m)}_{\sigma r}(s,b,r,\varphi)=
         \frac{\beta^{3/2}}{\pi a^{5/2}\epsilon_m}\sqrt{\frac {2c\gamma}
         {\pi \sigma}}\cos{m\varphi}
         \int\limits_0^\infty \frac{\sqrt{x}I_m(x b/a)
         I'_m(x r/a)}{I_m^2(x)}(\cos{\alpha x}+\sin{\alpha x})\,
         dx\,,\eqno(\mbox{\rm Д2}.41)$$
     $$\gv w^{(m)}_{\sigma\varphi}(s,b,r,\varphi)=
         \frac{\beta^{3/2}}{\pi  r a^{3/2}}\sqrt{\frac {2c\gamma}
         {\pi \sigma}}\,m\sin{m\varphi}\,
         \int\limits_0^\infty \frac{I_m(x b/a)
         I_m(x r/a)}{\sqrt{x}I_m^2(x)}(\cos{\alpha x}+
         \sin{\alpha x})\, dx\,,\eqno(\mbox{\rm Д2}.42)$$
где, как и  ранее,  $\alpha=s\gamma/a$,  интегрирование  ведётся  по
переменной  $x=\Gamma  a$,  а  при  выводе  этих  формул  использовано
соотношение $\rv E(\rv r,-\omega)=\rv E^*(\rv  r,\omega)$,  являющееся
прямым следствием того, что функция $\mbox{\boldmath$\gv E$}(\rv r,t)$
--- действительная.

     Для расчёта  ускорителей  основной  интерес  представляют малые
отклонения обоих зарядов от оси  ($b/a\ll  1,r/a\ll  1$),  так  что  в
разложениях  стоящих  под  интегралами  функций $I_m(x b/a)$ и~ $I_m(x
r/a)$ в ряд могут быть оставлены только первые члены. Основной вклад в
продольный   кильватерный   потенциал   даёт  член,  соответствующий
монополю:
     $$ \gv w_{\parallel}^{\sigma}(s)=-\frac 1 Q E_z^{\sigma}(s)
         =\frac {(\beta\gamma)^{3/2}}
         {\pi a^{5/2}}\sqrt{\frac c{2\pi \sigma}}\,F(\alpha)\,,
         \eqno(\mbox{\rm Д2}.43)$$
где
     $$ F(\alpha)=\int\limits_0^\infty \frac{\sqrt{x}}{I_0^2(x)}\,
         (\cos{\alpha x}-\sin{\alpha x})\,dx\,.\eqno(\mbox{\rm Д2}.44)$$

Вычислить этот   интеграл   в  аналитическом  виде  не
представляется возможным,  но в результате численного
интегрирования  на  компьютере легко   построить   график
функции   $F(\alpha)$,  приведённый  на рис.~{\mbox{\rm Д2}.2}.
Отметим,  что функция не является ни чётной,  ни нечётной,
обращается  в  нуль  при  $\alpha=0,65$ и обладает разной
асимптотикой при $\alpha\to\infty$ (поле за зарядом) и при
$\alpha\to -\,\infty$  (поле  перед зарядом).  Поле
$E_z^{\sigma}(s)$ перед зарядом (отрицательные $s$) убывает
экспоненциально с расстоянием от заряда  с тем же показателем
экспоненты, что и поле заряда в идеально проводящей камере, и
составляет малую добавку к нему.

     Для больших положительных значений $\alpha$ функция $F(\alpha)\to
-  \sqrt{\pi/2}\,\alpha^{-3/2}$,   и,   следовательно,   обусловленное
конечной   проводимостью   поле   $E_z^{\sigma}(s)$  при  больших  $s$
асимптотически убывает $\sim s^{-3/2}$ и не зависит от  $\gamma$.  Это
связано с тем, что шлейф поля, тянущийся за зарядом, вызван процессами
диффузии  и  затухания  поля  в  проводящей  среде,  скорость  которых
определяется проводимостью $\sigma$ и не зависит от энергии заряда.  В
конечном итоге в случае достаточно больших $s$ получаем, что
     $$\gv w_{\parallel}^\sigma (s)=-\frac 1{2\pi a s^{3/2}}\sqrt
         {\frac c{\sigma}}\,.\eqno(\mbox{\rm Д2}.45)$$
\vspace{-.2cm}
\begin{wrapfigure}[12]{l}{7.0cm}
\begin{picture}(80,45)
\put(0,45){\special{em:graph FigA2-2.bmp}}
\end{picture}
\hbox to 7.5cm{\hfil\footnotesize{Рис.~Д2.2.~Функция $F(\alpha)$.
}\hfil}
\end{wrapfigure}

Аналогичным способом   нетрудно  убедиться, что основной  вклад
в  поперечный  кильватерный потенциал  вблизи оси  дают   дипольные
составляющие    $(m=1)$.  Направление потенциала (напомним,  что он
вектор!) совпадает с направлением отклонения  первичного  заряда  от
оси  трубы (в нашем случае  по  оси  $x$). Оказывается,  что  для  достаточно
больших   $s$ потенциал однороден  вблизи оси  и не  зависит от $\gamma$:
     $$ \gv w^{\sigma}_x(s)=\frac b{\pi a^3}\sqrt{\frac c{\sigma s}}
         \,.\eqno(\mbox{\rm Д2}.46)$$

     Отрицательный знак продольного кильватерного потенциала говорит о
том,  что наведённое первичным зарядом поле при  этих  $s$  ускоряет
пробный  заряд.  Вблизи  же  заряда  $Q$ поле,  обусловленное конечной
проводимостью,  как видно из рис.~Д2.2,  является тормозящим. Хотя при
$s\to 0$ оно становится бесконечно малым по сравнению с полем заряда в
идеально проводящей камере,  тем не менее,  именно оно (и только  оно)
определяет  торможение  первичного  заряда,  приводящее  к потерям его
энергии,  идущим на омический  нагрев  стенок.  Мощность  этих  потерь
нетрудно     подсчитать     как     произведение    тормозящей    силы
$QE^{\sigma}_z(0)$ на скорость заряда $v\approx c$:
     $$ P=\frac{Q^2 \gamma^{3/2} c}{\pi a^2}\sqrt{\frac c
         {2\pi\sigma a}} \,F(0)\,.\eqno(\mbox{\rm Д2}.47)$$
(Рекомендуем читателям  самостоятельно убедиться,  что эта же
величина получается при вычислении потока энергии  через  любую
цилиндрическую поверхность,  охватывающую траекторию заряда;
удобнее всего вычислять поток через поверхность $r=a$,  где
тангенциальные составляющие  полей связаны соотношением (\mbox{\rm
Д2}.35)).

     Из сказанного выше следует, что продольный кильватерный потенциал
в  трубе  с конечной проводимостью при некотором положительном $s=s_0$
проходит через  нуль.  Типичное  значение  $s_0$  для  трубы радиуса
$a=5\,\mbox{см}$ с медными стенками при $\gamma=100$ соответствует $\alpha=2,5$ и
составляет  $0,13\,\mbox{см}$;  оно  увеличивается  с  ростом  $a$   и
уменьшается с $\gamma$.

     Отрицательное значение  кильватерного   потенциала   при   весьма
умеренных  значениях  $s$  позволяет в принципе говорить о возможности
{\it  кильватерного  ускорения}   в   гладкой   трубе   при   конечной
проводимости   стенок,   но   практически   этот   эффект   не  играет
сколько-нибудь значительной роли.  Заметим, что отрицательный максимум
кривой  $F(\alpha)$  на рис.~Д2.2 меньше по модулю её положительного
значения в нуле приблизительно в два раза --- а  именно  это  значение
$F(0)\approx   1,2$   определяет  величину  торможения  заряда  $Q$  и
соответствующие потери его  энергии.  Поэтому  наведённое  первичным
зарядом  ускоряющее  поле,  действующее  на  пробный  заряд,  даже при
оптимальном выборе $s$ меньше тормозящего первичный заряд.  На пробный
заряд  также  действует и поле,  наводимое им самим,  которое является
тормозящим. Если величина пробного заряда $q$ совпадает с $Q$, то даже
при оптимальном отставании пробного заряда полная сила, действующая на
него, остаётся тормозящей, хотя и приблизительно в два раза меньшей,
чем  на  заряд  $Q$.  При  $q\approx  Q/2$ пробный заряд не испытывает
действия возбуждаемого  первичным  зарядом  поля,  а  ускоряющая  сила
появляется  только  в том случае,  когда $q<Q/2$.  Очевидно,  что доля
энергии,  передаваемой  зарядом  $Q$  пробному  заряду  при   реальных
значениях   $a,\,\gamma$  и  $\sigma$,  составляет  лишь  малую  часть
омических потерь в стенках.

     Рассмотренными задачами  практически  исчерпываются  все  случаи,
когда $\delta$-- образный кильватерный потенциал может быть
выражен  в замкнутом аналитическом виде.  Вычисление наведённых
потенциалов для любой  локальной  нерегулярности  в  круглом
волноводе   требует   на завершающем  этапе  численных  расчётов
на  компьютере.  Однако для некоторых,  так называемых ступенчатых
нерегулярностей,  уже на  этапе предварительного    аналитического
анализа    можно   сделать   ряд определённых  выводов.   С
помощью   ступенчатых   нерегулярностей удаётся  с  хорошей
точностью  моделировать  большое число реальных элементов
ускорительного тракта. Рассмотрим простейшие нерегулярности,
изображённые схематически на рис.~\mbox{\rm Д2}.3:  а) уступ, б)
резонатор и в) диафрагма --- все в камере кругового сечения.

     Эти нерегулярные  структуры  естественным образом с помощью одной
или двух плоскостей,  нормальных к оси,  могут быть разбиты на две или
три  частичные  области,  каждая из которых представляет собой отрезок
регулярного круглого волновода.  Поле,  возбуждаемое движущимся по оси
первичным зарядом $Q$ с плотностью тока
     $$j_z(\rv r,\omega)=\frac Q{4\pi^2}\,\frac{\delta(r)}r
         \,e^{i\frac \omega v z}\,, \eqno(\mbox{\rm Д2}.48)$$
в области с номером $p$ (нумерация указана на рис.~{Д2}.3) ищем в виде
суммы  поля заряда в бесконечном регулярном волноводе соответствующего
радиуса $a_p$
     $$ E_{zp}^0(\rv r,\omega)=-i\frac{Q\omega}{\pi v^2\gamma^2}\left[
         K_0(\Gamma r)-I_0(\Gamma r)\frac{K_0(\Gamma a_p)}
         {I_0(\Gamma a_p)}\right] \,e^{i\frac \omega v z}
         \eqno(\mbox{\rm Д2}.49)$$
и совокупности $E$-волн:
     $$ E_{zp}(\rv r,\omega)= E_{zp}^0-\frac Q{a^2_p}\sum\limits_
         {n=1}^\infty \frac {\nu_{n}}{J_1(\nu_{n})}\,J_0(\nu_{n}
         r/a_p)\bigl [A_{np}^+\,e^{ih_{np}z}+A_{np}^-\,e^{-ih_{np}z}
         \bigr ]\,.\eqno(\mbox{\rm Д2}.50)$$
Здесь $h_{np}=\sqrt{k^2 -(\nu_n/a_p)^2}$,  а  нормировочные  множители
перед  амплитудами  $A_{np}^{\pm}$  выбраны  для упрощения последующих
формул,  прич\"ем везде далее  у  корней  $\nu_{0n}$  функции  Бесселя
$J_0(x)$ опущен первый индекс, равный нулю. Остальные отличные от нуля
компоненты поля $E_{r p}$ и $H_{\varphi  p}$  однозначно  определяются
выбором записи $E_{zp}$.

\begin{wrapfigure}[16]{l}{7.5cm}
\begin{picture}(80,60)
\put(0,60){\special{em:graph FigA2-3.bmp}}
\end{picture}
\hbox to 7.5cm{\hfil\footnotesize{Рис.~Д2.3.~Ступенчатые нергулярности
}\hfil}
\hbox to 7.5cm{\hfil\footnotesize{круглого волновода.}\hfil}
\end{wrapfigure}

     Одноступенчатая нерегулярность    представляет    собой   соосное
сочленение двух регулярных полубесконечных волноводов разных  радиусов
с  плоским  фланцем.  В  обоих  волноводах коэффициенты $A_{np}^{\pm}$
отличны от нуля только для волн, распространяющихся {\it от} плоскости
скачка и создающих поток мощности на $\pm \infty$;  поэтому $A_{n1}^+$
и $A_{n2}^-$ равны нулю. Уравнения для остальных коэффициентов следуют
из условия непрерывности тангенциальных компонент в плоскости скачка и
обращения в нуль тангенциальной составляющей  электрического  поля  на
торцевой  поверхности $z=0,\; a_1<r<a_2$.

     В результате преобразований,  тождественных с изложенными выше  в
разделе 13 для периодического волновода, приходим к бесконечной системе
линейных алгебраических уравнений вида
     $$ A_{n2}^+ h_{n2}+\sum\limits_{m=1}^\infty M_{nm}
         A_{m2}^+=f_n,\quad n=1,2,\dots\,,\eqno(\mbox{\rm Д2}.51)$$
с симметричной матрицей
     $$ M_{nm}= \alpha_n\alpha_m\sum\limits_{s=1}^\infty
         \frac {h_{s1}}{(\nu^2_s-y_n^2)(\nu^2_s-y_m^2)}\,,
         \eqno(\mbox{\rm Д2}.52)$$
где $\alpha_n=2(a_1/a_2)y_n    J_0(y_n)/J_1(\nu_n),\;    y_n=(a_1/a_2)
\nu_n$, и правой частью
     $$ f_n=\frac{i\alpha_n}{\pi c\beta}\Bigl\{\frac {2\Gamma
         a_1^2}{\gamma} I_0(\Gamma a_1)
         \Phi\sum\limits_{s=1}^\infty \frac{h_{s1}}{(\nu_s^2-y_n^2)
         [(\Gamma a_1)^2+\nu_s^2]}+\frac {(a_2/a_1)^2}{I_0(\Gamma a_1)
         [(\Gamma a_2)^2+\nu_n^2]}\Bigr\}\,,\eqno(\mbox{\rm Д2}.53)$$
причём
     $$ \Phi=\left [\frac{K_0(\Gamma a_2)}{I_0(\Gamma a_2)}-
         \frac{K_0(\Gamma a_1)}{I_0(\Gamma a_1)}\right]\,.
         \eqno(\mbox{\rm Д2}.54)$$
Необходимо также отметить,  что коэффициенты $A_{n1}^-$  и  $A_{n2}^+$
связаны следующим соотношением:
     $$ A^-_{n1}=\sum\limits_{s=1}^\infty\frac{\alpha_s A^+_{s2}}
         {\nu_n^2-y_s^2}-\frac{ka_1^2}{с\beta^2\gamma^2}\,
         \frac{I_0(\Gamma a_1) \Phi}{[(\Gamma a_1)^2+\nu_n^2]}\,.
         \eqno(\mbox{\rm Д2}.55)$$
Поэтому продольный кильватерный  потенциал,  обусловленный
излучением при  пролёте  нерегулярности,  в соответствии с
формулой (\mbox{\rm Д2}.3) после интегрирования по $z$ принимает
вид
     $$ \gv W_{\parallel}(s)=-\int\limits_{-\infty}^\infty
         e^{i\omega s/v}\left [\sum\limits_{n=1}^\infty
         \frac{\nu_n}{J_1(\nu_n)}\Bigl(\frac
         {A^-_{n1}/a_1^2}{h_{1n}+\omega/v}+\frac{A^+_{2n}/a_2^2}
         {h_{n2}-\omega/v}\Bigr)\right ]\,d\omega\,.
         \eqno(\mbox{\rm Д2}.56)$$

     На первый  взгляд  от  приведённых  громоздких  выражений  мало
пользы до тех пор, пока не будут проведены численные расчёты, но
это не  совсем так.  Левая часть системы уравнений (\mbox{\rm
Д2}.51) не зависит от параметров первичного заряда,  и,  в
частности, остаётся такой же при  расчёте рассеяния
собственных волн волновода на нерегулярности. С другой стороны,
правая часть существенно упрощается при  $\gamma\to \infty$:
     $$  f_n=\frac{i\alpha_n}{\pi c \nu_n^2}\,\Bigl(\frac{a_2}
         {a_1}\Bigr)^2\,.\eqno(\mbox{\rm Д2}.57)$$
Cистема уравнений (\mbox{\rm Д2}.51) с  этими  $f_n$  определяет
предельную зависимость коэффициентов $A_{np}^{\pm}$ от частоты
$\omega$,  которые оказываются функциями с резкими  изломами
вблизи  критических  частот обоих   волноводов.   Интеграл  же  по
$\omega$  в  (\mbox{\rm Д2}.56)  при предельных значениях
коэффициентов расходится  на  верхнем  и  нижнем пределе,
практически линейно возрастая как функции верхнего и нижнего
предела интегрирования. При больших,  но конечных значениях
$\gamma$, верхняя   граница спектра  излучаемых  частот
определяется  функцией $1/I_0(\Gamma a_1)\approx
\exp{(-ka_1/\gamma)}$.   В    результате кильватерный  потенциал
$\gv W_{\parallel}$ является линейной функцией $\gamma$.

     Количественная зависимость $\gv W_{\parallel}$ от $s$ может  быть
получена  только  в  результате  численных  расчётов,  однако  общие
закономерности излучения заряда  при  пролёте  мимо  нерегулярностей
(так  называемого  {\it  диффракционного излучения}) позволяют сделать
определённые  качественные  суждения.  При  релятивистском  движении
заряда  его  излучение  в основном направлено по ходу движения,  и оно
представляет собой волновой  пакет,  движущийся  вслед  за  зарядом  с
групповой  скоростью,  которая  в  волноводе всегда меньше $c$;  из-за
дисперсии  пакет  быстро   расплывается.   Пробный   заряд   $q$   при
отрицательных  $s$  (впереди  заряда  $Q$) не испытывает действия поля
излучения,  и кильватерный потенциал близок к нулю.  Отстающий пробный
заряд  всегда  обгоняет  волновой  пакет  и  испытывает  на  себе  его
тормозящее действие.  Чем больше $s$, тем с более расплывшимся пакетом
взаимодействует  пробный  заряд  и  тем меньше кильватерный потенциал.
Таким образом,  $\gv W_{\parallel}(s)$ близок к нулю при отрицательных
$s$,  резко  возрастает вблизи $s=0$ и плавно спадает с $s$.  Отметим,
что при больших $\gamma$ кильватерный потенциал практически не зависит
от  направления движения заряда,  хотя при малых скоростях действующая
на оба заряда сила вблизи плоскости скачка  существенно  разная  из-за
поля наведённых на торцевой поверхности зарядов.

     Отмеченные закономерности  в  поведении  кильватерного потенциала
сохраняются   и   для   двух   других   ступенчатых
нерегулярностей, представленных  на  рис.~\mbox{\rm Д2}.3.  Однако
в случае резонатора имеют место некоторые существенные отличия.
Спектр в области высоких частот менее   плотный,   и   поэтому
кильватерный  потенциал  и  остальные характеристики излучения
(такие,  например,  как полные потери энергии первичным зарядом)
растут с $\gamma$ по закону,  более медленному, чем линейный.
Другое важное отличие состоит в  существовании,  начиная  с
отношения  радиусов  $a_2/a_1$  порядка  двух,  конечного  числа
(тем большего,  чем  больше   это   отношение)   незатухающих
собственных колебаний,  частоты  которых лежат ниже наименьшей
критической частоты волновода,  определяемой  для  $E$-волн
условием   $ka_1=\nu_1$.

     В результате  после  пролёта  первичного  заряда  в  резонаторе
остаётся  запасённое  электромагнитное поле в виде суммы конечного
числа  собственных   колебаний.   Амплитуда   этих   колебаний   имеет
определённое  предельное  значение в случае $\gamma\to \infty$;  при
больших значениях $s$ пробный заряд взаимодействует практически только
с этим полем, которое определяет кильватерный потенциал вида
     $$\gv W_{\parallel}(s)=\sum\limits_n W_n\cos{\omega_n s}\,,
         \eqno(\mbox{\rm Д2}.58)$$
где коэффициенты $W_n$ при больших $\gamma$ не зависят от $\gamma$,  а
суммирование производится по конечному  числу  собственных  колебаний.
Важно, что собственные частоты и соответствующие им $W_n$ могут быть с
достаточно высокой точностью  вычислены  по  формулам  для  замкнутого
резонатора.

    Возбуждение кильватерного поля в резонаторе  в  виде  совокупности
собственных   колебаний,  поле  которых  остаётся  локализованным  в
объёме   резонатора,   по-новому   ставит   вопрос   о   возможности
кильватерного  ускорения.  Проще  всего  для  рассмотрения  и наиболее
очевидной возможность такого ускорения  представляется  в  случае  так
называемого  одномодового  резонатора.  Такие  резонаторы  в  природе,
естественно,  не существуют,  но такая  идеализация  приемлема  в  том
случае,  когда возбуждением остальных мод можно пренебречь.  Очевидно,
что если пробный заряд следует на  достаточно  большом  расстоянии  от
первичного,   то   ко   времени  его  влёта  в  резонатор  колебание
рассматриваемой моды установилось, и --- в зависимости от фазы влёта
---  пробный  заряд  испытывает  торможение  или  ускорение.  Встаёт
очевидный вопрос --- чему  равно  максимально  возможное  ускорение  и
максимально  возможное  торможение пробного заряда?  До тех пор,  пока
рассматривается одномодовое приближение,  ответ на эти  вопросы  может
быть получен из общих физических соображений.

     Представляется очевидным,   что   максимальный   прирост  энергии
пробный заряд получает в том случае,  когда он поглощает  всю  энергию
возбуждённого  колебания,  равную  потере энергии первичным зарядом,
который в пассивном резонаторе будет только тормозиться. Такое событие
может произойти лишь при вполне определённом отставании $s$ пробного
заряда,  равном  нечётному   числу   полуволн   колебания,   и   при
определённой  величине пробного заряда,  зависящей от траектории его
следования.

     Если траектории  обоих зарядов совпадают,  то необходимо равенство
зарядов $q=Q$;  в случае $q<Q$ пробный заряд ускоряется,  хотя и менее
эффективно,   то  есть  он  использует  на  сво\"е  ускорение  не  всю
запасённую  энергию  собственного  колебания  резонатора.  При  этом
необходимо  отметить,  что  ускоряющее  поле для пробного заряда будет
максимальным при $q\to 0$,  и  оно  в  {\it  два}  раза  превысит  эту
величину для оптимального случая $q=Q$. Пробный заряд ускоряется и при
$Q<q<2Q$, но всё менее эффективно; если $q=2Q$, то заряд пролетает
через  резонатор  без изменения своей энергии (фаза поля изменяется на
$\pi$),  и,  наконец,  при большем значении $q$ пробный  заряд  всегда
тормозится.

     Тот факт,  что  для  оптимального  (в  смысле поглощения энергии)
пробного заряда действующее на него ускоряющее поле в {\it два }  раза
меньше,  чем при $q\to 0$, составляет содержание {\it основной теоремы
кильватерного ускорения}.  Надо сказать,  что в литературе зачастую  в
формулировке   теоремы  отсутствует  упоминание  о  соотношении  между
зарядами, без чего утверждение теряет свой смысл. Снижение эффективной
величины  ускоряющего  поля  при увеличении заряда ускоряемого сгустка
(тока  пучка)  в  теории  ускорителей  носит  название  {\it  нагрузки
пучком}.

     Максимальное значение  силы  торможения   для   пробного   заряда
достигается  при  $s$,  равном  целому числу длин волн колебания.
При равных зарядах $q=Q$ оно в  три  раза  превышает  силу
<<обственного>> торможения  пробного  заряда,  амплитуда
собственного колебания после вылета  пробного  заряда  из
резонатора  возрастает  в  два  раза,  а запасённая   энергия
---  в  четыре.  Для  $q>Q$  сила  торможения, естественно,
растёт с увеличением $q$,  но роль первичного заряда в этом
становится всё меньше и меньше.

     К теории  кильватерного  ускорения   тесно   примыкает   проблема
двухпучковых ускорителей или,  как их ещё называют,  трансформаторов
поля.  Если  амплитуда   собственного   колебания   вдоль   траектории
первичного  заряда в резонаторе меньше,  чем вдоль траектории пробного
(для простоты изложения будем считать траектории зарядов в  резонаторе
параллельными  и  одинаковой  длины),  то  при оптимальном соотношении
между зарядами $q$ и  $Q$  (разумеется,  и  при  соответствующем  $s$)
энергия,  набираемая единичным пробным зарядом,  может быть во столько
же раз больше,  чем теряемая единичным зарядом первичного,  во сколько
раз  различаются  амплитуды  (кильватерный  потенциал,  при вычислении
которого не учитывается нагрузка пучком, ещё в два раза больше).

     Для двух одиночных  сгустков  этот  эффект  в  значительной  мере
подавляется  из-за  возбуждения большого числа собственных колебаний и
излучения в непрерывном спектре вдоль подводящих волноводов.  Если  же
работа  такого  ускорителя происходит в непрерывном режиме,  оба пучка
сгруппированы в короткие сгустки с частотой рабочей моды резонатора  и
вторичный  (ускоряемый)  пучок  впускается  в  резонатор  с  некоторой
задержкой,  необходимой для раскачки амплитуды колебания,  то возможен
эффективный  набор  энергии  на единицу заряда.  Естественно,  что ток
ускоряемого пучка (заряд сгустка) должен быть во столько же раз меньше
тока первичного,  во сколько соотносятся теряемая и набираемая энергии
единичными зарядами двух пучков.

     Фактически основная теорема  кильватерного  ускорения  есть  иная
формулировка  результата,  хорошо  известного  из  теории  возбуждения
собственных  колебаний  замкнутого  резонатора  пролетающим   точечным
зарядом.  Один  из  возможных  способов  расчёта  этого эффекта (так
называемый {\it гамильтонов метод}, который приводится в Дополнении 3)
рассматривает   возбуждение   как   нарастание  амплитуды  собственных
колебаний в течение времени пролёта зарядом резонатора.  При влёте
все  амплитуды  равны нулю,  при вылете достигают своих установившихся
значений.  Интегральное действие на  заряд  тормозящего  поля  каждого
собственного  колебания  точно  в  {\it два} раза меньше,  чем было бы
действие установившегося колебания  (оно  фактически  и  действует  на
пробный заряд, если $s>L$).

     В заключение  остановимся  на  важном  общем  соотношении   между
продольным   и  поперечным  кильватерными  потенциалами,  известном  в
литературе как {\it  теорема  Пановского-Венцеля};  она  формулируется
следующим образом:
       $$\frac{\partial{\mbox{\boldmath$\gv W$}_{\perp}}(s,b,b_0,
          \varphi)}{\partial s}=\nabla_{\perp} \gv W_{\parallel}
          (s,b,b_0,\varphi)\,.\eqno(\mbox{\rm Д2}.59)$$
Необходимо сказать, что приведённая современная формулировка теоремы
носит существенно более общий характер,  чем  соотношение,  полученное
авторами  в исходной работе для замкнутого резонатора,  опубликованной
на  самом  начальном  этапе  исследования  кильватерных  полей   и   в
значительной степени вызвавшей последующий интерес к проблеме.

     Как нетрудно  убедиться  прямым  дифференцированием,  соотношение
$(\mbox{\rm Д2}.59)$  выполняется для всех рассмотренных выше
частных задач. Оно  справедливо  для  любых   структур,
обладающих   трансляционной симметрией  вдоль  траектории
зарядов,  и  для аксиально симметричных структур произвольной
формы.  В общем случае теорема верна при  $s>L$, где   $L$   ---
характерный   размер   области,  в  которой  имеется
неоднородность  камеры.  Важно,  чтобы   первичный   заряд
находился достаточно  далеко от места локализации кильватерного
поля --- чтобы в нём  можно  было  пренебречь  вкладом  той
части   поля,   которая обусловлена скалярным потенциалом.

%\end{document}

\newpage
\oddsidemargin=-0.4mm \evensidemargin=-0.4mm
\topmargin=-0.4mm
\headsep=7mm
\textheight=231.875mm
\textwidth=160mm
\mathsurround=2.5pt
\unitlength=1mm
%\begin{document}
%\input{macr.tex}
\thispagestyle{empty}
%\addtocounter{page}{368}
\baselineskip=\normalbaselineskip
%\baselineskip=1.085\normalbaselineskip

\begin{center} \subsubsection*{\bf     Д3. Импедансы связи в теории
ускорителей}
\end{center}
\vspace*{0.5cm}

\markboth{Дополнения}{Д3. Импедансы связи в теории ускорителей}

\begin{center}\begin{minipage}[c]{0.75\textwidth}
\footnotesize{\parindent=0.5cm
         Определение импедансов  связи  и  примеры  их  вычисления для
         структур с трансляционной симметрией.  Разложение  импедансов
         по   мультиполям.   Свойства   симметрии.  Недостатки  теории
         кильватерных полей и импедансов для случая $v\equiv c$. Метод
         Гамильтона для расчета полей заряда,  движущегося в замкнутом
         резонаторе.
}\end{minipage}\end{center}
\vspace*{0.5cm}

     Наиболее последовательно и  строго  понятия  {\it  продольного  и
поперечного   импедансов  связи}  вводятся  как  преобразования  Фурье
соответствующих кильватерных потенциалов по переменной $s$,  а  именно
посредством формул
     $$ Z_{\parallel}(\omega,b,\rv r_{\perp})=\frac 1 v \int
         \limits_{-\infty}^\infty\gv W_{\parallel}(s,b,\rv r_{\perp})
         \,e^{i\omega s/v}\,ds\,,\eqno(\mbox{\rm Д3}.1)$$
     $$ \rv Z_{\perp}(\omega,b,\rv r_{\perp})=-\frac i v \int
         \limits_{-\infty}^\infty\mbox{\boldmath$\gv W$}_{\perp}
         (s,b,\rv r_{\perp}) \,e^{i\omega s/v}\,ds\,.
         \eqno(\mbox{\rm Д3}.2)$$
Настоящее Дополнение  базируется  на  материале Дополнения 2,  поэтому
здесь  и  в  дальнейшем  используются  введённые  там   обозначения.
Поперечный  импеданс  $\rv  Z_{\perp}$  (как и поперечный кильватерный
потенциал  $\mbox{\boldmath$\gv   W$}_{\perp}$)   является   двумерным
вектором,  лежащим  в  поперечной к оси $z$ плоскости;  в используемой
далее цилиндрической системе координат он  имеет  компоненты  $Z_r$  и
$Z_{\varphi}$.

     Поскольку по  своему  физическому  смыслу  $\gv   W_{\parallel}$,
$\mbox{\boldmath$\gv W$}_{\perp}$ --- величины действительные,  то для
импедансов имеют место сотношения
     $$\re Z_{\parallel}(-\omega)=\re Z_{\parallel}(\omega)\,,
         \qquad \im Z_{\parallel}(-\omega)=-\im Z_{\parallel}
         (\omega)\,,\eqno(\mbox{\rm Д3}.3)$$
     $$\re \rv Z_{\perp}(-\omega)=-\re \rv Z_{\perp}(\omega)\,,
         \qquad \im \rv Z_{\perp}(-\omega)=\im \rv Z_{\perp}
         (\omega)\,.\eqno(\mbox{\rm Д3}.4)$$
Как и  в случае кильватерных потенциалов,  следует различать импедансы
связи на единицу длины  (в  структурах  с  трансляционной  симметрией,
например,  в  гладкой  однородной  вдоль  оси  камере)  и интегральные
импедансы вдоль всего пути при пролёте мимо локальной нерегулярности
камеры.  Эти  величины,  очевидно,  имеют  разную размерность,  но для
упрощения записи далее для них используются одни и те же обозначения.

     Краткое дальнейшее рассмотрение импедансов ни  в  какой  мере  не
касается  вопросов  их  использования  для  расчета  динамики пучков в
ускорителях и  исследования  их  устойчивости  (эти  проблемы  целиком
относятся  к теории ускорителей),  а посвящено вычислению этих величин
методами   электродинамики   СВЧ.    Ниже    вычисляются    импедансы,
обусловленные  точечными  зарядами.  Обобщение  на  случай  сгустков с
распределённым зарядом тривиально и основано на теореме о свёртке.

     В Дополнении   2   было   показано,   что   вычисление   точечных
кильватерных   потенциалов  является  сложной  задачей  и  может  быть
доведено  до  конца  в  аналитическом  виде  только   для   нескольких
простейших  структур.  Поскольку импедансы связи определены выше через
кильватерные потенциалы,  то может создаться ложное  впечатление,  что
вычисление импедансов ещё более сложная задача. Это не так, и именно
этим   обстоятельством   объясняется   более   широкое   использование
импедансов,  чем  кильватерных потенциалов,  при расчётах конкретных
ускорителей.

     Действительно, в  соответствии  с  определением  (\mbox{\rm Д3}.1),
формулами  (\mbox{\rm Д2}.1)  и (\mbox{\rm Д2}.16)   продольный
импеданс связи   может  быть записан в виде  трёхкратного интеграла
     $$ Z_{\parallel}(\omega,b,\rv r_{\perp})=-\frac 1{2\pi Qv}\int
         \limits_{-\infty}^\infty ds\,e^{i\omega s/v}\int
         \limits_{-\infty}^\infty dz \int\limits_{-\infty}^\infty E_z
         (\omega',b,\rv r_{\perp},z)\, e^{-i\omega'(z+s)/v}\,d
         \omega'\,.\eqno(\mbox{\rm Д3}.5)$$
Изменив порядок  интегрирования,  начнём  с  вычисления интеграла по
$s$,  которое   даёт   $2\pi\delta(\omega-\omega')$;   после   этого
интегрирование  по  $\omega'$  становится  тривиальным  и в результате
получаем, что
     $$ Z_{\parallel}(\omega,b,\rv r_{\perp})=-\frac1 Q \int
         \limits_{-\infty}^\infty E_z(b,\rv r,\omega)\,e^
         {-i\omega z/v}\,dz\,.\eqno(\mbox{\rm Д3}.6)$$
Аналогичным способом   находим  выражение  для  поперечного  импеданса
связи:
      $$ \rv Z_{\perp}(\omega,b,\rv r_{\perp})=-\frac i Q \int
         \limits_{-\infty}^\infty \rv F_{\perp}(b,\rv r,\omega)\,
         e^{-i\omega z/v}\,dz\,,
         \eqno(\mbox{\rm Д3}.7)$$
где $\rv F_{\perp}=\rv E_{\perp}+\beta[\rv e_z \rv H]_{\perp}$,
$\rv e_z$ --- единичный вектор вдоль оси $z$.

     Для однородной вдоль оси $z$ структуры имеет место трансляционная
симметрия, так что
     $$ E_z(b,\rv r,\omega)=\tilde {E_z}(b,\rv r_{\perp},\omega)\,
         e^{i\frac \omega v z},\qquad \rv F_{\perp}(b,\rv r,
         \omega)= \tilde {\rv F}_{\perp}(b,\rv r_{\perp},\omega)\,e^
         {i\frac \omega v z}\,,
         \eqno(\mbox {\rm Д3}.8)$$
и для  импедансов  связи  {\it  на  единицу  длины} получаются простые
выражения:
     $$ Z_{\parallel}(\omega,b,\rv r_{\perp})=-\frac 1 Q
         \tilde E_z(b,\rv r_{\perp},\omega)\,, \eqno(\mbox{\rm Д3}.9)$$
     $$ \rv Z_{\perp}(\omega,b,\rv r_{\perp})=-\frac i Q
         \tilde{ \rv F}_{\perp}(b,\rv r_{\perp},\omega)\,.
         \eqno(\mbox{\rm Д3}.10)$$

     Представляя импедансы связи точечного заряда  в  круглой  гладкой
камере   со  стенками  конечной  проводимости  в  виде  разложения  по
мультиполям
     $$Z_{\parallel}(\omega,b,\rv r_{\perp})=\sum\limits_{m=0}
         ^\infty Z^{(m)}_{\parallel}(\omega,b,\rv r_{\perp})\,,
         \qquad \rv Z_{\perp}(\omega,b,\rv r_{\perp})=\sum\limits_
         {m=0}^\infty \rv Z^{(m)}_{\perp}(\omega,b,\rv r_{\perp})\,,
         \eqno(\mbox{\rm Д3}.11)$$
и учитывая, что $F_r=E_r/\gamma^2$,
$F_\varphi=E_\varphi/\gamma^2$, на основании формул
(3.26),~(\mbox{\rm Д2}.31) и (\mbox{\rm Д2}.39) находим импедансы
связи $m$-того мультиполя:

     $$Z_{\parallel}^{(m)}(\omega,b,\rv r_{\perp})=
         i\frac {\omega}{\pi v^2\gamma^2}\left \{\left [
         \begin{array}{l} K_m(\Gamma r)I_m(\Gamma b)\\[.3cm]
         I_m(\Gamma r) K_m(\Gamma b)\end{array}\right]
         +C_m I_m(\Gamma r) \right\}\,\frac {\cos{m\varphi}}
          {1+\delta_{0m}}\,,\eqno(\mbox{\rm Д3}.12)$$
     $$Z_r^{(m)}(\omega,b,\rv r_{\perp})=
         i\frac {\omega}{\pi v^2\gamma^3}\left \{\left [
         \begin{array}{l} K'_m(\Gamma r)I_m(\Gamma b)\\[.3cm]
         I'_m(\Gamma r) K_m(\Gamma b)\end{array}\right]
         +C_m I'_m(\Gamma r)\right\}\,\frac
         {\cos{m\varphi}}{1+\delta_{0m}}\,,\eqno(\mbox{\rm Д3}.13)$$
     $$Z_{\varphi}^{(m)}(\omega,b,\rv r_{\perp})=
         -\frac i{\pi v\gamma^2 r}\left \{\left [
         \begin{array}{l} K_m(\Gamma r)I_m(\Gamma b)\\[.3cm]
         I_m(\Gamma r) K_m(\Gamma b)\end{array}\right]
         +C_m I_m(\Gamma r)\right\}\,m\sin{m\varphi}
         \,,\eqno(\mbox{\rm  Д3}.14)$$
где $b$ --- отклонение первичного заряда от оси  трубы  в  направлении
оси  $x$,  верхняя  строчка  в квадратных скобках соответствует $r>b$,
нижняя --- $r<b$, а
     $$ C_m=-\frac{I_m(\Gamma b)}{I_m(\Gamma a)}\Bigl[
         {K_m(\Gamma a)}+\frac{iWc\beta^2\gamma^2}
         {\omega a I_m(\Gamma a)}\Bigr]\,.\eqno(\mbox{\rm Д3}.15)$$

     Как следствие  теоремы   Пановского-Венцеля   (\rm   {Д2}.59)   в
аксиально   симметричных  структурах  между  продольным  и  поперечным
импедансами связи мультиполей выполняется соотношение
     $$ \rv Z^{(m)}_{\perp}(\omega,\rv r)=\frac v \omega
         \nabla_{\perp} Z^{(m)}_{\parallel}(\omega,\rv r)\,.
         \eqno(\mbox{\rm Д3}.16)$$
Отметим ещё,  что  вблизи  оси  такой  структуры   импедансы   связи
ультрарелятивистского  точечного  заряда  имеют  следующую  радиальную
зависимость
     $$Z^{(m)}_{\parallel}(\omega,\rv r) \sim \Bigl(\frac r a
         \Bigr)^m\,,\qquad m\geqslant
         0\,;\qquad \rv Z^{(m)}_{\perp}(\omega,\rv r)
         \sim\Bigl(\frac r a \Bigr)^{m-1}\left\{\begin{array}{l}\rv e_r,
         \\[.2cm]\rv e_\varphi,\end{array}\right.\qquad m\geqslant
         1\,,\eqno(\mbox {\rm Д3}.17)$$
что непосредственно следует из разложения формул (\rm {Д3}.12) -- (\rm
{Д3}.15) вблизи оси по степеням $r$ (для $r<b$).

     Поскольку штатная работа ускорителя предполагает малые отклонения
от оси и первичного,  и пробного зарядов ($b/a\ll 1,\;r/a\ll  1$),  то
основной интерес представляют импедансы монополя $m=0$ и диполя $m=1$.
В ультрарелятивистском пределе $\gamma\to  \infty$  основной  вклад  в
продольный импеданс даёт монополь,  а в поперечный --- диполь. После
перехода к пределу получаем для продольного импеданса выражение
     $$ Z_{\parallel}(\omega,b)=\frac \omega{2\pi c^2}\bigl[\frac i
         {\gamma^2}\ln{\frac a b}+(1-i)\frac \delta {2a}\bigr]\,
         ,\eqno(\mbox{\rm Д3}.18)$$
где $\delta=c/\sqrt{2\pi\sigma\omega}$  ---  глубина  скин-слоя,  а  в
поперечном   импедансе  остаётся  только  компонента  в  направлении
отклонения первичного заряда
     $$ Z_x(\omega,b)=\frac b {2\pi c}\bigl[\frac i {\gamma^2}\bigl(
         \frac 1{b^2}-\frac 1{a^2}\bigr)+(1-i)\frac \delta {a^3}\bigr]
         \,.\eqno(\mbox{\rm Д3}.19)$$
Обе формулы  приведены  для  траекторий   пробного   заряда,   имеющих
отклонение  от  оси  меньшее,  чем  $b$  (в  этой области импедансы не
зависят от $\rv r_{\perp}$).

     В приведенных выше выражениях  для импедансов оставлены члены,
содержащие в знаменателе $\gamma^2$,  которые  в  ультрарелятивистском
пределе  $\gamma\to  \infty$ на первый взгляд могут быть опущены.  Эти
члены при любых конечных значениях $\gamma$ при стремлении  $b\to  0$,
то есть при движении первичного заряда строго по оси, расходятся, хотя
эта траектория, очевидно, ничем не выделена. Более внимательный анализ
показывает, что импедансы расходятся при любом значении $b$, но только
для таких траекторий пробного заряда,  для которых $r=b$. Расходимость
обусловлена  поведением  поля в непосредственной близости от точечного
заряда и требует аккуратного перехода к  пределу  $\gamma\to  \infty$.
Именно  этим  и  обусловлено присутствие указанных членов в предельных
формулах для импедансов.

     Вообще говоря,  предельный переход к ультрарелятивистскому случаю
требует  известной  осторожности.  Прежде  всего   подчеркнем,   что
введение  заряженных  частиц,  имеющих  конечную массу и движущихся со
скоростью $v=c$, принципиально недопустимо, поскольку оно противоречит
основным    положениям    классической    электродинамики   и   теории
относительности.  При  таком  формальном  переходе  нам   в   качестве
источника,   возбуждающего   в   исследуемой   структуре  наведённые
электромагнитные поля, приходится иметь странный точечный объект, поле
которого  строго  равно  нулю  во всем пространстве,  за исключением
бесконечно  тонкой  плоскости,  нормальной  к   направлению   движения
источника, во всех точках которой оно бесконечно велико. Очевидно, что
микроскопические уравнения Максвелла таких решений не имеют.

     Такой объект не может испускать дифракционное излучение.  В связи
с этим напомним, что излучение равномерно движущегося точечного заряда
при  пролёте  мимо  какого-нибудь  металлического  тела  обусловлено
возбуждением  полем  заряда  токов  на  поверхности  тела,  которые  и
излучают.   Этот   процесс   переизлучения  не  носит  энергетического
характера,  а связан с пространственной перестройкой  поля  ---  поток
энергии  через замкнутую поверхность,  охватывающую тело,  равен нулю.
Однако переизлучённое поле,  распространяясь  со  скоростью  $c$,  в
какой-то   момент  времени  достигает  движущегося  заряда,  поскольку
скорость последнего $v<c$.  Это поле,  в отличие от собственного  поля
заряда   в  пустом  пространстве,  имеет  в  точке  нахождения  заряда
тормозящую  продольную  составляющую  электрического   поля,   которая
работает над зарядом, приводя к потерям его энергии.

     Одновременно с возникновением  тормозящей  силы  в  перестроенном
вблизи  заряда поле возникает поток вектора Умова-Пойнтинга,
величина которого, вычисленная через замкнутую поверхность,
охватывающую заряд, равна  мощности  потерь.  Длительность
импульса излучения определяется временем взаимодействия
переизлучённого телом поля  с  зарядом.  При
ультрарелятивистском движении заряда возрастает время,  за которое
это поле <<догоняет>>  заряд, и увеличивается время и, следовательно,
и длина пути,   на   котором  импульс  поля  взаимодействует  с
зарядом.  Это расстояние в теории дифракционного излучения
называется  {\it  путём формирования  излучения};  для  больших
$\gamma$  длина  пути линейно растёт с $\gamma$, и
соответственно растёт излучённая энергия.

    Очевидно, что  реально  такое  возрастание  энергии  излучения не
наблюдается ни в ускорителях,  где к настоящему времени достигнуты для
электронов  значения  $\gamma$  порядка  $10^{5}$,  ни в природе,  где
встречаются частицы с энергией на много порядков большей.  Объясняется
это  тем,  что  при больших $\gamma$ поле заряда сосредоточено в узком
конусе вблизи плоскости,  проходящей через заряд и  нормальной  к  его
скорости.  Чем  больше  $\gamma$,  тем  уже  конус  конус и тем короче
импульс,  возбуждающий токи на поверхности тела,  и  тем  шире  спектр
частот  в  спектральном  разложении  импульса.  В  результате основной
участок спектра смещается в область таких частот,  для которых  нельзя
рассматривать  тело как металлическое,  приписывая ему хоть идеальную,
хоть  конечную  проводимость.  На  более   высоких   частотах   всякое
макроскопическое  тело  становится прозрачным и никакого переизлучения
поля не происходит.

     Нетрудно оценить  то  максимальное  значение $\gamma$,  вплоть до
которого  расчёт  кильватерных   потенциалов   и   импедансов
связи изложенным  выше  способом справедлив.  Пусть расстояние от
траектории заряда до ближайшей  части  проводящей  структуры  есть
$a$.  Верхняя граница  спектра  излучения  наведённого  тока
имеет порядок $\omega \approx \gamma c/a$.  Более высокие частоты
в спектральном  разложении поля  заряда  в  области  ближайшего
проводника сильно подавлены,  что следует из поведения при больших
значениях аргумента функции $K_0(x)$, определяющей  согласно
формуле  (\mbox{\rm Д2}.27)  комплексную амплитуду вектора Герца,
а, следовательно, и компоненты поля точечного заряда. Если
приведенная   верхняя   граница   частот   не  превышает  той
критической,  за  которой  рассматривать   структуру   стенок
камеры ускорителя как металлическую уже нельзя, то все расчёты
кильватерных потенциалов и импедансов верны.  Тем самым
устанавливается  предельное значение   $\gamma$,   до  которого  в
данной  конкретной  структуре, характеризуемой параметром $a$,
пригодна изложенная методика.

     Поэтому результаты,   полученные   в  многочисленных  работах  по
кильватерным потенциалам  и  импедансам  связи  как  прямое  следствие
устремления  $\gamma$  к  бесконечности,  в своей строгой формулировке
неверны. В первую очередь к ним следует отнести утверждение о том, что
$\mbox{\boldmath$\gv   W$}(s)\equiv   0$  при  $s<0$;  необоснованными
являются и все так называемые дисперсионные соотношения для импедансов
связи,  для  которых  принципиально  важно  знание  поведения  поля на
предельно  высоких  частотах  и  которые   являются   непосредственным
следствием  предыдущего  неверного утверждения.  Хотя,  несомненно,  в
какой-то  степени  (фактически  доведённой  до  абсурда)   последнее
утверждение   и  отражает  реальность:  кильватерный  потенциал  перед
первичным зарядом существенно меньше, чем за ним ($s>0$), но только не
в  ближайшей  окрестности  заряда.  С  ростом  $\gamma$  эта тенденция
усиливается,  но лишь до тех значений,  для которых ещё  справедлива
металлическая модель структуры.

     В общем случае необходимо  чётко  различать  две  ситуации  ---
во-первых,  когда  утверждение <<положим  скорость  заряда равной
скорости света ${c}\!$ >>  просто  соответствует  приравниванию  безразмерной
скорости $\beta$ единице,  что  вполне  допустимо  и  разумно,  если
$\beta$  входит в какое-то выражение в виде множителя,  и,
во-вторых,  когда на  основе этого   утверждения   разность  $1-\beta^2$
полагается равной нулю,  что соответствует устремлению энергии
заряженной частицы к бесконечности и может   привести   к
 ошибочным результатам.

     Ещё одной   распространённой  ошибкой  в  работах  по  теории
кильватерных  потенциалов  является   попытка   ввести   и   вычислить
кильватерный  потенциал точечного заряда в {\it замкнутом} резонаторе.
Естественно воспользоваться  для  этого  прямым способом  расчёта  полей
движущихся зарядов непосредственно во {\it временной} области,
вошедшим в  литературу  под  названием  {\it метода    Гамильтона}.    Ниже
кратко   излагается   {\it гамильтоновский} метод решения задачи о поле,
возбуждаемом в замкнутом резонаторе с идеально проводящими стенками
 движущимися внутри него зарядами. Рассмотрим простейший
случай, когда внутри резонатора вакуум.
Выпишем здесь ещё раз основные уравнения, приведенные в первых разделах.

     Уравнения Максвелла
     $$\left.\begin{array}{lcllcl}\rot\mbox{\boldmath$\gv H$}& = &
         \phantom{-}\displaystyle{\frac 1 c}{\frac{\partial
         \mbox{\boldmath$\gv  E$}}{\partial t}}+\displaystyle
         {\frac{4\pi} c}\rho \rv v\,,\qquad & \div\mbox{\boldmath$
         \gv E$}& = & 4\pi\rho\,,\\[.5cm]\rot\mbox{\boldmath$\gv E$}
         & = &-\displaystyle{\frac 1 c}\frac{\partial\mbox{\boldmath$
         \gv H$}}{\partial t}\,,\qquad & \div
         \mbox{\boldmath$\gv H$}& = & 0\,,\end{array}\right\}
         \eqno(\mbox{\rm Д3}.20)$$
путём введения  векторного  потенциала  $\mbox{\boldmath$\gv  A$}$ и
скалярного потенциала $\Phi$ посредством соотношений
     $$\mbox{\boldmath$\gv E$}=-\frac 1 c\frac{\partial\mbox
         {\boldmath$\gv A$}}{\partial t}-\grad\Phi,
         \qquad\mbox{\boldmath$\gv H$}=\rot\mbox{\boldmath$\gv A$}
         \eqno(\mbox{\rm Д3}.21)$$
сводятся к двум волновым уравнениям относительно этих потенциалов:
     $$\left.\begin{array}{l}\Delta\mbox{\boldmath$\gv A$}-
         \displaystyle{\frac 1 {c^2}\frac{\partial^2
         \mbox{\boldmath$\gv A$}}{\partial t^2}}-
         \grad\Bigl(\div\mbox{\boldmath$\gv A$}+\frac 1 c\frac
         {\partial\Phi}{\partial t}\Bigr)=-\frac{4\pi} c\rho\rv v\,,
         \\[.5cm]\Delta\Phi+\displaystyle{\frac 1 c\frac {\partial\div
         \mbox{\boldmath$\gv A$}}{\partial t}}=-4\pi\rho\,.
         \end{array}\right\}\eqno(\mbox{\rm Д3}.22)$$
Воспользуемся  кулоновской калибровкой $\div \mbox{\boldmath$\gv
A$} =0 $, что позволяет преобразовать эту систему уравнений к виду
     $$\left.\begin{array}{l}\Delta\mbox{\boldmath$\gv A$}-
         \displaystyle{\frac 1 {c^2}\frac{\partial^2
         \mbox{\boldmath$\gv A$}}{\partial t^2}}=-\frac{4\pi} c
         \rho\rv v+\frac 1 c \grad\frac{\partial\Phi}{\partial t}\,,
         \\[.5cm]\Delta\Phi =-4\pi\rho\,.\end{array}\right\}
         \eqno(\mbox{\rm Д3}.23)$$

     Напомним, что  в   результате   выбора   кулоновской   калибровки
векторный  потенциал  $\mbox{\boldmath$\gv A$}$ определяет только {\it
поперечное} электромагнитное поле.  Для скалярного потенциала  $\Phi$,
позволяющего   вычислить  {\it продольную}   часть   электрического   поля,
полученное  соотношение  представляет  собой   статическое   уравнение
Пуассона,  из  чего сразу следует,  что эта часть поля в данный момент
времени  $t$  в  данной  точке  пространства  $\rv   r$   определяется
положением  всех  зарядов  в {\it тот же} момент времени,  то есть
отсутствует эффект запаздывания.

     Решение второго уравнения системы (\mbox{\rm Д3}.34) ищем в виде ряда
     $$\Phi(\rv r,t)=\sum\limits_\lambda r_\lambda(t)\phi_\lambda
         (\rv r)\,,\eqno(\mbox{\rm Д3}.24)$$
где $\phi_\lambda(\rv r)$-- собственные функции краевой задачи
     $$\Delta\phi + \alpha^2 \phi=0\,,\qquad \phi=0
          \quad\mbox{на}\quad S\,,\eqno(\mbox{\rm Д3}.25)$$
причём $S$   ---   замкнутая  поверхность  (совпадающая  с  идеально
проводящими  стенками  резонатора),  а   $r_\lambda$--   коэффициенты,
зависящие от времени $t$ в том случае,  когда $\rho(\rv r,t)$ меняется
с $t$.  Аналогичная краевая задача (правда, двумерная) уже встречалась
в   разделе  8  при  исследовании  $E$-волн  в  волноводе.  Она  имеет
бесконечную  последовательность  действительных  собственных  значений
$\alpha_\lambda$,  каждому  из  них  соответствует собственная функция
$\phi_\lambda(\rv r)$.  Система этих функций является ортогональной  в
том смысле, что
     $$\int\limits_V \phi_\lambda(\rv r)\phi_\mu(\rv r)\,dV=0
         \qquad\mbox{при}\qquad \lambda\ne\mu\,,\eqno(\mbox{\rm Д3}.26)$$
и полной --- по ней может быть разложено произвольное решение
второго уравнения   (\mbox{\rm Д3}.23)  с  нулевым  значением  на
границе  области. Нормировку функций $\phi_\lambda$ удобно
определить условием
     $$\int\limits_V\Bigl(\nabla\phi_\lambda\Bigr)^2\,dV=8\pi .
         \eqno(\mbox{\rm Д3}.27)$$

     Собственные функции   $\phi_\lambda$   легко   находятся  методом
разделения переменных в том случае,  когда поверхность $S$
совпадает с одной  из  координатных  поверхностей  в системе
координат,  в которой переменные  в  уравнении  Гельмгольца
разделяются.  В   этом   случае $\lambda$  представляют собой
совокупность трёх индексов,  каждый из которых определяет число
вариаций поля по соответствующей  координате. Коэффициенты
$r_\lambda$ находятся в результате подстановки разложения
(\mbox{\rm Д3}.24) во второе уравнение системы (\mbox{\rm Д3}.23),
умножения  на $\phi_{\mu}$ и интегрирования по объёму
резонатора:
     $$r_\lambda=\frac 1 2 \int\limits_V\rho\phi_\lambda\,dV\,.
         \eqno(\mbox{\rm Д3}.28)$$
Отметим, что решение (\mbox{\rm Д3}.24) описывает,  в частности,
полное поле системы   покоящихся   зарядов  в  замкнутом
металлическими  стенками объёме.

     Если заряды   движутся  ($\rv  v\ne  0$),  то  необходимо  учесть
поперечное поле,  определяемое векторным потенциалом  $\mbox{\boldmath
$\gv A$}$, который следует искать в виде
     $$\mbox{\boldmath$\gv A$}(\rv r,t)=\sum\limits_\lambda
         q_\lambda(t)\rv a_\lambda (\rv r)\,,\eqno(\mbox{\rm Д3}.29)$$
где векторные функции $\rv a_\lambda(\rv r)$ суть собственные  функции
векторной краевой задачи
     $$ \Delta \rv a+\tilde\alpha^2 \rv a=0\,,\qquad \div \rv a=0\,,
         \qquad [\rv n \rv a]=0 \qquad\mbox{на}\qquad S
         \eqno(\mbox{\rm Д3}.30)$$
($\rv n$ --- нормаль к поверхности $S$), а уравнение для
коэффициентов $q_\lambda(t)$ находится путём подстановки
разложений (\mbox{\rm Д3}.24) и (\mbox{\rm Д3}.29) в первое
уравнение системы (\mbox{\rm Д3}.23).

     Собственные значения  $\tilde\alpha_\lambda$  векторной   краевой
задачи  принято обозначать как $\omega_\lambda/c$;  в общем случае
они не совпадают  с  собственными  значениями  $\alpha_\lambda$
скалярной задачи   (\mbox{\rm Д3}.25).  Собственные  функции  $\rv
a_\lambda$  помимо индекса $\lambda$,  представляющего собой
совокупность тех  же  трёх индексов,  что  и  в скалярной
задаче,  характеризуются дополнительным индексом,  определяющим
поляризацию  поля.  В  результате  подстановки приходим к
уравнению
     $$\sum\limits_\lambda\Bigl[(\ddot{q}_\lambda+\omega_\lambda^2q_
         \lambda)\rv a_\lambda+c\dot{r}_\lambda\,\nabla\phi_\lambda
         \Bigr]=4\pi c\rho\rv v\,,\eqno(\mbox{\rm Д3}.31)$$
домножая которое  на  $\rv a_\mu$ и интегрируя по объёму резонатора,
получим уравнениe для $q(t)$ вида
     $$\ddot{q}_\lambda+\omega_\lambda^2 q_\lambda=\frac{\omega^2_
         \lambda}{2c}\int\limits_V\rho\rv v\rv a_\lambda\,dV\,;
         \eqno(\mbox{\rm Д3}.32)$$
при выводе соотношения  (\mbox{\rm Д3}.32)  учтена
ортогональность  функций $\nabla   \phi_\lambda$   и   $\rv
a_\lambda$  (следствие  векторного равенства $\rv a_\lambda \cdot
\nabla \phi_\lambda=  \div(\phi_\lambda \cdot\rv a_\lambda$)) и
выбрана нормировка
     $$\int\limits_V\rv a_\lambda^2\,dV=\frac{8\pi c^2}{\omega_\lambda
         ^2}\,.\eqno(\mbox{\rm Д3}.33)$$

     Уравнение (\mbox{\rm Д3}.32) совпадает с обыкновенным  дифференциальным
уравнением,  описывающим   поведение   осциллятора  под  действием
вынуждающей силы. При заданной зависимости $\rho$ и $\rv v$ от
времени его  решение  легко  может  быть  найдено методом
вариации постоянных. Решения  $q_\lambda(t)$  и  $r_\lambda(t)$
при   известных   системах собственных   функций   $\rv a_\lambda$  и
$\phi_\lambda$  полностью определяют электромагнитные  поля  в
резонаторе,  которые   согласно формулам (\mbox{\rm Д3}.21) представляются в виде
     $$\left.\begin{array}{lcl}\mbox{\boldmath$\gv E$}(\rv r,t)& = &
         -\displaystyle{\sum\limits_\lambda}\Bigl(\displaystyle
         \frac{\dot{q}_\lambda(t)} c \rv a_\lambda(\rv r)
         +r_\lambda(t) \,\nabla\phi_\lambda(\rv r)\Bigr)\,,
         \\[.5cm]\mbox{\boldmath$\gv H$}(\rv r,t) & = &\phantom{-}
         \displaystyle{\sum\limits_\lambda}q_\lambda(t)\rot
         \rv a_\lambda(\rv r) \,.\end{array}\right\}
         \eqno(\mbox{\rm Д3}.34)$$
В случае  выбранной  выше  нормировки  собственных   функций   энергия
электромагнитного поля в объёме резонатора выражается формулой
     $$U(t)= \frac 1{8\pi}\int\limits_V(\mbox{\boldmath$\gv E$}^2+
         \mbox{\boldmath$\gv H^2$})\,dV=\sum\limits_\lambda\Bigl(
         \frac{\dot{q}_\lambda^2}{\omega_\lambda^2}+q_\lambda^2
         +r_\lambda^2\Bigr).\eqno(\mbox{\rm Д3}.35)$$

     Пусть в момент времени $t=0$ в  резонатор  влетает  заряд  $Q$  и
движется  со  скоростью  $v$  по  прямолинейной  траектории (оси $z$),
причём координата точки влёта --- $z=0$, а точки вылета --- $z=L$.
Подстановка соответствующей плотности заряда
     $$\rho(\rv r,t)=Q\delta(x)\delta(y)\delta(z-vt)
         \eqno(\mbox{\rm Д3}.36)$$
в уравнение (\mbox{\rm Д3}.32) даёт возможность преобразовать
его к виду
     $$\ddot q_\lambda+\omega_\lambda^2\,q_\lambda=
         \frac{Q\omega_\lambda^2 v}{2c}\,\left\{\begin{array}{lll}
         0 &\mbox{при}\quad &t<0\,,\\[.3cm]
         a_{z\lambda}(0,0,vt)\qquad &\mbox{при}\quad &0<t<L/v\,,\\
         [.3cm]0 &\mbox{при}&t>L/v\end{array}\right.
         \eqno(\mbox{\rm Д3}.37)$$
с начальными условиями
     $$q_\lambda(0)={\dot q}_\lambda(0)=0\,,
         \eqno(\mbox{\rm Д3}.38)$$
выражающими факт отсутствия  поля  в  резонаторе  до  момента
влёта заряда.   Решение   уравнения  (\mbox{\rm Д3}.37)  легко
находится  методом вариации постоянных:
     $$q_\lambda(t)=\frac{Q\omega_\lambda v}{2c}\int\limits_0^{\min
         (t,L/v)}\sin\omega_\lambda(t-t')a_{\lambda z}(0,0,vt')\,dt'.
         \eqno(\mbox{\rm Д3}.39)$$
Аналогичное выражение получается и для коэффициентов $r_\lambda$:
     $$  r_\lambda= \frac Q 2\,\left\{\begin{array}{lll}
         0 &\mbox{при}\quad &t<0\,,\\[.3cm]
         \phi_\lambda(0,0,vt)\qquad&\mbox{при}\quad &0<t<L/v\,,\\
         [.3cm]0 &\mbox{при}&t>L/v.\end{array}\right.
         \eqno(\mbox{\rm Д3}.40)$$

     Вычислим теперь энергию поля, которая осталась в резонаторе после
вылета заряда,  то есть при $t>L/v$,  когда $r_\lambda=0$.
Подставляя решение (\mbox{\rm Д3}.39) в формулу  (\mbox{\rm
Д3}.35)  в результате  несложных преобразований получаем для
энергии независящую от времени величину
         $$ U=Q^2\sum\limits_\lambda |V_\lambda|^2=
             Q^2\sum\limits_\lambda k_{\parallel \lambda}\,,
          \eqno(\mbox{\rm Д3}.41)$$
где
      $$V_\lambda=\frac {\omega_\lambda}{2c}\int\limits^L_0
           a_{z\lambda}(0,0,z)\,e^{i\omega_\lambda z/v}\,dz\,,
          \eqno(\mbox{\rm Д3}.42)$$
а $k_{\parallel\lambda}=|V_\lambda|^2$   ---  {\it  параметр  потерь},
названный так из-за того,  что он определяет убыль энергии  единичного
заряда   по  вылете  из  резонатора,  равную  потерям  на  возбуждение
$\lambda$-того собственного колебания.

     Не представляет   большого  труда  таким  же  способом  вычислить
распределение поля в резонаторе и кильватерные  потенциалы.  Не  будем
здесь этого делать в общем виде, а проанализируем полученный результат
для потерь энергии на конкретном  простейшем  примере  цилиндрического
резонатора кругового сечения радиуса $a$, длины $L$, по оси которого и
движется заряд.  В такой  азимутально  симметричной  структуре  и  при
выбранной  траектории  заряда  понадобятся только собственные функции,
зависящие от $r$ и $z$ цилиндрической системы координат и определяемые
поэтому двумя индексами:
     $$ \phi_{np}(\rv r)=\frac 4{\sqrt{L (1+\delta_{0p})}}\,
         \frac{c J_0(\nu_n r/a)}{a\omega_{np} J_1(\nu_n)}\,
         \sin\frac{\pi p z}L\,,\eqno(\mbox{\rm Д3}.43)$$
     $$ \rv a_{np}(\rv r)=\frac 4{\sqrt{L(1+\delta_{0p})}}\,
         \frac {c^2} {a\omega_{np}^2 J_1(\nu_n)}\,\left\{\begin
         {array}{l}\displaystyle{\frac{\pi p}L\,J_1(\nu_n r/a)\,
         \sin{\frac{\pi p z}L}\,\rv e_r},\\[.4cm]\displaystyle{
         \frac{\nu_n}a\,J_0(\nu_n r/a)\,\cos{\frac{\pi p z}L }
         \rv e_z},\end{array}\right.\eqno(\mbox{\rm Д3}.44)$$
где собственные значения
     $$ \alpha=\tilde\alpha=\frac{\omega_{np}}c=\sqrt
         {\Bigl(\frac{\nu_n}a\Bigr)^2+
         \Bigl(\frac{\pi p}L\Bigr)^2}\,,\eqno(\mbox{\rm Д3}.45)$$
а $\rv e_r,\,\rv e_z$ --- единичные векторы вдоль соответствующих осей.

     Используя приведенные  выше общие формулы,  получаем для функций,
определяющих зависимость от времени, следующие выражения:
     $$  r_{np}(t)=\frac{2Q}{\sqrt{L(1+\delta_{0p})}}\,\frac c
         {\omega_{np}a J_1(\nu_n)}\,\sin{\frac{\pi p vt}L}\qquad
         \mbox{при}\quad 0<t<L/v\eqno(\mbox{\rm Д3}.46)$$
и $r_{np}=0$ для всех других значений $t$,
     $$ q_{np}(t)\!=\!\frac{2Q\beta}{\sqrt{L(1\!+\!\delta_{0p})}\nu_n
        J_1(\nu_n)\Bigl[1\!+\!\Bigl(\displaystyle{\frac{\pi p a}{\nu_n L\gamma}}
        \Bigr)^2\Bigr]}\left\{\begin {array}{l}\cos{\displaystyle
        {\frac{\pi p v}L} t}-\cos{\omega_{np}t},\\[.5cm]
        (-1)^p\cos{\omega_{np}\displaystyle{(\frac L v\!-\!t)}}-\cos{\omega_{np}t},
        \end{array} \right.\eqno (\mbox{\rm Д3}.47)$$
верхняя строчка за фигурной скобкой соответствует моментам времени
$0\!<\!t\!<\!\ L/ v$, нижняя --- $t\!>\!\ L/ v$.

     Вычисляя $V_{np}$  по формуле (\mbox{\rm Д3}.42), получаем
        $$ V_{np}=i\frac{2\beta}{\sqrt{L(1+\delta_{np})}}\,
           \frac 1{\nu_n J_1(\nu_n)}\,
          \frac{1-(-1)^p e^{i\omega_{np}L/v}}
           {1+\Bigl(\displaystyle{\frac{\pi p a}{\nu_n L\gamma}\Bigr)^2}}\,,
           \eqno (\mbox{\rm Д3}.48)$$
что позволяет сразу выписать выражение для энергии поля,
оставшейся в резонаторе после вылета заряда:
      $$ U=Q^2\sum\limits_{p=0}^\infty\sum\limits_{n=1}^\infty |V_{np}|^2=
        \frac{8Q^2\beta^2}L\sum\limits_{p=0}^\infty\frac 1{1+\delta_{0p}}
        \sum\limits_{n=1}^\infty \frac{1-(-1)^p \cos{\displaystyle{
        \frac{\omega_{np}L}v}}}
        {\nu_n^2 J_1^2(\nu_n)\Bigl[1+\Bigl(\displaystyle{\frac{\pi p a}
        {\nu_n L\gamma}\Bigr)^2\Bigr]^2}}\,.\eqno (\mbox{\rm Д3}.49)$$

      Но этот результат, полученный безупречным в математическом
отношении способом, бессмыслен: дело в том, что входящая в него
двойная сумма {\it расходится}, причем расходится даже одинарная
сумма по $p$ при любом значении $n$ и наоборот. Такой
обескураживающий факт не останавливает многих авторов и они
продолжают последовательно, основываясь на формулах для
собственных функций  (\mbox{\rm Д3}.43) и (\mbox{\rm Д3}.44) и
формулах (\mbox{\rm Д3}.46) и (\mbox{\rm Д3}.47)  для функций
$r_{np}(t)$ и $q_{np}(t)$, определяющих зависимость от времени,
вычислять компоненты поля и кильватерный потенциал. Получающиеся в
результате выражения (не будем их здесь приводить) обладают тем же
существенным недостатком --- они имеют вид двойных сумм и
расходятся для всех точек пространства, моментов времени и $s$
(хотя и медленнее, чем выражение для $U$). Но и после этого
некоторые авторы пытаются извлечь физический смысл из расходящихся
сумм.

       Однако все такие попытки обречены на неудачу, так как математика
(при безупречном её применении) всегда жестоко <<мстит>>  за
некорректную физическую постановку задачи. Бессмысленно решать
задачу о вылете заряда из {\it идеально} проводящей среды с резкой
границей --- работа выхода в этом случае бесконечна, что в
какой-то мере оправдывает бесконечную энергию поля, возникающего в
резонаторе. Как и каждая наука, микроскопическая электродинамика (а
именно на ней и основывается  электродинамика СВЧ) имеет свою область
применимости. Она,  в частности, не может
рассматривать процесс возникновения <<голого>>  электрона и
<<одевание>>  его полем (хотя бы и со скоростью света), как, впрочем,
и процесс его исчезновения и соответствующего <<сбрасывания>>  поля
при вылете из замкнутого резонатора. Эти вопросы, если они и имеют
физический смысл, могут рассматриваться только в рамках квантовой
электродинамики. В данном же случае видна явная причина всех
трудностей --- она состоит в некорректности приближения идеальной
проводимости при высоких частотах. Заметим  также некорректность
попыток обойти расходимости путем <<размазывания>>  точечного заряда
по конечному объёму --- уже в начале 20-того века работами Лоренца была
доказана бесперспективность таких попыток устранения принципиальных
трудностей  классической электродинамики. Полученные таким
способом результаты при расчётах ускорителей могут оказаться как верными,
так и не верными. Единственный  последовательный путь, на
котором не возникают принципиальные трудности, состоит в том,
чтобы вся траектория частиц при расчетах проходила вне металла.

    Из сказанного выше напрашивается вывод о неприемлемости модели
замкнутого резонатора для расчёта динамики пучка в ускорителе и
необходимости использования для этой цели модели резонатора с
подводящими трубами, внутри которых и движется пучок. Но такой
поспешный вывод оказывается неверным.

    Основное достоинство модели замкнутого резонатора состоит в
сравнительной простоте расчёта и прозрачности результатов для
каждого собственного колебания. Более того, для резонаторов
простейших форм нетрудно получить, как это было показано выше на
примере цилиндрического резонатора, замкнутые выражения в
аналитическом виде для всех характеристик излучения.

     Последовательный анализ возбуждения резонатора с подводящими
волноводами, наоборот, очень громоздок и требует на завершающем этапе
обязательного использования трудоёмких численных расчётов даже
для резонаторов самой простой геометрии. Методика такого расчёта
для резонаторов, образованных ступенчатыми нерегулярностями круглого
волновода, кратко изложена в Дополнении 2. Наибольшего объёма
численного счёта требуют две операции: обращение бесконечной матрицы
для каждой частоты спектрального представления поля (что возможно
проделать только приближённо путем редукции матрицы) и вычисления
для любой физической величины --- будь то потери на излучение или
кильватерный потенциал --- интеграла по бесконечному интервалу
$\omega$ (что также возможно лишь с помощью компьютера). Однако
получающиеся при этом результаты корректны вплоть до тех значений
$\gamma$, при которых в спектре излучения реально важен вклад только того
диапазона частот, где стенки резонатора можно считать идеально
проводящими.

    Сопоставление результатов расчёта излучения точечного заряда
для замкнутого резонатора и резонатора той же формы, но с подводящими
волноводами, показывает, что они с достаточно хорошей точностью
совпадают в области низких частот, а именно таких, для которых и в модели
резонатора с волноводами имеются дискретные собственные колебания.
Верхняя граница спектра таких колебаний лежит ниже наименьшей критической
частоты подводящих волноводов; для области  высоких частот различие между
двумя этими моделями существенное. Важный для практических расчетов
ускорителей момент состоит в том, что в ультрарелятивистском пределе низкочастотная часть спектра
достигает насыщения и для неё можно положить $v=c$. Поэтому
модель замкнутого резонатора оказывается полезной при вычислении
импедансов связи, которые определяются свойствами структуры в достаточно
узкой области сравнительно низких частот.

%\end{document}

\newpage
\oddsidemargin=-0.4mm \evensidemargin=-0.4mm
\topmargin=-0.4mm
\headsep=7mm
\textheight=231.875mm
\textwidth=160mm
\mathsurround=2.5pt
\unitlength=1mm
%\begin{document}
%\input{macr.tex}
\thispagestyle{empty}
%\addtocounter{page}{379}

\begin{center}\subsubsection*{ЗАДАЧИ И УПРАЖНЕНИЯ}\end{center}
\vspace*{0.5cm}

\markboth{Задачи и упражнения}{Задачи и упражнения}

\begin{center}\begin{minipage}[c]{0.75\textwidth}
\footnotesize{\parindent=0.5cm
         Задачи и  упражнения  предназначены   для   более   глубокого
         изучения  и  закрепления  материала лекций по электродинамике
         СВЧ,  читаемых студентам четвёртого курса ФОПФ и ФПФЭ МФТИ;
         задачи   и   упражнения   подобраны   с   учётом  интересов
         соответствующих базовых кафедр.}
\end{minipage}\end{center}\vspace*{0.5cm}

     {\bf 1.}~Диэлектрическую проницаемость $\varepsilon$ морской воды
принято считать равной 80,  а её проводимость равна~5~Ом$^{-1}\cdot$
м$^{-1}$.  Найдите комплексную диэлектрическую проницаемость для длины
волны  10~см.  Почему  для   воды   $\varepsilon=80$,   а   показатель
преломления $n\approx1,33 \neq\sqrt {\varepsilon}$?\vspace*{0.13cm}

     $\mbox{\textit {Ответ}}$:~$\varepsilon=80+30i$; область   применения
формулы  $n= \sqrt {\varepsilon}$ ограничена.

     {\bf 2.}~Покажите,  что если в однородной и изотропной проводящей
среде   потенциалы  электромагнитного  поля  подчинить  калибровочному
соотношению
     $$\div\mbox{\boldmath$
\gv A$}+\frac{4\pi}{c}\sigma\mu\Phi+\frac{\varepsilon
         \mu}{c}\frac{\partial \Phi}{\partial t}=0,$$
то они в случае отсутствия свободных зарядов удовлетворяют уравнениям
     $$\nabla^2\Phi-\displaystyle{\frac
         {\varepsilon\mu}{c^2}}\displaystyle{\frac{\partial^2\Phi}
         {\partial t^2}}\!=\!\displaystyle{\frac{4\pi}{c^2}}\sigma
         \mu\displaystyle{\frac{\partial\Phi}{\partial t}},\qquad
         \nabla^2\mbox{\boldmath$
\gv A$}-\displaystyle{\frac{\varepsilon\mu}{c^2}}
         \displaystyle{\frac{\partial^2\mbox{\boldmath$
\gv A$}}{\partial t^2}}\!=\!
         \displaystyle{\frac{4\pi}{c^2}}\sigma\mu\displaystyle{\frac
         {\partial\mbox{\boldmath$
\gv A$}}{\partial t}}.$$

     {\bf 3.}~Обобщите волновые уравнения
     $$\nabla^2\mbox{\boldmath$
\gv E$}-\frac{\varepsilon\mu}{c^2}\frac{\partial^2\mbox{\boldmath$
\gv E$}}
         {\partial t^2}=0,\qquad\nabla^2\mbox{\boldmath$
\gv H$}-\frac{\varepsilon\mu}
         {c^2}\frac{\partial^2\mbox{\boldmath$
\gv H$}}{\partial t^2}=0$$
на случай неоднородной немагнитной ($\mu=1$) среды.\par
     Покажите, что    в    случае    неоднородной   среды,   магнитная
проницаемость которой есть функция пространственных координат,  вектор
напряжённости магнитного поля удовлетворяет уравнению
     $$\div\mbox{\boldmath$
\gv H$}=-\frac{1}{\mu}(\mbox{\boldmath$
\gv H$}\grad\mu).$$

     $\mbox{\textit {Ответ}}$:
     $$\nabla^2\mbox{\boldmath$
\gv E$}-\frac{\varepsilon}{c^2}\frac{\partial^2\mbox{\boldmath$
\gv E$}}
         {\partial t^2}-\grad\div\mbox{\boldmath$
\gv E$}=0,\qquad\nabla^2\mbox{\boldmath$
\gv H$}-\frac
         {\varepsilon}{c^2}\frac{\partial^2\mbox{\boldmath$
\gv H$}}{\partial t^2}+
         \frac{1}{\varepsilon}[\grad\varepsilon\rot\mbox{\boldmath$
\gv H$}]=0.$$

     {\bf 4.}~Опишите  распространение  одномерного  волнового  пакета
$f(\omega,k)=   \delta(\omega-\omega_0)   \cdot\exp(-(k-k_c)^2D^2/2)/2
\pi$ в среде с дисперсией $\omega_0  =\omega_c+v_g\cdot(k-k_c)+\beta_0
\cdot(k-k_c)^2$.       Рассмотрите       также      случай,      когда
$k_c=0$,~$\omega_c=0$,~$\beta_0=0$.

     $\mbox{\textit {Ответ}}$:~в общем случае
     $$f(x,t)=\Biggl(\frac{1}{4\pi(D^2/2+i\beta_0t)}\Biggr)^{1/2}
         \cdot\exp\Biggl(-\frac{(x-v_gt)^2}{4(D^2/2+i\beta_0t)}+i
         (k_cx-\omega_ct)\Biggr);$$
если $k_c=0$,~$\omega_c=0$,~$\beta_0=0$, то
     $$f(x,t)=\frac{1}{\sqrt{2\pi}D}\exp[-(v_gt-x)^2/2D^2].$$

     {\bf 5.}~Найдите выражение для фазовой скорости  и  разность  фаз
между  векторами  $\rv E$ и $\rv H$ однородной плоской волны в среде с
$\mu$,~$\varepsilon$ и~$\sigma$ в  случае  больших  и  малых  значений
отношения            $4\pi\sigma/\omega\varepsilon$,           равного
$\tg\delta_\varepsilon$.

     $\mbox{\textit {Ответ}}$:~общие выражения для $v_{\mbox{\footnotesize\textit{ф}}}$
и  тангенса  разности  фаз $\widetilde {\gamma} $ имеют следующий
вид:
     $$v_{\mbox{\footnotesize\textit{ф}}}=
         c\sqrt{\frac{2}{\varepsilon\mu}}\cdot\frac{1}
         {\sqrt{\sqrt{1+\tg^2\delta_\varepsilon}+1}};\qquad\tg
         \widetilde{\gamma}=\left(\frac{\sqrt{1+\tg^2\delta_
         \varepsilon}-1}{\sqrt{1+\tg^2\delta_\varepsilon}+1}
         \right)^{1/2}.$$

   {\bf 6.}~Оцените глубину $\delta$ проникновения СВЧ поля в медную
стенку   и  её  активное  поверхностное  сопротивление  $\rho_s$  на
частотах 0,5~ГГц  и  3~ГГц  (проводимость  меди  равна  $5,8\cdot10^7$
Ом$^{-1}\cdot$м$^{-1}$).  Почему  на  частоте  50~Гц  не  используется
сплошной медный провод диаметром более 1~см?\par
     Определите для  плоской  волны  с  амплитудой поля на поверхности
5~МВ/м величину мощности,  рассеиваемой на частоте 0,5~ГГц в меди  при
комнатной  температуре  и  в  ниобии  при температуре 4,2~K ($\rho_s=$
70~нОм);  рассмотрите  только  случай  нормального  падения  волны  на
поверхность.

     $\mbox{\textit Ответ}$:~$\delta$= 3~мкм, $\rho_s=5,8$~мОм при  0,5~ГГц;  $\delta$
=1,2~мкм,   $\rho_s$=14~мОм  при  $3$~ГГц;  $\delta_{Cu}(\mbox{50~Гц})
=0,95$~см; $P_{Nb}\approx6$ Вт/м$^2$, $P_{Cu}\approx0,5$~МВт/м$^2$.

     {\bf 7.}~Токоведущие  поверхности  многих устройств СВЧ покрывают
тонким слоем серебра.  Определите толщину  такого  слоя,  при  которой
плотность тока на его внутренней поверхности уменьшается в 1000 раз по
сравнению  с  плотностью  тока  на  границе  раздела   <<металл-воздух>>
($f=30$~ГГц, $\sigma=5,48\cdot10^{17}$~с$^{-1}$).\par
     При какой  толщине  экрана  обеспечивается  ослабление  амплитуды
электромагнитного поля в 1000 раз на этой же частоте, если он выполнен
из материала с $\sigma=5,48\cdot10^{17}$~с$^{-1}$ и~$\mu=10$?

     $\mbox{\textit {Ответ}}$:~2,57~мкм и 0,813~мкм.

     {\bf 8.}~Найдите   фазовую   и  групповую  скорости,  критическую
частоту  волны  типа  $E_{11}$,  распространяющейся  в   прямоугольном
волноводе  с размерами $4$~см$\times$2~см на частоте 9~ГГц;  то же для
волны типа $E_{01}$ в цилиндрическом волноводе с радиусом 2~см.

     $\mbox{\textit {Ответ}}$:~${v_{\mbox{\footnotesize\textit{ф}}}/c}=2,753$,
~${v_{\mbox{\footnotesize\textit{гр}}}/c}=0,363$, ~$f_{11}\approx
8,385 \hbox{ ГГц}$; $ {v_{\footnotesize\textit{ф}}/c}
\approx1,298$,~${v_{\mbox{\footnotesize\textit {гр}}}/c} \approx
0,770$,~$f_ {01} \approx5,741 \hbox{ ГГц}.$

     {\bf 9.}~Для  неискажённой передачи импульсов необходимо обеспечить
прохождение через  структуру колебаний  с   полосой   частот  не менее
$\Delta\omega=2\pi/\tau$,    где    $\tau$    ---     длительность импульса.
В   качестве   такой  структуры  рассмотрите волновод длиной 50~м  и  оцените
степень  искажения передаваемого  импульса с $\tau=0,05$ мкс (длина волны
генератора --- 10~см,  отношение рабочей  частоты  к  критической  --- 1.4).

     $\mbox{\textit {Ответ}}$:~расплывание импульса $\Delta\tau\approx0,33\cdot 10^{-8}
\hbox{ с}.$

     {\bf 10.}~Какие  типы  волн  могут  распространяться  на  частоте
10~ГГц в круглом волноводе с радиусом 1,1~см, заполненном диэлектриком
с $\varepsilon=3$?

     $\mbox{\textit {Ответ}}$:~$E_{01}$~(2,405), $E_{11}$~(3,832), $H_{11}$~(1,841),
$H_{21}$~(3,054), $H_{01}$~(3,832).

     {\bf 11.}~В   круглом   волноводе   радиуса  2,5~см,  заполненном
диэлектриком,  распространяется на частоте 3~ГГц волна типа  $H_{11}$
с  фазовой  скоростью,  равной  скорости  света.  Определите
диэлектрическую проницаемость вещества, заполняющего волновод.

     $\mbox{\textit {Ответ}}$:~$\varepsilon=2,374$.

     {\bf 12.}~Определите   размеры   прямоугольного   волновода,   по
которому должна передаваться только одна волна с частотой 3~ГГц, а для
круглого волновода диаметром 4~см --- диапазон одноволновости.\par
     Если в круглом волноводе волна типа $H_{11}$ не возбуждается,  то
в каком диапазоне радиусов $a$ на частоте 5~ГГц может распространяться
только волна типа $E_{01}$?

     $\mbox{\textit {Ответ}}$:~5~см$\ <a<\ $10~см, $b<\ $5~см; 4,395--5,741~ГГц; $2,296
\mbox{ см}<a<2,916\mbox{ см}$.

     {\bf 13.}~В прямоугольном волноводе,  в котором  распространяется
лишь  волна  типа  $H_{01}$,  имеется диэлектрическая вставка (пробка)
шириной    $d$    и     проницаемостью     вещества     $\varepsilon$.
Определите коэффициент прохождения $D$ волны
через вставку,  если ширина узкой  стенки  волновода  равна  $b$;  при
выполнении каких условий волна не отражается от вставки?

     $\mbox{\textit {Ответ}}$:~$D=2e^{-ih_1d}/\Delta$, где $\Delta=2\cos h_2d-i[(h_1^2+
h_2^2)/h_1h_2]\cdot\sin  h_2d$,  $h_1=  \sqrt{k^2-  \pi^2/b^2}$,~$h_2=
\sqrt{k^2\varepsilon   -   \pi^2/b^2}$;  отражение  отсутствует,  если
$h_1=h_2$ или $h_2d=n\pi$,~$n=1$,~$2$, ~$3$,~$ \ldots$.

     {\bf 14.}~Определите критические длины основных волн магнитного и
электрического типов в волноводе  полукруглого  сечения  радиуса  $a$;
изобразите силовые линии поля этих волн.

     $\mbox{\textit {Ответ}}$:~$H_{11}$ ($\lambda_{11}  =3,41a$)  и $E_{11}$ ($\lambda_
{11}= 1,64a$).

     {\bf 15.}~Волна  типа $TEM$ распространяется в коаксиальной линии
передачи с  размерами  проводников  0,9~см  и~2,1~см,  диэлектрик  ---
воздух.     Определите     амплитудные    значения    напряжённостей
электрического  и  магнитного  полей  $TEM$  волны   на   поверхностях
внутреннего и внешнего проводников и предельную передаваемую мощность,
если  пробой  происходит  при  напряжённости   электрического   поля
30~кВ/см.

     $\mbox{\textit {Ответ}}$:~$E_r=30\hbox{ кВ/см}$,~$H_  \varphi=8  \cdot10^3  \hbox{
А/м}\mbox{    и}$   $E_r=12,86\hbox{   кВ/см}$,~$   H_\varphi=   3,429
\cdot10^3\hbox{ А/м}$; $P=1,287\mbox{ МВт}$.

     {\bf 16.}~Найдите зависимость от длины волны генератора $\lambda$
давления,  которое оказывает электромагнитное поле на стенки структуры
в  случае:  а  --- волны типа $TEM$ в коаксиальной линии;  б --- волны
типа $H_{10}$ в прямоугольном волноводе.

     $\mbox{\textit {Ответ}}$:~а --- давление равно  нулю;  б  ---  давление  на  узкую
стенку пропорционально $-\lambda^2/ \lambda_{10}^2$,  а на широкую ---
$(1-\lambda^2/ \Lambda ^2) \sin^2(\pi x/a)- \lambda^2/  \lambda_{10}^2
\cos^2(\pi x/a)$.

     {\bf 17.}~Коаксиальная линия с радиусами проводников  $a=0,95$~см
и  $b=2$~см  служит для передачи мощности 10 кВт на длине волны 50 см;
диэлектриком      является      воздух,      проводимость       стенок
$\sigma=1,258\cdot10^{17}$~с$^{-1}$.   Определите   мощность,  которая
будет выделяться на участке длиной 1~м, прилегающем к генератору, если
затухание определяется соотношением

     $$h''=\frac{c}{8\pi\sigma\delta}\sqrt{\frac{\varepsilon}{\mu}}
         \cdot\frac{1/a+1/b}{\displaystyle {\ln(b/a)}}\,,$$

где $\delta$ --- толщина скин-слоя.

     $\mbox{\textit {Ответ}}$:~72~Вт.

     {\bf 18.}~Найдите  максимальное значение мощности,  которую можно
передать на частоте 3~ГГц  в  прямоугольном  волноводе  с  поперечными
размерами,   равными  9,9~см  и  4,9~см,  при  помощи  волны  $H_{10}$
(предельное  значение  напряжённости   электрического   поля   равно
$\sim30$~кВ/см).   Определите   максимальное   значение  поверхностной
плотности тока.

     $\mbox{\textit {Ответ}}$:~25~МВт;~6,7~кА/м.

     {\bf 19.}~В  круглом  волноводе  радиусом 2,5~см распространяется
волна типа $E_{01}$,  которая на частоте  6  ГГц  передаёт  мощность
20~кВт.     Определите     максимальное    значение    напряжённости
электрического поля в волноводе и  амплитуду  поверхностной  плотности
тока на стенках волновода.

     $\mbox{\textit {Ответ}}$:~161~кВ/м (на оси волновода);~290~А/м.

     {\bf 20.}~В  медном  ($\sigma=5\cdot10^{17}$~с$^{-1}$)  волноводе
квадратного сечения со стороной $a=2$~см распространяется только волна
типа $H_{11}$, затухание которой описывается выражением
     $$h''=\frac1a\sqrt{\frac{f_{11}}{\sigma}}\cdot\frac{\nu^2+1}
         {\sqrt{\nu}\sqrt{\nu^2-1}},\qquad\nu=\frac{f}{f_{11}}.$$
Определите частоту   поля,   при   которой   затухание  минимально,  и
минимальное значение затухания.  Какая мощность  будет  выделяться  на
участке   волновода   длиной   1~м,  прилегающем  к  генератору,  если
передаётся мощность 10~кВт на длине волны 1,17~см?

     $\mbox{\textit {Ответ}}$:~25,61~ГГц;~0,127~дБ/м;~292~Вт.

     {\bf 21.}~Диэлектрическая  пластинка  имеет  ширину  2~см,  длина
волны генератора --- 3.2~см. Определите фазовую скорость для магнитных
волн типа $H_{10}$ и~$H_{20}$, если известно, что поле на расстоянии 1~см от
пластинки  убывает соответственно в 11,658 и 4,865 раз;  испытывает ли
компонента электрического поля,  параллельная  поверхности  пластинки,
излом  в  функции  от  поперечной  координаты  на  границе пластинки и
вакуума?

     $\mbox{\textit {Ответ}}$:~$1,874\cdot10^{10}$~см/с и~$2,338 \cdot 10^ {10}
$~см/с;~нет.

     {\bf 22.}~Исходя из приближённого дисперсионного уравнения  для
гофры   $p=k\tg  kl$,  определите  фазовую  скорость  и  длину  волны,
распространяющейся на частоте  10~ГГц  вдоль  гофры  с  глубиной  паза
$l=0,62$~см. На каком расстоянии от гофры поле убывает в 10 раз?

     $\mbox{\textit {Ответ}}$:~$0,807\cdot 10^{10}$~см/с,~$0,807$~см;~0,307~см.

     {\bf 23.}~Получите   в    приближении    идеального    проводника
характеристическое   уравнение   для   основной  волны  в  замедляющей
структуре,  состоящей из  гофры  с  глубиной  паза  $l$  и  плоскости,
удалённой от края рёбер гофры на расстояние $d$; ширину паза можно
считать пренебрежимо малой.

     $\mbox{\textit {Ответ}}$:~$p\th pd=k\tg kl$.

     {\bf 24.}~Какую минимальную частоту  можно  возбудить  в  круглом
цилиндрическом  резонаторе,  радиус  которого равен 1.5~см и длина ---
1~см;  в прямоугольном резонаторе с размерами $2\hbox{ см}\times3\hbox
{ см}\times1\hbox{ см}$?

     $\mbox{\textit {Ответ}}$:~7,665~ГГц;~9,014~ГГц.

     {\bf 25.}~Перестраиваемый     резонатор     образован    отрезком
полубесконечного  прямоугольного   волновода
сечением  23~мм$\times$10~мм,  внутри  которого  перемещается поршень.
Определите границы перемещения поршня  для  перестройки  резонатора  в
пределах 8--12~ГГц на типе колебания $H_{101}$.

     $\mbox{\textit {Ответ}}$:~1,489--3,237~см.

     {\bf 26.}~Найдите  резонансные  частоты  структуры   из   четырёх
одинаковых     связанных     резонаторов     (цепочка    резонаторов),
воспользовавшись уравнением для одномодового приближения
     $$(\omega^2-\omega^2_0)\cdot A_n-K\cdot(A_{n-1}+A_{n+1})=0,$$
где $\omega_0$ --- резонансная частота отдельного резонатора,  $K$ ---
коэффициент  связи  между  смежными  резонаторами,  $A_n(\omega)$  ---
амплитуда поля в $n$-том резонаторе (дальние торцы крайних резонаторов
замкнуты накоротко).

     $\mbox{\textit {Ответ}}$:~$\omega^2_{1,2,3,4}=\omega^2_0\pm\Bigl((3\pm\sqrt{5})/2
         \Bigr)^{1/2}\cdot K$.

     {\bf 27.}~Оцените  для  объёмного  кубического  резонатора   со
стороной  $a$  и  проводимостью  стенок $\sigma$~$(\mu=1)$ добротность
произвольного типа колебаний.  Воспользовавшись формулой Релея-Джинса,
определите, при каких частотах резонансные свойства системы исчезнут?

     $\mbox{\textit {Ответ}}$:~$Q_n\approx\sqrt{\pi a^2\sigma\omega_n/2c^2}$, $\omega_n
\approx10^9 \sigma^{1/5} a^{-4/5}$.

     {\bf 28.}~Определите предельное значение энергии,  которая  может
быть  накоплена  в  коаксиальном  резонаторе длиной~8~см и с размерами
проводников 0,5~см и~2~см на  основном  типе  колебаний  (максимальная
напряжённость  электрического поля равна 30 кВ/см).

     $\mbox{\textit {Ответ}}$:~$3,466\cdot10^{-4}\mbox{ Дж}$.

     {\bf 29.}~Определите   максимальную  амплитуду  напряжённости
электрического поля в цилиндрическом резонаторе,  в  котором  запасена
энергия 0,01~Дж на типе колебаний $E_{010}$. Радиус резонатора равен 6~см,
длина --- 20~см.

     $\mbox{\textit {Ответ}}$:~$P=1,926\mbox { МВ/м}$.

     {\bf 30.}~Максимальная  амплитуда  напряжённости электрического
поля      в      прямоугольном      резонаторе       с       размерами
20~см$\times$10~см$\times$30~см  равна  $10^5$ В/м,  тип колебания ---
$H_{101}$. Определите запасённую энергию.

     $\mbox{\textit {Ответ}}$:~$6,63\cdot10^{-5}\mbox{ Дж}$.

     {\bf 31.}~В прямоугольном волноводе распространяется  волна
типа $H_{10}$, которая возбуждается элементарным электрическим диполем
$\rv j^e=J_0l_0\,\delta(x-x_0)\,\delta(y)\,\delta(z-z_0)\rv e_y$. Определите
комплексную   амплитуду   вынужденного   электрического    поля    для
бесконечного  волновода  и  для  волновода,  закороченного  проводящей
плоскостью  $z=0$.  При  каких  значениях  $x_0$   и~$z_0$   мощность,
отдаваемая в волновод, максимальна?

     $\mbox{\textit {Ответ}}$:
     $$\hspace*{-1.6cm}E_y=\frac{4\pi  kJ_0l_0}{chab}\cdot\sin(\pi x_0/a)
         \cdot\sin\frac{\pi x}{a}\cdot e^{\mp ih(z-z_0)};$$
     $$E_y=i\frac{8\pi kJ_0l_0}{chab}\cdot\sin\frac{\pi x_0}{a}\cdot\sin
         hz_0\cdot\sin\frac{\pi x}{a}\, e^{ihz},\quad z>z_0;$$
~$x_0=a/2$,~$z_0={(2n+1)\Lambda/4}$.

     $\bf 32.$~Цилиндрический   резонатор   возбуждается   петлей   на
резонансной частоте $\omega_{010}$  колебания  типа  $E_{010}$;  петля
расположена в плоскости $rz$ и создаёт магнитный ток
     $$j^m_\varphi=-i\frac{\omega_{010}I_0S_0}{2\pi cr}\,\delta(r-r_0)
         \,\delta(z-z_0),$$
причём заданными считаются ток $I_0$ в петле и её площадь $S_0$, а
также  добротность  колебания  $Q_{010}$  ($Q_{010}\gg1$).  Определите
комплексные амплитуды вынужденного электромагнитного поля в резонаторе
при   расположении   петли,   соответствующему   возбуждению  полей  с
максимальной амплитудой.

     $\mbox{\textit {Ответ}}$:~$r_0=a$,
     $$E_z=-4\frac{Q_{010}I_0S_0J_0(\nu_{01}r/a)}{ca^2d J_1(\nu_{01}
         )},\qquad H_\varphi=4i\frac{Q_{010}I_0S_0 J_1(\nu_{01}r/a)}
         {ca^2d J_1(\nu_{01})},$$
где $d$ --- длина резонатора.

     {\bf 33.}~Прямоугольный   резонатор   с  размерами  $a$,~$b$,~$d$
возбуждается щелью длиной $l_0$ на резонансной частоте  $\omega_{101}$
колебания  типа  $H_{101}$.  Комплексная  амплитуда  напряжения  между
краями  щели  равна  $U_0$,   добротность   резонатора   $Q_{101}\gg1$
известна,  влияние  щели  можно учесть при помощи магнитного тока
     $$\rv j^m=\rv e_x\frac{cU_0l_0}{4\pi}\,\delta(x-x_0)\,\delta(y-y_0)
        \,\delta(z).$$
Определите комплексную амплитуду электромагнитного поля в резонаторе.

     $\mbox{\textit {Ответ}}$:
     $$E_y=i\frac{4Q_{101}U_0l_0a}{\pi  b(a^2+d^2)}\cdot\sin\frac{\pi x_0}{a}
         \cdot\sin\frac{\pi x}{a}\cdot\sin\frac{\pi z}{d}.$$

     {\bf 34.}~В   широкой   стенке   полубесконечного  прямоугольного
волновода ($\lambda/2<a<\lambda$,~$b<\lambda/2$) прорезана  поперечная
щель  с длиной $l_0\ll\lambda$,  к которой приложено постоянное (вдоль
щели) напряжение $U_o$;  щель может быть описана при помощи магнитного
тока
     $$j_x^m=\frac{cU_0l_0}{4\pi}\,\delta(x-x_0)\,\delta(y)\,\delta(z-z_0).$$
Определите комплексную  амплитуду  напряжённости электрического поля
вдали от щели.  При каком расположении щели передаваемая по  волноводу
мощность максимальна?

     $\mbox{\textit {Ответ}}$:
     $$E_y=\frac{2U_0l_0}{ab}\cdot\sin\frac{\pi x_0}{a}\cdot\cos hz_0\cdot
         \sin\frac{\pi x}{a} \, e^{ihz};$$
$x_0=a/2$,~$z_0=\Lambda/2$.

     {\bf 35.}~Полубесконечный круглый волновод,  имеющий радиус  $a$,
возбуждается   тонким   штырём  длиной  $l_0<\lambda$,  по  которому
протекает переменный электрический ток с амплитудой  $J_0$  и  который
расположен  на  торце  резонатора $z=0$ параллельно оси $z$.  В случае
распространения только волны типа $E_{01}$
и  при оптимальном расположении штыря определите комплексную амплитуду
напряжённости магнитного поля в волноводе.

     $\mbox{\textit {Ответ}}$:
     $$H_\varphi=\frac{2i J_0 l_0\lambda\nu_{01}}{c\pi a^3J_1^2(\nu_
         {01})\sqrt{1- (\lambda \nu_{01} /2\pi a)^2}}\cdot J_1\Bigl(
         \nu_{01}\frac{r}{a}\Bigr)\,e^{ihz}.$$

     {\bf 36.}~С   одной   стороны   трёхдецибельного  направленного
ответвителя симметрично относительно   общей    стенки  расположены
механически спаренные закорачивающие поршни.  Покажите,  что   в такой
раздвижной   волноводной   линии (фазовращателе) можно как   угодно
регулировать   фазу выходного сигнала, не  нарушая согласования
прибора в целом.

     {\bf 37.}~На  сколько  изменится  частота  собственных колебаний
основного типа в прямоугольном резонаторе с размерами $a=5$~см,~$b=3$~
см,~$d=6$~см, если в середине верхней крышки ($x=a/2$,~$y=b$,~$z=d/2$)
вставить металлический подстроечный винт высотой $h=3$~мм и  диаметром
$D=5$~мм?

     $\mbox{\textit {Ответ}}$:~на~$\sim-5,11$~МГц.

     {\bf 38.}~Для немагнитных материалов ($\mu=1$,~$\varepsilon=4 \pi
i\sigma/\omega$) коэффициент отражения при нормальном падении  плоской
волны  можно  представить  в  виде $R\approx\exp[-(1-i)k\delta]$,  где
$\delta$ --- толщина скин-слоя.  Оцените в приближении  геометрической
оптики   добротность   открытых  резонаторов,  обусловленную  конечной
проводимостью  зеркал   и   излучением   через   боковую   поверхность
(характерный размер плоских параллельных зеркал равен $D$,  расстояние
между ними --- $2d$,  а собственное колебание представляет  собой  две
волны   с   $\lambda\ll2d,\;D$,   распространяющиеся   перпендикулярно
зеркалам навстречу друг другу и образующие стоячую волну).

     $\mbox{\textit {Ответ}}$:~$d/\delta$;~$\leqslant D^2\omega^2/2c^2$.

     {\bf 39.}~Для   открытого  резонатора,  описанного  в  предыдущей
задаче,  оцените  в  приближении  геометрической  оптики   добротность
колебания на частоте $\omega$,  обусловленную небольшой непараллельностью
плоских зеркал (угол между зеркалами равен $\theta$). Как найти полную
добротность  открытого  резонатора,  учитывающую  омические  потери  в
зеркалах,  излучение через боковую  поверхность  и  небольшой  перекос
зеркал?

     $\mbox{\textit {Ответ}}$:~$k\sqrt{2dD/\theta}$; по      правилу      параллельного
соединения.

%\end{document}

\newpage
\oddsidemargin=-0.4mm \evensidemargin=-0.4mm
\topmargin=-0.4mm
\headsep=7mm
\textheight=231.875mm
\textwidth=160mm
\mathsurround=2.5pt
\unitlength=1mm
%\begin{document}
%\input{macr.tex}
\thispagestyle{empty}
%\addtocounter{page}{386}

\begin{center}
   \subsubsection*{\rm П\,Р\,И\,Л\,О\,Ж\,Е\,Н\,И\,Я}
\end{center}\vspace{1cm}

\begin{center}{\bf П-1.  Перевод выражений и  уравнений  из  гауссовой
системы в СИ и наоборот; соотношения между единицами физических величин
}\end{center}

     Таблица 1  позволяет осуществить перевод выражений и уравнений из
гауссовой  (симметричной  СГС,  смешанной)  системы  единиц  в  СИ   и
наоборот;  масса, длина, время и другие неэлектродинамические величины
при  этом  не  изменяются.  Соотношения  между  некоторыми   единицами
физических величин приведены в Таблице 2;  множители 3 (кроме входящих
в показатели степени) следует при уточнённых расчётах заменить  на
2,997925. \vspace{0.5cm}\par

\hbox to \textwidth{\hfill\small{Таблица 1. Перевод из гауссовой
системы в СИ и наоборот.\hfil}}\vspace{-0.5cm}
     $$\begin{array}{|c|c|}
         \hline\vspace{-0.2cm}&\\\vspace{-0.2cm}
         \hspace{2.8cm}\hbox{Гауссова}\hspace{2.8cm}&
         \hspace{3.5cm}\hbox{СИ}\hspace{3.5cm}\\&\\
         \hline\vspace{-0.2cm}&\\\vspace{-0.2cm}
         c\quad\mbox{(скорость света)}&(\varepsilon_0\mu_0)^{-1/2}\\&\\
         \mbox{\boldmath
 $\gv E$}\quad(\Phi,V)&\sqrt{4\pi\varepsilon_0}\mbox{\boldmath$
 \gv E$}
           \quad(\Phi,V)\\&\\
         \mbox{\boldmath
 $\gv D$}&\sqrt{\displaystyle{\frac{4\pi}{\varepsilon_0}}}\mbox{\boldmath
 $\gv D$}\\&\\
         \rho\quad(q,\mbox{\boldmath
$\gv j$},J,\mbox{\boldmath
$\gv P$})&\displaystyle{\sqrt{
         \frac{1}{4\pi\varepsilon_0}}}\rho\quad(q,\mbox{\boldmath
$\gv j$},J,\mbox{\boldmath
$\gv P$})\\&\\
         \mbox{\boldmath
 $\gv B$}&\sqrt{\displaystyle{\frac{4\pi}{\mu_0}}}\mbox{\boldmath
 $\gv B$}\\&\\
         \mbox{\boldmath
 $\gv H$}&\sqrt{4\pi\mu_0}\mbox{\boldmath
 $\gv H$}\\&\\
         \mbox{\boldmath
 $\gv M$}&\sqrt{\displaystyle{\frac{\mu_0}{4\pi}}}\mbox{\boldmath
 $\gv M$}\\&\\
         \sigma\quad\mbox{(проводимость)}&\sigma/4\pi\varepsilon_0\\&\\
         \varepsilon&\varepsilon/\varepsilon_0\\&\\
         \mu&\mu/\mu_0\\&\\
         R\quad(L)&4\pi\varepsilon_0R\quad(L)\\&\\
         C&C/4\pi\varepsilon_0\\&\\
         \hline\end{array}$$
         \newpage
\hbox to  \textwidth{\hfill\small{Таблица   2.   Соотношения   между
некоторыми единицами физических величин.\hfil}}\vspace{-0.5cm}
$$\begin{array}{|l|c|c|}\hline\vspace{-0.2cm}&&\\\vspace{-0.2cm}
\hbox{Наименование величины}&\hbox{Гауссова система}&\hbox{СИ}\\&&\\
     \hline\vspace{-0.2cm}&&\\\vspace{-0.2cm}
     \mbox{   Длина}&\mbox{1 см}&10^{-2}\mbox{ м}\\&&\\
         \vspace{-0.25cm}
     \mbox{   Масса}&\mbox{1 г}&10^{-3}\mbox{ кг}\\&&\\
         \vspace{-0.25cm}
     \mbox{   Время}&\mbox{1 сек}&1\mbox{ с}\\&&\\\vspace{-0.25cm}
     \mbox{   Сила}&\mbox{1 дин(а)}&10^{-5}\mbox{ Н}\\&&\\
         \vspace{-0.25cm}
     \mbox{   Давление}&\mbox{1 дин/см}^2&10^{-1}\mbox{ Н/м}^2\\&&\\
         \vspace{-0.25cm}
     \mbox{   Работа, энергия}&\mbox{1 эрг}&10^{-7}\mbox{ Дж}\\&&\\
         \vspace{-0.25cm}
     \mbox{   Мощность}&\mbox{1 эрг/сек}&10^{-7}\mbox{ Вт}\\&&\\
         \vspace{-0.25cm}
     \mbox{   Сила электрического тока}&\mbox{1 статампер}&\displaystyle
         {\frac13}\cdot10^{-9}\mbox{ А}\\&&\\\vspace{-0.25cm}
     \mbox{   Количество электричества; электриче-}&\mbox{1 статкулон}&
         \displaystyle{\frac13}\cdot10^{-9}\mbox{ Кл}\vspace{-0.05cm}\\
         \mbox{ский заряд}&&\\&&\\\vspace{-0.25cm}
     \mbox{   Поверхностная плотность электриче-}&\mbox{1 статкулон/см}
         ^2&\displaystyle{\frac13}\cdot10^{-5}\mbox{ Кл/м}^2\vspace
         {-0.05cm}\\\mbox{ского заряда}&&\\&&\\\vspace{-0.25cm}
     \mbox{   Пространственная плотность электри-}&\mbox{1 статкулон/см}
         ^3&\displaystyle{\frac13}\cdot10^{-3}\mbox{ Кл/м}^3\vspace
         {-0.05cm}\\\mbox{ского заряда}&&\\&&\\\vspace{-0.25cm}
     \mbox{   Напряжённость электрического поля}&\mbox{1 статвольт/см}
         &3\cdot10^{4}\mbox{ В/м}\\&&\\\vspace{-0.25cm}
     \mbox{   Электрическое напряжение, электри-}&\mbox{1 статвольт}&
         3\cdot10^2\mbox{ В}\vspace{0.1cm}\\\mbox{ский потенциал,
         ЭДС}&&\\&&\\\vspace{-0.25cm}
     \mbox{   Поток электрического смещения}&\mbox{1 статкулон}&
         \displaystyle{\frac{1}{12\pi}}\cdot10^{-9}\mbox{ Кл}\\&&\\
         \vspace{-0.25cm}
     \mbox{   Диэлектрическая поляризация}&\mbox{1 статкулон/см}^2&
         \displaystyle{\frac13}\cdot10^{-5}\mbox{ Кл/м}^2\vspace
         {-0.05cm}\\&\mbox{(1 статвольт/см})&\\&&\\\vspace{-0.25cm}
     \mbox{   Электрическое смещение}&\mbox{1 статвольт/см}&
         \displaystyle{\frac{1}{12\pi}}\cdot10^{-5}\mbox{ Кл/м}^2\vspace
         {-0.05cm}\\&\mbox{(1 статкулон/см}^2)&\\&&\\\vspace{-0.25cm}
     \mbox{   Электрическая ёмкость}&\mbox{1 см}&\displaystyle
         {\frac19}\cdot10^{-11}\mbox{ Ф}\\&&\\\vspace{-0.25cm}
     \mbox{   Абсолютная диэлектрическая прони-}&\mbox{---}&8,854186
         \cdot10^{-12}\mbox{ Ф/м}\vspace{0.1cm}\\\mbox{цаемость}
         &&\\&&\\
         \hline\end{array}$$\newpage
\hbox to\textwidth{\hfill\small{Продолжение Таблицы 2.\hfil}}
\vspace{-0.5cm}
$$\begin{array}{|l|c|c|}\hline\vspace{-0.2cm}&&\\\vspace{-0.2cm}
\hbox{Наименование величины}&\hbox{Гауссова система}&\hspace{1.65cm}
\hbox{СИ}\hspace{1.65cm}\\&&\\
\hline\vspace{-0.2cm}&&\\
     \mbox{   Электрический момент}&\mbox{1 статкулон}\cdot\mbox{см}
         &\displaystyle{\frac13}\cdot10^{-11}\mbox{ Кл}\cdot\mbox{м}
         \\&&\\\vspace{-0.25cm}
     \mbox{   Плотность электрического тока}&\mbox{1 статампер/см}^2
         &\displaystyle{\frac13}\cdot10^{-5}\mbox{ А/м}^2\\&&\\\vspace
         {-0.25cm}
     \mbox{   Линейная плотность электрического}&\mbox{1 статампер/см}
         &\displaystyle{\frac13}\cdot10^{-7}\mbox{ А/м}\vspace{-0.05cm}
         \\\mbox{тока}&&\\&&\\\vspace{-0.25cm}
     \mbox{   Напряжённость магнитного поля}&\mbox{1 эрстед (Э)}
          &\displaystyle{\frac{1}{4\pi}}\cdot10^{3}\mbox{ А/м}\\&&\\
          \vspace{-0.25cm}
     \mbox{   Магнитодвижущая  сила;  разность}&\mbox{1 гильберт (Гб)}&
         \displaystyle{\frac{1}{4\pi}}\cdot10^1\mbox{ А}\vspace{-0.05cm}\\
         \mbox{магнитных потенциалов}&&\\&&\\\vspace{-0.25cm}
     \mbox{   Магнитная индукция}&\mbox{1 гаусс (Гс)}&10^{-4}\mbox{ Тл}
         \\&&\\\vspace{-0.25cm}
     \mbox{   Магнитный поток}&\mbox{1 максвелл (Мкс)}&10^{-8}\mbox{ Вб}
         \\&&\\\vspace{-0.25cm}
     \mbox{   Векторный потенциал}&\mbox{1 Гс}\cdot\mbox{см}&10^{-6}
         \mbox{ Тл}\cdot\mbox{м}\\&&\\\vspace{-0.25cm}
     \mbox{   Индуктивность, взаимная индуктив-}&\mbox{1 см}&10^{-9}
         \mbox{ Гн}\vspace{0.1cm}\\\mbox{ность}^{*)}&&\\&&\\
         \vspace{-0.25cm}
     \mbox{   Абсолютная магнитная проницаемость}&\mbox{---}&4\pi\cdot
         10^{-7}\mbox{ Гн/м}\\&&\\\vspace{-0.25cm}
     \mbox{   Магнитный момент (амперовский)}&\mbox{1 эрг/Гс}&
         10^{-3}\mbox{ А}\cdot\mbox{м}^2\\&&\\\vspace{-0.25cm}
     \mbox{   Магнитный момент (кулоновский)}&\mbox{1 Мкс}\cdot\mbox
         {см}&10^{-10}\mbox{ Вб}\cdot\mbox{м}\\&&\\\vspace{-0.25cm}
     \mbox{   Намагниченность}&\mbox{1 эрг/(Гс$\cdot$см$^3$)}&
         10^3\mbox{ А/м}\\&&\\\vspace{-0.25cm}
     \mbox{   Электрическое сопротивление}&\mbox{1 сек/см}&9\cdot10^
         {11}\mbox{ Ом}\\&&\\\vspace{-0.25cm}
     \mbox{   Электрическая проводимость}&\mbox{1 см/сек}&\displaystyle
         {\frac19}\cdot10^{-11}\mbox{ См}\\&&\\\vspace{-0.25cm}
     \mbox{   Удельное электрическое сопротивление}&\mbox{1 сек}&9
         \cdot10^{9}\mbox{ Ом}\cdot\mbox{м}\\&&\\\vspace{-0.25cm}
     \mbox{   Удельная электрическая проводимость}&\mbox{1 сек}^{-1}&
         \displaystyle{\frac19}\cdot10^{-9}\mbox{ См/м}\\&&\\\vspace
         {-0.25cm}
     \mbox{   Магнитное сопротивление}&\mbox{1 Гб/Мкс}&\displaystyle
         {\frac{1}{4\pi}}\cdot10^{9}\mbox{ А/Вб}\\&&\\\vspace{-0.25cm}
     \mbox{   Магнитная проводимость}&\mbox{1 Мкс/Гб}&4\pi\cdot10^{-9}
         \mbox{ Гн}\\&&\\
         \hline\end{array}$$\vspace*{0.15cm}\par
$^{*)}$ В (модифицированной) гауссовой системе единиц ток определяется
как  $I_m=dq/(c\cdot  dt)$  и  поэтому индуктивность имеет размерность
длины;  если  ток  измеряется  в   электростатических   единицах,   то
1~СГСЭ=$9\cdot10^{11}$~Гн.
\newpage

\begin{center}{\bf П-2. Некоторые часто встречающиеся постоянные}
\end{center}\begin{center}

\begin{tabular}{rl}
$c\ =$&$2,99792458\cdot10^{10}$~cм/c  --- скоpость света\\[0.5cm]
$\hbar\ =$&$1,05457\cdot10^{-27}$~эpг$\cdot$с  --- постоянная
Планка\\[0.5cm]
$e\ =$&$4,80286\cdot10^{-10}$~СГСЭ --- заpяд электpона\\[0.5cm]
$m_e=$&$9,10939\cdot10^{-28}$~г     --- масса электpона\\[0.5cm]
$m_p=$&$1,67262\cdot10^{-24}$~г    --- масса пpотона\\[0.5cm]
$\varepsilon_0=$&$8,85419\cdot10^{-12}$~Ф/м --- диэлектpическая
постоянная вакуума\\[0.5cm]
$\mu_0=$&$4\pi\cdot10^{-7}$~Гн/м --- магнитная постоянная вакуума\\[0.5cm]
$r_e=$&$2,81794\cdot10^{-13}$~см --- классический pадиус электpона
\\[0.5cm]\end{tabular} \end{center}
$$\pi=3,14159265;\qquad e=2,71828183;\qquad e^
C=1,78107242;$$
$$C=0,57721566\;\hbox{--- постоянная Эйлера}.$$
\newpage
\bigskip\bigskip\begin{center}{\bf П-3.    Ортогональные   координаты}
\medskip \end{center}

$$ds^2=h_1^2dq_1^2+h_2^2dq_2^2+h_3^2dq_3^2$$\smallskip
$$dV=h_1h_2h_3dq_1dq_2dq_3$$
\smallskip
$$h_i=\sqrt{\left(\frac{\partial x}{\partial q_1}\right)^2+\left(\frac
{\partial y}{\partial q_2}\right)^2+\left(\frac{\partial z}{
\partial q_3}\right)^2}$$\\\medskip
\hspace{3cm}Сфеpическая система: $h_r=1$, $h_\theta=r$,
$h_\varphi=r\sin\theta$.\\\smallskip
\hspace{3cm}Цилиндpическая система: $h_\rho=1$, $h_\varphi=\rho$,
$h_z=1$.\\\smallskip
\hspace{3cm}Пpямоугольная система: $h_x=h_y=h_z=1$.\\
\bigskip
$$
\begin{array}{ll}
x=r\sin\theta\cos\varphi & x=\rho\cos\varphi\\
y=r\sin\theta\sin\varphi & y=\rho\sin\varphi\\
z=r\cos\theta & z=z\\
ds^2=dr^2+r^{2}d\theta^2+r^{2}\sin^{2}\theta d\varphi^2 & ds^2=
d\rho^2+\rho^{2}d\varphi^2+dz^2
\end{array}
$$
\bigskip
$$
\begin{array}{ll}
F_r=F_x\sin\theta\cos\varphi+F_y\sin\theta\sin\varphi+F_z\cos
\theta& \qquad F_\rho=F_x\cos\varphi+F_y\sin\varphi\\
F_\theta=F_x\cos\theta\cos\varphi+F_y\cos\theta\sin
\varphi-F_z\sin\theta & \qquad F_\varphi=-F_x\sin\varphi+F_y\cos
\varphi\\F_\varphi=-F_x\sin\varphi+F_y\cos\varphi & \qquad F_z=F_z
\end{array}
$$
\bigskip
$$
\hspace{-0.4cm}\begin{array}{ll}
F_x=F_r\sin\theta\cos\varphi+F_\theta\cos\theta\cos\varphi-
        F_\varphi\sin\varphi\qquad& F_x=F_\rho\cos\varphi
        -F_\varphi\sin\varphi\\
F_y=F_r\sin\theta\sin\varphi+F_\theta\cos\theta\sin\varphi+F_\varphi
        \cos\varphi\qquad& F_y=F_\rho\sin\varphi+F_\varphi\cos\varphi\\
F_z=F_r\cos\theta-F_\theta\sin\theta\qquad& F_z=F_z
\end{array}
$$
\smallskip
$$(\grad\Phi)_i=\frac{1}{h_i}\frac{\partial \Phi}{\partial q_i}$$
\smallskip
$$\div\rv A=\frac{1}{h_1h_2h_3}{\left[\frac{\partial (h_2h_3A_1)}{
        \partial q_1}+\frac{\partial (h_1h_3A_2)}{\partial q_2}+
        \frac{\partial (h_1h_2A_3)}{\partial q_3}\right]}$$
\smallskip
$$\rot\rv A=\left|
    \begin{array}{ccc}
        \displaystyle{\frac{\vec\imath_1}{h_2h_3}}&
        \displaystyle{\frac{\vec\imath_2}{h_1h_3}}&
        \displaystyle{\frac{\vec\imath_3}{h_1h_2}}\\[0.5cm]
        \displaystyle{\frac{\partial}{\partial q_1}}&
        \displaystyle{\frac{\partial}{\partial q_2}}&
        \displaystyle{\frac{\partial}{\partial q_3}}\\[0.5cm]
        h_1A_1&h_2A_2&h_3A_3\\
    \end{array}
        \right|$$
\bigskip
$$
\Delta \Phi=\frac{1}{h_1h_2h_3}
{\left[{
        \frac{\partial}{\partial q_1}\, {{\left(\frac{h_2h_3}{h_1}\,
        \frac{\partial\Phi}{\partial q_1}\right)}}+
        \frac{\partial}{\partial q_2}\, {{\left(\frac{h_1h_3}{h_2}\,
        \frac{\partial\Phi}{\partial q_2}\right)}}+
        \frac{\partial}{\partial q_3}\, {{\left(\frac{h_1h_2}{h_3}\,
        \frac{\partial\Phi}{\partial q_3}\right)}}
}\right]}
$$ \bigskip
\vspace{0.5cm}
\begin{center}Проекции вектора  $\Delta\rv  A=\Delta\rv   A$   в
сферической   и цилиндрической\\
системах координат:\end{center}
\medskip
$$
\begin{array}{ll}
(\Delta\rv A)_r&=
\displaystyle{\Delta A_r-\frac{2 A_r}{r^2}-
\frac{2}{r^2\sin\theta}\,
\frac{\partial}{\partial\theta}(\sin\theta A_\theta)-
\frac{2}{r^2\sin\theta}\,
\frac{\partial A_\varphi}{\partial\varphi}},\\[0.5cm]
(\Delta\rv A)_\theta&=
\displaystyle{\Delta A_\theta-\frac{A_\theta}
{r^2\sin^2\theta}+\frac{2}{r^2}\,\frac{\partial A_r}{\partial\theta}-
\frac{2\cos\theta}{r^2\sin^2\theta}\,
\frac{\partial A_\varphi}{\partial\varphi}},\\[0.5cm]
(\Delta\rv A)_\varphi&=
\displaystyle{\Delta A_\varphi-\frac{A_\varphi}{
r^2\sin^2\theta}+\frac{2}{r^2\sin\theta}\,
\frac{\partial A_r}{\partial\varphi}+
\frac{2\cos\theta}{r^2\sin^2\theta}\,
\frac{\partial A_\theta}{\partial\varphi}},\\[0.9cm]
(\Delta\rv A)_\rho&=\displaystyle{\Delta A_\rho-\frac{A_\rho}{
\rho^2}-
\frac{2}{\rho^2}\,\frac{\partial A_\varphi}{\partial\varphi}},\\[0.5cm]
(\Delta\rv A)_\varphi&=\displaystyle{\Delta A_\varphi-\frac{A_\varphi
}{\rho^2}+\frac{2}{\rho^2}\,\frac{\partial A_\rho}{\partial\varphi}},
\\[0.5cm]
(\Delta\rv A)_z&=\Delta A_z.
\end{array}
$$\par\vspace{1cm}
\newpage
\vbox{\centering{\bf П-4. Некоторые векторные соотношения}}
\bigskip
$$\rv A[\rv B\rv C]= \left|
    \begin{array}{ccc}
        A_x&B_x&C_x\\[0.5cm]A_y&B_y&C_y\\[0.5cm]A_z&B_z&C_z\\
    \end{array}
        \right|=[\rv A\rv B\rv C],$$
\vspace{0.4cm}
$$\begin{array}{ll}
[\rv A[\rv B\rv C]]=\rv B(\rv A\rv C)-
\rv C(\rv A\rv B),&{}\\[0.5cm]
[\rv A\rv B][\rv C\rv D]=(\rv A\rv C)(\rv B\rv D)-
(\rv A\rv D)(\rv B\rv C),&{}\\[0.5cm]
[\rv A\rv B]^2=\rv A^2\rv B^2-(\rv A\rv B)^2,&{}\\[0.5cm]
[[\rv A\rv B][\rv C\rv B]]=[\rv A\rv C\rv D]\rv B-
[\rv B\rv C\rv D]\rv A=
[\rv A\rv B\rv D]\rv C- [\rv A\rv B\rv C]\rv D,&{}\\[0.5cm]
\rv A=\rv n(\rv A\rv n)+[\rv n[\rv A\rv n]],\qquad
\mbox{где $\rv n$ --- единичный вектоp.}&{}\\[0.2cm]
\end{array}$$
$$\begin{array}{ccc}
\rot\grad\Phi\equiv0\,,&\hspace{1cm}\div\rot\rv A\equiv0\,,\hspace{1cm}&
        \div\grad\Phi\equiv\Delta\Phi\,,
\end{array}$$
\vspace{0.1cm}
$$\begin{array}{ll}
\grad(\Phi\Psi)=\Phi\grad\Psi+\Psi\grad\Phi,&{}\\[0.5cm]
\div(\Phi\rv A)=\Phi\div\rv A+\rv A\grad\Phi,&{}\\[0.5cm]
\end{array}$$
$$\begin{array}{ll}
\rot(\Phi\rv A)=\Phi\rot\rv A-[\rv A\grad\Phi],&{}\\[0.5cm]
\div[\rv A\rv B]=\rv B\rot\rv A- \rv A\rot\rv B,&{}\\[0.5cm]
\rot[\rv A\rv B]=\rv A\div\rv B-\rv B\div\rv A
+(\rv B\nabla)\rv A-(\rv A\nabla)\rv B,&{}\\[0.5cm]
\grad(\rv A\rv B)=[\rv A\rot\rv B]+[\rv B\rot\rv A]
+(\rv B\nabla)\rv A+(\rv A\nabla)\rv B,&{}\\[0.25cm]
\end{array}$$
$$\begin{array}{ll}
\nabla\Phi(x,y,z)\equiv\grad\Phi(x,y,z),&{}\\[0.5cm]
\nabla\!\rv A(x,y,z)\equiv\div\rv A,&{}\\[0.5cm]
\left[\nabla\rv A(x,y,z)\right]\equiv\rot\rv A(x,y,z),&{}\\[0.5cm]
(\rv A\!\nabla)\Phi=A_x\displaystyle{\frac{\partial\Phi}{\partial x}}
+A_y\displaystyle{\frac{\partial\Phi}{\partial y}}
+A_z\displaystyle{\frac{\partial\Phi}{\partial z}},&{}\\[0.5cm]
(\rv A\!\nabla)\rv B=(\rv A\cdot\nabla B_x)\vec\imath
+(\rv A\cdot\nabla B_y)\vec\jmath
+(\rv A\cdot\nabla B_z)\vec k\,.
\end{array}$$
\newpage

{\it Полный диффеpенциал} $d\Phi=d\rv r\cdot
        \grad\Phi=(d\rv r\cdot\nabla)\Phi$, где $\Phi=\Phi(\rv r)$.\par
\smallskip

{\it Полная пpоизводная} $\Phi(\rv r)$ вдоль кpивой
        $\rv r=\rv r(t)$ pавна
$$
\frac{d\Phi}{dt}=\left(\displaystyle{\frac{d\rv r}{dt}}\nabla\right)\Phi
                +\displaystyle{\frac{\partial\Phi}{\partial t}}.
$$

{\it Пpоизводная по напpавлению}, заданному единичным вектоpом
$$
        \rv n=\cos\alpha_x\cdot\vec\imath
        +\cos\alpha_y\cdot\vec\jmath+\cos\alpha_z\cdot\vec k,
$$
pавна
$$
\frac{d\Phi}{ds}=(\rv n\nabla)\Phi,\qquad\rv n=\frac{d\rv r}{ds}\,.
$$
\noindent
Для вектоpной функции точки эти тpи величины опpеделяются
аналогично.
\medskip
$$   \Delta=\nabla\nabla\,,\qquad\Delta\Phi\equiv
        \left(\frac{\partial^2}{\partial x^2}
        +\frac{\partial^2}{\partial y^2}
        +\frac{\partial^2}{\partial z^2}\right)\Phi(x,y,z)\,$$
$$    \Delta\rv A=\Delta A_x\cdot\vec\imath
        +\Delta A_y\cdot\vec\jmath+\Delta A_z\cdot\vec k\,,$$

$$
        \Delta(\Phi\Psi)=\Psi\Delta\Phi
        +2(\nabla\Phi)(\nabla\Psi)+\Phi\Delta\Psi,
$$
\medskip
$$
\begin{array}{ll}
        \div\grad\Phi=\Delta\Phi,&{}\\[0.5cm]
        \grad\div\rv A=\Delta\rv A+\rot\rot\rv A,&{}\\[0.5cm]
        \rot\grad\Phi=0\,,&{}\\[0.5cm]
        \div\rot\rv A=0\,.&{}\\[0.5cm]
\end{array}
$$

Если скалярная функция $\Psi$ является решением уравнения
$\Delta\Psi+k^2\Psi=0$ и  $\rv  a$ --- некоторый постоянный вектор,
то векторные функции $\rv L=\grad\Psi$,  $\rv M=\rot(\rv a\Psi)$,
$\rv N=\rot\rv M$ удовлетворяют уравнению $$\Delta\rv A+k^2\rv A=0.$$
\newpage
\vbox{\centering{\bf П-5. Интегральные теоремы}}
$$
ds=\sqrt{dx^2+dy^2+dz^2}\equiv\sqrt{\dot a^2+\dot y^2+\dot z^2}\,dt
        \equiv\displaystyle{\frac{ds}{dt}}\,dt\,,
$$
$$
        d\rv r=\vec\imath\,dx+\vec\jmath\,dy+\vec k\,dz\,.
$$
\smallskip
$$
\begin{array}{ll}
     \displaystyle{\int\limits_C}\rv A(\rv r)\,d\rv r
        =\displaystyle{\int\limits_C}\left[
        A_x(x,y,z)\,dx+A_y(x,y,z)\,dy+A_z(x,y,z)\,dz\right],&{}\\[1cm]
     \displaystyle{\int\limits_C}\Phi\,d\rv r
        =\vec\imath\displaystyle{\int\limits_C}\Phi(x,y,z)\,dx
        +\vec\jmath\displaystyle{\int\limits_C}\Phi(x,y,z)\,dy
        +\vec k\displaystyle{\int\limits_C}\Phi(x,y,z)\,dz\,,&{}\\[1cm]
     \displaystyle{\int\limits_C}[\rv A\,d\rv r]=
        \vec\imath\displaystyle{\int\limits_C}(A_y\,dz-A_z\,dy)
        +\vec\jmath\displaystyle{\int\limits_C}(A_z\,dx-A_x\,dz)
        +\vec k\displaystyle{\int\limits_C}(A_x\,dy-A_y\,dx)\,.&{}\\[1cm]
\end{array}
$$

Регулярная  часть  поверхности  определяется  уравнением  $\rv
r=\rv r(u,v)$\,, где $u$,$v$ --- координаты на поверхности:
$$
     \rv r_u=\displaystyle{\frac{\partial\rv r}{\partial u}}\,,
     \qquad\qquad
     \rv r_v=\displaystyle{\frac{\partial\rv r}{\partial u}}\,.
$$
Нормаль к поверхности и элемент поверхности определяются
соотношениями:
$$
     \rv n=\displaystyle{\frac{[\rv r_u\rv r_v]}{|[\rv r_u\rv r_v]|},\qquad
      d\rv S=[\rv r_u\rv r_v]\,du\,dv=\rv n|d\rv S|=\rv n\,dS}\,,
$$
\medskip\par
Следующие интегралы --- как и многие другие --- удобно вычислять,
применяя известное соотношение
     $ d\rv S=[\rv r_u\rv r_v]\,du\,dv$:
$$
     \int\limits_S\Phi(\rv r)\,dS,\qquad
     \int\limits_S\rv A(\rv r)\,d\rv S,\qquad
     \int\limits_S\Phi(\rv r)\,d\rv S,\qquad
     \int\limits_S(\rv A(\rv r)\,d\rv S\,.
$$
\medskip
$$
      \displaystyle{\int\limits_S}\rv A(\rv r)\,d\rv S\!
     =\!\displaystyle{\int\!\!\int\limits_{\hspace{-0.19cm}S}}
     A_x[x(y,z),y,z]\,dy\,dz\!
     +\displaystyle{\int\!\!\int\limits_{\hspace{-0.19cm}S}}
     A_y[x,y(x,z),z]\,dx\,dz\!
     +\displaystyle{\int\!\!\int\limits_{\hspace{-0.19cm}S}}
     A_z[x,y,z(x,y)]\,dy\,dz,
$$\par\vspace{0.5cm}
\hbox{$
\displaystyle{\int\limits_S}\left[\displaystyle{\frac{\partial Q(x,y)
     }{\partial x}
     -\frac{\partial P(x,y)}{\partial y}}\right]\,dS
     =\oint\limits_C [P(x,y)\,dx+Q(x,y)\,dy]$,
\hfil}\par\vspace{0.5cm}
\hbox{$
     \displaystyle{\int\limits_S}(u\Delta v-v\Delta u)\,dS
     = \displaystyle{\oint\limits_C}\left(u\displaystyle
     {\frac{\partial v}{\partial n}}
     -v\displaystyle{\frac{\partial u}{\partial n}}\right)\,ds$ ---
двумерная вторая формула Грина.\hfil}\par\vspace{0.5cm}
В последнем соотношении используются следующие определения:
$$
     \Delta v\equiv\frac{\partial^2 v}{\partial x^2}
        +\frac{\partial^2 v}{\partial y^2}\,,\qquad\qquad
     \frac{\partial v}{\partial n}\,dS\equiv
        \frac{\partial v}{\partial x}\,dy
        -\frac{\partial v}{\partial y}\,dx\,.$$
$$\hbox{$
     \displaystyle{\int\limits_V}\rv A(\rv r)\,dV=
     \displaystyle{\int\!\!\int\!\!\int\limits_{\hspace{-0.35cm}V}}
     [A_x(x,y,z)\vec\imath+A_y(x,y,z)\vec\jmath+A_z(x,y,z)\vec k]
     \,dx\,dy\,dz\,,$\hfil}$$\vspace*{0.2cm}$$
\begin{array}{ll}
     \displaystyle{\int\limits_V}\div\rv A(\rv r)\,dV
     =\displaystyle{\int\limits_S}\rv A(\rv r)\,d\rv S&
     \hbox{ --- теорема о дивергенции,}\\[0.4cm]
     \displaystyle{\int\limits_V}\rot\rv A(\rv r)\,dV
     =\displaystyle{\int\limits_S}[\,d\rv S\rv A(\rv r)]&
     \hbox{ --- теорема о роторе,}\\[0.4cm]
     \displaystyle{\int\limits_V}\grad\Phi(\rv r)\,dV
     =\displaystyle{\int\limits_S}\Phi(\rv r)\,d\rv S&
\hbox{ --- теорема о градиенте.}\\[0.4cm]\end{array}$$\par
Первая и вторая формулы Грина:
%\vspace*{.5cm}
$$\begin{array}{ll}
     \displaystyle{\int\limits_V}\grad\Phi\grad\Psi\,dV
        +\displaystyle{\int\limits_V}\Psi\Delta\Phi\,dV
        =\displaystyle{\int\limits_S}(\Psi\grad\Phi)d\rv S
        =\displaystyle{\int\limits_S}
        \frac{\partial\Phi}{\partial n}\Psi\,dS\,,&{}\\[0.4cm]
\displaystyle{\int\limits_V}(\Psi\Delta\Phi-\Phi\Delta\Psi)\,dV
=\displaystyle{\int\limits_S}(\Psi\grad\Phi-\Phi\grad\Psi)\,d\rv S
        =\displaystyle{\int\limits_S}
        \Bigl(\Psi\frac{\partial\Phi}{\partial n}
        -\Phi\frac{\partial\Psi}{\partial n}\Bigr)\,dS\,,&{}\\[0.4cm]
\end{array}$$$$\begin{array}{ll}
     \displaystyle{\int\limits_V}\Delta\Phi\,dV
        =\displaystyle{\int\limits_S}\nabla\Phi\,d\rv S
        =\displaystyle{\int\limits_S}
        \frac{\partial\Phi}{\partial n}\,dS\quad\hbox{ --- теорема Гаусса\,,}
&\\[0.4cm]
     \displaystyle{\int\limits_V}|\grad\Phi|^2\,dV
        +\displaystyle{\int\limits_V}\Phi\Delta\Phi\,dV
        =\displaystyle{\int\limits_S}(\Phi\grad\Phi)\,d\rv S
        =\displaystyle{\int\limits_S}
        \Phi\frac{\partial\Phi}{\partial n}\,dS,\quad
\displaystyle{\frac{\partial\Phi}{\partial n}}\,dS\equiv
        (d\rv S\nabla)\Phi\,.&{}\\[0.4cm]
\end{array}$$\par
       Некоторые интегральные векторные формулы:
%\vspace*{0.5cm}
$$
\displaystyle{\int\limits_S}\rot\rv A(\rv r)\,d\rv S
=\displaystyle{\oint\limits_C}\rv A(\rv r)\,d\rv r\quad
\mbox{ --- теорема Стокса}\,,$$
%\vspace*{0.5cm}
    $$ \displaystyle{\int\limits_S}\left[[d\rv S\nabla]\rv  A(\rv r)\right]
        =\displaystyle{\oint\limits_C}  (d\rv r\times\rv A(\rv r))\,,$$
%\vspace*{0.5cm}
   $$  \displaystyle{\int\limits_S}[\grad\Phi\,d\rv S]
        =-\displaystyle{\int\limits_C}\Phi(\rv r)\,d\rv r\,,$$
%\vspace*{0.5cm}
    $$    \displaystyle{\int\limits_V}\grad\Phi\rot\rv A\,dV
        =\displaystyle{\oint\limits_S}[\rv A\grad\Phi]\,d\rv S
        =\displaystyle{\oint\limits_S}\Phi\rot\rv A\,d\rv S\,,$$
%\vspace*{0.5cm}
     $$   \displaystyle{\int\limits_V}(\rv A\rot\rot\rv B
        -\rv B\rot\rot\rv A)\,dV
        =\displaystyle{\oint\limits_S}\left([\rv B\rot\rv A]
        -[\rv A\rot\rv B]\right)\,d\rv S\,. $$
\newpage

\begin{center}{\bf П-6. Функции Бесселя}\end{center}
\medskip \par
Решениями   уравнения   $$z^2\frac{d^2Z}{dz^2}+z\frac
{dZ}{dz}+(z^2-\nu^2)Z=0$$  являются  функции  Бесселя   первого   рода
$J_{\pm\nu}(z)$,    второго    рода   $Y_\nu(z)$   и   третьего   рода
$H^{(1)}_\nu(z)$,~$H^{(2)}_\nu(z)$.  Через   $Z_\nu(z)$   обозначается
любая из этих функций.\par
     Функции $J_\nu(z)$ и~$J_{-\nu}(z)$ линейно независимы, кроме того
случая,  когда  $\nu$  --- целое.  Функция Вебера $Y_\nu(z)$ и функция
$J_\nu(z)$ линейно независимы;  ниже будет использоваться  принятое  в
физике обозначение функции $Y_\nu(z)$ через $N_\nu(z)$, где $N_\nu(z)$
--- функция Неймана.  Для всех значений $\nu$ функции Ганкеля
$H^{(1)}_\nu(z)$ и~$H^{(1)}_{-\nu}(z)$    линейно  независимы.
\vspace{0.2cm}
$$N_\nu(z)=\displaystyle{\frac{J_\nu(z)\cos\nu\pi-J_{-\nu}(z)}{\sin
\nu\pi  }};\qquad\vcenter{\vbox{\hbox{если $\nu$ --- целое или нуль,  то
правая    часть    этого}\hbox{соотношения    заменяется    предельным
значением.}}}$$

 $$H_\nu^{(1)}(z)=J_\nu(z)+iN_\nu(z)\,,\qquad   H_\nu^
{(2)}(z)=J_\nu(z)-iN_\nu(z)\,,$$\smallskip
$$J_{-n}(z)=(-1)^nJ_n(z)\,,\qquad    N_{-n}(z)=(-1)^nN_n(z)\,,$$\smallskip
$$H_{-\nu}^{(1)}(z)=e^{i\nu\pi}H_\nu^{(1)}(z)\,,\qquad
H_{-\nu}^{(2)}(z)=e^{-i\nu\pi}H_\nu^{(2)}(z)\,.$$\par\medskip
     При фиксированных значениях $\nu$ и малых значениях аргумента
\vspace{0.2cm}
$$\hspace{0.3cm}\begin{array}{ll}
   J_\nu(z)\approx\displaystyle{\frac{1}{\Gamma(\nu+1)}}\cdot\left(
\displaystyle{\frac{z}{2}}\right)^\nu\,,\qquad\nu\ne-1,-2,-3,\ldots;
&{}\\[0.9cm]
     N_0(z)\sim-iH^{(1)}_0(z)\sim iH^{(2)}_0(z)\approx-\displaystyle
{\frac{2}{\pi}}\ln\displaystyle{\frac{2}{\gamma z}};
\qquad\gamma\approx1,7811
\hbox{ --- постоянная Эйлера}\,,&{}\\[0.9cm]
     N_\nu(z)\sim-iH^{(1)}_\nu(z)\sim iH^{(2)}_\nu(z)\approx-
\displaystyle{\frac{\Gamma(\nu)}{\pi}}\left(\displaystyle{\frac2z}\right)
^{-\nu},\qquad\re \nu>0;&{}\\[0.2cm]
\end{array}$$
$$\lim\limits_{\alpha z\to0}\alpha^2H_0^{(1)}(\alpha z)=0;\quad
\lim\limits_{\alpha z\to0}\alpha H_0^{(1)}(\alpha z)=-\frac{2i}{\pi z}\,,
\quad H_0^{(1)}(\alpha z)\approx\frac{2}{\pi i}\ln{\frac{2}{\gamma z}}
\hbox{ при $\alpha z\ll1$\,.}$$
\vspace{0.2cm}

     Разложения в ряд:
$$\hspace{0.3cm}\begin{array}{ll}
J_\nu(z)=\left(\displaystyle{\frac{z}{2}}\right)^\nu\displaystyle{\sum
\limits_{k=o}^\infty}
\displaystyle{\frac{(-z^2/4)^k}{k!\Gamma(\nu+k+1)}};&{}\\[.5cm]
J_0(z)=1-\displaystyle{\frac{z^2}{4}}+\displaystyle
{\frac{(z^2/4)^2}{(2!)^2}}-
\displaystyle{\frac{(z^2/4)^3}{(3!)^2}}+\cdots\,;&{}\\[.5cm]
N_0(z)=-\displaystyle{\frac{2}{\pi}}\ln\displaystyle{\frac
{2}{\gamma z}}\cdot
J_0(z)+\displaystyle{\frac{2}{\pi}}\left\{\displaystyle
{\frac{z^2}{4}}-\Bigl(1+
\displaystyle{\frac12}\Bigr)\cdot\displaystyle{\frac{(z^2/4)^2}
{(2!)^2}}+
\cdots\,\,\right\}.&{}\\[1cm]\end{array}$$
\par  Вронскиан:
$$J_{\nu+1}(z)N_\nu(z)-J_\nu(z)N_{\nu+1}(z)=\frac{2}{\pi z}.$$
\par Интегральное представление:
$$J_n(z)=\frac{1}{\pi}\cdot\int\limits_0^\pi\cos(z\sin\theta-n\theta
)\,d\theta=\frac{i^{-n}}{\pi}\cdot\int\limits_0^\pi e^{iz\cos\theta}
\cos n\theta\,d\theta.$$
\par Рекуррентные соотношения:
$$\hspace{0.3cm}\begin{array}{ll}
Z_{\nu-1}(z)+Z_{\nu+1}(z)=\displaystyle{\frac{2\nu}{z}}
\cdot Z_\nu(z);&{}\\[1cm]
Z_{\nu-1}(z)-Z_{\nu+1}(z)=2Z'_\nu(z);&{}\\[.51cm]
J_0'(z)=-J_1(z),\qquad N'_0(z)=-N_1(z);&{}\\[.51cm]
z^{\nu+1}Z_{\nu+1}(z)=\displaystyle{\int} z^{\nu+1}Z_{\nu}(z)\,dz.
&{}\\[1cm]\end{array}$$
\par Аналитическое продолжение:
$$J_\nu(ze^{m\nu\pi i})=e^{m\nu\pi i}J_\nu(z),\quad
H^{(1)}_\nu(ze^{\pi i})=-e^{-\nu\pi i}H^{(2)}_\nu(z),\quad
H^{(2)}_\nu(ze^{-\pi i})=-e^{\nu\pi i}H^{(z)}_\nu(z).$$
\par При фиксированных значениях $\nu$  и  при
$|z|\to\infty$ имеет место следующая асимптотика:
$$\hspace{0.3cm}\begin{array}{ll}
J_\nu(z)\approx\sqrt{\displaystyle{\frac{2}{\pi z}}}\cdot\cos(z-\nu\pi/2-
\pi/4);\qquad
N_\nu(z)\approx\sqrt{\displaystyle{\frac{2}{\pi z}}}\cdot
\sin(z-\nu\pi/2-\pi/4);&{}\\[.51cm]
H^{(1)}_\nu(z)\approx\sqrt{\displaystyle{\frac{2}{\pi z}}}\cdot e^{i(z-
\nu\pi/2-\pi/4)};\quad\quad\quad\hspace{0.5cm}
H^{(2)}_\nu(z)\approx\sqrt{\displaystyle{\frac{2}{\pi z}}}\cdot e^
{-i(z-\nu\pi/2-\pi/4)}.&{}\\[1cm]\end{array}$$\vspace{-1.0cm}
\par Решениями   уравнения   $$z^2\frac{d^2Z}{dz^2}+z\frac{dZ}
{dz}-(z^2+\nu^2)Z=0$$ являются   модифицированные    функции    Бесселя
$I_{\pm\nu}(z)$ и~$K_\nu(z)$.  Функции $I_\nu(z)$ и~$I_{-\nu}$ линейно
независимы, если  $\nu$   не   является   целым   числом;   $I_\nu(z)$
и~$K_\nu(z)$, где   $K_\nu(z)$   ---   функция   Макдональда,  линейно
независимы для всех значений $\nu$.
$$K_\nu(z)=\displaystyle{\frac{\pi}{2}}\displaystyle{\frac
{I_{-\nu}(z)-I_\nu(z)}{\sin\nu\pi}};\qquad
\vcenter{\vbox{\hbox{если $\nu$ --- целое или
нуль, то правая часть
этого}\hbox{соотношения заменяется предельным значением.}}}$$
$$I_\nu^{(1)}(z)=i^{-\nu}J_\nu(iz),\quad K_\nu(z)=\frac{\pi}{2}i^{\nu+1}H_
{\nu}^{(1)}(iz);\qquad I_{-n}(z)=I_n(z),\quad K_{-n}(z)=K_n(z).$$
С помощью этих соотношений может  быть  получено  большинство  свойств
модифицированных функций    Бесселя    непосредственно    из   свойств
обыкновенных функций Бесселя, например, $$I'_0(z)=I_1(z),\qquad\qquad
 K'_0(z)=-K_1(z)$$.\par
     При малых значениях аргумента справедливы соотношения:
$$\begin{array}{ll}
I_\nu(z)\approx\displaystyle{\frac{1}{\Gamma(\nu+1)}}
\cdot\left(\displaystyle{\frac{z}{2}}\right)^\nu,\qquad\nu\neq -1,-2,-3,
\ldots\,;&{}\\[1cm]K_0(z)\approx\ln\displaystyle{\frac{2}{\gamma
z}};\qquad\qquad K_\nu(z)\approx\displaystyle{\frac{\Gamma(\nu)}{2}}\cdot
\left(\displaystyle{\frac2z}\right)^\nu,\quad\re \nu>0.&{}\\[1cm]
\end{array}$$\par При  фиксированном  $\nu$  и больших значениях $|z|$
справедливы следующие приближённые равенства: $$I_\nu(z)\approx\frac{e^z
}{\sqrt{2\pi z}},\qquad K_\nu(z)\approx\sqrt{\frac{\pi}{2z}}\cdot
e^{-z}.$$\par
\begin{center}
{\it Некоторые полезные
соотношения}\end{center}
$$
\displaystyle{\int\limits_0^z}uJ_0^2(u)\,du=\displaystyle{\frac{z^2}{2}}
\left[J_0^2(z)+J_1^2(z)\right];$$$$ \displaystyle{\int\limits_0
^a}J_1^2(\alpha u)u\,du=\displaystyle{\frac{a^2}{2}}\left[J_1^2(\alpha a)-
J_0(\alpha a)J_2(\alpha a)\right];$$$$ \displaystyle{\int
\limits_0^a}\rho J_\nu(x_{\nu n'}\rho/a)J_\nu(x_{\nu n}\rho/a)\,d
\rho=\displaystyle{\frac{a^2}{2}}\Bigl[J_{\nu+1}(x_{\nu n})\Bigr]^2\cdot
\delta_{nn'}\quad\hbox{где $J_\nu(x_{\nu n})=0$,~$n=1,2,\ldots\,.$}
$$$$ \displaystyle{\int\limits_0^a}xJ_m(kx)J_m(k'x)\,dx=
\displaystyle{\frac1k}\delta(k-k');$$$$ H_0^{(1)}(kD)=
\displaystyle
{\frac{1}{\pi i}}\displaystyle{\int\limits_{-\infty}^\infty}\displaystyle
\frac{e^{ik\sqrt{D^2+z^2}}}{\sqrt{D^2+z^2}}\,dz,\qquad %D>0,$$$$
\displaystyle{\frac12}\displaystyle{\int\limits_{-\infty}^\infty}
e^{iwz}H_0^{(1)}(k\sqrt{D^2+z^2})\,dz=\displaystyle{\frac{e^{ivD}}{v}},
$$$$ \displaystyle{\frac{e^{ik\sqrt{D^2+z^2}}}{\sqrt{D^2+z^2}}=
\displaystyle{\frac{i}{2}}\int\limits_{-\infty}^\infty}H_0^{(1)}(vD)e^
{iwz}\,dw,\quad D>0,\quad v=\sqrt{k^2-w^2},\quad\im v>0.$$
$$ e^{i\rho\cos\varphi}=\sum i^ne^{-in\varphi}\cdot
J_n(\rho),\qquad e^{i\rho\sin\varphi}=\sum e^{in\psi}\cdot J_n(\rho),
\qquad\psi=\displaystyle{\frac{\pi}{2}}-\varphi;$$\vspace{0.5cm}$$
\displaystyle{\frac{\delta(\rho)}{\rho}}=\displaystyle{\frac{2}{
a^2}\sum\limits_{s=1}^\infty}\displaystyle{\frac{J_0(\mu_s\rho/a)}{
J_1^2(\mu_s)}},\quad J_0(\mu_s)=0;\qquad\displaystyle{\frac{1}{\sqrt{
\rho^2+z^2}}}=\displaystyle{\int\limits_0^\infty}e^{-k|z|}J_0(k\rho)\,
dk.\\[1cm]$$
$$\mbox{Если }x=\nu_{01}=2,4048,\mbox{   то   }J_1(x)=0,51915,
\quad\frac{x^2}{2}J_1^2(x)=0,7793.$$$$
\max J_1(z)=J_1(1,8411)=0,5819.$$
\vspace*{1cm}
     \begin{center}
\begin{tabular}{|c|c|c|c|c|}\hline\multicolumn{5}{|c|}
{Корни уравнения $J_m(x)=0$}\\\hline
$n$&$m=0$&$m=1$&$m=2$&$m=3$\\\hline
1&\phantom{0}2,405&\phantom{0}3,832&\phantom{0}5,136&\phantom{0}6,380\\
2&\phantom{0}5,520&\phantom{0}7,016&\phantom{0}8,417&\phantom{0}9,761\\
3&\phantom{0}8,654&10,173&11,620&13,015\\
4&11,792&13,324&14,372&16,224\\
\hline\end{tabular}\qquad
\begin{tabular}{|c|c|c|c|c|}\hline\multicolumn{5}{|c|}
{Корни уравнения $J'_m(x)=0$}\\\hline
$n$&$m=0$&$m=1$&$m=2$&$m=3$\\\hline
1&\phantom{0}3,832&\phantom{0}1,841&\phantom{0}3,054&\phantom{0}4,201\\
2&\phantom{0}7,016&\phantom{0}5,331&\phantom{0}6,706&\phantom{0}8,015\\
3&10,173&\phantom{0}8,536&\phantom{0}9,969&11,346\\
4&13,324&11,706&13,170&14,586\\
\hline\end{tabular}
     \end{center}
\newpage
\begin{center}{\bf П-7. Уравнение Гельмгольца (в сферических
координатах)}\end{center} \medskip \par
     Уравнение Гельмгольца
$$\frac{1}{r}\cdot\frac{\partial^2}{\partial r^2}(rF)+\frac{1}{r^2\sin
\theta}\cdot\frac{\partial}{\partial\theta}\Bigl(\sin\theta\frac{\partial
F}{\partial\theta}\Bigr)+\frac{1}{r^2\sin^2\theta}\cdot\frac{\partial^2F
}{\partial\varphi^2}+k^2F=0$$ имеет решения вида
$$\begin{array}{ll}F(\rv r)=\displaystyle{\sum\limits_{l=0}^\infty\sum
\limits_{m=-l}^l}\left[A_{lm}\cdot j_l(kr)+B_{lm}\cdot n_l(kr)\right]
\cdot Y_{lm}(\theta,\varphi),&{}\\[1cm]F(\rv r)=\displaystyle{\sum
\limits_{l=0}^\infty\sum\limits_{m=-l}^l}\left[C_{lm}\cdot h_l^{(1)}
(kr)+D_{lm}\cdot h_l^{(2)}(kr)\right]\cdot Y_{lm}(\theta,\varphi).&{}
\\[1cm]\end{array}$$ В   последних   выражениях   сферические  функции
Бесселя определяются как
$$\begin{array}{ll} j_l(x)=\displaystyle{\sqrt{\frac{\pi}{2x}}}J_{l+1/2}(x)
=(-x)^l\displaystyle{{\left(\frac1x\cdot\frac{d}{dx}\right)^l\left
(\frac{\sin x}{x}\right)}},&{}\\[1cm]
n_l(x)=\displaystyle{\sqrt{\frac{\pi}{2x}}}N_{l+1/2}(x)
=-(-x)^l\displaystyle{\left(\frac{1}{x}\cdot\frac{d}{dx}\right)^l\left
(\frac{\cos x}{x}\right)},&{}\\[1cm]
h_l^{(1,2)}(x)=\displaystyle{\sqrt{\frac{\pi}{2x}}}\left[J_{l+1/2}(x)\pm
N_{l+1/2}(x)\right].&{}\\[1cm]\end{array}$$ Сферические гармоники
$$Y_{lm}(\theta,\varphi)=\left(\frac{2l+1}{4\pi}\cdot\frac{(l-m)!}{(l+m)!}
\right)^{1/2}\cdot P_l^m(\cos\theta)e^{im\varphi}$$ выражаются   через
присоедин\"енные полиномы Лежандра ($u=\cos\theta$) $$P_l^m(u)=
(1-u^2)^{m/2}\frac{d^mP_l(u)}{dx^m},$$ где полиномы Лежандра $P_l(u)$
равны $$P_l(u)=\frac{1}{2^l\cdot l!}\cdot\frac{d^l}{du^l}(u^2-1)^l.$$
\bigskip
\newpage

\begin{center}{\bf П-8.   Вычисление   коэффициентов    $R_n(u)$,~$V_n
(\zeta_0)$, ~$V_m^n( u)$ }\end{center}
\medskip \par
    Функциями Лежандра  (или   сферическими   функциями)   называются
решения дифференциального уравнения
     $$(1-z^2)\frac{d^2w}{dz^2}-2z\frac{dw}{dz}+\bigl[\nu(\nu+1)-\frac
         {\mu^2}{1-z^2}\bigr]w=0.\eqno(\mbox{П}8.1)$$
Здесь $z$   ---   комплексная  переменная,  $-1\leqslant  u=\re
z\leqslant 1$,~$u= \cos\theta$;  индексы  $\nu$  и~$\mu$  ---
произвольные   комплексные
постоянные  и  называются соответственно степенью и порядком,  а точки
$z=\pm1$ являются обыкновенными точками ветвления. Линейно независимые
решения   дифференциального  уравнения  (\mbox{П}8.1)  обозначаются  как
$P_\nu^\mu(z)$,~$Q_\nu^\mu(z)$    и    называются    присоедин\"енными
функциями   Лежандра   соответственно   первого  и  второго  рода  (на
действительной   оси   между   точками   $-1$   и~$+1$   эти   функции
действительны).  Для  функций  Лежандра  первого  рода  $P_\nu^\mu(z)$
справедливо соотношение
     $$P_{-\nu-1}^\mu(z)=P_\nu^\mu(z)\eqno(\mbox{П}8.2)$$
и рекуррентная формула вида
     $$(\nu-\mu+1)P_{\nu+1}^\mu(z)=(2\nu+1)zP_\nu^\mu(z)-(\nu+\mu)
         P_{\nu-1}^\mu(z).\eqno(\mbox{П8}.3)$$

     В частном   случае   $\nu=n$,~$\mu=0$  из  (\mbox{П8}.1)  получается
уравнение  для  полиномов   Лежандра   первого   рода   $P_n(z)$   ---
подразумевается, что $n$ являются действительным целым неотрицательным
числом; выражения для нескольких первых полиномов $P_n(z)$ имеют вид
     $$P_0(z)=1,\quad P_1(z)=z,\quad P_2(z)=\frac12(3z^2-1),\quad$$
     \vspace*{-0.75cm}$$\eqno(\mbox{П8}.4)$$\vspace*{-0.75cm}
     $$P_3(z)=\frac12(5z^3-3z),\qquad P_4(z)=\frac18(35z^4-30z^2+3).$$
Присоедин\"енные полиномы Лежандра первого рода $P_n^m(z)$,  где $m>0$
--- целое число, могут быть найдены согласно Приложению П-7.

     Функции Лежандра $P_\nu(\cos\theta)$ при всех  значениях  индекса
$\nu$ можно определить интегралом Мелера-Дирихле:
     $$P_\nu(\cos\theta)=\frac{1}{\pi}\int\limits_{-\theta}^\theta
         \frac{e^{i(\nu+1/2)\varphi}}{\sqrt{2(\cos\varphi+\cos\theta)
         }}\,d\varphi.\eqno(\mbox{П8}.5)$$
Из разложения производящей функции полиномов Лежандра
     $$\frac{1}{\sqrt{(t-\alpha)(t-\alpha^*)}}=\sum_{n=0}^\infty P_n
         (u)t^n,\quad|t|<1,\quad\alpha=e^{i\theta},\quad u=\cos\theta,
         \eqno(\mbox{П8}.6)$$
следует, что
     $$\sqrt{(t-\alpha)(t-\alpha^*)}=\left\{\begin{array}{lr}\phantom
         {-t}\displaystyle{\sum_{n=0}^\infty}\rho_n(u)t^n,&\quad|t|<1;
         \\[0.25cm]-t\displaystyle{\sum_{n=0}^\infty}\rho_n(u)t^{-n},
         &\quad|t|>1,\end{array}\right.\eqno(\mbox{П8}.7)$$
где
     $$\rho_0(u)=1,\quad\rho_1(u)=-u,\quad\rho_n(u)=P_n(u)-2uP_{n-1}
         (u)+P_{n-2}(u),\qquad n\geqslant 2.\eqno(\mbox{П8}.8)$$
В формулах  (\mbox{П8}.6),  (\mbox{П8}.7)  $\sqrt{(t-\alpha)(t-\alpha^*)}$
представляет ту ветвь корня, которая голоморфна во всей всей плоскости
с  разрезом вдоль дуги $L_1$,  соединяющим точки $\alpha$ и~$\alpha^*$
(рис.~25.3), и принимает в точке $t=0$ значение единицы.

     Ко\-эф\-фи\-ци\-ент $R_n(u)$  оп\-ре\-де\-ля\-ет\-ся фор\-му\-лой
(25.23) и в силу (\mbox{П8}.2), (\mbox{П8}.5) легко преобразуется к виду
     $$R_n(u)=\frac12P_n(u).\eqno(\mbox{П8}.9)$$
Чтобы вычислить главное значение интеграла,  определяемого  первой  из
формул (25.23), удобно сначала показать, что $V_n(\zeta_0)= Res(0)+Res
(\infty)$,~$\zeta\in L_1$.  Вычеты подынтегральной функции в нуле и на
бесконечности удаётся найти и тогда получается следующая формула для
коэффициентов $V_n(\zeta_0)$:
     $$V_n(\zeta_0)=\left\{\begin{array}{ll}\displaystyle{\sum\limits_
         {p=0}^{n+1}}\rho_{n-p+1}(u)\zeta_0^p,&\quad n\geqslant 0;\\
         [0.4cm]1-\zeta_0^{-1},&\quad n=-1;\\[0.2cm]\displaystyle{\sum
         \limits_{p=0}^{-n-1}}\rho_{-n-1-p}(u)\zeta_0^{-p-1},&\quad n
         <-1.\end{array}\right.\eqno(\mbox{П8}.10)$$

     Выражения для коэффициентов $V_m^n(u)$  через  полиномы  Лежандра
может  быть  получено  путём  подстановки  (\mbox{П8}.10)  в (25.23) и
последующего применения соотношений (\mbox{П8}.5) и (\mbox{П8}.2):
     $$V_m^n(u)=\left\{\begin{array}{lc}\phantom{-}\displaystyle{\frac
         12\sum\limits_{p=0}^{n+1}}\rho_{n+1-p}(u)P_{p-m-1}(u),\quad&n
         \geqslant 0;\\[0.4cm]\phantom{-}\displaystyle{\frac12}[P_m(u)-P_{m+
         1}(u)],\quad&n=-1;\\[0.25cm]\displaystyle{-\frac12\sum
         \limits_{p=0}^{-n-1}}\rho_{-n-1-p}(u)P_{p+m+1}(u),\quad&n<-
         1.\end{array}\right.\eqno(\mbox{П8}.11)$$
Эти формулы (при $n\ne-1$) могут быть упрощены в результате применения
известного рекуррентного соотношения (\mbox{П8}.3) и равенства
     $$\sum\limits_{j=0}^n\rho_{n-j}(u)P_{j+\nu}(u)=\left\{\begin
         {array}{lc}\displaystyle{\frac{\nu}{n+\nu}}[P_\nu(u)P_n(u)-P_
         {\nu-1}(u)P_{n-1}(u)],\quad&n\geqslant 1;\\[0.35cm]P_\nu(u),\quad&
         n=0;\end{array}\right.\eqno(\mbox{П8}.12)$$
при $\nu=-n$,~$n\geqslant 1$  неопределённость  в правой части (\mbox{П8}.12)
может быть раскрыта по правилу Лопиталя. Используя теперь (\mbox{П8}.12)
и (\mbox{П8}.2),  можно получить из (\mbox{П8}.11) следующее выражение для
$V_m^n$:
     $$V_m^n(u)=\frac{m+1}{2(m-1)}[P_m(u)P_{n+1}(u)-P_{m+1}(u)P_n(u)],
         \qquad m\ne n.\eqno(\mbox{П8}.13)$$

\newpage

\begin{center}{\bf П-9. Некоторые математические соотношения}\end{center}
$$\delta(ax)=\frac1a\delta(x),\quad a>0;\qquad\delta[f(x)]=\frac{1}{
f'_x}\delta(x-x_0),\quad f(x_0)=0;\\[0.5cm]$$$$\frac{1}{2\pi}\sum\limits_
{n=-\infty}^\infty e^{in(\varphi-\varphi')}=\delta(\varphi-\varphi')=
\frac{1}{\pi}\sum\limits_{m=0}^\infty\frac{\cos m(\varphi-\varphi')}{1+
\delta_{m0}};\\[0.5cm]$$$$ \frac{1}{2\pi}\int\limits_{h=-\infty}^\infty e
^{ih(z-z')}\,dh=\delta(z-z')=\frac{1}{\pi}\int\limits_0^\infty\cos h(z-z')
\,dh;\\[0.5cm]$$$$\frac{1}{2\pi}\int\limits_0^{2\pi}e^{i(m-n)}\,d\alpha=
\delta_{mn};\qquad\frac{1}{2\pi}\sum\limits_{n=-\infty}^\infty e^{inx}=
\sum\limits_{n=-\infty}^\infty\delta(x-2n\pi);\\[0.5cm]$$
$$\sin z=z\cdot\prod\limits_{n=1}^\infty\Bigl[1-\Bigl(\frac{z}{n\pi}
\Bigr)^2\Bigr];\qquad \cos z=z\cdot\prod\limits_{n=0}^\infty\Bigl\{1-
\Bigl[\frac{2z}{(2n+1)\pi}\Bigr]^2\Bigr\},\qquad |z|<\infty;\\[0.5cm]$$
$$\frac{1}{x+i0}=-\pi\delta(x)+V.p.\frac1x\quad\hbox{ ---
формула Сохоцкого;}\qquad\theta(\tau)=\frac{1}{2\pi i}\int
\limits_{\alpha-i\infty}^{\alpha
     +i\infty}\frac{e^{z\tau}}{z}\,dz,\;\;\alpha>0;$$
$$\int\limits_{-\infty}^\infty(ax^2+2bx+c)e^{-px^2-qx}\,dx=
\frac{\sqrt{\pi}}{4}p^{-5/2}[4cp^2-4pqb+a(q^2+2p)]e^{q^2/4p},\quad\re
p>0;\\[0.5cm]$$
$$\int\limits_{-\infty}^\infty e^{i(\lambda x^2+\mu x)}\,dx=\sqrt{
\frac{\pi}{\lambda}}\exp\left[\frac{i(\lambda\pi-\mu^2)}{4\lambda}\right],
\quad\re\lambda>0.$$\\[0.5cm]
Если преобразование  Фурье  задано
соотношением $$F[\varphi](k)=\int\varphi(x)e^{i(k,x)}\,dx,$$        то
справедливы следующие формулы:\vspace{0.5cm}
$$F[\delta(x-x_0)]=e^{i(k,x_0)};\qquad
F[\varphi(x-x_0)]=e^{i(k,x_0)}F[\varphi];\qquad F[\varphi](k+k_0)=F[
e^{i(k_0,x)}\varphi](k);\\[0.5cm]$$$$F[e^{-a^2x^2}]=\frac{\sqrt{\pi}}{a}
\cdot e^{-k^2/4a^2};\!\qquad\! F[e^{ix^2}]=\sqrt{\pi}\cdot e^{-i(k^2-
\pi)};\!\qquad\! F[\theta]=\pi\delta(k)+i V.p.\frac1k\,.$$

%\end{document}

\newpage
\oddsidemargin=-0.4mm \evensidemargin=-0.4mm
\topmargin=-0.4mm
\headsep=7mm
\textheight=231.875mm
\textwidth=160mm
%\begin{document}
%\input{macr.tex}
\thispagestyle{empty}
%\addtocounter{page}{404}

\begin{center}
\subsubsection*{\rm \,Л\,И\,Т\,Е\,Р\,А\,Т\,У\,Р\,А}
\end{center}
\parindent=1.5cm

\parskip=0.5em
   \hspace{1ex}1. Дж.~А.~Стрэттон. {\em Теория электромагнетизма.}
     М.-Л., Гостехиздат, \par\hspace{1.5em}1948.
\parskip=0em

   \hspace{1ex}2. И.~Е.~Тамм. {\em Основы теории электричества.}
     М., Наука, 1989.

   \hspace{1ex}3. Дж.~Джексон. {\em Классическая электродинамика.}
     М., Мир, 1965.

   \hspace{1ex}4. Л.~Д.~Ландау, Е.~М.~Лифшиц. {\em Электродинамика
     сплошных сред.} М.,\par
     \hspace{1.5em}Наука, 1982.

   \hspace{1ex}5. Л.~Д.~Ландау, Е.~М.~Лифшиц. {\em Теория поля.}
     М., Наука, 1967.

   \hspace{1ex}6. Б.~З.~Каценеленбаум. {\em Высокочастотная
     электродинамика (основы ма-\par
     \hspace{1.5em}тематического аппарата).} М., Наука, 1966.

   \hspace{1ex}7. Л.~А.~Вайнштейн. {\em Электромагнитные волны.}
     М., Сов. радио, 1957.

   \hspace{1ex}8. Г.~Т.~Марков, А.~Ф.~Чаплин. {\em Возбуждение
     электромагнитных волн.}\par
     \hspace{1.5em}М.--Л., Энергия, 1967.

   \hspace{1ex}9. А.~А.~Семёнов. {\em Теория электромагнитных волн.}
     М., МГУ, 1968.

   10. Л.~Левин. {\em Современная теория волноводов.} М., ИЛ, 1954.

   11. Р.~Миттра, С.~Ли. {\em Аналитические методы теории волноводов.}
     \par\hspace {1.5em} М., Мир, 1974.

   12. Сборник задач по курсу {\em <<Электродинамика и  распространение
     радио-}\par\hspace {1.5em}{\em волн>>.} Под редакцией
     С.~И.~Баскакова. М., Высшая школа, 1981.

   13. Л.~А.~Вайнштейн, В.~А.~Солнцев. {\em Лекции по
     сверхвысокочастотной\par
     \hspace{1.5em} электронике.} М., Сов. радио, 1973.

   14. Э.~Л.~Бурштейн, Г.~В.~Воскресенский.  {\em Линейные ускорители
     электронов\par
     \hspace{1.5em}с интенсивными пучками.} М., Атомиздат, 1970.

   15. Б.~М.~Болотовский, Г.~В.~Воскресенский.  {\em Излучение
     заряженных частиц\par
     \hspace{1.5em}в периодических структурах.} УФН, т.94,
     в.3, 1968, с.377--416. {\em Дифракци-\par
     \hspace{1.5em}онное излучение.} УФН, т.88, в.2,
     1966, с.209--251.

   16. Дж.~Л.~Альтман.  {\em Устройства сверхвысоких частот.} М., Мир,
     1968.

%\end{document}

\end{document}